\documentclass[reqno]{article}

\usepackage{amsthm}
\usepackage{amsmath}
\usepackage{amssymb}
\usepackage{graphicx}
\usepackage{color}
\usepackage{bbm}
\usepackage[top=2.5cm, bottom=2.5cm, left=2.5cm, right=2.5cm]{geometry}
\usepackage{faktor}
\usepackage{nicefrac}
\usepackage[nottoc]{tocbibind}
\usepackage{slashed}
\usepackage{authblk}
\usepackage[normalem]{ulem}
\usepackage{bm}
\usepackage{comment}
\usepackage{todonotes}
\usepackage{ulem}
\usepackage{marvosym}
\usepackage{multicol}
\usepackage{enumitem}
\usepackage{wrapfig}

\usepackage{authblk} 
\usepackage[hidelinks]{hyperref}

\usepackage[numbers,square]{natbib}
\usepackage[scr=euler]{mathalfa}

\def\mathclap#1{\text{\hbox to 0pt{\hss$\mathsurround=0pt#1$\hss}}}

\newcommand{\loc}{\mathrm{loc}}
\newcommand{\Z}{\mathbb{Z}}
\newcommand{\N}{\mathbb{N}}
\newcommand{\R}{\mathbb{R}}
\newcommand{\id}{\mathrm{id}}
\newcommand{\C}{\mathbb{C}}
\newcommand{\vol}{\mathrm{vol}}
\newcommand{\Rea}{\mathfrak{Re}}
\newcommand{\Ima}{\mathrm{Im}}
\newcommand{\Sp}{\mathbb{S}}

\newcommand{\vols}{\mathrm{vol}_{\mathbb{S}^2}}
\newcommand{\uu}{\underline{u}}

\newcommand{\chib}{\underline{\chi}}
\newcommand{\etab}{\underline{\eta}}
\newcommand{\omegab}{\underline{\omega}}

\newcommand{\alphab}{\underline{\alpha}}
\newcommand{\betab}{\underline{\beta}}
\newcommand{\be}{\pmb{e}}
\newcommand{\Co}{\mathscr{C}}
\newcommand{\Si}{\mathscr{S}}
\newcommand{\xib}{\underline{\xi}}
\def\ov {\overline}
\newcommand{\mb}{\overline{m}}
\def\doublestroke#1{\pdfliteral{1 Tr .3 w}#1\pdfliteral{0 Tr 0 w}}
\newcommand{\dD}{\doublestroke{D}}
\newcommand{\dDel}{\doublestroke{\Delta}}
\newcommand{\ddel}{\doublestroke{\delta}}
\newcommand{\rred}{r_\mathrm{red}}
\newcommand{\eai}{\underset{a.i.}{=}}
\newcommand{\swl}{\mathring{\slashed{\Delta}}_{[s]}}
\newcommand{\swlt}{\mathring{\slashed{\Delta}}_{[2]}}
\newcommand{\CH}{\mathcal{CH}}
\newcommand{\Hp}{\mathcal{H}^+}
\newcommand{\MK}{\mathcal{M}_{\mathrm{Kerr}}}
\newcommand{\phip}{{\varphi_+}}
\newcommand{\ee}{\hbox{\scriptsize \Ecommerce}}
\newcommand{\volg}{\vol_{\gamma}}
\newcommand{\psil}{\psi_0^{\mathrm{lin}}}

\newcommand{\pfstep}[1]{\smallskip \noindent {\it #1.}}

\newcommand{\f}{\frac}
\def\th {\theta}
\newcommand{\rd}{\partial}
\newcommand{\wtZ}{\widetilde{Z}}
\def\chib {\underline{\chi}}
\def\chih {\hat{\chi}}
\def\chibh {\hat{\underline{\chi}}}
\def\omegab {\underline{\omega}}
\def\etab {\underline{\eta}}
\def\xib {\underline{\xi}}
\def\betab {\underline{\beta}}
\def\alphab {\underline{\alpha}}

\def\hot{\widehat{\otimes}}

\def\alp {\alpha}

\def\bt {\beta}

\def\nab {\slashed{\nabla}}

\def\ep {\epsilon}
\def\om {\omega}
\def\omb {\underline{\omega}}

\def\f {\frac}

\def\rd {\partial}
\def\ls {\lesssim}
\def\de {\delta}
\def\i {\infty}
\def\Om {\Omega}
\def\Omg {\Omega}

\newcommand{\ud}{\mathrm{d}}
\def\ub {\underline{u}}
\def\trchb {\slashed{\mbox{tr}}\chib}
\def\trch{\slashed{\mbox{tr}}\chi}

\def\tpH {\widetilde{\psi}_H}
\def\tp {\widetilde{\psi}}
\def\tpHb {\widetilde{\psi}_{\Hb}}
\def\nab {\slashed{\nabla}}
\def\ep {\epsilon}
\def\Om {\Omega}

\def\tK {\widetilde{K}}

\def\tg {\widetilde{g}}
\def\tb {\widetilde{b}}
\def\Ke {\mathcal K}
\def\Hb {\underline{H}}

\def\pH {\psi_H}
\def\pHb {\psi_{\Hb}}

\DeclareFontFamily{U}{mathx}{}
\DeclareFontShape{U}{mathx}{m}{n}{<-> mathx10}{}
\DeclareSymbolFont{mathx}{U}{mathx}{m}{n}
\DeclareMathAccent{\widecheck}{0}{mathx}{"71}

\newcommand{\calA}{\mathcal A}
\newcommand{\calB}{\mathcal B}
\newcommand{\calC}{\mathcal C}
\newcommand{\calD}{\mathcal D}
\newcommand{\calE}{\mathcal E}

\newcommand{\calH}{\mathcal H}

\newcommand{\calM}{\mathcal M}
\newcommand{\calN}{\mathcal N}
\newcommand{\calO}{\mathcal O}

\newcommand{\calR}{\mathcal R}
\newcommand{\calS}{\mathcal S}
\newcommand{\calT}{\mathcal T}
\newcommand{\calU}{\mathcal U}
\newcommand{\calV}{\mathcal V}
\newcommand{\calW}{\mathcal W}

\newcommand{\bbR}{\mathbb R}
\newcommand{\bbS}{\mathbb S}

\newcommand{\eqs}{=_{\scalebox{.6}{\mbox{S}}}}
\newcommand{\eqrs}{=_{\scalebox{.6}{\mbox{RS}}}}

\newcommand{\stout}[1]
{\ifmmode\text{\sout{\ensuremath{#1}}}\else\sout{#1}\fi}

\makeatletter
\newcommand{\nrm}{\@ifstar{\nrmb}{\nrmi}}
\newcommand{\nrmi}[1]{\Vert{#1}\Vert}
\newcommand{\nrmb}[1]{\left\Vert{#1}\right\Vert}
\newcommand{\abs}{\@ifstar{\absb}{\absi}}
\newcommand{\absi}[1]{\vert{#1}\vert}
\newcommand{\absb}[1]{\left\vert{#1}\right\vert}
\newcommand{\brk}{\@ifstar{\brkb}{\brki}}
\newcommand{\brki}[1]{\langle{#1}\rangle}
\newcommand{\brkb}[1]{\left\langle{#1}\right\rangle}
\newcommand{\set}{\@ifstar{\setb}{\seti}}
\newcommand{\seti}[1]{\{#1\}}
\newcommand{\setb}[1]{\left\{ #1\right\}}
\makeatother

\renewcommand{\div}{\slashed{\mbox{div}}}
\newcommand{\curl}{\slashed{\mbox{curl}}}

\newcommand{\os}[2]{\overset{\scriptscriptstyle{\text{[{#2}]}}}{#1}}

\newcommand{\tr}{\slashed{\hbox{tr}}}

\NewDocumentCommand{\opnorm}{sO{}m}{%
  \IfBooleanTF{#1}{
    \left|\opnormkern\left|\opnormkern\left|
    #3
    \right|\opnormkern\right|\opnormkern\right|
  }{
    \mathopen{#2|\opnormkern #2|\opnormkern #2|}
    #3
    \mathclose{#2|\opnormkern #2|\opnormkern #2|}
  }%
}
\newcommand{\opnormkern}{\mkern-1.5mu\relax}

\begin{document}

\numberwithin{equation}{section}
\newtheorem{theorem}[equation]{Theorem}
\newtheorem{remark}[equation]{Remark}
\newtheorem{assumption}[equation]{Assumption}
\newtheorem{claim}[equation]{Claim}
\newtheorem{lemma}[equation]{Lemma}
\newtheorem{definition}[equation]{Definition}
\newtheorem{corollary}[equation]{Corollary}
\newtheorem{proposition}[equation]{Proposition}
\newtheorem{conjecture}[equation]{Conjecture}
\newtheorem{example}[equation]{Example}

\setcounter{tocdepth}{3}

\title{The formation of a weak null singularity \\in the interior of generic rotating black holes}
\author[1]{Jonathan Luk}
\author[2]{Jan Sbierski}

\affil[1]{Department of Mathematics, Stanford University, 450 Jane Stanford Way Building 380, Stanford CA 94305-2125, United States of America; email: \href{mailto:jluk@stanford.edu}{jluk@stanford.edu}}
\affil[2]{School of Mathematics and Maxwell Institute for Mathematical Sciences, University of Edinburgh, James Clerk Maxwell Building, Peter Guthrie Tait Road, Edinburgh, EH9 3FD, United Kingdom; email: \href{mailto:Jan.Sbierski@ed.ac.uk}{Jan.Sbierski@ed.ac.uk}}

\date{\today}

\maketitle

\begin{abstract}
    Given a characteristic initial value problem with smooth data representing a dynamical event horizon settling down to that of Kerr in the subextremal, strictly rotating range with suitable upper and lower bounds, we prove that a weak null singularity forms, across which the spacetime metric is continuously extendible but not Lipschitz extendible. The bulk of the proof is a stability argument showing that a dynamical Teukolsky field can be approximated by a linear Teukolsky field, whose linear instability was proved in previous works.
\end{abstract}

\tableofcontents

\section{Introduction}

In this paper, we prove the formation of a weak null singularity in the interior of dynamical black hole solutions to the vacuum Einstein equations
\begin{equation}\label{eq:EVE}
    \mathrm{Ric}(g) =0.
\end{equation}
Recall that \eqref{eq:EVE} admits an explicit two-parameter family of Kerr solutions $(\calM_{\mathrm{Kerr}}, \pmb{g})$, depending on parameters $(M,a)$ (see Section~\ref{SecKerrBackground}). In the subextremal, strictly rotating case $0< |a| <M$, these spacetimes represent rotating black hole solutions arising from complete asymptotically flat data, whose maximal Cauchy development contains a black hole region. The exterior regions of these solutions are stable \cite{Dafermos.ICM, Dafermos.extremal, DHRT, GKS, KS.Kerr, dwS2023} --- at least when $|a|\ll M$ but are conjectured to hold generally in the range $|a| < M$ --- and they are moreover expected to be relevant in describing the endstates of possibly large-data astrophysical systems \cite{Penroseunsolv}. However, inside these black holes, unlike their non-rotating counterpart the Schwarzschild solution ($a=0$, $M>0$), there is a smooth Cauchy horizon across which the metric can be extended smoothly --- but non-uniquely! --- as a solution to \eqref{eq:EVE}. 

In part to deal with the issues of determinism connected with these non-unique extensions, the strong cosmic censorship conjecture of Penrose asserts that such smooth extendibility is non-generic. In particular, under this conjecture, the smooth Kerr Cauchy horizon should be unstable under small perturbations. However, the strongest form of this conjecture, which posits that a crushing spacelike $C^0$-inextendible singularity like that of Schwarzschild would arise under small perturbations, turns out to be false and was disproved by Dafermos--Luk \cite{DafLuk17}. Instead, they proved that near timelike infinity, the Cauchy horizon must be stable in a weak sense, and the metric would at least be continuously extendible beyond the Cauchy horizon. Nevertheless, in their proof they need to contend with various potentially singular geometric quantities, which already strongly suggests that the Cauchy horizon may in fact be singular in some sense. In this paper, we rigorously establish that for a suitably defined generic subclass of initial data, the Cauchy horizon is a weak null singularity of \cite{LukWeakNull} and \cite{Sbie24}. In particular, the spacetime metric is not Lipschitz extendible beyond the Cauchy horizon.

More precisely, we consider a characteristic initial value problem with characteristic initial data representing a dynamical event horizon settling down to that of a subextremal rotating Kerr event horizon with suitable upper and lower bounds. The upper bound assumptions are similar to those imposed by Dafermos--Luk so that by their results the black hole interior terminates with a Cauchy horizon. The new lower bound assumptions guarantee that a weak null singularity indeed forms. The upper and lower bound assumptions that we impose allow for the behavior that are expected from \emph{generic} perturbations of Kerr Cauchy data, either from the class of initial data leading to a conformally regular null infinity, or from the Gajic--Kehrberger class of initial data (see Remark~\ref{rmk:are.the.rates.reasonable}).

Our results can be viewed as a generalization of results in the linear or spherically symmetric settings, where a weak null singularity forms under suitable upper and lower bound assumptions along the event horizon. We refer the reader to Section~\ref{sec:related.works} for a brief discussion of such results. In fact our result relies on existing linear instability results, and show that the nonlinear contribution --- given what is already established in \cite{DafLuk17} --- is \emph{perturbative}. The strategy is an adaptation of that in a simpler setting in \cite{LOSR}: 
\begin{enumerate}
    \item Obtain a nonlinear stability result in a weak topology (in this case done in \cite{DafLuk17}).
    \item Analyze the linear behavior to derive linear instability (in this case done in \cite{Sbie23}).
    \item Prove that the stability estimates are strong enough to show that the difference between linear and nonlinear solutions has a faster decay (which is carried out in this paper).
\end{enumerate}

In our case we use the linear result in \cite{Sbie23}, but it is also possible to formulate a result using \cite{Gurriaran.linear}. It is important to note that the linear result is used as a black box, and in particular the method for deriving the linear instability, e.g., whether it is robust or not, does not play a role here. We also observe, importantly, that the nonlinear stability result in Step~1 is completely independent of the more precise information about the linear field derived in the linear instability part in Step~2.


We state a version of the theorem where the assumptions are given informally. We refer the reader to Section~\ref{SecPreciseAssump} for the precise assumptions, which will be stated after we introduce the necessary background regarding Kerr geometry, and the NP and CK formalisms.

\begin{theorem}\label{thm:main}
Fix $M,a$ such that $0 < |a| < M$. 
Consider the characteristic initial value problem for the vacuum Einstein equations with smooth\footnote{Smoothness is assumed for convenience, but only finite regularity of order $I_0$ compatible with the precise assumptions in Section \ref{SecPreciseAssump} is needed.} initial data on two transversely intersecting null hypersurface $\calH^+ \equiv [1,\infty)\times \mathbb S^2$ and $\Hb_1 = [0,s_*)\times \mathbb S^2$. Assume that the characteristic initial data satisfy the following assumptions:
\begin{enumerate}[label=(\roman*)]
    \item (Upper bounds on $\calH^+$) The geometry on $\calH^+$  approaches with a suitable rate that of the event horizon of the Kerr spacetime with parameters $(M,a)$ and  is $\epsilon$-close to it with $I_0$ derivatives.  \label{item:main.thm.i}
    \item (Upper bounds on $\Hb_1$) The geometry on $\Hb_1$ is $\epsilon$-close with $I_0$ derivatives to that on a corresponding null hypersurface in the Kerr spacetime with parameters $(M,a)$. \label{item:main.thm.ii}
    \item (Assumptions on $\psi_0$ on $\calH^+$)\label{item:main.thm.iii}
    \begin{enumerate}
        \item (Lower bound on $\psi_0$) A weighted $L^2$-averaged lower bound holds for the $l =2$ angular modes of the dynamical $s=+2$ Teukolsky field $\psi_0$.
        \item (Auxiliary upper bounds on $\psi_0$) The $l >2$ angular modes of $\psi_0$ and the derivative of $\psi_0$ with respect to the background Killing vector field associated to time translation obey upper bounds with slightly improved decay rates.
    \end{enumerate}
\end{enumerate}

Then for $I_0$ big enough and $\epsilon>0$ small enough the following holds:
\begin{enumerate}
    \item (Existence) \label{item:main.thm.1} A solution $(\calM',g)$ to the vacuum Einstein equations with the following properties exists:
    \begin{enumerate}
        \item \label{item:main.thm.1.a} $(\calM',g)$ is the maximal globally hyperbolic future development of the data restricted to $\calH^+ \cup \widecheck{\Hb}_1$, where $\widecheck{\Hb}_1 =  \{ (s, \th_*,\varphi_*) \in \Hb_1: s< \widecheck{s}(\th_*,\varphi_*)\}$ (for a function  $\widecheck{s}:\mathbb S^2 \to (0,s_*)$ to be specified in Section~\ref{sec:proof.of.main.theorem.1.2}).
        \item \label{item:main.thm.1.b} In a suitable coordinate system $(u',\ub,\th_*,\varphi_*)$ on $\calM'$, the metric is $C^0$-close to the Kerr metric in Pretorius--Israel double null coordinates\footnote{See \cite[Section A.4]{DafLuk17}, \cite{fPwI1998}.} with parameters $(M,a)$ and converges to it in $C^0$ as $\min\{\ub,-u'\} \to \infty$.
        \item \label{item:main.thm.1.c} The null hypersurface $\widecheck{H}$ transversal to $\Hb_1$ emanating from $\widecheck{S} = \{ (s, \th_*,\varphi_*) \in \Hb_1: s= \widecheck{s}(\th_*,\varphi_*)\}$ is a future null boundary of $\calM'$ to which $g$ extends smoothly. Moreover, every future-directed future-inextendible timelike curve $\gamma:(a,b) \to \calM' \cup \widecheck{H}$ that does not intersect $\widecheck{H}$ satisfies $\lim_{\sigma\to b^-} \ub(\gamma(\sigma)) = \infty$.
    \end{enumerate}
    
    \item (Continuous extendibility of the metric) \label{item:main.thm.2} Define $\ub_{\CH^+}: [1, \infty) \to (-\infty,0)$ by
    \begin{equation}\label{eq:ubCH.thm}
        \ub_{\CH^+}(\ub) = - \f{r_-^2+a^2}{r_+-r_-} e^{-\f{r_+-r_-}{r_-^2+a^2} \ub} ,
    \end{equation}
    where $r_+ >r_-$ are the roots of $\Delta = r^2 -2Mr + a^2$. 
    
    Then, for appropriately defined coordinate functions $u$, $\th^1_{(i),\CH^+}$ and $\th^2_{(i),\CH^+}$ (for $i=1,2$, see Sections~\ref{sec:M2}, \ref{sec:CH.coord}), we can attach the boundary $\CH^+ :=\{\ub_{\CH^+} = 0\}$ to $\calM'$ so that the metric extends continuously to $\overline{\calM'} = \calM' \cup \CH^+$ in the $(u,\ub_{\CH^+}, \th^1_{(i),\CH^+}, \th^2_{(i),\CH^+})$ coordinate systems. Moreover, in a neighborhood of $\CH^+$, $\widecheck{H}$ corresponds to $\{ u = u_f\}$ for some $u_f \leq -1$. 
    \item (Lipschitz inextendibility of the metric) \label{item:main.thm.3} There is no $C^{0,1}_{\mathrm{loc}}$ extension $\iota:\calM' \to \widehat{\calM'}$ with the property that there is an affinely parameterized, future-directed and future-inextendible timelike geodesic $\tau: (-t_0,0)  \to \calM'$ (for some $t_0>0$) with the properties that 
    \begin{itemize}
        \item $\lim_{\sigma \to 0^-}  \ub_{\CH^+}(\tau(\sigma)) = 0$, for the function $\ub_{\CH^+}$ in \eqref{eq:ubCH.thm},
        \item $\lim_{\sigma \to 0^-} u(\tau(\sigma)) < u_f$ for the coordinate function $u$ as in part 2, and 
        \item $\lim_{\sigma \to 0^-} \iota \circ \tau(\sigma) \in \widehat{\calM'}$ exists.
    \end{itemize}
\end{enumerate}
\end{theorem}

By \cite{DafLuk17} and \cite{DafLuk26}, the assumptions \ref{item:main.thm.i} and \ref{item:main.thm.ii} of Theorem~\ref{thm:main} imply\footnote{The work \cite{DafLuk26} treats a small neighborhood of the event horizon, and the work \cite{DafLuk17} treats the remaining region, including, importantly, the region near the Cauchy horizon. We note that the region near the event horizon is sufficiently stable and can also be treated as part of the \emph{exterior} region of the black hole; see for instance \cite{pHaV2018, KS.Kerr}. See also the related works \cite{xtCsK2024, pH2024} on the construction of event horizon in a gauge that is not event horizon-normalized.} the conclusions \ref{item:main.thm.1} and \ref{item:main.thm.2}. Theorem~\ref{thm:main} 
\begin{wrapfigure}{r}{0.45\textwidth}
 \centering
  \def\svgwidth{5.5cm}
\begingroup%
  \makeatletter%
  \providecommand\color[2][]{%
    \errmessage{(Inkscape) Color is used for the text in Inkscape, but the package 'color.sty' is not loaded}%
    \renewcommand\color[2][]{}%
  }%
  \providecommand\transparent[1]{%
    \errmessage{(Inkscape) Transparency is used (non-zero) for the text in Inkscape, but the package 'transparent.sty' is not loaded}%
    \renewcommand\transparent[1]{}%
  }%
  \providecommand\rotatebox[2]{#2}%
  \newcommand*\fsize{\dimexpr\f@size pt\relax}%
  \newcommand*\lineheight[1]{\fontsize{\fsize}{#1\fsize}\selectfont}%
  \ifx\svgwidth\undefined%
    \setlength{\unitlength}{334.2699114bp}%
    \ifx\svgscale\undefined%
      \relax%
    \else%
      \setlength{\unitlength}{\unitlength * \real{\svgscale}}%
    \fi%
  \else%
    \setlength{\unitlength}{\svgwidth}%
  \fi%
  \global\let\svgwidth\undefined%
  \global\let\svgscale\undefined%
  \makeatother%
  \begin{picture}(1,0.68009886)%
    \lineheight{1}%
    \setlength\tabcolsep{0pt}%
    \put(0,0){\includegraphics[width=\unitlength,page=1]{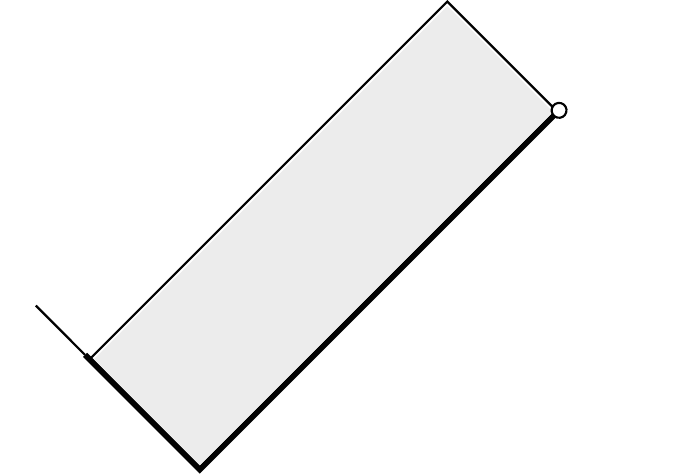}}%
    \put(0.57391202,0.22948347){\color[rgb]{0,0,0}\makebox(0,0)[lt]{\lineheight{1.25}\smash{\begin{tabular}[t]{l}$\mathcal{H}^+$\end{tabular}}}}%
    \put(0.28788403,0.41616061){\color[rgb]{0,0,0}\makebox(0,0)[lt]{\lineheight{1.25}\smash{\begin{tabular}[t]{l}$\widecheck{H}$\end{tabular}}}}%
    \put(0.71915933,0.63567081){\color[rgb]{0,0,0}\makebox(0,0)[lt]{\lineheight{1.25}\smash{\begin{tabular}[t]{l}$\mathcal{CH}^+$\end{tabular}}}}%
    \put(0.46301485,0.03174116){\color[rgb]{0,0,0}\makebox(0,0)[lt]{\lineheight{1.25}\smash{\begin{tabular}[t]{l}$\mathcal{M}'$\end{tabular}}}}%
    \put(0.12759405,0.03660403){\color[rgb]{0,0,0}\makebox(0,0)[lt]{\lineheight{1.25}\smash{\begin{tabular}[t]{l}$\widecheck{\underline{H}}_1$\end{tabular}}}}%
    \put(-0.00055253,0.15623808){\color[rgb]{0,0,0}\makebox(0,0)[lt]{\lineheight{1.25}\smash{\begin{tabular}[t]{l}$\underline{H}_1$\end{tabular}}}}%
    \put(0,0){\includegraphics[width=\unitlength,page=2]{FigMainThm.pdf}}%
  \end{picture}%
\endgroup%
 
   \caption{Penrose-style diagram illustrating Theorem \ref{thm:main}.}
   \label{FigMainThm}
\end{wrapfigure}
states that if we add assumption \ref{item:main.thm.iii}, then conclusion \ref{item:main.thm.3} also holds. In order to establish this conclusion, we use the result of \cite{Sbie24}, which shows that the desired Lipschitz inextendibility follows from a curvature blow-up condition. For the \emph{linearized} equation, the desired curvature blow-up condition needed in \cite{Sbie24} was in turn established in \cite{Sbie23}. The main contribution in this paper is to show that the linearized curvature blow-up of \cite{Sbie23} persists for the full \emph{nonlinear} vacuum Einstein equations, which then implies conclusion~\ref{item:main.thm.3} of Theorem~\ref{thm:main}. See Section~\ref{sec:proof.of.main.theorem.1.2} and Section~\ref{sec:main.result.proven.in.this.paper} for further details on the main steps of the proof and the reduction to the proof of a curvature blow-up condition.

\subsection{Discussion of the proof}

\subsubsection{Main strategy}

As we have already mentioned, the upper bounds in the main theorem give a solution to the Einstein equations with a Cauchy horizon at which the metric extends continuously \cite{DafLuk17, DafLuk26}. In a neighborhood of the Cauchy horizon, the construction in \cite{DafLuk17} gives the metric in a double null coordinate system $(u,\ub,\vartheta_*^1,\vartheta_*^2)$: 
\begin{equation}\label{eq:metric.intro}
    g=-2\Omega^2(\ud u\otimes \ud\ub + \ud\ub\otimes \ud u)+\gamma_{AB}(\ud\vartheta_*^A-b^A \ud\ub)\otimes (\ud\vartheta_*^B-b^B \ud\ub),
\end{equation}
where $u,\ub$ are null variables and $\{\ub = \infty\}$ formally corresponds to the Cauchy horizon. The metric coefficient $\Omg^2$ in these coordinates satisfies
\begin{equation}\label{eq:Omg.centered}
    \Omg^2 \sim e^{-\f{r_+-r_-}{r_-^2+a^2}(u+\ub)}. 
\end{equation} 
It was also proven in \cite{DafLuk17} that upon changing to the $(u,\ub_{\CH^+},\vartheta^1_{\CH^+},\vartheta^2_{\CH^+})$ coordinate system, which is different but also double null, with $\ub_{\CH^+}(\ub) = -\f{r_-^2+a^2}{r_+-r_-} e^{-\f{r_+-r_-}{r_-^2+a^2} \ub}$ as in \eqref{eq:ubCH.thm} (and $(\vartheta^1_{\CH^+},\vartheta^2_{\CH^+})$ a suitable change of angular coordinates), the metric in the new coordinates extends continuously to the Cauchy horizon $\CH^+ = \{\ub_{\CH^+} = 0\}$. 

In the regular coordinate system $(u,\ub_{\CH^+},\vartheta^1_{\CH^+},\vartheta^2_{\CH^+})$, the bounds proven in \cite{DafLuk17} are consistent with the connection coefficient and the curvature components blowing up  with powers of $\Omg^{-2}$ (see \eqref{eq:Omg.centered}). However, the estimates in \cite{DafLuk17} are only upper bounds, and the goal of this paper is to show that such a blowup indeed occurs.

In order to use the Lipschitz inextendibility result of the second author \cite{Sbie24}, we will show that in any tubular neighborhood of a constant-$(u,\vartheta^1_{\CH^+},\vartheta^2_{\CH^+})$ curve, the integral of suitable components of the Riemann curvature tensor blows up. More precisely, for $\overline{V}$ denoting such a tubular neighborhood and $\vol_g$ the Lorentzian volume form, \cite{Sbie24} requires us to show that for some choice of vector fields $\overline{X}^{(j)}_i$ ($i=1,2,3,4$, $j=1,2$) which are continuous up to the Cauchy horizon (in the $(u,\ub_{\CH^+},\vartheta^1_{\CH^+},\vartheta^2_{\CH^+})$ coordinates), the following holds for all $\widehat{X}^{(j)}_i$ sufficiently close (in $C^0$) to $\overline{X}^{(j)}_i$
\begin{equation}\label{eq:intro.goal}
    \lim_{\ub_{\CH^+,k} \to 0^-} \Big| \int_{\overline{V}\cap \{\ub_{\CH^+} < \ub_{\CH^+,k}\}} \Big{\{}R(\widehat{X}^{(1)}_1, \widehat{X}^{(1)}_2, \widehat{X}^{(1)}_3, \widehat{X}^{(1)}_4) + i R(\widehat{X}^{(1)}_1, \widehat{X}^{(1)}_2, \widehat{X}^{(1)}_3, \widehat{X}^{(1)}_4) \Big{\}} \vol_g \Big| = \infty.
\end{equation}

The double null coordinate system in \cite{DafLuk17} comes with a null frame\footnote{We will actually use a slightly different notation in the text. In the text the frame will be denoted as $e_1'^{(\cdot)},e_2'^{(\cdot)},\os{\ee}{2}{}_3, \os{\ee}{2}{}_4$, but in the introduction, we will slightly lighten the notation without explaining all the different frames.} $\{\ee_1,\ee_2,\ee_3,\ee_4\}$ which is continuous up to the the Cauchy horizon $\CH^+$. Here, $\ee_3, \ee_4$ are future-directed, null, and $\{\ee_1,\ee_2,\ee_3\}$ are tangential to $\CH^+$. The analysis in \cite{DafLuk17} already suggest that in this frame, the most singular curvature components are 
$$\alpha(\ee_1,\ee_2):= R(\ee_1,\ee_4,\ee_2,\ee_4),\quad \alpha(\ee_1,\ee_1) = -\alpha(\ee_2,\ee_2) := R(\ee_1,\ee_4,\ee_1,\ee_4)$$
(and in fact a crucial insight in \cite{DafLuk17} is to avoid these curvuture components completely in the main estimates). Thus we aim at establishing \eqref{eq:intro.goal} with 
\begin{equation}\label{eq:intro.goal.X}
    \overline{X}^{(j)}_2, \overline{X}^{(j)}_4 = \ee_4 \quad \hbox{for $j=1,2$}, \quad (\overline{X}^{(1)}_1, \overline{X}^{(1)}_3) = (\ee_1, \ee_2), \quad (\overline{X}^{(2)}_1, \overline{X}^{(2)}_3) = (\ee_1, \ee_1). 
\end{equation}

In order to prove \eqref{eq:intro.goal}, we connect the components of $\alp$  with the dynamical Teukolsky field $\psi_0$, which is a complex scalar function capturing two curvature components associated with a \emph{different} null frame $(e_1,e_2,e_3,e_4)$, related to the principal null frame on a fixed Kerr background. The reason for using $\psi_0$ is two-fold: First, like $\alp$, the components of $\psi_0$ are expected to be the most singular; that some suitably understood ``linear combination'' of $\alp$ and $\psi_0$ would be more regular. Second, and more importantly, the \emph{linear analysis} for $\psi_0$ is easier. It has been observed since the work of Teukolsky \cite{Teu73} that in linear theory, $\psi_0$ satisfies a \emph{decoupled} equation; to see such a decoupling it is important to use the principal null frame. This turns out to be very useful for the linear instability analysis: Interestingly, while the known proofs for the linear stability for systems of wave equations in the black hole interior appear to be very robust, those for linear instability are much more sensitive to the precise linear structure. Here we rely on the linear instability result for the Teukolsky equation proven in \cite{Sbie23}.

Let us already emphasize that the frame field $(e_1,e_2,e_3,e_4)$ we introduce is normalized to \emph{not} be a $C^0$-regular frame field at the Cauchy horizon (in contrast to $\{\ee_1,\ee_2,\ee_3,\ee_4\}$). In particular, with this normalization, $\psi_0$ is bounded. A \emph{polynomial} in $\ub$ lower bound in $\psi_0$ will then translate to the desired blowup in the regular coordinate system.

Our overall strategy will thus be the following:
\begin{enumerate}
    \item Introduce a suitable principal null frame on the dynamical spacetime so as to define $\psi_0$.
    \item After an identification with Kerr, and choosing a suitable comparison linear Teukolsky field $\psil$, we prove energy estimates for $\psi_0 - \psil$ up to a hypersurface $\Gamma$, defined to be ``logarithmically closer'' to the Cauchy horizon than a constant-Boyer--Lindquist $r$ hypersurface (cf.~\cite{D2}). This shows that $\psi_0 - \psil$ decays faster than the $L^2$-averaged lower bound for $\psil$ obtained from linear theory in \cite{Sbie23}, thus inducing an $L^2$-averaged lower bound for $\psi_0$ along $\Gamma$.
    \item Prove that the lower bound for $\psi_0$ gives a lower bound on $\alp$ along $\Gamma$ and then propagate the lower bound for $\alp$ to $\ov{V}$, concluding \eqref{eq:intro.goal}.
\end{enumerate}
We will further elaborate on each step below.

\subsubsection{Principal null frame on dynamical spacetime}\label{sec:intro.frames}

We define a map $\Phi:\calM\to \calM_{\mathrm{Kerr}}$, which is a diffeomorphism onto its image, a subset of the Kerr interior in a neighborhood of timelike infinity. In order to define $\psi_0$, we define an analog of the principal null frame $(e_1,e_2,e_3,e_4)$ on $\calM$. The principal null frame should satisfy the following properties:
\begin{enumerate}
    \item $(e_1,e_2,e_3,e_4)$ is a null frame, i.e., $g(e_A,e_B) = \de_{AB}$, $g(e_3,e_3)= g(e_4,e_4) = g(e_3,e_A)= g(e_4,e_A) = 0$ (for $A,B=1,2$) and $g(e_3,e_4) = -2$.
    \item $(e_1,e_2,e_3,e_4)$ has the right regularity properties, i.e., $(e_3,e_4)$ are smooth and that after a suitable rotation near the poles, $(e_1,e_2)$ are smooth.
    \item Denoting by $(\pmb{e}_1,\pmb{e}_2,\pmb{e}_3,\pmb{e}_4)$ as the push-forward of the Kerr principal null frame under $\Phi^{-1}$, we require $(e_1,e_2,e_3,e_4) \to (\pmb{e}_1,\pmb{e}_2,\pmb{e}_3,\pmb{e}_4)$ in a strong enough topology with a sufficiently fast rate, consistent with the decay of the geometric quantities.
\end{enumerate}

Our approach will be to construct \emph{in coordinates} a global double null frame $(e_1',e_2',e_3',e_4')$, which approaches $(\pmb{e}_1',\pmb{e}_2',\pmb{e}_3',\pmb{e}_4')$, the push-forward of the (Pretorius--Israel) Kerr double null frame. We then construct the principal null frame $(e_1,e_2,e_3,e_4)$ on the dynamical background by imposing the algebraic relation between the transformation $(\pmb{e}_1,\pmb{e}_2,\pmb{e}_3,\pmb{e}_4) \leftrightarrow (\pmb{e}_1',\pmb{e}_2',\pmb{e}_3',\pmb{e}_4')$.

Importantly, we will show that the principal null frame so defined obeys a null structure, manifested as follows:
\begin{itemize}
    \item In terms of the principal null frame, different connection coefficients have different behaviors near the Cauchy horizon. In particular, only some but not all components are singular. The same is true for curvature components.
    
    This structure of the connection coefficients is inherited for the null structure in the double null frame in \cite{DafLuk17} through the specific form of the transformation between the double null frame and the principal null frame on exact Kerr spacetimes.
    \item In the nonlinear terms that appear in the nonlinear Teukolsky equation (see Section~\ref{sec:intro.ee}), the most singular terms never appear quadratically.  See  Proposition~\ref{prop:nonlinear.Teukolsky.error.1} and Proposition~\ref{prop:nonlinear.Teukolsky.error.2}.
\end{itemize}

\subsubsection{Energy estimates for $\protect\psi_0 - \protect\psil$}\label{sec:intro.ee}

Using the Einstein equations, it follows that $\psi_0$ satisfies a nonlinear wave equation
$\mathfrak{T}_{[2]} \psi_0 = \mathfrak{N}$, where $\mathfrak{T}_{[2]}$ is a second order wave operator with coefficients depending on the dynamical background, and $\mathfrak{N}$ is a nonlinear term that is at least quadratic in quantities that vanish in the background; see \eqref{EqNonlinearTeuk}.

We design a comparison solution $\psil$ solving the linear Teukolsky equation $\pmb{\mathfrak{T}}_{[2]} \psil = 0$, where the bold $\pmb{\mathfrak{T}}_{[2]}$ denotes the operator on the background Kerr spacetime, and bound $\psi = \psi_0 - \psil$. The solution $\psil$ is chosen to have the same data as $\psi_0$ on the event horizon $\mathcal H^+$, but such that its transversal derivatives also obey good asymptotic properties.

To control $\psi$, we consider $\pmb{\mathfrak{T}}_{[2]} \psi = (\pmb{\mathfrak{T}}_{[2]} - \mathfrak{T}_{[2]})\psi_0 + \mathfrak{N}$, and 
think of it as a linear inhomogeneous Teukolsky equation on the fixed Kerr background. The error terms $(\pmb{\mathfrak{T}}_{[2]} - \mathfrak{T}_{[2]})\psi_0 + \mathfrak{N}$ are at least quadratic, and are to be controlled using the stability estimates. In particular, since all the estimates have been closed, we can allow for the loss of derivatives on the right-hand side. Our goal will be to have sufficient control of the error terms to obtain an estimate
\begin{equation}\label{eq:psi.diff.intro.goal}
    \int_{\Gamma} \ub^q |\psi_0 - \psil|^2 \vols \ud\ub <\infty,
\end{equation}
where $\vols$ should be thought of as the standard volume form on the sphere in $\vartheta_*$ coordinates and $\Gamma$ is roughly\footnote{The precise definition is given in a different coordinate system; see Section~\ref{sec:Dafermos}.} given by 
\begin{equation}\label{eq:Gamma.intro}
    \Big\{ u + \ub -\f{\sigma_q (r_-^2+a^2)}{r_+-r_-} \log \ub = 0 \Big\},
\end{equation}
where $\sigma_q$ is a parameter that needs to be carefully chosen, as we discuss below.

In \eqref{eq:psi.diff.intro.goal}, $q \in \mathbb Z_{\geq 7}$ is a parameter quantifying the decay rate along the event horizon in the assumptions of the theorem. Importantly, as a consequence of the linear theory
\begin{equation}\label{eq:intro.linear.estimates}
    \int_{\Gamma} \ub^q |(\psil)_{l=2}|^2 \vols \ud \ub = \infty,\qquad \int_{\Gamma} \ub^q |(\psil)_{l>2}|^2 \vols \ud \ub < \infty,
\end{equation}
where the subscript $l$ denotes the projection on the corresponding angular modes. When combined with the linear estimates, \eqref{eq:psi.diff.intro.goal} gives an $L^2$-averaged polynomial lower bound for the $l=2$ angular mode of $\psi_0$ on $\Gamma$, roughly corresponding to decay being no faster than $\ub^{-\f{q+1}{2}}$.

In order to control the error terms in the energy estimates to achieve \eqref{eq:psi.diff.intro.goal}, we use the nonlinear estimates established in \cite{DafLuk17, DafLuk26}, which corresponds to the difference between any geometric quantities on the dynamical spacetime and on the background Kerr spacetime decays with an upper bound of $\ub^{-\f{q_--3}{2}}$ or $|u|^{-\f{q_--3}{2}}$ (depending on the precise quantity) for $q_- < q$. We mention a few ideas relevant to this analysis.

\begin{itemize}
    \item (Translation to principal null frame) To handle the error terms, we will translate all the estimates in \cite{DafLuk17, DafLuk26} in terms of the principal null frame defined in Section~\ref{sec:intro.frames}. Using, in addition, the bounds about the frame transformation that we derive, we will treat as error terms both the original nonlinear term $\mathfrak{N}$ in the Teukolsky equation, and the term $(\pmb{\mathfrak{T}}_{[2]} - \mathfrak{T}_{[2]})\psi_0$ that involves the difference between the dynamical and the background Teukolsky operator. This requires careful identification with the background Kerr spacetime, especially in a transition region between the regions in \cite{DafLuk26} and the regions in \cite{DafLuk17}.
    \item (Singular terms, null structure, the use of $\Gamma$) When dealing with the error terms, some terms are singular and we need to handle an additional power of $\Omg^{-2}$. An important observation, already mentioned above, is a null structure with respect to the terms in the principal null frame, and that there is never a term where more than one factor is singular. To deal with the singularity, it is helpful to prove estimates only up to $\Gamma$. By \eqref{eq:Gamma.intro}, $\Omg^{-2}$ only grows polynomially in $\ub$ to the past of $\Gamma$, and thus with a careful choice of $\sigma_q$, one can use decay in the nonlinear terms to offset the singular polynomial $\ub$-weight. Note that this imposes an upper bound for $\sigma_q$. See \cite{D2} for the introduction of a similar $\Gamma$ already in the spherically symmetric setting.
    \item (Improving estimates in \cite{DafLuk17}) Even with the above observations, the bounds in \cite{DafLuk17} are not enough and need to be sharpened. One necessary improvement is to also control higher derivatives, which is relatively straightforward using the ideas in \cite{DafLuk17}. More importantly, one needs to improve the weights in the estimates. In \cite{DafLuk17}, there are three types of estimates, which we caricature by bounding the following by $\ls \ep$:
    \begin{equation}\label{eq:baby.model.estimates}
        \begin{split}
            \underbrace{\| \widetilde{\phi} \|_{L^\i_u L^\i_{\ub} L^2(S)}}_{=:I},\quad  \underbrace{\| |u|^{\f{q_--2}2} \Omg \widetilde{\phi} \|_{L^2_u L^2_{\ub} L^2(S)},\,\| \ub^{\f{q_--2}2} \Omg \widetilde{\phi} \|_{L^2_u L^2_{\ub} L^2(S)}}_{=:II},\\
            \underbrace{\| \ub^{\f{q_--2}2} \widetilde{\phi} \|_{L^2_u L^\i_{\ub} L^2(S)}, \| |u|^{\f{q_--2}2} \widetilde{\phi} \|_{L^2_u L^\i_{\ub} L^2(S)}, \| \ub^{\f{q_--2}2} \widetilde{\phi} \|_{L^2_{\ub} L^\i_{u} L^2(S)}}_{=:III},
        \end{split}
    \end{equation}
    where $\widetilde{\phi}$ here should be thought of as being ``normalized'' by the suitable $\Omg^2$ weights (like $\psi_0$) so that it is bounded inverse polynomially in $|u|$ or $\ub$. In each group, exactly whether one has a $|u|$ or a $\ub$ weight depends on the exact component in question.

    We will prove the following improvements:
    \begin{itemize}
        \item Estimate $I$ is improved to $\| |u|^{\f{q_--3}{2}} \widetilde{\phi} \|_{L^\i_u L^\i_{\ub} L^2(S)}$ and $\| \ub^{\f{q_--3}{2}} \widetilde{\phi} \|_{L^\i_u L^\i_{\ub} L^2(S)}$, which is consistent with the total decay rate in $III$ but is now pointwise in $(u,\ub)$. A further improvement turns out to be possible to the past of $\Gamma$, using the bound of the $\ub$-length and $|u|$-length in this region so as to bound $\| |u|^{\f{q_{--}-2}{2}} \widetilde{\phi} \|_{L^\i_u L^\i_{\ub} L^2(S)}$ and $\| \ub^{\f{q_{--}-2}{2}} \widetilde{\phi} \|_{L^\i_u L^\i_{\ub} L^2(S)}$ for $q_{--} < q_-$.
        \item One observes that the $\Omg$-weight in $II$ is highly degenerate  (see \eqref{eq:Omg.centered}) near the Cauchy horizon, as compared to the other estimates. This estimate can be improved so that the $\Omg^2$ weights are improved to $\brk{u+\ub}^{-\f 34}$ weights. This type of improved spacetime terms are inspired by similar estimates in \cite[Theorem~3.7]{LOSR}, \cite[Proposition~9.2]{LukOh2017one} (see also \cite{mDiR2009}).
    \end{itemize}
\end{itemize}

\subsubsection{Propagation from $\Gamma$ to $\ov{V}$}\label{sec:intro.propagation}

In order to obtain \eqref{eq:intro.goal}, we start with $\ov{V}$ and project the set along the integral curves of $e_3'$ onto the hypersurface $\Gamma$. The vector field $e_3'$ (the same $e_3'$ as in Section~\ref{sec:intro.frames}) is adapted to the double null foliation in \cite{DafLuk17}, and the $\alp$ component satisfies, along integral curves of $e_3'$,
\begin{equation}\label{eq:intro.transport.for.alp}
    \f{\ud }{\ud u} |(\det \gamma_{\CH^+})^{\frac{1}{4}} \Omega^4 \alpha|^2 =   (\det \gamma_{\CH^+})^{\frac{1}{4}} \Omega^4 \alpha \cdot (\det \gamma_{\CH^+})^{\frac{1}{4}} \Omega^4 \Big[ \cdots \Big],
\end{equation}
where $(\det \gamma_{\CH^+})\sim 1$ but importantly $\Omg$ degenerates as in \eqref{eq:Omg.centered}. 

To use \eqref{eq:intro.transport.for.alp}, we need a lower bound on $\alp$ on $\Gamma$, by relating it to the lower bound of $\psi_0$ above, as well as upper bounds to show that terms in $\Big[ \cdots \Big]$ are perturbative. In both of these estimates, we need to use (a) the estimates in \cite{DafLuk17} showing that the contribution from the differences of all other components of the geometric quantities are less singular than the lower bound for $\alp$ that we aim at, and that (b) with $\sigma_q$ sufficiently large, all the background contributions are sufficiently small. For (b), notice in particular that for the background contributions, the terms in $\Big[\cdots \Big]$ in \eqref{eq:intro.transport.for.alp} are only $O(1)$, but we can use the $\ub$-decay coming from $\Omg^4$ to the future of $\Gamma$ (see \eqref{eq:Omg.centered}, \eqref{eq:Gamma.intro}). 
The above argument thus imposes a lower bound on $\sigma_q$. It turns out that there is a choice such that both this lower bound and the upper bound in Section~\ref{sec:intro.ee} are satisfied simultaneously.

Finally, the linear instability result we obtain only gives a lower bound on a weighted $L^2$ norm analogous to the linear bound \eqref{eq:intro.linear.estimates}, as one would expect from only a lower bound on the weighted $L^2$ norm in the initial data assumption. On the other hand, the inextendibility result in \cite{Sbie24} relies on the blow-up of averages in every small $\ov{V}$ (see \eqref{eq:intro.goal}) without taking absolute values --- we thus need to exclude cancellations. To rule out cancellations in the angular direction, we use that the blow-up only occurs in the $l =2$ modes (see \eqref{eq:intro.linear.estimates}), which in particular is finite dimensional. Along the direction of the null curve, we use a functional inequality (see Lemma~\ref{LemFinal}) which shows that because of the rapid blowup of curvature in the weighted $L^2$ sense, the very weak control we have for its derivative is already sufficient to rule out that too much oscillation occurs.

\subsection{Related works}\label{sec:related.works}

Our work follows a long tradition both in the mathematics and in the physics literature. We will in particular discuss some works in the simpler linear and spherically symmetric settings, where the interplay between stability and instability, important for the present work, was already apparent. Our discussions will be brief, but we refer the reader to the introduction of \cite{DafLuk17} and to \cite{Maxime.survey}, respectively, for a detailed overview of the more classical and the more recent results.

\subsubsection{Linear stability and instability results in black hole interiors}

Linear stability and instability of Cauchy horizons have been a subject of much discussions in the physics literature. See for instance \cite{CH, GSNS, McN, McN.2, SP} and the references therein.

The linear stability and instability are by now very well-understood mathematically, at least concerning subextremal (non-Schwarzschild) Reissner--Nordstr\"om and Kerr spacetimes. In fact, there are now multiple approaches for such rigorous results. In terms of stability, it is known that for the linear scalar wave equation, the scalar field itself remains continuously extendible up to the Cauchy horizon, and that some degenerate energy estimates hold. These estimates for both Reissner--Nordstr\"om and Kerr, which can be viewed as a linearized statement of the $C^0$-stability of the Cauchy horizon, have been established for instance in \cite{Franzen1, Franzen2, pH2017, LukSbi15}.

In addition to linear stability, there is a linear instability mechanism associated to the blue-shift effect showing that higher norms blow up. For Reissner--Nordstr\"om, instability can be established by restricting to the spherically symmetry mode. Assuming a pointwise polynomial lower bound on the spherically symmetric mode on the event horizon (which is later proven to hold, see below), instability in the sense of blow-up of the derivatives follows from adapting the argument of Dafermos \cite{D2} to the linear case. The first unconditional instability result was proven by Luk--Oh \cite{LukOhpub} which established the (non-degnerate) energy blow-up at the Cauchy horizon for smooth and localized Cauchy data. See also \cite{Gleeson} which shows that, at least in a subrange of parameters, blow-up occurs even in the Sobolev space $W^{1,p}$ for all $p>1$. An alternative proof was later given by Luk--Oh--Shlapentokh-Rothman \cite{LOSR} based on scattering theory.

For Kerr, we established a blow up result assuming an energy lower bound on the event horizon \cite{LukSbi15}. A different proof relying on more precise asymptotics in the exterior was later given by Ma--Zhang \cite{syMlZ2023}. There are other types of approaches and results capturing the global blue-shift instability: A Gaussian beam construction of solutions with infinite energy at the Cauchy horizon was given in \cite{jS2015}, while a different construction using the time-translation invariance of the scattering map and the scattering theory of \cite{mDiRySR2018}, which in addition allows for rapidly decaying solutions along past null infinity was carried out in \cite{mDySR2017}.

More relevant to the present paper are the instability results for the spin $+2$ Teukolsky equation. An instability result was first proven by the second author \cite{Sbie23} with a proof inspired by the scattering approach of \cite{LOSR}. Subsequently, Gurriaran \cite{Gurriaran.linear} gives an alternative proof using the approach of \cite{syMlZ2023}. We refer the reader also to \cite{Gurriaran.mass} for results on the spin $-2$ Teukolsky equation, which is expected to be relevant to the generic blow-up of curvature scalars, a question that can be viewed as an analog of mass inflation (see Section~\ref{sec:intro.ss}).

As the discussion above already indicated, the blow-up at the Cauchy horizon is closely related to the global behavior of the solutions, including in the black hole exterior. Indeed, some of the linear instability results above are conditioned on suitable lower bounds, or sometimes even on precise asymptotic behavior, in the black hole exterior. In these settings, an averaged lower bound was first established in \cite{LukOhpub} for Reissner--Nordstr\"om, and the precise asymptotics --- often known as Price's law \cite{Price} --- for the linear wave equation on Reissner--Nordstr\"om and Kerr were first proven with different methods by Angelopoulos--Aretakis--Gajic \cite{AAG2020, AAGKerr, AAGPrice} and Hintz \cite{HintzPriceLaw}. (See also the earlier results \cite{DSS1, DSS2, MTT, Ta} on upper bounds consistent with Price's law.) The corresponding sharp asymptotics results for the Teukolsky field have been proven by different approaches in \cite{syMlZ2022.2, pM2023}. More recently, there are new works concerning corrections to the Price's law asymptotics in various dynamical settings; see \cite{dGlK2025, dGlmaK2022, lmaK2022, LO}.

For results on the linear scalar wave equation in the interior of other black holes, we refer the reader to \cite{gFjS2020} for the Schwarzschild case, and to \cite{Gajic:2015csa, Gajic:2015hyu} for the extremal case. 

Finally, we mention that linear stability and instability of Cauchy horizons in the cases of non-vanishing cosmological constants also present new and interesting challenges; see for instance \cite{PhysRevLett.120.031103, MR3952830, MR4487911, mDySR2018, MR4776522,MR3882684, MR3892262, pHaV2017} for the case of positive cosmological constant and \cite{cK2020, cK2021, cK2022} for the case of negative cosmological constant.

\subsubsection{Nonlinear spherically symmetric models}\label{sec:intro.ss}

The generic formation of weak null singularities in black hole interiors was first studied in various spherically symmetric models. Hiscock \cite{Hiscock} first showed that such singularities can arise in the black hole interior for spherically symmetric solutions to the Einstein--Maxwell--null dust system with an incoming dust. Poisson--Israel later showed that after adding an outgoing dust, the Hawking mass becomes infinite at the Cauchy horizon \cite{PI1, PI2}, demonstrating a phenomenon that is known as mass inflation.

Many of the mathematical ideas important for the (in)stability of the Cauchy horizon were first developed in the study of the Einstein--Maxwell--(real) scalar field model in spherical symmetry. It is important in particular because the system has a dynamical degree of freedom given by a wave equation, and is thus a better model for the vacuum equations outside symmetry. In a breakthrough work \cite{D1}, Dafermos first identified the stability and instability mechanisms in the black hole interior near the Cauchy horizon in this spherically symmetric setting, constructing examples for which the metric is continuously extendible through the Cauchy horizon but that the Hawking mass blows up. He later proved that assuming suitable upper and lower bounds on the (derivative of the) scalar field on the event horizon, one has a similar weak null singularity at the Cauchy horizon \cite{D2}. The necessary upper bound was subsequently proven in the full black hole exterior region by Dafermos--Rodnianski \cite{DRPL}, which implies unconditionally that as long as the charge is non-zero and that the exterior does not settle down to an extremal black hole (even for large data!), the black hole interior has a Cauchy horizon across which the metric is $C^0$-extendible. In fact, Dafermos also showed that if a global smallness assumption is imposed, then the black hole interior has a global bifurcate $C^0$-Cauchy horizon \cite{D3}.

The above already hinted at an instability of the Cauchy horizon, but the necessary pointwise lower bound along the event horizon was not known at the time. Instead, the strong cosmic censorship conjecture in $C^2$ --- a statement that the maximal Cauchy developments arising from a generic, i.e., open and dense,  subset of initial data are $C^2$-inextendible --- was proven by Luk--Oh \cite{LukOh2017one, LukOh2017two} by establishing a weaker $L^2$-averaged lower bound along the event horizon. Still with the same analytic instability result, the second author \cite{Sbierski.C1} introduced a new geometric inextendibility argument (see Section~\ref{sec:intro.inextendible}) to show that the solutions arising from the generic set of data of Luk--Oh are in fact Lipschitz-inextendible. More recently, the original needed lower bound in \cite{D2} (and in fact more precise asymptotics!) has been established by combining the works of Gautam \cite{Gau} and Luk--Oh \cite{LukOh2017two, LO}, which also proves mass inflation at the Cauchy horizon, in addition to strong cosmic censorship.  (See also \cite{LOSR}, which can be combined with \cite{Gau} to give a different proof of mass inflation.)

Some of the results above have been extended to the Einstein--Maxwell--charged scalar field model \cite{VDM, VDM4}, where the scalar field is moreover allowed to be massive. Unlike the case of uncharged scalar field, this model allows for the study of gravitational collapse from one-ended data. It turns out that in such a setting, weak null singularities can be proven to break down to give way to a coexisting spacelike part, and the precise structure of the singularity, as well as global structure of spacetime, were analyzed in the works of Van de Moortel \cite{VDM3, VDM.coexistence, VDM.coexistence.global}. There are in addition a large number of other interesting phenomena for the Einstein--Maxwell--charged scalar field model; see, for instance, \cite{mVdM2024, wLmVdM2025, cKmVdM2024}. Many of these will be very interesting to understand outside symmetry!

\subsubsection{Nonlinear results without symmetry assumptions}

The only full nonlinear result without symmetry assumptions for the black hole interior is that of Dafermos--Luk \cite{DafLuk17}, mentioned above, showing that the Kerr Cauchy horizon is $C^0$ stable. 

The weak stability of the Kerr Cauchy horizon has been previously conjectured in the literature, together with heuristics suggesting that the perturbed Cauchy horizon is singular \cite{prBcmC1995, aO1992, aO1997}. That the perturbed Cauchy horizon becomes singular is also suggested by the estimates in the proof of \cite{DafLuk17}. However, it was not known prior to this work (and \cite{Gurriaran.nonlinear}, see below). On the other hand, weak null singularities in vacuum without symmetry assumptions were first constructed in the analytic class in \cite{FO}, and more generally in \cite{LukWeakNull}. The estimates used in \cite{LukWeakNull} were in particular important in the design of the norms for \cite{DafLuk17}, which need to account for the fact that there might be a singularity. The construction of \cite{LukWeakNull} has recently been extended by Song \cite{Song} to the Einstein--Euler system, where he moreover showed that the fluid variables remain continuous at the singularity (see also \cite{Mancheva} for (uncoupled) fluid equations on a background with a weak null singularity). 

\subsubsection{Geometric inextendibility results}\label{sec:intro.inextendible}

An important ingredient of our proof is the geometric Lipschitz inextendibility result of \cite{Sbie24}, which relies on uniqueness results for Lipschitz extensions established in \cite{Sbie24a}. (Such uniqueness results for extensions fail to hold for merely continuous extensions to weak null singularities, see \cite{CaSbi25}.)  Low-regularity geometric inextendibility results have in fact only been developed very recently; even the proof of strong cosmic censorship conjecture for the Einstein--Maxwell--(real) scalar field system was first only established as a $C^2$-inextendibility result \cite{LukOh2017one, LukOh2017two}. The first low-regularity geometric inextendibility result below $C^2$ was obtained in \cite{Sbi.Schwarzschild}. The first Lipschitz inextendibility result for weak null singularities was proven by the second author in \cite{Sbierski.C1}, already mentioned above, for the class of spherically symmetric spacetimes considered in \cite{D2, LukOh2017one}.
See also \cite{pCpK2018, GalLinSbi17, gjGeL2017,  mGmvdBS2025, GraKuSa19, Le25, Le25a, eL2024, bM2024, MinSuhr19, Racz10,   jS2023} for related results in the field of geometric inextendibility.

\subsubsection{The paper \cite{Gurriaran.nonlinear}}

While finalizing the current paper, Gurriaran has posted a paper on the arXiv \cite{Gurriaran.nonlinear} proving a very similar result.

\subsection{Organization of the paper} The remainder of the paper is structured as follows. In \textbf{Section~\ref{SecKerrBackground}}, we will discuss the geometry of Kerr. In \textbf{Section~\ref{sec:NP.CK}}, we discuss some notations related to horizontal tensors corresponding to a null pair, and in particular recall the Newman--Penrose and Christodoulou--Klainerman formalisms. After these preliminaries, we will state our main theorem in \textbf{Section~\ref{sec:statement}}. 

The remainder of the paper proves the main theorem. In \textbf{Section~\ref{SecInhomLinTeuk}}, we reproduce some of the estimates for the linear Teukolsky equation on fixed Kerr background in \cite{Sbie23}, now allowing for inhomogeneous terms. In \textbf{Section~\ref{sec:main.est}}, we discuss the main estimates on the dynamical spacetime. Here we obtain estimates improving those in \cite{DafLuk17}, as well as derive all the estimates in terms of the principal null frame. In \textbf{Section~\ref{sec:prop.up.to.Gamma}}, we put together the linear analysis in Section~\ref{SecInhomLinTeuk} and the nonlinear estimates in Section~\ref{sec:main.est} to control the difference between the curvature component $\psi_0$ and the linear Teukolsky field $\psil$ up to the hypersurface $\Gamma$.

Finally, in \textbf{Section~\ref{sec:final}}, we propagate the lower bound up to the Cauchy horizon and conclude the proof using \cite{Sbie24}.

\subsection*{Acknowledgements} 

Both authors thank the Institute for Mathematical Sciences in Singapore for its hospitality during a stay in January 2025. J.~Luk thanks the University of Edinburgh for its hospitality during a visit in August 2025.

J.~Luk acknowledges the support of the National Science Foundation through the grant DMS-2304445, and J.~Sbierski acknowledges the support through the Royal Society University Research Fellowship URF\textbackslash R1\textbackslash 211216.

\section{The explicit Kerr background} \label{SecKerrBackground}

In this section, we introduce the geometry of Kerr. We first introduce the standard Boyer--Linquist coordinates in Section~\ref{sec:BL}, but in Sections~\ref{sec:Kerr.star}, \ref{sec:Kerr.double.null}, and \ref{sec:Kerr.s.ub}, we will need a few different coordinate systems, useful in particular near the event horizon and Cauchy horizons and have been used in the analysis of \cite{DafLuk17, DafLuk26, Sbie23}. We then introduce the notions important for the analysis of the Teukolsky equation, including the Kerr principal null frame, spin weighted functions, and the Teukolsky equation itself; see Sections~\ref{sec:Kerr.frames}--\ref{sec:Teukolsky.first.mention}. Finally, in Section~\ref{sec:Dafermos}, we introduce the important hypersurface $\Gamma$, already mentioned in the introduction, which plays a crucial role in the analysis. 

\subsection{Kerr geometry in Boyer--Lindquist coordinates}\label{sec:BL}

We consider the standard $(t,r,\theta, \varphi)$ coordinates on the smooth manifold $\mathcal{M}_{\mathrm{Kerr}} = \R \times (r_-,r_+) \times \mathbb{S}^2$, where $r_-$ and $r_+$ will be defined momentarily. A Lorentzian metric $\pmb{g}$ on $\mathcal{M}_{\mathrm{Kerr}}$ is defined by
\begin{equation}\label{EqKerrMetric}
\pmb{g} = \pmb{g}_{tt} \, \ud t^2 + \pmb{g}_{t\varphi}\,(\ud t \otimes \ud \varphi + \ud \varphi \otimes \ud t) + \frac{\Sigma}{\Delta} \, \ud r^2 + \Sigma \, \ud \theta^2 + \pmb{g}_{\varphi \varphi} \, \ud \varphi^2 \;,
\end{equation}
where
\begin{equation*}
\begin{aligned}
&\Sigma = r^2 + a^2 \cos^2\theta\;,  \qquad \qquad &&\pmb{g}_{tt} = -1 + \frac{2Mr}{\Sigma}\;{,} \\
&\Delta = r^2 -2Mr + a^2\;,  &&\pmb{g}_{t\varphi} = -\frac{2Mra\sin^2\theta}{\Sigma}\;{,} \\
&R^2 = r^2 + a^2 + \frac{2Mr a^2  \sin^2 \theta}{\Sigma}\;{,}\qquad  &&\pmb{g}_{\varphi \varphi} = \big[ r^2 + a^2 +\frac{2Mra^2 \sin^2\theta}{\Sigma}\big] \sin^2\theta = R^2 \sin^2 \theta\;.
\end{aligned}
\end{equation*}
Here, $a$ and $M$, which are required to satisfy $0 <  |a| < M$, are constants representing the angular momentum per unit mass and the mass of the black hole, respectively. We now define $r_- <  r_+$  to be the roots of $\Delta$. \textbf{We lay out the convention, already alluded to in Sections \ref{sec:intro.frames}, \ref{sec:intro.ee}, that geometrical objects defined on the explicit Kerr background, which do have a corresponding analogue with the same label on the dynamical spacetime arising in Theorem \ref{thm:main}, are labeled in bold in order to distinguish them later on.} So, for example the background Kerr metric is labeled with $\pmb{g}$, the dynamical metric in Theorem \ref{thm:main} is labeled with $g$; and the background $\Sigma$ does not have a dynamical analogue in this paper and for this reason it is not written in bold.

For later reference we note that the inverse metric $\pmb{g}^{-1}$ in the \emph{Boyer--Lindquist coordinates} $(t, \varphi, r, \theta)$ is given by
\begin{equation}
\label{gInverse}
\pmb{g}^{-1} = \begin{pmatrix}
-\frac{\pmb{g}_{\varphi \varphi}}{\Delta \sin^2 \theta} & \frac{\pmb{g}_{t \varphi}}{\Delta \sin^2 \theta} & 0 & 0 \\
\frac{\pmb{g}_{t \varphi}}{ \Delta \sin^2 \theta} & -\frac{\pmb{g}_{tt}}{\Delta \sin^2 \theta} & 0 & 0 \\
0 & 0 & \frac{\Delta}{\Sigma} &0 \\
0 & 0 & 0& \frac{1}{\Sigma}
\end{pmatrix} \;.
\end{equation}

For convenience we introduce the abbreviations $\mathscr{S} = \sin \theta$ and $\mathscr{C}=\cos \theta$. We also note the following identity for later reference:
\begin{equation} \label{EqMagicFormula}
    \Sigma R^2 = (r^2 + a^2 \Co^2)(r^2 + a^2) + 2Ma^2r \Si^2 = (r^2+a^2)^2 - a^2\Delta \Si^2.
\end{equation}
Moreover, using the Boyer--Lindquist coordinates, we define
\begin{equation} \label{EqDefVW}
V= (r^2 + a^2) \frac{\partial}{\rd t}\Big|_{BL} + a \frac{\partial}{\partial \varphi}\Big|_{BL}  \qquad \textnormal{ and } \qquad W= \frac{\partial}{\partial \varphi}\Big|_{BL} + a \Si^2 \frac{\partial}{\partial t}\Big|_{BL} \;.
\end{equation}
\textbf{The notation $\big|_{\ldots}$ attached to a coordinate vector field is used in this paper to indicate the reference coordinate system whenever there is an ambiguity due to the use of several coordinate systems.} The indication may be in terms of a shorthand ($\big|_{BL}$ for the Boyer-Lindquist coordinates) or in terms of the explicit coordinate system ($\big|_{(t,r,\theta, \varphi)}$). If, however, only a single coordinate system is used or the reference coordinate system is stated explicitly in the text, we drop the $\big|_{\ldots}$. 

Moreover, we define 
\begin{equation}\label{eq:kappa.def}
    \kappa_\pm :=  \frac{r_+ - r_-}{2(r_\pm^2 + a^2)} >0
\end{equation} and note that
\begin{equation} \label{EqAsympDelta}
   |\Delta| \sim  e^{- 2 \kappa_- r^*(r)} \qquad \textnormal{ for } r \to r_- \;.
\end{equation}

\subsection{Kerr-star coordinates}\label{sec:Kerr.star}

Let $r^*(r)$ be a function on $(r_-,r_+)$ satisfying $\frac{dr^*}{dr} = \frac{r^2 + a^2}{\Delta}$ and $\overline{r}(r)$ a function on $(r_-,r_+)$ satisfying $\frac{d\overline{r}}{dr} = \frac{a}{\Delta}$. We now define the following functions on $\mathcal M_{\mathrm{Kerr}}$:
\begin{equation} \label{EqDefKerrStarCoord}
\begin{aligned}
v_+ := t + r^* \quad &, \qquad \varphi_+ := \varphi + \overline{r}\quad  \mathrm{mod}\; 2\pi \\
v_-:= r^* - t \quad &, \qquad \varphi_- := \varphi - \overline{r} \quad \mathrm{mod} \; 2\pi \;.
\end{aligned}
\end{equation}
It is easy to check that $(v_+, \varphi_+, r, \theta)$  is a  coordinate system for $\mathcal M$. The metric $\pmb{g}$ in these coordinates takes the form
\begin{equation*}
\begin{split}
\pmb{g}&= \pmb{g}_{tt} \, \ud v_+^2 + \pmb{g}_{t\varphi} \, \big( \ud v_+ \otimes \ud \varphi_+ + \ud \varphi_+ \otimes \ud v_+\big) + \pmb{g}_{\varphi \varphi} \, \ud \varphi^2_+  +\big(\ud v_+ \otimes \ud r + \ud r \otimes \ud v_+\big) \\[2pt] &\qquad - a\sin^2\theta \, \big( \ud r \otimes \ud \varphi_+ + \ud \varphi_+ \otimes \ud r\big) + \rho^2 \, \ud \theta^2 \;.
\end{split}
\end{equation*}
In these coordinates the metric extends smoothly to the right event horizon $\mathcal{H}^+ := \{ r = r_+\}$. We denote with $\mathcal{M}_{\mathrm{Kerr},+}$ the manifold with boundary which arises from attaching $\mathcal{H}^+$ to $\mathcal{M}_{\mathrm{Kerr}}$. In the $(v_-, \varphi_-, r, \theta)$ coordinates a similar computation shows that the metric extends smoothly to the right Cauchy horizon $\CH^+:= \{r = r_-\}$. The manifold arising from attaching $\CH^+$ to $\calM_{\mathrm{Kerr}}$ is denoted by $\calM_{\mathrm{Kerr},-}$. And finally the manifold arising from attaching $\CH^+$ and $\Hp$ to $\calM_{\mathrm{Kerr}}$ is denoted by $\calM_{\mathrm{Kerr},\pm}$.

We express the Boyer--Lindquist coordinate vector fields on the left in terms of the $\{v_+, \varphi_+, r, \theta\}$ coordinate vector fields on the right, indicated by $\big|_+$:
\begin{equation} \label{DefPlusCoordVectorFields}
\begin{aligned}
\frac{\partial}{\partial r} \Big|_{BL}&= \frac{r^2 + a^2}{\Delta} \frac{\rd}{\partial {v_+}}\Big|_+ + \frac{a}{\Delta} \frac{\rd}{\partial {\varphi_+}}\Big|_+ + \frac{\rd}{\partial r}\Big|_+ \qquad \quad &&\frac{\partial}{\partial t}\Big|_{BL} = \frac{\rd}{\partial {v_+}}\Big|_+ \\
\frac{\partial}{\partial \varphi}\Big|_{BL} &= \frac{\rd}{\partial {\varphi_+}}\Big|_+ &&\frac{\partial}{\partial \theta}\Big|_{BL} = \frac{\rd}{\partial \theta}\Big|_+ \;.
\end{aligned}
\end{equation}

We also note that the volume form in $\{v_+, \varphi_+, r, \theta\}$-coordinates is given by $\vol = \Sigma^2 \sin \theta \, dv_+ \wedge dr \wedge d\theta \wedge d\varphi_+ = \Sigma^2 dv_+ \wedge dr \wedge \vols$, where we have defined $\vols := \sin \theta d \theta \wedge d \varphi_+$.

\subsection{The $(u,\ub,\th_*,\varphi_*)$ coordinate system and the associated null frame}\label{sec:Kerr.double.null}

In \cite{DafLuk17}, the Pretorius--Israel construction \cite{fPwI1998} of double null coordinates was adapted to the Kerr interior. We briefly recall the key formulas here, but refer the reader to \cite{DafLuk17} for details.

Before introducing the null coordinates, we first define the Pretorius--Israel \cite{fPwI1998} transformation $(r,\th)\mapsto (r_*,\th_*)$. First, $\th_*(r,\th)$ is defined implicitly by
$$F(\th_*,r,\th):= \int_{\th_*}^{\th} \frac{\ud\th'}{a\sqrt{\sin^2\th_*-\sin^2\th'}}+\int_{r}^{r_+}\frac{\ud r'}{\sqrt{(r'^2+a^2)^2-a^2 \sin^2\th_*\Delta'}}=0.$$
Then define $r_*(r,\th) = \rho(r,\th,\th_*(r,\th))$, where 
\begin{equation}\label{eq:rho.def}
\begin{split}
\rho=&\int \frac{r^2+a^2}{\Delta}\,dr+\int^{r_+}_r \frac{\left((r')^2+a^2-\sqrt{((r')^2+a^2)^2-a^2\sin^2\th_*\Delta'}\right)}{\Delta'}dr'\\
&+\int_{\th_*}^{\th}a\sqrt{\sin^2\th_*-\sin^2\th'}\,d\th'.
\end{split}
\end{equation}

It can be shown that $(r,\th)\mapsto (r_*,\th_*)$ is a well-defined diffeomorphism away from the axis $\th = 0, \pi$. The map also fixes the axis, i.e., $\th_*(r_*,0) = 0$, $\th_*(r_*,\pi) = \pi$. Moreover, $\f{\rd\th}{\rd\th_*}$ is a smooth function in $r_*$ and $\cos\th_*$. The partial derivatives of $(r,\th)$ with respect to $(r_*,\th_*)$ can be computed from the definition, see \cite[Proposition~A.5]{DafLuk17}.
\begin{lemma}\label{lem:partials}
The following identities hold:
    \begin{align}
    \f{\rd r}{\rd r_*}=&\: \f{\Delta \sqrt{(r^2+a^2)^2-a^2\sin^2\th_*\Delta}}{(r^2+a^2)^2-a^2\sin^2\th\Delta},\label{eq:partials.1} \\
    \f{\rd r}{\rd \th_*}= &\:\f{a^2\Delta (\sin^2\th_*-\sin^2\th)\sqrt{(r^2+a^2)^2-a^2\sin^2\th_*\Delta}G}{(r^2+a^2)^2-a^2\sin^2\th\Delta}, \\
    \f{\rd \th}{\rd r_*}=&\: \f{a\Delta \sqrt{\sin^2\th_*-\sin^2\th}}{(r^2+a^2)^2-a^2\sin^2\th\Delta}, \\
    \f{\rd \th}{\rd \th_*}=&\:-\f{a \sqrt{\sin^2\th_*-\sin^2\th}((r^2+a^2)^2-a^2\sin^2\th_*\Delta)G}{(r^2+a^2)^2-a^2\sin^2\th\Delta}.\label{eq:partials.4}
\end{align}
where 
\begin{equation}\label{eq:G.def}
\begin{split}
G= \:&\f{\rd}{\rd\th_*}\restriction_{r, \th fixed} F(\th_*;r,\th)\\
=\:& -\f{1}{a\sin\th_*\cos\th_*}\int_\th^{\th_*} \sqrt{\sin^2\th_*-\sin^2\th'}d\th'-\f{\sin (2\th)}{a\sin (2\th_*)\sqrt{\sin^2\th_*-\sin^2\th}}\\
\:&+\int_{r}^{r_+}\frac{a^2\Delta'\sin 2\th_*\,dr'}{2\big((r'^2+a^2)^2-a^2 \sin^2\th_*\Delta'\big)^{\f32}};
\end{split}
\end{equation}

Moreover, the Jacobian determinant satisfies
\begin{equation}\label{eq:Jac.det.PI}
   \f{\rd r}{\rd r_*} \f{\rd \th}{\rd \th_*} - \f{\rd r}{\rd \th_*} \f{\rd \th}{\rd r_*} = \f{\Delta \ell}{(r^2+a^2)^2-a^2\sin^2\th\Delta},
\end{equation}
where for $G$ as in \eqref{eq:G.def}, $\ell$ is given by 
\begin{equation}\label{eq:ell.def}
    \ell= -G a\sqrt{\sin^2\th_*-\sin^2\th}\sqrt{(r^2+a^2)^2-a^2\sin^2\th_*\Delta}.
\end{equation}
\end{lemma}


\begin{lemma} \label{LemPartialsLimit}
\begin{enumerate}
    \item $\Delta \f{\rd r}{\rd r_*}$, $\Delta^{-1} \sin\th_*\f{\rd r}{\rd \th_*}$, $\Delta^{-1} \f 1{\sin\th_*} \f{\rd \th}{\rd r_*}$, $\f{\rd \th}{\rd \th_*}$, $\sin\th_* G$, $\ell$ are smooth functions with respect to the $|\Delta|^{-1} \rd_{r_*}$ and $\f{1}{\sin\th_*} \rd_{\th_*}$ derivatives, up to the axis and up to the horizons.
    \item Moreover, at $r = r_{\pm}$, 
    \begin{align}
        \Delta^{-1} \f{\rd r}{\rd r_*} \restriction_{r = r_\pm} = &\: \f{1}{r_\pm^2+a^2} \\
        \Delta^{-1} \f{\rd r}{\rd \th_*} \restriction_{r=r_\pm} = &\: \f{a^2 (\sin^2\th_* - \sin^2 \th)G}{r_\pm^2+a^2} \\
        \Delta^{-1} \f{\rd\th}{\rd r_*} \restriction_{r=r_\pm} = &\: \f{a\sqrt{\sin^2\th_* - \sin^2\th}}{(r_\pm^2+a^2)^2}\\
        \f{\rd\th}{\rd\th_*} \restriction_{r=r_\pm} = &\: -a\sqrt{\sin^2\th_* - \sin^2\th} G \label{eq:dth.dth*}
    \end{align}
\end{enumerate}

\end{lemma}


\begin{definition}[Kerr double null coordinates]\label{def:Kerr.double.null.coordinates}
    On $\calM_{\mathrm{Kerr}}$, define the coordinates\footnote{Observe that we used a slightly different notation as \cite{DafLuk17, DafLuk26}, where $\varphi_*$ is denoted as $\phi_*$ and $\mathfrak h$ is denoted as $h$. } $\ub = \f 12 (r_* + t)$, $u = \f 12 (r_* - t)$, $\varphi_* = \varphi - \mathfrak h(r_*,\th_*)$, where $\f{\rd \mathfrak h}{\rd r_*}(r_*,\th_*) = -\f{2Mar}{\Sigma R^2}$, $\mathfrak{h}(r_* = 0, \theta_*) = 0$. The shorthand for denoting partial derivatives with respect to this $(u, \ub, \theta_*, \varphi_*)$ double null coordinate system is $|_{DN}$.   
\end{definition}

In the $(u,\ub,\th_*, \varphi_*)$ coordinates of Definition~\ref{def:Kerr.double.null.coordinates}, the Kerr metric then takes the following form:
\begin{equation}\label{Kerr.doublenull}
\pmb{g} = -2{\bf \Omega}^2 (\ud u\otimes \ud \ub+\ud \ub\otimes \ud u)+{\boldsymbol \gamma}_{AB}(\ud \vartheta_*^A-{\boldsymbol b}^A \ud \ub)\otimes(\ud \vartheta_*^B-{\boldsymbol b}^B \ud \ub).
\end{equation}
Here, $\vartheta_*^1=\th_*$ and $\vartheta_*^2=\varphi_*$ and the metric components take the following values\footnote{Recall that since $\mathfrak h$ is independent of $\varphi_*$, $\frac{\rd \mathfrak h}{\rd\th_*}$ can be defined unambiguously independent of whether $\f{\rd}{\rd\th_*}$ is defined with respect to the $(u,\ub,\th_*,\varphi)$ or the $(u,\ub,\th_*,\varphi_*)$ coordinate system.}:
\begin{equation}\label{Kerr.metric.comp}
\begin{split}
{\boldsymbol \Om}^2=-\frac{\Delta}{R^2},\quad {\boldsymbol b}^{\varphi_*}&=\frac{4Mar}{\Sigma R^2},\quad {\boldsymbol b}^{\th_*}=0,\\
{\boldsymbol \gamma}_{\varphi_*\varphi_*}=R^2\sin^2\th,\quad &{\boldsymbol \gamma}_{\th_*\th_*}=\frac{\ell^2}{R^2}+(\frac{\rd \mathfrak h}{\rd\th_*})^2 R^2\sin^2\th,\\
{\boldsymbol \gamma}_{\th_*\varphi_*}=&\:{\boldsymbol \gamma}_{\varphi_*\th_*}=(\frac{\rd \mathfrak h}{\rd\th_*}) R^2\sin^2\th,
\end{split}
\end{equation}
where $\ell$ is as in \eqref{eq:ell.def}. See \cite[Section~A4]{DafLuk17}.

The inverse of ${\boldsymbol \gamma}$ is given by
\begin{equation}\label{eq:gamma.Ke.inverse}
\begin{split}
({\boldsymbol \gamma}^{-1})^{\varphi_*\varphi_*}=\frac{1}{R^2\sin^2\th}+(\frac{\rd \mathfrak h}{\rd\th_*})^2 &\frac{R^2}{\ell^2},\quad ({\boldsymbol \gamma}^{-1})^{\th_*\th_*}=\frac{R^2}{\ell^2},\\
({\boldsymbol \gamma}^{-1})^{\th_*\varphi_*}=({\boldsymbol \gamma}^{-1})^{\varphi_*\th_*}&=-(\frac{\rd \mathfrak h}{\rd\th_*})\frac{R^2}{\ell^2}.
\end{split}
\end{equation}

It will be useful later to compare the Kerr double null coordinates introduced here and Kerr-star coordinates in Section~\ref{sec:Kerr.star}. In particular, we will use the following lemma:
\begin{lemma} \label{LemRelationStarAndNullCoord}
    There exists a constant $C>0$ which only depends on $a$ and $M$ such that $|r^* - r_*| \leq C$, $|2\ub - v_+| \leq C$, and $|2u - v_-| \leq C$.
\end{lemma}

\begin{proof}
    We recall the definition of $r_*$ in \eqref{eq:rho.def}. The first term on the right-hand side of \eqref{eq:rho.def} corresponds, after possible addition of a constant, exactly to $r^*$. The last two integrals on the right-hand side of \eqref{eq:rho.def} are clearly uniformly bounded, which proves the first claim. The remaining estimates now follow directly from \eqref{EqDefKerrStarCoord} and Definition~\ref{def:Kerr.double.null.coordinates}.
\end{proof}

Associated with the metric \eqref{Kerr.doublenull} in the $(u,\ub,\th_*,\varphi_*)$ coordinate system, we introduce the following double null frame on Kerr. 
\begin{definition}\label{def:Kerr.double.null}
    Define the \textbf{double null frame} on Kerr by
    \begin{equation}
        \begin{aligned}
            &\pmb{e}'_1 := \f{R}{\ell} \Big( \rd_{\th_*} - (\f{\rd \mathfrak h}{\rd\th_*}) \rd_{\varphi_*} \Big) \;,  \qquad \qquad && \pmb{e}'_3 := -\f{R^2}{\Delta} \rd_u \;,\\
            &\pmb{e}_2' := \f{1}{R\Si} \rd_{\varphi_*}\;, \qquad \qquad
            &&\pmb{e}_4' := \rd_{\ub} + \f{4Mar}{\Sigma R^2} \rd_{\varphi_*} \;.
        \end{aligned}
    \end{equation}    
\end{definition}

It will also be useful to make the following definitions:
\begin{definition}
    Given $\pmb{e}'_3$, $\pmb{e}'_4$ as above, define a differently rescaled version
    \begin{equation}\label{eq:background.ee.BS}
        \os{\pmb{\ee}}{2}_3 = \pmb{\Omg}^2 \pmb{e}'_3,\quad \os{\pmb{\ee}}{2}_4 = \pmb{\Omg}^{-2} \pmb{e}'_4.
    \end{equation}
\end{definition}

\begin{definition}\label{def:Kerr.double.null.rotation}
    Given $\pmb{e}'_1$, $\pmb{e}_2'$ as above, define $\pmb{e}'^{(N)}_1$, $\pmb{e}'^{(N)}_2$, $\pmb{e}'^{(S)}_1$, $\pmb{e}'^{(N)}_2$ by
    \begin{align}
    \pmb{e}_1'^{(N)} = \cos\varphi_* \pmb{e}_1' - \sin \varphi_* \pmb{e}_2',\quad \pmb{e}_2'^{(N)} = \sin\varphi_* \pmb{e}_1' + \cos\varphi_* \pmb{e}_2', \\
    \pmb{e}_1'^{(S)} = \cos\varphi_* \pmb{e}_1' + \sin \varphi_* \pmb{e}_2',\quad \pmb{e}_2'^{(S)} = -\sin\varphi_* \pmb{e}_1' + \cos\varphi_* \pmb{e}_2', 
\end{align}
\end{definition}

We will collect some properties about the frame introduced in Definition~\ref{def:Kerr.double.null}. Before we proceed, note that the $(\th_*,\varphi_*)$ coordinates give a natural differentiable structure on the $2$-sphere. The following standard fact describes the differentiable structure, which in particular captures the subtlety near the axis.

\begin{lemma}\label{lem:smoothness}
    \begin{enumerate}
        \item $f(\th_*,\varphi_*)$ is a smooth function on the sphere if and only if there exists $\widetilde{f}$ which is smooth in its variables such that $f(\th_*,\varphi_*) = \widetilde{f}(\sin\th_* \cos\varphi_*, \sin\th_* \sin\varphi_*,\cos\th_*)$.
        \item A vector field on the sphere is smooth if and only if it is a linear combination of $Z_{1,*}$, $Z_{2,*}$ and $Z_{3,*}$ with smooth coefficients, where
        \begin{equation}
\begin{aligned}
Z_{1,*} &= - \sin \varphi_* \, \partial_{\theta_*} -  \frac{\cos \varphi_*\cos \th_*}{\sin \theta_*} \partial_{\varphi_*} \\
Z_{2,*} &= -\cos \varphi_* \, \partial_{\theta_*} +  \frac{\sin \varphi_*\cos \th_*}{\sin \theta_*} \partial_{\varphi_*} \\
Z_{3,*} &= \partial_{\varphi_*} 
\end{aligned}
\end{equation}
    \end{enumerate}
\end{lemma}

\begin{proposition}\label{prop:smmothness.background.e}
    \begin{enumerate}
        \item The vector field $\pmb{e}_3'$ is geodesic and satisfies $\pmb{\nabla}_{\pmb{e}_3'}\pmb{e}_3' = 0$.
        \item $(\pmb{e}_3',\pmb{e}_4')$ are smooth null vector fields in $\calM_{\mathrm{Kerr}}$ orthogonal to the tangent space of the constant-$(u,\ub)$ spheres. Moreover $\pmb{g}(\pmb{e}_3',\pmb{e}_4') = -2$.
        \item Each pair of $(\pmb{e}_1',\pmb{e}_2')$, $(\pmb{e}_1'^{(N)},\pmb{e}_2'^{(N)})$ and $(\pmb{e}_1'^{(S)},\pmb{e}_2'^{(S)})$ form smooth orthonormal vector fields away from the axis $\th_* = 0, \pi$. Moreover, $(\pmb{e}_1'^{(N)},\pmb{e}_2'^{(N)})$ extends smoothly to $\th_* = 0$ and $(\pmb{e}_1'^{(S)},\pmb{e}_2'^{(S)})$ extends smoothly to $\th_* = \pi$.
    \end{enumerate}
    
\end{proposition}
\begin{proof}
    Parts 1 and 2 are proven in \cite{DafLuk17}.
    
    For part 3, it is clear that all the vector fields are orthonormal and smooth away from the axis. To check the smooth extensions to the axis, we will only verify that
    $$\pmb{e}_1'^{(N)}= \f{R \cos \varphi_*}{\ell} \Big( \rd_{\th_*} - (\f{\rd \mathfrak h}{\rd\th_*}) \rd_{\varphi_*} \Big) - \f{\sin \varphi_*}{R\Si} \rd_{\varphi_*}$$ is smooth away from the south pole ($\th_* = \pi$). The other vector fields can be checked in the same way. For this we use the criterion in Lemma~\ref{lem:smoothness}. 
    
   Since $R$, $\ell$, $\f{1}{\sin\th_*}\f{\rd \mathfrak h}{\rd \th_*}$, $\cos \varphi_* \sin \th_*$ are smooth functions, it follows that $$\f{R\cos\varphi_*}{\ell}\f{\rd \mathfrak h}{\rd\th_*} \rd_{\varphi_*} =  \f{R}{\ell \sin\th_*}\f{\rd \mathfrak h}{\rd\th_*} (\cos\varphi_* \sin\th_*) Z_{3,*} $$ is a smooth vector field. It therefore suffices to consider 
    \begin{equation}
        \begin{split}
             \f{R \cos \varphi_*}{\ell} \rd_{\th_*}  - \f{\sin \varphi_*}{R\Si} \rd_{\varphi_*} 
            = &\: - \f{R}{\ell} Z_{2,*} + \f{1}{\sin^2\th_*}\Big(\f{R\cos\th_*}{\ell} - \f{1}{R} \cdot \f{\sin \th_*}{\sin \th}\Big) \Big(\sin\varphi_* \sin\th_* \rd_{\varphi_*}\Big).
        \end{split}
    \end{equation}
    Now $\f{R}{\ell} Z_{2,*}$ is manifestly a smooth vector field. For the second term, $\rd_{\varphi_*}$ is a smooth vector field, $\sin\varphi_* \sin\th_*$ is a smooth function, and $\f{R\cos\th_*}{\ell} - \f{1}{R}\cdot \f{\sin \th_*}{\sin \th}$ is an analytic function of $\sin^2 \th_*$ (and $r$). It thus remains to check that $(\f{R\cos\th_*}{\ell} - \f{1}{R}) |_{\th_* = 0} = 0$ so that we can divide by $\sin^2\th_*$ to get a smooth function. A direct computation shows that
    $$(\f{R\cos\th_*}{\ell} - \f{1}{R}\cdot\f{\sin \th_*}{\sin \th}) |_{\th_* = 0} = r^2 + a^2 - \f{\rd \th}{\rd \th_*}|_{\th_* = 0} (r^2 + a^2),$$
    which indeed vanishes since $\f{\rd \th}{\rd \th_*}|_{\th_* = 0} = 1$. 
\end{proof}

\subsection{The $(s,\ub,\th_*,\varphi_*)$ coordinate system and the associated null frame}\label{sec:Kerr.s.ub}

We now introduce another coordinate system on the background Kerr spacetime, where the null variable $u$ is replaced by a variable $s$ that is defined as the affine parameter of the incoming geodesic vector field.
\begin{definition}\label{def:s.Kerr}
    Let $(u,\ub,\th_*,\varphi_*)$ be as in Definition~\ref{def:Kerr.double.null.coordinates}. For every $\ub \in \bbR$ and $\vartheta_* = (\th_*,\varphi_*) \in \bbS^2$, define
    \begin{equation}\label{eq:s.Kerr.def}
        s(u,\ub,\th_*,\phi_*) = \int_{-\infty}^u \pmb{\Omg}^{2}(\mathfrak u,\ub,\th_*,\phi_*)\, \ud \mathfrak u,
    \end{equation}
    where $\pmb{\Omg}$ is as in \eqref{Kerr.metric.comp}.
\end{definition}
In the $(s,\ub,\th_*,\varphi_*)$ coordinate system, to which we assign the shorthand $|_s$, the Kerr metric takes the form
\begin{equation}
\label{eq:Kerr.s}
\pmb{g} =- 2 (\ud s\otimes \ud\ub+ \ud\ub\otimes \ud s)+ 4\pmb{f} \ud\ub\otimes \ud\ub + \pmb{\gamma}_{AB}(\ud\vartheta_*^A-\pmb{h}^A \ud\ub)\otimes (\ud\vartheta_*^B-\pmb{h}^B \ud\ub),
\end{equation}
where $\gamma$ is as in \eqref{Kerr.metric.comp}, and $\pmb{f}$, $\pmb{h}$ are given by
\begin{equation}\label{eq:h.f.Kerr}
\pmb{h}^A =  \pmb{b}^A - 2 (\pmb{\gamma}^{-1})^{AB} \f{\rd s}{\rd \vartheta_*^B}|_{(u,\ub,\vartheta_*)},\quad \pmb{f} = \f{\rd s}{\rd \ub}|_{(u,\ub,\vartheta_*)} - |\nab_{|_{(u,\ub,\vartheta_*)}} s|^2_{\pmb{\gamma}} + \pmb{b}^A \f{\rd s}{\rd \vartheta_*^A}|_{(u,\ub,\vartheta_*)},
\end{equation}
where $\pmb{b}$, $\pmb{\gamma}^{-1}$ are as in \eqref{Kerr.metric.comp}--\eqref{eq:gamma.Ke.inverse}.

We will also introduce null vectors associated to the coordinate system \eqref{eq:Kerr.s}. One could complete it to a null frame as in Definition~\ref{def:Kerr.double.null}, but we will not do this since it will not be used.
\begin{definition}
    Define $(\os{\pmb{\ee}}{1}_3, \os{\pmb{\ee}}{1}_4)$ in the $(s,\ub,\vartheta_*)$ coordinates by
    \begin{equation}\label{eq:background.RS.ee.3.4.def}
        \os{\pmb{\ee}}{1}_3 = \f{\rd}{\rd s},\quad \os{\pmb{\ee}}{1}_4 = \f{\rd}{\rd \ub} + \pmb{f} \f{\rd}{\rd s} + \pmb{h}^A \f{\rd}{\rd\vartheta_*^A}.
    \end{equation}
\end{definition}

The following is easy to check:
\begin{proposition}
    \begin{enumerate}
        \item The vector field $\os{\pmb{\ee}}{1}_3$ is geodesic and satisfies $\pmb{\nabla}_{\os{\pmb{\ee}}{1}_3} \os{\pmb{\ee}}{1}_3 = 0$.    
        \item $(\os{\pmb{\ee}}{1}_3, \os{\pmb{\ee}}{1}_4)$ are null vector fields orthogonal to the tangent space of the constant-$(s,\ub)$ spheres. Moreover $\pmb{g}(\os{\pmb{\ee}}{1}_3, \os{\pmb{\ee}}{1}_4) = -2$.
    \end{enumerate}
    
\end{proposition}

\subsection{The Kerr principal null frame and the relation with the Kerr double null frame}\label{sec:Kerr.frames}

Recalling \eqref{EqDefVW}, the \emph{principal null frame} is given with respect to Boyer--Lindquist coordinates by
\begin{equation}\label{eq:Teukolsky.in.BL}
\begin{aligned}
&\pmb{e}_1 := \frac{1}{\sqrt{\Sigma}} \partial_\theta \;,  && \pmb{e}_3 := - \frac{1}{\Sigma}\partial_r + \frac{1}{\Delta \Sigma} V \;,\\
&\pmb{e}_2 := \frac{W}{|W|} = \frac{1}{\sqrt{\Sigma} \Si}(\partial_\varphi + a \Si^2 \partial_t)\;, \qquad \qquad
&&\pmb{e}_4 :=  \Delta \partial_r + V \;.
\end{aligned}
\end{equation}
The vector fields $\pmb{e}_3$ and $\pmb{e}_4$ are regular at the  event horizon $\mathcal{H}^+$, while at the Cauchy horizon $\be_4$ degenerates. We have $\pmb{g}(\be_3,\be_4) = -2$. 
We will also need the complex vector field
\begin{equation} \label{EqDefMBackground}
    \pmb{m}:=  \frac{1}{\sqrt{2}}\cdot \frac{\sqrt{\Sigma}}{r + i a \cos \theta}(\pmb{e}_1 + i \cdot \pmb{e}_2).
\end{equation}


We now compute the transformation between the double null frame introduced in Definition~\ref{def:Kerr.double.null} and the principal null frame introduced in \eqref{eq:Teukolsky.in.BL}.


\begin{proposition}\label{prop:transformation}
    The double null frame and the principal null frame on Kerr spacetime satisfy the following relations:
    \begin{align}
        \pmb{e}'_1 = &\: \f{R}{\ell}\f{\rd \th}{\rd \th_*} \sqrt{\Sigma} \pmb{e}_1 -\f 12 \f{\Sigma R}{\ell} \f{\rd r}{\rd \th_*} \pmb{e}_3 + \f{1}{2 \Delta} \f{R}{\ell} \f{\rd r}{\rd \th_*}\pmb{e}_4, \label{eq:principal.to.double.null.1}\\
        \pmb{e}'_2 = &\: \f{r^2+a^2 }{R\sqrt{\Sigma}} \pmb{e}_2 - \f{a \Delta \Si}{2 R} \pmb{e}_3 - \f{a \Si}{2 \Sigma R} \pmb{e}_4, \label{eq:principal.to.double.null.2}\\
        \pmb{e}'_3 = &\: -\f{R^2}{\Delta} \f{\rd \th}{\rd r_*} \sqrt{\Sigma} \pmb{e}_1  - \f{a \Si}{\sqrt{\Sigma} } \pmb{e}_2 + \Big(\f{r^2+a^2}{2} + \f {\Sigma R^2}{2\Delta} \f{\rd r}{\rd r_*} \Big) \pmb{e}_3 + \Big(\f {r^2+a^2}{2\Sigma \Delta} - \f{\rd r}{\rd r_*} \f {R^2}{2\Delta^2} \Big) \pmb{e}_4 , \label{eq:principal.to.double.null.3}\\
        \pmb{e}'_4 = &\:  \f{\rd \th}{\rd r_*} \sqrt{\Sigma} \pmb{e}_1  - \f{a\Delta\Si}{\sqrt{\Sigma} R^2 } \pmb{e}_2  + \Big(\f{\Delta(r^2+a^2)}{2 R^2} - \f {\Sigma }2 \f{\rd r}{\rd r_*} \Big) \pmb{e}_3 + \Big(  \f {r^2+a^2}{2\Sigma R^2} + \f{\rd r}{\rd r_*} \f {1}{2\Delta} \Big) \pmb{e}_4.\label{eq:principal.to.double.null.4}
    \end{align}
    Moreover, the inverse transformation is given as follows:
    \begin{align}
    \pmb{e}_1 = &\: \f{R}{\ell}\f{\rd \th}{\rd \th_*} \sqrt{\Sigma} \pmb{e}_1' - \f{\sqrt{\Sigma}}{2} \f{\rd \th}{\rd r_*} \pmb{e}_3' + \f{R^2\sqrt{\Sigma}}{2\Delta }\f{\rd \th}{\rd r_*} \pmb{e}_4', \label{eq:principal.null.in.terms.of.double.null.1} \\
    \pmb{e}_2 = &\: \f{r^2+a^2}{R\sqrt{\Sigma}}\pmb{e}_2' + \f 12 \f{a\Si \Delta}{\sqrt{\Sigma}R^2}\pmb{e}_3' + \f{a \Si}{2\sqrt{\Sigma}} \pmb{e}_4', \label{eq:principal.null.in.terms.of.double.null.2}\\
    \pmb{e}_3 
    = &\: -\f{R}{\Delta \ell} \f{\rd r}{\rd \th_*} \pmb{e}_1' + \f{a \Si}{\Sigma R}\pmb{e}_2' + \Big(  \f {r^2+a^2}{2\Sigma R^2} + \f{\rd r}{\rd r_*} \f {1}{2\Delta} \Big) \pmb{e}_3' + \Big(\f{r^2+a^2}{2\Sigma \Delta} - \f{\rd r}{\rd r_*} \f {R^2}{2\Delta^2} \Big) \pmb{e}_4', \label{eq:principal.null.in.terms.of.double.null.3} \\
    \pmb{e}_4 
    = &\: \f{\Sigma R}\ell \f{\rd r}{\rd \th_*} \pmb{e}_1' + \f{a \Delta \Si}{R}\pmb{e}_2' +  \Big(\f{\Delta(r^2+a^2)}{2 R^2} - \f {\Sigma }2 \f{\rd r}{\rd r_*} \Big) \pmb{e}_3' + \Big(\f{r^2+a^2}{2} + \f {\Sigma R^2}{2\Delta} \f{\rd r}{\rd r_*} \Big) \pmb{e}_4'. \label{eq:principal.null.in.terms.of.double.null.4}
\end{align}

\end{proposition}


\begin{proof}
In the following computations, it is convenient to keep \eqref{EqMagicFormula} in mind. 
We also compute that
\begin{equation}\label{eq:BL.derivatives.in.terms.of.e}
    \begin{split}
         \rd_{\theta} |_{(t,r,\th,\varphi)} = \sqrt{\Sigma} \pmb{e}_1,\quad \rd_{\varphi}|_{(t,r,\th,\varphi)} = \f{(r^2+a^2) \Si}{\sqrt{\Sigma}} \pmb{e}_2 - \f{a\Delta\Si^2}{2} \pmb{e}_3 - \f{a \Si^2}{2\Sigma} \pmb{e}_4,\\
        \rd_r |_{(t,r,\th,\varphi)}= \f 12 (\Delta^{-1} \pmb{e}_4 - \Sigma \pmb{e}_3),\quad  \rd_t |_{(t,r,\th,\varphi)} = -\f {a \Si}{\sqrt{\Sigma}} \pmb{e}_2 + \f 12 \Delta  \pmb{e}_3 + \f 1{2\Sigma} \pmb{e}_4 .
    \end{split}
\end{equation}

We start with the proof of the first four identities. Observe that $\rd_{\th_*} |_{(t,r_*,\th_*,\varphi_*)}- \f{\rd \mathfrak h}{\rd \th_*} \rd_{\varphi_*} |_{(t,r_*,\th_*,\varphi_*)} = \f{\rd\th}{\rd\th_*}\rd_\th |_{(t,r,\th,\varphi)}+ \f{\rd r}{\rd\th_*} \rd_r|_{(t,r,\th,\varphi)}$ since $\f{\rd \varphi}{\rd\th_*} = \f{\rd \mathfrak h}{\rd \th_*}$. Hence,
\begin{equation*}
    \begin{split}
        \pmb{e}'_1 = &\: \f{R}{\ell}\f{\rd \th}{\rd \th_*} \sqrt{\Sigma} \pmb{e}_1 -\f 12 \f{\Sigma R}{\ell} (\f{\rd r}{\rd \th_*}) \pmb{e}_3 + \f{1}{2 \Delta} \f{R}{\ell} (\f{\rd r}{\rd \th_*})\pmb{e}_4.
    \end{split}
\end{equation*}
Next, we observe $\rd_{\varphi_*} |_{(t,r_*,\th_*,\varphi_*)}= \rd_\varphi |_{(t,r,\th,\varphi)}$ to obtain
\begin{equation*}
    \begin{split}
        \pmb{e}'_2 = &\: \f{1}{R} \Big( \f{(r^2+a^2) }{\sqrt{\Sigma}} \pmb{e}_2 - \f{a \Delta \Si}{2 } \pmb{e}_3 - \f{a \Si}{2 \Sigma} \pmb{e}_4 \Big).
    \end{split}
\end{equation*}
For $\pmb{e}'_3$, we first observe $\rd_{r_*}|_{(t,r_*,\th_*,\varphi_*)} = \f{\rd r}{\rd r_*} \rd_r|_{(t,r,\th,\varphi)} + \f{\rd \th}{\rd r_*} \rd_\th |_{(t,r,\th,\varphi)}+ \f{\rd\varphi}{\rd r_*} \rd_\varphi |_{(t,r,\th,\varphi)}$ and then use $\rd_u|_{(u,\ub,\th_*,\varphi_*)} = \rd_{r_*}|_{(t,r_*,\th_*,\varphi_*)} - \rd_t|_{(t,r_*,\th_*,\varphi_*)}$ to obtain
\begin{equation*}
    \begin{split}
        \pmb{e}'_3 = &\: -\f{R^2}{\Delta}\Big( \f{\rd r}{\rd r_*} (\f {1}{2\Delta} \pmb{e}_4 - \f {\Sigma}2 \pmb{e}_3) + \f{\rd \th}{\rd r_*} \sqrt{\Sigma} \pmb{e}_1 - \f{2Mar}{\Sigma R^2}(\f{(r^2+a^2)\Si }{\sqrt{\Sigma} } \pmb{e}_2 - \f{a \Delta \Si^2}{2} \pmb{e}_3 - \f{a \Si^2}{2 \Sigma} \pmb{e}_4)\\
        &\: \qquad - \f{1}{2\Sigma} \pmb{e}_4 - \f{\Delta}{2} \pmb{e}_3 + \f{a \Si}{\sqrt{\Sigma}} \pmb{e}_2\Big) \\
        = &\: -\f{R^2}{\Delta}\Big( (\f{\rd r}{\rd r_*} \f {1}{2\Delta} - \f 1{2\Sigma} + \f{Ma^2r\Si^2}{\Sigma^2 R^2}) \pmb{e}_4 - (\f{\Delta}{2} + \f {\Sigma }2 \f{\rd r}{\rd r_*} - \f{Ma^2r\Delta\Si^2}{\Sigma R^2}) \pmb{e}_3 \\
        &\: \qquad + \f{\rd \th}{\rd r_*} \sqrt{\Sigma} \pmb{e}_1  + (\f{a \Si}{\sqrt{\Sigma}} - \f{2Mar}{\Sigma R^2}\f{(r^2+a^2)\Si }{\sqrt{\Sigma} } )\pmb{e}_2\Big) \\
        = &\: -\f{R^2}{\Delta}\Big( (\f{\rd r}{\rd r_*} \f {1}{2\Delta} - \f {r^2+a^2}{2\Sigma R^2}) \pmb{e}_4 - (\f{\Delta(r^2+a^2)}{2 R^2} + \f {\Sigma}2 \f{\rd r}{\rd r_*} ) \pmb{e}_3+ \f{\rd \th}{\rd r_*} \sqrt{\Sigma} \pmb{e}_1  + \f{a\Delta\Si}{\sqrt{\Sigma} R^2 } \pmb{e}_2\Big).
    \end{split}
\end{equation*}
For $\pmb{e}'_3$, we again use $\rd_{r_*}|_{(t,r_*,\th_*,\varphi_*)} = \f{\rd r}{\rd r_*} \rd_r|_{(t,r,\th,\varphi)} + \f{\rd \th}{\rd r_*} \rd_\th |_{(t,r,\th,\varphi)}+ \f{\rd\varphi}{\rd r_*} \rd_\varphi |_{(t,r,\th,\varphi)}$, but this time we use that $\rd_{\ub}|_{(u,\ub,\th_*,\varphi_*)} = \rd_{r_*}|_{(t,r_*,\th_*,\varphi_*)} + \rd_t |_{(t,r_*,\th_*,\varphi_*)}$. We then obtain
\begin{equation*}
    \begin{split}
        \pmb{e}'_4 = &\:  \f{\rd r}{\rd r_*} (\f {1}{2\Delta} \pmb{e}_4 - \f {\Sigma}2 \pmb{e}_3) + \f{\rd \th}{\rd r_*} \sqrt{\Sigma} \pmb{e}_1 + \f{2Mar}{\Sigma R^2}(\f{(r^2+a^2)\Si }{\sqrt{\Sigma} } \pmb{e}_2 - \f{a \Delta \Si^2}{2} \pmb{e}_3 - \f{a \Si^2}{2\Sigma} \pmb{e}_4) + \f{1}{2\Sigma} \pmb{e}_4 + \f{\Delta}{2} \pmb{e}_3 - \f{a \Si}{\sqrt{\Sigma}} \pmb{e}_2 \\
        = &\:  (\f{\rd r}{\rd r_*} \f {1}{2\Delta} + \f 1{2\Sigma} - \f{Ma^2r\Si^2}{\Sigma^2 R^2}) \pmb{e}_4 + (\f{\Delta}{2} - \f {\Sigma}2 \f{\rd r}{\rd r_*} - \f{Ma^2r\Delta\Si^2}{\Sigma R^2}) \pmb{e}_3\\
        &\: + \f{\rd \th}{\rd r_*} \sqrt{\Sigma} \pmb{e}_1  - (\f{a \Si}{\sqrt{\Sigma}} - \f{2Mar}{\Sigma R^2}\f{(r^2+a^2)\Si }{\sqrt{\Sigma} } ) \pmb{e}_2 \\
        = &\:  (\f{\rd r}{\rd r_*} \f {1}{2\Delta} + \f {r^2+a^2}{2\Sigma R^2}) \pmb{e}_4 + (\f{\Delta(r^2+a^2)}{2 R^2} - \f {\Sigma }2 \f{\rd r}{\rd r_*} ) \pmb{e}_3+ \f{\rd \th}{\rd r_*} \sqrt{\Sigma} \pmb{e}_1  - \f{a\Delta\Si}{\sqrt{\Sigma} R^2 } \pmb{e}_2.
    \end{split}
\end{equation*}
This concludes the proof of \eqref{eq:principal.to.double.null.1}--\eqref{eq:principal.to.double.null.4}.

Finally, to obtain the inverse map, it suffices to notice that $(\pmb{e}_1,\cdots, \pmb{e}_4)$ and $(\pmb{e}_1',\cdots, \pmb{e}_4')$ are both null frames so that we can obtain the inverse map by computing $g(\pmb{e}_\mu,\pmb{e}_\nu')$. \qedhere
\end{proof}

The vector fields $(\pmb{e}_3, \pmb{e}_4)$ are smooth but $(\pmb{e}_1, \pmb{e}_2)$ are not regular near the axis. We thus make the following definition in analogy with Definition~\ref{def:Kerr.double.null.rotation}:
\begin{definition}\label{def:Kerr.principal.null.rotation}
    Given $\pmb{e}_1$, $\pmb{e}_2$ as above, define $\pmb{e}^{(N)}_1$, $\pmb{e}^{(N)}_2$, $\pmb{e}^{(S)}_1$, $\pmb{e}^{(S)}_2$ by
    \begin{align}
    \pmb{e}_1^{(N)} = \cos\varphi_* \pmb{e}_1 - \sin \varphi_* \pmb{e}_2,\quad \pmb{e}_2^{(N)} = \sin\varphi_* \pmb{e}_1 + \cos\varphi_* \pmb{e}_2, \\
    \pmb{e}_1^{(S)} = \cos\varphi_* \pmb{e}_1 + \sin \varphi_* \pmb{e}_2,\quad \pmb{e}_2^{(S)} = -\sin\varphi_* \pmb{e}_1 + \cos\varphi_* \pmb{e}_2. 
\end{align}
\end{definition}

We also introduce the following shorthand:
\begin{definition}\label{def:B}
Define $\pmb{\calB}_\mu^\nu$ and $\pmb{\calB}_\mu'^\nu$ by 
\begin{equation}\label{eq:calB.def}
\pmb{e}_\mu = \pmb{\calB}_\mu^\nu \pmb{e}'_\nu,\quad \pmb{e}'_\mu = \pmb{\calB}_\mu'^\nu \pmb{e}_\nu,
\end{equation}
where the components can be read off from Proposition~\ref{prop:transformation}.
For $^{(\cdot)} = ^{(N)}, ^{(S)}$, we also write 
\begin{equation}\label{eq:pmbB().def}
\pmb{e}_\mu^{(\cdot)} = (\pmb{\calB}^{(\cdot)})_\mu^\nu \pmb{e}'^{(\cdot)}_\nu,\quad \pmb{e}'^{(\cdot)}_\mu = (\pmb{\calB}'^{(\cdot)})_\mu^\nu \pmb{e}_\nu^{(\cdot)}.
\end{equation}
The components $(\pmb{\calB}^{(\cdot)})_\mu^\nu$ and $(\pmb{\calB}'^{(\cdot)})_\mu^\nu$ can be computed explicitly using Proposition~\ref{prop:transformation} and Definition~\ref{def:Kerr.principal.null.rotation} and \eqref{def:Kerr.double.null.rotation}.
\end{definition}

We will later use the estimates regarding the transformation of the frames. It will be convenient to separate out the following computation, which shows an algebraic cancellation.
\begin{lemma}
    \begin{equation}\label{eq:surprise.estimate}
        \Big| \f {r^2+a^2}{2\Sigma \Delta} - \f{\rd r}{\rd r_*} \f {R^2}{2\Delta^2} \Big|\ls 1,\quad 
        \Big| \f{\Delta(r^2+a^2)}{2 R^2} - \f {\Sigma }2 \f{\rd r}{\rd r_*} \Big| \ls |\Delta|^2,
    \end{equation}
\end{lemma}
\begin{proof}
    First notice that $\f {r^2+a^2}{2\Sigma } - \f{\rd r}{\rd r_*} \f {R^2}{2\Delta}$ and $\f{(r^2+a^2)}{2 R^2} - \f {\Sigma }{2\Delta} \f{\rd r}{\rd r_*}$ are smooth functions up to $r= r_\pm$ by \eqref{eq:partials.1}. Noting $\Sigma$ is bounded above and away from $0$, it thus remains to show that
        \begin{equation}\label{eq:surprise.estimate.need.to.show}
            \Big(\f {r^2+a^2}{2 } - \f{\rd r}{\rd r_*} \f {R^2\Sigma}{2\Delta}\Big)_{r=r_\pm} = \Big(\f{(r^2+a^2)}{2 R^2 \Sigma} - \f {1 }{2\Delta} \f{\rd r}{\rd r_*}\Big)_{r=r_\pm} = 0.
        \end{equation}
        Now \eqref{eq:surprise.estimate.need.to.show} is true after observing
        $$\Big(\f 1{\Delta} \f{\rd r}{\rd r_*}\Big)_{r=r_\pm} = \frac{1}{r_\pm^2 + a^2}, \quad R^2\Sigma_{r=r_\pm} = (r_\pm^2 + a^2)^2,$$
        which follow from \eqref{eq:partials.1}, \eqref{EqMagicFormula}.
\end{proof}

The following proposition consists of the main estimates for the change of frames.
\begin{proposition}\label{prop:B}
The following estimates hold for any $i_1,i_2,i_3,i_4 \in \mathbb Z_{\geq 0}$, with implict constants depending on $i_1,i_2,i_3,i_4$, for $\th_* \in [0,\f{3\pi}4]$ if $^{(\cdot)} = ^{(N)}$ and for $\th_* \in [\f{\pi}4,\pi]$ if $^{(\cdot)} = ^{(S)}$:
\begin{enumerate}
    \item The following bounds hold for all $\mu$, $\nu$:
    \begin{equation}\label{eq:calB.most.basic}
        |(\pmb{e}_1'^{(\cdot)})^{i_1} (\pmb{e}_2'^{(\cdot)})^{i_2} (\pmb{e}_3')^{i_3} (\pmb{e}_4')^{i_4} (\pmb{\calB}^{(\cdot)})_{\mu}^\nu|\ls 1
    \end{equation}
    \item The following improved estimate holds for some components:
    \begin{align}
        |(\pmb{e}_1'^{(\cdot)})^{i_1} (\pmb{e}_2'^{(\cdot)})^{i_2} ( \pmb{e}_3')^{i_3} (\pmb{e}_4')^{i_4}(\pmb{\calB}^{(\cdot)})_A^3|, |(\pmb{e}_1'^{(\cdot)})^{i_1} (\pmb{e}_2'^{(\cdot)})^{i_2} ( \pmb{e}_3')^{i_3} ( \pmb{e}_4')^{i_4}(\pmb{\calB}^{(\cdot)})_4^A|\ls &\: |\Delta|^{1-i_3}, \label{eq:calB.good.components.1}\\
        |(\pmb{e}_1'^{(\cdot)})^{i_1} (\pmb{e}_2'^{(\cdot)})^{i_2} ( \pmb{e}_3')^{i_3} (\pmb{e}_4')^{i_4}(\pmb{\calB}^{(\cdot)})_4^3| \ls &\: |\Delta|^{2-i_3}.\label{eq:calB.good.components.2}
    \end{align}
    \item In particular, it follows from the above that the following holds: 
    \begin{align}
        |(\pmb{e}_1'^{(\cdot)})^{i_1} (\pmb{e}_2'^{(\cdot)})^{i_2} (|\Delta| \pmb{e}_3')^{i_3} (\pmb{e}_4')^{i_4} (\pmb{\calB}^{(\cdot)})_{\mu}^\nu|\ls &\: 1\quad \hbox{all $\mu,\nu$}, \label{eq:calB.BS.1}\\
        |(\pmb{e}_1'^{(\cdot)})^{i_1} (\pmb{e}_2'^{(\cdot)})^{i_2} (|\Delta| \pmb{e}_3')^{i_3} (\pmb{e}_4')^{i_4}(\pmb{\calB}^{(\cdot)})_A^3|, |(\pmb{e}_1'^{(\cdot)})^{i_1} (\pmb{e}_2'^{(\cdot)})^{i_2} (|\Delta| \pmb{e}_3')^{i_3} (\pmb{e}_4')^{i_4}(\pmb{\calB}^{(\cdot)})_4^A|\ls &\: |\Delta|, \label{eq:calB.BS.2}\\
        |(\pmb{e}_1'^{(\cdot)})^{i_1} (\pmb{e}_2'^{(\cdot)})^{i_2} (|\Delta| \pmb{e}_3')^{i_3} (\pmb{e}_4')^{i_4}(\pmb{\calB}^{(\cdot)})_4^3| \ls &\: |\Delta|^2.\label{eq:calB.BS.3}
    \end{align}
\end{enumerate}
\end{proposition}
\begin{proof}
   We note that $\pmb{\calB}_{\mu}^\nu$ are functions of $(r_*,\th_*)$ alone and that the components satisfy 
   \begin{equation*}|(|\Delta|^{-1} \rd_{r_*})^{i_1} (\f{1}{\sin\th_*}\rd_{\th_*})^{i_2} \f{\pmb{\calB}_{\mu}^\nu}{\sin \th_*}| \ls 1 \textnormal{ if exactly $\mu = 1,2$ or $\nu =1,2$, but not $(\mu,\nu) \in \{1,2\}^2$.}
   \end{equation*}
   All other components satisfy $|(|\Delta|^{-1} \rd_{r_*})^{i_1} (\f{1}{\sin\th_*}\rd_{\th_*})^{i_2} \pmb{\calB}_{\mu}^\nu| \ls 1$. Furthermore, we note that $\pmb{\calB}_1^1(\theta= 0, \pi) = \pmb{\calB}_2^2(\th = 0, \pi)$, which follows in the same way as in the proof of Proposition \ref{prop:smmothness.background.e}. These properties directly ensure \eqref{eq:calB.most.basic}. To see this, we compute $\pmb{\calB}^{(\cdot)}$. Consider for example
   \begin{equation*}
       \begin{split}
           \pmb{e}_1^{(N)} &= \Big(\cos^2 \varphi_* \cdot \f{R}{\ell} \f{\rd \theta}{\rd \theta_*} \sqrt{\Sigma} + \sin^2 \varphi_* \f{r^2 + a^2}{R\sqrt{\Sigma}} \Big) \pmb{e}_1'^{(N)} + \cos \varphi_* \sin \varphi_* \Big( \f{R}{\ell} \f{\rd \theta}{\rd \theta_*} \sqrt{\Sigma} -  \f{r^2 + a^2}{R\sqrt{\Sigma}} \Big) \pmb{e}_2'^{(N)} \\
           &-\Big(\cos \varphi_*  \f{\sqrt{\Sigma}}{2} \f{\rd \theta}{\rd r_*} + \sin \varphi_* \f{a \Si \Delta}{2\sqrt{\Sigma} R^2} \Big)  \pmb{e}_3' + \Big( \cos \varphi_* \f{R^2 \sqrt{\Sigma}}{2 \Delta} \f{\rd \theta}{\rd r_*} - \sin \varphi_* \f{a \Si}{2 \sqrt{\Sigma}} \Big) \pmb{e}_4'.
       \end{split}
   \end{equation*}
Using the above we see that each coefficient is a smooth function for $\th_* \in [0, \f{3 \pi}{4}]$ and has uniformly bounded derivatives in $(|\Delta|^{-1} \rd_{r_*})^{i_1}$. The other components of $\pmb{\calB}^{(\cdot)}$ follow similarly, which shows \eqref{eq:calB.most.basic}.

    For the remaining estimates, it suffices to note that the components $(\pmb{\calB}^{(\cdot)})_A^3$,  $(\pmb{\calB}^{(\cdot)})_4^A$, $(\pmb{\calB}^{(\cdot)})_4^3$ have one or two extra powers of $|\Delta|$, which again follows from Lemma~\ref{LemPartialsLimit} and \eqref{eq:surprise.estimate}.
    When differentiating with $\pmb{e}_3'$, one may lose the extra powers of $|\Delta|$, giving  \eqref{eq:calB.good.components.1}--\eqref{eq:calB.good.components.2}. If, instead, one differentiates with $|\Delta|\pmb{e}_3'$ or $\pmb{e}_4'$, then the extra powers of $|\Delta|$ persist, giving \eqref{eq:calB.BS.2}--\eqref{eq:calB.BS.3}.
    \qedhere    
\end{proof}
We point out that the analogous estimates hold for $(\pmb{\calB}')^{(\cdot)}$ in place of $\pmb{\calB}^{(\cdot)}$, since the matrices $\pmb{\calB}'$ satisfy the same structure as is clear from \eqref{eq:principal.to.double.null.1} - \eqref{eq:principal.to.double.null.4}.

\subsection{Spin $s$-weighted functions}\label{sec:spin.weighted}

\begin{definition} \label{DefSpinWeightedFunction}
    Let $s \in \Z$ and let $k \in \N \cup \{\infty\}$. A $C^k$-regular spin $s$-weighted function on an open subset $\mathcal{A} \subseteq \mathcal{M}_{\mathrm{Kerr},+}$ is a function $f \in C^{k}(\mathcal{A} \setminus \{\theta = 0, \pi\}, \C)$ such that $e^{si \varphi_+}f \in C^{k}(\mathcal{A} \setminus \{\theta =  \pi\}, \C)$ and $e^{-si \varphi_+}f \in C^{k}(\mathcal{A} \setminus \{\theta =  0\}, \C)$. The space of $C^k$-regular spin $s$-weighted functions on $\mathcal{A}$ is denoted by $\mathscr{I}^k_{[s]}(\mathcal{A})$. 
\end{definition}

Recall that the vector fields
\begin{equation}\label{EqDefVectorFieldZ}
\begin{split}
Z_{1,+} &:= - \sin \varphi_+ \, \partial_\theta - \cos \varphi_+ \frac{\cos \theta}{\sin \theta} \, \partial_{\varphi_+} \\
Z_{2,+} &:= - \cos \varphi_+ \, \partial_\theta + \sin \varphi_+ \frac{\cos \theta}{\sin \theta} \, \partial_{\varphi_+} \\
Z_{3,+} &:= \partial_\phip
\end{split}
\end{equation}
are smooth on $\mathcal{M}_{\mathrm{Kerr},+}$, they are tangent to the Boyer--Lindquist spheres  $\{v_+ = \mathrm{const}\} \cup \{r = \mathrm{const}\} \simeq \Sp^2$, and they span the tangent space of the Boyer--Lindquist spheres at each point. Moreover, they satisfy the commutation relations
\begin{equation} \label{EqCommRelationsZ}
[Z_{1,+}, Z_{2,+}] = Z_{3,+}\,,\qquad [Z_{2,+}, Z_{3,+}] = Z_{1,+}\,, \qquad [Z_{3,+}, Z_{1,+}] = Z_{2,+}\,.
\end{equation}

\begin{lemma} \label{LemAPMSmooth}
    Consider the functions $a_{\pm}(\theta, \phip) := e^{i \phip} \cdot \frac{1 \mp \cos \theta}{\sin \theta}$. Then $a_+ \in C^\infty \big((\mathcal{M}_{\mathrm{Kerr},+})\setminus \{\theta = \pi\}; \C\big)$ and $a_- \in C^\infty \big((\mathcal{M}_{\mathrm{Kerr},+})\setminus \{\theta = 0\}; \C\big)$.
\end{lemma}
In other words, $a_+$ is smooth away from the south pole and $a_-$ is smooth away from the north pole.
\begin{proof}
    We consider the embedding of the coordinate Boyer--Lindquist sphere $\Sp^2$ into $\R^3$ given by
    \begin{equation} \label{EqDefEmbeddingSphere}
        x = \sin \theta \cos \phip\,, \qquad \qquad y = \sin \theta \sin \phip\,, \qquad \qquad z = \cos \theta \,.
    \end{equation}
    We then have $a_+ = e^{i \phip} \frac{1 - \cos \theta}{\sin \theta} = (x + i y) \frac{1-z}{1 - z^2} = \frac{x+ iy}{1 + z}$ and $a_- = e^{i \phip} \frac{1 + \cos \theta}{\sin \theta} = (x + i y) \frac{1+z}{1 - z^2} = \frac{x+ iy}{1 - z}$, which proves the lemma.
\end{proof}

\begin{lemma} 
    Recall the vector field $\pmb{m}$ from \eqref{EqDefMBackground}. Then, the vector field $e^{i \varphi_+} \pmb{m}$ is smooth on $\mathcal{M}_{\mathrm{Kerr},+}$ away from the south pole $\{\theta = \pi\}$ and $e^{-i \varphi_+} \pmb{m}$ is smooth on $\mathcal{M}_{\mathrm{Kerr},+}$ away from the north pole $\{\theta = 0\}$.
\end{lemma}

\begin{proof}
    We only prove the case $e^{i \varphi_+} \pmb{m}$, the other case is similar. It suffices to show that $$e^{i \varphi_+} \big(\rd_\theta + i \frac{1}{\sin \theta}(\rd_{\phip} + a \sin^2 \theta \cdot \rd_{v_+}) \big)$$ is smooth away from the south pole.
    A direct computation gives $Z_{2,+} + i Z_{1,+} = - e^{i \phip}(\rd_{\theta} + i \frac{\cos \theta}{\sin \theta} \rd_\phip )$ so that we obtain
    \begin{equation*}
    \begin{split}
        e^{i \varphi_+} \big(\rd_\theta + i \frac{1}{\sin \theta}(\rd_{\phip} + a \sin^2 \theta \cdot \rd_{v_+}) \big) &= -Z_{2,+} - i Z_{1,+} - e^{i \varphi_+} i \frac{\cos \theta}{\sin \theta} \rd_{\phip} +e^{i \phip} i \frac{1}{\sin \theta} \rd_\phip + ia e^{i \phip} \sin \theta \cdot \rd_{v_+} \\
        &= -Z_{2,+} - i Z_{1,+} + i a_+ Z_{3,+} + ia e^{i \phip} \sin \theta \cdot \rd_{v_+} \;.
        \end{split}
    \end{equation*}
    The function $a_+$ is smooth away from the south pole by Lemma \ref{LemAPMSmooth} and, using \eqref{EqDefEmbeddingSphere}, $e^{i \phip} \sin \theta = x +iy$ is smooth throughout.
\end{proof}

We also recall the definition of the vector fields
\begin{equation}\label{EqZTilde}
\begin{aligned}
\widetilde{Z}_{1_+} &:= - \sin \varphi_+ \, \partial_\theta + \cos \varphi_+(-is \frac{1}{\sin \theta} - \frac{\cos \theta}{\sin \theta} \partial_{\varphi_+}) \\
\widetilde{Z}_{2_+} &:= -\cos \varphi_+ \, \partial_\theta - \sin \varphi_+(-is \frac{1}{\sin \theta} - \frac{\cos \theta}{\sin \theta} \partial_{\varphi_+}) \\
\widetilde{Z}_{3,+} &:= \partial_{\varphi_+}  \,.
\end{aligned}
\end{equation}
For $f \in C^1\big((\mathcal{M}_{\mathrm{Kerr},+}) \setminus\{\theta = 0, \pi\}; \C\big)$ we also recall the relations (see \cite[Lemma 2.23]{Sbie23})
\begin{equation} \label{EqCommuteExpZ}
\begin{aligned}
Z_{1,+}(e^{\pm is\phip}f) &= e^{\pm is\phip}\big(\widetilde{Z}_{1,+} f + is\underbrace{ \frac{\cos \phip}{\sin \theta}( 1 \mp \cos \theta)}_{= \Rea (a_\pm)} f\big) \\
Z_{2,+}(e^{\pm is \varphi_+}f) &= e^{\pm is \phip} \big(\widetilde{Z}_{2,+} f - is\underbrace{ \frac{\sin \phip}{\sin \theta}(1 \mp \cos \theta)}_{= \Ima (a_\pm)}f\big) \\
Z_{3,+}(e^{\pm is \phip}f) &= e^{\pm is \phip}\big(\widetilde{Z}_{3,+} f \pm is f\big) \;.
\end{aligned}
\end{equation}
It in particular follows that the vector fields $\widetilde{Z}_{i,+}$ act smoothly on spin $s$-weighted functions.

\subsection{The Teukolsky equation and spin-weighted spherical harmonics}\label{sec:Teukolsky.first.mention}

The Teukolsky operator $\pmb{\mathcal{T}}_{[2]}$
in $(v_+, \varphi_+, r, \theta)$ coordinates takes the form
\begin{equation}
\label{EqTeukolskyStarCoordinates}
\begin{split}
\pmb{\mathcal{T}}_{[s]} \psi := & a^2 \sin^2 \theta \,\partial_{v_+}^2\psi + 2a \,\partial_{v_+}\partial_{\varphi_+} \psi + 2(r^2 + a^2)\, \partial_{v_+}\partial_r \psi 
+2 a\, \partial_{\varphi_+}\partial_r \psi + \Delta \,\partial_r^2 \psi + \mathring{\slashed{\Delta}}_{[s]} \psi\\
&+ 2\Big( r(1-2s) - isa\cos \theta\Big)\, \partial_{v_+} \psi 
+2(r-M)(1-s) \,\partial_r \psi  - 2s \psi 
\end{split}
\end{equation}
with $s = +2$. Here, $$\mathring{\slashed{\Delta}}_{[s]} := \frac{1}{\sin\theta} \partial_\theta\big(\sin \theta \, \partial_\theta \psi\big) + \frac{1}{\sin^2 \theta} \partial_{\varphi_+}^2 \psi + 2si\frac{\cos \theta}{\sin^2 \theta}\partial_{\varphi_+} \psi -\big( s^2\frac{\cos^2\theta}{\sin^2\theta} -s\big)\psi$$ is the spin-weighted Laplacian on $\mathbb{S}^2$.

For $\psi \in \mathscr{I}^k_{[2]}(\mathcal{A})$ with $\mathcal{A} \subseteq \mathcal{M}_{\mathrm{Kerr},+}$ we recall the definition of the projection onto the $ml$ spin $2$-weighted spherical harmonic $Y_{ml}^{[2]}(\theta, \varphi_+; 0) = S_{ml}^{[2]}(\cos \theta;0) e^{im \varphi_+}$ (see \cite[Section 5.1]{Sbie23}) by 
\begin{equation} \label{DefP}
(\mathbb{S}_{ml}\psi) (v_+, r, \theta, \varphi_+) := \underbrace{\int_{\Sp^2} \psi(v_+,r,\theta', \varphi_+') \overline{Y_{ml}^{[2]}(\theta',\varphi_+';0)} \, \vols }_{=: \psi_{ml}(v_+,r)}\cdot Y_{ml}^{[2]}(\theta, \varphi_+;0) \;.
\end{equation}
Note that in \cite{Sbie23}, \cite{Sbie26} the projection is denoted by $\mathbb{P}_{S(ml)}\psi$. 
We also define $\psi_{l = 2} := \sum_{m = -2}^2  \mathbb{S}_{m2}\psi$ and $\psi_{l > 2} := \psi - \psi_{l=2}$. Note that the projection is defined with respect to the spin $2$-weighted spherical harmonics on the Boyer--Lindquist spheres in the $(\theta, \varphi_+)$-coordinates.

\subsection{The hypersurface $\Gamma$}\label{sec:Dafermos}

The natural number $q \geq 7$ determines the rate of decay of the dynamical geometry to the background geometry along the event horizon, see Section \ref{SecPreciseAssump}.  We set $\sigma_q := \frac{1}{4}(q+3)$  
and define on $\mathcal{M}_{\mathrm{Kerr}} \cap \{v_+ \geq 1\}$ the function $f_\Gamma(v_+, v_-) := v_+ + v_- - \frac{\sigma_q}{\kappa_-} \log (v_+)$ and, for $v_+ \geq 1$, the hypersurface  $$\mathcal{M}_{\mathrm{Kerr}} \cap \{v_+ \geq 1\} \supseteq \Gamma := \{ f_\Gamma(v_+,v_-) = 0\} \;.$$ 
Note that the hypersurface $\Gamma$ depends on $q$, which however is a fixed number $\geq 7$. We compute 
$$\ud f_\Gamma = \frac{r^2 + a^2}{\Delta}(2 - \frac{\sigma_q}{\kappa_-} \frac{1}{v_+} ) \ud r - \frac{\sigma_q}{\kappa_-} \frac{1}{v_+} \ud t$$
and thus, using \eqref{EqMagicFormula}, we obtain
$$\Sigma \cdot g^{-1}(df_\Gamma, df_\Gamma) = a^2 \sin^2 \theta \frac{\sigma_q^2}{\kappa_-^2} \frac{1}{v_+^2} + \frac{4}{\Delta} (r^2 + a^2)^2 (1- \frac{\sigma_q}{\kappa_- v_+}) \;.$$
Note that we have $v_+ + v_- = 2r^*$ and thus $r \to r_-$ on $\Gamma$ for $v_+ \to \infty$. Also recall that $\Delta \nearrow 0$ for $r \to r_-$. This shows directly that $\Gamma$ is spacelike for large $v_+$.

Moreover, we obtain from \eqref{EqAsympDelta} for any $r_- < r_0 < r_+$
\begin{align}
    \textnormal{ on } \{v_+ \geq 1\} \cap \{f_\Gamma \geq 0\} \; &\textnormal{ the bound } |\Delta| \lesssim v_+^{- \sigma_q}  \label{EqDeltaFutureGamma} \\
    \textnormal{ on } \{v_+ \geq 1\} \cap \{f_\Gamma \leq 0\} \cap \{r \leq r_0\} \; &\textnormal{ the bound } |\Delta|^{-1} \lesssim v_+^{ \sigma_q} \;. \label{EqDeltaPastGamma}
\end{align}
The implicit constant in the second bound depends on the choice of $r_0$.

Note that on $\Gamma$ we have $2 r^* = \frac{\sigma_q}{\kappa_-} \log (v_+) 
\geq 0$ and thus $\Gamma \subseteq \{r \leq r_1\}$ for some $r_- < r_1 < r_+$. As a consequence we have 
\begin{equation} \label{EqUVFutureGamma}
\frac{1}{v_+} \lesssim \frac{1}{\sqrt{1 + v_-^2}} \qquad \textnormal{ in } \{f_\Gamma \geq 0\} \cap \{v_+ \geq 1\} \cap \{u \leq -1\}\;,
\end{equation}
where we have used Lemma \ref{LemRelationStarAndNullCoord} to infer that $v_-$ is bounded from above in this region. 

\section{The NP and the CK formalisms}\label{sec:NP.CK}

\subsection{The NP formalism}   \label{SecNP}

Consider a Lorentzian manifold $(M,g)$ with a Ricci-flat metric which is of signature $(-,+,+,+)$. Let $(l,n,f_3,f_4)$ be a real frame field, possibly only defined on some subset of $M$, such that with respect to this frame the metric takes the form
\begin{equation*}
    \begin{pmatrix}
        0 & -1 & 0 & 0 \\ -1 & 0 & 0 & 0 \\ 0 & 0 & 1 & 0 \\ 0 & 0 & 0 & 1
    \end{pmatrix} \;.
\end{equation*}
We now complexify the tensor bundles over $M$ and extend the metric $g$, Levi-Civita connection $\nabla$, and curvature tensor $R$ to those complex bundles such that they are complex-linear. Define the complex vector $m := \frac{1}{ \sqrt{2}}(f_3 + if_4)$ and the frame field $z_1 := l$, $z_2 := n$, $z_3 := m$, $z_4 := \overline{m}$. The overline stands for complex conjugation. We thus obtain
\begin{equation} \label{EqNormalisationNullFrameNP}
    g(z_a, z_b) = \begin{pmatrix}
        0 & -1 & 0 & 0 \\ -1 & 0 & 0 & 0 \\ 0 & 0 & 0 & 1 \\ 0 & 0 & 1 & 0
    \end{pmatrix} \;.
\end{equation}
When the frame fields act as derivatives on scalar functions we denote them by
\begin{equation*}
    l =: \dD \;, \qquad n =: \dDel \;, \qquad m =: \ddel \;, \qquad \overline{m} =: \ov{\ddel} \;.
\end{equation*}
Our convention for the Riemann curvature tensor is the same as in \cite{NP62}:
\begin{equation*}
    R_{abcd} = g\big( \nabla_{z_c}(\nabla_{z_d} z_b) - \nabla_{z_d} ( \nabla_{z_c} z_b) - \nabla_{[z_c,z_d]} z_b, z_a\big) \;.
\end{equation*}
\emph{Since, however, our signature of $g$ is $-1$ times that of \cite{NP62}, we define the connection coefficients and curvature components with an additional minus sign when compared to \cite{NP62} -- in this way the null structure equations and Bianchi equations from \cite{NP62} keep their validity.} We have
\begin{equation} \label{EqRicciCoeff}
\begin{aligned}
&\rho = - g(\nabla_{\ov{m}} l, m) \qquad &&\mu = g(\nabla_m n , \ov{m})\\
&\sigma = - g(\nabla_m l, m) && \lambda = g(\nabla_{\ov{m}} n, \ov{m}) \\
&\kappa = -g(\nabla_l l, m) \qquad &&\nu = g(\nabla_n n, \ov{m}) \\
&\tau = - g(\nabla_nl,m)  &&\pi = g(\nabla_ln, \ov{m}) \\
&\varepsilon = \frac{1}{2} \big[ g(\nabla_lm, \ov{m}) - g(\nabla_ll,n) \big] \qquad &&\gamma = \frac{1}{2} \big[g(\nabla_n m, \ov{m}) - g(\nabla_n l, n) \big] \\
&\alpha = \frac{1}{2} \big[ g(\nabla_{\ov{m}} m, \ov{m}) - g(\nabla_{\ov{m}} l, n) \big] &&\beta = \frac{1}{2}\big[ g(\nabla_m m, \ov{m}) - g(\nabla_ml,n) \big] 
\end{aligned}
\end{equation}
and
\begin{equation} \label{EqCurvatureNP}
\begin{aligned}
    &\psi_0 = R(l,m,l,m) \qquad &&\psi_1 = R(l,n,l,m) \qquad &&&\psi_2 = R(l,m,\ov{m},n) \\
    &\psi_3 = R(l,n,\ov{m},n) &&\psi_4 = R(n,\ov{m}, n, \ov{m}) \;.
\end{aligned}
\end{equation}

\subsubsection{Relating the NP formalism to the covariant horizontal formalism}

In this section we augment the NP formalism by GHP-like operators to emulate the covariant horizontal formalism (see Section \ref{SecCK}) within the complex NP formalism.

\begin{definition}
    Let $k \in \N$. A $k$-covector field $\varpi$ on $M$ is called \emph{horizontal} iff  $\varpi( \ldots, l, \ldots) = 0 = \varpi(\ldots, n, \ldots)$. In other words, $\varpi$ is horizontal if and only if it is annihilated by inserting $l$ or $n$ into any of its slots.
\end{definition}

The following relations are a direct consequence of the definition \eqref{EqRicciCoeff} of the Ricci coefficients in the NP-formalism:\footnote{The last two relations in the right column are just complex conjugates of the first two.}
\begin{equation} \label{EqRelRicci}
\begin{aligned}
    &g(\nabla_l m, \ov{m}) = \varepsilon - \ov{\varepsilon} \qquad  &&g(\nabla_{\ov{m}} m , \ov{m}) = \alpha - \ov{\beta} \\
    &g(\nabla_l l, n) = -(\varepsilon + \ov{\varepsilon}) &&g(\nabla_{\ov{m}} l, n) = -(\alpha + \ov{\beta}) \\
    &g(\nabla_n m, \ov{m}) = \gamma - \ov{\gamma} &&g(\nabla_m m, \ov{m}) = \beta - \ov{\alpha} \\
    &g(\nabla_n l, n) = -( \gamma + \ov{\gamma}) &&g(\nabla_m l ,n) = - ( \ov{\alpha} + \beta) \;.
\end{aligned}
\end{equation}

\begin{definition}\label{def:Dr}
    Let $r \in \Z$. We define
    \begin{align*}
        &\dD_r := \dD - r(\varepsilon - \ov{\varepsilon}) \qquad &&\dDel_r := \dDel - r(\gamma - \ov{\gamma}) \\
        &\ddel_r := \ddel - r(\beta - \ov{\alpha}) && \ov{\ddel}_r := \ov{\ddel} - r(\alpha - \ov{\beta}) \;.
    \end{align*}
\end{definition}

\begin{remark}[Relation to GHP formalism]
    The above defined operators exactly correspond to the thorn and eth operators defined in the GHP formalism ((2.14) in \cite{GHP73}) \emph{if they act on quantities of type $(r,-r)$ in that language.} If the above defined operators act on quantities of a different type, they do not agree with the thorn and eth operators of the GHP formalism. 
    
    The above definitions correspond to the framework of horizontal covector fields (see Lemma \ref{LemNPHorizontal} below) and are in this sense intermediate between the NP and the GHP formalism. Indeed, one may have denoted the Ricci coefficients in a more suggestive way with underlines (see for example the appendix of \cite{IoKlai09}) to make this correspondence even clearer. Here, however, we have decided to stick to the original terminology of Newman and Penrose so that we can directly refer to their set of equations.
\end{remark}

\begin{lemma}
    \label{LemNPHorizontal}
    Let $\varpi$ be a horizontal $k$-covector field and let $ k_1 \in \N$, $0 \leq k_1 \leq k$. Let $(\mathfrak{m}_1, \ldots, \mathfrak{m}_k)$ denote any permutation of $(\underbrace{\ov{m}, \ldots, \ov{m}}_{k_1}, \underbrace{m, \ldots, m}_{k - k_1})$ and set $r := k - 2k_1$. We then have
    \begin{align*}
        &\dD_r \big[ \varpi(\mathfrak{m}_1, \ldots, \mathfrak{m}_k)\big] = (\nabla_l \varpi)(\mathfrak{m}_1, \ldots, \mathfrak{m}_k) \qquad &&\dDel_r\big[ \varpi(\mathfrak{m}_1, \ldots, \mathfrak{m}_k)\big] = (\nabla_n \varpi)(\mathfrak{m}_1, \ldots, \mathfrak{m}_k) \\
        &\ddel_r \big[ \varpi (\mathfrak{m}_1, \ldots, \mathfrak{m}_k) \big]= (\nabla_m \varpi)(\mathfrak{m}_1, \ldots, \mathfrak{m}_k) && \ov{\ddel}_r\big[\varpi(\mathfrak{m}_1, \ldots, \mathfrak{m}_k)\big] = (\nabla_{\ov{m}}\varpi)(\mathfrak{m}_1, \ldots, \mathfrak{m}_k) \;.
    \end{align*}
\end{lemma}

\begin{proof}
    We compute
    \begin{equation*}
    \begin{split}
        (\nabla_l \varpi)(\mathfrak{m}_1, \ldots, \mathfrak{m}_k) &= l \big( \varpi(\mathfrak{m}_1, \ldots, \mathfrak{m}_k) \big) - \varpi( \nabla_l \mathfrak{m}_1, \mathfrak{m}_2, \ldots, \mathfrak{m}_k) - \ldots - \varpi(\mathfrak{m}_1, \ldots, \mathfrak{m}_{k-1}, \nabla_l \mathfrak{m}_k) \\
        &= \dD \big[ \varpi(\mathfrak{m}_1, \ldots, \mathfrak{m}_k) \big] - \Big(k_1 \cdot g(\nabla_l \ov{m},m) + (k-k_1) \cdot g(\nabla_l m ,\ov{m}) \Big) \varpi(\mathfrak{m}_1, \ldots, \mathfrak{m}_k) \\
        &= \dD_r \big[ \varpi(\mathfrak{m}_1, \ldots, \mathfrak{m}_k)\big] \;,
    \end{split}
    \end{equation*}
    where we have used that $\varpi$ is horizontal, $g(\nabla_l m, m) = 0 = g(\nabla_l \ov{m}, \ov{m})$, and $g(\nabla_l m, \ov{m}) = - g(\nabla_l \ov{m}, m)$.
    The proof of the other identities is analogous.
\end{proof}

\subsubsection{Derivation of nonlinear Teukolsky equation} \label{SecNonLinearTeukolsky}

\begin{proposition} \label{PropDeriNonLinearTeuk}
    The following equation holds for solutions to the vacuum Einstein equations:
    \begin{equation} \label{EqNonlinearTeuk}
        \begin{split}
            &(\ddel - 3 \beta - \ov{\alpha} + \ov{\pi} - 4 \tau)(\ov{\ddel} - 4 \alpha + \pi) \psi_0 -(\dD - 3 \varepsilon + \ov{\varepsilon} - 4 \rho - \ov{\rho})(\dDel - 4 \gamma + \mu) \psi_0 + 3 \psi_0 \psi_2 \\
&\quad + 4 \Big( \ov{\ddel}(\sigma \psi_1) - 3 \sigma \psi_1 ( \alpha - \ov{\beta}) \Big)  - 4 \Big( \dDel ( \kappa \psi_1) - 2 \kappa \psi_1 ( \gamma - \ov{\gamma}) \Big) + \sigma \psi_1 \big[ -8( \alpha + \ov{\beta}) - 4 \ov{\tau} \big] \\ 
&\quad + \kappa \psi_1 \big[ 12(\gamma + \ov{\gamma}) - 4 \ov{\mu} \big]   + 3\psi_0 (\nu \kappa - \lambda \sigma)  - 10 \psi_1^2 \\
&= 0.
        \end{split}
    \end{equation}
\end{proposition}

It might be helpful for the reader to already keep in mind that on exact Kerr, in the algebraically special frame,  the analogs of $\kappa$, $\sigma$, $\nu$, $\lambda$, $\psi_0$, $\psi_1$, $\psi_3$, and $\psi_4$ all vanish identically. Thus the second and third lines in \eqref{EqNonlinearTeuk} will become error terms. The terms $\alpha - \ov{\beta}$ and $\gamma - \ov{\gamma}$ have been added to the derivatives so that those terms correspond to horizontal covariant derivatives, see Lemma \ref{LemNPHorizontal}. Note that all other Ricci coefficients appearing in the second and third lines arise from horizontal covector fields.

\begin{proof}
We follow the original derivation of the (linear) Teukolsky equation in \cite{Teu73} but keeping nonlinear terms.
We need the following null-structure equations
\begin{align}
    \dD \tau - \dDel \kappa &= (\tau + \overline{\pi}) \rho + \sigma( \overline{\tau} + \pi) + \tau (\varepsilon - \overline{\varepsilon}) - \kappa ( 3 \gamma + \overline{\gamma}) + \psi_1 \label{EqNS1} \\
    \dD \beta - \ddel \varepsilon &= \sigma ( \alpha + \pi) + \beta ( \overline{\rho} - \overline{\varepsilon}) - \kappa(\mu + \gamma) - \varepsilon(\overline{\alpha} - \overline{\pi}) + \psi_1 \label{EqNS2} \\
    \ddel \rho - \overline{\ddel} \sigma &= \rho ( \overline{\alpha} + \beta) - \sigma(3 \alpha - \overline{\beta}) +( \rho - \overline{\rho}) \tau + ( \mu - \ov{\mu}) \kappa - \psi_1 \label{EqNS3} 
\end{align}
which are equations (4.2c), (4.2e), and (4.2k) from \cite{NP62}. We also need the commutation relation
\begin{equation}
    \label{EqComRel}
    (\ddel \dD - \dD \ddel) = (\ov{\alpha} + \beta - \ov{\pi})\dD + \kappa \dDel - \sigma \ov{\ddel} - (\ov{\rho} + \varepsilon - \ov{\varepsilon}) \ddel
\end{equation}
which is equation (4.4) in \cite{NP62}. These equations together imply the modified commutation relation 
\begin{equation} \label{EqFirstStepTeuk}
    \begin{split}
        (\dD - &3 \varepsilon + \ov{\varepsilon} - 4 \rho - \ov{\rho})(\ddel - 2 \beta - 4 \tau) - ( \ddel - 3 \beta - \overline{\alpha} + \overline{\pi} - 4 \tau)(\dD - 2 \varepsilon - 4 \rho) \\
        &= - \kappa \dDel + \sigma \ov{\ddel} + \sigma( -14 \alpha - 6 \pi - 4 \ov{\tau} + 4 \ov{\beta}) + \kappa(6 \mu + 14 \gamma + 4 \ov{\gamma} - 4 \ov{\mu}) - 4(\dDel \kappa) + 4 (\ov{\ddel} \sigma) - 10 \psi_1 \;.
    \end{split}
\end{equation}

Furthermore we recall the Bianchi equations
\begin{align}
    \dD \psi_2 - \ov{\ddel}\psi_1 &= - 2 \kappa \psi_3 + 3 \rho \psi_2 - (-2 \pi + 2 \alpha) \psi_1 - \lambda \psi_0 \label{EqBi1} \\
    \dDel \psi_1 - \ddel \psi_2 &= \nu \psi_0 + (2 \gamma - 2 \mu) \psi_1 - 3 \tau \psi_2 + 2 \sigma \psi_3 \label{EqBi2}
\end{align}
as well as
\begin{align}
    (\ov{\ddel} - 4 \alpha + \pi) \psi_0 - (\dD - 4 \rho - 2 \varepsilon) \psi_1 - 3 \kappa \psi_2 &=0 \label{EqBi3} \\
    (\dDel - 4 \gamma + \mu) \psi_0 - ( \ddel - 4 \tau - 2 \beta) \psi_1 - 3 \sigma \psi_2 &= 0 \label{EqBi4}
\end{align}
which are equations (4.5) in \cite{NP62} and the null structure equation
\begin{equation}
    \label{EqNS4}
    (\dD - \rho - \ov{\rho} - 3 \varepsilon + \ov{\varepsilon}) \sigma - ( \ddel - \tau + \ov{\pi} - \ov{\alpha} - 3 \beta) \kappa - \psi_0 = 0
\end{equation}
which is equation (4.2b) in \cite{NP62}.
We now multiply equation \eqref{EqNS4} by $\psi_2$ and use equations \eqref{EqBi1} and \eqref{EqBi2} to bring $\psi_2$ inside the $\dD$ and $\ddel$ derivatives to obtain
\begin{equation} \label{EqSecondStepTeuk}
    \begin{split}
        (\dD - &4 \rho - \ov{\rho} - 3 \varepsilon + \ov{\varepsilon}) \psi_2 \sigma - ( \ddel - 4 \tau + \ov{\pi} - \ov{\alpha} - 3 \beta) \psi_2 \kappa - \psi_2 \psi_0 \\
        &\quad - \big[ (\ov{\ddel} \psi_1) - 2 \kappa \psi_3 - (-2 \pi + 2 \alpha) \psi_1 - \lambda \psi_0 \big] \sigma + \big[ (\dDel \psi_1) - \nu \psi_0 + (2 \mu - 2 \gamma) \psi_1 - 2 \sigma \psi_3  \big] \kappa \\
        &=0 \;.
    \end{split}
\end{equation}
Next, we act with $(\ddel - 3 \beta - \ov{\alpha} + \ov{\pi} - 4\tau)$ on \eqref{EqBi3} and with $(\dD - 3 \varepsilon + \ov{\varepsilon} - 4 \rho - \ov{\rho})$ on \eqref{EqBi4} and subtract these equations to obtain
\begin{equation}
    \begin{split}
        0 &= (\ddel - 3 \beta - \ov{\alpha} + \ov{\pi} - 4\tau) (\ov{\ddel} - 4 \alpha + \pi) \psi_0 - \uwave{(\ddel - 3 \beta - \ov{\alpha} + \ov{\pi} - 4\tau)(\dD - 4 \rho - 2 \varepsilon) \psi_1} \\
        &\qquad -\dashuline{(\ddel - 3 \beta - \ov{\alpha} + \ov{\pi} - 4\tau)(3 \kappa \psi_2)} \\
        &\quad - (\dD - 3 \varepsilon + \ov{\varepsilon} - 4 \rho - \ov{\rho})(\dDel - 4 \gamma + \mu) \psi_0 + \uwave{(\dD - 3 \varepsilon + \ov{\varepsilon} - 4 \rho - \ov{\rho}) ( \ddel - 4 \tau - 2 \beta) \psi_1} \\
        &\qquad + \dashuline{(\dD - 3 \varepsilon + \ov{\varepsilon} - 4 \rho - \ov{\rho})(3 \sigma \psi_2)}
    \end{split}
\end{equation}
We now use \eqref{EqFirstStepTeuk} applied to $\psi_1$ to replace the wavily underlined terms and \eqref{EqSecondStepTeuk} to replace the terms with a dashed underline. This yields
\begin{equation*} 
        \begin{split}
            0&=(\ddel - 3 \beta - \ov{\alpha} + \ov{\pi} - 4 \tau)(\ov{\ddel} - 4 \alpha + \pi) \psi_0 -(\dD - 3 \varepsilon + \ov{\varepsilon} - 4 \rho - \ov{\rho})(\dDel - 4 \gamma + \mu) \psi_0 + 3 \psi_0 \psi_2 \\
&\quad + 3 \Big( \big[ (\ov{\ddel} \psi_1) - 2 \kappa \psi_3 -(-2 \pi + 2 \alpha) \psi_1 - \lambda \psi_0 \big] \sigma - \big[(\dDel \psi_1) - \nu \psi_0 + (2 \mu - 2 \gamma) \psi_1 - 2 \sigma \psi_3 \big] \kappa \Big) \\
&\quad - \kappa \dDel \psi_1 + \sigma \ov{\ddel} \psi_1 + \sigma (-14 \alpha - 6 \pi - 4 \ov{\tau} + 4 \ov{\beta}) \psi_1 + \kappa(6 \mu + 14 \gamma + 4 \ov{\gamma} - 4 \ov{\mu}) \psi_1 -4(\dDel \kappa) \psi_1 + 4(\ov{\ddel} \sigma) \psi_1 - 10 \psi_1^2 
        \end{split}
    \end{equation*}
Reordering the terms and adding and subtracting $8 \sigma \psi_1 \ov{\beta}$ as well as $8 \kappa \psi_1 \ov{\gamma}$ gives \eqref{EqNonlinearTeuk}.
\end{proof}


Finally, using the notation introduced in Definition \ref{def:Dr},  we define
\begin{equation}\label{eq:mathfrak.T.def}
    \mathfrak{T}_{[2]} := \Big(\ddel_1 - 2 (\beta + \ov{\alpha}) + \ov{\pi} - 4 \tau\Big)\Big(\ov{\ddel}_2 - 2 (\alpha + \ov{\beta}) + \pi\Big)  -\Big(\dD_2 -  (\varepsilon + \ov{\varepsilon}) - 4 \rho - \ov{\rho}\Big)\Big(\dDel_2 - 2( \gamma + \ov{\gamma}) + \mu\Big)   + 3  \psi_2
\end{equation}
and
\begin{equation}\label{eq:mathfrak.N.def}
    \mathfrak{N} := 4  \ov{\ddel}_3(\sigma \psi_1)   - 4  \dDel_2 ( \kappa \psi_1) + \sigma \psi_1 \big[ -8( \alpha + \ov{\beta}) - 4 \ov{\tau} \big]  + \kappa \psi_1 \big[ 12(\gamma + \ov{\gamma}) - 4 \ov{\mu} \big]   + 3\psi_0 (\nu \kappa - \lambda \sigma)  - 10 \psi_1^2 
\end{equation}
so that the nonlinear Teukolsky equation \eqref{EqNonlinearTeuk} takes the form 
\begin{equation} \label{EqShortHandNonLinearTeuk}
0 =  \mathfrak{T}_{[2]} \psi_0 + \mathfrak{N} \;.
\end{equation}
Already keeping the application to perturbation of Kerr in mind, the term $\mathfrak{N}$ corresponds to error terms.


\subsubsection{The background quantities}

Recalling the definition of the principal null frame \eqref{eq:Teukolsky.in.BL} on the background Kerr spacetime $(\mathcal{M}_{\mathrm{Kerr}}, \pmb{g})$ we define the complex NP principal null frame
\begin{equation}\label{eq:principal.null.NP}
        \pmb{l}:= \pmb{e}_4, \qquad  \pmb{n}:= \frac{1}{2} \pmb{e}_3, \qquad 
        \pmb{m}:=  \frac{1}{\sqrt{2}}\cdot \frac{\sqrt{\Sigma}}{r + i a \cos \theta}(\pmb{e}_1 + i \cdot \pmb{e}_2) \;.
\end{equation}
Here, and throughout the paper we use the convention that background quantities are denoted in bold. We also define $\mathfrak{c} := \frac{\sqrt{\Sigma}}{r + i a \cos \theta}$. 
Note that the null frame $(\pmb{l}, \pmb{n}, \pmb{m}, \ov{\pmb{m}})$ satisfies the normalisation \eqref{EqNormalisationNullFrameNP} of the NP formalism. The Ricci coefficients \eqref{EqRicciCoeff} of the background $(\mathcal{M}_{\mathrm{Kerr}}, \pmb{g})$ with respect to this null frame take the values
\begin{align*}
    &\pmb{\rho} = - \frac{\Delta}{r - ia \Co}     \qquad &&\pmb{\mu} =- \frac{1}{2 \Sigma (r - ia \Co)}  \qquad 
    &&&\pmb{\tau} = - \frac{ia \Si}{\sqrt{2} \cdot \Sigma}    \\
    &\pmb{\pi} = \frac{ia \Si}{\sqrt{2} \cdot (r -ia \Co)^2} \qquad &&\pmb{\beta} = \frac{\cot \theta}{2 \sqrt{2} \cdot (r + ia \Co)}  \qquad &&&\pmb{\alpha} = \pmb{\pi} - \pmb{\ov{\beta}}    \\
    &\pmb{\gamma} = - \frac{1}{2 \Sigma \cdot (r -ia \Co)} \qquad &&\pmb{\varepsilon} = r - M
\end{align*} and
\begin{equation}\label{eq:vanishing.background.NP.quantities}
    \pmb{\kappa}= \pmb{\sigma} = \pmb{\lambda} = \pmb{\nu} = 0 \;,
\end{equation}
where the expressions are given in Boyer--Lindquist coordinates. These values can be inferred from those given in \cite{ChandBlackHole} by taking the null rotation with a factor $\Delta$ of the frame into account. Furthermore, the curvature components \eqref{EqCurvatureNP} take the values
\begin{equation}\label{eq:background.NP.curvature}
    \pmb{\psi}_2 = - \frac{M}{(r - ia \Co)^3}  \quad \textnormal{and} \quad \pmb{\psi}_0 = \pmb{\psi}_1 = \pmb{\psi}_3 = \pmb{\psi}_4 = 0 \;.
\end{equation}
The Teukolsky operators $\pmb{\mathcal{T}}_{[2]}$ and $\pmb{\mathfrak{T}}_{[2]}$ on Kerr are defined by
\begin{equation} \label{EqDefTeukOperator}
\begin{split}
    \frac{1}{2 \Sigma}\pmb{\mathcal{T}}_{[2]} := \pmb{\mathfrak{T}}_{[2]} :=   &\Big(\pmb{\ddel}_1 - 2 (\pmb{\beta} + \ov{\pmb{\alpha}}) + \ov{\pmb{\pi}} - 4 \pmb{\tau}\Big)\Big(\pmb{\ov{\ddel}}_2 - 2 (\pmb{\alpha} + \ov{\pmb{\beta}}) + \pmb{\pi}\Big) \\
    &\quad -\Big(\pmb{\dD}_2 -  (\pmb{\varepsilon} + \ov{\pmb{\varepsilon}}) - 4 \pmb{\rho} - \ov{\pmb{\rho}}\Big)\Big(\pmb{\dDel}_2 - 2( \pmb{\gamma} + \ov{\pmb{\gamma}}) + \pmb{\mu}\Big) 
     + 3 \pmb{\psi}_2 \;.
\end{split}
\end{equation}
In Boyer--Lindquist coordinates, and acting on a spin 2-weighted  function $f$, it takes the form
\begin{equation} \label{EqTeukolskyEquationBL}
\begin{split}
\pmb{\mathcal{T}}_{[s]} f := &-\Big[\frac{(r^2 + a^2)^2}{\Delta} - a^2 \sin^2 \theta\Big] \partial_t^2 f - \frac{4Mar}{ \Delta} \partial_t \partial_\varphi f - \Big[\frac{a^2}{\Delta} - \frac{1}{\sin^2 \theta}\Big] \partial^2_\varphi f \\ 
&+ \Delta^{-s} \partial_r (\Delta^{s+1} \partial_r f) + \frac{1}{ \sin \theta}\partial_\theta(\sin \theta \partial_\theta f) + 2s\Big[\frac{a(r-M)}{\Delta} + \frac{i \cos\theta}{\sin^2\theta}\Big] \partial_\varphi f \\ 
&+ 2s\Big[\frac{M(r^2 -a^2)}{\Delta} - r - ia\cos \theta\Big] \partial_t f -\Big[\frac{s^2 \cos^2 \theta}{\sin^2\theta} +s\Big] f -4s(r-M) \partial_r f = 0 
\end{split}
\end{equation}
with $s = +2$. This agrees with (2.12) in \cite{Sbie23} and with \eqref{EqTeukolskyStarCoordinates} in $(v_+, r, \theta, \varphi_+)$ coordinates.

\subsubsection{Transformation of NP quantities under change of frame}

\begin{proposition} \label{PropTrafoNPQuant}
    Let $(l,n,m, \ov{m})$ be an NP-frame defined on a subset $D \subseteq M$, let $\varphi \in C^\infty(D,\R)$ and let $(l' := l, n' := n, m':= e^{i \varphi} m, \ov{m}':= e^{-i \varphi} \ov{m})$ be another NP-frame in $D$. The NP quantities with respect to the second frame are denoted with a prime. Let $f \in C^\infty(D, \C)$ and $r \in \Z$. We then have
    \begin{equation*}
    \begin{aligned}
        &e^{(r-1)i \varphi} \overline{\ddel}_r f = \overline{\ddel}_r'(e^{ r i \varphi} f) \qquad \qquad &&e^{r i \varphi} \dD_r f = \dD_r' (e^{r i \varphi} f)    \\
        &e^{(r+1)i \varphi} \ddel_r f = \ddel_r' (e^{r i \varphi} f) &&e^{r i \varphi}\dDel_r f = \dDel_r' (e^{r i \varphi} f) \;.
        \end{aligned}
    \end{equation*}
\end{proposition}

\begin{proof}
    We only prove the first identity, the others follow similarly.
    \begin{equation*}
        \begin{split}
            \overline{\ddel}_r' (e^{r i \varphi} f) &= \big(\ov{m}' - r g(\nabla_{\ov{m}'}m', \ov{m}') \big) (e^{r i \varphi}f) = e^{- i \varphi} \big( \ov{m} - r g(\nabla_{\ov{m}} m, \ov{m}) - r i \ov{m}(\varphi) \big) (e^{r i \varphi} f) \\
            &= e^{(r-1) i \varphi} \big( \ov{m} - r g(\nabla_{\ov{m}} m , \ov{m}) \big) f = e^{(r-1)i \varphi} \overline{\ddel}_r f \;.
        \end{split}
    \end{equation*}
\end{proof}

\subsection{The CK formalism} \label{SecCK}

We briefly recall the horizontal formalism \cite{CK}. Assume that $(M,g)$ is oriented. We start with two null vectors $e_3$ and $e_4$ which satisfy $g(e_3, e_4) = -2$. The distribution spanned by these two vector fields is denoted by $\Pi := \mathrm{span}\{e_3, e_4\}$. The orthogonal distribution $\Pi^\perp$ is referred to as the horizontal space. The induced metric on $\Pi^\perp$ is denoted by $\slashed{g}$ and the induced volume form by $\slashed{\varepsilon}(\cdot, \cdot) := \frac{1}{2}\varepsilon (\cdot, \cdot, e_3, e_4)$, where $\varepsilon$ denotes the Lorentzian volume form on $M$. For ease of presentation, complement the two null vectors by an arbitrary choice of basis vectors $e_1$ and $e_2$ (denoted by $e_A$, $A \in \{1,2\})$ spanning $\Pi^\perp$ such that $(e_1, e_2, e_3, e_4)$ is an oriented frame field. (The vectors $(e_1, e_2)$ will be orthonormal in some settings, but we will not be always imposing that.) The connection coefficients are defined by 
\begin{equation}\label{eq:Ricci.def}
        \begin{aligned}
\chi_{AB}&= g(\nabla_{e_A} e_4,e_B),   & \chib_{AB}&=g(\nabla_{e_A} e_3,e_B),\\
\omega&= -\frac{1}{4} g(\nabla_{e_4} e_3,e_4),  & \omegab&=-\frac{1}{4} g(\nabla_{e_3} e_4,e_3),\\
\eta_A&=\frac{1}{2}g(\nabla_{e_3} e_4,e_A),    & \etab_A&=\frac{1}{2}g(\nabla_{e_4} e_3,e_A),\\
\xi_A&=\frac{1}{2}g(\nabla_{e_4} e_4,e_A),  &  \xib_A&=\frac{1}{2}g(\nabla_{e_3} e_3,e_A), \\
\zeta_A&= \frac{1}{2}g(\nabla_{e_A} e_4,e_3).
\end{aligned}
\end{equation}
and the curvature components are defined by
\begin{equation}\label{eq:curvature.def}
        \begin{aligned}
    \alpha_{AB} &= R(e_A, e_4, e_B, e_4), & \alphab_{AB}&= R(e_A, e_3, e_B, e_3), \\
    \beta_A &= \frac{1}{2} R(e_A, e_4, e_3, e_4), & \betab_{A}&= \frac{1}{2} R(e_A, e_3, e_3, e_4), \\
    \rho&= \frac{1}{4} R(e_4, e_3, e_4, e_3), & \sigma &= \frac{1}{4} {}^*R(e_4, e_3, e_4, e_3),
\end{aligned}
\end{equation}
where ${}^*R_{\mu \nu \lambda \sigma} = \varepsilon_{\mu \nu \alpha \beta}R^{\alpha \beta}_{\; \; \; \lambda \sigma}$ denotes the Hodge dual of $R$.

In what follows, when discussing a null frame in the CK formalism, we will use upper case Latin indices $A,B=1,2$ for the frame elements in $\Pi^\perp$, and lower case Greek indices $\mu,\nu=1,2,3,4$ for the full set of frame elements.

\subsubsection{The horizontal derivatives}\label{sec:horizontal.derivatives}

The quantities in \eqref{eq:Ricci.def} and \eqref{eq:curvature.def} are all defined so as to be tensorial in $(e_1,e_2)$. For such quantities, we define the horizontal derivatives $\nab_3 = \nab_{e_3}$, $\nab_4 = \nab_{e_4}$ and $\nab_A = \nab_{e_A}$ to be, respectively, the projection of the covariant derivative $\nabla_{e_3}$, $\nabla_{e_4}$, $\nabla_{e_A}$ to $\Pi^\perp$.

\subsubsection{All derivatives of the frame fields}

Observe that \eqref{eq:Ricci.def} only includes connection coefficients which are tensorial in $e_1,e_2$. In general, we also need the derivatives of $(e_1,e_2)$. We collect some computations here.
\begin{lemma}\label{lem:computation.nablaslashed}
    The following identities hold for $A,B=1,2$:
    \begin{align}
        g(\nabla_{e_3} e_A, e_B) = &\:  g([e_3, e_A], e_B) + \chib(e_A, e_B),\label{eq:nablaslashed.chib} \\
        g(\nabla_{e_4} e_A, e_B) = &\: g([e_4, e_A], e_B) + \chi(e_A, e_B), \label{eq:nablaslashed.chi}\\
        g(\nabla_{e_A} e_B, e_C) = &\: \mathfrak G_{ABC},\label{eq:nablaslashed.angular} 
    \end{align}
    where, assuming $A\neq B$,
    \begin{align}
        \mathfrak G_{AAA} =\f 12 e_A(g(e_A,e_A)),\quad  \mathfrak G_{BAA} = \f 12 e_B(g(e_A,e_A)), \label{eq:frkG.easy}\\
        \mathfrak G_{ABA} =  g\Big([e_A,e_B],e_A \Big) + \f 12 e_B(g(e_A,e_A)),\quad 
        \mathfrak G^{(\cdot)}_{AAB} = e_A(g(e_A,e_B)) - \mathfrak G_{ABA}.\label{eq:frkG.less.easy}
    \end{align}
\end{lemma}
\begin{proof}  
    The identities \eqref{eq:nablaslashed.chib} and \eqref{eq:nablaslashed.chi} follows from the fact that $\nabla$ is torsion-free. Next, \eqref{eq:frkG.easy} follows from the fact that $\nabla$ is metric compatible. We then compute
    \begin{align}
        g(\nabla_{e_A} e_B, e_A) = g([e_A,e_B],e_A) + g(\nabla_{e_B} e_A, e_A), \quad 
        g(\nabla_{e_A} e_A, e_B) = e_A(g(e_A,e_B))- g( \nabla_{e_A} e_B , e_A ),
    \end{align}
     which gives \eqref{eq:frkG.less.easy} after using computations above. \qedhere
\end{proof}

\subsection{The dictionary}

Let $(l,n,m, \ov{m})$ be the NP double null frame and let $e_3 = 2n$ and $e_4 = l$ be the double null frame of the horizontal formalism. Note that the vector field $m$ is a horizontal vector field.
Let $e_1, e_2$ be as in the CK formalism, which we now assume in addition to be \emph{orthonormal}. We assume without loss of generality that the orientation of $(e_1, e_2)$ is such that that there exists a complex scalar function $\mathfrak{c}$ of unit norm on $M$   such that   $m = \frac{1}{\sqrt{2}}\mathfrak{c}(e_1 +i e_2)$. We have $\tr \chi = \chi_{11} + \chi_{22}$ and $\slashed{\varepsilon} \cdot \chi = \chi_{12} - \chi_{21}$; and similarly for $\chib$.
 Quantities on the left-hand side denote the NP quantities defined in Section \ref{SecNP} while quantities on the right-hand side stand for those defined in Section \ref{SecCK}. The connection coefficients are related by
\begin{align*}
    \rho &= - \frac{1}{2} (\tr \chi + i \slashed{\varepsilon} \cdot \chi) & \mu &= \frac{1}{4}(\tr \chib - i \slashed{\varepsilon} \cdot \chib) \\
    \sigma &= - \chih(m,m) & \lambda &= \frac{1}{2} \chibh(\ov{m}, \ov{m}) \\
    \kappa &= -2 \xi(m) & \nu &= \frac{1}{2} \xib (\ov{m}) \\
    \tau &= - \eta(m) & \pi &= \etab(\mb) \\
    \varepsilon + \ov{\varepsilon} &= -2 \omega & \gamma + \ov{\gamma} &= \omegab \\
    \alpha + \ov{\beta} &= - \zeta(\mb) \;.
\end{align*}
while the curvature components are related by
\begin{align*}
    \psi_0 &= \alpha(m,m) & \psi_1 &= \beta(m) & \psi_2 &= \rho - \frac{i}{2} \sigma \\
    \psi_3 &= - \frac{1}{2} \betab (\mb) & \psi_4 &= \frac{1}{4} \alphab(\mb, \mb) \;.
\end{align*}

\section{Precise assumptions of Theorem \ref{thm:main} and reduction to main result proven in this paper}\label{sec:statement}

\subsection{Gauge conditions for the initial data and precise assumptions of the main theorem} \label{SecPreciseAssump}

We consider a characteristic initial value problem for the vacuum Einstein equations in the following gauge
\begin{equation*}
    g =- 2 (\ud s\otimes \ud\ub+ \ud\ub\otimes \ud s)+ 4f \ud\ub\otimes \ud\ub + \gamma_{AB}(\ud\vartheta_*^A-h^A \ud\ub)\otimes (\ud\vartheta_*^B-h^B \ud\ub)\;,
\end{equation*}
where $(s,\ub, \vartheta_*)$ are standard coordinates on $[0,s_f) \times [1, \infty) \times \Sp^2$, $f(s, \ub, \vartheta_*)$ is a  scalar function, $\gamma_{AB}$ a $(s, \ub)$-dependent Riemannian metric on $\Sp^2$, and $h$ a  $(s, \ub)$-dependent vector field on $\Sp^2$. The Riemannian volume form on $\Sp^2$ induced by $\gamma$ is denoted by $\volg$. We introduce the two null vectors
\begin{equation*}
    e_3:= \frac{\rd}{\rd s}, \qquad e_4 := \frac{\rd}{\rd \ub} + f \frac{\rd}{\rd s} + h^A \frac{\rd}{\rd \vartheta_*^A},
\end{equation*}
which satisfy $g(e_3,e_4) = -2$ and which are both orthogonal to $\frac{\rd}{\rd \vartheta_*^A}$ (see \cite{DafLuk26} for the computational details). Consider now the CK quantities \eqref{eq:Ricci.def} and \eqref{eq:curvature.def} with respect to this null frame. In this gauge one has
\begin{equation*}
    \omegab = 0, \qquad \zeta = \eta = - \etab, \qquad \xib = 0, \qquad \chi \textnormal{ and } \chib \textnormal{ are symmetric.}
\end{equation*}
Let $f|_{\{s=0\}} = 0$ and fix $0 < |a| < M$. We now assume that smooth characteristic initial data for the vacuum Einstein equations is given on the null hypersurfaces $\{\ub = 1\}$ and $\{s=0\}$ in the \emph{event horizon gauge} of \cite{DafLuk26}, i.e., with
\begin{equation*}
  h^{\theta_*}|_{\{s=0\}} = 0,\qquad    h^{\varphi_*}|_{\{s=0\}} = \frac{a}{Mr_+},\qquad  \omega|_{\{s=0\}} = - \kappa_+\;.
\end{equation*}
The exact Kerr metric in this coordinate system has been given in Section \ref{sec:Kerr.s.ub} and, as usual, we denote the geometric quantities of the background in bold\footnote{Note that on the event horizon $s=0$ we have (see \cite{DafLuk26})
\begin{equation*}
  \pmb{h}^{\theta_*}|_{\{s=0\}} = 0,\qquad    \pmb{h}^{\varphi_*}|_{\{s=0\}} = \frac{a}{Mr_+},\qquad  \pmb{\omega}|_{\{s=0\}} = - \kappa_+\;.
\end{equation*}} and difference quantities by a tilde, i.e., $\tilde{\phi} = \phi - \pmb{\phi}$. Furthermore, we introduce the following shorthand 
\begin{equation*}
\calS := \{ \widetilde{\gamma}, \widetilde{\gamma^{-1}}, \,\widetilde{f},\,\widetilde{h},\, \widetilde{\chib},\widetilde{\eta},\widetilde{\om},\, \widetilde{\chi},\widetilde{\xi}\}.
\end{equation*}
Let $I_0 \in \N$ be large, $q \in \mathbb Z_{\geq 7}$, and $q_- < q$ arbitrarily close to $q$, but fixed. We assume that the characteristic initial data is such that the following holds on $\{s=0\}$: First the stability estimate
\begin{equation} \label{EqAsDecRest}
                \sum_{i+j+k \leq I_0} \sum_{\widetilde{\phi} \in \calS} \int\limits_1^\infty \int\limits_{\Sp^2} \ub^{q_--2} |\nab_3^j \nab_4^k \nab^i \widetilde{\phi} |_\gamma^2\volg \ud \ub \leq \ep
\end{equation}
on $\Hp$, where we refer the reader to Section \ref{SecNorms} for the definition of the norm $| \cdot |_\gamma$.
Furthermore, our instability estimates 
\begin{equation} \label{EqAsDecAlpha}
                \sum_{i_1 + i_2 + i_3 +j+k \leq I_0}  \int\limits_1^\infty \int\limits_{\Sp^2} \ub^{q_-} |\rd_r^j \rd_{v_+}^k \wtZ_{1,+}^{i_1} \wtZ_2^{i_2} \wtZ_{3,+}^{i_3} \big(R(\pmb{e}_4, \pmb{m}, \pmb{e}_4, \pmb{m})\big)|^2 \volg \ud \ub \ls 1,
\end{equation}
\begin{equation}  \label{EqAsInst2}
            \begin{split}
                \int\limits_1^\infty \int\limits_{\Sp^2}  \ub^{q} |\Sp_{(2)}R(\pmb{e}_4, \pmb{m}, \pmb{e}_4, \pmb{m})|^2 \volg \ud \ub  = \infty\;,
            \end{split}
\end{equation}
\begin{equation} \label{EqAsInst3}
 \sum_{i_1 + i_2 + i_3 +j+k \leq I_0}  \int\limits_1^\infty \int\limits_{\Sp^2} \ub^{q} |\rd_r^j \rd_{v_+}^k \wtZ_{1,+}^{i_1} \wtZ_{2,+}^{i_2} \wtZ_{3,+}^{i_3} \Sp_{(>2)}\big(R(\pmb{e}_4, \pmb{m}, \pmb{e}_4, \pmb{m})\big)|^2 \volg \ud \ub \ls 1,
\end{equation}
\begin{equation} \label{EqAsInst4}
  \sum_{i_1 + i_2 + i_3 +j+k \leq I_0}  \int\limits_1^\infty \int\limits_{\Sp^2} \ub^{q} |\rd_r^j \rd_{v_+}^k \wtZ_{1,+}^{i_1} \wtZ_{2,+}^{i_2} \wtZ_{3,+}^{i_3}\rd_{v_+}\big(R(\pmb{e}_4, \pmb{m}, \pmb{e}_4, \pmb{m})\big)|^2 \volg \ud \ub \ls 1,
\end{equation}
where $R$ denotes the Riemann curvature of the initial data, the bold vector fields $\pmb{e}_4$ and $\pmb{m}$  are defined with respect to the background by \eqref{eq:Teukolsky.in.BL} and \eqref{EqDefMBackground}, the vector fields $\rd_r$, $\rd_{v_+}$, $\wtZ_{i,+}$ are with respect to background $(v_+,r, \theta, \varphi_+)$ coordinates, where $\wtZ_{i,+}$ are defined in \eqref{EqZTilde}, and  $\mathbb{S}_{(2)}$, $\mathbb{S}_{(>2)}$ 
denote the projection onto the $l = 2$ and $l >2$ spin $2$-weighted spherical harmonics with respect to the $(v_+,r, \theta, \varphi_+)$ coordinate system (cf.~Section \ref{sec:Teukolsky.first.mention}), respectively. 

And the following holds on $\{\ub = 1\}$:
\begin{equation} \label{EqAssumptionsTransversal}
            \begin{split}
                \sum_{i+j+k \leq I_0} \sum_{\widetilde{\phi} \in \calS} \int\limits_0^{s_f} \int\limits_{\Sp^2} | \nab_3^j \nab_4^k \nab^i \widetilde{\phi}|_\gamma^2 \volg \ud s \leq \ep.
            \end{split}
\end{equation}

\begin{remark}
To understand the compatibility of the decay rates assumed in \eqref{EqAsDecRest} and \eqref{EqAsDecAlpha} we observe the following: To begin with we assume \eqref{EqAsDecAlpha} and focus on the implied decay rate of $\chi$. By virtue of the event horizon gauge we have $|R(\pmb{e}_4, \pmb{m}, \pmb{e}_4, \pmb{m})| \sim |\alpha|_\gamma$ and $\xi = 0$ on $\Hp$. The latter implies that the null structure equations $\nab_4 \tr \chi + \frac{1}{2} (\tr \chi)^2 = - |\chih|^2 - 2 \omega \tr \chi$ and $\nab_4 \chih + \tr \chi \chih = - 2 \omega \chih - \alpha$ hold on $\Hp$. Recalling that $\omega = - \kappa_+$ in the event horizon gauge and $\os{\ee}{1}_4 = \frac{\rd}{\rd \ub} + \frac{a}{Mr_+}\frac{\rd}{\rd \varphi_*}$, we thus obtain
\begin{equation*}
    \frac{\rd}{\rd \ub}\underbrace{ \int_{S_{\ub}} \Big( |\chih|^2_\gamma + |\tr \chi|^2 \Big) \volg}_{=: \mathfrak{f}^2(\ub)} =  4 \kappa_+ \int_{S_{\ub}} \Big( |\chih|^2_\gamma + |\tr \chi|^2 \Big) \volg \underbrace{- 3 \int_{S_{\ub}} \tr \chi |\chih|^2_\gamma \, \volg  - 2 \int_{S_{\ub}} \langle \alpha, \chih \rangle_\gamma \, \volg}_{=:\mathfrak{G}} \;.
\end{equation*}
This gives
\begin{equation*}
    \mathfrak{f}^2(\ub) = e^{4 \kappa_+ \ub} \Big( e^{- 4\kappa_+} \mathfrak{f}^2(1) + \int_1^{\ub} e^{- 4 \kappa_+ \ub'} \mathfrak{G} (\ub') \, \ud \ub' \Big)\;.
\end{equation*}
Under the assumption that $\mathfrak{f}^2$ remains bounded, we thus obtain
\begin{equation*}
    \mathfrak{f}^2(\ub) = - e^{4 \kappa_+ \ub} \Big(\int_{\ub}^\infty e^{- 4 \kappa_+ \ub'} \mathfrak{G}(\ub') \, \ud \ub' \Big) \;.
\end{equation*}
Assuming further that $\sup\limits_{\Hp} |\tr \chi| < \epsilon$, Cauchy--Schwarz gives
\begin{equation*}
    \mathfrak{f}^2(\ub) \ls \int_{\ub}^\infty e^{4 \kappa_+(\ub - \ub')} \Big( \epsilon  \cdot \mathfrak{f}^2(\ub')  + \epsilon^{-1} \underbrace{\int_{S_{\ub'}} |\alpha|_\gamma^2 \volg}_{=:A^2(\ub')} \Big) \ud \ub'\,.
\end{equation*}

 We now compute with a similar integration by parts argument as in Lemma \ref{LemBlueDecayPoly}
 \begin{equation*}
     \begin{split}
    \int_1^\infty \ub^{q_-} \mathfrak{f}^2(\ub) \, \ud \ub &\ls \int_1^\infty \ub^{q_-} \frac{1}{4 \kappa_+} \big(\rd_{\ub} e^{4 \kappa_+ \ub}\big) \int_{\ub}^\infty e^{- 4 \kappa_+ \ub'}\big( \epsilon \cdot \mathfrak{f}^2(\ub') + \epsilon^{-1} A^2(\ub') \big)\, \ud \ub' \ud \ub \\
    &\leq \int_1^\infty \ub^{q_-} \frac{1}{4\kappa_+} \big(\epsilon \cdot\mathfrak{f}^2(\ub) + \epsilon^{-1} A^2(\ub) \big)\, \ud \ub \;.
     \end{split}
 \end{equation*}
 For $\epsilon>0$ small enough, we can thus absorb the first term on the right-hand side and we obtain
 \begin{equation*}
     \int_1^\infty \ub^{q_-} \mathfrak{f}^2(\ub) \, \ud \ub \ls \int_1^\infty \ub^{q_-} A^2(\ub) \, \ud \ub\;.
 \end{equation*}
Hence, we conclude that \emph{if $|\chi|_\gamma$ decays along $\Hp$}, then it inherits the same decay rate from $|\alpha|_\gamma$ --- we do not lose a power. However, when we consider for example the implied decay rate for $\widetilde{\gamma}$, we need to integrate $(\rd_{\ub} + \frac{a}{Mr_+} \rd_{\varphi_*}) \widetilde{\gamma}_{AB} = 2 \chi_{AB}$, which loses one power, cf.\ Lemma \ref{LemNoShiftDecay}, thus yielding the power $q_- -2$ in \eqref{EqAsDecRest}. The case for $\widetilde{\eta}$ is analogous, using the $\nab_4 \widetilde{\eta}$ null structure equation in combination with the Codazzi equation to replace $\tilde{\beta}$. When considering the decay rate of $\widetilde{\chib}$, we use the null structure equation $\nab_4 \widetilde{\chib} = 2 \omega \widetilde{\chib} + \ldots$, express $\tilde{\rho}$ in terms of the Gauss curvature $\tilde{K}$ on the spheres and note the presence of the red-shift term $\omega = - \kappa_+<0$ on the right-hand side, such that the same polynomial decay of the right-hand side is inherited. We omit the details.
\end{remark}

\begin{remark}\label{rmk:are.the.rates.reasonable}
    Conjecturally, when $q= 11$, our assumptions correspond to the decay rate for generic solutions within the class which are conformally regular at future null infinity. (Notice that this rate is slower than that predicted by Price's law, but is instead the modified rate suggested by work \cite{LO}.) When $q = 7$, they correspond to the decay rate for Gajic--Kehrberger type data; see \cite{dGlmaK2022}.
\end{remark}

\begin{remark}
    For initial data in the exterior region which satisfy suitable vector field bounds, one also expects that for the upper bound of $R(\pmb{e}_4, \pmb{m}, \pmb{e}_4, \pmb{m})$, every $\rd_{v_+}$ derivative gains a power of $\ub$ decay, i.e., 
\begin{equation*}
    \sum_{i_1 + i_2 + i_3 +j+k \leq I_0}  \int\limits_1^\infty \int\limits_{\Sp^2} \ub^{q_- + 2p} |\rd_r^j \rd_{v_+}^k \wtZ_{1,+}^{i_1} \wtZ_{2,+}^{i_2} \wtZ_{3,+}^{i_3}\rd_{v_+}^p\big(R(\pmb{e}_4, \pmb{m}, \pmb{e}_4, \pmb{m})\big)|^2 \volg \ud \ub \ls 1
\end{equation*}
    We do not need the full strength of this assumption, but note that it in particular justifies \eqref{EqAsInst4}.
\end{remark}

\subsection{The dynamical spacetime}\label{sec:dynamical.spacetime}

We recall the spacetimes $\os{\calM}{1}$, $\os{\calM}{2}$ (to be defined below) constructed in \cite{DafLuk17, DafLuk26} and then define the combined manifold $\calM$, which is a union of $\overset{\scriptscriptstyle{\text{[1]}}}{\calM}$ and $\overset{\scriptscriptstyle{\text{[2]}}}{\calM}$ with a suitable identification. We also recall the $C^0$ extension of the metric to the Cauchy horizon. 

Since we will define and use various frame fields in the discussion, the reader may find the glossary in Section~\ref{sec:glossary.frame} helpful.

\subsubsection{The red-shift region $\protect\overset{\scriptscriptstyle{\text{[1]}}}{\calM}$}\label{sec:M1}

Let $\overset{\scriptscriptstyle{\text{[1]}}}{\calM}$ be the manifold given by
$$\overset{\scriptscriptstyle{\text{[1]}}}{\calM} = \{(s,\ub,\vartheta_*): s \in (0,s_f), \ub \in [1, \infty), \vartheta_* \in \bbS^2\}$$
where $0< s_f \ll 1$. As in the case on Kerr, we will often denote the coordinates on the spheres by $\vartheta_* = (\vartheta_*^1, \vartheta_*^2) = (\th_*,\varphi_*)$.
Define $\calH^+ = \{s=0\}$ to be the event horizon. Denote $\os{\calM}{1}_{+}= \os{\calM}{1} \cup \calH^+$. 

On $\overset{\scriptscriptstyle{\text{[1]}}}{\calM}$, there is a Lorentizan metric $\overset{\scriptscriptstyle{\text{[1]}}}{g}$ which solves $\mathrm{Ric}(\overset{\scriptscriptstyle{\text{[1]}}}{g}) = 0$ and takes the form 
\begin{equation}\label{eq:metric.form.1}
    \overset{\scriptscriptstyle{\text{[1]}}}{g}=- 2 (\ud s\otimes \ud\ub+ \ud\ub\otimes \ud s)+ 4f \ud\ub\otimes \ud\ub+\gamma_{AB}(\ud\vartheta_*^A - h^A \ud\ub)\otimes (\ud\vartheta_*^B - h^B \ud\ub).    
\end{equation}
The metric is smooth and smoothly extends to $\calH^+$. In particular, the coordinate system $(s,\ub,\vartheta_*^A)$ obeys the following gauge conditions:
\begin{equation}\label{eq:transport.gauge.geometric}
    \overset{\scriptscriptstyle{\text{[1]}}}{g}{}^{-1}(\ud\ub,\ud\ub) = 0,\quad \overset{\scriptscriptstyle{\text{[1]}}}{g}{}^{-1}(\ud\ub,\ud\vartheta_*^A) = 0, \quad \overset{\scriptscriptstyle{\text{[1]}}}{g}{}^{-1}(\ud\ub, \ud s) = -\f 12.    
\end{equation}

The solution \eqref{eq:metric.form.1} is constructed in \cite{DafLuk26} by solving the vacuum Einstein equations starting from the event horizon $\calH^+ = \{s=0\}$ up till a spacelike hypersurface $\{s=s_f\}$. As in the setup in \cite{DafLuk26}, we require $f \geq 0$ with $f=0$ if and only if $s = 0$ so that $\{s=0\}$ is a null hypersurface and that the other constant-$s$ hypersurfaces with $s>0$ are spacelike. (Note that $\overset{\scriptscriptstyle{\text{[1]}}}{g}{}^{-1}(\ud s,\ud s) = -\f 12 f$.) The upper bound assumptions in Section~\ref{SecPreciseAssump} guarantee that \cite{DafLuk26} can be applied and that a solution exists.

We will use $|_s$ to denote derivatives in the $(s,\ub,\vartheta_*)$ coordinates. Associated to the metric $\overset{\scriptscriptstyle{\text{[1]}}}{g}$ in \eqref{eq:metric.form.1} is a null pair $(\overset{\scriptscriptstyle{\text{[1]}}}{\ee}_3,\overset{\scriptscriptstyle{\text{[1]}}}{\ee}_4)$ given in coordinates by
\begin{align}
    \overset{\scriptscriptstyle{\text{[1]}}}{\ee}_3 = \f{\rd}{\rd s}\Big|_s, \quad \overset{\scriptscriptstyle{\text{[1]}}}{\ee}_4 = \f{\rd}{\rd \ub}\Big|_s + f\f{\rd}{\rd s}\Big|_s + h^A\f{\rd}{\rd\vartheta_*^A}\Big|_s \label{eq:dynamical.ee.3.4.def}
\end{align}
so that $\overset{\scriptscriptstyle{\text{[1]}}}{\ee}_3 = -2 (\ud \ub)^\sharp$, $\overset{\scriptscriptstyle{\text{[1]}}}{g}(\overset{\scriptscriptstyle{\text{[1]}}}{\ee}_3, \overset{\scriptscriptstyle{\text{[1]}}}{\ee}_4)= -2$, and that both $\overset{\scriptscriptstyle{\text{[1]}}}{\ee}_3$ and $\overset{\scriptscriptstyle{\text{[1]}}}{\ee}_4$ are orthogonal to the constant-$(s,\ub)$ spheres.

\subsubsection{The blue-shift region $\protect\overset{\scriptscriptstyle{\text{[2]}}}{\calM}$}\label{sec:M2}

Let $\overset{\scriptscriptstyle{\text{[2]}}}{\calM}$ be a manifold with boundary defined by
$$\overset{\scriptscriptstyle{\text{[2]}}}{\calM} = \{(u,\ub,\vartheta_*): u \in (-\infty,u_f], \ub \in [1, \infty), u+ \ub \geq C_R, \vartheta_* \in \bbS^2 \},$$
where $u_f \leq -1$. Again, we denote $\vartheta_* = (\vartheta_*^1, \vartheta_*^2) = (\th_*,\varphi_*)$.

On $\overset{\scriptscriptstyle{\text{[2]}}}{\calM}$, the metric $\overset{\scriptscriptstyle{\text{[2]}}}{g}$ takes the form
\begin{equation}\label{eq:metric.form.2}
    \overset{\scriptscriptstyle{\text{[2]}}}{g}=-2\Omega^2(\ud u\otimes \ud\ub + \ud\ub\otimes \ud u)+\gamma_{AB}(\ud\vartheta_*^A-b^A \ud\ub)\otimes (\ud\vartheta_*^B-b^B \ud\ub).
\end{equation}
In particular, $(u,\ub,\vartheta_*^A)$ obeys the following gauge conditions:
\begin{equation}
    \overset{\scriptscriptstyle{\text{[2]}}}{g}{}^{-1}(\ud\ub,\ud\ub) = 0,\quad \os{g}{2}{}^{-1}(\ud\ub,\ud\vartheta_*^A) = 0, \quad \os{g}{2}{}^{-1}(\ud u,\ud u) = 0.    
\end{equation}

$(\overset{\scriptscriptstyle{\text{[2]}}}{\calM},\overset{\scriptscriptstyle{\text{[2]}}}{g})$ corresponds to the solution to the vacuum Einstein equations with initial data induced on a spacelike hypersurface $\Sigma \subset \os{\calM}{1}$. The solution is then shown to exist up till the Cauchy horizon. We specify the hypersurface $\Sigma \subset \os{\calM}{1}$ as follows:
\begin{itemize}
    \item Let $\pmb{u}: (0,\infty)\times (-\infty,\infty) \times \bbS^2$ be the function such that on the Kerr background, $\pmb{u}(s,\ub,\vartheta_*)$ corresponds to the double null $u$ coordinate in the $(u,\ub,\vartheta_*)$ coordinate system.
    \item Define the function $u_1:\os{\calM}{1}\to \bbR$ by setting 
    \begin{equation}\label{eq:u1.def}
        u_1 = \pmb{u}(s,\ub,\vartheta_*).
    \end{equation}
    \item Let $\Sigma = \{(s,\ub,\vartheta_*) \in \os{\calM}{1}: u_1 + \ub = C_R\}$, where we fix $C_R \in \bbR$ to be a sufficiently negative constant such that $\Sigma \subset \{ (s,\ub,\vartheta_*) \in \os{\calM}{1}: s \in (0,\f{s_f}2)\}$. That such a $C_R$ exists by is proven in Lemma~\ref{lem:can.initialize} below.
\end{itemize}

\begin{lemma}\label{lem:can.initialize}
    Let $\pmb{s}:\bbR\times \bbR\times \bbS^2 \to \bbR_{>0}$ be the inverse of $\pmb{u}$ for fixed $(\ub,\vartheta_*)$, i.e., $$\pmb{s}(\pmb{u}(s,\ub,\vartheta_*),\ub,\vartheta_*) = s.$$
    Then on the background Kerr spacetime, given any $s_f >0$, there exists $C_R \in \bbR$ sufficiently negative such that 
    $$ u+ \ub = C_R \implies 0< \inf_{\vartheta_* \in \bbS^2} \pmb{s}(u,\ub,\vartheta_*) \leq  \sup_{\vartheta_* \in \bbS^2} \pmb{s}(u,\ub,\vartheta_*) < \f{s_f}2 .$$
\end{lemma}
\begin{proof}
    Note that $\f{\rd}{\rd u}|_{(u,\ub,\vartheta_*)} + \f{\rd}{\rd \ub}|_{(u,\ub,\vartheta_*)}$ is tangential to both $\{u+\ub = \hbox{constant}\}$ and $\{\pmb{s} = \hbox{constant}\}$. (To show that $\f{\rd \pmb{s}}{\rd u}|_{(u,\ub,\vartheta_*)} + \f{\rd \pmb{s}}{\rd \ub}|_{(u,\ub,\vartheta_*)}$, we use Definition~\ref{def:s.Kerr} and the fact that $\f{\rd\pmb{\Omg}}{\rd u}|_{(u,\ub,\vartheta_*)} + \f{\rd\pmb{\Omg}}{\rd \ub}|_{(u,\ub,\vartheta_*)}=0$.) We thus only need to check the claim for any fixed value of $\ub$. Now for $\ub = 1$ and any $\vartheta_*$, we can take $C_R$ more negative so that $\pmb{s}(u,\ub,\vartheta_*) \in (0, \f{s_f}2)$. The desired claim, which is for all $\vartheta_* \in \bbS^2$, therefore follows from the compactness of $\bbS^2$.
\end{proof}

With the estimates proven in $\os{\calM}{1}$ (see Theorem~\ref{thm:DL.2} below), the data induced on $\Sigma$ approach that of a corresponding Kerr hypersurface sufficiently fast so that the main theorem of \cite{DafLuk17} apply; see \cite{DafLuk26}. This implies the existence of a solution in $\os{\calM}{1}$.

We will use $|_{DN}$ to denote derivatives in the $(u,\ub,\vartheta_*)$ coordinates. Associated to the metric $\overset{\scriptscriptstyle{\text{[2]}}}{g}$ in \eqref{eq:metric.form.2} is a null pair $(\overset{\scriptscriptstyle{\text{[2]}}}{\ee}_3,\overset{\scriptscriptstyle{\text{[2]}}}{\ee}_4)$ given in coordinates by
\begin{align}
    \overset{\scriptscriptstyle{\text{[2]}}}{\ee}_3 = \f{\rd}{\rd u}\Big|_{DN}, \quad \overset{\scriptscriptstyle{\text{[2]}}}{\ee}_4 = \Omg^{-2} \Big(\f{\rd}{\rd \ub}\Big|_{DN} + b^A\f{\rd}{\rd\vartheta_*^A} \Big|_{DN}\Big) \label{eq:dynamical.ee.2.3.4.def}
\end{align}
so that $\overset{\scriptscriptstyle{\text{[2]}}}{\ee}_3 = -2 \Omg^2 (\ud \ub)^\sharp$, $\overset{\scriptscriptstyle{\text{[2]}}}{g}(\overset{\scriptscriptstyle{\text{[2]}}}{\ee}_3, \overset{\scriptscriptstyle{\text{[2]}}}{\ee}_4)= -2$, and that both $\overset{\scriptscriptstyle{\text{[2]}}}{\ee}_3$ and $\overset{\scriptscriptstyle{\text{[2]}}}{\ee}_4$ are orthogonal to the constant-$(u,\ub)$ spheres.

\subsubsection{Identification and the combined manifold $\calM$}\label{sec:combined.manifold}

We now introduce $\calM$. To this end, we identify suitable subsets of  $\overset{\scriptscriptstyle{\text{[1]}}}{\calM}$ and $\overset{\scriptscriptstyle{\text{[2]}}}{\calM}$, which is carried out with the $(s,\ub,\vartheta_*)$ coordinates and thus we first need to introduce such coordinates on $\overset{\scriptscriptstyle{\text{[2]}}}{\calM}$.

Before we proceed, it is convenient to introduce $s_\Sigma:\bbR\times \bbS^2 \to \bbR_{>0}$ by the implicit relation 
\begin{equation}\label{eq:sSigma.def}
    \pmb{u}(s_\Sigma(\ub,\vartheta_*), \ub,\vartheta_*) = C_R -\ub,
\end{equation}
where $\pmb{u}$ and $C_R$ are as in Section~\ref{sec:M2}. Thus $s_\Sigma$ corresponds to the Kerr $s$-value for points on $\Sigma$. Notice that since $\pmb{u}$ is strictly increasing in $s$ for every fixed $(\ub,\vartheta_*)$, $s_\Sigma$ is well-defined. It is also easy to check that $s_\Sigma$ is a smooth function.

On $\overset{\scriptscriptstyle{\text{[2]}}}{\calM}$, for every fixed $(\ub,\vartheta_*)$, define a new variable $s$ by integrating the following ordinary differential equation
\begin{equation}\label{eq:s.def.in.M2}
    \f{\ud s}{\ud u}(u;\ub,\vartheta_*) = \Omg^{2}(u,\ub,\vartheta_*)
\end{equation}
with the initial condition for $s$ on $\Sigma$:
\begin{equation}\label{eq:initial.condition.for.s}
    s(C_R-\ub;\ub,\vartheta_*) = s_\Sigma(\ub,\vartheta_*).
\end{equation}

Consider now the subset $\overset{\scriptscriptstyle{\text{[2]}}}{\calU} \subset \overset{\scriptscriptstyle{\text{[2]}}}\calM$ given by
$$\overset{\scriptscriptstyle{\text{[2]}}}{\calU} = \{(u,\ub,\vartheta_*) \in \overset{\scriptscriptstyle{\text{[2]}}}\calM: s_\Sigma< s < s_f\}.$$
Define $\Psi: \overset{\scriptscriptstyle{\text{[2]}}}{\calU} \to \overset{\scriptscriptstyle{\text{[1]}}}\calM$ to be the mapping $(u,\ub,\vartheta_*)\mapsto (s(u,\ub,\vartheta_*),\ub,\vartheta_*)$. Due to \eqref{eq:s.def.in.M2} and the positivity of $\Omg^2$, the map $\Psi$ is injective. Moreover, noticing that
\begin{equation}
    \begin{pmatrix}
    \frac{\rd s}{\rd u} & \frac{\rd s}{\rd \ub} & \frac{\rd s}{\rd \theta_*} & \frac{\rd s}{\rd \varphi_*} \\
     \frac{\rd \ub}{\rd u} & \frac{\rd \ub}{\rd \ub} & \frac{\rd \ub}{\rd \theta_*} & \frac{\rd \ub}{\rd \varphi_*} \\  
     \frac{\rd \theta_*}{\rd u}   &\frac{\rd \theta_*}{\rd \ub}   &\frac{\rd \theta_*}{\rd \theta_*} &\frac{\rd \theta_*}{\rd \varphi_*} \\
     \frac{\rd \varphi_*}{\rd u}  &\frac{\rd \varphi_*}{\rd \ub}    &\frac{\rd \varphi_*}{\rd \theta_*} & \frac{\rd \varphi_*}{\rd \varphi_*} 
 \end{pmatrix} = \begin{pmatrix}
    \Omg^2 & \star & \star & \star \\
     0 & 1 & 0 & 0 \\  
     0   & 0    & 1 & 0 \\
     0  &0    &0 & 1 
 \end{pmatrix},
\end{equation} 
\begin{wrapfigure}{r}{0.45\textwidth}
 \centering
  \def\svgwidth{5.5cm}
   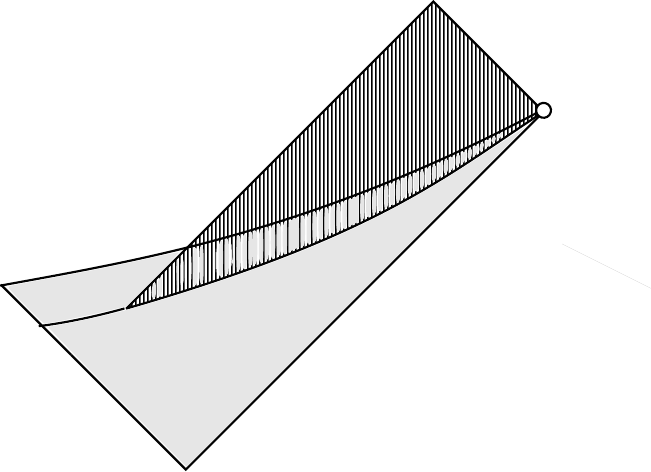 
   \caption{A Penrose-style diagram illustrating the regions $\calM$, $\os{\calM}{1}$, $\os{\calM}{2}$.}
   \label{FigRegions}
\end{wrapfigure}
(and that $\Omg^2>0$), it follows from the inverse function theorem that $\Psi$ is a diffeomorphism onto its image. Denote the image by $\overset{\scriptscriptstyle{\text{[1]}}}{\calU} \subset \overset{\scriptscriptstyle{\text{[1]}}}\calM$. Finally, notice that $s$ is chosen so that $\f{\rd}{\rd s}$ in the $(s,\ub,\vartheta_*)$ coordinates agree with $-2(\ud u)^\sharp$. Thus, we have put the metric in exactly the same gauge as \eqref{eq:metric.form.1}. Given that the metric also achieve the same geometric data on $\Sigma$, it follows that $\Psi^* \os{g}{1} = \os{g}{2}$.

Define now $\calM = \overset{\scriptscriptstyle{\text{[1]}}}\calM \cup \overset{\scriptscriptstyle{\text{[2]}}}\calM/\sim$, where $p_1 \in \overset{\scriptscriptstyle{\text{[1]}}}\calM$ and $p_2 \in \overset{\scriptscriptstyle{\text{[2]}}}\calM$ are identified if and only if $p_1 \in \overset{\scriptscriptstyle{\text{[1]}}}\calU$ and $p_2 \in \overset{\scriptscriptstyle{\text{[2]}}}\calU$, and $p_1 = \Psi(p_2)$.  Moreover, by considerations above, we can define $g$ on $\calM$, where $g\restriction_{\overset{\scriptscriptstyle{\text{[1]}}}\calM} = \overset{\scriptscriptstyle{\text{[1]}}}{g}$ and $g\restriction_{\overset{\scriptscriptstyle{\text{[2]}}}\calM} = \overset{\scriptscriptstyle{\text{[2]}}}{g}$. Denote also $\calM_+ = \calM \cup \calH^+$.

By the above, in the overlapping region $\overset{\scriptscriptstyle{\text{[1]}}}\calU = \Psi(\overset{\scriptscriptstyle{\text{[2]}}}\calU)$, which we will denote by $\calU$ from now on, we have both $(u,\ub,\th_*,\varphi_*)$ and $(s,\ub,\th_*,\varphi_*)$ coordinates. We compute the change of variables map in the following proposition:
\begin{proposition}\label{prop:change.s.to.u}
    In the region $\calU\subset \calM$, the change of coordinate maps and the metrics in the $(s,\ub,\th_*,\varphi_*)$ coordinates in \eqref{eq:metric.form.1} and in the $(u,\ub,\th_*,\varphi_*)$ coordinates in \eqref{eq:metric.form.2} relate as follows. The metric $\gamma$ agrees in both coordinate systems, and
    \begin{align}
    \f{\rd s}{\rd u} = &\: \Omg^2, \\
    \f{\rd s}{\rd \vartheta_*^A} = &\: \f 12 \gamma_{AB} (b^B - h^B) ,\\
    \f{\rd s}{\rd \ub} = &\: f - |\nab s|^2_\gamma - h^A \f{\rd s}{\rd \vartheta_*^A} = f - \f 14 |b - h|_\gamma^2 - \f 12 \langle b,h\rangle_\gamma + \f 12 |h|_\gamma^2.
\end{align}
\end{proposition}
\begin{proof}
    The expression for $\f{\rd s}{\rd u}$ follows from the definition of $s$. For the other derivatives, we first write
    $$\ud s = \Omg^2 \ud u + \f{\rd s}{\rd \ub} \ud \ub + \f{\rd s}{\rd \vartheta_*^A} \ud \vartheta_*^A.$$
We then compute the transformation of the metric from the $(s,\ub,\th_*,\varphi_*)$ coordinates to the $(u,\ub,\th_*,\varphi_*)$ coordinates:
\begin{equation}
    \begin{split}
        &\: - 2 (\ud s\otimes \ud\ub+ \ud\ub\otimes \ud s)+ 4f \ud\ub\otimes \ud\ub+\gamma_{AB}(\ud\vartheta_*^A - h^A \ud\ub)\otimes (\ud\vartheta_*^B - h^B \ud\ub) \\
        = &\: - 2 \Omg^2 (\ud u\otimes \ud\ub+ \ud\ub\otimes \ud u) - 4 \f{\rd s}{\rd \ub} \ud \ub \otimes \ud \ub - 2 \f{\rd s}{\rd \vartheta_*^A} (\ud \ub \otimes \ud \vartheta_*^A + \ud \vartheta_*^A \otimes \ud \ub) \\
        &\: + 4f \ud\ub\otimes \ud\ub+\gamma_{AB}(\ud\vartheta_*^A - h^A \ud\ub)\otimes (\ud\vartheta_*^B - h^B \ud\ub) \\
        = &\: - 2 \Omg^2 (\ud u\otimes \ud\ub+ \ud\ub\otimes \ud u) + 4 \Big(f- \f{\rd s}{\rd \ub} - |\nab s|_\gamma^2 - h^A \f{\rd s}{\rd \vartheta_*^A}\Big) \ud \ub \otimes \ud \ub  \\
        &\: +\gamma_{AB} \Big(\ud\vartheta_*^A - \Big(h^A + 2 (\gamma^{-1})^{AC} \f{\rd s}{\rd\vartheta_*^C} \Big)  \ud\ub \Big)\otimes \Big(\ud\vartheta_*^B - \Big(h^B + 2 (\gamma^{-1})^{BD} \f{\rd s}{\rd\vartheta_*^D}\Big) \ud\ub \Big).
    \end{split}
\end{equation}
Comparing this and the form of the metric in \eqref{eq:metric.form.2} yields the desired conclusion. \qedhere
\end{proof}

\subsubsection{The global coordinates $(u',\ub,\vartheta_*)$}\label{sec:u'def}

We continue to view $\overset{\scriptscriptstyle{\text{[1]}}}\calM,\, \overset{\scriptscriptstyle{\text{[2]}}}\calM \subset \calM$, and $\calU = \overset{\scriptscriptstyle{\text{[1]}}}\calM \cap \overset{\scriptscriptstyle{\text{[2]}}}\calM$ (see Section~\ref{sec:combined.manifold}). Define $u':\calM \to \bbR$ by
\begin{equation}\label{eq:u'.def}
    u' = 
    \begin{cases}
        u_1 &\hbox{on $\overset{\scriptscriptstyle{\text{[1]}}}\calM\setminus \calU$}\\
        u &\hbox{on $\overset{\scriptscriptstyle{\text{[2]}}}\calM \setminus \calU$}\\
        \chi(s) u_1 + (1-\chi)(s) u &\hbox{on $\calU$},
    \end{cases}
\end{equation}
where $u:\overset{\scriptscriptstyle{\text{[2]}}}\calM \to \bbR$ is the original coordinate function in \eqref{eq:metric.form.2}, $u_1:\overset{\scriptscriptstyle{\text{[1]}}}\calM\to \bbR$ is as in \eqref{eq:u1.def} and $\chi:[0,\infty) \to [0,1]$ is a cut-off function where $\chi(s) = 1$ for $s \leq \sup_{\ub,\vartheta_*} s_\Sigma(\ub,\vartheta_*)$ and $\chi(s) = 0$ for $s \geq \f{s_f}{2}$. (That such a cutoff function exists follows from Lemma~\ref{lem:can.initialize}.)

We emphasize that $u'$ is \underline{not} a null variable. The reader should think that $u'$ is close to both $u$ and $u_1$, and moreover that their differences decay as $\ub \to \infty$. See Lemma~\ref{lem:u.diff.main} for a precise statement.

\subsubsection{The Cauchy horizon coordinates}\label{sec:CH.coord}

We now introduce the coordinates $(u,\ub_{\CH^+}, \th^1_{(1),\CH^+}, \th^2_{(1),\CH^+})$ and $(u,\ub_{\CH^+}, \th^1_{(2),\CH^+}, \th^2_{(2),\CH^+})$ from \cite[Section~16]{DafLuk17} with which the metric extends continuously to the Cauchy horizon. When there is no need to specify the coordinate patch, we also use the notation $(u,\ub_{\CH^+}, \th^1_{\CH^+}, \th^2_{\CH^+}) = (u, \ub_{\CH^+}, \vartheta_{\CH^+})$ to mean either one of them.

First, $\ub_{\CH^+}$ is given by \eqref{eq:ubCH.thm}, i.e., $\ub_{\CH^+} = -\f 1{2\kappa_-} e^{-2\kappa_-\ub}$. 
To define the angular coordinates, we first recall the usual stereographic projection. On $\Sp^2 \subseteq \R^3$ we consider the standard $(\theta, \varphi)$ coordinates which are given by $(x = \sin \theta \cos \varphi, y = \sin \theta \sin \varphi, z= \cos \theta)$. Define $\mathcal{V}_1 := \Sp^2 \setminus \{\theta = 0\}$ and $\mathcal{V}_2 := \Sp^2 \setminus \{\theta = \pi\}$. On $\mathcal{V}_2$ we consider the stereographic projection from the south pole of $\Sp^2 \subseteq \R^3$, which associates to a point $\vartheta \in \Sp^2 \setminus \{\theta = \pi\}$ the Euclidean coordinates of the point of intersection of $\{z=0\} \subseteq \R^3$ with the straight line in $\R^3$ from the south pole to $\vartheta$. These stereographic coordinates we denote with $(\theta^1_{(2)}, \theta^2_{(2)}) \in \R^2$. Similarly one defines the stereographic projection from the north pole of $\Sp^2 \subseteq \R^3$ in the subset $\mathcal{V}_1$ with stereographic coordinates $(\theta^1_{(1)}, \theta^2_{(1)}) \in \R^2$. The coordinate transformation between $(\theta, \varphi)$ and $(\theta^1_{(i)}, \theta^2_{(i)})$ coordinates is given by the map $\mathscr{S}_{(i)}$ with the expressions
\begin{align}
    \mathscr{S}_{(2)}(\theta, \varphi) = &\: \Big( \th^1_{(2)} = \tan \frac{\theta}{2} \cos \varphi,  \th^2_{(2)} =\tan \frac{\theta}{2} \sin \varphi\Big),\label{eq:def.scrS.2} \\
    \mathscr{S}_{(1)}(\theta, \varphi) = &\: \Big( \th^1_{(1)} = \cot\frac{\theta}{2} \cos \varphi,  \th^2_{(1)} = \cot \frac{\theta}{2} \sin \varphi \Big).\label{eq:def.scrS.1}
\end{align}

The angular coordinates are defined as follows\footnote{Notice that some computation issues regarding this change of variables were introduced in the published version of \cite{DafLuk17}. The citations below refer to the arXiv version of the paper.}. First, let\footnote{Notice that the Kerr coordinates were denoted by $\th_{(i),\CH^+,\Ke}^A$ in \cite{DafLuk17}, but in this paper we use bold face instead of the subscript ${}_{\Ke}$} 
\begin{equation}\label{eq:Kerr.CH.coord}
    (\pmb{\th}_{*,\CH^+},\pmb{\varphi}_{*,\CH^+}) = \Big(\th_*, \phi_* -\f{4Mar_-}{(r_-^2 +a^2)^2
}\ub \Big).    
\end{equation}
For $i=1,2$, define 
\begin{equation}\label{eq:Kerr.CH.stereo}
    (\th_{(i)}^1,\th_{(i)}^2) = \mathscr{S}_{(i)} (\th_*,\varphi_*),\quad (\pmb{\th}_{(i),\CH^+}^1,\pmb{\th}_{(i),\CH^+}^2) = \mathscr{S}_{(i)} (\pmb{\th}_*,\pmb{\varphi}_*).
\end{equation}
Define then $(\th^1_{(i),\CH^+}, \th^2_{(i),\CH^+})$ by the following conditions:
\begin{enumerate}
    \item On $\{u = u_f\}$, we impose
    \begin{equation}\label{eq:reg.coord.def.1}
        \Big(\f{\rd\th^A_{(i),\CH^+}}{\rd \ub}+ b_{(i)}^B\f{\rd\th^A_{(i),\CH^+}}{\rd\th_{(i)}^B}\Big)(u,\ub,\th_{(i)}^1,\th_{(i)}^2)= \underline{\pmb{b}}_{(i)}^A,
    \end{equation}
    with initial data $(\th^1_{(i),\CH^+}, \th^2_{(i),\CH^+})= (\th^1_{(i)},\th^2_{(i)})$ on $S_{u_f,\ub_0}$ for some fixed $\ub_0$ satisfying $\f{4Mar_-}{r_-^2+a^2}\ub_0 \in 2\pi \mathbb Z$, where $b_{(i)}^A$ is the component of $b$ (see \eqref{eq:metric.form.2}) in the $(\th^1_{(i)},\th^2_{(i)})$ coordinates, $\underline{\pmb{b}}$ is the vector field $\underline{\pmb{b}}^{\th_*} =0$, $\underline{\pmb{b}}^{\varphi_*} = -\f{4Mar_-}{r_-^2+a^2}+ \f{4Mar}{\Sigma R^2}$ in $(\th_*,\varphi_*)$ coordinates, and $\underline{\pmb{b}}_{(i)}^A$
 are the components of the vector field $\underline{\pmb{b}}$ in the $(\th^1_{(i)},\th^2_{(i)})$ coordinates. 
    \item To the past of $\{u=u_f\}$, we require
    \begin{equation}\label{eq:reg.coord.def.2}
        \f{\rd\th^A_{(i),\CH^+}}{\rd u} = 0.
    \end{equation}
\end{enumerate}
    It is shown in \cite[Section~16.4]{DafLuk17} that for some fixed 
    \begin{equation*}
    \calV_1'\subset \calV''_1 \subset \{\frac{\pi}{4} \leq \theta_* \leq \pi\} \subseteq \calV_1\quad  \textnormal{ and } \quad \calV_2'\subset \calV''_2 \subset \{0 \leq \theta_* \leq \frac{3\pi}{4} \} \subseteq \calV_2
    \end{equation*}
    satisfying $\calV_1' \cup \calV_2' =\mathbb S^2$, the new coordinates $(\th^1_{(i),\CH^+}, \th^2_{(i),\CH^+})$ as functions of $(\th^1_{(i)}, \th^2_{(i)})$ is well-defined for $(\th^1_{(i)}, \th^2_{(i)}) \in \calV''_i$. On the overlapping region $\calV''_1 \cap \calV''_2$, the coordinates $(\th^1_{(1),\CH^+}, \th^2_{(1),\CH^+})$ and $(\th^1_{(2),\CH^+}, \th^2_{(2),\CH^+})$ agree up to the necessary transformation\footnote{This is the content of \cite[Lemma~16.8]{DafLuk17}, even though the transformation formula was not computed there.}, i.e., $(\th^1_{(2),\CH^+},\th^2_{(2),\CH^+}) = (\mathscr{S}_{(2)} \circ \mathscr{S}_{(1)}^{-1})(\th^1_{(1),\CH^+},\th^2_{(1),\CH^+})$. Moreover, the image of $\calV_i''$ under the map $(\th^1_{(i)}, \th^2_{(i)}) \mapsto (\th^1_{(i),\CH^+}, \th^2_{(i),\CH^+})$ covers $\calV_i'$, and $(\th^1_{(i),\CH^+}, \th^2_{(i),\CH^+})$ forms a coordinate system on $\calV_i'$ in the constant $(u,\ub)$-sphere if 
    \begin{equation}\label{eq:Winfty.def}
        (u,\ub) \in \calW_{\infty} := \{(u,\ub): u< u_f,\, -u+C_R\leq \ub\}.
    \end{equation}

    The following was proven in \cite[Section~16.5]{DafLuk17}:
\begin{theorem}\label{thm:C0.extendibility}
    Let $\calW_{\infty}$ be defined by \eqref{eq:Winfty.def}. For $i=1,2$, the metric constructed in \cite{DafLuk17} on $\calW_{\infty}\times \calV'_{(i)} \subset \os{\calM}{2}$ in the $(u,\ub_{\CH^+}, \th^1_{(i),\CH^+}, \th^2_{(i),\CH^+})$ coordinates extends continuously up to the Cauchy horizon $\CH^+$.
\end{theorem}
Note that the metric remains in the double null form
\begin{equation*}
    g=-2\Omega_{\CH^+}^2(\ud u\otimes \ud\ub_{\CH^+} + \ud\ub_{\CH^+}\otimes \ud u)+(\gamma_{\CH^+})_{AB}(\ud\vartheta_{\CH^+}^A-b_{\CH^+}^A \ud\ub_{\CH^+})\otimes (\ud\vartheta_{\CH^+}^B-b_{\CH^+}^B \ud\ub_{\CH^+}).
\end{equation*}
The determinant is computed to be 
\begin{equation} \label{EqDetG}\det g = - 4 \Omega_{\CH^+}^4 \det \gamma_{\CH^+}
\end{equation}
and we have 
\begin{equation} \label{Eqe4ExtCont}
\overset{\scriptscriptstyle{\text{[2]}}}{\ee}_4 = \f{1}{\Omega_{\CH^+}^2} \Big(\f{\rd}{\rd \ub}\Big|_{\CH^+} + b_{\CH^+}^A\f{\rd}{\rd\vartheta_{\CH^+}^A} \Big|_{\CH^+}\Big)\;,
\end{equation}
where we have introduced the abbreviation $|_{\CH^+}$ for the $(u, \ub_{\CH^+}, \vartheta_{\CH^+})$ coordinate system.

\subsection{Proof of Parts~\ref{item:main.thm.1} and \ref{item:main.thm.2} of Theorem~\ref{thm:main}}\label{sec:proof.of.main.theorem.1.2}

\begin{proof}[Proof of Parts~\ref{item:main.thm.1} and \ref{item:main.thm.2} of Theorem~\ref{thm:main}]
The existence of $\os{\calM}{1}$ and $\os{\calM}{2}$ (and thus $\calM$ as constructed in Section~\ref{sec:combined.manifold}) follow from the assumed upper bounds and the results in \cite{DafLuk17, DafLuk26}.

The manifold $\calM'$ in the statement of Theorem~\ref{thm:main} can be realized as a subset $\calM'\subset \calM$. To construct $\calM'$, first consider the sphere $S_{u_f,C_R-u_f}$ with $(u,\ub)= (u_f, C_R - u_f)$ in $\Sigma$. Let $\widecheck{H}$ be the null hypersurface emanating from this sphere transversal to the constant-$\ub$ hypersurfaces. To the future of $S_{u_f,C_R-u_f}$, this necessarily coincide with the $\{u=u_f\}$ hypersurface in the $(u,\ub,\vartheta_*)$ coordinates in $\os{\calM}{2}$. To the past of $S_{u_f,C_R-u_f}$, by choosing $\ep$ smaller if necessary, we know from the estimates in Theorem~\ref{thm:DL.2} below that the dynamical spacetime is close to Kerr in the $(s,\ub,\vartheta_*)$ coordinates. Since $\widecheck{H}$ is contained in a compact subset, we can choose $\ep$ smaller so that by Cauchy stability for the geodesic equation, it remains a smooth hypersurface until it intersects the initial hypersurface $\underline{H}_1$ at a $2$-sphere. Define that $2$-sphere of intersection as $\widecheck{S}$, and $\widecheck{s}:\bbS^2 \to \mathbb R$ as the function such that $\widecheck{S} = \{ (s, \th_*,\varphi_*) \in \Hb_1: s= \widecheck{s}(\th_*,\varphi_*)\}$. Finally, define $\calM''\subset \calM$ to be the subset to the past of $\widecheck{H}$, and let $\widecheck{\Hb}_1 =  \{ (s, \th_*,\varphi_*) \in \Hb_1: s< \widecheck{s}(\th_*,\varphi_*)\}$. 

    Now the properties \ref{item:main.thm.1.a}--\ref{item:main.thm.1.c} in Theorem~\ref{thm:main} follow easily when we also use the estimates proven in \cite{DafLuk17, DafLuk26}.
    
    For Part~\ref{item:main.thm.1.a}, the fact that $(\calM',g)$ is the maximal globally hyperbolic future development of the data restricted to $\calH^+ \cup \widecheck{\Hb}_1$ can be proven similarly as \cite[Proposition~16.16]{DafLuk17}. For Part~\ref{item:main.thm.1.b}, we use the coordinates $(u',\ub,\th_*,\varphi_*)$, where $(\ub,\th_*,\varphi_*)$ are as in Section~\ref{sec:M1} and Section~\ref{sec:M2}, and $u'$ is as in Section~\ref{sec:u'def}. The proof of the convergence of the metric in these coordinates require some estimates that will be stated and proven later. We postpone the proof to Proposition~\ref{prop:end.of.1.b}. 
    
    For Part~\ref{item:main.thm.1.c}, $\widecheck{H}$ is by construction a future null boundary of $\calM'$. The fact that the metric extends smoothly to $\widecheck{H}$ is a consequence of \cite{DafLuk17, DafLuk26}. For the last assertion, first note that $\calM'\cap \os{\calM}{1}\cap \{ s \leq s_{\Sigma}\}$ is bounded to the future by $\Sigma$ and a portion of $\widecheck{H}$. Thus, a future-directed future-inextendible timelike curve in $\calM'\cap \os{\calM}{1}\cap \{ s \leq s_{\Sigma}\}$ must either intersect $\widecheck{H}$ or enter $\os{\calM}{2}$. Now since any future-directed timelike vector $X$ in $\os{\calM}{2}$ satisfies $X u>0$ and $X\ub >0$, it follows that along any future-directed timelike curve $\gamma$ in $\os{\calM}{2}$, $u$ and $\ub$ must both be strictly increasing. If the curve $\gamma:(a,b)\to \calM' \cap \widecheck{H}$ is also future-inextendible, then it must either intersect $\widecheck{H}$ or $\lim_{\sigma \to b^-} \ub(\gamma(\sigma)) = \infty$.

    Finally, the continuous extendibility of the metric in Part~\ref{item:main.thm.2} of Theorem~\ref{thm:main} follows from Theorem~\ref{thm:C0.extendibility}. As already mentioned above, $\widecheck{H}$ corresponds to $\{u = u_f\}$ in $\os{\calM}{2}$. \qedhere
\end{proof}

\subsection{Main result proven in this paper}\label{sec:main.result.proven.in.this.paper}

As outlined in the introduction, Part~\ref{item:main.thm.3} of Theorem \ref{thm:main} is established by proving that the assumptions of the $C^{0,1}_{\loc}$-inextendibility result in \cite{Sbie24} are satisfied. The key requirement of \cite{Sbie24} is an integrated blowup of curvature, the verification of which is the content of Theorem \ref{ThmInextConditionVerified} below. This is the main result of this paper. In addition to the integrated curvature blowup, a second auxiliary upper bound on various connection coefficients  is needed in \cite{Sbie24}. This is the content of Proposition \ref{PropAuxiliaryBounds} below.
We denote the Lorentzian volume on $(\calM,g)$ by $\vol_g$ and define the auxiliary metric  $h := \ud u^2 + \ud\ub_{\CH^+}^2 + \gamma_{\CH^+}$, which is referred to in the theorem below. Here, $\gamma_{\CH^+} = \gamma$ is the induced metric on the spheres $\{u = \mathrm{const}\} \cap \{\ub = \mathrm{const}\}$ in $(\calM, g)$.

\begin{theorem} \label{ThmInextConditionVerified}
    Under the assumptions of Theorem \ref{thm:main} from Section \ref{SecPreciseAssump} and with respect to the double null coordinate system $(u, \ub_{\CH^+}, \theta^1_{(i), \CH^+}, \theta^2_{(i), \CH^+})$ introduced in Section \ref{sec:CH.coord}  the following holds: Given any compact neighborhoods $K_u \subseteq (-\infty, u_f)$ and $K_{\Sp^2} \subseteq \Sp^2$ such that we have\footnote{This is not an `either or'.} $K_{\Sp^2} \subseteq \calV_1'$ or $K_{\Sp^2} \subseteq \calV_2'$, let  $K := K_u \times K_{\Sp^2} \subseteq (-\infty, u_f) \times \Sp^2 \simeq \{\ub_{\CH^+} = 0\}$ be a compact neighborhood  within the Cauchy horizon. Let $\ee_4 := \os{\ee}{2}_4$ be as in \eqref{eq:dynamical.ee.2.3.4.def}. For $A=1,2$ there exist continuous vector fields $E_A^{(\cdot)}$ which extend continuously to $\{\ub_{\CH} = 0\}$  and an 
$\hat{\varepsilon}>0$ such that for all continuous vector fields $\hat{\ee}_4$, $\hat{E}_A^{(\cdot)}$ on $(-\delta_0,0] \times K$ with  $||\hat{\ee}_4 - \ee_4||_h < \hat{\varepsilon}$ and $||\hat{E}^{(\cdot)}_A - E^{(\cdot)}_A||_h < \hat{\varepsilon}$ on $(-\delta_0, 0] \times K$ we have
\begin{equation} \label{EqInextVerified}
    \int\limits_{-\delta_0}^0 \Big| \int\limits_{-\delta_0}^{\ub_{\CH^+}'}\int_{K} \big( R(\hat{\ee}_4, \hat{E}^{(\cdot)}_1, \hat{\ee}_4, \hat{E}^{(\cdot)}_1) + i R(\hat{\ee}_4, \hat{E}^{(\cdot)}_1, \hat{\ee}_4, \hat{E}^{(\cdot)}_2) \big) \vol_g \Big|^2 \ud \ub_{\CH^+}' = \infty \;.
\end{equation}
Here, $\delta_0 >0$ is so small that $(-\delta_0,0) \times K$ is contained within the domain of the $(u, \ub_{\CH^+}, \theta^1_{(i), \CH^+}, \theta^2_{(i), \CH^+})$ coordinates, but otherwise arbitrary.
    \end{theorem}

\begin{proposition} \label{PropAuxiliaryBounds}
Under the subset\footnote{I.e., without the need to assume \eqref{EqAsDecAlpha}, \eqref{EqAsInst2}, \eqref{EqAsInst3}, \eqref{EqAsInst4}.} of assumptions \eqref{EqAsDecRest}, \eqref{EqAssumptionsTransversal} of Theorem \ref{thm:main} it was proven in \cite{DafLuk17} (combined with \cite{DafLuk26}) that for each compact $K \subseteq \{\ub_{\CH} = 0\}$ the following bounds on the connection hold near the Cauchy horizon:
\begin{equation} \label{EqBoundsConnectionDafLuk}
\sup\limits_{(-\delta_0,0) \times K} \{|\omb| + |\eta|_\gamma + |\etab|_\gamma + |\chib|_\gamma\} \leq C \quad \textnormal{ and } \quad \int_{-\delta_0}^0 \Big(\sup\limits_{(u, \vartheta_{\CH^+}) \in K} |\chi|_\gamma(u, \uu_{\CH^+}, \vartheta_{\CH^+}) \Big)\, \ud \uu_{\CH^+} \leq C \;.
\end{equation}
Moreover, under the same assumptions, the following bound also holds for each compact $K_u \subseteq (-\infty, u_f]$ and $i=1,2$:
    \begin{equation}\label{EqBoundConnectionB}
\int\limits_{-\delta_0}^0 \sup_{(u, \vartheta_{(i), \CH^+}) \in K_u \times \calV'_{i}} \big| \frac{\rd}{\rd \theta^B_{(i), \CH^+}} b^A_{(i), \CH^+} (u, \ub_{\CH^+}, \vartheta_{(i),\CH^+})\big| \, \ud \uu_{\CH^+}  \leq C\;.
\end{equation}
Here, the constants on the right-hand side are allowed to depend on $K$ and $K_u$ --- and $\delta_0 >0$ is again such that $(-\delta_0,0) \times K$, $(-\delta_0,0) \times K_u \times \calV'_i$, respectively, are contained within the domain of the $(u, \ub_{\CH^+}, \theta^1_{(i), \CH^+}, \theta^2_{(i), \CH^+})$ coordinates, but otherwise arbitrary.
\end{proposition}    
The proof of \eqref{EqBoundConnectionB}, and thus of Proposition \ref{PropAuxiliaryBounds}, is carried out in Section \ref{SecPfAuxEstimate}, where an even stronger statement is proven in Proposition \ref{PropAuxEst}. The proof of Theorem \ref{ThmInextConditionVerified} is concluded in Section \ref{SecProofMainThm}. Given these two statements, it is straightforward to conclude the proof of Theorem \ref{thm:main}.

\begin{proof}[Proof of Part~\ref{item:main.thm.3} of Theorem \ref{thm:main}]
    We note that \eqref{EqInextVerified} in particular implies the existence of a sequence $\ub_{\CH^+, k} \nearrow 0$ for $k \to \infty$ with 
    $$ \Big| \int\limits_{-\delta_0}^{\ub_{\CH^+,k}}\int_{K} \big( R(\hat{\ee}_4, \hat{E}^{(\cdot)}_1, \hat{\ee}_4, \hat{E}^{(\cdot)}_1) + i R(\hat{\ee}_4, \hat{E}^{(\cdot)}_1, \hat{\ee}_4, \hat{E}^{(\cdot)}_2) \big) \vol_g  \Big| \to \infty$$
    for $k \to \infty$. We now conclude by \cite[Theorem 3.14 and Remark 3.18]{Sbie24}.
\end{proof}

\section{Estimates for the inhomogeneous linear Teukolsky equation} \label{SecInhomLinTeuk}

In this section we consider the exact Kerr background $(\mathcal{M}_{\mathrm{Kerr}}, \pmb{g})$ introduced in Section \ref{SecKerrBackground} and prove estimates for the inhomogeneous linear Teukolsky equation. The estimates (and the notation) follow those from \cite{Sbie23} where they were carried out for the homogeneous Teukolsky equation. However, here we have taken a more streamlined approach with just one multiplier vector field.

We define the function $f_+ := v_+ -r + r_+$. An easy computation gives 
\begin{equation}
\label{NormF}
\langle \ud f_+, \ud f_+ \rangle =  \frac{a^2 \sin^2 \theta}{\rho^2} + \frac{\Delta}{\rho^2} - \frac{2(r^2 + a^2)}{\rho^2}
\end{equation}
which shows that the level sets of $f_+$ are spacelike hypersurfaces. Moreover, it is immediate that the level sets of $r$ are spacelike hypersurfaces.
We also introduce the energy densities
\begin{equation*}
\begin{split}
    &d_\Delta(\psi) := \sum_{0 \leq i_1 + i_2 + i_3 + j + k \leq 1} |\Delta|^k |\widetilde{Z}_{1,+}^{i_1}\widetilde{Z}_{2,+}^{i_2}\widetilde{Z}_{3,+}^{i_3}\rd_{v_+}^{j}\rd_r^{k}\psi|^2 \\
    &d(\psi):= \sum_{0 \leq i_1 + i_2 + i_3 + j + k \leq 1} |\widetilde{Z}_{1,+}^{i_1}\widetilde{Z}_{2,+}^{i_2}\widetilde{Z}_{3,+}^{i_3}\rd_{v_+}^{j}\rd_r^{k}\psi|^2
    \end{split} \;,
\end{equation*}
where the coordinate vector fields  and the fields \eqref{EqZTilde} are defined with respect to the $(v_+, r, \theta, \varphi_+)$ coordinate system, which will be used throughout this section.
Note that in particular $|\psi|^2$ is controlled by both energy densities while, for $r = r_+$,  $|\partial_r \psi|^2$ is not controlled by $d_\Delta(\psi)$.

\begin{proposition} \label{PropInhomTeukI}
    Consider the inhomogeneous linear Teukolsky equation $F = \pmb{\mathcal{T}}_{[2]} \psi$, where $\psi$ and $F$ are $C^{10}$-regular\footnote{We are not keeping track of the exact required regularity in this paper.} spin $2$-weighted functions on $\mathcal{M}_{\mathrm{Kerr},+}$. Let  $q > 0$. Then there exist $\rred \in (r_-, r_+)$ and a $C>0$ such that for $v_1 \geq 1 + r_+ - r_-$ big enough the following estimate holds:
\begin{equation}
\label{EqPropRightH}
\begin{split}
    \int\limits_{\{r=\rred\} \cap \{f_+ \geq v_1\}} v_+^{q} d(\psi) \,\vols \ud v_+ &\leq C \Big( \sum_{k=0}^2 \int\limits_{\mathcal{H}^+ \cap \{f_+ \geq v_1\}}  v_+^{q} d_\Delta(\rd_r^k \psi)\, \vols \ud v_+ + C_{\{f_+ = v_1\}}(\psi) \\
    &\qquad + \int\limits_{\{\rred \leq r \leq r_+\} \cap \{f_+ \geq v_1\}} v_+^{q} |\rd_r^2F|^2 \,\vols\, \ud v_+ \ud r \Big) \;.
\end{split}
\end{equation}

Here, $C_{\{f_+ = v_1\}}(\psi) >0$ is a constant which only depends on up to three derivatives of $\psi$ on $\{f_+ = v_1\} \cap \{\rred \leq r \leq r_+\}$.
\end{proposition}

\begin{proposition}\label{PropInhomTeukII}
    Consider the inhomogeneous linear Teukolsky equation $F = \pmb{\mathcal{T}}_{[2]} \psi$,  where $\psi$ and $F$ are $C^{10}$-regular spin $2$-weighted functions on $\mathcal{M}_{\mathrm{Kerr},+}$. Let $\rred \in (r_-, r_+)$ as in Proposition \ref{PropInhomTeukI} and $q >0$. Then there exists a $C>0$ such that for $v_1 \geq 1 + r_+ - r_-$ big enough we have
\begin{equation} \label{EqPropGamma}
\begin{split}
    \int\limits_{\Gamma \cap \{f_+ \geq v_1\} } v_+^q d_\Delta(\psi) \, \vols \ud v_+ &\leq C \Big( \int\limits_{\{r= \rred\} \cap \{f_+ \geq v_1\}} v_+^q d(\psi) \, \vols \ud v_+ + C_{\{f_+ = v_1\}}(\psi) \\
    &\qquad + \int\limits_{\substack{\{f_\Gamma \leq 0\} \cap \{r \leq \rred\} \\ \cap \{f_+ \geq v_1\}} }v_+^q|F|^2 \, \vols \ud v_+ \ud r \Big) \;.
    \end{split}
\end{equation}
    Here, $C_{\{f_+ = v_1\}}(\psi)>0$ is a constant which only depends on up to one derivative of $\psi$ on $\{f_+ = v_1\} \cap \{r \leq \rred\} \cap \{f_\Gamma \leq 0\}$.
\end{proposition}

For the proof of both propositions we use $\chi(v_+) e^{\lambda r} \overline{N \psi}$ with $\lambda >0$ as a multiplier, where $\chi(v_+) = v_+^q$ and $N:= \rd_{v_+} - \frac{M}{r} \rd_r$. Note that this vector field is uniformly timelike and future directed in the interior of Kerr. We also introduce the notation $\Phi:= \frac{1}{\sin \theta} (is \cos \theta + \rd_{\varphi_+})$, cf.\ beginning of Appendix \ref{AppendixEstimates}. We will prove Proposition \ref{PropInhomTeukII} first.

\begin{proof}[Proof of Proposition \ref{PropInhomTeukII}]


\pfstep{Step 1: The multiplier} We start out from the following multiplier identity, where $\lambda, \mu >0$ are constants to be chosen and $\chi(v_+) = v_+^q$:
\begin{equation}\label{EqPropCauchyMultiplier}
\begin{split}
\Rea\big( F \cdot \chi(v_+) e^{\lambda r} \overline{N \psi}\big) &= \Rea\Big(\mathcal{T}_{[s]} \psi \cdot \chi(v_+) e^{\lambda r} \overline{(\rd_{v_+} - \frac{M}{r} \rd_r) \psi}\Big) \\
&\qquad + \underbrace{\rd_r(\chi(v_+) \mu e^{\lambda r}| \psi|^2) - \chi(v_+)\mu \lambda e^{\lambda r} | \psi|^2 - 2\chi(v_+) \mu e^{\lambda r} \Rea(\overline{\psi}\rd_r \psi)}_{=0} \;.
\end{split}
\end{equation}
After integration over the spheres, the right-hand side of \eqref{EqPropCauchyMultiplier} consists of the sum of the following terms
\begin{enumerate}
\item the sum of all the terms on the right-hand sides of Appendix \ref{AppendixCommutator}, i.e., the terms $$ \rd_{v_+} [A] + \rd_r [B] + \uwave{D} + D_{\mathrm{rem}}$$ as defined in Appendix \ref{AppendixEstimates}
\item the real parts of the terms
\begin{equation*}
\begin{split}
2\big(r(1-2s) &- isa \cos \theta\big) \rd_{v_+} \psi \cdot \chi(v_+) e^{\lambda r} \overline{(\rd_{v_+} - \frac{M}{r} \rd_r) \psi}
\;\, \dashuline{-\,\chi(v_+)e^{\lambda r}\frac{M}{r} 2(r-M)(1-s)|\rd_r \psi|^2} \\[4pt]
&+ \chi(v_+)e^{\lambda r}2(r-M)(1-s) \rd_r \psi (\overline{\rd_{v_+} \psi})
-2s\psi \cdot \chi(v_+)e^{\lambda r}\overline{(\rd_{v_+} - \frac{M}{r} \rd_r) \psi}
\end{split}
\end{equation*}
\item the underbraced terms in \eqref{EqPropCauchyMultiplier}.
\end{enumerate}
The desired boundedness statement requires all the bulk terms on the right-hand side to yield a negative definite contribution. In the following Steps 2 and 3 we only consider the right-hand side.

\pfstep{Step 2: Estimating the bulk term without the underbraced term in \eqref{EqPropCauchyMultiplier}}
We only consider the terms $\uwave{D}$, $D_{\mathrm{rem}}$, and those from 2. We first consider all those terms that are quadratic in $\partial_r \psi$, but which do not have a factor of $\Delta$. There are three such terms. The dashed term from 2.\ above and the dashed term from Appendix \ref{AppendixCommutator}, contained in $D_{\mathrm{rem}}$, sum to
\begin{equation*}
    \chi(v_+)e^{\lambda r} \frac{M}{r} (r-M) \underbrace{\big(1 - 2(1-s) \big)}_{= 3 \textnormal{ for } s = +2} |\rd_r \psi|^2 \;,
\end{equation*}
which is negative for $r \in (r_-, M)$. The third term is $\chi'(v_+) e^{\lambda r} \frac{M}{r}(r^2 + a^2) |\rd_r \psi|^2$. Recalling that $\chi(v_+) = v_+^q$, we now fix an $r_0 \in (r_-,M)$ and choose $v_1$ large enough such that 
\begin{equation} \label{EqControlRrminus}
    \textnormal{the sum of those three terms is } \leq - c_0 \cdot  \chi(v_+) e^{\lambda r} |\rd_r \psi|^2 \qquad \textnormal{ for }  \{r_- < r < r_0\} \cap \{f_+ \geq v_1\}
\end{equation}
for some $c_0 >0$.

Lemma \ref{LemmaAppendixDUWave} gives us
\begin{equation} \label{EqControlDuwave}\uwave{D} \leq -c \lambda \chi(v_+) e^{\lambda r} \Big( - \Delta |\rd_r \psi|^2 + |\rd_{v_+} \psi|^2 + |\rd_\theta \psi|^2 + |\Phi \psi|^2 \Big) + C \lambda \chi(v_+) e^{\lambda r} |\psi|^2
\end{equation}
for $r \in [r_-, r_+]$.

Now, the important structure to observe is that none of the remaining terms in $D_{\mathrm{rem}}$ and those from 2.\ have a factor of $\lambda$ in it and furthermore, terms of the form $|\rd_r \psi|^2$ come with a factor of $\Delta$. (The sum of the dashed terms restricted to $\{r_0  \leq r \leq \rred\}$ is trivially also of this form since $\Delta$ is bounded away from zero in this region.) We can thus apply a weighted Cauchy--Schwarz inequality to the remaining terms of the form $\Rea(\psi \ov{\rd_r \psi})$, $\Rea(\rd_{\varphi_+} \psi \ov{\rd_r \psi})$, and $\Rea(\rd_{v_+} \psi \ov{\rd_r \psi})$ to absorb $|\rd_r \psi|^2$ into \eqref{EqControlRrminus} at the expense of a large term of the form $|\rd_{v_+} \psi|^2$, $|\rd_{\varphi_+} \psi|^2$, $|\psi|^2$. We can now choose $\lambda_0>0$ big enough and use Cauchy--Schwarz on all the remaining mixed first (and zeroth) derivative terms to obtain for all $\lambda \geq \lambda_0$
\begin{equation} \label{EqPfPropGammaStep2}
\begin{split}
    \textnormal{RHS\eqref{EqPropCauchyMultiplier}}  &\underset{a.i.}{\leq} \rd_{v_+} [A] + \rd_r [B] - c \chi(v_+) e^{\lambda r} \Big[ |\rd_r \psi|^2 + \lambda \Big( |\Delta| |\rd_r \psi|^2 + |\rd_{v_+} \psi|^2 + |\rd_\theta \psi|^2 + |\Phi \psi|^2 \Big)\Big] \\ 
    &\qquad + C_0 \lambda \chi(v_+)e^{\lambda r} |\psi|^2 
     + \underbrace{\rd_r(\chi(v_+) \mu e^{\lambda r}| \psi|^2) - \chi(v_+)\mu \lambda e^{\lambda r} | \psi|^2 - 2\chi(v_+) \mu e^{\lambda r} \Rea(\overline{\psi}\rd_r \psi)}_{=0} 
    \end{split}
\end{equation}
in $r \in [r_-, \rred]$.

\pfstep{Step 3: Estimating the boundary terms and the remaining bulk terms}
We estimate the last of the underbraced terms in \eqref{EqPfPropGammaStep2} by
\begin{equation} \label{EqEstUnderb}
    |2\chi(v_+) \mu e^{\lambda r} \Rea(\overline{\psi}\rd_r \psi)| \leq \frac{1}{2} \chi(v_+) \mu \lambda e^{\lambda r} |\psi|^2 + 2 \chi(v_+) \frac{\mu}{\lambda} e^{\lambda r} |\rd_r \psi|^2
\end{equation}
so that the underbraced term is estimated by
\begin{equation} \label{EqPfPropUnderb}
    \underbrace{ \ldots}_{=0} \leq \rd_r(\chi(v_+) \mu e^{\lambda r}| \psi|^2) - \frac{1}{2}\chi(v_+)\mu \lambda e^{\lambda r} | \psi|^2 + 2 \chi(v_+) \frac{\mu}{\lambda} e^{\lambda r} |\rd_r \psi|^2 \;.
\end{equation}

Furthermore, the Lemmas  \ref{LemAppendixBA}, \ref{LemApendixGamma} give us
\begin{align*}
 B - A &\geq c \chi(v_+) e^{\lambda r} \Big( |\rd_r \psi|^2 + |\rd_{v_+} \psi|^2 + |\rd_\theta \psi|^2 + |\Phi \psi|^2 \Big) - C_1 \chi(v_+) e^{\lambda r} |\psi|^2  \textnormal{ for } r \in [r_-, r_+] \\
 B + \frac{\sigma_q}{\kappa_- v_+} \cdot \frac{|\Delta|}{2(r^2 + a^2)} A &\geq  c \chi(v_+) e^{\lambda r} \Big( - \Delta |\rd_r \psi|^2 + |\rd_{v_+} \psi|^2 + |\rd_\theta \psi|^2 + |\Phi \psi|^2 \Big)  \\
 &\qquad \qquad\qquad \qquad \qquad \qquad \qquad\qquad - C_2 \chi(v_+) e^{\lambda r} |\psi|^2 \textnormal{ on } \Gamma \cap \{r_- < r \leq r_G\}\;,
\end{align*}
where $r_G \in (r_-, r_+)$. 
The first term on the right-hand side of \eqref{EqPfPropUnderb} gives an additional boundary term. All the arising boundary terms in the next step can now be made positive definite also in $|\psi|^2$ if we choose $\mu > \max\{C_1, C_2\}$. Furthermore, if we also choose $\mu > 2C_0$, then the bulk term in $|\psi|^2$ on the right-hand side of \eqref{EqPfPropGammaStep2} can be absorbed by the second term on the right-hand side of \eqref{EqPfPropUnderb}. Finally, we choose $\lambda> \lambda_0$ even larger if necessary such that $\frac{2 \mu}{\lambda} < c$ such that the last term on the right-hand side of \eqref{EqPfPropUnderb} can be absorbed by the third term on the right-hand side of \eqref{EqPfPropGammaStep2}. In summary, this gives us in $r \in [r_-, \rred]$
\begin{equation} \label{EqPfPropGammaStep3}
\begin{split}
    \textnormal{RHS\eqref{EqPropCauchyMultiplier}}  &\underset{a.i.}{\leq} \rd_{v_+} [A] + \rd_r [B + \chi(v_+) \mu e^{\lambda r}| \psi|^2)]\\
    &\qquad \qquad - c \chi(v_+) e^{\lambda r} \Big[ |\rd_r \psi|^2 + \lambda \Big( |\Delta| |\rd_r \psi|^2 + |\rd_{v_+} \psi|^2 + |\rd_\theta \psi|^2 + |\Phi \psi|^2 + |\psi|^2 \Big)\Big] \;.
    \end{split}
\end{equation}

\pfstep{Step 4: Putting it all together and estimating the LHS of \eqref{EqPropCauchyMultiplier}} We now choose $v_1$ even bigger if necessary such that the range of the $r$-coordinate on $\Gamma \cap \{f_+ \geq v_1\}$ is contained in $(r_-, r_G]$. We then integrate \eqref{EqPfPropGammaStep3} over the region $V(v'):=\{r \leq \rred\} \cap \{v_1 \leq f_+ \leq v'\} \cap \{f_\Gamma \leq 0\}$ with respect to $dv_+ \wedge dr \wedge \vols = \frac{1}{\Sigma}\vol$ to obtain
\begin{figure}[h]
\centering
 \def\svgwidth{4cm}
\begingroup%
  \makeatletter%
  \providecommand\color[2][]{%
    \errmessage{(Inkscape) Color is used for the text in Inkscape, but the package 'color.sty' is not loaded}%
    \renewcommand\color[2][]{}%
  }%
  \providecommand\transparent[1]{%
    \errmessage{(Inkscape) Transparency is used (non-zero) for the text in Inkscape, but the package 'transparent.sty' is not loaded}%
    \renewcommand\transparent[1]{}%
  }%
  \providecommand\rotatebox[2]{#2}%
  \newcommand*\fsize{\dimexpr\f@size pt\relax}%
  \newcommand*\lineheight[1]{\fontsize{\fsize}{#1\fsize}\selectfont}%
  \ifx\svgwidth\undefined%
    \setlength{\unitlength}{275.92472347bp}%
    \ifx\svgscale\undefined%
      \relax%
    \else%
      \setlength{\unitlength}{\unitlength * \real{\svgscale}}%
    \fi%
  \else%
    \setlength{\unitlength}{\svgwidth}%
  \fi%
  \global\let\svgwidth\undefined%
  \global\let\svgscale\undefined%
  \makeatother%
  \begin{picture}(1,0.97661233)%
    \lineheight{1}%
    \setlength\tabcolsep{0pt}%
    \put(0,0){\includegraphics[width=\unitlength,page=1]{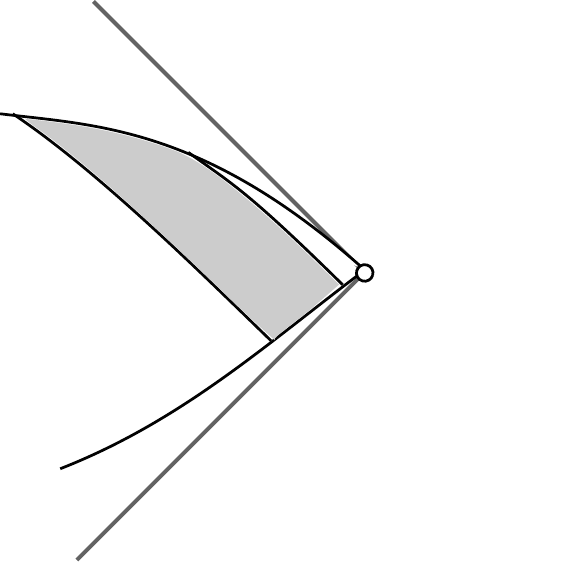}}%
    \put(0.40367464,0.75684508){\color[rgb]{0,0,0}\makebox(0,0)[lt]{\lineheight{1.25}\smash{\begin{tabular}[t]{l}$\mathcal{CH}^+$\end{tabular}}}}%
    \put(0.35471639,0.13903859){\color[rgb]{0,0,0}\makebox(0,0)[lt]{\lineheight{1.25}\smash{\begin{tabular}[t]{l}$\mathcal{H}^+$\end{tabular}}}}%
    \put(0.00501474,0.28824474){\color[rgb]{0,0,0}\makebox(0,0)[lt]{\lineheight{1.25}\smash{\begin{tabular}[t]{l}$r=\rred$\end{tabular}}}}%
    \put(0.13090733,0.78248943){\color[rgb]{0,0,0}\makebox(0,0)[lt]{\lineheight{1.25}\smash{\begin{tabular}[t]{l}$\Gamma$\end{tabular}}}}%
    \put(0.76270152,0.62395825){\color[rgb]{0,0,0}\makebox(0,0)[lt]{\lineheight{1.25}\smash{\begin{tabular}[t]{l}$f_+ = v'$\end{tabular}}}}%
    \put(0.63914042,0.33254016){\color[rgb]{0,0,0}\makebox(0,0)[lt]{\lineheight{1.25}\smash{\begin{tabular}[t]{l}$f_+=v_1$\end{tabular}}}}%
    \put(0,0){\includegraphics[width=\unitlength,page=2]{EnEstGamma.pdf}}%
  \end{picture}%
\endgroup%

      \caption{The region of integration} \label{FigEnEstGamma}
\end{figure}
\begin{equation*}
    \begin{split}
        \int\limits_{V(v')} &\Rea \big( F \cdot \chi(v_+) e^{\lambda r} \ov{N\psi} \big) \,\vols \ud v_+ \ud r \leq \int\limits_{V(v')} -c \chi(v_+) e^{\lambda r} \big[ |\rd_r \psi|^2 + \lambda d_\Delta(\psi)] \, \vols \ud v_+ \ud r \\
        & + \int\limits_{\{r = \rred\} \cap V(v')} \big( B + \chi(v_+) \mu e^{\lambda r} |\psi|^2 \big) \, \vols \ud v_+  - \int\limits_{\{f_+ = v'\} \cap V(v')} \big[\big( B + \chi(v_+) \mu e^{\lambda r} |\psi|^2\big) \ud v_+ - A \ud r\big] \vols \\
        &-\int\limits_{\Gamma \cap V(v')} \big[ \big(B+ \chi(v_+) \mu e^{\lambda r} |\psi|^2 \big) \ud v_+ + A\ud r\big] \vols + \int\limits_{\{f_+ = v_1\}\cap V(v')} \big[\big(B + \chi(v_+) \mu e^{\lambda r} |\psi|^2\big)\ud v_+ - A\ud r\big] \vols \;.
    \end{split}
\end{equation*}
We note that we have $\ud v_+ = \ud r$ on $\{f_+ = \mathrm{const}\}$ and $\ud r = \frac{\sigma_q}{\kappa_- v_+} \cdot \frac{ |\Delta|}{2(r^2 + a^2)} \ud v_+ $ on $\Gamma$. Using this together with our choice of $\mu$ in Step 3, and letting $v' \to \infty$, yields
\begin{equation*}
    \begin{split}
        \sup_{v' \geq v_1} &\int\limits_{\{f_+ = v'\} \cap V(\infty)} v_+^q d(\psi) \, \vols \ud v_+ + \int\limits_{\Gamma \cap V(\infty)} v_+^q d_\Delta(\psi) \, \vols \ud v_+ + \int\limits_{V(\infty)} v_+^q d(\psi) \, \vols \ud v_+ \ud r \\
        &\leq C \Big( \int\limits_{V(\infty)}|F \cdot v_+^q \ov{N\psi}| \, \vols \ud v_+ \ud r + \int\limits_{\{r = \rred\} \cap V(\infty)} v_+^q d(\psi) \vols \ud v_+ + \int\limits_{\{f_+ = v_1\} \cap V(\infty)} v_+^q d(\psi) \, \vols \ud v_+ \Big) \;.
    \end{split}
\end{equation*}
For the first term on the right-hand side we use a weighted Cauchy--Schwarz inequality to absorb the $|N\psi|^2$ term by the third term on the LHS. This concludes the proof. \qedhere

\end{proof}

\begin{proof}[Proof of Proposition \ref{PropInhomTeukI}]
As in \cite{Sbie23} (Proposition 4.11) we differentiate the inhomogeneous Teukolsky equation twice in $r$ to obtain an equation which admits a red-shift estimate for the energy. Using \eqref{EqTeukolskyStarCoordinates} this gives
\begin{equation} \label{EqTeukTwiceDiff}
    \begin{split}
\rd_r^2 F &=  \rd_r^2 \mathcal{T}_{[s]} \psi \\
&= a^2 \sin^2 \theta \rd_{v_+}^2 \rd_r^2 \psi + 2a \rd_{v_+} \rd_{\varphi_+} \rd_r^2 \psi + 2(r^2 + a^2) \rd_{v_+} \rd_r^3 \psi + 2a\rd_{\varphi_+} \rd_r^3 \psi + \Delta \rd_r^4 \psi + \swl \rd_r^2 \psi\\
&\quad  +2\big(r(5-2s) - isa \cos \theta\big) \rd_{v_+} \rd_r^2 \psi + 2(r-M)(3-s) \rd_r^3 \psi   +6(1-s) \rd_r^2 \psi +8(1-s)\rd_{v_+}\rd_r \psi  \\
&=  a^2 \sin^2 \theta \rd_{v_+}^2 \Psi + 2a \rd_{v_+} \rd_{\varphi_+} \Psi + 2(r^2 + a^2) \rd_{v_+} \rd_r\Psi + 2a\rd_{\varphi_+} \rd_r\Psi + \Delta \rd_r^2 \Psi + \swl \Psi\\
&\quad  +2\big(r(5-2s) - isa \cos \theta\big) \rd_{v_+} \Psi + 2(r-M)(3-s) \rd_r \Psi   +6(1-s) \Psi +8(1-s)\rd_{v_+}\rd_r \psi \;,
\end{split}
\end{equation}
where we have introduced the shorthand $\Psi := \rd_r^2 \psi$. 

\pfstep{Step 1: The multiplier} We start out from the following multiplier identity, where $\lambda, \mu >0$ are constants to be chosen and $\chi(v_+) = v_+^q$:
\begin{equation} \label{EqMultiplierH}
\begin{split}
    \Rea \big( \rd_r^2 F \cdot \chi(v_+) e^{\lambda r} &\ov{N \Psi} \big) = \Rea \big( \rd_r^2 \mathcal{T}_{[s]} \psi \cdot  \chi(v_+) e^{\lambda r} \ov{N \Psi} \big) \\
    & + \underbrace{\rd_r(\chi(v_+) \mu e^{\lambda r}| \Psi|^2) - \chi(v_+)\mu \lambda e^{\lambda r} | \Psi|^2 - 2\chi(v_+) \mu e^{\lambda r} \Rea(\overline{\Psi}\rd_r\Psi)}_{=0} \\
    & + \underbrace{\rd_r \big(\chi(v_+) \mu e^{\lambda r} | \rd_{v_+} \rd_r \psi|^2 \big) - \chi(v_+) \mu \lambda e^{\lambda r} |\rd_{v_+} \rd_r \psi|^2 - 2 \chi(v_+) \mu e^{\lambda r} \Rea ( \ov{\rd_{v_+} \rd_r \psi} \rd_{v_+} \Psi)}_{=0} \;.
\end{split}
\end{equation}
Note that we have used the same multiplier vector field $N$ as in the proof of Proposition \ref{PropInhomTeukII}, just applied to $\Psi$ instead of to $\psi$. The second underbraced modification has been added to control the terms arising from the last term on the right-hand side of \eqref{EqTeukTwiceDiff}, which is the only term not containing two $r$-derivatives of $\psi$. After integration over the spheres, the right-hand side of \eqref{EqMultiplierH} is the sum of
\begin{enumerate}
    \item the sum of all the terms on the right-hand side of Appendix \ref{AppendixCommutator} with $\psi$ replaced by $\Psi$, i.e., the terms
    \begin{equation*}
        \rd_{v_+}[A(\Psi)] + \rd_r[B(\Psi)] + \uwave{D(\Psi)} + D_{\mathrm{rem}}(\Psi)
    \end{equation*}
    as defined in Appendix \ref{AppendixEstimates}
    \item the real parts of the terms
    \begin{equation*}
        \begin{split}
            &\chi(v_+) e^{\lambda r} 2\big(r(5-2s) - isa \cos \theta\big) \rd_{v_+} \Psi \cdot \ov{N\Psi} - \dashuline{\chi(v_+) e^{\lambda r}\frac{M}{r} 2(r-M)(3-s) |\rd_r\Psi|^2} \\ 
            &+ \chi(v_+) e^{\lambda r} 2(r-M)(3-s) \rd_r\Psi \cdot \ov{\rd_{v_+} \Psi} 
            +\chi(v_+) e^{\lambda r} 6(1-s) \Psi \cdot \ov{N \Psi} \\
            &+ \uuline{\chi(v_+) e^{\lambda r} 8(1-s)\rd_{v_+}\rd_r \psi \cdot \ov{N \Psi}}
        \end{split}
    \end{equation*}
    \item the two underbraced terms in \eqref{EqMultiplierH}.
\end{enumerate}
Again, the desired boundedness statement requires all the bulk terms on the right-hand side to yield a negative definite contribution. The following Steps 2-4 follow closely those of the proof of Proposition \ref{PropInhomTeukII}. Step 5 is new.  In the next two steps we only consider the right-hand side of \eqref{EqMultiplierH}.

\pfstep{Step 2: Estimating the bulk term without the underbraced terms in \eqref{EqMultiplierH}} We consider the three terms in $\uwave{D(\Psi)}$, $D_{\mathrm{rem}}(\Psi)$, and in 2.\ which are quadratic in $\rd_r\Psi$ and do not contain a factor of $\Delta$. The dashed term from 2.\ and the dashed term from $D_{\mathrm{rem}}(\Psi)$ sum to 
\begin{equation*}
    \chi(v_+) e^{\lambda r} \frac{M}{r}(r-M) \underbrace{\big(-2(3-s) + 1 \big)}_{=-1 \textnormal{ for } s =+2} |\rd_r\Psi|^2 \;,
\end{equation*}
which is negative definite for $r \in (M, r_+)$. The third term is $\chi'(v_+) e^{\lambda r} \frac{M(r^2 + a^2)}{r} |\rd_r\Psi|^2$.  We now fix an $\rred \in (M, r_+)$ and choose $v_1$ big enough such that
\begin{equation} \label{EqControlRH}
    \textnormal{ the sum of those three terms is} \leq -c_0 \cdot \chi(v_+) e^{\lambda r} |\rd_r\Psi|^2 \qquad \textnormal{ for } \{\rred \leq r \leq r_+\} \cap \{f_+ \geq v_1\} 
\end{equation}
for some $c_0 >0$. Furthermore, Lemma \ref{LemmaAppendixDUWave} gives
\begin{equation} \label{EqControlDuwaveH}\uwave{D(\Psi)} \leq -c \lambda \chi(v_+) e^{\lambda r} \Big( - \Delta |\rd_r\Psi|^2 + |\rd_{v_+} \Psi|^2 + |\rd_\theta \Psi|^2 + |\Phi \Psi|^2 \Big) + C \lambda \chi(v_+) e^{\lambda r} |\Psi|^2
\end{equation}
for $r \in [r_-, r_+]$. Note that all the remaining terms in $D_{\mathrm{rem}}$ (and in 2.) which are quadratic in $\rd_r\Psi$ come with a factor of $\Delta$. We apply a weighted Cauchy--Schwarz inequality to the remaining terms of the form $\Rea(\Psi \ov{\rd_r\Psi}) $, $\Rea(\rd_{\varphi_+} \Psi  \ov{\rd_r\Psi}) $, $\Rea(\rd_{v_+} \Psi \ov{\rd_r\Psi})$, $\Rea(\rd_{v_+} \rd_r \psi \ov{\rd_r\Psi})$ to absorb $|\rd_r\Psi|^2$ into \eqref{EqControlRH} at the expense of a large term of the form $|\rd_{v_+} \Psi|^2$, $|\rd_{\varphi_+} \Psi|^2$, $|\Psi|^2$, $|\rd_{v_+} \rd_r \psi|^2$. We can now choose $\lambda_0 >1$ big enough and use Cauchy--Schwarz on all the remaining terms quadratic in derivatives of $\psi$ to obtain for all $\lambda \geq \lambda_0$
\begin{equation} \label{EqStep2H}
    \begin{split}
        RHS \eqref{EqMultiplierH} &\underset{a.i.}{\leq} \rd_{v_+}[A(\Psi)] + \rd_r[B(\Psi)] -c \chi(v_+) e^{\lambda r} \Big[ |\rd_r\Psi|^2  + \lambda \Big( |\Delta| |\rd_r\Psi|^2 + |\rd_{v_+} \Psi|^2 + |\rd_\theta \Psi|^2 + |\Phi \Psi|^2 \Big) \Big] \\
        &\quad + C_0 \lambda \chi(v_+) e^{\lambda r} |\Psi|^2 + C_0 \chi(v_+) e^{\lambda r} |\rd_{v_+} \rd_r \psi|^2 \\
        &\quad + \underbrace{\rd_r(\chi(v_+) \mu e^{\lambda r}| \Psi|^2) - \chi(v_+)\mu \lambda e^{\lambda r} | \Psi|^2 - 2\chi(v_+) \mu e^{\lambda r} \Rea(\overline{\Psi}\rd_r\Psi)}_{=0} \\
    &\quad + \underbrace{\rd_r \big(\chi(v_+) \mu e^{\lambda r} | \rd_{v_+} \rd_r \psi|^2 \big) - \chi(v_+) \mu \lambda e^{\lambda r} |\rd_{v_+} \rd_r \psi|^2 - 2 \chi(v_+) \mu e^{\lambda r} \Rea ( \ov{\rd_{v_+} \rd_r \psi} \rd_{v_+} \Psi)}_{=0}
    \end{split}
\end{equation}
in $r \in [\rred, r_+]$.

\pfstep{Step 3: Estimating the boundary terms and the remaining bulk terms} As in \eqref{EqEstUnderb} the two underbraced terms in \eqref{EqStep2H} can be estimated by
\begin{equation} \label{EqEstUnderbH}
    \begin{split}
    \underbrace{\ldots}_{=0} + \underbrace{\ldots}_{=0} \leq &\: \rd_r \big(\underbrace{ \chi(v_+) \mu e^{\lambda r} (|\Psi|^2 + |\rd_{v_+} \rd_r \psi|^2)}_{=:B_{\mathrm{mod}}} \big) - \frac{1}{2} \chi(v_+) \mu \lambda e^{\lambda r}(|\Psi|^2 +|\rd_{v_+} \rd_r \psi|^2 ) \\
    &\:+ 2 \chi(v_+) \frac{\mu}{\lambda} e^{\lambda r} (|\rd_r \Psi|^2 + |\rd_{v_+} \Psi|^2).
    \end{split}
\end{equation}
Furthermore, the Lemmas \ref{LemAppendixB}, \ref{LemAppendixBA} give
\begin{align*}
     B(\Psi) - A(\Psi) &\geq c \chi(v_+) e^{\lambda r} \Big( |\rd_r \Psi|^2 + |\rd_{v_+} \Psi|^2 + |\rd_\theta \Psi|^2 + |\Phi \Psi|^2 \Big) - C_1 \chi(v_+) e^{\lambda r} |\Psi|^2  \textnormal{ for } r \in [r_-, r_+] \\
     B(\Psi)&\geq c \chi(v_+) e^{\lambda r} \Big( |\Delta| \cdot |\rd_r \Psi|^2 + |\rd_{v_+} \Psi|^2 + |\rd_\theta \Psi|^2 + |\Phi \Psi|^2 \Big) - C_2 \chi(v_+) e^{\lambda r} |\Psi|^2  \textnormal{ for } r \in [r_-, r_+] \;.
\end{align*}
Choosing $\mu > \max\{C_1, C_2\}$ ensures that $B(\Psi) + B_{\mathrm{mod}}$ and $B(\Psi) - A(\Psi) + B_{\mathrm{mod}}$ are coercive also in $|\Psi|^2$ (and indeed also in $|\rd_{v_+} \rd_r \psi|^2$, but this will not be used). If we also choose $\mu > 2C_0$, then the two bulk terms right before the underbraced terms in the right-hand side of \eqref{EqStep2H} can be absorbed by the second term on the right-hand side of \eqref{EqEstUnderbH} (recall that $\lambda \geq \lambda_0 >1$). Finally, we choose $\lambda \geq \lambda_0$ even bigger if necessary such that $\frac{2\mu}{\lambda} < c$ such that the last term on the right-hand side of \eqref{EqEstUnderbH} can be absorbed by the third term on the right-hand side of \eqref{EqStep2H}. In summary, we have obtained for $r \in [\rred, r_+]$
\begin{equation} \label{EqStep3H}
    \begin{split}
        RHS \eqref{EqMultiplierH} &\underset{a.i.}{\leq} \rd_{v_+}[A(\Psi)] + \rd_r[B(\Psi) + B_{\mathrm{mod}}] \\
        &\qquad \quad -c \chi(v_+) e^{\lambda r} \Big[ |\rd_r\Psi|^2   + \lambda \Big( |\Delta| |\rd_r\Psi|^2 + |\rd_{v_+} \Psi|^2 + |\rd_\theta \Psi|^2 + |\Phi \Psi|^2 + |\Psi|^2 + |\rd_{v_+} \rd_r \psi|^2 \Big) \Big] \;.
    \end{split}
\end{equation}

\pfstep{Step 4: Putting it all together and estimating the LHS of \eqref{EqMultiplierH}} We integrate \eqref{EqStep3H} over the region $V(v'):= \{\rred \leq r \leq r_+\} \cap \{v_1 \leq f_+ \leq v'\}$ with respect to $\ud v_+ \wedge \ud r \wedge \vols = \frac{1}{\Sigma} \vol$ and use $\ud v_+ = \ud r$ on $\{f_+ = \mathrm{const}\}$ to obtain
\begin{equation*}
    \begin{split}
        &\int\limits_{\{f_+ = v'\} \cap V(v')} [B(\Psi) + B_{\mathrm{mod}} - A] \, \vols \ud v_+ + \int\limits_{\{r = \rred\} \cap V(v')} [B(\Psi) + B_{\mathrm{mod}}] \, \vols \ud v_+ \\
        &\qquad \qquad  + c \int\limits_{V(v')} \chi(v_+) e^{\lambda r} \Big[ |\rd_r \Psi|^2 + \lambda (d_\Delta (\Psi) +|\rd_{v_+} \rd_r \psi|^2) \Big] \, \vols \ud v_+ \ud r \\
        &\quad \leq \int\limits_{V(v')} \Rea ( \rd_r^2 F \cdot \chi(v_+) e^{\lambda r} \ov{N\Psi}) \, \vols \ud v_+ \ud r + \int\limits_{\{f_+ = v_1\} \cap V(v')} [B(\Psi) + B_{\mathrm{mod}} - A ] \, \vols \ud v_+ \\
        &\qquad \qquad  + \int\limits_{\Hp \cap V(v')} [B(\Psi) + B_{\mathrm{mod}}] \, \vols \ud v_+ \;.
    \end{split}
\end{equation*}
We apply a weighted Cauchy--Schwarz inequality to the first term on the right-hand side to absorb the $\ov{N \Psi}$ contribution into the bulk term on the LHS. We also let $v' \to \infty$ and, recalling the estimates on the boundary terms from Step 3,  drop some positive terms on the LHS to obtain
\begin{equation} \label{EqStep4H}
    \begin{split}
        \int\limits_{V(\infty)} v_+^q d(\rd_r^2 \psi) \, \vols \ud v_+ \ud r  &\leq C\Big(\int\limits_{V(\infty)}v_+^q |\rd_r^2 F|^2 \, \vols \ud v_+ \ud r + \int\limits_{\{f_+ = v_1\} \cap V(\infty)}[d(\rd_r^2 \psi) + d(\rd_r \psi)] \, \vols \ud v_+ \\
        &\qquad + \int\limits_{\Hp \cap V(\infty)} v_+^q[d_\Delta(\rd_r^2 \psi) + d_\Delta(\rd_r \psi)] \, \vols \ud v_+ \Big) \;.
    \end{split}
\end{equation}
The upper bound $B(\Psi) \leq C d_\Delta(\Psi)$ used on $\Hp$ follows directly from \eqref{EqUpBoundB}, recalling that $B = - \frac{1}{\lambda} \uwave{D}$.

\pfstep{Step 5: Estimating $\psi$} It follows from $\rd_r(v_+^q e^r |\phi|^2) - v_+^q e^r |\phi|^2 - 2v_+^q e^r \Rea( \ov{\phi} \cdot \rd_r \phi) = 0$ and Cauchy--Schwarz on the last term that $- \rd_r (v_+^q e^r |\phi|^2) \leq v_+^q e^r |\rd_r \phi|^2$. Here, $\phi$ is a $C^2$-regular spin $2$-weighted function. Integrating this over $\{v_1 \leq f_+ \leq v'\} \cap \{r' \leq r \leq r_+\}$ and letting $v' \to \infty$ gives
\begin{equation*}
\begin{split}
    \sup_{\rred \leq r' \leq r_+} \int\limits_{\{r = r'\} \cap V(\infty)} v_+^q e^r |\phi|^2 \, \vols \ud v_+ &\leq \int\limits_{\{f_+ = v_1\} \cap V(\infty)} v_+^q e^r |\phi|^2 \, \vols \ud v_+ + \int\limits_{\Hp \cap V(\infty)} v_+^q e^r |\phi|^2 \, \vols \ud v_+ \\
    &\quad + \int\limits_{V(\infty)} v_+^q e^r |\rd_r \phi|^2 \, \vols \ud v_+ \;.
    \end{split}
\end{equation*}
Applying this not only to $\phi$ but also to $\rd_r \phi$, $\rd_{v_+} \phi$, and $\wtZ_{i,+} \phi$ for $i=1,2,3$ gives
\begin{equation} \label{EqStep5H}
\begin{split}
    \sup_{\rred \leq r' \leq r_+} \int\limits_{\{r = r'\} \cap V(\infty)} v_+^q e^r d(\phi) \, \vols \ud v_+ &\leq \int\limits_{\{f_+ = v_1\} \cap V(\infty)} v_+^q e^r d(\phi) \, \vols \ud v_+ + \int\limits_{\Hp \cap V(\infty)} v_+^q e^r d(\phi) \, \vols \ud v_+ \\
    &\quad + \int\limits_{V(\infty)} v_+^q e^r d(\rd_r \phi) \, \vols \ud v_+ \;.
    \end{split}
\end{equation}
We now use \eqref{EqStep5H} with $\phi = \rd_r \psi$ to infer that 
\begin{equation} \label{EqStep5aH}
\sup_{\rred \leq r' \leq r_+} \int\limits_{\{r = r'\} \cap V(\infty)} v_+^q d(\rd_r \psi) \, \vols \ud v_+  \leq C \cdot RHS \eqref{EqStep4H} \;.
\end{equation}
Applying \eqref{EqStep5H} now with $\phi = \psi$ and using \eqref{EqStep5aH} concludes the proof.
\end{proof}




\section{Estimates for dynamical interior spacetime}\label{sec:main.est}

We will first recall the estimates proven in \cite{DafLuk17, DafLuk26} for the solution $(\calM,g)$. We will then prove stronger estimates that improve those in \cite{DafLuk26}, as well as control the global coordinates introduced in Section~\ref{sec:u'def}. After that we introduce a global dynamical principal null frame and prove estimates in the dynamical spacetime of \cite{DafLuk17, DafLuk26} in terms of this principal null frame. This will be used in the next section as an input to control the error terms arising in the nonlinear Teukolsky equation.

\subsection{Notational and analytic preliminaries}

\subsubsection{Sets}

In $\os{\calM}{1}$, we denote the constant-$(s,\ub)$ $2$-spheres by $S_{s,\ub}$. In $\os{\calM}{2}$, we denote the constant-$(u,\ub)$ $2$-spheres by $S_{u,\ub}$. 


\subsubsection{Dynamical quantities, Kerr quantities, and difference quantities}\label{sec:difference}

Consistent with the notation so far, for any geometric quantity, we use regular face to denote the quantity on the dynamic spacetime and bold face to denote the quantity on the background Kerr spacetime. (Note that this is different from the convention in \cite{DafLuk17, DafLuk26}, where background quantities were denoted with a subscript ${}_\Ke$.) 

Using the identification by the coordinates ($(s,\ub,\vartheta_*)$ coordinates in $\overset{\scriptscriptstyle{\text{[1]}}}{\calM}$ and $(u,\ub,\vartheta_*)$ coordinates in $\overset{\scriptscriptstyle{\text{[2]}}}{\calM}$), we also use $\,\,{}\widetilde{ }\,\,$ to denote difference quantities, i.e.,
\begin{equation}\label{eq:diff.schematic}
    \widetilde{\phi} = \phi - \pmb{\phi}.    
\end{equation}

For $S$-tangent tensors ($S_{s,\ub}$-tangent in $\os{\calM}{1}$ and $S_{u,\ub}$-tangent in $\os{\calM}{2}$) corresponding to the metric components in \eqref{Kerr.doublenull}, \eqref{eq:Kerr.s}, \eqref{eq:metric.form.1}, \eqref{eq:metric.form.2}, the Ricci coefficients in \eqref{eq:Ricci.def} or curvature components in \eqref{eq:curvature.def}, when the notation \eqref{eq:diff.schematic} is used, the differences are taken as $S$-tangent tensors. For instance, in $\os{\calM}{1}$,
$$\widetilde{\chi} = \chi - \pmb{\chi}$$
is understood as the difference of two $S_{s,\ub}$-tangent $2$-tensors, where $\chi$ is defined with respect to the dynamical null pair $(\os{\ee}{1}{}_3, \os{\ee}{1}{}_4)$ in \eqref{eq:dynamical.ee.3.4.def} and $\pmb{\chi}$ is defined with respect to the background null pair $(\os{\pmb{\ee}}{1}{}_3, \os{\pmb{\ee}}{1}{}_4)$ in \eqref{eq:background.RS.ee.3.4.def}.

We will only use \eqref{eq:diff.schematic} when there is no ambiguity: Later on, we will also identify the full $\calM$ with a subset of $\calM_{\mathrm{Kerr}}$; see Section~\ref{sec:identification}. In the transition region (which will be contained in $\os{\calM}{1}$), it will be important to distinguish the different identifications and we will explicitly write the identification maps (see Definition~\ref{def:Phi}).

\subsubsection{Schematic notation}\label{sec:schematic}

We will often use a schematic notation to simultaneously denote multiple quantities. We will introduce the sets of schematic difference quantities in \eqref{eq:S.def.RS}, \eqref{S.def}. In the schematic notation, $\tp$ would represent a quantity in $\calS_{\tp}$, etc.

We will also often write equations schematically, where the left-hand side of the equation is exact, but the right-hand side is only schematic. We use $\eqs$ when the equation is to be understood only schematically. We will also use the reduced schematic notation (see \cite[Section~7.1]{DafLuk17}), indicated by $\eqrs$, where we moreover ignore factors in the nonlinear terms which are bounded (in $L^\i$). We further use brackets to denote terms with one of the components in the brackets.
For instance, the notation $\tp (\tp, \tpH)$ denotes the sum of all terms of the form $\tp \tp$ or $\tp \tpH$.

\subsubsection{Norms} \label{SecNorms}

\begin{definition}[Norms]\label{def:norms}
    \begin{enumerate}
        \item (Volume form on $2$-spheres) In $\overset{\scriptscriptstyle{\text{[1]}}}{\calM}$ (respectively $\overset{\scriptscriptstyle{\text{[2]}}}{\calM}$), we use $\volg$ to denote the volume form induced by the Riemmanian metric $\gamma$ on the constant-$(s,\ub)$ (respectively constant-$(u,\ub)$) $2$-spheres.
        \item (Norms on $2$-spheres) In $\overset{\scriptscriptstyle{\text{[1]}}}{\calM}$ (respectively $\overset{\scriptscriptstyle{\text{[2]}}}{\calM}$), we define the $L^p(S_{s,\ub})$ (respectively $L^p(S_{u,\ub})$) norm for $p \in [1,\infty]$ with respectively to $\volg$.
        \item (Mixed norms) Introduce that $\overset{\scriptscriptstyle{\text{[1]}}}{\calM}$, introduce the $L^p_s L^q_{\ub} L^r(S)$ and $L^q_{\ub}L^p_s  L^r(S)$ norms for $p,q,r\in [1,\infty)$ as follows:
        \begin{align}
            \|\phi\|_{L^p_s L^q_{\ub} L^r(S)}:= &\: \Bigg( \int_{0}^{s_f} \Big(\int_{1}^{\ub_f} \|\phi\|_{L^r(S_{s,\ub})}^q \ud \ub \Big)^{\frac pq} \ud s\Bigg)^{\frac 1p}, \\
            \|\phi\|_{L^q_{\ub}L^p_s  L^r(S)}:=&\: \Bigg(\int_1^{\ub_f}\left(\int_{0}^{s_f} \|\phi\|_{L^r(S_{s,\ub})}^p \ud s\right)^{\frac qp} \ud\ub\Bigg)^{\frac 1q},
        \end{align}
        and with obvious changes when $p,q = \infty$. In particular, norms on the right are taken first. On $\os{\calM}{2}$, we likewise introduce the $L^p_u L^q_{\ub} L^r(S)$ and $L^q_{\ub}L^p_u  L^r(S)$ norms in a similar manner.
        \item (Norms for $S$-tangent tensors) If $\phi$ is not a scalar but an $S$-tangent tensor, then introduce $|\phi|_\gamma = \sqrt{(\gamma^{-1})^{A_1B_1} \cdots (\gamma^{-1})^{A_r B_r} \phi_{A_1,\cdots,A_r} \phi_{B_1,\cdots B_r}}$. The $L^p_s L^q_{\ub} L^r(S)$ norm of $\phi$ is then defined as the $L^p_s L^q_{\ub} L^r(S)$ norm of $|\phi|_\gamma$.
    \end{enumerate}
\end{definition}

\begin{remark}[Notations for unified treatment]
    Let us already mention that while in some of the arguments it is important to use a variety of mixed norms as in Definition~\ref{def:norms}, once all the estimates are obtained, it will be convenient below to only consider more simple and unified norms. The corresponding notations $\mathfrak w(\ub)$, $\opnorm{\cdot}$ and $\opnorm{\cdot}_*$ will be defined in Definition~\ref{def:RS.weight} and Definition~\ref{def:BS.general.norm}.
\end{remark}

\subsubsection{Frames and derivatives}

We will use different frame fields, including those introduced in Section~\ref{SecKerrBackground} and Section~\ref{sec:dynamical.spacetime}, as well as some other ones to be defined below. The reader may find it helpful to refer to the glossary in Section~\ref{sec:glossary.frame}.

We will also differentiate with respect to the coordinate vector fields in different coordinate systems. We use the convention that $\Big|_s$ denotes derivatives in the $(s,\ub,\vartheta_*)$ coordinate system (in $\os{\calM}{1})$, $\Big|_{DN}$ denotes derivatives in the $(u,\ub,\vartheta_*)$ coordinate system (in $\os{\calM}{2}$) and $\Big|_{DN'}$ denotes derivatives in the $(u',\ub,\vartheta_*)$ coordinate system (in all of $\calM$). Throughout, we use the usual convention $\vartheta_* = (\th_*,\varphi_*)$.

\subsubsection{Dependence of constants}

For the rest of this section, we use $A \ls B$ to denote that $A \leq CB$ for some constant $C>0$. \textbf{All implicit constants will depend only on $M$, $a$, $q$, $q_-$ and $q_{--}$}, where $q_{--}<q_-$ will be introduced in Proposition~\ref{prop:pointwise.improved.2} below.

\textbf{We will from now on freely choose $\ep$ smaller if necessary. }

\subsection{Estimates in the red-shift region}\label{sec:old.estimates.1}

\subsubsection{Main estimates in the red-shift region}

In $\overset{\scriptscriptstyle{\text{[1]}}}\calM$, all the geometric quantities satisfy similar bounds (in contrast to $\overset{\scriptscriptstyle{\text{[2]}}}\calM$). (Notice that the estimates for different geometric quantities were closed with different levels of regularity in \cite{DafLuk26}, but here we do not need ``top order'' estimates and simply collect the estimates we need after losing a finite number of derivatives.)

We describe the estimates that were proven in \cite{DafLuk26} with $\de$ set to $\f{q_--3}2$. We will consider the following quantities:
\begin{equation}\label{eq:S.def.RS}
\calS := \{ \widetilde{\gamma}, \widetilde{\gamma^{-1}}, \,\widetilde{f},\,\widetilde{h},\, \widetilde{\chib},\widetilde{\eta},\widetilde{\om},\, \widetilde{\chi},\widetilde{\xi}\}.
\end{equation}
Here, $\,\widetilde{}\,$ denotes the differences in the sense of described in Section~\ref{sec:difference}. The dynamical metric quantities $\widetilde{\gamma}, \widetilde{\gamma^{-1}}, \,\widetilde{f},\,\widetilde{h}$ are as in \eqref{eq:metric.form.1}, and the dynamical connection coefficients are defined with respect to the null pair $(\overset{\scriptscriptstyle{\text{[1]}}}{\ee}_3, \overset{\scriptscriptstyle{\text{[1]}}}{\ee}_4)$ in \eqref{eq:dynamical.ee.3.4.def}. The corresponding background quantities are defined with respect to \eqref{eq:Kerr.s} and background null pair $(\os{\pmb{\ee}}{1}{}_3, \os{\pmb{\ee}}{1}{}_4)$ in \eqref{eq:background.RS.ee.3.4.def}.

\begin{theorem}\label{thm:DL.2}
    For every $I_{\mathrm{red}} \in \mathbb Z_{\geq 0}$, there exists $I_0$ (denoting the total number of derivatives for the initial data in Section~\ref{SecPreciseAssump}) sufficiently large such that the following holds in $\overset{\scriptscriptstyle{\text{[1]}}}\calM\subset \calM$: 
    \begin{enumerate}
        \item The following pointwise estimates hold:
        \begin{equation}
            \begin{split}
                \sum_{i+j+k \leq I_{\mathrm{red}}} \sum_{\widetilde{\phi} \in \calS} \| \ub^{\f{q_--2}2} \os{\nab}{1}{}_{\os{\ee}{1}_3}^j \os{\nab}{1}{}_{\os{\ee}{1}_4}^k \os{\nab}{1}{}^i \widetilde{\phi} \|_{L^\i_s L^\i_{\ub}L^\i(S)} \ls \ep.
            \end{split}
        \end{equation}
        \item The following spacetime $L^2$ bounds hold:
        \begin{equation}
            \begin{split}
                \sum_{i+j+k \leq I_{\mathrm{red}}} \sum_{\widetilde{\phi} \in \calS} \| \ub^{\f{q_--2}2}  \os{\nab}{1}{}_{\os{\ee}{1}_3}^j \os{\nab}{1}{}_{\os{\ee}{1}_4}^k \os{\nab}{1}{}^i \widetilde{\phi} \|_{L^2_s L^2_{\ub}L^2(S)} \ls \ep.
            \end{split}
        \end{equation}
    \end{enumerate}
    Here, the notation denotes that the derivatives are understood as horizontal derivatives (Section~\ref{sec:horizontal.derivatives}) with respect to the $(\overset{\scriptscriptstyle{\text{[1]}}}{\ee}_3, \overset{\scriptscriptstyle{\text{[1]}}}{\ee}_4)$ and the dynamical metric $\os{g}{1}$.
\end{theorem}

\begin{remark}[Equivalence of norms]\label{rmk:eq.of.norm.RS}
    In Theorem~\ref{thm:DL.2}, we have fixed a particular way to write the norms. Nonetheless, as we pass to different coordinate systems and introduce new frames, it will be useful to notice equivalent ways to write the norms:
    \begin{enumerate}
        \item Since we control $\gamma - \pmb{\gamma}$, when we use the norms on the spheres (Definition~\ref{def:norms}) we can take them with respect to $\gamma$ or $\pmb{\gamma}$. 
        \item Moreover, using the estimates in Theorem~\ref{thm:DL.2} itself, we can control $\os{\nab}{1}{}_{\os{\ee}{1}_3} - \os{\pmb{\nab}}{1}{}_{\os{\pmb{\ee}}{1}_3}$, $\os{\nab}{1}{}_{\os{\ee}{1}_4} - \os{\pmb{\nab}}{1}{}_{\os{\pmb{\ee}}{1}_4}$ and $\os{\nab}{1} - \os{\pmb{\nab}}{1}$ and so all the derivatives $\os{\nab}{1}{}_{\os{\ee}{1}_3}^j \os{\nab}{1}{}_{\os{\ee}{1}_4}^k \os{\nab}{1}{}^i$ in Theorem~\ref{thm:DL.2} can be replaced by $\os{\pmb{\nab}}{1}{}_{\os{\pmb{\ee}}{1}_3}^j \os{\pmb{\nab}}{1}{}_{\os{\pmb{\ee}}{1}_4}^k \os{\pmb{\nab}}{1}{}^i$.
        \item Since commuting the derivatives would give rise to connection coefficient terms (and their derivatives), which in turn are controlled by Theorem~\ref{thm:DL.2}, the derivatives in Theorem~\ref{thm:DL.2} can be taken in any order.
    \end{enumerate}
     
\end{remark}

\begin{remark}[Sobolev embedding]\label{rmk:Sobolev.RS}
    The estimates in Theorem~\ref{thm:DL.2} in particular give strong control over the difference $\widetilde{\gamma} = \gamma - \pmb{\gamma}$ to control the areas and the isoperimetric constants of the spheres. As a result, using for instance \cite[Lemma~5.1, 5.2]{Chr}, the following estimate holds for arbitrary $S_{s,\ub}$-tangent tensor fields:
    \begin{equation}\label{eq:Sobolev.RS}
        \|\phi\|_{L^\i(S_{s,\ub})} \ls \sum_{i\leq 2} \| \os{\nab}{1}{}^i \phi\|_{L^2(S_{s,\ub})}.
    \end{equation}
    
\end{remark}

\subsubsection{Notations for general weight functions}

In view of the form of estimates in Theorem~\ref{thm:DL.2}, it will be convenient to introduce the following weight function which will be used throughout when considering $\os{\calM}{1}$.
\begin{definition}\label{def:RS.weight}
    We use $\mathfrak w(\ub)$ to denote a general weight such that 
    \begin{equation}
        \| \ub^{\f{q_--2}2}\mathfrak w(\ub) \|_{L^2_{\ub}} + \| \ub^{\f{q_--2}2}\mathfrak w(\ub) \|_{L^\i_{\ub}} \ls 1.
    \end{equation}
\end{definition}

Note that (after losing a finite number of derivatives for Sobolev embedding), the estimates in Theorem~\ref{thm:DL.2} exactly say that the difference of the geometric quantities and their derivatives are bounded above by a weight $\mathfrak w(\ub)$.

\subsubsection{Estimates in terms of $(\os{\ee}{1}_1, \os{\ee}{1}_2, \os{\ee}{1}_3, \os{\ee}{1}_4)$}

\begin{definition}
    We define the vector fields
        \begin{align}
            \os{\ee}{1}_1:= \os{\pmb{\ee}}{1}_1 := &\: \f{R}{\ell} \Big( \rd_{\th_*}\Big|_s - (\f{\rd \mathfrak h}{\rd\th_*}) \rd_{\varphi_*} \Big|_s \Big), \label{eq:fat.ee.1}\\
            \os{\ee}{1}_2 :=  \os{\pmb{\ee}}{1}_2:= &\: \f{1}{R\Si} \rd_{\varphi_*}\Big|_s.\label{eq:fat.ee.2}
        \end{align}
    Note that while $(\os{\ee}{1}_3, \os{\ee}{1}_4)$ by definition forms a null pair (with $\os{\ee}{1}_3$, $\os{\ee}{1}_4$ null, and $g(\os{\ee}{1}_3, \os{\ee}{1}_4)=-2$), the frame $(\os{\ee}{1}_1, \os{\ee}{1}_2, \os{\ee}{1}_3, \os{\ee}{1}_4)$ in general does not form a null frame.

    Define also the north and south versions in analogy with Definition~\ref{def:Kerr.double.null.rotation}:
    \begin{align}
    \os{\ee}{1}{}_1^{(N)} = \cos\varphi_* \os{\ee}{1}_1 - \sin \varphi_* \os{\ee}{1}_2,\quad \os{\ee}{1}{}_2^{(N)} = \sin\varphi_* \os{\ee}{1}_1 + \cos\varphi_* \os{\ee}{1}_2, \\
    \os{\ee}{1}{}_1^{(S)} = \cos\varphi_* \os{\ee}{1}_1 + \sin \varphi_* \os{\ee}{1}_2,\quad \os{\ee}{1}{}_2^{(S)} = -\sin\varphi_* \os{\ee}{1}_1 + \cos\varphi_* \os{\ee}{1}_2,
\end{align}
    and analogously, $\os{\pmb{\ee}}{1}{}_A^{(N)}$ and $\os{\pmb{\ee}}{1}{}_A^{(S)}$.
\end{definition}
An argument as in Proposition~\ref{prop:smmothness.background.e} shows that $(\os{\ee}{1}{}_1^{(N)}, \os{\ee}{1}{}_2^{(N)})$ is smooth away from the south pole and $(\os{\ee}{1}{}_1^{(S)}, \os{\ee}{1}{}_2^{(S)})$ is smooth away from the north pole. To unify notations, we will denote $\os{\ee}{1}{}_\mu^{(\cdot)}= \os{\ee}{1}{}_\mu$ if $\mu \in \{3,4\}$.

\begin{lemma}\label{lem:fat.e.full.frame.est}
    In $\os{\calM}{1}$, the following holds, where all estimates are understood to hold for $\th_* \in [0,\f{3\pi}4]$ if $^{(\cdot)} = ^{(N)}$ and for $\th_* \in [\f{\pi}4,\pi]$ if $^{(\cdot)} = ^{(S)}$:
    \begin{enumerate}
        \item The vectors $(\os{\ee}{1}{}_1^{(\cdot)}, \os{\ee}{1}{}_2^{(\cdot)}, \os{\ee}{1}_3, \os{\ee}{1}_4)$ is almost a null frame in the sense that
        $$g(\os{\ee}{1}{}_\nu'^{(\cdot)}, \os{\ee}{1}{}_\lambda'^{(\cdot)} )= \begin{cases}
        -2 & \hbox{if $\{\lambda,\nu\} = \{3,4\}$} \\
        \gamma(\os{\pmb{\ee}}{1}{}_\nu'^{(\cdot)}, \os{\pmb{\ee}}{1}{}_\lambda'^{(\cdot)} ) & \hbox{if $\lambda,\nu \in \{1,2\}$} \\
        0 & \hbox{otherwise}
    \end{cases}$$
        \item The vectors $(\os{\ee}{1}{}_1, \os{\ee}{1}{}_2)$ is almost an orthonormal basis on the tangent space $S_{s,\ub}$ in the sense that
        \begin{equation}\label{eq:transition.almost.orthonormal}
            \sum_{i_1+i_2+i_3+i_4 \leq I_{\mathrm{red}}} \Big| (\os{\ee}{1}{}_1^{(\cdot)})^{i_1}
         (\os{\ee}{1}{}_2^{(\cdot)})^{i_2}
         (\os{\ee}{1}{}_3^{(\cdot)})^{i_3}
         (\os{\ee}{1}{}_4^{(\cdot)})^{i_4}
         \Big( g(\os{\ee}{1}{}_A^{(\cdot)}, \os{\ee}{1}{}_B^{(\cdot)} ) - \de_{AB} \Big) \Big| \ls \ep \mathfrak w(\ub).
        \end{equation}
        \item It holds that
        $$[\os{\ee}{1}{}_A, \os{\ee}{1}{}_B^{(\cdot)}] = [\os{\pmb{\ee}}{1}{}_A, \os{\pmb{\ee}}{1}{}_B^{(\cdot)}], \quad [\os{\ee}{1}{}_3, \os{\ee}{1}{}_A^{(\cdot)}] = [\os{\pmb{\ee}}{1}{}_3, \os{\pmb{\ee}}{1}{}_A^{(\cdot)}]$$
        and, moreover, writing
        $$[\os{\ee}{1}{}_4, \os{\ee}{1}{}_A^{(\cdot)}] - [\os{\pmb{\ee}}{1}{}_4, \os{\pmb{\ee}}{1}{}_A^{(\cdot)}] = (\mathfrak j_A)^{\mu} \os{\pmb{\ee}}{1}_\mu,\quad [\os{\ee}{1}{}_4, \os{\ee}{1}{}_3^{(\cdot)}] - [\os{\pmb{\ee}}{1}{}_4, \os{\pmb{\ee}}{1}{}_3^{(\cdot)}] = (\mathfrak j_3)^{\mu} \os{\pmb{\ee}}{1}_\mu, $$
        it holds for $\mu = 1,2,3,4$ that
        \begin{equation}\label{eq:transition.commutators}
            \sum_{i_1+i_2+i_3+i_4 \leq I_{\mathrm{red}}} \Big| (\os{\ee}{1}{}_1^{(\cdot)})^{i_1}
         (\os{\ee}{1}{}_2^{(\cdot)})^{i_2}
         (\os{\ee}{1}{}_3^{(\cdot)})^{i_3}
         (\os{\ee}{1}{}_4^{(\cdot)})^{i_4}
         \Big( (\mathfrak j_{\cdot})^{\mu}\Big) \Big| \ls \ep \mathfrak w(\ub).
        \end{equation}
    \end{enumerate}
        
\end{lemma}
\begin{proof}
    The first part is a consequence of the fact that $(\os{\ee}{1}_3, \os{\ee}{1}_4)$ forms a null pair orthogonal to the $S_{s,\ub}$-spheres and that $(\os{\ee}{1}_1, \os{\ee}{1}_2)$ are tangent to the $S_{s,\ub}$-spheres. The second part follows from the fact that $\pmb{g}(\os{\ee}{1}{}_A, \os{\ee}{1}{}_A) = \pmb{\gamma}(\os{\pmb{\ee}}{1}{}_A, \os{\pmb{\ee}}{1}{}_A) = \de_{AB}$ and the estimates for $\gamma - \pmb{\gamma}$ in Theorem~\ref{thm:main.DL.est}.

    For the third part, we use that
    \begin{equation}\label{eq:ee.1.explcit}
        \os{\ee}{1}{}_A= \os{\pmb{\ee}}{1}{}_A,\quad \os{\ee}{1}{}_3 = \os{\pmb{\ee}}{1}{}_3, \quad \os{\ee}{1}{}_4 = \os{\pmb{\ee}}{1}{}_4 + (f - \pmb{f}) \os{\pmb{\ee}}{1}{}_3 + (h^A - \pmb{h}^A) \f{\rd}{\rd \vartheta_*^A}\Big|_s,
    \end{equation}
    and combine with part 1 and the estimates for $f-\pmb{f}$ and $h-\pmb{h}$ in Theorem~\ref{thm:main.DL.est}.
\end{proof}

\begin{proposition}\label{prop:RS.original.gauge.full.frame}
    In $\os{\calM}{1}$, the following estimates hold for $\th_* \in [0,\f{3\pi}4]$ if $^{(\cdot)} = ^{(N)}$ and for $\th_* \in [\f{\pi}4,\pi]$ if $^{(\cdot)} = ^{(S)}$:
    \begin{equation}
        \sum_{i_1+i_2+i_3+i_4 \leq I_{\mathrm{red}}-1} \Big| (\os{\ee}{1}{}_1^{(\cdot)})^{i_1}
         (\os{\ee}{1}{}_2^{(\cdot)})^{i_2}
         (\os{\ee}{1}{}_3^{(\cdot)})^{i_3}
         (\os{\ee}{1}{}_4^{(\cdot)})^{i_4}
         \Big( g(\nabla_{\os{\ee}{1}{}_\mu^{(\cdot)}} \os{\ee}{1}{}_\nu^{(\cdot)}, \os{\ee}{1}{}_\lambda^{(\cdot)} ) - \pmb{g}(\pmb{\nabla}_{\os{\pmb{\ee}}{1}{}_\mu^{(\cdot)}} \os{\pmb{\ee}}{1}{}_\nu^{(\cdot)}, \os{\pmb{\ee}}{1}{}_\lambda^{(\cdot)} ) \Big) \Big| \ls \ep \mathfrak w(\ub).
    \end{equation}
\end{proposition}
\begin{proof}
    Suppose $\nu \notin \{1,2\}$, then $g(\nabla_{\os{\ee}{1}{}_\mu^{(\cdot)}} \os{\ee}{1}{}_\nu^{(\cdot)}, \os{\ee}{1}{}_\lambda^{(\cdot)} ) - \pmb{g}(\pmb{\nabla}_{\os{\pmb{\ee}}{1}{}_\mu^{(\cdot)}} \os{\pmb{\ee}}{1}{}_\nu^{(\cdot)}, \os{\pmb{\ee}}{1}{}_\lambda^{(\cdot)} )$ can be expressed as the difference of one of the connection coefficients in \eqref{eq:Ricci.def} and then controlled by Theorem~\ref{thm:DL.2}. If $\nu \in \{ 1,2\}$ and $\lambda \notin \{1,2\}$, we can write
     \begin{equation}
         \begin{split}
             &\: g(\nabla_{\os{\ee}{1}{}_\mu'^{(\cdot)}} \os{\ee}{1}{}_\nu'^{(\cdot)}, \os{\ee}{1}{}_\lambda'^{(\cdot)} ) - \pmb{g}(\pmb{\nabla}_{\os{\pmb{\ee}}{1}{}_\mu'^{(\cdot)}} \os{\pmb{\ee}}{1}{}_\nu'^{(\cdot)}, \os{\pmb{\ee}}{1}{}_\lambda'^{(\cdot)} ) \\
             = &\: \os{\ee}{1}{}_\mu'^{(\cdot)}\Big(g(\os{\ee}{1}{}_\nu'^{(\cdot)}, \os{\ee}{1}{}_\lambda'^{(\cdot)} )\Big) - \os{\pmb{\ee}}{1}{}_\mu'^{(\cdot)}\Big(\pmb{g}( \os{\pmb{\ee}}{1}{}_\nu'^{(\cdot)}, \os{\pmb{\ee}}{1}{}_\lambda'^{(\cdot)} )\Big) -g( \os{\ee}{1}{}_\nu'^{(\cdot)}, \nabla_{\os{\ee}{1}{}_\mu'^{(\cdot)}} \os{\ee}{1}{}_\lambda'^{(\cdot)} ) + \pmb{g}( \os{\pmb{\ee}}{1}{}_\nu'^{(\cdot)}, \pmb{\nabla}_{\os{\pmb{\ee}}{1}{}_\mu'^{(\cdot)}} \os{\pmb{\ee}}{1}{}_\lambda'^{(\cdot)} ).
         \end{split}
     \end{equation}
     Since $\nu \in \{ 1,2\}$ and $\lambda \notin \{1,2\}$, we have $g(\os{\ee}{1}{}_\nu'^{(\cdot)}, \os{\ee}{1}{}_\lambda'^{(\cdot)} ) =0$ and $\pmb{g}(\os{\pmb{\ee}}{1}{}_\nu'^{(\cdot)}, \os{\pmb{\ee}}{1}{}_\lambda'^{(\cdot)} )= 0$ by Lemma~\ref{lem:fat.e.full.frame.est}. We are thus left with the last two terms, which can be bounded by $\ep \mathfrak w(\ub)$ by Theorem~\ref{thm:DL.2}.

     Finally, we are left with the case $\nu,\lambda \in \{1,2\}$, which can be computed with Lemma~\ref{lem:computation.nablaslashed}. Now we either have terms expressed as connection coefficients in \eqref{eq:Ricci.def}, which can be controlled by Theorem~\ref{thm:DL.2}, or otherwise we have terms that can be bounded by \eqref{eq:transition.almost.orthonormal} or \eqref{eq:transition.commutators} in Lemma~\ref{lem:fat.e.full.frame.est}. \qedhere
\end{proof}

\subsection{Estimates in the blue-shift region and their refinements}\label{sec:DL.refined}

We first describe the estimates that were proven in \cite{DafLuk17}. 

Define, as in\footnote{Observe that $\widetilde{\zeta}$ is not part of $\mathcal S_{\tp}$ in \cite{DafLuk17}. However, using $\zeta = \eta$ (see \cite[Proposition~2.3]{DafLuk17}), we can trivially add $\widetilde{\zeta}$. This is convenient as we change frames later.} \cite{DafLuk17},
\begin{equation}\label{S.def}
\mathcal S_{\tg} := \{\widetilde{\gamma},\,\widetilde{\gamma^{-1}},\,\widetilde{\log\Om},\,\Om^{-2}\widetilde{\Om^2}\}, \quad \mathcal S_{\tp} := \{\widetilde{\eta},\,\widetilde{\etab},\, \widetilde{\zeta}\},\quad\mathcal S_{\tpHb} := \{\widetilde{\slashed{tr}\chib},\,\widetilde{\chibh}\}, \quad \mathcal S_{\tpH}:= \{\widetilde{\slashed{tr}\chi},\,\widetilde{\chih}\},
\end{equation}
where we used the convention for differences in Section~\ref{sec:difference}. In particular, the dynamical $\gamma$, $\Omg$ (and later on also $b$) are understood as in \eqref{eq:metric.form.2}, and the dynamical connection coefficients are all understood with respect to the $(\os{\ee}{2}_3, \os{\ee}{2}_4)$ null pair in \eqref{eq:dynamical.ee.2.3.4.def}, and the background quantities correspond to \eqref{Kerr.doublenull} and the $(\os{\pmb{\ee}}{2}_3, \os{\pmb{\ee}}{2}_4)$ null pair in \eqref{eq:background.ee.BS}. We will also use the convention as in Section~\ref{sec:schematic} that when we write $\tg$, it schematically represents one of the quantities $\tg \in \mathcal S_{\tg}$. We also define $K$ to be the Gauss curvature of the sphere $S_{u,\ub}$, and define $\pmb{K}$, $\widetilde{K}$ analogously using the convention in Section~\ref{sec:difference}. In addition to these quantities, we also control $\tb$ and $\widetilde{\om}$.

\textbf{For the rest of this subsection until Section~\ref{SecPfAuxEstimate}, we will fix the gauge in $\os{\calM}{2}$ as in Section~\ref{sec:M2} and the null pairs} $(\os{\ee}{2}_3, \os{\ee}{2}_4)$ \textbf{and} $(\os{\pmb{\ee}}{2}_3, \os{\pmb{\ee}}{2}_4)$\textbf{. We will therefore drop the oversets in our notations,} For instance, $\nab_3 = \os{\nab}{1}_{\os{\ee}{1}_3}$, $\pmb{\nab_3}\ = \os{\pmb{\nab}}{1}_{\os{\pmb{\ee}}{1}_3}$ etc.

The following estimates were established in \cite{DafLuk17}, with the choice of parameter such that $\de$ in \cite{DafLuk17} is set to $\f{q_--3}2$. Indeed, the quantities on the left-hand sides of \eqref{NS.def}, \eqref{NH.def} and \eqref{NI.def} are controlled by (the square root of) $\calN_{\mathrm{sph}}$, $\calN_{\mathrm{hyp}}$ and $\calN_{\mathrm{int}}$ in \cite[(4.24)]{DafLuk17}, \cite[(4.23)]{DafLuk17}, and \cite[(4.22)]{DafLuk17}, respectively, and these are shown to be bounded as part of the conclusion in \cite[Theorem~4.24]{DafLuk17}. (Note that there were other bounds that were established and needed in the argument in \cite{DafLuk17}, but since they are not relevant here, we will not cite those estimates.)
\begin{theorem}\label{thm:main.DL.est}
    For $I_0$ sufficiently large, the following estimates hold in $\os{\calM}{2}$:
    \begin{enumerate}
        \item The following fixed-sphere bounds hold:
        \begin{equation}\label{NS.def}
\begin{split}
&\sum_{i\leq 3}\sum_{\tg \in \mathcal S_{\tg} }\|\nab^i\tg\|_{L^\infty_{\ub}L^\infty_uL^2(S)}+\sum_{i\leq 2}\sum_{\substack{ \tpHb\in \mathcal S_{\tpHb} \\ \tp \in \mathcal S_{\tp}}}\|\nab^i(\tpHb,\widetilde{\omb},\tp,\tb)\|_{L^\infty_{\ub}L^\infty_uL^2(S)}\\
&+\sum_{i\leq 2}\sum_{\tpH\in \mathcal S_{\tpH}}\|\pmb{\Om}^2\nab^i\tpH\|_{L^\infty_{\ub}L^\infty_uL^2(S)}+\sum_{i\leq 1}\|\nab^{i}\tK\|_{L^\infty_{\ub}L^\infty_uL^2(S)} \ls \ep.
\end{split}
\end{equation}
    \item The following mixed norms estimates hold:
    \begin{equation}\label{NH.def}
        \begin{split}
\sum_{i\leq 2}\sum_{\tpHb\in \mathcal S_{\tpHb}}\||u|^{\f{q_--2}{2}}\nab^i\tpHb\|_{L^2_u L^\infty_{\ub}L^2(S)}+\sum_{i\leq 2}\sum_{\tpH\in \mathcal S_{\tpH}}\|\ub^{\f{q_--2}{2}}\pmb{\Om}^2\nab^i\tpH\|_{L^2_{\ub} L^\infty_uL^2(S)} \ls \ep.
        \end{split}
    \end{equation}
    \item The following spacetime $L^2$ bounds hold:
    \begin{equation}\label{NI.def}
\begin{split}
&\sum_{i\leq 3}\Big(\sum_{\tpHb\in \mathcal S_{\tpHb}}\||u|^{\f{q_--2}{2}} \pmb{\Om}\nab^i(\tpHb,\widetilde{\omb})\|_{L^2_uL^2_{\ub}L^2(S)}\\
&\qquad+\sum_{\substack{\tg \in \mathcal S_{\tg} \\ \tp \in \mathcal S_{\tp}}}\|\ub^{\f{q_--2}{2}} \pmb{\Om}(\nab^i(\tp,\tg,\tb),\nab^{\min\{i,2\}}\tK)\|_{L^2_uL^2_{\ub}L^2(S)}\\
&\qquad+\sum_{\tpH\in \mathcal S_{\tpH}}\|\ub^{\f{q_--2}{2}} \pmb{\Om}^3\nab^i\tpH\|_{L^2_uL^2_{\ub}L^2(S)} \Big) \ls \ep.
\end{split}
\end{equation}
    \end{enumerate}
\end{theorem}

\begin{remark}[Sobolev embedding]\label{rmk:Sobolev.BS}
    Like in $\os{\calM}{1}$ (see Remark~\ref{rmk:Sobolev.RS}), we also have in $\os{\calM}{2}$ sufficient control of the geometry to have Sobolev embedding for arbitrary $S_{u,\ub}$-tangent tensor fields (see \cite[Proposition~5.5]{DafLuk17}):
    \begin{equation}\label{eq:Sobolev.BS}
        \|\phi\|_{L^\i(S_{u,\ub})} \ls \sum_{i\leq 2} \| \os{\nab}{1}{}^i \phi\|_{L^2(S_{u,\ub})}.
    \end{equation}
\end{remark}
We need some improvements over the bounds in \cite{DafLuk17}. 
\begin{enumerate}
    \item We need to improve the decay rate in the pointwise estimates in \eqref{NS.def}. These follow from the proofs in \cite{DafLuk17}. They were not explicitly stated because they were not needed in \cite{DafLuk17}. This will be carried out in Proposition~\ref{prop:pointwise.improved}.
    \item We will improve the weights in the spacetime estimates \eqref{NI.def}, replacing $\Omg$ by $\brk{u+\ub}^{-\f 34}$ is some instances. This will be carried out in Proposition~\ref{prop:imp.int}.
    \item Finally, we need estimates for higher derivatives. The scheme in \cite{DafLuk17} in particular was designed so that estimates for $\alpha$ and $\alphab$ are not necessary to close the argument. We prove nonetheless that all higher derivative estimates can be derived a posteriori. This will be carried out in Proposition~\ref{prop:derivatives.all.directions}.
\end{enumerate}

\subsubsection{Improved fixed-sphere estimates}

We improve some of the estimates in \eqref{NS.def}. Compared to the energies defined in \cite[(4.24)]{DafLuk17}, the main difference is the additional $|u|^{\f{q_--3}2}$ or $\ub^{\f{q_--3}2}$ weights.

\begin{proposition}[Improved fixed-sphere estimates]\label{prop:pointwise.improved}
The following estimates hold in $\os{\calM}{2}$:
   \begin{equation}\label{eq:NS.improved}
\begin{split}
&\sum_{i\leq 2} \Big( \sum_{\substack{ \tpHb\in \mathcal S_{\tpHb}\\ \tp \in \mathcal S_{\tp}}}\||u|^{\f{q_--3}2} \nab^i(\tpHb,\widetilde{\omb},\tp,\tb)\|_{L^\infty_{\ub}L^\infty_uL^2(S)} + \sum_{\tpH\in \mathcal S_{\tpH}}\|\ub^{\f{q_--3}2} \pmb{\Om}^2 \nab^i \tpH\|_{L^\infty_{\ub}L^\infty_uL^2(S)}\Big) \\
& +\sum_{i\leq 3}\sum_{\tg \in \mathcal S_{\tg} }\||u|^{\f{q_--3}2}  \nab^i\tg\|_{L^\infty_{\ub}L^\infty_uL^2(S)} +\sum_{i\leq 1}\||u|^{\f{q_--3}2} \nab^{i}\tK\|_{L^\infty_{\ub}L^\infty_uL^2(S)}  \ls \ep.
\end{split}
\end{equation}
\end{proposition}
\begin{proof}

    We revisit the transport equations for the geometric quantities in \cite[Section~13]{DafLuk17}. In \cite[Section~13]{DafLuk17}, the pointwise estimate is obtained by using
    \begin{equation}\label{eq:u.integrate.wasteful}
        \Big(\int_{-\ub+C_R}^u |\mathfrak u|^{-1-(q_--3)} \, \ud \mathfrak u\Big)^{\f 12} \ls_{C_R} 1,\quad \Big(\int_{-u+C_R}^{\ub} \underline{\mathfrak{u}}^{-1-(q_--3)} \, \ud \underline{\mathfrak{u}}\Big)^{\f 12} \ls_{C_R} 1
    \end{equation}
    when $\ub +u \geq C_R$. It is straightforward to replace this by
    \begin{equation}\label{eq:u.integrate.better}
        \Big(\int_{-\ub+C_R}^u |\mathfrak u|^{-1-(q_--3)} \, \ud \mathfrak u\Big)^{\f 12} \ls_{C_R} |u|^{-\f{q_--3}2},\quad \Big(\int_{-u+C_R}^{\ub} \underline{\mathfrak{u}}^{-1-(q_--3)} \, \ud \underline{\mathfrak{u}}\Big)^{\f 12} \ls_{C_R} |u|^{-\f{q_--3}2}.
    \end{equation}
    We also need a slight variation of \eqref{eq:u.integrate.better}, namely, that
     \begin{equation}\label{eq:u.integrate.better.2}
        \Big(\int_{-\ub+C_R}^u \pmb{\Omg} |\mathfrak u|^{-1-(q_--3)} \, \ud \mathfrak u\Big)^{\f 12} \ls_{C_R} \ub^{-\f{q_--3}2}.
    \end{equation}
    To prove \eqref{eq:u.integrate.better.2}, we observe that $\pmb{\Omg} \ls e^{-c(u+\ub)}$ (by \cite[Proposition~A.15]{DafLuk17}), and so we need to bound
        $$\Big(\int_{-\ub+C_R}^u e^{-c(\mathfrak u+\ub)} |\mathfrak u|^{-1-(q_--3)} \, \ud \mathfrak u\Big)^{\f 12}.$$
    It is then straightforward to check, e.g., after integrating by parts, that \eqref{eq:u.integrate.better.2} holds. (In fact a stronger estimate, with $\ub^{-\f{q_--3}2}$ replaced by $\ub^{-\f 12 -\f{q_--3}2}$, holds.)

     We start considering specific estimates. In the proof of \cite[Proposition~13.3]{DafLuk17}, we use (the first estimate of) \eqref{eq:u.integrate.better} instead of \eqref{eq:u.integrate.wasteful} in \cite[(13.3)]{DafLuk17} and similarly for the remaining estimates in the proof. We thus obtain an additional weight in $|u|^{-\f{q_--3}2}$ and therefore obtain the desired estimates for 
     \begin{equation}
         \tg,\quad \widetilde{\etab},\quad\tb,\quad \tK.
     \end{equation}

     Turning to $\tpH$, we revisit the proof of \cite[Proposition~13.4]{DafLuk17}. We now put in an additional $\ub^{\f{q_--3}2}$ weight, which does not affect the integration in $u$. Notice that there are two types of terms in \cite[(13.6)]{DafLuk17}, namely those which contain a factor of $\nab^{i} \tpH$ (for which we use Gr\"onwall's inequality), and those which do not. For the terms which do not contain a factor of $\nab^{i} \tpH$, the key is to notice that there is at least an additional factor of $\pmb{\Omg}$ which can be used for integration. For instance, whereas in the proof of \cite[(13.6)]{DafLuk17}, only the bound $\|\pmb{\Omg}^3 \nab^3 \widetilde{\eta} \|_{L^1_uL^2(S)} \ls \ep$ was needed, in the present setting, we need to control the term $\|\ub^{\f{q_--3}2} \pmb{\Omg}^3 \nab^3 \widetilde{\eta} \|_{L^1_uL^2(S)}$. For this, we use Cauchy--Schwarz (noting $\varpi \geq 1$, $\pmb{\Omg}\ls 1$) to obtain
     $$\|\ub^{\f{q_--3}2} \pmb{\Omg}^3 \nab^3 \widetilde{\eta} \|_{L^1_u L^2(S)} \ls \| (\f{\ub}{|u|})^{\f{q_--3}2} |u|^{-\f 12} \pmb{\Omg} \|_{L^2_u L^2(S)} \| |u|^{\f{q_--2}2} \pmb{\Omg} \varpi^N \nab^3 \widetilde{\eta} \|_{L^2_u L^2(S)}.$$ 
     Now using \eqref{eq:u.integrate.better.2}, we thus have
     $$\| (\f{\ub}{|u|})^{\f{q_--3}2} |u|^{-\f 12} \pmb{\Omg} \|_{L^2_u L^2(S)} \ls \ub^{\f{q_--3}2} \Big(\int_{-\ub+C_R}^u \pmb{\Omg}^2 |\mathfrak u|^{-1-(q_--3)} \, \ud \mathfrak u\Big)^{\f 12} \ls 1.$$
     Thus altogether we obtain
     $$\|\ub^{\f{q_--3}2} \pmb{\Omg}^3 \nab^3 \widetilde{\eta} \|_{L^1_u L^2(S)} \ls \ep$$
     using the bound for $\calN_{\mathrm{hyp}}$ (see \cite[(4.23)]{DafLuk17}). The other terms which do not contain a factor of $\nab^{i} \tpH$ all have an additional factor of $\pmb{\Omg}$ and can thus be treated in a similar fashion. The terms that contain a factor of $\nab^{i} \tpH$ can be dealt with using Gr\"onwall's inequality, in the same way as in \cite[Proposition~13.4]{DafLuk17}, even in the presence of an additional $\ub^{\f{q_--3}2}$ weight. We have thus obtained the desired bound for 
     \begin{equation}
         \tpH.
     \end{equation}

     It remains to consider 
     $$\widetilde{\eta},\quad \tpHb.$$ 
     For $\widetilde{\eta}$, we argue as in \cite[Proposition~13.5]{DafLuk17}, except for using (the second estimate of) \eqref{eq:u.integrate.better} instead of \eqref{eq:u.integrate.wasteful}. Similarly, we also use \eqref{eq:u.integrate.better} for the bounds for $\tpHb$ in \cite[Proposition~13.6]{DafLuk17}. As for $\tpH$, there are terms that are treated with Gr\"onwall's inequality, but one can also put $|u|^{\f{q_--3}2}$ weights in the estimates. This thus concludes the proof. \qedhere

\end{proof}
    It will later on be useful to obtain an improvement when restricted to the past of $\Gamma$. For this purpose, observe the following:
    \begin{lemma}\label{lem:CGamma}
        Let $\Gamma$ be the pullback of the hypersurface in Kerr defined in Section~\ref{sec:Dafermos}. Then, when restricted to the past of $\Gamma$ in  $\os{\calM}{2}$, there exists a constant $C_\Gamma$ such that
        $$\os{\calM}{2} \cap \{f_\Gamma \leq 0\} \subset \os{\calM}{2} \cap \{u+\ub \leq \f 12 \f{\sigma_q}{\kappa_-}\log \ub + C_\Gamma\}.$$
    \end{lemma}
    \begin{proof}
        By Lemma~\ref{LemRelationStarAndNullCoord}, there exists $C>0$ (which is allowed to change in every step) such that to the past of $\Gamma$,
        \begin{equation}\label{eq:Gamma.in.u.ub}
            |u+\ub| \leq \f 12 |v_+ + v_-| + C \leq \f 12 \f{\sigma_q}{\kappa_-}\log(v_+)+C\leq \f 12 \f{\sigma_q}{\kappa_-}\log(\ub) + C.
        \end{equation}
        Label the last constant in \eqref{eq:Gamma.in.u.ub} as $C_{\Gamma}$. \qedhere
    \end{proof}
    We now show that Proposition~\ref{prop:pointwise.improved} can be improved when restricted to $\{u+\ub \leq \f 12 \f{\sigma_q}{\kappa_-}\log \ub + C_\Gamma\}$:
    \begin{proposition}\label{prop:pointwise.improved.2}
         When restricted to $\os{\calM}{2} \cap \{u+\ub \leq \f 12 \f{\sigma_q}{\kappa_-}\log \ub + C_\Gamma\}$, the estimates in Proposition~\ref{prop:pointwise.improved} can be further improved so that the $|u|^{\f{q_--3}2}$ and $\ub^{\f{q_--3}2}$ weights can be replaced by $|u|^{\f{q_{--}-2}2}$ and $\ub^{\f{q_{--}-2}2}$, i.e., 
        \begin{equation}\label{eq:NS.improved.2}
            \begin{split}
                &\sum_{i\leq 2} \Big( \sum_{\substack{ \tpHb\in \mathcal S_{\tpHb}\\ \tp \in \mathcal S_{\tp}}}\||u|^{\f{q_{--}-2}2} \nab^i(\tpHb,\widetilde{\omb},\tp,\tb)\|_{L^\infty_{\ub}L^\infty_uL^2(S)} + \sum_{\tpH\in \mathcal S_{\tpH}}\|\ub^{\f{q_{--}-2}2} \pmb{\Om}^2 \nab^i \tpH\|_{L^\infty_{\ub}L^\infty_uL^2(S)}\Big) \\
                & +\sum_{i\leq 3}\sum_{\tg \in \mathcal S_{\tg} }\||u|^{\f{q_{--}-2}2}  \nab^i\tg\|_{L^\infty_{\ub}L^\infty_uL^2(S)} +\sum_{i\leq 1}\||u|^{\f{q_{--}-2}2} \nab^{i}\tK\|_{L^\infty_{\ub}L^\infty_uL^2(S)}  \ls \ep.
            \end{split}
        \end{equation}
        where $q_{--}< q_-$ is arbitrarily close to $q_-$, but fixed.
    \end{proposition}
    \begin{proof}
         We reconsider the integrals \eqref{eq:u.integrate.better} and \eqref{eq:u.integrate.better.2}, which was the main point of the proof of Proposition~\ref{prop:pointwise.improved}, but with the restricted range of $(u,\ub)$ in \eqref{eq:Gamma.in.u.ub}. It follows that the length of the integrals is at most $\f 12 \f{\sigma_q}{\kappa_-}\log(\ub) + C_R+ C_\Gamma$, and thus (suppressing dependence on $C_R$, $C_\Gamma$, $\sigma_q$, $\kappa_-$)
        \begin{equation}\label{eq:u.integrate.better.b4.Gamma}
        \Big(\int_{-\ub+C_R}^u |\mathfrak u|^{-1-(q_--3)} \, \ud \mathfrak u\Big)^{\f 12} \ls |u|^{-\f{q_--2}2} \log (1+\ub) ,\quad \Big(\int_{-u+C_R}^{\ub} \underline{\mathfrak{u}}^{-1-(q_--3)} \, \ud \underline{\mathfrak{u}}\Big)^{\f 12} \ls \ub^{-\f{q_--2}2} \log (1+\ub) ,
    \end{equation}
     \begin{equation}\label{eq:u.integrate.better.b4.Gamma.2}
        \Big(\int_{-\ub+C_R}^u \pmb{\Omg} |\mathfrak u|^{-1-(q_--3)} \, \ud \mathfrak u\Big)^{\f 12} \ls_{C_R} |u|^{-\f{q_--2}2} \log (1+\ub).
    \end{equation}
    The condition $C_R \leq u+\ub \leq \f 12 \f{\sigma_q}{\kappa_-}\log \ub + C_\Gamma$ moreover enforces that $u$ and $\ub$ are comparable, and that the log-loss can be absorbed by replacing $\f{q_--2}2$ by $\f{q_{--}-2}2$ above. The remainder of the proof proceeds as in Proposition~\ref{prop:pointwise.improved}. \qedhere
    \end{proof}

\subsubsection{Improved integrated estimates}

\begin{proposition}[Improved integrated estimates]\label{prop:imp.int}
    \begin{equation}\label{eq:NI.strengthened}
\begin{split}
&\sum_{i\leq 3}\Big(\sum_{\tpHb\in \mathcal S_{\tpHb}}\||u|^{\f{q_--2}{2}} \varpi'^N \brk{u+\ub}^{-\f 34} \nab^i(\tpHb,\widetilde{\omb})\|_{L^2_uL^2_{\ub}L^2(S)}\\
&\qquad+\sum_{\substack{\tg \in \mathcal S_{\tg} \\ \tp \in \mathcal S_{\tp}}}\|\ub^{\f{q_--2}{2}}\varpi'^N\pmb{\Om}(\nab^i(\tp,\tg,\tb),\nab^{\min\{i,2\}}\tK)\|_{L^2_uL^2_{\ub}L^2(S)}\\
&\qquad+\sum_{\tpH\in \mathcal S_{\tpH}}\|\ub^{\f{q_--2}{2}}\varpi'^N \brk{u+\ub}^{-\f 34} \pmb{\Om}^2\nab^i\tpH\|_{L^2_uL^2_{\ub}L^2(S)} \Big) \ls N^{\f 12} 2^N \ep,
\end{split}
\end{equation}
where $\varpi'$ will be introduced in \eqref{eq:varpi'.def} and the estimate holds for any\footnote{Notice that in the final estimate in \cite{DafLuk17}, $N$ is fixed and the dependence of the estimate on $N$ is not written explicitly. Nevertheless, if one traces through the estimates in \cite[Section~12]{DafLuk17}, one finds the dependence of $N^{\f 12} 2^{2N}$.} $N \in \mathbb Z_{\geq 1}$ large.
\end{proposition}
\begin{proof}
    In \cite{DafLuk17}, a weight function 
    \begin{equation}
        \varpi  = 1+e^{-\f{r_+-r_-}{r_-^2+a^2}(u+\ub-C_R)}
    \end{equation}
    was used. The relevant properties of the weight $\varpi$ for the proof (see \cite[(4.19)--(4.21)]{DafLuk17}) in the region $u + \ub \geq C_R$ are as follows:
    \begin{enumerate}
        \item \begin{equation}\label{varpi.bd}
            1\leq \varpi\leq 2,
        \end{equation}
    \item \begin{equation}\label{varpi.34}
\f{\rd}{\rd u}\varpi=\f{\rd}{\rd \ub}\varpi\ls -\pmb{\Om}^2,
\end{equation}
    \item \begin{equation}\label{varpi.ang}
        \nab \varpi=0.
    \end{equation}
    \end{enumerate}

    Here, we define a different weight
    \begin{equation}\label{eq:varpi'.def}
        \varpi' := 1+(1+u+\ub-C_R)^{-\f 12}.
    \end{equation}
    The properties \eqref{varpi.bd} and \eqref{varpi.ang} obviously still hold. To see that \eqref{varpi.34} also holds, we compute that
    \begin{equation}\label{eq:varpi.34.improved}
        \f{\rd}{\rd u}\varpi'=\f{\rd}{\rd \ub}\varpi' = -\f{3}{2}(1+u+\ub-C_R)^{-\f 32}
    \end{equation}
    and recall from \cite[Proposition~A.15]{DafLuk17} that
    $$e^{-\f{r_+-r_-}{r_-^2+a^2}(u+\ub)} \ls_{C_R} \pmb{\Om}^2\ls_{C_R} e^{-\f{r_+-r_-}{r_-^2+a^2}(u+\ub)}.$$
    It thus follows that all the estimates in \cite{DafLuk17} hold with $\varpi$ replaced by $\varpi'$. (In particular, changing these weights do not change the error terms, except for replacing $\varpi$ by $\varpi'$, since whenever a derivative hits the weight, the term is considered as a good main term instead of an error term.)

    The upshot of using $\varpi'$ instead of $\varpi$ is that when differentiated, the derivatives of $\varpi'$ provide better control (weights of $(1+u+\ub-C_R)^{-\f 32}$ instead of $\pmb{\Om}^2$). Revisiting the proofs in \cite[Section~9]{DafLuk17}, all the integrated terms with $N\pmb{\Omg}^2$ can be replaced by $N (1+u+\ub-C_R)^{-\f 32}$.
\end{proof}

\begin{remark}
    It is possible to have an improvement that controls $$\sum_{\substack{\tg \in \mathcal S_{\tg} \\ \tp \in \mathcal S_{\tp}}}\|\ub^{\f{q_--2}{2}}\varpi'^N \brk{u+\ub}^{-\f 34} (\nab^i(\tp,\tg,\tb),\nab^{\min\{i,2\}}\tK)\|_{L^2_uL^2_{\ub}L^2(S)} \ls N^{\f 12} 2^N\ep$$ improving the weights above. However, this is not needed in our main theorem and requires slightly more changes to the argument in \cite{DafLuk17}. Therefore, we do not pursue the improved estimates here.
\end{remark}

\subsubsection{Higher order estimates}

We now take both the improved fixed-sphere estimates in Proposition~\ref{prop:pointwise.improved} and the improved integrated estimates in Proposition~\ref{prop:imp.int} and prove versions of them with higher order derivatives.

Let us make two remarks before we proceed. First, note that some form of higher derivative estimates were derived in \cite[Section~14]{DafLuk17}, but those bounds degenerate as $\ub \to \infty$. This is because in \cite{DafLuk17}, there is no assumptions about the behavior of the initial data as $\ub \to \infty$. In our case, since we impose such assumptions, we can therefore obtain the desired estimates for all higher derivatives. Second, in the statement of Proposition~\ref{prop:derivatives.all.directions} below, we allow ourselves to lose many derivatives, i.e., we will assume that the initial data obey bounds for many more derivatives (corresponding to $I_{\mathrm{blue}} \ll I_0$) and we will not keep track of the exact count.

\begin{proposition}[Estimates for derivatives in all directions]\label{prop:derivatives.all.directions}
    For every $I_{\mathrm{blue}} \in \mathbb Z_{\geq 1}$, there exists $I_0$ (denoting the total number of derivatives for the initial data in Section~\ref{SecPreciseAssump}) sufficiently large such that all the estimates in Proposition~\ref{prop:pointwise.improved} and Proposition~\ref{prop:imp.int} can be replaced by analogous versions with higher derivatives as follows when $N$ is sufficiently large. 
    \begin{enumerate}
        \item The following fixed-sphere bounds hold:
        \begin{equation}\label{eq:NS.higher}
\begin{split}
&\sum_{i+j+k\leq I_{\mathrm{blue}}} \Big( \sum_{\substack{ \tpHb\in \mathcal S_{\tpHb}, \tp \in \mathcal S_{\tp} \\ \tg \in \mathcal S_{\tg}}}\||u|^{\f{q_--3}2} \varpi'^N \nab_3^j (\Om^2\nab_4)^k\nab^i(\tpHb,\widetilde{\omb},\tp,\tb,\tg,\tK)\|_{L^\infty_{\ub}L^\infty_uL^2(S)} \\
&\qquad +\sum_{\tpH\in \mathcal S_{\tpH}}\|\ub^{\f{q_--3}2} \varpi'^N \pmb{\Om}^2 \nab_3^j (\Om^2\nab_4)^k\nab^i \tpH\|_{L^\infty_{\ub}L^\infty_uL^2(S)}  \Big) \ls N^{\f 12} 2^N \ep,
\end{split}
\end{equation}
    with the improvement that $\f{q_--3}2$ is replaced by $\f{q_{--}-2}2$ when restricted to $\os{\calM}{2} \cap \{u+\ub \leq \f 12 \f{\sigma_q}{\kappa_-}\log \ub + C_\Gamma\}$.
    \item The following integrated estimates hold:
    \begin{equation}\label{eq:NI.higher}
\begin{split}
&\sum_{i+j+k\leq I_{\mathrm{blue}}}\Big(\sum_{\tpHb\in \mathcal S_{\tpHb}}\||u|^{\f{q_--2}{2}}\varpi'^N \brk{u+\ub}^{-\f 34}\nab_3^j (\Om^2\nab_4)^k\nab^i(\tpHb,\widetilde{\omb})\|_{L^2_uL^2_{\ub}L^2(S)}\\
&\qquad+\sum_{\substack{\tg \in \mathcal S_{\tg} \\ \tp \in \mathcal S_{\tp}}}\|\ub^{\f{q_--2}{2}}\varpi'^N\pmb{\Om}\nab_3^j (\Om^2\nab_4)^k\nab^i(\tp,\tg,\tb,\tK)\|_{L^2_uL^2_{\ub}L^2(S)}\\
&\qquad+\sum_{\tpH\in \mathcal S_{\tpH}}\|\ub^{\f{q_--2}{2}}\varpi'^N \brk{u+\ub}^{-\f 34}\pmb{\Om}^2\nab_3^j (\Om^2\nab_4)^k\nab^i\tpH\|_{L^2_uL^2_{\ub}L^2(S)} \Big) \ls N^{\f 12} 2^N \ep.
\end{split}
\end{equation}
    \end{enumerate}
\end{proposition}
\begin{proof}
    In obtaining the estimates \eqref{eq:NS.higher} and \eqref{eq:NI.higher}, we also need to simultaneously obtain auxiliary bounds for the geometric quantities in some $L^2$, $L^\i$ mixed norms in $u$, $\ub$ (cf.~the $\calN_{hyp}$ norms in \cite[(4.23)]{DafLuk17} though we only need a much simpler version here):
    \begin{equation}\label{eq:NH.higher}
        \begin{split}
            & \sum_{\tpHb\in \mathcal S_{\tpHb}} \Big( \||u|^{\f{q_--2}{2}}\varpi'^N\nab_3^j\nab^i(\tpHb,\widetilde{\omb})\|_{L^2_uL^\infty_{\ub}L^2(S)}\\
            &\qquad \qquad \qquad \qquad +\sup_{\ub} |u|^{\f{q_--3}2} \int_{-\ub+C_R}^{u} \|\varpi'^N\nab_3^j (\Om^2\nab_4)^k\nab^i(\tpHb,\widetilde{\omb}) \|_{L^2(S_{\mathfrak u,\ub})} \ud \mathfrak u\Big) \\
            &+\sum_{\tpH\in \mathcal S_{\tpH}}\Big(\|\ub^{\f{q_--2}{2}}\varpi'^N\pmb{\Om}^2(\Om^2\nab_4)^k\nab^i\tpH\|_{L^2_{\ub}L^\infty_uL^2(S)} + \| \varpi'^N\pmb{\Om}^2\nab_3^j (\Om^2\nab_4)^k\nab^i\tpH\|_{L^\infty_uL^1_{\ub}L^2(S)} \Big)\ls N^{\f 12} 2^N \ep,
        \end{split}
    \end{equation}
    where the range of $i$, $j$ and $k$ is to be specified below. We also prove an improvement to \eqref{eq:NH.higher} for the second term, namely that when restricted to $\os{\calM}{2} \cap \{u+\ub \leq \f 12 \f{\sigma_q}{\kappa_-}\log \ub + C_\Gamma\}$, we have the improved estimate
    \begin{equation}\label{eq:NH.higher2}
        \begin{split}
            & \sum_{\tpHb\in \mathcal S_{\tpHb}} \sup_{\ub} |u|^{\f{q_{--}-2}2} \int_{-\ub+C_R}^{u} \|\varpi'^N\nab_3^j (\Om^2\nab_4)^k\nab^i(\tpHb,\widetilde{\omb}) \|_{L^2(S_{\mathfrak u,\ub})} \ud \mathfrak u \ls N^{\f 12} 2^N \ep.
        \end{split}
    \end{equation}.
    
    We make three remarks regarding \eqref{eq:NH.higher}: 
    \begin{enumerate}
        \item For the estimates of $\tpHb$ and $\widetilde{\omb}$ in $L^2_uL^\infty_{\ub}L^2(S)$, we only require $\nab_3$ and $\nab$ derivatives (but not $\Om^2\nab_4$ derivatives) on $(\tpHb,\widetilde{\omb})$. (Similarly for $\tpH$ after switching $\os{\ee}{2}{}_3\leftrightarrow \Om^2 \os{\ee}{2}{}_4$ and $u\leftrightarrow \ub$.)
        \item The norms in the second and fourth terms of \eqref{eq:NH.higher} are quite different. This has taken into account the difference between integrating towards increasing $\ub$ and decreasing $|u|$.
        \item When considering the same quantity, the weighted $L^2_uL^\infty_{\ub}L^2(S)$ estimate is stronger than the weighted $L^\infty_{\ub}L^1_u L^2(S)$ estimate (and similarly for $L^2_{\ub}L^\infty_u L^2(S)$ and $L^\infty_u L^1_{\ub} L^2(S)$). Indeed, by the Cauchy--Schwarz inequality,
        \begin{equation}\label{eq:compare.flux.bounds}
            \begin{split}
                &\: \sup_{\ub} |u|^{\f{q_--3}2} \int_{-\ub+C_R}^{u} \|\varpi'^N\nab_3^j \nab^i(\tpHb,\widetilde{\omb}) \|_{L^2(S_{\mathfrak u,\ub})} \ud \mathfrak u \\
                \ls &\: \sup_{\ub} |u|^{\f{q_--3}2} \Big(\int_{-\ub+C_R}^{u}  |\mathfrak u|^{-1-(q_--3)} \ud \mathfrak u\Big)^{\f 12}\||u|^{\f{q_--2}2}\varpi'^N\nab_3^j (\Om^2\nab_4)^k\nab^i(\tpHb,\widetilde{\omb}) \|_{L^\i_{\ub} L^2_u L^2(S)} \\
                \ls &\: \||u|^{\f{q_--2}2}\varpi'^N\nab_3^j (\Om^2\nab_4)^k\nab^i(\tpHb,\widetilde{\omb}) \|_{L^\i_{\ub} L^2_u L^2(S)}
            \end{split}
        \end{equation}
        and
        \begin{equation}
            \begin{split}
                \|\varpi'^N \pmb{\Omg}^2(\Om^2\nab_4)^k \nab^i \tpH \|_{L^\infty_u L^1_{\ub} L^2(S)} \ls &\:\|\ub^{\f{q_--2}{2}}\varpi'^N \pmb{\Omg}^2(\Om^2\nab_4)^k \nab^i \tpH \|_{L^\infty_u L^2_{\ub} L^2(S)} \| \ub ^{-\frac 12-\f{q_--3}2}\|_{L^2_{\ub}} \\
                \ls &\:\|\ub^{\f{q_--2}{2}}\varpi'^N \pmb{\Omg}^2(\Om^2\nab_4)^k \nab^i \tpH \|_{L^\infty_u L^2_{\ub} L^2(S)},
            \end{split}
        \end{equation}
        where in the last bound we used $\f{q_--3}2>0$. 
        \item When restricted to $u+\ub \leq \f 12 \f{\sigma_q}{\kappa_-}\log \ub + C_\Gamma$, if $k=0$, the bound \eqref{eq:NH.higher2} also follows from that for $\||u|^{\f{q_--2}2}\varpi'^N\nab_3^j (\Om^2\nab_4)^k\nab^i(\tpHb,\widetilde{\omb}) \|_{L^\i_{\ub} L^2_u L^2(S)}$ by the following improvement of \eqref{eq:compare.flux.bounds} in this subregion:
    \begin{equation}\label{eq:compare.flux.bounds.2}
            \begin{split}
                &\: \sup_{\ub} |u|^{\f{q_{--}-2}2} \int_{-\ub+C_R}^{-\ub+\f 12 \f{\sigma_q}{\kappa_-}\log \ub + C_\Gamma} \|\varpi'^N\nab_3^j \nab^i(\tpHb,\widetilde{\omb}) \|_{L^2(S_{\mathfrak u,\ub})} \ud \mathfrak u \\
                \ls &\: \sup_{\ub} |u|^{\f{q_{--}-2}2} \Big(\int_{-\ub+C_R}^{-\ub+\f 12 \f{\sigma_q}{\kappa_-}\log \ub + C_\Gamma}  |\mathfrak u|^{-(q_{--}-2)} \ud \mathfrak u\Big)^{\f 12}\||u|^{\f{q_{--}-2}2}\varpi'^N\nab_3^j (\Om^2\nab_4)^k\nab^i(\tpHb,\widetilde{\omb}) \|_{L^\i_{\ub} L^2_u L^2(S)} \\
                \ls &\: \||u|^{\f{q_--2}2}\varpi'^N\nab_3^j (\Om^2\nab_4)^k\nab^i(\tpHb,\widetilde{\omb}) \|_{L^\i_{\ub} L^2_u L^2(S)},
            \end{split}
        \end{equation}
        where now $|u|^{\f{q_{--}-2}2} \Big(\int_{-\ub+C_R}^{-\ub+\f 12 \f{\sigma_q}{\kappa_-}\log \ub + C_\Gamma}  |\mathfrak u|^{-(q_{--}-2)} \ud \mathfrak u\Big)^{\f 12}$ is bounded in this subregion for $q_{--} < q_-$.
    \end{enumerate}

    We now begin the proof. By choosing $I_0$ larger if necessary, we can assume $I_{\mathrm{red}}$ in Theorem~\ref{thm:DL.2} to be as large as we wish. As a result, we can assume that for the data for $\os{\calM}{2}$, posed on $\Sigma \subset \os{\calM}{1}$, all the difference quantities satisfy both $L^\i_{\ub}$ and $L^2_{\ub}$ bounds with a weight $\ub^{\f{q_--3}{2}}$ for $i+j+k \leq I_{\mathrm{red}}$, consistent with the estimates obtained in Theorem~\ref{thm:DL.2}. 

    In the following, we will also freely choose $N$ larger as needed. This parameter changes the $\varpi'$ weights in the norms and is used to absorb some bulk terms to the left-hand sides.
    
    \pfstep{Step~1: Higher angular derivatives} The first step is to derive estimates analogous to \eqref{eq:NS.higher},  \eqref{eq:NI.higher} and \eqref{eq:NH.higher}, but with $j,k=0$ and $i\leq I'$ for some $I_{\mathrm{blue}} \ll I' \ll I_{\mathrm{red}}  \ll I_0$.

    The estimates in Theorem~\ref{thm:main.DL.est}, Proposition~\ref{prop:pointwise.improved}, Proposition~\ref{prop:pointwise.improved.2}, and Proposition~\ref{prop:imp.int} above correspond to the $I' = 1$ case of what we need to prove (with up to $2$ or $3$ angular derivatives for some components).\footnote{Here, we used the remark above that the weighted $L^2_u L^\i_{\ub} L^2(S)$ and weighted $L^2_{\ub} L^\i_u L^2(S)$ bounds in \eqref{eq:NH.higher} are stronger, and so for \eqref{eq:NH.higher} we only need those bounds, which follow from \eqref{NH.def}.}  Moreover, in Theorem~\ref{thm:main.DL.est}, Proposition~\ref{prop:pointwise.improved}, Proposition~\ref{prop:pointwise.improved.2}, we trace through the dependence on $N$ in the argument to see that the right-hand sides of these estimates are bounded above by $N^{\f 12} 2^{N} \ep$.
    
    Now we make an observation that taking more angular derivatives does not change the argument in \cite{DafLuk17} (as well as in the above improvements). We thus obtain the desired higher angular derivative bounds in an identical manner.


    \pfstep{Step~2: Higher $\nab_3$ derivatives} From this point onwards, we start to gain $\nab_3$ and $\nab_4$ derivatives, but in the process, we allow ourselves to lose a finite number of angular derivatives. 
    
    Our goal in this step is to derive the $k=0$ case of the estimates \eqref{eq:NS.higher}--\eqref{eq:NH.higher2} for any $i$, $j$, as long as we allow for a loss of finitely many derivatives. More precisely, we induct in $J=1,\cdots, I_{\mathrm{blue}}$ (the number of $\nab_3$ derivatives). For every $J$, introduce $I_J$ satisfying 
    \begin{equation}\label{eq:sequence.of.I}
        I_{\mathrm{blue}} \ll I_{I_{\mathrm{blue}}} \ll I_{I_{\mathrm{blue}}-1} \ll \cdots \ll I_1 \ll I'
    \end{equation} such that we will prove the estimates \eqref{eq:NS.higher}--\eqref{eq:NH.higher2} with $i \leq I_J$, $j\leq J$, $k=0$.
    
    To make the exposition clearer, in the estimates below, we first focus on explaining how to control one $\nab_3$ derivatives for all the quantities in \eqref{eq:NS.higher}--\eqref{eq:NH.higher2}, i.e., we will consider the $J=1$ case with $i=0$. Since we can take a large number of angular derivatives in Step~1, the will then allow us to prove the estimates for $j\leq 1$ and $i \leq I_1$, for $I_1$ as in \eqref{eq:sequence.of.I}. In Step~2(d), we will then induct in the number of $\nab_3$ derivatives. 

    \pfstep{Step~2(a): Estimates for $\nab_3\nab^i\tg$, $\nab_3\nab^i\tb$, $\nab_3\nab^i\widetilde{\etab}$, $\nab_3\nab^i\tpH$, and $\nab_3\nab^i\tK$} We first consider quantities satisfying a $\nab_3$ equation. These include $\tg$, $\tb$, $\widetilde{\etab}$, $\tpH$, and $\tK$. For these quantities, it suffices to directly bound the right-hand sides of the equations. Because of the $L^\i$ control that we have (due to Step~1 and Sobolev embedding \eqref{eq:Sobolev.BS}), we can treat all the quadratic or higher order terms as linear, i.e., (recalling the $\eqrs$ notation in Section~\ref{NH.def})\footnote{One needs some care in deriving the reduced schematic equation for $\nab_3 \tpH$. In particular, there is a $\tpH \tpHb$ term, but by \eqref{NS.def} (and Sobolev embedding), only $\tpHb$, but not $\tpH$, is in $L^\i_u L^\i_{\ub} L^\i(S)$. This term will therefore be written as $\tpH$ instead of $\tpHb$.}
    \begin{equation}\label{eq:higher.order.nab3.schematic}
        \nab_3 (\tg,\tb,\widetilde{\etab},\tK) \eqrs \sum_{i\leq 2} \nab^i(\tpHb,\widetilde{\omb},\tg,\tb,\tp),\quad \nab_3 \tpH \eqrs (\tpH, \tpHb) + \sum_{i\leq 1} \nab^i(\tp,\tg, \tK).
    \end{equation}
    See\footnote{The reduced schematic equation in \cite{DafLuk17} was derived for $\nab_3 (\pmb{\Om}^2\tpH)$ instead of $\nab_3 \tpH$, but is easily shown to imply the reduced schematic equation above after noting $|\nab_3 \log \pmb{\Omg}| \sim 1$. } \cite[Propositions~7.8, 7.9, 7.10, 7.13, 7.15, 7.22]{DafLuk17}. It is straightforward to use the bounds that we have so far (which do not have $\nab_3$ derivatives) to bound the quantities on the left-hand side of \eqref{eq:higher.order.nab3.schematic}:
    \begin{enumerate}
        \item For the needed $L^\i_u L^\i_{\ub}L^2(S)$ and $L^2_u L^2_{\ub}L^2(S)$ estimates corresponding to those in \eqref{eq:NS.higher}--\eqref{eq:NI.higher}, including the needed improvement when restricted to $\os{\calM}{2} \cap \{u+\ub \leq \f 12 \f{\sigma_q}{\kappa_-}\log \ub + C_\Gamma\}$, we control the left-hand sides of \eqref{eq:higher.order.nab3.schematic} in these norms by bounding the right-hand sides in the same norms. Since the right-hand sides only have angular derivatives but not $\nab_3$ (or $\nab_4$) derivatives, this reduces to estimates that we have already derived in Step~1. Notice that for some of the terms, in order to obtain the desired bound, we will need to change between $|u|$ and $\ub$ weights, but observe that we have $|u| \ls \ub$ and $\ub \pmb{\Omg}^\alp \ls |u|$ (for $\alp>0$), where the latter follows from the fact that $\pmb{\Omg}\sim e^{-2\kappa_-(u+\ub)}$, and that $x^q e^{-x}$ is decreasing for $x \geq q >0$.
        \item For $\nab_3 \tpH$, we will also need to bound $\|\pmb{\Omg}^2 \nab_3 \tpH\|_{L^\i_u L^1_{\ub} L^2(S)}$; see \eqref{eq:NH.higher}. For this purpose, we control the $\|\pmb{\Omg}^2 \cdot \|_{L^\i_u L^1_{\ub} L^2(S)}$ norm of the terms on the right-hand side of the second equation in \eqref{eq:higher.order.nab3.schematic}. 
        We already have such a bound for $\tpH$ using \eqref{eq:NH.higher} in the $j,k=0$ case. On the other hand, while $\tpHb$ and $\tp$ do not have an explicitly stated $L^1_{\ub} L^2(S)$ bound, using the $L^\i_u L^\i_{\ub} L^2(S)$ estimate and the extra $\pmb{\Omg}^2$ factor, we have
        \begin{equation}\label{eq:L1.using.Omg}
            \begin{split}
                &\: \|\varpi'\pmb{\Omg}^2\nab^i (\tp, \tpHb,\tg,\tK)\|_{L^\i_u L^1_{\ub} L^2(S)} \\
                \ls &\: \sup_u \Big(\int_{-u+C_R}^\infty e^{-c(u+\ub)} \, \ud \ub\Big)^{\f 12} \| \varpi' (\tp, \tpHb,\tg,\tK)\|_{L^\i_u L^\i_{\ub} L^2(S)} \ls N^{\f 12} 2^N \ep.
            \end{split}    
        \end{equation}
        \item Observe also, importantly, that \eqref{eq:NH.higher} is defined so that we do not need to obtain an estimate for $\|\ub^{\f{q_--2}{2}}\varpi'^N\pmb{\Om}^2\nab_3\tpH\|_{L^2_{\ub}L^\infty_uL^2(S)}$.
        \item Note also that \eqref{eq:NH.higher2} is irrelevant here because it only concerns $\tpHb$ and $\widetilde{\omegab}$.
    \end{enumerate}
    We have thus proven all the needed bounds for $\nab_3\tg$, $\nab_3\tb$, $\nab_3\widetilde{\etab}$, $\nab_3\tpH$, and $\nab_3\tK$. At this point, observe also that since Step~1 allows us to take a large number of angular derivatives, we can control $\nab_3\nab^i\tg$, $\nab_3\nab^i\tb$, $\nab_3\nab^i\widetilde{\etab}$, $\nab_3\nab^i\tpH$, and $\nab_3\nab^i\tK$ for $i\leq I_1+3$, where $I_1$ is as in \eqref{eq:sequence.of.I}.

    \pfstep{Step~2(b): Estimates for $\nab_3\nab^i \widetilde{\eta}$} We now turn to the only quantities not satisfying $\nab_3$ equations, namely $\widetilde{\eta}$, $\tpHb$ and $\widetilde{\omb}$. For $\widetilde{\eta}$, we use (see \cite[(2.27)]{DafLuk17})
    $$\eta = 2 \nab \log\Om - \etab,\quad \nab_3 \log \Om = -\omb$$
    and the $\nab_3 \etab$ equation (see \cite[(3.3)]{DafLuk17}) to obtain a schematic equation
    $$\nab_3 \eta \eqs \sum_{i_1 + i_2 =1} \psi^{i_1} \nab^{i_2}(\psi_{\Hb},\omb).$$ 
    Taking difference with the corresponding equation on Kerr background, we thus obtain a $\nab_3\widetilde{\eta}$ equation and we can control $\nab_3 \widetilde{\eta}$ as the quantities in Step~2(a).

    In other words, as in Step~2(a), we have the estimates for $\nab_3\nab^i \widetilde{\eta}$ corresponding to\footnote{This includes, as above, the needed improvement when restricted to $\os{\calM}{2} \cap \{u+\ub \leq \f 12 \f{\sigma_q}{\kappa_-}\log \ub + C_\Gamma\}$.} \eqref{eq:NS.higher} and \eqref{eq:NI.higher} for $i \leq I_1+1$, where $I_1$ is as in \eqref{eq:sequence.of.I}.

    \pfstep{Step~2(c): Estimates for $\nab_3\nab^i \tpHb$ and $\nab_3\nab^i \widetilde{\omb}$} For $\tpHb$ and $\widetilde{\omb}$, we rely on the $\nab_4$ equations that they satisfy.
    We now use the transport estimate in \cite[Proposition~6.5]{DafLuk17} for $\widetilde{\phi} = \widetilde{\omb},\, \tpHb$, but with $\varpi'$ in place of $\varpi$ (see \eqref{eq:varpi'.def}, \eqref{eq:varpi.34.improved}):
    \begin{equation}\label{eq:transport.est.for.higher.order.est.1}
\begin{split}
&\||u|^{\f{q_--2}{2}}\varpi'^N\nab_3 \widetilde{\phi}\|_{L^2_uL^\infty_{\ub}L^2(S)}^2+N\||u|^{\f{q_--2}{2}}\varpi'^N \brk{u+\ub}^{-\f 34}\nab_3 \widetilde{\phi}\|_{L^2_uL^2_{\ub}L^2(S)}^2\\
\ls &\||u|^{\f{q_--2}{2}}\varpi'^N\nab_3 \widetilde{\phi}\|_{L^2_uL^2(S_{u,-u+C_R})}^2+\| |u|^{q_--2}\varpi'^{2N}\pmb{\Om}^2\nab_3 \widetilde{\phi}\nab_4\nab_3 \widetilde{\phi}\|_{L^1_{\ub}L^1_uL^1(S)}.
\end{split}
\end{equation}
    We also have the following variation of \eqref{eq:transport.est.for.higher.order.est.1} for every fixed $u$, which follows\footnote{The estimates in \cite{DafLuk17} technically correspond only to the $|u|^{\f{q_--2}{2}}$ weight, but the corresponding estimates with the weaker weights $|u|^{\f{q_{--}-2}{2}}$ and $|u|^{\f{q_--3}2}$ follow with the same proof.} from the last line of the proof of \cite[Proposition~6.5]{DafLuk17}::
    \begin{equation}\label{eq:transport.est.for.higher.order.est.2}
        \begin{split}
            &\||u|^{p}\varpi'^N\nab_3 \widetilde{\phi}\|_{L^\infty_{\ub}L^2(S)}^2+N\||u|^{p}\varpi'^N \brk{u+\ub}^{-\f 34}\nab_3 \widetilde{\phi}\|_{L^2_{\ub}L^2(S)}^2\\
            \ls &\||u|^{p}\varpi'^N\nab_3 \widetilde{\phi}\|_{L^2(S_{u,-u+C_R})}^2+\| |u|^{2p}\varpi'^{2N}\pmb{\Om}^2\nab_3 \widetilde{\phi}\nab_4\nab_3 \widetilde{\phi}\|_{L^1_{\ub}L^1(S)},\quad \hbox{$p = \f{q_{--}-2}{2}$ or $ \f{q_--3}2$}.
        \end{split}
    \end{equation}

    To control $\nab_4 \nab_3 \widetilde{\omb}$ and $\nab_4 \nab_3 \tpHb$ in \eqref{eq:transport.est.for.higher.order.est.1}--\eqref{eq:transport.est.for.higher.order.est.2}, we start with the equations $\nab_4 (\widetilde{\omb},\tpHb)$ (derived by taking difference of \cite[(3.14)]{DafLuk17} and \cite[(3.9)]{DafLuk17} with the corresponding background equations) 
    and then commute $\nab_3$ using 
    \begin{equation}\label{eq:34.commutator}
        \begin{split}
            \nab_4(\nab_3 \widetilde{\phi}) = \nab_3 (\nab_4 \widetilde{\phi}) - [\nab_3, \nab_4 ] \widetilde{\phi}
            = &\: \nab_3 (\nab_4 \widetilde{\phi}) - 2\omb \nab_4 \widetilde{\phi}-2(\gamma^{-1})^{BC}(\eta_B-\etab_B)\nab_C \widetilde{\phi},
        \end{split}
    \end{equation}
    we then obtain 
    \begin{equation}\label{eq:higher.nab3.commuted.eq}
        \begin{split}
            \nab_4 (\nab_3\widetilde{\omb},\nab_3\tpHb) \eqrs &\: \nab_3 \tpHb + \sum_{j_1+j_2\leq 1} \nab_3^{j_1}\tpHb\nab_3^{j_2}\tpH + \sum_{i\leq 1} \nab^i\tpHb + \sum_{\substack{ j\leq 1 \\ i\leq 1}} \nab_3^j \nab^i (\pmb{\Om}^2\tpH,\tp,\tg,\tK)\\
            =:&\:I + II+ III+ IV.
        \end{split}
    \end{equation}
    Using the estimates we already obtained (including the ones we just obtained in Steps 2(a) and 2(b)), we have
    \begin{equation}\label{eq:higher.nab3.better}
        \begin{split}
             \||u|^{\f{q_--2}2} \varpi'^N \pmb{\Omg} (|III| + |IV|) \|_{L^2_uL^2_{\ub}L^2(S)} ,\,\||u|^{\f{q_--3}2} \varpi'^N \pmb{\Omg}(|III| + |IV|)\|_{L^\i_uL^1_{\ub}L^2(S)} &\ls N^{\f 12} 2^N \ep.
        \end{split}
    \end{equation}
    Notice in particular that the $L^\i_uL^1_{\ub}L^2(S)$ estimates are obtained from integrating the $L^\i_uL^\i_{\ub}L^2(S)$ bound in a similar manner as \eqref{eq:L1.using.Omg} (using the $\pmb{\Omg}$ factor for integration). 
    We now return to the transport estimate \eqref{eq:transport.est.for.higher.order.est.1} for $\widetilde{\phi} = (\psi_{\Hb},\widetilde{\omb})$. For the last term on the right-hand side, we use \eqref{eq:higher.nab3.commuted.eq}, \eqref{eq:higher.nab3.better}, H\"older's inequality, and the estimates we have obtained so far (including that for $\| \pmb{\Om}^2 \nab_3\tpH\|_{L^\infty_uL^1_{\ub}L^\i(S)}$ that we just derived in Step~2(a) with Sobolev embedding) to get
    \begin{equation}\label{eq:higher.order.est.main.nab3.1}
        \begin{split}
            &\: \| |u|^{q_--2}\varpi'^{2N}\pmb{\Om}^2\nab_3 \widetilde{\phi} \nab_4\nab_3  \widetilde{\phi} \|_{L^1_{\ub}L^1_uL^1(S)} \\
            \ls &\:  N^{\f 12} 2^N \ep \| |u|^{\f{q_--2}2} \varpi'^N \pmb{\Omg} \nab_3 (\widetilde{\omb},\tpHb) \|_{L^2_u L^2_{\ub} L^2(S)} + \| |u|^{\f{q_--2}2} \varpi'^N \pmb{\Omg} \nab_3 (\widetilde{\omb},\tpHb) \|_{L^2_u L^2_{\ub} L^2(S)}^2 \\
            &\: + \sum_{j_1+j_2\leq 1}\| |u|^{\f{q_--2}2} \varpi'^N \nab_3 (\widetilde{\omb},\tpHb)\|_{L^2_uL^\i_{\ub}L^2(S)}  \| |u|^{\f{q_--2}2} \varpi'^N \nab_3^{j_1} \tpHb \|_{L^2_uL^\i_{\ub}L^2(S)}  \| \pmb{\Om}^2\nab_3^{j_2} \tpH\|_{L^\i_u L^1_{\ub} L^\i(S)}  \\
            \ls &\: N 2^{2N}\ep^2 + \| |u|^{\f{q_--2}2} \varpi'^N \pmb{\Omg} \nab_3 (\widetilde{\omb},\tpHb) \|_{L^2_u L^2_{\ub} L^2(S)}^2 + \ep N^{\f 12} 2^N\sum_{j \leq 1} \| |u|^{\f{q_--2}2} \varpi'^N \nab_3^j (\widetilde{\omb},\tpHb)\|_{L^2_uL^\i_{\ub}L^2(S)}^2.
        \end{split}
    \end{equation}
    Plugging \eqref{eq:higher.order.est.main.nab3.1} into \eqref{eq:transport.est.for.higher.order.est.1}, choosing $N$ large, and then $\ep$ small (so that $\ep N^{\f 12} 2^N \ll 1$), we can absorb terms to the left (noting $\pmb{\Omg} \ls \brk{u+\ub}^{-\f 34}$). We thus obtain
    $$\||u|^{\f{q_--2}{2}}\varpi'^N\nab_3 (\tpHb, \widetilde{\omb})\|_{L^2_uL^\infty_{\ub}L^2(S)}^2+N\||u|^{\f{q_--2}{2}}\varpi'^N \brk{u+\ub}^{-\f 34}\nab_3 (\tpHb, \widetilde{\omb})\|_{L^2_uL^2_{\ub}L^2(S)}^2\ls N 2^{2N}\ep^2.$$
    We have thus obtained the $(i,j,k) = (0,1,0)$ case for the estimates \eqref{eq:NI.higher} and \eqref{eq:NH.higher} of $\widetilde{\omb}$ and $\tpHb$.

    When restricted to $u+\ub \leq \f 12 \f{\sigma_q}{\kappa_-}\log \ub + C_\Gamma$, we need the bound for $\nab_3(\tpHb,\widetilde{\omb})$ corresponding to \eqref{eq:NH.higher2}. As we observed already, this follows from \eqref{eq:compare.flux.bounds.2} and the bounds we have achieved above.
        
    To obtain the fixed-sphere bound \eqref{eq:NS.higher}, we use \eqref{eq:transport.est.for.higher.order.est.2} with $p = \f{q_--3}{2}$. We control the right-hand side of \eqref{eq:transport.est.for.higher.order.est.2}. Using H\"older's inequality, \eqref{eq:higher.nab3.commuted.eq}, \eqref{eq:higher.nab3.better}, and the part of \eqref{eq:NH.higher} for $\tpH$ that has been established, we obtain
    \begin{equation}\label{eq:higher.order.est.main.nab3.2}
        \begin{split}
            &\: \| |u|^{q_--3}\varpi'^{2N}\pmb{\Om}^2\nab_3 (\tpHb, \widetilde{\omb})\nab_4\nab_3 (\tpHb, \widetilde{\omb})\|_{L^\i_u L^1_{\ub}L^1(S)}\\
            \ls &\: N^{\f 12} 2^N \ep \| |u|^{\f{q_--3}2} \varpi'^N \nab_3(\widetilde{\omb},\tpHb)\|_{L^\i_uL^\i_{\ub}L^2(S)} + \| |u|^{\f{q_--3}2} \varpi'^N \pmb{\Om} \nab_3(\widetilde{\omb},\tpHb)\|_{L^\i_uL^2_{\ub}L^2(S)}^2\\
            &\: + \sum_{j_1+j_2\leq 1}\| |u|^{\f{q_--3}2} \varpi'^N \nab_3 (\widetilde{\omb},\tpHb)\|_{L^\i_uL^\i_{\ub}L^2(S)}  \| |u|^{\f{q_--3}2} \varpi'^N \nab_3^{j_1} \tpHb \|_{L^\i_uL^\i_{\ub}L^2(S)}  \| \pmb{\Om}^2\nab_3^{j_2} \tpH\|_{L^\i_u L^1_{\ub} L^\i(S)} \\
            \ls &\: N^{\f 12} 2^N \ep \| |u|^{\f{q_--3}2} \varpi'^N \nab_3(\widetilde{\omb},\tpHb)\|_{L^\i_uL^\i_{\ub}L^2(S)} + \| |u|^{\f{q_--3}2} \varpi'^N \pmb{\Om} \nab_3(\widetilde{\omb},\tpHb)\|_{L^\i_uL^2_{\ub}L^2(S)}^2\\
            &\: + N^{\f 12} 2^N \ep \sum_{j\leq 1}\| |u|^{\f{q_--3}2} \varpi'^N \nab_3 (\widetilde{\omb},\tpHb)\|_{L^\i_uL^\i_{\ub}L^2(S)}  \| |u|^{\f{q_--3}2} \varpi'^N \nab_3^{j} \tpHb \|_{L^\i_uL^\i_{\ub}L^2(S)}.
        \end{split}
    \end{equation}
    Here, for $\| \pmb{\Om}^2\nab_3^{j_2} \tpH\|_{L^\i_u L^1_{\ub} L^\i(S)}$, we used the bounds in Steps~1 and 2(a) together with Sobolev embedding to get $L^\i(S)$ control. Plugging \eqref{eq:higher.order.est.main.nab3.2} into \eqref{eq:transport.est.for.higher.order.est.2}, choosing $N$ large and then $\ep$ small to absorb terms to the left, we obtain the desired estimate \eqref{eq:NS.higher} for $\nab_3 (\widetilde{\omb},\tpHb)$. This also completes the proof of all estimates when $(i,j,k) = (0,1,0)$.

    We also need the improved fixed-sphere bound, corresponding to changing $\f{q_--3}2$ in \eqref{eq:NS.higher} to $\f{q_{--}-2}2$ when $u+\ub \leq \f 12 \f{\sigma_q}{\kappa_-}\log \ub + C_\Gamma$. For this we use \eqref{eq:transport.est.for.higher.order.est.2} with $p = \f{q_{--}-2}{2}$, and restrict to this subregion. The key observation is that in this subregion, the second term in \eqref{eq:higher.nab3.better} obeys a better bound
    \begin{equation}
        \||u|^{\f{q_{--}-2}2} \varpi'^N \pmb{\Omg}(|III| + |IV|)\|_{L^\i_uL^1_{\ub}L^2(S)} \ls N^{\f 12} 2^N \ep.
    \end{equation}
    This is because the proof of \eqref{eq:higher.nab3.better} only rely on fixed sphere estimates, and those estimates are better in this subregion from the results in Step~1. Thus, repeating the same argument as before gives us the needed improvement when $u+\ub \leq \f 12 \f{\sigma_q}{\kappa_-}\log \ub + C_\Gamma$.

    Now we observe as in the end of Step~2(a) that Step~1 allows us to take higher angular derivatives. Thus, after losing finitely many angular derivatives, exactly the same argument works to obtain \eqref{eq:NS.higher}--\eqref{eq:NH.higher} with $i \leq I_1$, $j \leq 1$ and $k=0$, where $I_1$ is as in \eqref{eq:sequence.of.I}. (We observe also that we need to lose angular derivatives compared to the estimates in Steps~2(a) and 2(b) because of the term \eqref{eq:higher.nab3.commuted.eq} and angular Sobolev embedding. For this reason, we have allowed $i\leq I_1+3$, instead of $i\leq I_1$, in those estimates.)
    
    \pfstep{Step~2(d): Higher $\nab_3$ derivatives} We now derive estimates for higher $\nab_3$ derivatives by induction in the number of $\nab_3$ derivatives, i.e., assuming the estimates for $j\leq J$, $i \leq I_J$, $k=0$ and prove the estimates for $j \leq J+1$, $i \leq I_{J+1}$, $k=0$. In the process, we again allow ourselves to lose a large but finite number of angular derivatives. Now notice that the above argument in Steps~2(a)--2(c) is the special case where $J=0$. However, the exact same argument works for all $J$ to get estimates for one additional $\nab_3$ derivative, as long as we account for the loss of a finite number of derivatives as compared to data.
    
    \pfstep{Step~3: Higher $\Om^2\nab_4$ derivatives} We now consider higher derivatives in $\Om^2 \nab_4$, allowing also a large number of angular derivatives, but with $j=0$. (In particular, at this point we do not consider mixed $\nab_3$ and $\Om^2 \nab_4$ derivatives.) The basic strategy is the same as in the case of $\nab_3$ in Step~2, and our goal is to prove \eqref{eq:NS.higher}--\eqref{eq:NH.higher2} with $i \leq I_K$, $j=0$, $k\leq K$ for $K=0,\cdots, I_{\mathrm{blue}}$, with $I_1,\cdots,I_{\mathrm{blue}}$ as in \eqref{eq:sequence.of.I}.  For simplicity, we will again first discuss the estimates in the case of one $\Om^2 \nab_4$ and no angular derivatives.

    \pfstep{Step~3(a): Estimates for $\Om^2 \nab_4 \nab^i \widetilde{\eta}$, $\Om^2 \nab_4 \nab^i \tpHb $, $\Om^2 \nab_4 \nab^i \widetilde{\omb}$, $\Om^2 \nab_4 \tK$} We begin with quantities that satisfy a $\nab_4$ equation, i.e., $\widetilde{\eta}$, $\tpHb$, $\widetilde{\omb}$, and $\tK$. Arguing analogously as Step~2, for the $L^\i_u L^\i_{\ub}L^2(S)$ and $L^2_u L^2_{\ub}L^2(S)$ estimates\footnote{Again, as in Step~2(a), this includes the improved when $u+\ub\leq \f 12 \f{\sigma_q}{\kappa_-} \log \ub + C_\gamma$.} of the $\Omg^2 \nab_4$ derivatives of these quantities corresponding to those in \eqref{eq:NS.higher}--\eqref{eq:NI.higher}, we directly look at the $\nab_4$ equation, multiply by $\Omg^2$, and control the $L^\i_u L^\i_{\ub}L^2(S)$ and $L^2_u L^2_{\ub}L^2(S)$ norms of the right-hand sides. Notice that this multiplication by $\Omg^2$ is important: Unlike in the $\nab_3$ equations, the $\nab_4$ equations contain quantities that are only controlled in a space that with worse $\pmb{\Omg}$ weights.\footnote{For instance, $\tpH$ appears on the right-hand side of the equation $\nab_4\widetilde{\eta}$ so that only $\Omg^2 \nab_4 \widetilde{\eta}$ --- not $\nab_4 \widetilde{\eta}$ itself --- can be put into the desired space.} It is only after multiplying by $\Omg^2$ that we can control the right-hand sides of these equations in the necessary space.  

    For $\Omg^2\nab_4\tpHb$ and $\Omg^2\nab_4\widetilde{\omb}$, we in addition need to bound them in norms corresponding to the second term in \eqref{eq:NH.higher}. For this, first note the reduced schematic equation (see \cite[Proposition~7.18]{DafLuk17})
    \begin{equation}\label{eq:schematic.nab4.tpHb}
        \Omg^2\nab_4 (\tpHb, \widetilde{\omb}) \eqrs (\tpHb, \widetilde{\omb}) + \Omg^2(\Om^2 \tpH) + \sum_{i\leq 1} \Omg^2 \nab^{i} (\tp,\tg,\tb,\tK) =:I+II+III.
    \end{equation}
    The term $I$ already satisfies the desired bound by Step~1. For the terms $II$ and $III$, we have an extra $\Omg^2$ factor so that after using the $L^\i_u L^\i_{\ub}L^2(S)$ bounds (from Step~1), we have
    \begin{equation}\label{eq:arguing.for.nab4.tpHb.NH.norm}
        \begin{split}
            &\: \sup_{\ub} |u|^{\f{q_--3}2} \int_{-\ub+C_R}^{u} \|\varpi'^N (|II|+|III|) \|_{L^2(S_{\mathfrak u,\ub})} \ud \mathfrak u \\
            \ls &\: N 2^N \ep \sup_{\ub} |u|^{\f{q_--3}2} \Big(\int_{-\ub+C_R}^{u} e^{-c(\mathfrak u+\ub)} |\mathfrak u|^{-\f{q_--3}2} \ud \mathfrak u\Big) \ls N 2^N \ep,
        \end{split}
    \end{equation}
    where we used that the integral can be bounded by $|u|^{-\f{q_--3}{2}}$ after integration by parts. This gives the necessary bound for \eqref{eq:NH.higher}.

    When restricted to $u+\ub \leq \f 12 \f{\sigma_q}{\kappa_-}\log \ub + C_\Gamma$ we need to prove the improved estimate \eqref{eq:NH.higher2}. We return to the equation \eqref{eq:schematic.nab4.tpHb}. For $I$, the needed bound is already proven in Step~1. For the terms $II$ and $III$, we argue as in \eqref{eq:arguing.for.nab4.tpHb.NH.norm} except for using the fact that we have improved $L^\i_u L^\i_{\ub}L^2(S)$ in the subregion. This gives the needed bound corresponding to \eqref{eq:NH.higher2}.

    We have thus completed all the needed estimates for $\Omg^2\nab_4\widetilde{\eta}$, $\Omg^2\nab_4\tpHb$, $\Omg^2\nab_4\widetilde{\omb}$, and $\Omg^2\nab_4\tK$. Similarly as in Step~2, we can add additional angular derivatives and obtain the needed estimates for $(\Omg^2\nab_4)\nab^i\widetilde{\eta}$, $(\Omg^2\nab_4)\nab^i\tpHb$, $(\Omg^2\nab_4)\nab^i\widetilde{\omb}$, and $(\Omg^2\nab_4)\nab^i\tK$ for $i\leq I_1+4$.
    
    \pfstep{Step~3(b): Estimates for $\Om^2 \nab_4 \nab^i \tg$, $\Om^2 \nab_4 \nab^i \tb$, $\Om^2 \nab_4 \nab^i \widetilde{\etab}$} We next consider the quantities $\tg$, $\tb$, $\widetilde{\etab}$, and $\tpH$, which do not obey a $\nab_4$ equation. We argue in a similar manner as in Step~2, except for noticing that we have more quantities to consider in this case.

    We use the commutator formula in \eqref{eq:34.commutator} as above, but notice that the weight $\Omg^2$ generates a $\omb$ contribution since $\nab_3 \log \Omg = - \omb$ (see \cite[(2.27)]{DafLuk17}). Thus, we have 
    \begin{equation}\label{eq:34.commutator.2}
        \begin{split}
            \nab_3(\Omg^2 \nab_4 \widetilde{\phi}) = &\: \Omg^2\nab_4 (\nab_3 \widetilde{\phi}) + (\nab_3 \log \Omg) \Omg^2 \nab_4 \widetilde{\phi}+ \Omg^2[\nab_3, \nab_4 ] \widetilde{\phi} \\
            = &\: \Omg^2 \nab_3 (\nab_4 \widetilde{\phi}) +2\Omg^2 (\gamma^{-1})^{BC}(\eta_B-\etab_B)\nab_C \widetilde{\phi}.
        \end{split}
    \end{equation}
    The commuted equations take the following reduced schematic form (see \cite[Propositions~7.8, 7.9, 7.10, 7.13]{DafLuk17}):
    \begin{align}
        \nab_3 (\Omg^2 \nab_4 (\tg,\tb,\widetilde{\etab})) \eqrs &\: \Omg^2 (\Omg^2\nab_4)(\tg,\tb,\widetilde{\etab})+ \Omg^2 (\Omg^2\nab_4)\widetilde{\eta}+ \Omg^2 \sum_{i \leq 1} \nab^i (\tg,\tb,\tp) +  \sum_{\substack{k\leq 1 \\ i\leq 1}} (\Omg^2\nab_4)^k \nab^i \tpHb \nonumber \\
        =: &\: I + \cdots + IV. \label{eq:higher.order.main.nab4.1} 
    \end{align}

        We use the following transport estimates:
    \begin{equation}\label{eq:higher.order.transport.nab3.1}
        \begin{split}
            &\:\|\ub^{\f{q_--2}{2}}\varpi'^N\pmb{\Om}\Om^2 \nab_4\widetilde{\phi}\|_{L^2_{\ub}L^\infty_uL^2(S)}^2+N\|\ub^{\f{q_--2}{2}}\varpi'^N\pmb{\Om}\brk{u+\ub}^{-\f 34}\Om^2 \nab_4\widetilde{\phi}\|_{L^2_uL^2_{\ub}L^2(S)}^2\\
            &\: +\|\ub^{\f{q_--2}{2}}\varpi'^N\pmb{\Om}\Om^2 \nab_4\widetilde{\phi}\|_{L^2_uL^2_{\ub}L^2(S)}^2\\
            \ls &\:\|\ub^{\f{q_--2}{2}}\varpi'^N \pmb{\Om}\Om^2 \nab_4\widetilde{\phi}\|_{L^2_{\ub}L^2(S_{-\ub+C_R,\ub})}^2+\| \ub^{q_--2}\varpi'^{2N}\pmb{\Om}^2\Om^2 \nab_4\widetilde{\phi}\nab_3(\Om^2 \nab_4\widetilde{\phi})\|_{L^1_{\ub}L^1_uL^1(S)},
        \end{split}
    \end{equation}
        \begin{equation}\label{eq:higher.order.transport.nab3.2}
            \begin{split}
                &\:\|\ub^{\f{q_--2}{2}}\varpi'^N\Om^2 \nab_4\widetilde{\phi}\|_{L^2_{\ub}L^\infty_uL^2(S)}^2+N\|\ub^{\f{q_--2}{2}}\varpi'^N\brk{u+\ub}^{-\f 34}\Om^2 \nab_4\widetilde{\phi}\|_{L^2_uL^2_{\ub}L^2(S)}^2\\
                \ls &\:\|\ub^{\f{q_--2}{2}}\varpi'^N\Om^2 \nab_4\widetilde{\phi}\|_{L^2_{\ub}L^2(S_{-\ub+C_R,\ub})}^2+\| \ub^{q_--2}\varpi'^{2N}\Om^2 \nab_4\widetilde{\phi}\nab_3(\Om^2 \nab_4\widetilde{\phi})\|_{L^1_{\ub}L^1_uL^1(S)},
            \end{split}
        \end{equation}
        and, for every fixed $u$, $\ub$,
        \begin{equation}\label{eq:higher.order.transport.nab3.3}
            \begin{split}
                &\:\|\varpi'^N\Om^2 \nab_4\widetilde{\phi}\|_{L^2(S_{u,\ub})}^2+N\int_{-\ub+C_R}^u \|\varpi'^N\brk{u+\ub}^{-\f 34}\Om^2 \nab_4\widetilde{\phi}\|_{L^2(S_{\mathfrak u,\ub})}^2\,\ud \mathfrak u \\
                \ls &\:\|\varpi'^N\Om^2 \nab_4\widetilde{\phi}\|_{L^2(S_{-\ub+C_R,\ub})}^2+ \int_{-\ub+C_R}^u \| \varpi'^{2N}\Om^2 \nab_4\widetilde{\phi}\nab_3(\Om^2 \nab_4\widetilde{\phi})\|_{L^1(S_{\mathfrak u,\ub})}\ud \mathfrak u.
            \end{split}
        \end{equation}
        After accounting for the use of $\varpi'$ instead of $\varpi$, the estimate \eqref{eq:higher.order.transport.nab3.1} is taken from \cite[Proposition~6.3]{DafLuk17}, \eqref{eq:higher.order.transport.nab3.1} is taken from \cite[Proposition~6.2]{DafLuk17}, while the final estimate \eqref{eq:higher.order.transport.nab3.3} is a consequence of the proof of the same proposition \cite[Proposition~6.2]{DafLuk17}.

        The terms $II$, $III$ and $IV$ in \eqref{eq:higher.order.main.nab4.1} can be bounded directly by\footnote{Notice that we can put in an extra $\brk{u+\ub}^{\f 34}$ weight because there are enough powers of $\Omg$. In particular, we use that in the $L^2_u L^2_{\ub}L^2(S)$ estimate for $(\Omg^2\nab_4)^k \nab^i \tpHb$ (see \eqref{eq:NI.higher}), we have improved the weights to $\brk{u+\ub}^{-\f 34}$. Finally, we also note that applying $\ub \pmb{\Om}^\alp \ls_\alp |u|$,  $\forall \alp >0$ allows us to change the $\ub$-weights to $|u|$-weights.} $\|\ub^{\f{q_--2}2} \varpi'^N \pmb{\Om}\brk{u+\ub}^{\f 34} (|II|+ |III| + |IV|)\|_{L^2_{\ub}L^2_uL^2(S)} \ls N^{\f 12}2^N \ep$, using the estimates we have derived so far, including those from Step~1 and Step~3(a). Hence, using \eqref{eq:higher.order.transport.nab3.1} with $\widetilde{\phi} = \tg, \tb, \widetilde{\etab}$, we obtain
        \begin{equation}\label{eq:higher.order.nab4.est.1}
            \begin{split}
                &\:N\|\ub^{\f{q_--2}{2}}\varpi'^N\pmb{\Om}\brk{u+\ub}^{-\f 34}\Om^2 \nab_4(\tg,\tb,\widetilde{\etab})\|_{L^2_uL^2_{\ub}L^2(S)}^2\\
                \ls &\:N 2^{2N} \ep^2 + N^{\f 12} 2^N \ep\|\ub^{\f{q_--2}{2}}\varpi'^N\pmb{\Om}^2\Om^2 \nab_4(\tg,\tb,\widetilde{\etab})\|_{L^2_uL^2_{\ub}L^2(S)} + \|\ub^{\f{q_--2}{2}}\varpi'^N\pmb{\Om}^2\Om^2 \nab_4(\tg,\tb,\widetilde{\etab})\|_{L^2_uL^2_{\ub}L^2(S)}^2,
            \end{split}
        \end{equation}
        which implies the desired bound needed for \eqref{eq:NI.higher} after choosing $N$ large to absorb terms to the left-hand side. 
        
        For the estimate needed for \eqref{eq:NS.higher}, we use \eqref{eq:higher.order.transport.nab3.3}. First observe that using the $L^\i_uL^\i_{\ub}L^2(S)$ estimate for $II$ and $III$ in \eqref{eq:higher.order.main.nab4.1} from Step~1, we have 
        \begin{equation}\label{eq:higher.order.nab4.est.2}
            \begin{split}
                \int_{-\ub+C_R}^u \|\varpi'^N(|II|+ |III|)\|_{L^2(S_{\mathfrak u,\ub})}\ud \mathfrak u \ls N^{\f 12} 2^N \ep |u|^{-\f{q_--3}2} \int_{-\ub+C_R}^u \pmb{\Om}^2(\mathfrak u,\ub)\ud \mathfrak u \ls N^{\f 12} 2^N \ep |u|^{-\f{q_--3}2}.
            \end{split}
        \end{equation}
        For $IV$ in \eqref{eq:higher.order.main.nab4.1}, we use the estimate in \eqref{eq:NH.higher} derived in Steps~1 and 3(a),
        \begin{equation}\label{eq:higher.order.nab4.est.3}
            \begin{split}
                \int_{-\ub+C_R}^u \|\varpi'^{N} |IV|\|_{L^2(S_{\mathfrak u,\ub})}\ud \mathfrak u \ls N^{\f 12} 2^N \ep |u|^{-\f{q_--3}2}.
            \end{split}
        \end{equation}
        Thus, plugging \eqref{eq:higher.order.nab4.est.2}, \eqref{eq:higher.order.nab4.est.3} into \eqref{eq:higher.order.transport.nab3.3}, choosing $N$ large and absorbing terms to the left, we obtain
        \begin{equation}
            \begin{split}
                &\:\|\varpi'^N\Om^2 \nab_4 (\tg,\tb,\widetilde{\etab}) \|_{L^2(S_{u,\ub})}^2+N\int_{-\ub+C_R}^u \|\varpi^N\brk{u+\ub}^{-\f 34}\Om^2 \nab_4(\tg,\tb,\widetilde{\etab})\|_{L^2(S_{\mathfrak u,\ub})}^2\,\ud \mathfrak u \\
                \ls &\:\ep^2 N 2^{2N} |u|^{-(q_--3)} + \int_{-\ub+C_R}^u \|\varpi^N\Om^2 \nab_4(\tg,\tb,\widetilde{\etab})\|_{L^2(S_{\mathfrak u,\ub})}^2\,\ud \mathfrak u\\
                &\: + \Big( \sup_{\mathfrak u\in [-\ub+C_R, u]} \|\varpi'^N\Om^2 \nab_4 (\tg,\tb,\widetilde{\etab}) \|_{L^2(S_{u,\ub})} \Big) \int_{-\ub+C_R}^u \| \varpi'^{N} (|II| + |III|+ |IV|)\|_{L^1(S_{\mathfrak u,\ub})}\ud \mathfrak u \\
                \ls &\: \ep^2 N 2^{2N} |u|^{-(q_--3)}.
            \end{split}
        \end{equation}
        Multiplying by $|u|^{q_--3}$ and then taking supremum in $u$ and $\ub$, we obtain \eqref{eq:NS.higher} for $\Om^2\nab_4 (\tg,\tb,\widetilde{\etab})$.

        Observe also that we have the needed improvement of \eqref{eq:NS.higher} when $u+\ub \leq \f 12 \f{\sigma_q}{\kappa_-}\log \ub + C_\Gamma$. This is because under this restriction:
        \begin{itemize}
            \item The right-hand side of \eqref{eq:higher.order.nab4.est.2} improves to $N^{\f 12} 2^N \ep |u|^{-\f{q_{--}-2}2}$ by using the improved estimates in this subregion from Steps~1 and 3(a).
            \item To improve \eqref{eq:higher.order.nab4.est.3}, we use \eqref{eq:NH.higher2} instead of \eqref{eq:NH.higher}, which is valid in this subregion.
        \end{itemize}

        We have thus proven all the needed estimates for $\Om^2 \nab_4 \tg$, $\Om^2 \nab_4 \tb$, $\Om^2 \nab_4 \widetilde{\etab}$. As before, we can then get estimates for $\Om^2 \nab_4 \nab^i \tg$, $\Om^2 \nab_4 \nab^i \tb$, $\Om^2 \nab_4 \nab^i \widetilde{\etab}$ for $i \leq I_1+3$ (with the loss of one angular derivatives compared to Step~3(a) due to the last term in \eqref{eq:higher.order.main.nab4.1}.)

        \pfstep{Step~3(c): Estimates for $\Om^2\nab_4 \nab^i \tpH$} We now use \eqref{eq:34.commutator.2} and \cite[Proposition~7.15]{DafLuk17} to derive
        \begin{align}
            \nab_3 (\Omg^2 \nab_4 (\pmb{\Omg}^2\tpH)) \eqrs &\: \Omg^6\nab_4 \tpH + \Omg^2 \sum_{k_1+k_2\leq 1} (\Omg^2\nab_4)^{k_1}(\widetilde{\omb},\tpHb) (\Omg^2\nab_4)^{k_2}\tpH + \Omg^4\sum_{i\leq 1} \nab^i\tpH \nonumber\\
            &\: + \Omg^2\sum_{\substack{k\leq 1 \\ i \leq 1}} (\Omg^2\nab_4)^k\nab^i (\tpHb,\widetilde{\omb},\tp,\tg,\tK)=: V + \cdots + VIII. \label{eq:higher.order.main.nab4.2}
        \end{align}
            Note that in equation \eqref{eq:higher.order.main.nab4.2}, we first put in an $\pmb{\Omg}^2$ weight before commuting with $\Omg^2 \nab_4$ to remove an undesirable linear term; see details in \cite[Proposition~7.15]{DafLuk17}. The estimate for $\Om^2\nab_4 \tpH$ is analogous to the estimates for \eqref{eq:higher.nab3.commuted.eq} in Step~2(c). In this step, we will in particular use the bounds for $\Om^2 \nab_4(\tg,\tb,\widetilde{\eta})$ that we have just derived in Step~3(b). First, we have
        \begin{equation}\label{eq:higher.nab4.better}
        \begin{split}
             \|\ub^{\f{q_--2}2} \varpi'^N \pmb{\Omg} (|VII| + |VIII|) \|_{L^2_uL^2_{\ub}L^2(S)} ,\,\|\ub^{\f{q_--3}2} \varpi'^N \pmb{\Omg}(|VII| + |VIII|)\|_{L^\i_{\ub} L^1_u L^2(S)} &\ls N^{\f 12} 2^N \ep.
        \end{split}
        \end{equation}
        As in Step~2(c), the $L^2_uL^2_{\ub}L^2(S)$ estimate directly follows from the estimates we already have, while $L^\i_{\ub} L^1_u L^2(S)$ can be derived from integrating the $L^\i_u L^\i_{\ub} L^2(S)$ estimate and noting that 
        \begin{equation}
            \begin{split}
                \ub^{\f{q_--3}2} \Big( \int_{-\ub+C_R}^u \pmb{\Omg} |\mathfrak u|^{-\f{q_--3}2}\ud \mathfrak u\Big) \ls 1.
            \end{split}
        \end{equation}
        We now use \eqref{eq:higher.order.transport.nab3.2} for $\tpH$. When dealing with the weights, it will be convenient to note that
        \begin{equation}\label{eq:weights.commute}
            \Omg^2 \ls \pmb{\Omg}^2 \ls \Omg^2,\quad |\pmb{\Om}^2\nab_4\tpH|_\gamma \ls |\nab_4(\pmb{\Om}^2\tpH)|_\gamma + |\pmb{\Om}^2\tpH|_\gamma,\quad |\nab_4(\pmb{\Om}^2\tpH)|_\gamma \ls |\pmb{\Om}^2\nab_4\tpH|_\gamma + |\pmb{\Om}^2\tpH|_\gamma
        \end{equation}
        Thus, applying H\"older's inequality, using the first bound in \eqref{eq:higher.nab4.better} and other estimates already derived, and absorbing terms to the left (after choosing $N$ large and then $\ep$ small), we have
        \begin{equation}\label{eq:int.est.for.nab4.tpH}
            \begin{split}
                &\:\|\ub^{\f{q_--2}{2}}\varpi'^N\Om^2 \nab_4(\pmb{\Omg}^2\tpH)\|_{L^2_{\ub}L^\infty_uL^2(S)}^2+N\|\ub^{\f{q_--2}{2}}\varpi'^N\brk{u+\ub}^{-\f 34}\Om^2 \nab_4(\pmb{\Omg}^2\tpH)\|_{L^2_uL^2_{\ub}L^2(S)}^2\\
                \ls &\:N 2^{2N}\ep^2 + N^{\f 12} 2^N \ep\|\ub^{\f{q_--2}{2}}\varpi'^N\pmb{\Om} \Om^2 \nab_4(\pmb{\Omg}^2\tpH)\|_{L^2_uL^2_{\ub}L^2(S)} + \|\ub^{\f{q_--2}{2}}\varpi'^N\pmb{\Om} \Om^2 \nab_4(\pmb{\Omg}^2\tpH)\|_{L^2_uL^2_{\ub}L^2(S)}^2 \\
                &\: + \sum_{k_1+k_2\leq 1} \|\ub^{\f{q_--2}{2}}\varpi'^N\Om^2 \nab_4(\pmb{\Omg}^2\tpH)\|_{L^2_{\ub}L^\infty_uL^2(S)} \|\ub^{\f{q_--2}{2}}\varpi'^N(\Om^2 \nab_4)^{k_2}(\pmb{\Omg}^2\tpH)\|_{L^2_{\ub}L^\infty_uL^2(S)}\\
                &\: \qquad\qquad \times \| (\Omg^2 \nab_4)^{k_1}\tpHb \|_{L^\i_{\ub}L^1_uL^\i(S)} \ls N 2^{2N} \ep^2,
            \end{split}
        \end{equation}
        where we have noted that $\| (\Omg^2 \nab_4)^{k_1}\tpHb \|_{L^\i_{\ub}L^1_uL^\i(S)}$ is controlled by \eqref{eq:NH.higher}, after using the bounds derived in Steps~1 and 3(a), together with Sobolev embedding.
        This gives \eqref{eq:NI.higher} and \eqref{eq:NH.higher} for $\nab_4\tpH$ after noting \eqref{eq:weights.commute} again. 
        
        Finally, arguing similarly as \eqref{eq:int.est.for.nab4.tpH} except for using \eqref{eq:higher.order.transport.nab3.3} and the second bound in \eqref{eq:higher.nab4.better}, we obtain
        \begin{equation}\label{eq:higher.nab4.fixed.sphere.final}
            \begin{split}
                &\:\|\ub^{\f{q_--3}2}\varpi'^N\Om^2 \nab_4(\pmb{\Omg}^2\tpH)\|_{L^2(S_{u,\ub})}^2+N  \int_{-\ub+C_R}^u \|\ub^{\f{q_--3}2}\varpi^N\brk{u+\ub}^{-\f 34}\Om^2 \nab_4(\pmb{\Omg}^2\tpH)\|_{L^2(S_{\mathfrak u,\ub})}^2\,\ud \mathfrak u \\
                \ls &\:N 2^{2N} \ep^2+ N^{\f 12} 2^N \ep \|\ub^{\f{q_--3}2}\varpi'^N\Om^2 \nab_4(\pmb{\Omg}^2\tpH)\|_{L^\i_{\ub} L^\i_u L^2(S)} + \|\ub^{\f{q_--3}2}\varpi'^N\Om^3 \nab_4(\pmb{\Omg}^2\tpH)\|_{L^\i_{\ub} L^\i_u L^2(S)}^2 \\
                &\: + \sum_{k_1+k_2\leq 1} \|\ub^{\f{q_--3}2}\varpi'^N\Om^2 \nab_4(\pmb{\Omg}^2\tpH)\|_{L^\i_uL^\i_{\ub} L^2(S)} \|\ub^{\f{q_--3}2}\varpi'^N(\Om^2 \nab_4)^{k_2}(\pmb{\Omg}^2\tpH)\|_{L^\i_uL^\i_{\ub} L^2(S)} \\
                &\: \qquad \qquad \times \| (\Om^2 \nab_4)^{k_1}\tpHb\|_{L^\i_{\ub} L^1_uL^\i(S)}\\
                \ls &\: N 2^{2N}\ep^2,
            \end{split}
        \end{equation}
        which gives the desired estimate \eqref{eq:NS.higher} for $\Omg^2\nab_4 \tpH$ after using \eqref{eq:weights.commute}. For the needed improved estimate when $u+\ub \leq \f 12 \f{\sigma_q}{\kappa_-}\log \ub + C_\Gamma$, we observe that under the restriction, we can repeat the above argument but using the improved $L^\i_uL^i_{\ub}L^2(S)$ bounds (both in \eqref{eq:higher.nab4.better} and \eqref{eq:higher.nab4.fixed.sphere.final}) as input.

        \pfstep{Step~3(d): Higher $\Om^2 \nab_4$ derivatives} As in Step~2, we now use an induction argument in the number of $\Om^2\nab_4$ derivatives to conclude this step.
    
    \pfstep{Step~4: Mixed $\nab_3$ and $\Om^2\nab_4$ derivatives} The above argument does not bound quantities when differentiated by $\nab_3^j (\Om^2 \nab_4)^k\nab^i$ when $j$ and $k$ both $\neq 0$. Nonetheless, this case is in fact easier: Each quantity must satisfy either a $\nab_3$ or a $\nab_4$ equation. If it satisfies a $\nab_3$ equation, we differentiate it by $\nab_3^{j-1} (\Om^2 \nab_4)^k \nab^i$ and if it satisfies a $\nab_4$ equation, then we multiply by $\Om^2$ and then differentiate by $\nab_3^{j-1} (\Om^2 \nab_4)^{k-1} \nab^i$. Denoting $\widetilde{\phi}$ as the quantity to be controlled, we thus obtained one of the following:
    \begin{align}
        \nab_3^{j-1} (\Om^2 \nab_4)^k \nab^i \nab_3 \widetilde{\phi} = &\: \cdots, \\
        \nab_3^{j} (\Om^2 \nab_4)^{k-1} \nab^i \nab_4 \widetilde{\phi} = &\: \cdots.
    \end{align}
    To control $\nab_3^{j} (\Om \nab_4)^{k} \nab^i \widetilde{\phi}$, we thus need to commute the derivatives on the left-hand side and control the terms on the right-hand side. These terms either have a smaller total number of $\nab_3$ and $\Om \nab_4$ derivatives, or else has the same total number of $\nab_3$ and $\Om \nab_4$ derivatives but with fewer $\nab$ derivatives. One can therefore repeat this procedure inductively to reduce to terms that have been controlled in Steps~1--3. \qedhere



\end{proof}

From this point on, we fix $N$ and also remove the weights $\varpi'^N$ from all our estimates. All the dependence on $N$ can now be absorbed into the implicit constants in the estimates.

\textbf{In addition, since we have now closed all the higher order derivative estimates, from this point onward, we introduce the notation that $I \leq I_{\mathrm{blue}}$ denotes the total number of derivatives that we control. The number $I$ could decrease by a finite amount from line to line, with the only requirement that by the end we still have }
\begin{equation}\label{eq:I.convention}
    10^{10} \leq I \leq I_{\mathrm{blue}} \ll I_{\mathrm{red}} \ll I_0.
\end{equation}.

\subsubsection{Estimates for the curvature components}\label{sec:curv.est.BS}

In Proposition~\ref{prop:curvature.original} below, we will collect some estimates for the differences of the curvature components. These bounds follow quite straightforwardly from Proposition~\ref{prop:derivatives.all.directions} since we have already controlled all derivatives of the connection coefficients.

\begin{proposition}\label{prop:curvature.original}
    The following estimates hold for the differences of the curvature components decomposed with respect to the double null frame in $\os{\calM}{1}$:
    \begin{enumerate}
        \item The following fixed-sphere bounds hold:
        \begin{equation}\label{eq:curvature.double.null.final.pointwise}
        	\begin{split}
            &\: \sum_{i+j+k\leq I} \Big( \| |u|^{ \f{q_--3}2} \nab_3^j (\pmb{\Omg}^2 \nab_4)^k \nab^i (\widetilde{\alphab}, \widetilde{\betab}) \|_{L^\i_{\ub} L^\i_u L^2(S)} \\
            &\: \qquad + \| \ub^{\f{q_--3}2} \nab_3^j (\pmb{\Omg}^2 \nab_4)^k \nab^i (\pmb{\Omg}^2 \widetilde{\rho}, \pmb{\Omg}^2 \widetilde{\sigma}, \pmb{\Omg}^2 \widetilde{\beta}, \pmb{\Omg}^4 \widetilde{\alpha}) \|_{L^\i_{\ub} L^\i_u L^2(S)} \Big) \ls \ep,
        \end{split}
        \end{equation}
        with the improvement that $\f{q_--3}2$ is replaced by $\f{q_{--}-2}2$ when restricted to $\os{\calM}{2} \cap \{u+\ub \leq \f 12 \f{\sigma_q}{\kappa_-}\log \ub + C_\Gamma\}$.
        \item The following integrated estimates hold:
        \begin{equation}\label{eq:curvature.double.null.final.integrated}
        \begin{split}
            &\: \sum_{i+j+k\leq I} \Big(\| |u|^{\f{q_--2}2} \brk{u+\ub}^{-\f 34} \nab_3^j (\pmb{\Omg}^2 \nab_4)^k \nab^i (\widetilde{\alphab}, \widetilde{\betab}) \|_{L^2_{\ub} L^2_u L^2(S)} \\
            &\: \qquad + \| \ub^{\f{q_--2}2} \brk{u+\ub}^{-\f 34} \nab_3^j (\pmb{\Omg}^2 \nab_4)^k \nab^i (\pmb{\Omg}^2 \widetilde{\rho}, \pmb{\Omg}^2 \widetilde{\sigma}, \pmb{\Omg}^2 \widetilde{\beta}, \pmb{\Omg}^4 \widetilde{\alpha}) \|_{L^2_{\ub} L^2_u L^2(S)}\Big) \ls \ep.
        \end{split}
        \end{equation}
    \end{enumerate}
\end{proposition}
\begin{proof}
We prove this using the bounds of the connection coefficients and their derivatives in Proposition~\ref{prop:derivatives.all.directions}. Since the vacuum Einstein equations are satisfied, we use \cite[(3.2), (3.3), (3.5)]{DafLuk17} to relate the curvature components with the connection coefficients:
    \begin{align}
\beta =&\: -\div\chih + \frac 12 \slashed{\nabla} \trch - \zeta \cdot (\chi - \trch\gamma), \label{eq:Bianchi.for.curvature.1} \\
\betab  = &\: \div\chibh - \frac 12 \slashed{\nabla} \trchb - \zeta\cdot (\chibh-\trchb\gamma) ,\\
 \sigma = &\:  \curl\eta- \frac 1 2\chibh \wedge\chih, \label{eq:Bianchi.for.curvature.3}\\
\rho = &\: -K+\frac 1 2 \chih\cdot\chibh-\frac 1 4 \trch \trchb, \label{eq:Bianchi.for.curvature.4}\\
\alpha= &\: -\nab_4\chih - \trch \chih -2 \omega \chih, \\
\alphab =&\:  -\nab_3\chibh - \trchb\,  \chibh -2\omegab \chibh.\label{eq:Bianchi.for.curvature.6}
\end{align}
Here, $\div$, $\curl$ are angular operators (defined in \cite[(2.16), (2.17)]{DafLuk17}), but here we only need that they are angular operators. Notice also that the difference of the Gauss curvatures $K - \pmb{K}$ can be computed as angular derivatives of the difference $\gamma - \pmb{\gamma}$ (see \cite[Proposition~7.11]{DafLuk17}).

    To obtain the estimates \eqref{eq:curvature.double.null.final.pointwise}--\eqref{eq:curvature.double.null.final.integrated}, we directly consider the equations \eqref{eq:Bianchi.for.curvature.1}--\eqref{eq:Bianchi.for.curvature.6}, take the difference between the dynamical spacetime and the background Kerr spacetime, and then use the estimates in Proposition~\ref{prop:derivatives.all.directions}.

    There are two useful observations for this procedure:
    \begin{enumerate}
        \item There are estimates for which we need a $\ub^{\f{q_--3}2}$ or $\ub^{\f{q_--2}2}$ weight in \eqref{eq:curvature.double.null.final.pointwise}--\eqref{eq:curvature.double.null.final.integrated}, but the bound in Proposition~\ref{prop:derivatives.all.directions} for the terms on the right-hand side  only has a $|u|^{\f{q_--3}2}$ or $|u|^{\f{q_--2}2}$ weight. However, in all such cases where is extra $\pmb{\Om}$ power and we can use 
        \begin{equation}\label{eq:Omg.switch.u.ub}
            \ub \pmb{\Om}^\alp \ls_\alp |u|\quad \forall \alp >0.
        \end{equation}
        \item In \eqref{eq:NI.higher}, the integrated estimates of some components (namely $\tp$, $\tg$, $\tb$ and $\tK$) require a weight of $\pmb{\Omg}$ instead of $\brk{u+\ub}^{-\f 34}$. Nonetheless, when these terms arise there is always an additional power of $\pmb{\Omg}^2$. 
    \end{enumerate}
    We take the equation \eqref{eq:Bianchi.for.curvature.1} as an example, for which both observations are important. We will only write out the $i=j=k=0$ since the estimates in Proposition~\ref{prop:derivatives.all.directions} easily allow us to take higher $\nab_3^j (\pmb{\Omg}^2 \nab_4)^k \nab^i$ derivatives. Taking differences and using pointwise bounds (which follow from \eqref{eq:NS.higher} and Sobolev embedding \eqref{eq:Sobolev.BS}) for the nonlinear terms, we have, schematically,
    \begin{equation}
        \widetilde{\bt} \eqrs \sum_{i\leq 1}\nab^i(\tpH,\tg,\tp),
    \end{equation}
    cf.~\cite[(3.15)]{DafLuk17}. This immediately gives
    \begin{equation}\label{eq:DL.improved.bt.pointwise}
        \| \ub^{\f{q_--3}2} \pmb{\Omg}^2 \widetilde{\beta} \|_{L^\i_{\ub} L^\i_u L^2(S)} \ls   \sum_{i\leq 1} \Big(\|\ub^{\f{q_--3}2} \pmb{\Omg}^2 \nab^i \tpH \|_{L^\i_{\ub} L^\i_u L^2(S)} + \|\ub^{\f{q_--3}2} \pmb{\Omg}^2 \nab^i(\tg,\tp)\|_{L^\i_{\ub} L^\i_u L^2(S)}\Big),
    \end{equation}
    and 
    \begin{equation}\label{eq:DL.improved.bt.integrated}
        \begin{split}
        &\:\| \ub^{\f{q_--3}2} \brk{u+\ub}^{-\f 34}\pmb{\Omg}^2 \widetilde{\beta} \|_{L^2_{\ub} L^2_u L^2(S)} \\
        \ls &\: \sum_{i\leq 1} \Big(\|\ub^{\f{q_--3}2} \brk{u+\ub}^{-\f 34}\pmb{\Omg}^2 \nab^i \tpH \|_{L^2_{\ub} L^2_u L^2(S)} + \|\ub^{\f{q_--3}2} \brk{u+\ub}^{-\f 34} \pmb{\Omg}^2 \nab^i (\tg,\tp)\|_{L^2_{\ub} L^2_u L^2(S)}\Big).
        \end{split}
    \end{equation}
    In both \eqref{eq:DL.improved.bt.pointwise} and \eqref{eq:DL.improved.bt.integrated}, the $\tpH$ terms can be controlled directly using Proposition~\ref{prop:derivatives.all.directions}. The $\tg,\tp$ terms has an $\pmb{\Omg}^2$ factor so that they can be bounded Proposition~\ref{prop:derivatives.all.directions} after using \eqref{eq:Omg.switch.u.ub}. \qedhere
    
\end{proof}

\subsubsection{An auxiliary estimate in the blue-shift region} \label{SecPfAuxEstimate}

Here we establish an auxiliary estimate in the blue-shift region near the Cauchy horizon which is needed in the argument of \cite{Sbie24} (see \eqref{EqBoundConnectionB} in Proposition~\ref{PropAuxiliaryBounds}). 

Before that, we first need a lemma. Notice that the lemma is not already implied by Proposition~\ref{prop:derivatives.all.directions} because we have a $\ub^{\f{q_--3}{2}}$ weight here instead of a $|u|^{\f{q_--3}{2}}$ weight.
\begin{lemma}\label{lem:b.improve}
    For $I_0$ sufficiently large, the following estimates hold in $\os{\calM}{2}$:
    \begin{equation}
        \sum_{i\leq 2}\|\ub^{\f{q_--3}{2}} \nab^i \widetilde{b} \|_{L^\i_u L^\i_{\ub} L^\i(S)} \ls \ep.
    \end{equation}
\end{lemma}
\begin{proof}
    Observe that the estimate with $\ub^{\f{q_--3}{2}}$ replaced by $|u|^{\f{q_--3}{2}}$ already follows from Proposition~\ref{prop:derivatives.all.directions} (after using Sobolev embedding \eqref{eq:Sobolev.BS}). In particular the desired estimate holds for $|u| \geq \f 12 \ub$. We now consider the estimates for a point in $\os{\calM}{2}$ with $(u^{0},\ub^{0},\vartheta_*^{0})$ coordinates such that $|u^0|\leq \f 12 \ub^0$. Without loss of generality, assume also that $\ub^{0} \geq C_R$ for otherwise the estimate also follows from Proposition~\ref{prop:derivatives.all.directions}.

    Consider the reduced schematic equation satisfied by $\widetilde{b}$ (cf.~\cite[(7.10)]{DafLuk17}) for $i\leq 2$:
    \begin{equation}\label{eq:b.final.schematic}
        \nab_3 \nab^i \widetilde{b} \eqrs \pmb{\Omg}^2 \sum_{i' \leq i} \nab^{i'}(\tp,\tg,\tb) + \sum_{i_1+i_2 \leq i} \nab^{i_1} \tpHb \nab^{i_2} \tb,
    \end{equation}
    where we have used the $L^\i$ bounds in Proposition~\ref{prop:derivatives.all.directions} (after using Sobolev embedding \eqref{eq:Sobolev.BS}) to control many of the nonlinear terms. Using moreover the pointwise bounds for $\sum_{i' \leq i} \nab^{i'}(\tp,\tg,\tb)$ and $\nab^{i_1} \tpHb$ from Proposition~\ref{prop:derivatives.all.directions} and \eqref{eq:Sobolev.BS}, we know that 
    $$|\hbox{RHS of \eqref{eq:b.final.schematic}}|_\gamma \ls |u|^{-\f{q_--3}{2}} \Big(\pmb{\Omg}^2  + \sum_{i'\leq i} |\nab^{i'} \widetilde{b}|_\gamma\Big) $$

    Now given the point as above with $(u^{0},\ub^{0},\vartheta_*^{0})$ coordinates, consider the integral curve of $\os{\ee}{2}{}_3$ connecting this point with $(u,\ub,\vartheta_*) = (-\f 12 \ub^0,\ub^0,\vartheta_*)$. (Here we have noted that $\os{\ee}{2}{}_3 \ub = \os{\ee}{2}{}_3 \vartheta_*^A = 0$.) Along this integral curve, we have $\pmb{\Omg}^2 \ls e^{-c(u+\ub)} \ls e^{-c'\ub^0}$. Parametrizing the integral curve of $\os{\ee}{2}{}_3$ by $u$ (noting $\os{\ee}{2}{}_3 u = 1$), it follows from \eqref {eq:b.final.schematic} that along this integral curve
    \begin{equation}
        \sum_{i\leq 2} \Big| \f{\ud}{\ud u} (\ub^{q_--3}|\nab^i b|_\gamma^2) \Big| \ls  e^{-c'\ub} \ub^{q_--3} |u|^{-\f{q_--3}{2}} + |u|^{-\f{q_--3}{2}} \sum_{i \leq 2} (\ub^{q_--3}|\nab^{i'} \widetilde{b}|_\gamma^2).
    \end{equation}
    By Gr\"onwall's inequality, using $q_--3 >1$,  it follows that 
    $$\sum_{i\leq 2} (\ub^{q_--3}|\nab^i b|_\gamma^2)(u^{0},\ub^{0},\vartheta_*^{0}) \ls \sum_{i\leq 2} (\ub^{q_--3}|\nab^i b|_\gamma^2)(-\f 12 \ub^{0},\ub^{0},\vartheta_*^{0})+ e^{-c'\ub^0} (\ub^0)^{q_--3} \ls 1,$$
    where the final inequality follows from the observation in the beginning of the proof. Since $(u^{0},\ub^{0},\vartheta_*^{0})$ is an arbitrary point with $|u^0|\leq \f 12 \ub^0$, the bound follows.
\end{proof}

We now prove the main auxiliary estimate needed for \cite{Sbie24}.
\begin{proposition} \label{PropAuxEst}
    In the coordinate system $(u,\ub_{\CH^+},\th^1_{(i),\CH^+},\th^2_{(i),\CH^+})$ (for $i=1,2$) in $\os{\calM}{2}$ introduced in Section~\ref{sec:CH.coord}, the following estimates hold:
    \begin{equation}\label{eq:sbi.cond.goal}
        |\f{\rd b^A_{(i),\CH^+}}{\rd \th^B_{(i),\CH^+}}| \ls_u 1,
    \end{equation}
    uniformly in $\ub_{\CH^+}$ up to the Cauchy horizon, but with a constant that depends on $u$ (which $\to \infty$ as $u \to -\infty$).
\end{proposition}
\begin{proof}
    Let us recall the definition of $\th_{(i),\CH^+}$ in \eqref{eq:reg.coord.def.1} and \eqref{eq:reg.coord.def.2}. 

    We first control the second angular derivatives of $\th^A_{\CH^+}$. In view of \eqref{eq:reg.coord.def.2}, it suffices to establish the bound on $\{u=u_f\}$. Following the proof of \cite[Proposition~16.11]{DafLuk17}, but taking an additional angular derivative (and with the usual convention that background quantities are bold instead of denoted by ${}_{\Ke}$, we obtain
    \begin{equation}
        \begin{split}
            &\: \Big(\f{\rd}{\rd \ub}+ b_{(i)}^D\f{\rd}{\rd \th_{(i)}^D}\Big)\Big(\f{\rd^2\th_{(i),\CH^+}^A}{\rd\th_{(i)}^C \rd\th_{(i)}^B}-\f{\rd^2\pmb{\th}_{(i),\CH^+}^A}{\rd\th_{(i)}^C\rd\th_{(i)}^B}\Big) \\
            = &\: - \f{\rd b_{(i)}^D}{\rd \th_{(i)}^C}\Big(\f{\rd^2\th_{(i),\CH^+}^A}{\rd\th_{(i)}^D \rd\th_{(i)}^B}-\f{\rd^2\pmb{\th}_{(i),\CH^+}^A}{\rd\th_{(i)}^D\rd\th_{(i)}^B} \Big) \\
            &\:-\f{\rd}{\rd\th_{(i)}^C}\Big(\Big(\f{\rd b_{(i)}^D}{\rd\th_{(i)}^B}-\f{\rd \pmb{b}_{(i)}^D}{\rd\th_{(i)}^B}\Big)\f{\rd \pmb{\th}_{(i),\CH^+}^A}{\rd\th_{(i)}^D}+\f{\rd b_{(i)}^D}{\rd\th_{(i)}^B}\Big(\f{\rd\th_{(i),\CH^+}^A}{\rd\th_{(i)}^D}-\f{\rd\pmb{\th}_{(i),\CH^+}^A}{\rd\th_{(i)}^D}\Big)  - (\pmb{b}_{(i)}^D - b_{(i)}^D) \f{\rd^2\pmb{\th}_{(i),\CH^+}^A}{\rd \th_{(i)}^D \rd\th_{(i)}^B}\Big).
        \end{split}
    \end{equation}
    We now apply the same argument as in \cite[Proposition~16.11]{DafLuk17}, noting that estimates for $b-\pmb{b}$ obtain above in Lemma~\ref{lem:b.improve}, together with the higher order derivative estimates for the metric in Proposition~\ref{prop:derivatives.all.directions}, imply that we also control the $L^\i_u L^1_{\ub}L^\i(S)$ norm of second coordinate angular derivatives of $b- \pmb{b}$. Thus the same argument as in \cite[Proposition~16.11]{DafLuk17} with Gr\"onwall's inequality gives 
    \begin{equation}\label{eq:angular.second.derivative.bound}
        \begin{split}
            \Big|\f{\rd^2\th_{(i),\CH^+}^A}{\rd\th_{(i)}^C \rd\th_{(i)}^B}-\f{\rd^2\pmb{\th}_{(i),\CH^+}^A}{\rd\th_{(i)}^C\rd\th_{(i)}^B}\Big| \ls \ep.
        \end{split}
    \end{equation}
    
    By \cite[Theorem~16.14]{DafLuk17}, in the $(u,\ub_{\CH^+},\th^1_{(i),\CH^+},\th^2_{(i),\CH^+})$ coordinate system, $b_{(i),\CH^+}^A$ takes the form:
    $$b_{(i),\CH^+}^A = e^{\f{r_+-r_-}{r_-^2 + a^2}\ub} \Big( \f{\rd \th^A_{(i),\CH^+}}{\rd \ub} + b_{(i)}^B \f{\rd \th^A_{(i),\CH^+}}{\rd \th_{(i)}^B} \Big).$$
    Now, it has already been observed that for the change of variables map on the sphere, $\f{\rd \th_{(i),\CH^+}^A}{\rd \th_{(i)}^B}$ and its inverse are both continuous up to the Cauchy horizon. Therefore, in order to obtain \eqref{eq:sbi.cond.goal}, it suffices to take derivatives of $b_{(i),\CH^+}^A$ with respect to $\f{\rd}{\rd \th_{(i)}}$.
    
    On $\{u=u_f\}$, the right-hand side $= e^{\f{r_+-r_-}{r_-^2 + a^2}\ub} \underline{\pmb{b}}_{(i)}^A$. Importantly, the Kerr vector field $\underline{\pmb{b}}$ is smooth and decays as $e^{-\f{r_+-r_-}{r_-^2 + a^2}\ub}$ in norm, with angular derivatives also decaying as $e^{-\f{r_+-r_-}{r_-^2 + a^2}\ub}$. Taking angular derivatives of $e^{\f{r_+-r_-}{r_-^2 + a^2}\ub} \underline{\pmb{b}}_{(i)}^A$ immediately gives the desired bound \eqref{eq:sbi.cond.goal} on $\{u=u_f\}$.
    
    Away from $\{u=u_f\}$, we use \cite[(16.58)]{DafLuk17} (which is a consequence of \eqref{eq:reg.coord.def.2}) to obtain
    \begin{equation}\label{eq:transporting.b}
        \begin{split}
            \f{\rd}{\rd u} \Big( \f{\rd \th^A_{(i),\CH^+}}{\rd \ub} + b^B \f{\rd \th^A_{(i),\CH^+}}{\rd \th_{(i)}^B} \Big) = 2\Omg^2 (\eta_{(i)}^B - \etab_{(i)}^B) \f{\rd \th^A_{(i),\CH^+}}{\rd \th_{(i)}^B}.
        \end{split}
    \end{equation}
    Using the bounds in Proposition~\ref{prop:derivatives.all.directions} (and Sobolev embedding \eqref{eq:Sobolev.BS}) and \eqref{eq:angular.second.derivative.bound}, the angular derivative of the right-hand side of \eqref{eq:transporting.b} is bounded. Integrating in $u$ (and recalling that the implicit constant in the inequality is allowed to depend on $u$), and using the bound on $\{u=u_f\}$ that we have derived above, we obtain \eqref{eq:sbi.cond.goal} for all $u < u_f$. \qedhere
\end{proof}

\subsubsection{General norm}

Just as in the red-shift region (cf.~Definition~\ref{def:RS.weight}), it is convenient to introduce a notation to simultaneously capture the different types of norms we consider. Here it is less convenient to consider a single weight function, as the weight in the integrated norm is also important. Instead we make the following definition.
\begin{definition}\label{def:BS.general.norm}
    Define the norms $\opnorm{\cdot}$ and $\opnorm{\cdot}_*$ to measure the size of difference quantities in $\os{\calM}{2}$:
    \begin{align}
        \opnorm{\phi - \pmb{\phi}}:= &\: \| |u|^{\f{q_--3}2} (\phi - \pmb{\phi}) \|_{L^\i(\os{\calM}{2}, \vol_{\gamma}\ud u\ud \ub)} + \| |u|^{\f{q_{--}-2}2} (\phi - \pmb{\phi}) \|_{L^\i(\os{\calM}{2}\cap \{f_\Gamma \leq 0\}, \vol_{\gamma}\ud u\ud \ub)} \nonumber \\
        &\: + \| |u|^{\f{q_--2}2} \pmb{\Omg}^2 (\phi - \pmb{\phi}) \|_{L^2(\os{\calM}{2}, \vol_{\gamma}\ud u\ud \ub)}, \label{eq:triple.norm.def.1}\\
        \opnorm{\phi - \pmb{\phi}}_* := &\:  \| |u|^{\f{q_--3}2} (\phi - \pmb{\phi}) \|_{L^\i(\os{\calM}{2}, \vol_{\gamma}\ud u\ud \ub)} + \| |u|^{\f{q_{--}-2}2} (\phi - \pmb{\phi}) \|_{L^\i(\os{\calM}{2}\cap \{f_\Gamma \leq 0\}, \vol_{\gamma}\ud u\ud \ub)} \nonumber \\
        &\: + \| |u|^{\f{q_--2}2} \brk{u+\ub}^{-\f 34} (\phi - \pmb{\phi}) \|_{L^2(\os{\calM}{2}, \vol_{\gamma}\ud u\ud \ub)}.\label{eq:triple.norm.def.2}
    \end{align}
    
\end{definition}

\begin{remark}
    In view of Lemma~\ref{lem:CGamma}, the improved $L^\i$ estimates we proved in the region $\os{\calM}{2} \cap \{u+\ub \leq \f 12 \f{\sigma_q}{\kappa_-}\log \ub + C_\Gamma\}$ implies improved estimates in $\os{\calM}{2}\cap \{f_\Gamma \leq 0\}$. The definitions \eqref{eq:triple.norm.def.1}--\eqref{eq:triple.norm.def.2} capture this improvement in $\os{\calM}{2}\cap \{f_\Gamma \leq 0\}$.
\end{remark}

\begin{remark}
    Observe that $\opnorm{\cdot}_*$ is a stronger norm for which the integrated part has the weight $\brk{u+\ub}^{-\f 34}$ instead of the weight $\pmb{\Om}^2$. Observe also that in Definition~\ref{def:BS.general.norm}, we only keep the $|u|$ weights. Some terms indeed obey stronger estimates with $|u|$-weights replaced by $\ub$-weights, but these improvements will not be relevant for most of the remainder of the paper.
\end{remark}

\subsubsection{Switching from the $(\os{\ee}{2}_3, \os{\ee}{2}_4)$ pair to the $(e'_3, e'_4)$ pair}

Define on $\os{\calM}{2}\setminus (\calU \cap \{s \leq \f{s_f}{2}\})$ the vector fields 
\begin{equation}\label{eq:e3'.e4'.def.restricted}
    e_3' = \Omg^{-2} \os{\ee}{2}_3,\quad e_4' = \Omg^2 \os{\ee}{2}_4.
\end{equation}
(We will later define $e_3'$, $e_4'$ in all of $\calM$; see Definition~\ref{def:e3'.e4'}. The definitions coincide when restricted to $\os{\calM}{2}\setminus (\calU \cap \{s \leq \f{s_f}{2}\})$; see \eqref{eq:e3.e4.in.coordinates.M2}.)

We will now use $\chi$, $\chib$, etc.~to denote the quantities with respect to the frame $(e_3',e_4')$. In this gauge, we have $\omegab=0$, but $\om \neq 0$ (and $\xi=\xib = 0$). 

The bounds in Proposition~\ref{prop:derivatives.all.directions} and Proposition~\ref{prop:curvature.original} imply the following estimates $\os{\calM}{2}\setminus (\calU \cap \{s \leq \f{s_f}{2}\})$:
\begin{proposition}\label{prop:DL.rescaled}
    In $\os{\calM}{2}$, the following estimates hold, where the sets $\calS_{\cdot}$ are as in \eqref{S.def}, but the horizontal quantities \eqref{eq:Ricci.def}--\eqref{eq:curvature.def} in the CK formalism are now understood with respect to the $(e_3',e_4')$ null pair in \eqref{eq:e3'.e4'.def.restricted}:
    \begin{align}
        \sum_{i+j+k\leq I} \sum_{\substack{\tg \in \mathcal S_{\tg} \\ \tp \in \mathcal S_{\tp}}} \opnorm{(\pmb{\Omg}^2\os{\nab}{2}_{e_3'})^j \os{\nab}{2}{}_{e_4'}^k \os{\nab}{2}{}^i (\tg,\tb,\tK,\tp)} \ls &\: \ep, \label{eq:final.double.null.1}\\
        \sum_{i+j+k\leq I} \sum_{\substack{\tpHb\in \mathcal S_{\tpHb} \\ \tpH\in \mathcal S_{\tpH}}} \opnorm{(\pmb{\Omg}^2\os{\nab}{2}_{e_3'})^j \os{\nab}{2}{}_{e_4'}^k \os{\nab}{2}{}^i (\pmb{\Omg}^2\tpHb,\tpH)}_* \ls &\: \ep, \label{eq:final.double.null.2} \\
        \sum_{i+j+k\leq I}  \opnorm{(\pmb{\Omg}^2\os{\nab}{2}_{e_3'})^j \os{\nab}{2}{}_{e_4'}^k \os{\nab}{2}{}^i (\pmb{\Omg}^4\widetilde{\alphab}, \pmb{\Omg}^2\widetilde{\betab}, \pmb{\Omg}^2 \widetilde{\rho}, \pmb{\Omg}^2 \widetilde{\sigma}, \widetilde{\beta}, \widetilde{\alpha}) }_* \ls &\: \ep. \label{eq:final.double.null.3}
    \end{align}
\end{proposition}
\begin{proof}
        With the change from $(\os{\ee}{2}_3,\os{\ee}{2}_4)$ to $(e_3',e_4')$, the metric components $\Omg$, $b^A$ and $\gamma_{AB}$ remain unchanged. By Proposition~\ref{prop:derivatives.all.directions}, we thus have the desired estimates for the metric components. Since $K$ depends only on $\gamma$, we also have the desired estimate for $K$.

        We now turn to the frame coefficients. One checks that because $e'_3$ is geodesic (i.e., $\nabla_{e'_3} e'_3 = 0$), $\omegab$ and $\xib$ vanish. Since $\nabla_{e'_4} e'_4$ is parallel to $e'_4$, it also follows that $\xi =0$. The transformation of the remaining frame coefficients can be easily computed as follows:
        \begin{align*}
            g(\nabla_{\f{\rd}{\rd\vartheta_*^A}} e'_4,\f{\rd}{\rd\vartheta_*^B}) = \Omg^2 g(\nabla_{\f{\rd}{\rd\vartheta_*^A}} \os{\ee}{2}_4,\f{\rd}{\rd\vartheta_*^B}), \quad g(\nabla_{\f{\rd}{\rd\vartheta_*^A}} e'_3,\f{\rd}{\rd\vartheta_*^B}) = \Omg^{-2} g(\nabla_{\f{\rd}{\rd\vartheta_*^A}} \os{\ee}{2}_3,\f{\rd}{\rd\vartheta_*^B}), \\
            g(\nabla_{e'_4} e'_3,e'_4) =  4 \Omg^2 \os{\ee}{2}_4 \log \Omg,\quad  g(\nabla_{\f{\rd}{\rd\vartheta_*^A}} e'_4,e'_3) = g(\nabla_{\f{\rd}{\rd\vartheta_*^A}} \os{\ee}{2}_4,\os{\ee}{2}_3) - 4 \f{\rd}{\rd\vartheta_*^A}\log \Omg,\\
            g(\nabla_{e'_3} e'_4,\f{\rd}{\rd\vartheta_*^A}) = g(\nabla_{e'_3} \os{\ee}{2}_4,\f{\rd}{\rd\vartheta_*^A}),\quad g(\nabla_{e'_4} e'_3,\f{\rd}{\rd\vartheta_*^A}) = g(\nabla_{e'_4} \os{\ee}{2}_3,\f{\rd}{\rd\vartheta_*^A}).
        \end{align*}
        The desired estimates for $\psi$,$\pH$ and $\pHb$ are then consequences of Proposition~\ref{prop:derivatives.all.directions} after noting that $\pmb{\Omg}^2 \ls \Omg^2 \ls \pmb{\Omg}^2$, $|\f{\rd}{\rd\vartheta_*^A} \log \Omg|_\gamma \ls 1$. 

        Finally, we turn to the curvature components $\alphab, \betab,\rho,\sigma,\bt,\alp$. Since the curvature components are tensorial, the estimate \eqref{eq:final.double.null.3} follows directly from Proposition~\ref{prop:curvature.original}.
\end{proof}

\begin{remark}[Equivalence of norms]\label{rmk:norm.equivalence}
    In a similar manner as in $\os{\calM}{1}$ (see Remark~\ref{rmk:eq.of.norm.RS}), it is useful to note the equivalent ways of applying Proposition~\ref{prop:DL.rescaled}:
    \begin{enumerate}
        \item Since we control $\gamma - \pmb{\gamma}$, when we use the norms $\opnorm{\cdot}$ or $\opnorm{\cdot}_*$ in \eqref{eq:triple.norm.def.1}--\eqref{eq:triple.norm.def.2}, the $L^\i(S)$ can be taken with respect to $\gamma$ or $\pmb{\gamma}$. In particular, later on after we define $e_A'^{(\cdot)}$ (unit with respect to $\gamma$) or $\pmb{e}_A'^{(\cdot)}$ (unit with respect to $\pmb{\gamma}$), we can bound e.g.~either $\widetilde{\eta}(e_A'^{(\cdot)})$ or $\widetilde{\eta}(\pmb{e}_A'^{(\cdot)})$.
        \item Using the estimates in Proposition~\ref{prop:DL.rescaled} itself, we can control $\pmb{\Omg}^2 (\nab_3 - \pmb{\nab}_{\pmb{3}})$, $\nab_4 - \pmb{\nab}_{\pmb{4}}$ and $\nab - \pmb{\nab}$ and so all the derivatives $(\pmb{\Omg}^2\nab_3)^j \nab_4^k \nab^i$ in Proposition~\ref{prop:DL.rescaled} can be replaced by $(\pmb{\Omg}^2 \pmb{\nab}_{\pmb{3}})^j \pmb{\nab}_{\pmb{4}}^k \pmb{\nab}^i$.
        \item The estimates also allow us to commute the order of the derivatives. Indeed, the commutators can be computed using \cite[Proposition~7.1]{DafLuk17}, and then controlled by the bounds for the connection coefficients and curvature components in Proposition~\ref{prop:DL.rescaled}.
    \end{enumerate}
\end{remark}

\subsection{Estimates for global coordinates in the dynamical spacetime $\calM$}\label{sec:global.coordinates}

Our goal in this subsection is to control the global coordinates $(u',\ub,\vartheta_*)$ on $\calM$ defined in Section~\ref{sec:u'def}, again with $\vartheta_* =  (\th_*,\varphi_*)$. 


\subsubsection{Estimates for $u-u_1$}

We now show that $u-u_1$ is small in $\calU$, so that recalling \eqref{eq:u'.def}, $u$, $u_1$ and $u'$ are all close to each other. This will allow us to consider $(u',\ub,\th_*,\varphi_*)$ as global coordinates.

\begin{lemma}\label{lem:u.diff.main}
In the region $\calU \subset \calM$,
    \begin{equation}\label{eq:u.diff.main}
        \sum_{i+j+k\leq I} |\os{\nab}{1}{}^i \os{\ee}{1}{}_3^j \os{\ee}{1}{}_4^k (u-u_1)|_\gamma(s,\ub,\vartheta_*) \ls \ep \mathfrak w(\ub).
    \end{equation}
\end{lemma}
\begin{proof}
    By the definition of $u$ and $u_1$, we have
    \begin{equation}\label{eq:u-u1.diff.ode}
        \begin{split}
            \f{\ud}{\ud s}(u-u_1)(s) =&\:  \Omg^2(u(s),\ub,\vartheta_*) - \pmb{\Omg}^2(u_1(s),\ub,\vartheta_*)\\
            = &\: (\Omg^2-\pmb{\Omg}^2)(u(s),\ub,\vartheta_*) + \pmb{\Omg}^2(u(s),\ub,\vartheta_*)- \pmb{\Omg}^2(u_1(s),\ub,\vartheta_*),
        \end{split}
    \end{equation}
    where we used the shorthand $u(s) = u(s,\ub,\vartheta_*)$, $u_1(s) = u_1(s,\ub,\vartheta_*)$

    The difference $(\Omg^2-\pmb{\Omg}^2)(u(s),\ub,\vartheta_*)$ can be bounded by $\ep \mathfrak w(\ub)$ using \eqref{eq:NI.higher} and Sobolev embedding. Notice that in the region $\calU$, $\pmb{\Omg}^2$ is uniformly bounded below, and so there is no degeneration in either the integration or in the derivatives $\os{\nab}{2}{}_3$, $\os{\nab}{2}{}_4$, $\os{\nab}{1}$. Thus Sobolev embedding implies both $\|\ub^{\f{q_--2}2}(\Omg^2-\pmb{\Omg}^2) \|_{L^2_{\ub}L^\i_s L^\i(S)} \ls \ep$ (with Sobolev embedding in angular directions by \eqref{eq:Sobolev.BS} and a simple $1$-dimensional Sobolev embedding in $u$) and $\|\ub^{\f{q_--2}2}(\Omg^2-\pmb{\Omg}^2) \|_{L^\i_{\ub}L^\i_s L^\i(S)} \ls \ep$ (with Sobolev embedding in all directions).
    
    For the second term in \eqref{eq:u-u1.diff.ode}, by the fundamental theorem of calculus
    $$\pmb{\Omg}^2(u(s),\ub,\vartheta_*)- \pmb{\Omg}^2(u_1(s),\ub,\vartheta_*) = (u-u_1)(s) \int_0^1 \f{\rd}{\rd u}\pmb{\Omg}^2 (u_1(s) + \tau (u-u_1)(s), \ub, \vartheta_*) \, \ud \tau. $$
    Thus, combining the above estimates, and using the bounds for the background Kerr solution, we have
    $$|\f{\ud}{\ud s}(u-u_1)(s,\ub,\vartheta_*) |\ls \ep \mathfrak w(\ub) + |(u-u_1)(s,\ub, \vartheta_*)|.$$
    By \eqref{eq:sSigma.def} and \eqref{eq:initial.condition.for.s}, we know that $u(s_\Sigma(\ub,\vartheta_*),\ub,\vartheta_*) = u_1(s_\Sigma(\ub,\vartheta_*),\ub,\vartheta_*)$. Thus, a standard Gr\"onwall argument gives the desired estimates.

    We now move on to higher derivative estimates, noting that Proposition~\ref{prop:derivatives.all.directions} also controls higher derivatives of $\Omg^2 - \pmb{\Omg}^2$. Notice that the estimates in Proposition~\ref{prop:derivatives.all.directions} are with respect to $\os{\nab}{2}{}^i \os{\ee}{2}{}_3^j \os{\ee}{2}{}_4^k$ derivatives\footnote{Proposition~\ref{prop:derivatives.all.directions} are stated by $\os{\nab}{2}{}^i \os{\ee}{2}{}_3^j \os{\ee}{2}{}_4^k$ derivatives, but recall that the commutators can be controlled as discussed in Remark~\ref{rmk:eq.of.norm.RS}.}, while we want to control the $\os{\nab}{1}{}^i \os{\ee}{1}{}_3^j \os{\ee}{1}{}_4^k$ derivative. For this purpose, we compute the difference of the derivatives. First, the easiest to compute is
    \begin{align}
        \os{\ee}{1}_3 (\phi (u(s,\ub,\vartheta_*),\ub,\vartheta_*))
        = &\: \Omg^{-2} \os{\ee}{2}_3 \phi(u(s,\ub,\vartheta_*),\ub,\vartheta_*) \label{eq:u.est.e3.1} \\
        \os{\ee}{1}_3 (\phi (u_1(s,\ub,\vartheta_*),\ub,\vartheta_*))
        = &\: \pmb{\Omg}^{-2} \os{\ee}{2}_3 \phi(u_1(s,\ub,\vartheta_*),\ub,\vartheta_*) \label{eq:u.est.e3.2}
    \end{align}        
    where we used that $\f{\rd u}{\rd s}\Big|_s = \Omg^{-2}$ and $\f{\rd u_1}{\rd s}\Big|_s = \pmb{\Omg}^{-2}$.

    For $\os{\ee}{1}_4$, we have
    \begin{equation}\label{eq:u.est.e4.1}
        \begin{split}
            &\: \os{\ee}{1}_4 (\phi(u(s),\ub,\vartheta_*)) \\
            = &\: \Big[\Big(\f{\rd}{\rd \ub} \Big|_{DN} + h^A \f{\rd}{\rd \vartheta_*^A}\Big|_{DN} + \Big( \f{\rd u}{\rd \ub} + f \f{\rd u}{\rd s} + h^A \f{\rd u}{\rd \vartheta_*^A} \Big)\Big|_s  \f{\rd}{\rd u} \Big|_{DN}\Big) \phi \Big] (u(s),\ub,\vartheta_*) \\
            = &\: \Big[ \Big( \Omg^2 \os{\ee}{2}_4 + (h^A -b^A) \f{\rd}{\rd \vartheta_*^A}\Big|_{DN} + \Big( \f{\rd u}{\rd \ub} + f \f{\rd u}{\rd s} + h^A \f{\rd u}{\rd \vartheta_*^A} \Big)\Big|_s \os{\ee}{2}_3 \Big) \phi \Big] (u(s),\ub,\vartheta_*) 
        \end{split}
    \end{equation}
    Similarly, 
    \begin{equation}\label{eq:u.est.e4.2}
        \begin{split}
            &\: \os{\ee}{1}_4 (\phi(u_1(s),\ub,\vartheta_*)) \\
            = &\: \Big[ \Big( \Omg^2 \os{\ee}{2}_4 + (h^A -b^A) \f{\rd}{\rd \vartheta_*^A}\Big|_{DN} + \Big( \f{\rd u_1}{\rd \ub} + f \f{\rd u_1}{\rd s} + h^A \f{\rd u_1}{\rd \vartheta_*^A} \Big)\Big|_s \os{\ee}{2}_3 \Big) \phi \Big] (u(s),\ub,\vartheta_*) 
        \end{split}
    \end{equation}
    
    For the angular derivatives, we will need to compute with covariant derivatives. First observe that
    \begin{equation*}
        \begin{split}
            &\: \f{\rd}{\rd \vartheta_*^B}|_{(s,\ub, \vartheta_*)}\phi_{A_1\cdots A_r}(u(s), \ub,\vartheta_*) \\
            = &\: (\f{\rd}{\rd \vartheta_*^B}\Big|_{DN}\phi_{A_1\cdots A_r})(u(s,\ub,\vartheta_*), \ub,\vartheta_*) + (\f{\rd}{\rd u} \Big|_{DN}\phi_{A_1\cdots A_r})(u(s), \ub,\vartheta_*) \f{\rd u}{\rd \vartheta^B}\Big|_s ,
        \end{split}
    \end{equation*}
    which in particular implies that 
    \begin{equation*}
        \begin{split}
            \os{\slashed{\Gamma}}{1}{}_{DE}^C = &\: \os{\slashed{\Gamma}}{2}{}_{DE}^C + \f 12 (\gamma^{-1})^{CF}(\f{\rd u}{\rd \vartheta_*^D}\Big|_s\f{\rd\gamma_{EF}}{\rd u}\Big|_{DN}+\f{\rd u}{\rd\vartheta_*^E}\Big|_s \f{\rd\gamma_{DF}}{\rd u}\Big|_{DN} -\f{\rd u}{\rd\vartheta_*^F}\Big|_s \f{\rd\gamma_{DE}}{\rd u}\Big|_{DN}) \\
            = &\: \os{\slashed{\Gamma}}{2}{}_{DE}^C + (\gamma^{-1})^{CF}(\f{\rd u}{\rd \vartheta_*^D}\Big|_s\os{\chib}{2}_{EF}+\f{\rd u}{\rd\vartheta_*^E}\Big|_s \os{\chib}{2}_{DF} -\f{\rd u}{\rd\vartheta_*^F}\Big|_s \os{\chib}{2}_{DE}).
        \end{split}
    \end{equation*}
    Together, we thus have
    \begin{equation}\label{eq:u.est.eB.1}
        \begin{split}
            &\: \os{\nab}{1}_B (\phi_{A_1\cdots A_r}(u(s,\ub,\vartheta_*),\ub,\vartheta_*)) \\
        = &\: \Big[ \os{\nab}{2}_B \phi_{A_1\cdots A_r} + \f{\rd u}{\rd \vartheta_*^B}\Big|_s \os{\nab}{2}_3 \phi_{A_1\cdots A_r} \\
        &\: - \sum_{i=1}^r (\gamma^{-1})^{CD} \Big(\f{\rd u}{\rd \vartheta_*^{A_i}}\Big|_s\os{\chib}{2}_{BD} -\f{\rd u}{\rd\vartheta_*^D}\Big|_s \os{\chib}{2}_{A_i B}\Big) \phi_{A_1\dots \hat{A_i}C\dots A_r} \Big](u(s,\ub,\vartheta_*),\ub,\vartheta_*). 
        \end{split}
    \end{equation}
    Similarly,
    \begin{equation}\label{eq:u.est.eB.2}
        \begin{split}
            &\: \os{\nab}{1}_B (\phi_{A_1\cdots A_r}(u_1(s,\ub,\vartheta_*),\ub,\vartheta_*)) \\
        = &\: \Big[ \os{\nab}{2}_B \phi_{A_1\cdots A_r} + \f{\rd u_1}{\rd \vartheta_*^B}\Big|_s \os{\nab}{2}_3 \phi_{A_1\cdots A_r} \\
        &\: - \sum_{i=1}^r (\gamma^{-1})^{CD} \Big(\f{\rd u_1}{\rd \vartheta_*^{A_i}}\Big|_s\os{\chib}{2}_{BD} -\f{\rd u_1}{\rd\vartheta_*^D}\Big|_s \os{\chib}{2}_{A_i B}\Big) \phi_{A_1\dots \hat{A_i}C\dots A_r} \Big](u_1(s,\ub,\vartheta_*),\ub,\vartheta_*). 
        \end{split}
    \end{equation}

     We now start with equation \eqref{eq:u-u1.diff.ode}, differentiate by $\os{\nab}{1}{}^i \os{\ee}{1}{}_3^j \os{\ee}{1}{}_4^k$, and use the formulas \eqref{eq:u.est.e3.1}--\eqref{eq:u.est.eB.2} to re-express all the derivatives on the right-hand side in terms of $\os{\nab}{2}{}^i \os{\ee}{2}{}_3^j \os{\ee}{2}{}_4^k$ of the geometric quantities. We now use Proposition~\ref{prop:derivatives.all.directions} to inductively control higher derivatives of $u-u_1$.

    Observe that unlike the undifferentiated case, the derivatives of $(u_1-u)$ need not vanish on $\Sigma$; it is only the tangential derivatives to $\Sigma$ that vanish. Nonetheless, using the equation \eqref{eq:u-u1.diff.ode} and inductively in the number of derivatives, one can show that the data for the higher derivatives $\ls \ep \mathfrak w(\ub)$ on $\Sigma$. The remainder of the argument then proceeds as before. \qedhere
\end{proof}

\subsubsection{Computations in the $(u',\ub,\th_*, \varphi_*)$ coordinates}\label{sec:computations.u'}

After taking after taking $\ep$ smaller if necessary, the estimates in \eqref{eq:u.diff.main} in particular shows that the first derivatives of $u-u_1$ are small. In particular, it follows that $(u',\ub,\th_*, \varphi_*)$ forms a coordinate system on $\calU$.

The following proposition computes the dynamical metric in the $(u',\ub,\vartheta_*)$ coordinates.

\begin{proposition}\label{prop:metric.in.global.coordinates}
    \begin{enumerate}
        \item In $\overset{\scriptscriptstyle{\text{[2]}}}\calM \setminus (\calU\cap \{s \leq \f{s_f}2\})$, the metric in the $(u',\ub,\th_*,\varphi_*)$ coordinate system takes the following form:
        \begin{equation}\label{eq:metric.in.global.coordinates.2}
            g=-2\Omega^2(\ud u'\otimes \ud\ub + \ud\ub\otimes \ud u')+\gamma_{AB}(\ud\vartheta_*^A-b^A \ud\ub)\otimes (\ud\vartheta_*^B-b^B \ud\ub).
        \end{equation}
        \item In $\overset{\scriptscriptstyle{\text{[1]}}}\calM$, the metric in the $(u',\ub,\th_*,\varphi_*)$ coordinate system takes the following form:
        \begin{equation}\label{eq:metric.in.global.coordinates}
    \begin{split}
        g = &\: -  \f{2}{\f{\rd u'}{\rd s}\Big|_s} \Big(\ud u' \otimes \ud\ub+ \ud\ub\otimes \ud u' \Big) \\
        &\: + 4\Big(f + \f{\f{\rd u'}{\rd \ub}\Big|_s}{\f{\rd u'}{\rd s}\Big|_s} + \f{h^A \f{\rd u'}{\rd \vartheta_*^A}\Big|_s}{\f{\rd u'}{\rd s}\Big|_s} - \f{|\nab u'\Big|_s|_{\gamma}^2}{(\f{\rd u'}{\rd s}\Big|_s)^2}\Big) \ud \ub\otimes \ud \ub  \\
        &\: +\gamma_{AB}\Big(\ud\vartheta_*^A - \Big(h^A - 2 \f{\gamma^{AA'}\f{\rd u'}{\rd \vartheta_*^{A'}}\Big|_s}{\f{\rd u'}{\rd s}\Big|_s}\Big) \ud\ub \Big)\otimes \Big(\ud\vartheta_*^B - \Big(h^B - 2 \f{\gamma^{BB'}\f{\rd u'}{\rd \vartheta_*^{B'}}\Big|_s}{\f{\rd u'}{\rd s}\Big|_s}\Big) \ud\ub \Big),
    \end{split}
\end{equation}
    where $\f{\rd u'}{\rd s}\Big|_s$, $\f{\rd u'}{\rd \ub}\Big|_s$ and $\f{\rd u'}{\rd \vartheta_*^A}\Big|_s$ are derivatives in the $(s,\ub,\th_*,\varphi_*)$ coordinate system and $(f,h,\gamma)$ are the metric components in \eqref{eq:metric.form.1}.

        The derivatives $\f{\rd u'}{\rd s}\Big|_s$, $\f{\rd u'}{\rd \ub}\Big|_s$ and $\f{\rd u'}{\rd \vartheta_*^A}\Big|_s$ are computed as follows:
        \begin{align}
            \f{\rd u'}{\rd s}\Big|_s = &\: \chi'(s) (u_1(s) - u(s)) + \chi(s) \pmb{\Omg}^{-2} + (1-\chi(s)) \Omg^{-2}, \label{eq:derivatives.tu.1}\\
            \f{\rd u'}{\rd \ub}\Big|_s = &\: -\chi(s) \pmb{\Omg}^{-2} \Big(\pmb{f} - \f 14 |\pmb{b} - \pmb{h}|_{\pmb{\gamma}}^2 - \f 12 \langle \pmb{b},\pmb{h}\rangle_{\pmb{\gamma}} + \f 12 |\pmb{h}|_{\pmb{\gamma}}^2\Big) \nonumber \\
            &\: - (1- \chi(s)) \Omg^{-2} \Big(f - \f 14 |b - h|_\gamma^2 - \f 12 \langle b,h\rangle_\gamma + \f 12 |h|_\gamma^2\Big), \label{eq:derivatives.tu.2}\\
            \f{\rd u'}{\rd \vartheta_*^A}\Big|_s = &\: -\f 12 \Big( \chi(s) \pmb{\Omg}^{-2}\pmb{\gamma}_{AB} (\pmb{b}^B - \pmb{h}^B) + (1- \chi(s)) \Omg^{-2}\gamma_{AB} (b^B - h^B) \Big). \label{eq:derivatives.tu.3} 
        \end{align}
        Here, the bold-faced double null quantities $\pmb{\gamma}$, $\pmb{b}$ and $\pmb{\Omg}$ are evaluated at $(u_1(s),\ub,\vartheta_*)$ and the regular $\gamma$, $b$ and $\Omg$ are evaluated at $(u(s),\ub,\vartheta_*)$. The quantities $\pmb{f}, \pmb{h}, f, h$ are all evaluated at $(s,\ub,\vartheta_*)$.
    \end{enumerate}
\end{proposition}
\begin{proof}
    Part 1 is immediate from the fact that $u' = u$ on $\overset{\scriptscriptstyle{\text{[2]}}}\calM \setminus (\calU\cap \{s \leq \f{s_f}2\})$.

    To compute the metric in the $(u',\ub,\th_*,\varphi_*)$ coordinate system, we first use $\ud u' = \f{\rd u'}{\rd s}\Big|_s \ud s  + \f{\rd u'}{\rd \ub}\Big|_s \ud \ub + \f{\rd u'}{\rd \vartheta_*^A}\Big|_s \ud \vartheta_*^A$ to obtain
$$\ud s = \f{1}{\f{\rd u'}{\rd s}\Big|_s} \ud u' - \f{\f{\rd u'}{\rd \ub}\Big|_s}{\f{\rd u'}{\rd s}\Big|_s} \ud \ub  - \f{\f{\rd u'}{\rd \vartheta_*^A}\Big|_s}{\f{\rd u'}{\rd s}\Big|_s} \ud \vartheta_*^A.$$
    Then, starting with \eqref{eq:metric.form.1} (and writing angular derivatives as $\nab$), we compute
    \begin{equation}
    \begin{split}
        g= &\: - 2 (\ud s\otimes \ud\ub+ \ud\ub\otimes \ud s)+ 4f \ud\ub\otimes \ud\ub+\gamma_{AB}(\ud\vartheta_*^A - h^A \ud\ub)\otimes (\ud\vartheta_*^B - h^B \ud\ub) \\
        = &\: - 2 \f{1}{\f{\rd u'}{\rd s}\Big|_s} \Big(\ud u' \otimes \ud\ub+ \ud\ub\otimes \ud u' \Big) + 4\Big(f + \f{\f{\rd u'}{\rd \ub}\Big|_s}{\f{\rd u'}{\rd s}\Big|_s} + \f{h^A\nab_Au'}{\f{\rd u'}{\rd s}\Big|_s} - \f{|\nab u'|_{\gamma}^2}{(\f{\rd u'}{\rd s}\Big|_s)^2}\Big) \ud \ub\otimes \ud \ub  \\
        &\: +\gamma_{AB}\Big(\ud\vartheta_*^A - \Big(h^A - 2 \f{\nab^A u'}{\f{\rd u'}{\rd s}\Big|_s}\Big) \ud\ub \Big)\otimes \Big(\ud\vartheta_*^B - \Big(h^B - 2 \f{\nab^B u'}{\f{\rd u'}{\rd s}\Big|_s}\Big) \ud\ub \Big),
    \end{split}
\end{equation}
obtaining the desired formula \eqref{eq:metric.in.global.coordinates}.

It remains to compute \eqref{eq:derivatives.tu.1}--\eqref{eq:derivatives.tu.3}. For this purpose, first notice that
\begin{equation}
    \ud u' = \chi'(s) (u_1 - u) \, \ud s + \chi(s) \ud u_1 + (1- \chi(s)) \ud u.
\end{equation}
By Proposition~\ref{prop:change.s.to.u}, we have
\begin{equation}
    \ud s = \Omg^2 \ud u + \f 12 \gamma_{AB} (b^B - h^B) \ud \vartheta_*^A + \Big( f - \f 14 |b - h|_\gamma^2 - \f 12 \langle b,h\rangle_\gamma + \f 12 |h|_\gamma^2\Big) \ud \ub.
\end{equation}
Now notice that the argument in Proposition~\ref{prop:change.s.to.u} also applies to the derivatives of $s$ in the $(u_1,\ub,\th_*,\varphi_*)$ coordinate system, except that by \eqref{eq:u1.def}, we have the ordinary diffential equation $\f{\ud s}{\ud u_1} = \pmb{\Omg}^2$ (in contrast to $\f{\ud s}{\ud u} = \Omg^2$). Hence, in the setting, one adapts the computation in Proposition~\ref{prop:change.s.to.u} but replaces the dynamical metric by the Kerr metric. Thus,
\begin{equation}
    \ud s = \pmb{\Omg}^2 \ud u_1 + \f 12 \pmb{\gamma}_{AB} (\pmb{b}^B - \pmb{h}^B) \ud \vartheta_*^A + \Big( \pmb{f}  - \f 14 |\pmb{b} - \pmb{h}|_{\pmb{\gamma}}^2 - \f 12 \langle \pmb{b},\pmb{h}\rangle_{\pmb{\gamma}} + \f 12 |\pmb{h}|_{\pmb{\gamma}}^2\Big) \ud \ub.
\end{equation}
Combining, we thus obtain
\begin{equation}
    \begin{split}
        \ud u' = &\: \chi'(s) (u_1 - u) \, \ud s + \chi(s) \ud u_1 + (1- \chi(s)) \ud u  \\
        = &\: \Big( \chi'(s) (u_1 - u) + \chi(s) \pmb{\Omg}^{-2} + (1-\chi(s)) \Omg^{-2} \Big) \, \ud s \\
        &\: - \f 12 \Big( \chi(s) \pmb{\Omg}^{-2}\pmb{\gamma}_{AB} (\pmb{b}^B - \pmb{h}^B) + (1- \chi(s)) \Omg^{-2}\gamma_{AB} (b^B - h^B) \Big)\ud \vartheta_*^A  \\
        &\: - \chi(s) \pmb{\Omg}^{-2} \Big(\pmb{f} - \f 14 |\pmb{b} - \pmb{h}|_{\pmb{\gamma}}^2 - \f 12 \langle \pmb{b},\pmb{h}\rangle_{\pmb{\gamma}} + \f 12 |\pmb{h}|_{\pmb{\gamma}}^2\Big)  \ud \ub \\ 
        &\: - (1- \chi(s)) \Omg^{-2} \Big(f - \f 14 |b - h|_\gamma^2 - \f 12 \langle b,h\rangle_\gamma + \f 12 |h|_\gamma^2\Big)  \ud \ub,
    \end{split}
\end{equation}
which implies \eqref{eq:derivatives.tu.1}--\eqref{eq:derivatives.tu.3}. 
\end{proof}


\subsubsection{Completion of Part~\ref{item:main.thm.1.b} of Theorem~\ref{thm:main}}

\begin{proposition}\label{prop:end.of.1.b}
    Part~\ref{item:main.thm.1.b} of Theorem~\ref{thm:main} holds.
\end{proposition}
\begin{proof}
    The metric in $(u',\ub,\th_*,\varphi_*)$ coordinates is computed in \eqref{eq:metric.in.global.coordinates.2} and \eqref{eq:metric.in.global.coordinates}. Using the computations for $u'$ in \eqref{eq:derivatives.tu.1}--\eqref{eq:derivatives.tu.3}, the estimates of the differences on $\os{\calM}{1}$ in \cite{DafLuk26} (see Theorem~\ref{thm:DL.2}), the estimates of the differences on $\os{\calM}{2}$ in \cite{DafLuk17} (see Theorem~\ref{thm:main.DL.est}) and the estimates for $u_1-u$ (see Lemma~\ref{lem:u.diff.main}). \qedhere
\end{proof}

\subsection{Identification with the Kerr spacetime}\label{sec:identification}

Having shown that $(u',\ub,\vartheta_*)$ is a coordinate system in \eqref{sec:global.coordinates}, we now use it to globally define an identification of $\calM$ with a subset of $\calM_{\mathrm{Kerr}}$. We define this identification $\Phi$, together with $\os{\Phi}{1}$, $\os{\Phi}{2}$ which are implicitly used in \cite{DafLuk17, DafLuk26} as follows:
\begin{definition}\label{def:Phi}
    Define the maps $\Phi:\calM \to \calM_{\mathrm{Kerr}}$, $\os{\Phi}{1}:\os{\calM}{1} \to \calM_{\mathrm{Kerr}}$ and $\os{\Phi}{2}:\os{\calM}{2} \to \calM_{\mathrm{Kerr}}$ by
    \begin{itemize}
        \item The map $\os{\Phi}{1}:\os{\calM}{1} \to \calM_{\mathrm{Kerr}}$ is defined by \textbf{identifying the $(s,\ub,\vartheta_*)$ coordinates} on $\os{\calM}{1}$ from Section~\ref{sec:M1} with the $(s,\ub,\vartheta_*)$ coordinates on $\calM_{\mathrm{Kerr}}$ from Section~\ref{sec:Kerr.s.ub}.
        \item The map $\os{\Phi}{2}:\os{\calM}{2} \to \calM_{\mathrm{Kerr}}$ is defined by \textbf{identifying the $(u,\ub,\vartheta_*)$ coordinates}
        on $\os{\calM}{2}$ from Section~\ref{sec:M2} with the $(u,\ub,\vartheta_*)$ coordinates on $\calM_{\mathrm{Kerr}}$ from Section~\ref{sec:Kerr.double.null}.
        \item The map $\Phi:\calM \to \calM_{\mathrm{Kerr}}$ is defined by \textbf{identifying the $(u',\ub,\vartheta_*)$ coordinates} on $\calM$ from Section~\ref{sec:u'def}, with those on $\calM_{\mathrm{Kerr}}$, after setting $u' = u$ on $\calM_{\mathrm{Kerr}}$, for $(u,\ub,\vartheta_*)$ as in Section~\ref{sec:Kerr.double.null}.
    \end{itemize}
\end{definition}

Each of these maps is a diffeomorphism with its image. Since $u'|_{\os{\calM}{2}\setminus (\calU\cap \{s\geq s_f\})} = u$, it follows that
\begin{equation}\label{eq:Phi.agree.in.2}
    \Phi|_{\os{\calM}{2}\setminus (\calU\cap \{s\geq s_f\})} = \os{\Phi}{2}|_{\os{\calM}{2}\setminus (\calU\cap \{s\geq s_f\})}.
\end{equation}
In general, however, $\Phi|_{\os{\calM}{1}}$ is different from $\os{\Phi}{1}$ (though $\Phi|_{\os{\calM}{1}\setminus\calU}= \os{\Phi}{1}|_{\os{\calM}{1} \setminus \calU}$). Nonetheless, these maps are close in a sense that we will make precise below. 

We will consider the estimates for the $^{(N)}$ and $^{(S)}$ variant separately: All the estimates below for the rest of the subsection hold for $\th_* \in [0,\f{3\pi}4]$ if $^{(\cdot)} = ^{(N)}$ and for $\th_* \in [\f \pi 4, \pi]$ if $^{(\cdot)} = ^{(S)}$. For convenience with indexing, it will be useful to adapt the notation $\os{\ee}{1}{}_\mu^{(\cdot)}= \os{\ee}{1}{}_\mu$ for $\mu =3,4$, similarly for $\os{\pmb{\ee}}{1}{}_\mu^{(\cdot)}$ when $\mu = 3,4$.

\begin{lemma}\label{lem:function.pull.back}
    Suppose $\phi:\Phi(\os{\calM}{1})\cap \os{\Phi}{1}(\os{\calM}{1})\subset \calM_{\mathrm{Kerr}} \to \mathbb R$ is a smooth scalar function. If $\phi$ satisfies
    \begin{equation}\label{eq:function.pull.back.general.assumption}
        \sum_{i_1+i_2+i_3+i_4 \leq I+1} \Big| (\os{\pmb{\ee}}{1}{}_1^{(\cdot)})^{i_1}
         (\os{\pmb{\ee}}{1}{}_2^{(\cdot)})^{i_2}
         (\os{\pmb{\ee}}{1}{}_3)^{i_3}(\os{\pmb{\ee}}{1}{}_4)^{i_4} \os{\Phi}{1}{}^* \phi \Big| \ls 1,
    \end{equation}
    then, on $\Phi^{-1} (\Phi(\os{\calM}{1}) \cap \os{\Phi}{1}(\os{\calM}{1}))\cap \os{\Phi}{1}{}^{-1} (\Phi(\os{\calM}{1}) \cap \os{\Phi}{1}(\os{\calM}{1}))$,
    \begin{equation}
        \sum_{i_1+i_2+i_3+i_4 \leq I} \Big| (\os{\pmb{\ee}}{1}{}_1^{(\cdot)})^{i_1}
         (\os{\pmb{\ee}}{1}{}_2^{(\cdot)})^{i_2}
         (\os{\pmb{\ee}}{1}{}_3)^{i_3}(\os{\pmb{\ee}}{1}{}_4)^{i_4} (\Phi^* \phi - \os{\Phi}{1}{}^* \phi ) \Big| \ls \ep \mathfrak w(\ub).
    \end{equation}
\end{lemma}
\begin{proof}
    This is a consequence of the definition of $u'$ in \eqref{eq:u'.def}, the estimates in Lemma~\ref{lem:u.diff.main}, and the mean value theorem, after noting that in Lemma~\ref{lem:u.diff.main}, we can change the derivatives between the background and the dynamical frame (see Remark~\ref{rmk:eq.of.norm.RS}). \qedhere    
\end{proof}

Next we compute and estimate the transformation of vector fields under $\os{\Phi}{1}{}^{-1} \Phi$:
\begin{lemma}\label{lem:comparing.Kerr.vector.fields}
    For any $p \in \os{\calM}{1}$ such that $\Phi(p) \in \os{\Phi}{1}(\os{\calM}{1})$, the following holds on $T_{\Phi^{-1} \os{\Phi}{1}(p)} \calM$:
    \begin{align}
        \Phi_*^{-1} \os{\Phi}{1}{}_*  \rd_s \Big|_s =&\:  \f{\pmb{\Omg}^{-2}(\os{\Phi}{1}) }{\f{\rd u'}{\rd s}\Big|_s(\Phi\circ\os{\Phi}{1}(p))} \rd_s \Big|_s , \\
        \Phi_*^{-1} \os{\Phi}{1}{}_*  \rd_{\ub} \Big|_s =&\: \rd_{\ub} \Big|_s  - \Big( \f{\f{\rd u'}{\rd \ub}\Big|_s}{\f{\rd u'}{\rd s}\Big|_s}(\Phi\circ\os{\Phi}{1}(p)) + \f{\pmb{\Omg}^{-2} (\pmb{f} -\f 14 |\pmb{b}-\pmb{h}|_{\pmb{\gamma}}^2 - \f 12 \langle \pmb{b},\pmb{h}\rangle + \f 12 |\pmb{h}|^2)(\os{\Phi}{1}(p))}{\f{\rd u'}{\rd s}\Big|_s(\Phi\circ\os{\Phi}{1}(p))}\Big) \rd_s\Big|_s, \\
        \Phi_*^{-1} \os{\Phi}{1}{}_* \rd_{\vartheta_*^A} \Big|_s =&\: \rd_{\vartheta_*^A} \Big|_s 
        - \Big(\f{\f{\rd u'}{\rd \vartheta^A_*}\Big|_s}{\f{\rd u'}{\rd s}\Big|_s} (\Phi\circ\os{\Phi}{1}(p)) + \f{\f 12 \pmb{\gamma}_{AB} (\pmb{b}^B - \pmb{h}^B)(\os{\Phi}{1}(p))}{\f{\rd u'}{\rd s}\Big|_s(\Phi\circ\os{\Phi}{1}(p))} \Big) \rd_s\Big|_s.
    \end{align}
    As a result, on $\Phi^{-1} (\Phi(\os{\calM}{1}) \cap \os{\Phi}{1}(\os{\calM}{1}))\cap \os{\Phi}{1}{}^{-1} (\Phi(\os{\calM}{1}) \cap \os{\Phi}{1}(\os{\calM}{1}))$, writing $\Phi_*^{-1} \os{\Phi}{1}{}_* \os{\pmb{\ee}}{1}{}_\mu^{(\cdot)} = (\mathcal D^{(\cdot)})_{\mu}^\nu  \os{\pmb{\ee}}{1}{}_\mu^{(\cdot)}$, it holds that
    \begin{equation}
        \begin{split}
            \sum_{i_1+i_2+i_3+i_4 \leq I} \Big| (\os{\pmb{\ee}}{1}{}_1^{(\cdot)})^{i_1}
         (\os{\pmb{\ee}}{1}{}_2^{(\cdot)})^{i_2}
         (\os{\pmb{\ee}}{1}{}_3)^{i_3}(\os{\pmb{\ee}}{1}{}_4)^{i_4} (\mathcal D^{(\cdot)})_{\mu}^\nu\Big| \ls \ep \mathfrak w(\ub).
        \end{split}
    \end{equation}
\end{lemma}
\begin{proof}
    This is a direct change of variables computation using the definitions, except that we plug the Kerr values in the relevant places (which correspond to the computations \eqref{eq:derivatives.tu.1}--\eqref{eq:derivatives.tu.3}, but with $\chi\equiv 1$). The desired bound is a consequence of Theorem~\ref{thm:DL.2}, Proposition~\ref{prop:DL.rescaled} and Lemma~\ref{lem:u.diff.main} (and we used Remark~\ref{rmk:eq.of.norm.RS} so as to apply these results with derivatives in terms of the $\os{\pmb{\ee}}{1}_\mu$ frame fields). \qedhere
\end{proof}

On $\os{\calM}{1}$, define the pull-back metrics $\Phi^* \pmb{g}$ and $\os{\Phi}{1}{}^* \pmb{g}$, where $\pmb{g}$ is the metric on $\calM_{\mathrm{Kerr}}$. By definition, $\{\os{\pmb{\ee}}{1}{}_{1}, \os{\pmb{\ee}}{1}{}_{2}, \os{\pmb{\ee}}{1}{}_{3}, \os{\pmb{\ee}}{1}{}_{4}\}$ is a null frame with respect to $\os{\Phi}{1}{}^* \pmb{g}$. 
\begin{lemma}\label{lem:almost.null.different.pullback}
    On $\Phi^{-1} (\Phi(\os{\calM}{1}) \cap \os{\Phi}{1}(\os{\calM}{1}))\cap \os{\Phi}{1}{}^{-1} (\Phi(\os{\calM}{1}) \cap \os{\Phi}{1}(\os{\calM}{1}))$, $\{\os{\pmb{\ee}}{1}{}_{1}, \os{\pmb{\ee}}{1}{}_{2}, \os{\pmb{\ee}}{1}{}_{3}, \os{\pmb{\ee}}{1}{}_{4}\}$ is almost a null frame with respect to $\Phi^* \pmb{g}$ in the sense that the following holds for all $\mu$, $\nu$:
        \begin{equation}
            \sum_{i_1+i_2+i_3+i_4 \leq I} \Big| (\os{\pmb{\ee}}{1}{}_1^{(\cdot)})^{i_1}
         (\os{\pmb{\ee}}{1}{}_2^{(\cdot)})^{i_2}
         (\os{\pmb{\ee}}{1}{}_3)^{i_3}(\os{\pmb{\ee}}{1}{}_4)^{i_4} \Big( (\Phi^* \pmb{g})(\os{\pmb{\ee}}{1}{}^{(\cdot)}_{\mu}, \os{\pmb{\ee}}{1}{}^{(\cdot)}_{\nu}) - (\os{\Phi}{1}{}^* \pmb{g})(\os{\pmb{\ee}}{1}{}^{(\cdot)}_{\mu}, \os{\pmb{\ee}}{1}{}_{\nu}^{(\cdot)})\Big) \Big| \ls \ep \mathfrak w(\ub).
        \end{equation}
\end{lemma}
\begin{proof}
    For fixed $\mu$, $\nu$, $\pmb{g}(\os{\Phi}{1}{}_* \os{\pmb{\ee}}{1}{}^{(\cdot)}_\mu, \os{\Phi}{1}{}_* \os{\pmb{\ee}}{1}{}^{(\cdot)}_\nu)$ is a constant function on $\calM_{\mathrm{Kerr}}$ and so
    $$(\os{\Phi}{1}{}^* \pmb{g})(\os{\pmb{\ee}}{1}{}^{(\cdot)}_{\mu}, \os{\pmb{\ee}}{1}{}_{\nu}^{(\cdot)}) = \os{\Phi}{1}{}^*\Big( \pmb{g}(\os{\Phi}{1}{}_* \os{\pmb{\ee}}{1}{}^{(\cdot)}_\mu, \os{\Phi}{1}{}_* \os{\pmb{\ee}}{1}{}^{(\cdot)}_\nu)\Big) = \Phi^* \Big( \pmb{g}(\os{\Phi}{1}{}_* \os{\pmb{\ee}}{1}{}^{(\cdot)}_\mu, \os{\Phi}{1}{}_* \os{\pmb{\ee}}{1}{}^{(\cdot)}_\nu)\Big) = (\Phi^* \pmb{g})(\Phi{}^{-1}_*\os{\Phi}{1}{}_* \os{\pmb{\ee}}{1}{}^{(\cdot)}_\mu, \Phi{}^{-1}_* \os{\Phi}{1}{}_* \os{\pmb{\ee}}{1}{}^{(\cdot)}_\nu). $$
    It thus reduces to controlling $(\Phi^* \pmb{g})(\Phi{}^{-1}_*\os{\Phi}{1}{}_* \os{\pmb{\ee}}{1}{}^{(\cdot)}_\mu, \Phi{}^{-1}_* \os{\Phi}{1}{}_* \os{\pmb{\ee}}{1}{}^{(\cdot)}_\nu) - (\Phi^* \pmb{g})(\os{\pmb{\ee}}{1}{}^{(\cdot)}_{\mu}, \os{\pmb{\ee}}{1}{}^{(\cdot)}_{\nu})$, and the desired bound follows from Lemma~\ref{lem:function.pull.back}. \qedhere
\end{proof}

Next, we consider the transformation of the connections. 
\begin{lemma}\label{lem:pull.back.comparison}
    On $\Phi^{-1} (\Phi(\os{\calM}{1}) \cap \os{\Phi}{1}(\os{\calM}{1}))\cap \os{\Phi}{1}{}^{-1} (\Phi(\os{\calM}{1}) \cap \os{\Phi}{1}(\os{\calM}{1}))$, the following holds for all $\mu$, $\nu$, $\lambda$:
        \begin{equation}\label{eq:pull.back.comparison}
            \begin{split}
                \Big| (\os{\Phi}{1}{}^* \pmb{g})(\pmb{\nabla}^{\os{\Phi}{1}{}^* \pmb{g}}_{\os{\pmb{\ee}}{1}{}^{(\cdot)}_{\mu}} \os{\pmb{\ee}}{1}{}^{(\cdot)}_{\nu}, \os{\pmb{\ee}}{1}{}^{(\cdot)}_{\lambda}) - ({\Phi}{}^* \pmb{g})(\pmb{\nabla}^{\Phi^* \pmb{g}}_{\Phi{}_* ^{-1}\os{\Phi}{1}{}_* \os{\pmb{\ee}}{1}{}^{(\cdot)}_{\mu}} (\Phi{}_* ^{-1}\os{\Phi}{1}{}_*\os{\pmb{\ee}}{1}{}^{(\cdot)}_{\nu}), \Phi{}_* ^{-1}\os{\Phi}{1}{}_*\os{\pmb{\ee}}{1}{}^{(\cdot)}_{\lambda})  \Big| \ls \ep \mathfrak w(\ub),
            \end{split}   
        \end{equation}
        where $\pmb{\nabla}^{\os{\Phi}{1}{}^* \pmb{g}}$, $\pmb{\nabla}^{\Phi^* \pmb{g}}$ are the Levi-Civita connections for the pull-back metrics $\os{\Phi}{1}{}^* \pmb{g}$ and $\Phi^* \pmb{g}$, respectively. Moreover, higher $(\os{\pmb{\ee}}{1}{}_1^{(\cdot)})^{i_1}
         (\os{\pmb{\ee}}{1}{}_2^{(\cdot)})^{i_2}
         (\os{\pmb{\ee}}{1}{}_3)^{i_3}(\os{\pmb{\ee}}{1}{}_4)^{i_4}$ derivatives of the quantity in \eqref{eq:pull.back.comparison} obeys the same estimate when $i_1+i_2+i_3+i_4 \leq I$.
\end{lemma}
\begin{proof}
This holds due to Lemma~\ref{lem:function.pull.back} since the two expressions represent the pull back of the same function 
\begin{equation}\label{eq:the complicated.expression.on.MKerr}
\pmb{g}(\pmb{\nabla}^{\pmb{g}}_{\os{\Phi}{1}{}_*\os{\pmb{\ee}}{1}{}^{(\cdot)}_{\mu}} (\os{\Phi}{1}{}_*\os{\pmb{\ee}}{1}{}^{(\cdot)}_{\nu}), \os{\Phi}{1}{}_* \os{\pmb{\ee}}{1}{}^{(\cdot)}_{\lambda})
\end{equation}
on $\calM_{\mathrm{Kerr}}$ by $\os{\Phi}{1}$ and $\Phi$. The function in \eqref{eq:the complicated.expression.on.MKerr} is moreover a background quantity satisfying \eqref{eq:function.pull.back.general.assumption}. \qedhere
    
\end{proof}

At this point, we also note that by virtue of Lemma~\ref{lem:u.diff.main}, we have an estimate of the distance (say, in the $s$-coordinate of $\calM_{\mathrm{Kerr}}$) between $\Phi(p)$ and $\os{\Phi}{1}(p)$ for $p \in \os{\calM}{1}$. As a result:
\begin{lemma}\label{lem:to.find.restricting.set}
    Choosing $\ep$ smaller if necessary, 
    $$\os{\calM}{1} \cap \Big\{s \leq \f{3 s_f}{4} \Big\} \subset \Phi^{-1} (\Phi(\os{\calM}{1}) \cap \os{\Phi}{1}(\os{\calM}{1}))\cap \os{\Phi}{1}{}^{-1} (\Phi(\os{\calM}{1}) \cap \os{\Phi}{1}(\os{\calM}{1})).$$
\end{lemma}
From now on, we will use that, as a consequence of Lemma~\ref{lem:to.find.restricting.set}, all the estimates in Lemma~\ref{lem:comparing.Kerr.vector.fields}, Lemma~\ref{lem:almost.null.different.pullback}, and Lemma~\ref{lem:pull.back.comparison} hold on $\os{\calM}{1} \cap \{s \leq \f{3 s_f}{4}\}$.

\subsection{The principal null frame on dynamical spacetime}

In this subsection, we introduce a globally defined principal null frame on the dynamical spacetime. The way we define the principal null frame is as follows: First, we define a null frame $(e_1',e_2',e_3',e_4')$, which is close to the Kerr double null frame in a suitable sense, even though it is not associated with a global double null foliation. We then introduce the global dynamical principal null frame by imposing the Kerr transformation to $(e_1',e_2',e_3',e_4')$ (see Section~\ref{sec:Kerr.frames}).

When dealing with frames, we will use the $^{(N)}$ and $^{(S)}$ variants for the $e_A'$, $\pmb{e}_A'$, $e_A$, $\pmb{e}_A$, $\os{\ee}{1}{}_A$ frames when $A=1,2$. As in the previous subsection, all the estimates in this subsection are understood to valid for $\th_* \in [0,\f{3\pi}4]$ if $^{(\cdot)} = ^{(N)}$ and for $\th_* \in [\f \pi 4, \pi]$ if $^{(\cdot)} = ^{(S)}$. Again, for convenience in indexing, we denote $e_\mu'^{(\cdot)} = e_\mu'$ for $\mu = 3,4$, and similarly for $e_\mu^{(\cdot)}$, $\pmb{e}_\mu'^{(\cdot)}$ and $\pmb{e}_\mu^{(\cdot)}$.

\subsubsection{Definition of the global $(e_1',e_2',e_3',e_4')$ null frame}\label{sec:e'.frame}

Recall that we have a global coordinate system $(u',\ub, \th_*,\varphi_*)$ (see Section~\ref{sec:global.coordinates}). Slightly abusing notation\footnote{Later on in this section, when we need to consider the background double null frame on Kerr, we will explicitly write $\Phi_* \pmb{e}'_\mu$.}, we first define the background double null frame $(\pmb{e}_1', \pmb{e}_2', \pmb{e}_3', \pmb{e}_4')$ as the push-forward of those in Definition~\ref{def:Kerr.double.null} by $\Phi^{-1}$, i.e., 
\begin{align}
        \pmb{e}_3' = \pmb{\Omg}^{-2} \f{\rd}{\rd u'}\Big|_{DN'}, \quad 
        \pmb{e}_4' = \f{\rd}{\rd \ub}\Big|_{DN'} + \pmb{b}^A \f{\rd}{\rd \vartheta_*^A}\Big|_{DN'},\label{eq:pmbe3e4.def} \\
        \pmb{e}_1' = \f{R }{\ell} \Big( \rd_{\th_*}\Big|_{DN'} - (\f{\rd \mathfrak h}{\rd\th_*}) \rd_{\varphi_*} \Big|_{DN'} \Big),\quad \pmb{e}_2' = \f{1}{R\Si} \rd_{\varphi_*} \Big|_{DN'}, \label{EqDefBolde_A}
    \end{align}
    where $\pmb{\Omg}$ and $\pmb{b}$, $\f{R }{\ell}$ and $(\f{\rd \mathfrak h}{\rd\th_*})$ are background quantities evaluated at $(u',\ub,\vartheta_*)$. Define also the rotated versions $\pmb{e}_1'^{(N)}$, $\pmb{e}_2'^{(N)}$, $\pmb{e}_1'^{(S)}$, $\pmb{e}_2'^{(S)}$ as the push-forward of those in Definition~\ref{def:Kerr.double.null.rotation} under $\Phi^{-1}$.

We also define a dynamical frame $(e_1',e_2',e_3',e_4')$. We begin with $e_3'$ and $e_4'$:
\begin{definition}\label{def:e3'.e4'}
    Define $e_3'$ and $e_4'$ as unique vector fields on $\calM$ satisfying the following conditions:
\begin{enumerate}
    \item Let $e_3' = -2(g^{-1})^{\alp\bt} \rd_\alp \ub \rd_\bt$.
    \item Let $e_4'$ be such that $e_4'$ is orthogonal to the constant $(u',\ub)$-spheres, and $g(e_4',e_4') = 0$, $g(e_4', e_3') = -2$.
\end{enumerate}
(By \eqref{eq:e3.e4.in.coordinates.M2} below, this is consistent with \eqref{eq:e3'.e4'.def.restricted}.)
\end{definition}
For $e_1'$ and $e_2'$, we use Gram--Schmidt to ensure that they are orthogonal. 
\begin{definition} \label{DefDefe_A'}
    For $\th_* \in [0,\f{3\pi}{4}]$, normalize $(\pmb{e}_1'^{(N)}, \pmb{e}_2'^{(N)})$ by Gram--Schmidt, i.e., define
\begin{align}
    d_1'^{(N)} = \f{1}{\sqrt{g(\pmb{e}_1'^{(N)},\pmb{e}_1'^{(N)})}} \pmb{e}_1'^{(N)}, \quad 
    d_2'^{(N)} = \f{\pmb{e}_2'^{(N)} - g(d_1'^{(N)}, \pmb{e}_2'^{(N)}) d_1'^{(N)}}{\sqrt{g(\pmb{e}_2'^{(N)}, \pmb{e}_2'^{(N)}) - (g(d_1'^{(N)}, \pmb{e}_2'^{(N)}))^2 }},
\end{align}
so that $d_1'^{(N)}$, $d_2'^{(N)}$ are orthonormal with respect to $g$. Notice that $(\pmb{e}_1'^{(N)}, \pmb{e}_2'^{(N)})$ are orthonormal with respect to the Kerr metric. In what follows, $g-\pmb{g}$ will be small, and thus $g(\pmb{e}_1'^{(N)},\pmb{e}_1'^{(N)})$ and $g(\pmb{e}_2'^{(N)}, \pmb{e}_2'^{(N)}) - (g(d_1'^{(N)}, \pmb{e}_2'^{(N)}))^2$ are positive and bounded away from $0$, say, when $\th_* \in [0,\f{3\pi}{4}]$. In particular, $d_1'^{(N)}$, $d_2'^{(N)}$ are well-defined for $\th_* \in [0,\f{3\pi}{4}]$; see Lemma~\ref{lem:basic.double.null}.

Introduce a similar frame which is regular away from the north pole. Define
\begin{align}
    d_1'^{(S)} = \f{1}{\sqrt{g(\pmb{e}_1'^{(S)},\pmb{e}_1'^{(S)})}} \pmb{e}_1'^{(S)},\quad 
    d_2'^{(S)} = \f{\pmb{e}_2'^{(S)} - g(d_1'^{(S)}, \pmb{e}_2'^{(S)}) d_1'^{(S)}}{\sqrt{g(\pmb{e}_2'^{(S)}, \pmb{e}_2'^{(S)}) - (g(d_1'^{(S)}, \pmb{e}_2'^{(S)}))^2 }}.
\end{align}
The vector fields $d_1'^{(S)}$ and $d_2'^{(S)}$ are well-defined when $\th_* \in [\f{\pi}4, \pi]$ by similar considerations as above.

Now let $\upsilon:\mathbb S^2 \to [0,1]$ be an axisymmetric cutoff function such that $\upsilon \equiv 1$ for $\th_* \in [0,\f{\pi}{3}]$ (near the north pole) and $\upsilon \equiv 0$ for $\th_* \in [\f{2\pi}3, \pi]$ (near the south pole). Define
\begin{align}
    \mathfrak d_1' = &\: \upsilon \Big(\cos\varphi_* d_1'^{(N)} + \sin \varphi_* d_2'^{(N)} \Big) + (1 - \upsilon) \Big( \cos\varphi_* d_1'^{(S)} - \sin\varphi_* d_2'^{(S)}\Big), \label{eq:f1'} \\
    \mathfrak d_2' = &\: \upsilon \Big(-\sin\varphi_* d_1'^{(N)} + \cos \varphi_* d_2'^{(N)} \Big) + (1 - \upsilon) \Big( \sin\varphi_* d_1'^{(S)} + \cos\varphi_* d_2'^{(S)}\Big), \label{eq:f2'}
\end{align}
and then use Gram--Schimdt to define
\begin{align}
    e_1' =  \f{\mathfrak d_1'}{\sqrt{g(\mathfrak d_1',\mathfrak d_1')}}, \quad 
    e_2' =  \f{\mathfrak d_2' - g(e_1',\mathfrak d_2')e_1' }{\sqrt{g(\mathfrak d_2',\mathfrak d_2') - (g(e_1',\mathfrak d_2'))^2}}.\label{eq:e'}
\end{align}
\end{definition}

Define also the northern/southern versions given as follows in complex notations (cf.~Definition~\ref{def:Kerr.double.null.rotation}):
\begin{equation}\label{eq:e'.NS}
    e_1'^{(N)} + i e_2'^{(N)} = e^{i\varphi_*} (e_1' + i e_2'),\quad e_1'^{(S)} + i e_2'^{(S)} = e^{-i\varphi_*} (e_1' + i e_2').
\end{equation}

\subsubsection{Estimates for the frame in the red-shift region}

Before we proceed, it is useful to write down the transformation for the angular derivatives.
\begin{lemma}\label{lem:angular.transform}
    \begin{equation}
        \f{\rd}{\rd\vartheta_*^A}\Big|_{DN'} = \f{\rd}{\rd\vartheta_*^A}\Big|_s - \f{\f{\rd u'}{\rd\vartheta_*^A}\Big|_s}{\f{\rd u'}{\rd s}\Big|_s}  \os{\ee}{1}_3.
    \end{equation}
\end{lemma}

In the next two lemmas, we compute the vector fields $(\pmb{e}_3', \pmb{e}_4')$ in terms of the $(\os{\pmb{\ee}}{1}_3, \os{\pmb{\ee}}{1}{}_4, \f{\rd}{\rd \vartheta_*^A}\Big|_s)$ basis.
\begin{proposition}\label{prop:global.background.e3'e4'.comp}
The following holds in $\os{\calM}{1}$:
        \begin{align}
            \pmb{e}_3' = &\: \f{\pmb{\Omg}^{-2}}{\f{\rd u'}{\rd s}\Big|_s}\os{\pmb{\ee}}{1}_3, \label{eq:pmbe3'.comp}\\
            \pmb{e}_4' = &\: \os{\pmb{\ee}}{1}{}_4 + (\pmb{b}^A - \pmb{h}^A ) \f{\rd}{\rd \vartheta_*^A}\Big|_s -  \f{1}{\f{\rd u'}{\rd s}\Big|_s}\Big( \pmb{f}\f{\rd u'}{\rd s}\Big|_s + \f{\rd u'}{\rd \ub}\Big|_s + \pmb{b}^A \f{\rd u'}{\rd\vartheta_*^A}\Big|_s \Big) \os{\pmb{\ee}}{1}{}_3 . \label{eq:pmbe4'.comp}
        \end{align}
        Here, in all quantities except for $\pmb{\Omg}$, $\pmb{b}$ (which are quantities associated to double null coordinates) are evaluated at $(s,\ub, \vartheta_*)$, and $\pmb{\Omg}$, $\pmb{b}$ are evaluated at $(u'(s,\ub,\vartheta_*), \ub,\vartheta_*)$.
\end{proposition}
\begin{proof}
    Written in the $(\f{\rd}{\rd u'}\Big|_{DN'}, \f{\rd}{\rd \ub}\Big|_{DN'}, \f{\rd}{\rd \vartheta_*^A}\Big|_{DN'})$ basis,
    \begin{align}
        \os{\pmb{\ee}}{1}{}_3 = &\: \f{\rd u'}{\rd s}\Big|_s \f{\rd}{\rd u'}\Big|_{DN'}, \\
        \os{\pmb{\ee}}{1}{}_4 = &\: \f{\rd}{\rd \ub}\Big|_{DN'} + \pmb{h}^A \f{\rd}{\rd \vartheta_*^A}\Big|_{DN'} +  \Big( \pmb{f}\f{\rd u'}{\rd s}\Big|_s + \f{\rd u'}{\rd \ub}\Big|_s + \pmb{h}^A \f{\rd u'}{\rd \vartheta_*^A}\Big|_s \Big) \f{\rd}{\rd u'} \Big|_{DN'}.
    \end{align}

    Using the first equation and \eqref{eq:pmbe3e4.def}, we obtain \eqref{eq:pmbe3'.comp}.
    
    Rearranging the second equation, and using the first equation and Lemma~\ref{lem:angular.transform}, we obtain
    \begin{equation}
        \begin{split}
            \f{\rd}{\rd \ub}\Big|_{DN'} = &\: \os{\pmb{\ee}}{1}{}_4 - \pmb{h}^A \f{\rd}{\rd \vartheta_*^A}\Big|_s  -  \f{1}{\f{\rd u'}{\rd s}\Big|_s}\Big( \pmb{f}\f{\rd u'}{\rd s}\Big|_s + \f{\rd u'}{\rd \ub}\Big|_s \Big) \os{\pmb{\ee}}{1}{}_3 .
        \end{split}
    \end{equation}
    Going back to the definition of $\pmb{e}_4'$ and using Lemma~\ref{lem:angular.transform} again, we obtain \eqref{eq:pmbe4'.comp}. \qedhere
\end{proof}

The vector fields $e_3'$ and $e_4'$ defined in \eqref{def:e3'.e4'} can be expressed as follows:
\begin{proposition}\label{prop:global.e3'e4'.comp}
    In $\overset{\scriptscriptstyle{\text{[1]}}}\calM$, the vector fields $(e'_3,e'_4)$ in the $(u',\ub,\vartheta_*)$ coordinate system take the following form:
        \begin{align}
            e'_3 = &\:\f{\rd u'}{\rd s}\Big|_s \f{\rd}{\rd u'}\Big|_{DN'}, \label{eq:e3'.global} \\
    e'_4 = &\: \f{\rd}{\rd \ub}\Big|_{DN'} + \Big(h^A - 2 \f{\gamma^{AB} \f{\rd u'}{\rd \vartheta_*^B}\Big|_s}{\f{\rd u'}{\rd s}\Big|_s}\Big) \f{\rd}{\rd\vartheta_*^A}\Big|_{DN'}  + \Big(\f{\rd u'}{\rd s}\Big|_s f + \f{\rd u'}{\rd \ub}\Big|_s + h^A\f{\rd u'}{\rd \vartheta_*^A}\Big|_s-\f{|\nab u'|_s|_{\gamma}^2}{\f{\rd u'}{\rd s}\Big|_s}\Big) \f{\rd}{\rd u'}\Big|_{DN'}, \label{eq:e4'.global}
        \end{align}
        where $\f{\rd u'}{\rd s}\Big|_s$, $\f{\rd u'}{\rd \ub}\Big|_s$ and $\f{\rd u'}{\rd\vartheta_*^A}\Big|_s$ are computed as in Proposition~\ref{prop:metric.in.global.coordinates}, and $\nab u'|_s$ also denotes the angular derivatives.
    Moreover, 
     \begin{align}
        e'_3 = &\:\overset{\scriptscriptstyle{\text{[1]}}}\ee_3,\label{eq:e3'.in.terms.of.ee} \\
        e_4' =&\: \os{\ee}{1}_4 - \f{2 \gamma^{AB} \f{\rd u'}{\rd \vartheta_*^B}\Big|_s}{\f{\rd u'}{\rd s}\Big|_s} \f{\rd}{\rd\vartheta_*^A}\Big|_s -  \f{|\nab u'|_s|_{\gamma}^2}{(\f{\rd u'}{\rd s}\Big|_s)^2} \os{\ee}{1}_3.\label{eq:e4'.in.terms.of.ee}
    \end{align}
\end{proposition}
\begin{proof}

    Taking the inverse of \eqref{eq:metric.in.global.coordinates}, we obtain
\begin{equation}
    \begin{split}
        g^{-1} = &\: - \f 12 e_3'\otimes e_4' - \f 12  e_4' \otimes e_3' + \gamma^{-1},
    \end{split}
\end{equation}
if $e_3'$ and $e_4'$ take the form \eqref{eq:e3'.global}--\eqref{eq:e4'.global}. This verifies the formula for $(e'_3,e'_4)$. 
    
    To proceed, observe that 
    \begin{equation}\label{eq:du'ds.inverse}
        \f{\rd u'}{\rd s}\Big|_s \f{\rd s}{\rd u'} \Big|_{DN'} =1,    
    \end{equation}
    which follows from standard considerations about derivatives of inverse in one dimension (as $(\ub,\vartheta_*)$ is fixed). 

    The formula \eqref{eq:e3'.in.terms.of.ee} for $e_3'$ then follows from \eqref{eq:e3'.global} and \eqref{eq:du'ds.inverse}. For $e_4'$, we use \eqref{eq:dynamical.ee.3.4.def}
    \begin{equation}
        \begin{split}
            \os{\ee}{1}_4 = &\: \f{\rd}{\rd \ub}\Big|_s + f\f{\rd}{\rd s}\Big|_s +  h^A \f{\rd}{\rd\vartheta_*^A}\Big|_s \\
            = &\: \f{\rd}{\rd \ub}\Big|_{DN'} + h^A \f{\rd}{\rd\vartheta_*^A}\Big|_s + \Big( f\f{\rd u'}{\rd s}\Big|_s + \f{\rd u'}{\rd \ub} |_s \Big) \f{\rd}{\rd u'}\Big|_{DN'}.
        \end{split}
    \end{equation}
    Hence, rearranging and using \eqref{eq:e3'.global}, \eqref{eq:e3'.in.terms.of.ee}, we obtain
    \begin{equation}
        \begin{split}
            \f{\rd}{\rd \ub}\Big|_{DN'} = &\:\os{\ee}{1}_4 - h^A \f{\rd}{\rd\vartheta_*^A}\Big|_s - \Big( f + \f{\f{\rd u'}{\rd \ub}\Big|_s }{\f{\rd u'}{\rd s}\Big|_s} \Big) \os{\ee}{1}_3 .
        \end{split}
    \end{equation}
    Returning to \eqref{eq:e4'.global}, and using the above with Lemma~\ref{lem:angular.transform}, we thus obtain \eqref{eq:e4'.in.terms.of.ee}. \qedhere
\end{proof}

\begin{lemma}\label{lem:basic.double.null.RS}
    The following holds in $\os{\calM}{1}\cap \{s \leq \f{3s_f}{4} \}$: The vectors $(e_1', e_2')$ are well-defined and smooth away from the poles, $(e_1'^{(N)}, e_2'^{(N)})$ (respectively, $(e_1'^{(S)}, e_2'^{(S)})$) extend smoothly to the north pole (respectively, south pole). Moreover, for $^{(\cdot)} = ^{(N)}, ^{(S)}$ and denoting 
        \begin{equation}\label{eq:double.null.transform.RS}
            e_A'^{(\cdot)} = (\calA^{(\cdot)})_A^{B} \pmb{e}_B'^{(\cdot)},\quad \pmb{e}_A'^{(\cdot)} = (\pmb{\calA}^{(\cdot)})_A^B e_B'^{(\cdot)},
        \end{equation}
        the components $(\calA^{(\cdot)})_A^{B}$ and $(\pmb{\calA}^{(\cdot)})_A^B$ satisfy
        \begin{align}
            \sum_{i_1+i_2+i_3+i_4\leq I} \Big| (\os{\pmb{\ee}}{1}{}_1^{(\cdot)})^{i_1}
         (\os{\pmb{\ee}}{1}{}_2^{(\cdot)})^{i_2}
         (\os{\pmb{\ee}}{1}{}_3)^{i_3}(\os{\pmb{\ee}}{1}{}_4)^{i_4} ((\calA^{(\cdot)})_A^{B} - \de_A^B) \Big| \ls &\: \ep \mathfrak w(\ub), \label{eq:frame.diff.RS.1}\\
            \sum_{i_1+i_2+i_3+i_4\leq I} \Big| (\os{\pmb{\ee}}{1}{}_1^{(\cdot)})^{i_1}
         (\os{\pmb{\ee}}{1}{}_2^{(\cdot)})^{i_2}
         (\os{\pmb{\ee}}{1}{}_3)^{i_3}(\os{\pmb{\ee}}{1}{}_4)^{i_4} ((\pmb{\calA}^{(\cdot)})_A^B - \de_A^B) \Big| \ls &\: \ep \mathfrak w(\ub).\label{eq:frame.diff.RS.2}
        \end{align}
\end{lemma}
\begin{proof}

    \pfstep{Step~1: Smoothness of $d_1'^{(\cdot)}$ and $d_2'^{(\cdot)}$} Observe that $(\pmb{e}_1'^{(\cdot)}, \pmb{e}_2'^{(\cdot)})$ satisfies the orthonormality condition $(\Phi^* \pmb{g})(\pmb{e}_A'^{(\cdot)}, \pmb{e}_B'^{(\cdot)})= \pmb{g}(\Phi_*\pmb{e}_A'^{(\cdot)}, \Phi_*\pmb{e}_B'^{(\cdot)}) = \de_{AB}$. Hence,
    \begin{align}
        d_1'^{(\cdot)} - \pmb{e}_1'^{(\cdot)} = &\: \Big(\f{1}{\sqrt{g(\pmb{e}_1'^{(\cdot)},\pmb{e}_1'^{(\cdot)})}} -\f{1}{\sqrt{(\Phi^*\pmb{g})(\pmb{e}_1'^{(\cdot)},\pmb{e}_1'^{(\cdot)})}} \Big) \pmb{e}_1'^{(\cdot)}, \label{eq:most.annoying.1} \\
        d_2'^{(\cdot)} - \pmb{e}_2'^{(\cdot)} = &\: \f{\pmb{e}_2'^{(\cdot)} - g(d_1'^{(\cdot)}, \pmb{e}_2'^{(\cdot)}) d_1'^{(\cdot)}}{\sqrt{g(\pmb{e}_2'^{(\cdot)}, \pmb{e}_2'^{(\cdot)}) - (g(d_1'^{(\cdot)}, \pmb{e}_2'^{(\cdot)}))^2 }}  - \f{\pmb{e}_2'^{(\cdot)} - (\Phi^*\pmb{g})(\pmb{e}_1'^{(\cdot)}, \pmb{e}_2'^{(\cdot)}) d_1'^{(\cdot)}}{\sqrt{(\Phi^*\pmb{g})(\pmb{e}_2'^{(\cdot)}, \pmb{e}_2'^{(\cdot)}) - ((\Phi^*\pmb{g})( \pmb{e}_1'^{(\cdot)}, \pmb{e}_2'^{(\cdot)}))^2 }}. \label{eq:most.annoying.2}
    \end{align} 
    
    By Proposition~\ref{prop:smmothness.background.e}, $(\pmb{e}_1'^{(\cdot)},\pmb{e}_2'^{(\cdot)})$ are smooth away from the south pole (respectively, north pole) when $^{(\cdot)} = ^{(N)}$ (respectively $^{(\cdot)} = ^{(S)}$). Thus, well-definedness and smoothness of $(d_1'^{(\cdot)}, d_2'^{(\cdot)})$ follows from showing that the right-hand sides of \eqref{eq:most.annoying.1}--\eqref{eq:most.annoying.2} and their derivatives are bounded by $\ep \mathfrak w(\ub)$. For this, we make use of the fact that they can be written in terms of $g-\Phi^*\pmb{g}$. To consider the quantity $(g  - \Phi^*\pmb{g})(\pmb{e}_{A}'^{(\cdot)},\pmb{e}_{B}'^{(\cdot)})$, observe that by Lemma~\ref{lem:angular.transform}, $\pmb{e}_{A}'^{(\cdot)} = \os{\pmb{\ee}}{1}{}_{A}^{(\cdot)} -\f{\os{\pmb{\ee}}{1}{}_{A}^{(\cdot)} u'}{\f{\rd u'}{\rd s}|_s}  \os{\ee}{1}_3$. It then follows, since $\os{\ee}{1}_3$ is null and orthogonal to the constant-$\ub$ hypersurface for both $g$ and $\Phi^*\pmb{g}$, that 
    \begin{equation}\label{eq:writing.g.diff.in.terms.of.projected}
        \begin{split}
            (g  - \Phi^*\pmb{g})(\pmb{e}_{A}'^{(\cdot)},\pmb{e}_{B}'^{(\cdot)}) = (g  - \Phi^*\pmb{g})(\os{\pmb{\ee}}{1}{}_{A}^{(\cdot)},\os{\pmb{\ee}}{1}{}_{B}^{(\cdot)}).
        \end{split}
    \end{equation}
    Noting $(\os{\Phi}{1}{}^*\pmb{g})(\os{\pmb{\ee}}{1}{}_{A}^{(\cdot)},\os{\pmb{\ee}}{1}{}_{B}^{(\cdot)}) = \de_{AB}$, we now write
    \begin{equation}\label{eq:writing.g.diff.in.terms.of.projected.2}
        \begin{split}
             (\Phi^*\pmb{g})(\os{\pmb{\ee}}{1}{}_{A}^{(\cdot)},\os{\pmb{\ee}}{1}{}_{B}^{(\cdot)}) 
        = &\: \de_{AB} + (\Phi^*\pmb{g})(\os{\pmb{\ee}}{1}{}_{A}^{(\cdot)},\os{\pmb{\ee}}{1}{}_{B}^{(\cdot)}) - (\os{\Phi}{1}{}^*\pmb{g})(\os{\pmb{\ee}}{1}{}_{A}^{(\cdot)},\os{\pmb{\ee}}{1}{}_{B}^{(\cdot)}).
        \end{split}    
    \end{equation}
    Plugging this back into \eqref{eq:writing.g.diff.in.terms.of.projected}, we have
    \begin{equation}
        (g  - \Phi^*\pmb{g})(\pmb{e}_{A}'^{(\cdot)},\pmb{e}_{B}'^{(\cdot)}) = \Big(g(\os{\pmb{\ee}}{1}{}_{A}^{(\cdot)},\os{\pmb{\ee}}{1}{}_{B}^{(\cdot)}) - \de_{AB}\Big) + \Big((\Phi^*\pmb{g})(\os{\pmb{\ee}}{1}{}_{A}^{(\cdot)},\os{\pmb{\ee}}{1}{}_{B}^{(\cdot)}) - (\os{\Phi}{1}{}^*\pmb{g})(\os{\pmb{\ee}}{1}{}_{A}^{(\cdot)},\os{\pmb{\ee}}{1}{}_{B}^{(\cdot)})\Big) =:I+II.
    \end{equation}
    We bound $I$ and its derivatives by\footnote{Notice that using \eqref{eq:ee.1.explcit} and the estimates in Theorem~\ref{thm:main.DL.est}, we can change the derivatives in \eqref{eq:transition.almost.orthonormal} to $(\os{\pmb{\ee}}{1}{}_1'^{(\cdot)})^{i_1} (\os{\pmb{\ee}}{1}{}_2'^{(\cdot)})^{i_2} (\os{\pmb{\ee}}{1}{}_3')^{i_3} (\os{\pmb{\ee}}{1}{}_4')^{i_4}$ derivatives.} \eqref{eq:transition.almost.orthonormal}, and bound $II$ and its derivatives by Lemma~\ref{lem:almost.null.different.pullback}. As a result, returning to \eqref{eq:most.annoying.1}--\eqref{eq:most.annoying.2}, it follows that
    \begin{equation}\label{eq:d-e.expression.RS.1}
        d_A'^{(\cdot)} - \pmb{e}_A'^{(\cdot)} = (\widetilde{\mathfrak{A}}^{(\cdot)})_A^B \pmb{e}'^{(\cdot)}_B,\quad A=1,2,
    \end{equation}
    for some $(\widetilde{\mathfrak{A}}^{(\cdot)})_A^B$ satisfying
    \begin{equation}\label{eq:d-e.bd.RS.1}
        \Big| (\os{\pmb{\ee}}{1}{}_1'^{(\cdot)})^{i_1} (\os{\pmb{\ee}}{1}{}_2'^{(\cdot)})^{i_2} (\os{\pmb{\ee}}{1}{}_3')^{i_3} (\os{\pmb{\ee}}{1}{}_4')^{i_4} (\widetilde{\mathfrak{A}}^{(\cdot)})_A^B \Big| \ls \ep\mathfrak w(\ub).
    \end{equation}
    %

    \pfstep{Step~2: Proof of \eqref{eq:frame.diff.RS.1}--\eqref{eq:frame.diff.RS.2}} 
    Recall that $(d_1'^{(N)}, d_2'^{(N)})$, $(d_1'^{(S)}, d_2'^{(S)})$ are $g$-orthonormal. It therefore follows from \eqref{eq:f1'}, \eqref{eq:f2'}, \eqref{eq:e'} that $e_A' = \mathfrak d_A'$ ($A=1,2$) when $\upsilon = 0$ or $\upsilon =1$. Hence,  $e_A'^{(N)} = d_A'^{(N)}$ on $[0,\f{\pi}3]$ (outside the support of $1-\upsilon$), and $e_A'^{(S)} = d_A'^{(S)}$ on $[\f{2\pi}3,\pi]$ (outside the support of $\upsilon$). 
    
    We now consider different cases. It will be convenient to use complex notations and write $\mathfrak d' = \mathfrak d_1' + i \mathfrak d_2'$, $d'^{(N)} =d_1'^{(N)} + i d_2'^{(N)}$, and similarly for $d'^{(S)}$, $\pmb{e}'$, $\pmb{e}'^{(N)}$, $\pmb{e}'^{(S)}$, etc. In the following steps, we prove the bound for the $^{(N)}$ version with $\th_* \in [0,\f{3\pi}{4}]$. The $^{(S)}$ version can be argued similarly.
    \begin{enumerate}
        \item On $[0,\f{\pi}3]$, the bound \eqref{eq:frame.diff.RS.1}--\eqref{eq:frame.diff.RS.2} in this region is immediate from $e_A'^{(N)} = d_A'^{(N)}$ and \eqref{eq:d-e.expression.RS.1}--\eqref{eq:d-e.bd.RS.1}.
        \item On $[\f{2\pi}{3},\f{3\pi}4]$, since $\upsilon = 0$, we have, by \eqref{eq:f1'}--\eqref{eq:f2'}, that $e' = \mathfrak d' = e^{-i\varphi_*} d'^{(S)}$.
    Thus, 
            \begin{equation}
                e'^{(N)} = e^{-i\varphi_*}e' = e^{-2i\varphi_*} d'^{(S)}.
            \end{equation}
            Observing also that the Kerr double null frame satisfies
            \begin{equation}
                \pmb{e}'^{(N)} = e^{-2i\varphi_*} \pmb{e}'^{(S)},
            \end{equation}
            we thus obtain $e'^{(N)} - \pmb{e}'^{(N)} = e^{-2i\varphi_*} (d'^{(S)} - \pmb{e}'^{(S)})$. The desired estimate \eqref{eq:frame.diff.RS.1}--\eqref{eq:frame.diff.RS.2} thus follows from \eqref{eq:d-e.expression.RS.1}--\eqref{eq:d-e.bd.RS.1} and the smoothness of $e^{-2i\varphi_*}$ in this region.
        \item We now consider $[\f{\pi}3, \f{2\pi}{3}]$. We have by \eqref{eq:f1'} and \eqref{eq:f2'} that
            \begin{equation}
                \mathfrak d'  = \upsilon e^{-i\varphi_*} d'^{(N)}   + (1 - \upsilon) e^{i\varphi_*} d'^{(S)}.  
            \end{equation}
            Writing $\pmb{e}' = \upsilon e^{-i\varphi_*} \pmb{e}'^{(N)} + (1-\upsilon) e^{i\varphi_*} \pmb{e}'^{(S)}$
            and using \eqref{eq:d-e.expression.RS.1}, we have
    \begin{align}
        \mathfrak d' - \pmb{e}' =\upsilon e^{-i\varphi_*}(d'^{(N)} - \pmb{e}'^{(N)}) + (1-\upsilon) e^{i\varphi_*} (d'^{(S)} - \pmb{e}'^{(S)})  =: \widetilde{\mathfrak{a}} \,\pmb{e} = e^{i\varphi_*} \widetilde{\mathfrak{a}} \pmb{e}'^{(N)} = e^{-i\varphi_*} \widetilde{\mathfrak{a}} \pmb{e}'^{(S)}
    \end{align}
    where by \eqref{eq:d-e.expression.RS.1}--\eqref{eq:d-e.bd.RS.1}, for complex-valued functions $\widetilde{a} = \mathfrak{a}, e^{i\varphi_*} \widetilde{\mathfrak{a}}, e^{-i\varphi_*} \widetilde{\mathfrak{a}}$ it holds that
    \begin{equation}\label{eq:near.equator.complex.RS}
        \sum_{i_1+i_2+i_3+i_4\leq I} \Big|(\os{\pmb{\ee}}{1}{}_1^{(\cdot)})^{i_1}
         (\os{\pmb{\ee}}{1}{}_2^{(\cdot)})^{i_2}
         (\os{\pmb{\ee}}{1}{}_3)^{i_3}(\os{\pmb{\ee}}{1}{}_4)^{i_4} (\widetilde{\mathfrak{a}})_A^B \Big| \ls \ep \mathfrak w(\ub),\quad \th_* \in [\f{\pi}3,\f{2\pi}3],
    \end{equation}
    where we used that $e^{\pm i\varphi_*}$ is smooth when $\th_* \in [\f{\pi}3, \f{2\pi}{3}]$.
    Finally, since $(e_1',e_2')$ is obtained from $(\mathfrak d_1',\mathfrak d_2')$ via Gram--Schmidt (see \eqref{eq:e'}), for $\ep$ sufficiently small, the desired estimates \eqref{eq:frame.diff.RS.1}--\eqref{eq:frame.diff.RS.2} follow from \eqref{eq:near.equator.complex.RS}. \qedhere
    \end{enumerate}

\end{proof}

The following is the main corollary about the change of frames that we will need later.
\begin{corollary}\label{cor:RS.e'.to.ee}
    The following holds in $\os{\calM}{1}\cap \{s \leq \f{3s_f}{4} \}$: Writing
    \begin{equation}\label{eq:bbA.def}
        e_\mu'^{(\cdot)} = (\mathbb A^{(\cdot)})_\mu^\nu \os{\ee}{1}{}_\nu^{(\cdot)}, \quad \pmb{e}_\mu'^{(\cdot)} = (\pmb{\mathbb A}^{(\cdot)})_\mu^\nu \os{\pmb{\ee}}{1}{}_\nu^{(\cdot)},
    \end{equation}
    it holds that 
    \begin{equation}\label{eq:RS.e'.to.ee}
        \sum_{i_1+i_2+i_3+i_4\leq I} \Big| (\os{\pmb{\ee}}{1}{}_1^{(\cdot)})^{i_1}
         (\os{\pmb{\ee}}{1}{}_2^{(\cdot)})^{i_2}
         (\os{\pmb{\ee}}{1}{}_3)^{i_3}
         (\os{\pmb{\ee}}{1}{}_4)^{i_4} \Big((\mathbb A^{(\cdot)})_\mu^\nu - (\pmb{\mathbb A}^{(\cdot)})_\mu^\nu\Big)\Big|\ls \ep \mathfrak w(\ub).
    \end{equation}
    Moreover, all components of $(\mathbb A^{(\cdot)})_\mu^\nu$ and $(\pmb{\mathbb A}^{(\cdot)})_\mu^\nu$, and all components of their inverses, as well as their derivatives by $(\os{\pmb{\ee}}{1}{}_1^{(\cdot)})^{i_1}
         (\os{\pmb{\ee}}{1}{}_2^{(\cdot)})^{i_2}
         (\os{\pmb{\ee}}{1}{}_3)^{i_3}
         (\os{\pmb{\ee}}{1}{}_4)^{i_4}$, are bounded. 
    
    Furthermore, defining
    \begin{equation}\label{eq:bbF}
        \os{\ee}{1}{}_\mu^{(\cdot)} = (\mathbb F^{(\cdot)})_\mu^\nu \os{\pmb{\ee}}{1}{}_\nu^{(\cdot)},\quad \os{\pmb{\ee}}{1}{}_\mu^{(\cdot)} = (\pmb{\mathbb F}^{(\cdot)})_\mu^\nu \os{\ee}{1}{}_\nu^{(\cdot)},
    \end{equation}
    it holds that
     \begin{align}
        \sum_{i_1+i_2+i_3+i_4\leq I} \Big| (\os{\pmb{\ee}}{1}{}_1^{(\cdot)})^{i_1}
         (\os{\pmb{\ee}}{1}{}_2^{(\cdot)})^{i_2}
         (\os{\pmb{\ee}}{1}{}_3)^{i_3}
         (\os{\pmb{\ee}}{1}{}_4)^{i_4} ((\mathbb F^{(\cdot)})_\mu^\nu - \de_\mu^\nu)\Big|\ls \ep \mathfrak w(\ub), \label{eq:RS.ee.diff.to.ee.1} \\
         \sum_{i_1+i_2+i_3+i_4\leq I} \Big| (\os{\pmb{\ee}}{1}{}_1^{(\cdot)})^{i_1}
         (\os{\pmb{\ee}}{1}{}_2^{(\cdot)})^{i_2}
         (\os{\pmb{\ee}}{1}{}_3)^{i_3}
         (\os{\pmb{\ee}}{1}{}_4)^{i_4} ((\pmb{\mathbb F}^{(\cdot)})_\mu^\nu - \de_\mu^\nu)\Big|\ls \ep \mathfrak w(\ub), \label{eq:RS.ee.diff.to.ee.2}
    \end{align}
\end{corollary}
\begin{proof}
    We first prove \eqref{eq:RS.e'.to.ee}. Recall throughout that $\os{\ee}{1}{}_A= \os{\pmb{\ee}}{1}{}_A$ and $\os{\ee}{1}{}_3 = \os{\pmb{\ee}}{1}{}_3$. 
    
    Suppose $\mu = A= 1,2$. Then since $\pmb{e}_A'^{(\cdot)} = \os{\ee}{1}{}_A'^{(\cdot)} - \f{1}{\f{\rd u'}{\rd s}|_s} (\os{\ee}{1}{}_A'^{(\cdot)} u') \os{\ee}{1}{}_3$, the desired estimate follows from Lemma~\ref{lem:basic.double.null.RS}.

    Suppose $\mu = 3$. Then, $(\mathbb A^{(\cdot)})_3^\nu$ and $(\pmb{\mathbb A}^{(\cdot)})_3^\nu$ are only non-vanishing when $\nu=3$, in which case $(\pmb{\mathbb A}^{(\cdot)})_3^3 = \f{\pmb{\Omg}^{-2}(u'(s,\ub,\vartheta_*), \ub,\vartheta_*)}{\f{\rd u'}{\rd s}|_s}$ (by \eqref{eq:pmbe3'.comp}) $(\mathbb A^{(\cdot)})_3^3 = 1$ (by \eqref{eq:e3'.in.terms.of.ee}). The difference estimate follows from \eqref{eq:derivatives.tu.1}, Lemma~\ref{lem:u.diff.main} and Proposition~\ref{prop:DL.rescaled}. (Notice that in the process we need to control $\pmb{\Omg}^2(u',\cdots) - \pmb{\Omg}^2(u_1,\cdots)$, $\pmb{\Omg}^2(u_1,\cdots) -\pmb{\Omg}^2(u,\cdots)$ and $\Omg^{-2}(u,\cdots) - \pmb{\Omg}^{-2}(u,\cdots)$; for the first two one uses the mean value theorem together with Lemma~\ref{lem:u.diff.main}.)

    Finally, when $\mu = 4$, we have by \eqref{eq:pmbe4'.comp} and \eqref{eq:e4'.in.terms.of.ee} that $(\mathbb A^{(\cdot)})_4^4 = (\pmb{\mathbb A}^{(\cdot)})_4^4 = 1$, 
    \begin{equation}
        (\mathbb A^{(\cdot)})_4^3 = - \f{|\nab u'|_{\gamma}^2}{(\f{\rd u'}{\rd s}\Big|_s)^2},\quad (\pmb{\mathbb A}^{(\cdot)})_4^3 = -  \Big( \pmb{f} + \f{\f{\rd u'}{\rd \ub}\Big|_s}{\f{\rd u'}{\rd s}\Big|_s} + \f{\pmb{b}^A \f{\rd u'}{\rd\vartheta_*^A}\Big|_s}{\f{\rd u'}{\rd s}\Big|_s} \Big),
    \end{equation}
    and 
    \begin{equation}\label{eq:mathbb.A.diff.to.invert}
       (\mathbb A^{(\cdot)})_4^B  \os{\ee}{1}{}_B^{(\cdot)} = - \f{2 \gamma^{AB} \f{\rd u'}{\rd \vartheta_*^B}\Big|_s}{\f{\rd u'}{\rd s}\Big|_s} \f{\rd}{\rd\vartheta_*^A}\Big|_s,\quad (\pmb{\mathbb A}^{(\cdot)})_4^B  \os{\ee}{1}{}_B^{(\cdot)} = (\pmb{b}^A - \pmb{h}^A ) \f{\rd}{\rd \vartheta_*^A}\Big|_s,
    \end{equation}
    where $\pmb{b}$ is evaluated at $(u'(s,\ub,\vartheta_*),\ub,\vartheta_*)$ and all the other quantities are evaluated at $(s,\ub,\vartheta_*)$. 
    For $\mathbb A_4^3 - \pmb{\mathbb A}_4^3$, we have the desired bound from the computations \eqref{eq:derivatives.tu.1}--\eqref{eq:derivatives.tu.3}, as well as the bounds for $u-u_1$ in Lemma~\ref{lem:u.diff.main}, for $f-\pmb{f}$, $h-\pmb{h}$ in Proposition~\ref{thm:DL.2}, and for $b-\pmb{b}$ in Proposition~\ref{prop:DL.rescaled}. (For $b-\pmb{b}$, we need to control $\pmb{b}(u',\cdots) - \pmb{b}^2(u_1,\cdots)$, $\pmb{b}^2(u_1,\cdots) -\pmb{b}^2(u,\cdots)$ in the process, which can again be done using Lemma~\ref{lem:u.diff.main} as for $\pmb{\Omg}^{-2}$ above.)
    
    Since $(\os{\ee}{1}{}_1^{(\cdot)}, \os{\ee}{1}{}_2^{(\cdot)})$ is almost an orthonormal basis (in the sense of \eqref{eq:transition.almost.orthonormal}), we can invert the linear system \eqref{eq:mathbb.A.diff.to.invert} to solve for $(\mathbb A^{(\cdot)})_4^B - (\pmb{\mathbb A}^{(\cdot)})_4^B$. Each component can then be bounded by the norm of $(\f{2 \gamma^{AB} \f{\rd u'}{\rd \vartheta_*^B}|_s}{\f{\rd u'}{\rd s}|_s} + \pmb{b}^A - \pmb{h}^A ) \f{\rd}{\rd \vartheta_*^A}\Big|_s$, which, as before, can be bounded using \eqref{eq:derivatives.tu.1}--\eqref{eq:derivatives.tu.3}, Lemma~\ref{lem:u.diff.main}, Proposition~\ref{thm:DL.2}, and Proposition~\ref{prop:DL.rescaled}. 

    Given \eqref{eq:RS.e'.to.ee}, and the boundedness of the background quantities,  we obtain the boundedness of the components of $\pmb{\mathbb A}$, $\mathbb A$, and their derivatives. By the above calculations, $\pmb{\mathbb A}$ takes the form
    \begin{equation*}
    \begin{pmatrix}
        1 & 0 & \star & 0 \\ 0 & 1 & \star & 0 \\ 0 & 0 & 1 & 0 \\ \star & \star & 0 & 1
    \end{pmatrix},
\end{equation*}
    giving $\det \pmb{\mathbb A} = 1$ and thus the inverses and their derivatives are also bounded.
    
    Finally, for \eqref{eq:RS.ee.diff.to.ee.1}--\eqref{eq:RS.ee.diff.to.ee.2}, we use \eqref{eq:ee.1.explcit}, and then conclude using \eqref{eq:transition.almost.orthonormal}, Lemma~\ref{lem:u.diff.main}, Proposition~\ref{thm:DL.2}, and Proposition~\ref{prop:DL.rescaled} as above. \qedhere

\end{proof}

\subsubsection{Estimates for the frame in the blue-shift region}

We now turn to the blue-shift region. In fact, from now on, we will restrict our estimates to $\os{\calM}{2}\setminus (\calU \cap \{s \leq \f{s_f}2\})$ (instead of all of $\os{\calM}{2}$) so that $u' \equiv u$. The complementary region will always be treated as part of the blue-shift region.

First, the analog of Proposition~\ref{prop:global.background.e3'e4'.comp} is much simpler in this case because $u' \equiv u$, and is a direct computation:

\begin{lemma}
    In $\os{\calM}{2}\setminus (\calU \cap \{s \leq \f{s_f}2\})$, 
        \begin{equation}\label{eq:e3.e4.in.coordinates.M2}
        \begin{aligned}
            \pmb{e}_3' = &\: \pmb{\Omg}^{-2} \os{\pmb{\ee}}{2}_3, \quad 
            \pmb{e}_4' = \pmb{\Omg}^2 \os{\pmb{\ee}}{2}_4, \quad
            e'_3 = \Omg^{-2} \f{\rd}{\rd u'} = \Omg^{-2} \overset{\scriptscriptstyle{\text{[2]}}}{\ee}_3 ,\quad e'_4 = \f{\rd}{\rd \ub} + b^A \f{\rd}{\rd \vartheta_*^A}= \Omg^2 \overset{\scriptscriptstyle{\text{[2]}}}{\ee}_4.
        \end{aligned}
        \end{equation}
\end{lemma}

The following is an analog of Lemma~\ref{lem:basic.double.null.RS} and Corollary~\ref{cor:RS.e'.to.ee}. We will simultaneously control the transformation of all frame fields. Note that unlike the red-shift region, these estimates see some degeneration in $|\Delta|$.

\begin{lemma}\label{lem:basic.double.null}
    The following holds in $\os{\calM}{2}\setminus (\calU \cap \{s \leq \f{s_f}2\})$: The vectors $(e_1', e_2')$ are well-defined and smooth away from the poles, $(e_1'^{(N)}, e_2'^{(N)})$ (respectively, $(e_1'^{(S)}, e_2'^{(S)})$) extend smoothly to the north pole (respectively, south pole). The vectors $(e_3',e_4')$ are smooth. 

    Denoting the transformation by
        \begin{equation}\label{eq:double.null.transform}
            e_\mu'^{(\cdot)} = (\calA^{(\cdot)})_\mu^{\nu} \pmb{e}_\nu'^{(\cdot)},\quad \pmb{e}_\mu'^{(\cdot)} = (\pmb{\calA}^{(\cdot)})_\mu^\nu e_\nu'^{(\cdot)},
        \end{equation}
        it holds that $(\calA^{(\cdot)})_\mu^{\nu}$ and $(\pmb{\calA}^{(\cdot)})_\mu^\nu$ satisfy
        \begin{align}
            \sum_{i_1+i_2+i_3+i_4\leq I}\opnorm[\Big]{(\pmb{e}_1'^{(\cdot)})^{i_1} (\pmb{e}_2'^{(\cdot)})^{i_2} (|\Delta| \pmb{e}_3')^{i_3} (\pmb{e}_4')^{i_4} ((\calA^{(\cdot)})_\mu^{\nu} - \de_\mu^\nu) } \ls &\: \ep,\label{eq:frame.diff.1}\\
            \sum_{i_1+i_2+i_3+i_4\leq I}  \opnorm[\Big]{(\pmb{e}_1'^{(\cdot)})^{i_1} (\pmb{e}_2'^{(\cdot)})^{i_2} (|\Delta| \pmb{e}_3')^{i_3} (\pmb{e}_4')^{i_4}((\pmb{\calA}^{(\cdot)})_\mu^\nu - \de_\mu^\nu)} \ls &\: \ep.\label{eq:frame.diff.2}
        \end{align}
        Moreover $(\calA^{(\cdot)})_{3}^\mu = (\calA^{(\cdot)})_\mu^3 = (\pmb{\calA}^{(\cdot)})_3^\mu = (\pmb{\calA}^{(\cdot)})_\mu^3 = 0$ if $\mu \neq 3$, and $ \calA_A^4  = (\pmb{\calA}^{(\cdot)})_A^4 = 0$ if $A = 1,2$, and $(\calA^{(\cdot)})_{4}^4 = (\pmb{\calA}^{(\cdot)})_4^4 = 1$.
\end{lemma}
\begin{proof}
    The proof is similar to Lemma~\ref{lem:basic.double.null.RS} and we will only point out the differences. The formulas \eqref{eq:most.annoying.1}--\eqref{eq:most.annoying.2} are the same, but in order to control $(g  - \pmb{g})(\pmb{e}_{A}'^{(\cdot)},\pmb{e}_{B}'^{(\cdot)}) = (\gamma  - \pmb{\gamma})(\pmb{e}_{A}'^{(\cdot)},\pmb{e}_{B}'^{(\cdot)})$, it is now important to note that \eqref{eq:final.double.null.1} only allows us to differentiate with $\pmb{\Omg}^2\pmb{e}_3' \sim |\Delta| \pmb{e}_3'$ and thus we need a degeneration for such derivatives, i.e., after writing as before
    \begin{equation}\label{eq:d-e.expression.1}
        d_A'^{(\cdot)} - \pmb{e}_A'^{(\cdot)} = (\widetilde{\mathfrak{A}}^{(\cdot)})_A^B \pmb{e}'^{(\cdot)}_B,\quad A=1,2,
    \end{equation}
    the $(\widetilde{\mathfrak{A}}^{(\cdot)})_A^B$ only satisfy
    \begin{equation}\label{eq:d-e.bd.1}
        \opnorm{(\pmb{e}_1'^{(\cdot)})^{i_1} (\pmb{e}_2'^{(\cdot)})^{i_2} (|\Delta| \pmb{e}_3')^{i_3} (\pmb{e}_4')^{i_4} (\widetilde{\mathfrak{A}}^{(\cdot)})_A^B } \ls \ep.
    \end{equation}
    Here, the notation $\opnorm{\cdot}$ is understood after restricting to the set of $\th_*$ depending on whether $^{(\cdot)} = ^{(N)},^{(S)}$. Starting with \eqref{eq:d-e.expression.1} and \eqref{eq:d-e.bd.1}, we then argue as in Step~2 of the proof of Lemma~\ref{lem:basic.double.null.RS} to obtain \eqref{eq:frame.diff.1}--\eqref{eq:frame.diff.2} when $\mu,\nu \in \{1,2\}$.

Finally, for the other components in \eqref{eq:frame.diff.1}--\eqref{eq:frame.diff.2}, recall now that 
\begin{align} \pmb{e}'_3 := \pmb{\Omg}^{-2} \rd_u \quad \pmb{e}_4' := \rd_{\ub} + \pmb{b}^A \rd_{\vartheta_*^A},\quad e_3' = \Om^{-2} \rd_u ,\quad e_4' = \rd_{\ub} + b^A \rd_{\vartheta_*^A}
\end{align}
where $\pmb{\Omg}^{-2} = -\f{R^2}{\Delta}$, $\pmb{b}^{\th_*} = 0$, and $\pmb{b}^{\varphi_*} = \f{4Mar}{\Sigma R^2}$. From this, it is easy to read of the estimates \eqref{eq:frame.diff.1}--\eqref{eq:frame.diff.2} when $\mu = 3,4$ (and $\nu$ arbitrary) using the bounds for $\widetilde{b}$ and $\widetilde{g}$ in Proposition~\ref{prop:DL.rescaled}. \qedhere
\end{proof}

\begin{lemma}\label{lem:commutators}
The following holds in $\os{\calM}{2}\setminus (\calU \cap \{s < \f{s_f}{2}\})$. For $^{(\cdot)} = ^{(N)},^{(S)}$, we write the commutators as follows
%
\begin{align}
[e_A'^{(\cdot)}, e_B'^{(\cdot)}] - [\pmb{e}_A'^{(\cdot)}, \pmb{e}_B'^{(\cdot)}] = &\: (\overline{\mathfrak c}^{(\cdot)})_{AB}^C \pmb{e}_C'^{(\cdot)}, \\
[e_A'^{(\cdot)}, e'_4] - [\pmb{e}_A'^{(\cdot)}, \pmb{e}'_4] = &\: (\overline{\mathfrak c}^{(\cdot)})_{A4}^C \pmb{e}_C'^{(\cdot)} \\
[e_A'^{(\cdot)}, e'_3] - [\pmb{e}_A'^{(\cdot)}, \pmb{e}'_3]= &\: (\overline{\mathfrak c}^{(\cdot)})_{A3}^C \pmb{e}_C'^{(\cdot)} + \overline{\mathfrak c}_{A3}^3 \pmb{e}_3',
\end{align}
Then the following holds:
    $$\sum_{i_1+i_2+i_3+i_4\leq I} \opnorm{(\pmb{e}_1'^{(\cdot)})^{i_1} (\pmb{e}_2'^{(\cdot)})^{i_2} (|\Delta| \pmb{e}_3')^{i_3} (\pmb{e}_4')^{i_4}(\overline{\mathfrak c}_{AB}^C, \overline{\mathfrak c}_{A4}^C, |\Delta| \overline{\mathfrak c}_{A3}^C,\overline{\mathfrak c}_{A3}^3)} \ls \ep.$$

\end{lemma}
\begin{proof}
    This is an immediate consequence of Lemma~\ref{lem:basic.double.null}. \qedhere
\end{proof}

\subsubsection{Principal null frame on dynamical background}\label{sec:principal.null.first.def}

Finally, define the background principal null frame $(\pmb{e}_1,\pmb{e}_2,\pmb{e}_3,\pmb{e}_4)$ and the dynamical principal null frame $(e_1,e_2,e_3,e_4)$ by relating them with $(e_1',e_2',e_3',e_4')$ through the Kerr relations, i.e., 
\begin{equation} \label{EqDefBackgroundPNFrame}
    \pmb{e}_\mu = \pmb{\calB}_\mu^\nu \pmb{e}'_\nu,\quad \pmb{e}'_\mu = \pmb{\calB}_\mu'^\nu \pmb{e}_\nu,
\end{equation}
and
\begin{equation} \label{EqDefDynamicalPNFrame}
    e_\mu = \pmb{\calB}_\mu^\nu e'_\nu,\quad e'_\mu = \pmb{\calB}_\mu'^\nu e_\nu,
\end{equation}
where $\pmb{\calB}$ and $\pmb{\calB}'$ are the functions on $\calM_{\mathrm{Kerr}}$ defined in Definition~\ref{def:B}, now pulled back to $\calM$ via the map $\Phi$ (Definition~\ref{def:Phi}).

We will also define 
\begin{equation}\label{eq:e.NS}
    e_1^{(N)} + i e_2^{(N)} = e^{i\varphi_*} (e_1 + i e_2),\quad e_1^{(S)} + i e_2^{(S)} = e^{-i\varphi_*} (e_1 + i e_2)
\end{equation}
so that for $^{(\cdot)} = ^{(N)}, ^{(S)}$, the following holds (with the usual convention that $e_\mu^{(\cdot)} = e_\mu$ and $e_\mu'^{(\cdot)} = e_\mu'$ if $\mu =3,4$):
\begin{equation}\label{eq:B().def}
e_\mu^{(\cdot)} = (\pmb{\calB}^{(\cdot)})_\mu^\nu e'^{(\cdot)}_\nu,\quad e'^{(\cdot)}_\mu = (\pmb{\calB}'^{(\cdot)})_\mu^\nu e_\nu.
\end{equation}

 Notice that the frames $\pmb{e}_\mu^{(N)}$ and $\pmb{e}_\mu'^{(N)}$ on Kerr are smooth in $\th_* \in [0,\f{3\pi}4]$. Hence, the transformations $(\pmb{\calB}^{(N)})_\mu^\nu$ and $(\pmb{\calB}'^{(N)})_\mu^\nu$ are also smooth in $\th_* \in [0,\f{3\pi}4]$. (This is because $(\pmb{\calB}^{(N)})_1^1 = \pmb{g}(\pmb{e}_1^{(N)},\pmb{e}_1'^{(N)})$, etc.) As a result, $e_\mu^{(N)}$ are smooth vector fields when $\th_* \in [0,\f{3\pi}4]$. Similar comments apply to $e_\mu^{(S)}$ for $\th_* \in [\f \pi 4, \pi]$. 

\subsection{Estimates in the red-shift region in principal null frame} \label{SecEstRedShiftPrincipalNull}

We now write the estimates in the red-shift region in terms of different frames that we defined. We will also restrict to $s\leq \f{3s_f}{4}$ to take advantage of Lemma~\ref{lem:to.find.restricting.set}. We will also use the $^{(N)}$ and $^{(S)}$ variant of the frames. As before, all estimates are to be understood for $\th_* \in [0,\f{3\pi}4]$ if $^{(\cdot)} = ^{(N)}$ and for $\th_* \in [\f{\pi}4,\pi]$ if $^{(\cdot)} = ^{(S)}$. We continue to use $e_\mu'^{(\cdot)} = e_\mu'$, etc.~for $\mu = 3,4$.

\subsubsection{Estimates in terms of the $(e_1'^{(\cdot)}, e_2'^{(\cdot)}, e_3', e_4')$ null frame}

We first write the estimates of Proposition~\ref{prop:RS.original.gauge.full.frame} in terms of the $(e_1'^{(\cdot)}, e_2'^{(\cdot)}, e_3', e_4')$ null frame. This is an auxiliary step towards proving estimates in the principal null frame in Section~\ref{sec:RS.est.principal.null}.

\begin{remark}\label{rmk:RS.change.how.to.differentiate}
    For the results to be used later, it is more convenient to state Proposition~\ref{prop:RS.double.null.frame.main} and Proposition~\ref{prop:RS.doublle.null.curvature} with $(\pmb{e}_1'^{(\cdot)})^{i_1} (\pmb{e}_2'^{(\cdot)})^{i_2} ( \pmb{e}_3')^{i_3} (\pmb{e}_4')^{i_4}$. Nonetheless, using Corollary~\ref{cor:RS.e'.to.ee}, the statement will be equivalent to those where $(\pmb{e}_1'^{(\cdot)})^{i_1} (\pmb{e}_2'^{(\cdot)})^{i_2} ( \pmb{e}_3')^{i_3} (\pmb{e}_4')^{i_4}$ is replaced by $(\os{\pmb{\ee}}{1}{}_1^{(\cdot)})^{i_1}
         (\os{\pmb{\ee}}{1}{}_2^{(\cdot)})^{i_2}
         (\os{\pmb{\ee}}{1}{}_3^{(\cdot)})^{i_3}
         (\os{\pmb{\ee}}{1}{}_4^{(\cdot)})^{i_4}$. We will sometimes pass between these statements without further comments.
\end{remark}

\begin{proposition}\label{prop:RS.double.null.frame.main}
    The following holds in $\os{\calM}{1}\cap \{s \leq \f{3s_f}{4} \}$:
    \begin{equation}\label{eq:RS.double.null.frame.main}
        \sum_{i_1+i_2+i_3+i_4\leq I} \Big| (\pmb{e}_1'^{(\cdot)})^{i_1} (\pmb{e}_2'^{(\cdot)})^{i_2} ( \pmb{e}_3')^{i_3} (\pmb{e}_4')^{i_4} \Big( g(\nabla_{e_\mu'^{(\cdot)}} e_\nu'^{(\cdot)}, e_\lambda'^{(\cdot)} ) - \Phi^* \Big(\pmb{g}(\pmb{\nabla}_{\Phi_* \pmb{e}_\mu'^{(\cdot)}} (\Phi_* \pmb{e}_\nu'^{(\cdot)}), \Phi_* \pmb{e}_\lambda'^{(\cdot)} )\Big) \Big) \Big| \ls \ep \mathfrak w(\ub).
    \end{equation}
\end{proposition}
\begin{proof}
    We will suppress the higher derivatives in this discussion but just observe that all the estimates used allow us to take higher derivatives.
    
    First, let us rephrase the estimate in Proposition~\ref{prop:RS.original.gauge.full.frame}. In Section~\ref{sec:old.estimates.1}, the identification with Kerr is via the $(s,\ub,\vartheta_*)$ coordinates. Thus, in the notation of Section~\ref{sec:identification}, we control (the derivatives of)
    \begin{equation}\label{eq:RS.original.gauge.full.frame.rephrased}
        \begin{split}
            &\: g(\nabla_{\os{\ee}{1}{}_\mu^{(\cdot)}} \os{\ee}{1}{}_\nu^{(\cdot)}, \os{\ee}{1}{}_\lambda^{(\cdot)} ) - \os{\Phi}{1}{}^* \Big( \pmb{g}( \pmb{\nabla}_{\os{\Phi}{1}{}_* \os{\pmb{\ee}}{1}{}_\mu^{(\cdot)}} (\os{\Phi}{1}{}_* \os{\pmb{\ee}}{1}{}_\nu^{(\cdot)}), \os{\Phi}{1}{}_* \os{\pmb{\ee}}{1}{}_\lambda^{(\cdot)} )\Big) \\
            = &\: g(\nabla_{\os{\ee}{1}{}_\mu^{(\cdot)}} \os{\ee}{1}{}_\nu^{(\cdot)}, \os{\ee}{1}{}_\lambda^{(\cdot)} ) - (\os{\Phi}{1}{}^* \pmb{g})( \pmb{\nabla}^{\os{\Phi}{1}{}^* \pmb{g}}_{\os{\pmb{\ee}}{1}{}_\mu^{(\cdot)}} \os{\pmb{\ee}}{1}{}_\nu^{(\cdot)}, \os{\pmb{\ee}}{1}{}_\lambda^{(\cdot)} ).
        \end{split}
    \end{equation}

    For the terms in \eqref{eq:RS.double.null.frame.main}, we now use \eqref{eq:bbA.def} to write $e_\mu'^{(\cdot)}$, $e_\nu'^{(\cdot)}$, $e_\lambda'^{(\cdot)}$ (and their bold counterpart) as a linear combination of $(\os{\ee}{1}{}_1^{(\cdot)}, \os{\ee}{1}{}_2^{(\cdot)}, \os{\ee}{1}{}_3, \os{\ee}{1}{}_4)$ (and their bold counterpart). 
    \begin{equation}\label{eq:RS.double.null.frame.main.1}
        \begin{split}
            &\: g(\nabla_{e_\mu'^{(\cdot)}} e_\nu'^{(\cdot)}, e_\lambda'^{(\cdot)} ) \\
            = &\: (\mathbb A^{(\cdot)})^{\bar{\mu}}_{\mu} (\mathbb A^{(\cdot)})^{\bar{\lambda}}_{\lambda} (\mathbb A^{(\cdot)})^{\bar{\nu}}_{\nu} g(\nabla_{\os{\ee}{1}{}_{\bar{\mu}^{(\cdot)}}} \os{\ee}{1}{}_{\bar{\nu}}^{(\cdot)}, \os{\ee}{1}{}_{\bar{\lambda}}^{(\cdot)} ) + (\mathbb A^{(\cdot)})^{\bar{\mu}}_{\mu} (\mathbb A^{(\cdot)})^{\bar{\lambda}}_{\lambda} (\os{\ee}{1}{}_{\bar{\mu}^{(\cdot)}} (\mathbb A^{(\cdot)})^{\bar{\nu}}_{\nu}) g( \os{\ee}{1}{}_{\bar{\nu}}^{(\cdot)}, \os{\ee}{1}{}_{\bar{\lambda}}^{(\cdot)} ).
        \end{split}
    \end{equation}
    and
    \begin{equation}\label{eq:RS.double.null.frame.main.2}
        \begin{split}
            &\: \Phi^* \Big(\pmb{g}(\pmb{\nabla}_{\Phi_* \pmb{e}_\mu'^{(\cdot)}} (\Phi_* \pmb{e}_\nu'^{(\cdot)}), \Phi_* \pmb{e}_\lambda'^{(\cdot)} )\Big) 
            =  (\Phi^* \pmb{g})(\pmb{\nabla}^{\Phi^* \pmb{g}}_{\pmb{e}_\mu'^{(\cdot)}} \pmb{e}_\nu'^{(\cdot)}, \pmb{e}_\lambda'^{(\cdot)}) \\
            = &\: (\pmb{\mathbb A}^{(\cdot)})^{\bar{\mu}}_{\mu} (\pmb{\mathbb A}^{(\cdot)})^{\bar{\nu}}_{\nu} (\pmb{\mathbb A}^{(\cdot)})^{\bar{\lambda}}_{\lambda}  (\Phi^* \pmb{g})(\nabla^{\Phi^*\pmb{g}}_{\os{\pmb{\ee}}{1}{}_{\bar{\mu}^{(\cdot)}}}  \os{\pmb{\ee}}{1}{}_{\bar{\nu}}^{(\cdot)}, \os{\pmb{\ee}}{1}{}_{\bar{\lambda}}^{(\cdot)} ) + (\pmb{\mathbb A}^{(\cdot)})^{\bar{\mu}}_{\mu} (\pmb{\mathbb A}^{(\cdot)})^{\bar{\lambda}}_{\lambda} (\os{\pmb{\ee}}{1}{}_{\bar{\mu}^{(\cdot)}} (\pmb{\mathbb A}^{(\cdot)})^{\bar{\nu}}_{\nu}) (\Phi^* \pmb{g})( \os{\pmb{\ee}}{1}{}_{\bar{\nu}}^{(\cdot)}, \os{\pmb{\ee}}{1}{}_{\bar{\lambda}}^{(\cdot)} ).
        \end{split}
    \end{equation}

    To obtain the desired bound, we take the difference of \eqref{eq:RS.double.null.frame.main.1} and \eqref{eq:RS.double.null.frame.main.2}. The difference $\mathbb A - \pmb{\mathbb A}$, $\os{\ee}{1} - \os{\pmb{\ee}}{1}$ are controlled in Corollary~\ref{cor:RS.e'.to.ee}. The difference $g( \os{\ee}{1}{}_{\bar{\nu}}^{(\cdot)}, \os{\ee}{1}{}_{\bar{\lambda}}^{(\cdot)} )-(\Phi^* \pmb{g})( \os{\pmb{\ee}}{1}{}_{\bar{\nu}}^{(\cdot)}, \os{\pmb{\ee}}{1}{}_{\bar{\lambda}}^{(\cdot)} )$ is bounded by Lemma~\ref{lem:fat.e.full.frame.est} and Lemma~\ref{lem:almost.null.different.pullback}. It thus remains to bound
    \begin{equation}
        \begin{split}
            &\: g(\nabla_{\os{\ee}{1}{}_{\bar{\mu}'^{(\cdot)}}} \os{\ee}{1}{}_{\bar{\nu}}^{(\cdot)}, \os{\ee}{1}{}_{\bar{\lambda}}^{(\cdot)} ) - (\Phi^* \pmb{g})(\nabla^{\Phi^*\pmb{g}}_{\os{\pmb{\ee}}{1}{}_{\bar{\mu}'^{(\cdot)}}}  \os{\pmb{\ee}}{1}{}_{\bar{\nu}}^{(\cdot)}, \os{\pmb{\ee}}{1}{}_{\bar{\lambda}}^{(\cdot)} ) \\
        \end{split}
    \end{equation}
    By the triangle inequality, this can be achieved by combining Proposition~\ref{prop:RS.original.gauge.full.frame} (with the quantity expressed as in \eqref{eq:RS.original.gauge.full.frame.rephrased}), Lemma~\ref{lem:comparing.Kerr.vector.fields}, and Proposition~\ref{prop:RS.original.gauge.full.frame}. \qedhere

\end{proof}

    We also control the curvature components, essentially by reducing to Proposition~\ref{prop:RS.double.null.frame.main}:
    \begin{proposition}\label{prop:RS.doublle.null.curvature}
        Denote by $R$ and $\Phi^*\pmb{R}$ the Riemann curvature tensor on $\os{\calM}{1}\cap \{s \leq \f{3s_f}{4} \}$ for $g$ and $\Phi^*\pmb{g}$, respectively. The following holds on $\os{\calM}{1}\cap \{s \leq \f{3s_f}{4} \}$:
        \begin{equation}\label{eq:main.curvature.RS}
            \begin{split}
                &\: \sum_{i_1+i_2+i_3+i_4\leq I} \Big| (\pmb{e}_1'^{(\cdot)})^{i_1} (\pmb{e}_2'^{(\cdot)})^{i_2} (\pmb{e}_3')^{i_3} (\pmb{e}_4')^{i_4}\Big(R(e'^{(\cdot)}_{\mu_1}, e'^{(\cdot)}_{\mu_2}, e'^{(\cdot)}_{\mu_3}, e'^{(\cdot)}_{\mu_4}) - (\Phi^*\pmb{R})(\pmb{e}'^{(\cdot)}_{\mu_1}, \pmb{e}'^{(\cdot)}_{\mu_2}, \pmb{e}'^{(\cdot)}_{\mu_3}, \pmb{e}'^{(\cdot)}_{\mu_4}) \Big) \Big| \ls \ep \mathfrak w(\ub). 
            \end{split}            
        \end{equation}
    \end{proposition}
    \begin{proof}
        The curvature term $R(e'^{(\cdot)}_{\mu_1}, e'^{(\cdot)}_{\mu_2}, e'^{(\cdot)}_{\mu_3}, e'^{(\cdot)}_{\mu_4})$ can be expressed as
        \begin{equation}
            \begin{split}
                &\: e'^{(\cdot)}_{\mu_2}\Big( g(\nabla_{e'^{(\cdot)}_{\mu_1}} e'^{(\cdot)}_{\mu_3}, e'^{(\cdot)}_{\mu_4})\Big)  - g( \nabla_{e'^{(\cdot)}_{\mu_1}} e'^{(\cdot)}_{\mu_3}, \nabla_{e'^{(\cdot)}_{\mu_2}} e'^{(\cdot)}_{\mu_4}) - e'^{(\cdot)}_{\mu_1} \Big(g( \nabla_{e'^{(\cdot)}_{\mu_2}}  e'^{(\cdot)}_{\mu_3}, e'^{(\cdot)}_{\mu_4})\Big) \\
                &\: + g( \nabla_{e'^{(\cdot)}_{\mu_2}}  e'^{(\cdot)}_{\mu_3}, \nabla_{e'^{(\cdot)}_{\mu_1}} e'^{(\cdot)}_{\mu_4}) - g\Big(\nabla_{\nabla_{e'^{(\cdot)}_{\mu_2}} e'^{(\cdot)}_{\mu_1} - \nabla_{e'^{(\cdot)}_{\mu_1}} e'^{(\cdot)}_{\mu_2}, } e'^{(\cdot)}_{\mu_3}, e'^{(\cdot)}_{\mu_4} \Big).
            \end{split}
        \end{equation}
        We now need to control the differences between these terms and their background equivalence. For the derivative terms, the difference, e.g., takes the following form:
        \begin{equation}
            \begin{split}
                &\: e'^{(\cdot)}_{\mu_2} \Big( g(\nabla_{e'^{(\cdot)}_{\mu_1}} e'^{(\cdot)}_{\mu_3}, e'^{(\cdot)}_{\mu_4})\Big) - \pmb{e}'^{(\cdot)}_{\mu_2} \Big( (\Phi^*\pmb{g})(\nabla_{\pmb{e}'^{(\cdot)}_{\mu_1}} \pmb{e}'^{(\cdot)}_{\mu_3}, \pmb{e}'^{(\cdot)}_{\mu_4})\Big) \\
                = &\: e'^{(\cdot)}_{\mu_2} \Big( g(\nabla_{e'^{(\cdot)}_{\mu_1}} e'^{(\cdot)}_{\mu_3}, e'^{(\cdot)}_{\mu_4}) -  (\Phi^*\pmb{g})(\nabla_{\pmb{e}'^{(\cdot)}_{\mu_1}} \pmb{e}'^{(\cdot)}_{\mu_3}, \pmb{e}'^{(\cdot)}_{\mu_4})\Big) + (e'^{(\cdot)}_{\mu_2}  - \pmb{e}'^{(\cdot)}_{\mu_2} )\Big( (\Phi^*\pmb{g})(\nabla_{\pmb{e}'^{(\cdot)}_{\mu_1}} \pmb{e}'^{(\cdot)}_{\mu_3}, \pmb{e}'^{(\cdot)}_{\mu_4})\Big),
            \end{split}
        \end{equation}
        which can be controlled by Proposition~\ref{prop:RS.double.null.frame.main} after using the bound on the difference of the frames in Corollary~\ref{cor:RS.e'.to.ee}, and the fact that we can change the vectors we use to differentiate (see Remark~\ref{rmk:RS.change.how.to.differentiate}).
        
        For the quadratic terms, we can expand using the null frame, e.g., 
        \begin{equation*}
            \begin{split}
                g( \nabla_{e'^{(\cdot)}_{\mu_1}} e'^{(\cdot)}_{\mu_3}, \nabla_{e'^{(\cdot)}_{\mu_2}} e'^{(\cdot)}_{\mu_4}) =&\:  \sum_{A=1}^2 g( \nabla_{e'^{(\cdot)}_{\mu_1}} e'^{(\cdot)}_{\mu_3}, e'^{(\cdot)}_A) g( \nabla_{e'^{(\cdot)}_{\mu_2}} e'^{(\cdot)}_{\mu_4}, e'^{(\cdot)}_A) \\
                &\: -\f 12 g( \nabla_{e'^{(\cdot)}_{\mu_1}} e'^{(\cdot)}_{\mu_3}, e'^{(\cdot)}_3) g( \nabla_{e'^{(\cdot)}_{\mu_2}} e'^{(\cdot)}_{\mu_4}, e'^{(\cdot)}_4) -\f 12 g( \nabla_{e'^{(\cdot)}_{\mu_1}} e'^{(\cdot)}_{\mu_3}, e'^{(\cdot)}_4) g(  \nabla_{e'^{(\cdot)}_{\mu_2}} e'^{(\cdot)}_{\mu_4}, e'^{(\cdot)}_3).
            \end{split}
        \end{equation*}
        We then need to control (the derivatives of) some term of the form, say,
        \begin{equation}
            g( \nabla_{e'^{(\cdot)}_{\mu_1}} e'^{(\cdot)}_{\mu_3}, e'^{(\cdot)}_A) g( \nabla_{e'^{(\cdot)}_{\mu_2}} e'^{(\cdot)}_{\mu_4}, e'^{(\cdot)}_A) - (\Phi^* \pmb{g})( (\Phi^* \pmb{\nabla})_{\pmb{e}'^{(\cdot)}_{\mu_1}} \pmb{e}'^{(\cdot)}_{\mu_3}, \pmb{e}'^{(\cdot)}_A) (\Phi^* \pmb{g})( (\Phi^*\pmb{\nabla})_{\pmb{e}'^{(\cdot)}_{\mu_2}} \pmb{e}'^{(\cdot)}_{\mu_4}, \pmb{e}'^{(\cdot)}_A),
        \end{equation}
        which can easily be estimated by Proposition~\ref{prop:RS.double.null.frame.main} using the boundedness of the background quantities. \qedhere
    \end{proof}

\subsubsection{Estimates in terms of the principal null frame}\label{sec:RS.est.principal.null}

\begin{proposition}\label{prop:RS.principal.null}
    Let $e_\mu^{(\cdot)} = (\calE^{(\cdot)})_\mu^\nu \pmb{e}_\nu^{(\cdot)}$ and $\pmb{e}_\mu^{(\cdot)} = (\pmb{\calE}^{(\cdot)})_\mu^\nu {e}_\nu^{(\cdot)}$. The following estimates hold in $\os{\calM}{1}\cap \{s \leq \f{3s_f}{4} \}$:
    \begin{align}
    &\sum_{i_1+i_2+i_3+i_4\leq I} \Big|(\pmb{e}_1^{(\cdot)})^{i_1} (\pmb{e}_2^{(\cdot)})^{i_2} ( \pmb{e}_3)^{i_3} (\pmb{e}_4)^{i_4} \Big((\calE^{(\cdot)})_\mu^\nu - \delta_\mu^\nu \Big) \Big| \ls \epsilon \mathfrak{w}(\ub)\,, \label{EqTrafoFrame1}\\
     &\sum_{i_1+i_2+i_3+i_4\leq I} \Big|(\pmb{e}_1^{(\cdot)})^{i_1} (\pmb{e}_2^{(\cdot)})^{i_2} ( \pmb{e}_3)^{i_3} (\pmb{e}_4)^{i_4} \Big((\pmb{\calE}^{(\cdot)})_\mu^\nu - \delta_\mu^\nu \Big) \Big| \ls \epsilon \mathfrak{w}(\ub)\,, \label{EqTraforFrame2} \\
        &\sum_{i_1+i_2+i_3+i_4\leq I} \Big| (\pmb{e}_1^{(\cdot)})^{i_1} (\pmb{e}_2^{(\cdot)})^{i_2} ( \pmb{e}_3)^{i_3} (\pmb{e}_4)^{i_4} \Big( g(\nabla_{e_\mu^{(\cdot)}} e_\nu^{(\cdot)}, e_\lambda^{(\cdot)} ) - \Phi^* \Big( \pmb{g}(\pmb{\nabla}_{\Phi_*\pmb{e}_\mu^{(\cdot)}} (\Phi_*\pmb{e}_\nu^{(\cdot)}), \Phi_* \pmb{e}_\lambda^{(\cdot)} ) \Big) \Big) \Big| \ls \ep \mathfrak w(\ub). \label{EqTrafoRicciProp}
        \end{align}
\end{proposition}
\begin{proof}
Recall from \eqref{eq:bbA.def}, \eqref{eq:bbF}, \eqref{EqDefBackgroundPNFrame}, \eqref{EqDefDynamicalPNFrame} that 
\begin{align}
    e_\mu^{(\cdot)} = &\: (\pmb{\calB}^{(\cdot)})_\mu^\nu (\mathbb{A}^{(\cdot)})_\nu^\rho (\mathbb{F}^{(\cdot)})_\rho^\sigma ((\pmb{\mathbb A}^{(\cdot)})^{-1})_{\sigma}^\de (\pmb{\calB}'^{(\cdot)})_\de^\eta \pmb{e}_\eta^{(\cdot)}.
\end{align} 
Since $\pmb{\calB}^{(\cdot)}$ and $\pmb{\calB}'^{(\cdot)}$ are inverses of each other, and $(\pmb{\mathbb A}^{(\cdot)})^{-1}$ is bounded, the first estimate \eqref{EqTrafoFrame1} reduces to bounding $(\mathbb{A}^{(\cdot)})_\nu^\rho (\mathbb{F}^{(\cdot)})_\rho^\sigma - (\pmb{\mathbb{A}}^{(\cdot)})_\nu^\rho \de_\rho^\sigma$ (and its derivatives), which in turn follows from Corollary~\ref{cor:RS.e'.to.ee}. The proof of \eqref{EqTraforFrame2} is similar.

    To prove \eqref{EqTrafoRicciProp} we compute the transformation:
    \begin{equation}\label{eq:RC.main.transformation.RS}
        \begin{split}
            &\: g(\nabla_{e_\mu^{(\cdot)}} e_\nu^{(\cdot)}, e_\lambda^{(\cdot)} ) - \Phi^* \Big( \pmb{g}(\pmb{\nabla}_{\Phi_*\pmb{e}_\mu^{(\cdot)}} (\Phi_*\pmb{e}_\nu^{(\cdot)}), \Phi_* \pmb{e}_\lambda^{(\cdot)} ) \Big) \\
            = &\: \underbrace{(\pmb{\calB}^{(\cdot)})_\mu^{\mu'} (\pmb{\calB}^{(\cdot)})_\lambda^{\lambda'}  (\pmb{\calB}^{(\cdot)})_\nu^{\nu'} \Big( g(\nabla_{e_{\mu'}'^{(\cdot)}} e_{\nu'}'^{(\cdot)}, e_{\lambda'}'^{(\cdot)} ) - (\Phi^*\pmb{g})(\pmb{\nabla}_{\pmb{e}_{\mu'}'^{(\cdot)}} \pmb{e}_{\nu'}'^{(\cdot)}, \pmb{e}_{\lambda'}'^{(\cdot)} )\Big)}_{=:I} \\ 
            &\: + \underbrace{(\pmb{\calB}^{(\cdot)})_\mu^{\mu'} (\pmb{\calB}^{(\cdot)})_\lambda^{\lambda'}\Big( (e_{\mu'}'^{(\cdot)} - \pmb{e}_{\mu'}'^{(\cdot)}) (\pmb{\calB}^{(\cdot)})_\nu^{\nu'} \Big) g(e_{\nu'}'^{(\cdot)}, e_{\lambda'}'^{(\cdot)})}_{=:II} ,
        \end{split}
    \end{equation}
    where we used $g(e_{\nu'}^{(\cdot)}, e_{\lambda'}^{(\cdot)}) = (\Phi^*\pmb{g})(\pmb{e}_{\nu'}^{(\cdot)}, \pmb{e}_{\lambda'}^{(\cdot)})$, and repeated indices are summed over.

    For the term $I$ in \eqref{eq:RC.main.transformation.RS} and its derivatives, we use the bounds for $\calB$ in Proposition~\ref{prop:B} and the bounds for $g(\nabla_{e_{\mu'}'^{(\cdot)}} e_{\nu'}'^{(\cdot)}, e_{\lambda'}'^{(\cdot)} ) - (\Phi^*\pmb{g})(\pmb{\nabla}^{\Phi^* \pmb{g}}_{\pmb{e}_{\mu'}'^{(\cdot)}} \pmb{e}_{\nu'}'^{(\cdot)}, \pmb{e}_{\lambda'}'^{(\cdot)} )$ just established above in Proposition~\ref{prop:RS.double.null.frame.main}.

    For the term $II$ in \eqref{eq:RC.main.transformation.RS} and its derivatives, we combine the estimates for $\calB$ in Proposition~\ref{prop:B} with the bounds for the differences of the frames in Corollary~\ref{cor:RS.e'.to.ee}. \qedhere

\end{proof}

\begin{proposition}\label{prop:RS.curvature.principal.null}
    The following estimates hold in $\os{\calM}{1}\cap \{s \leq \f{3s_f}{4} \}$:
    \begin{equation}
        \sum_{i_1+i_2+i_3+i_4\leq I} \Big| (\pmb{e}_1^{(\cdot)})^{i_1} (\pmb{e}_2^{(\cdot)})^{i_2} (\pmb{e}_3')^{i_3} (\pmb{e}_4)^{i_4}\Big(R(e^{(\cdot)}_{\mu_1}, e^{(\cdot)}_{\mu_2}, e^{(\cdot)}_{\mu_3}, e^{(\cdot)}_{\mu_4}) - (\Phi^*\pmb{R})(\pmb{e}^{(\cdot)}_{\mu_1}, \pmb{e}^{(\cdot)}_{\mu_2}, \pmb{e}^{(\cdot)}_{\mu_3}, \pmb{e}^{(\cdot)}_{\mu_4}) \Big)\Big| \ls \ep \mathfrak w(\ub).
    \end{equation}
\end{proposition}
\begin{proof}
    Starting from Proposition~\ref{prop:RS.doublle.null.curvature}, we introduce the change of frame as in the proof of Proposition~\ref{prop:RS.principal.null} to obtain the desired estimates. We omit the details. \qedhere
\end{proof}

\subsection{Estimates in the blue-shift region in principal null frame} \label{SecEstBlueShiftPN}

We now rewrite the estimates in the blue-shift region in Section~\ref{sec:DL.refined} in terms of the principal null frame. We will restrict our estimates to $\os{\calM}{2}\setminus (\calU \cap \{s \leq \f{s_f}{2}\})$. In this region, \eqref{eq:Phi.agree.in.2} holds and there is no ambiguity about $\Phi^*$ versus $\os{\Phi}{2}{}^*$. For the rest of the subsection, we will just write $\pmb{g}(\pmb{\nabla}_{\pmb{e}'^{(\cdot)}_\mu} \pmb{e}'^{(\cdot)}_\nu, \pmb{e}'^{(\cdot)}_\lambda )$ for $\Phi^*\pmb{g}((\Phi^*\pmb{\nabla})_{\pmb{e}'^{(\cdot)}_\mu} \pmb{e}'^{(\cdot)}_\nu, \pmb{e}'^{(\cdot)}_\lambda )$, and similarly for the curvature tensor.

As in previous subsections, all estimates are to be understood for $\th_* \in [0,\f{3\pi}4]$ if $^{(\cdot)} = ^{(N)}$ and for $\th_* \in [\f{\pi}4,\pi]$ if $^{(\cdot)} = ^{(S)}$. We continue to use $e_\mu'^{(\cdot)} = e_\mu'$, etc.~for $\mu = 3,4$.

\subsubsection{Rewriting the estimates}\label{sec:rewrite.double.null}


\begin{proposition}\label{prop:Ricci.double.null}
The following estimates hold in $\os{\calM}{2}\setminus (\calU\cap \{s \leq \f{s_f}2\})$: 
\begin{enumerate}
    \item If $(\mu,\nu,\lambda)$ contains two $3$, or $(\mu,\nu,\lambda)$ is a permutation of $(4,4,A)$, or $\nu = \lambda$, then
    \begin{equation}\label{eq:RC.double.null.main.1}
        g(\nabla_{e'^{(\cdot)}_\mu} e'^{(\cdot)}_\nu, e'^{(\cdot)}_\lambda ) - \pmb{g}(\pmb{\nabla}_{\pmb{e}'^{(\cdot)}_\mu} \pmb{e}'^{(\cdot)}_\nu, \pmb{e}'^{(\cdot)}_\lambda ) = 0.
    \end{equation}
    \item If $(\mu,\nu,\lambda) = (A,B,C)$ or if it is a permutation of $(3,4,A)$, $(4,A,B)$ or $(4,4,3)$, then
    \begin{equation}\label{eq:RC.double.null.main.2}
        \opnorm[\Big]{g(\nabla_{e'^{(\cdot)}_\mu} e'^{(\cdot)}_\nu, e'^{(\cdot)}_\lambda ) - \pmb{g}(\pmb{\nabla}_{\pmb{e}'^{(\cdot)}_\mu} \pmb{e}'^{(\cdot)}_\nu, \pmb{e}'^{(\cdot)}_\lambda ) } \ls \ep.
    \end{equation}
    \item If $(\mu,\nu,\lambda)$ is a permutation of $(3,A,B)$, then
\begin{equation}\label{eq:RC.double.null.main.3}
        \opnorm[\Big]{ |\Delta| \Big(g(\nabla_{e'^{(\cdot)}_\mu} e'^{(\cdot)}_\nu, e'^{(\cdot)}_\lambda ) - \pmb{g}(\pmb{\nabla}_{\pmb{e}'^{(\cdot)}_\mu} \pmb{e}'^{(\cdot)}_\nu, \pmb{e}'^{(\cdot)}_\lambda )\Big) } \ls \ep.
    \end{equation}
    \item Moreover, all the above estimates still hold after taking higher derivatives in $\pmb{e}_1'^{(\cdot)}$, $\pmb{e}_2'^{(\cdot)}$, $|\Delta| \pmb{e}_3'$, or $\pmb{e}_4'$, i.e, if $(\mu,\nu,\lambda)$ satisfies the assumptions of point 2 above, then 
    \begin{equation*}
        \sum_{i_1+i_2+i_3+i_4\leq I}\opnorm[\Big]{(\pmb{e}_1'^{(\cdot)})^{i_1} (\pmb{e}_2'^{(\cdot)})^{i_2} (|\Delta| \pmb{e}_3'^{(\cdot)})^{i_3} (\pmb{e}_4'^{(\cdot)})^{i_4}\Big( g(\nabla_{e'^{(\cdot)}_\mu} e'^{(\cdot)}_\nu, e'^{(\cdot)}_\lambda ) - \pmb{g}(\pmb{\nabla}_{\pmb{e}'^{(\cdot)}_\mu} \pmb{e}'^{(\cdot)}_\nu, \pmb{e}'^{(\cdot)}_\lambda )\Big)  } \ls \ep.
    \end{equation*}
    Similarly, if $(\mu,\nu,\lambda)$ satisfies the assumptions for point 3 above, then a similar higher derivative bound holds (with an additional $|\Delta|$ as in \eqref{eq:RC.double.null.main.3}).
\end{enumerate}
\end{proposition}
\begin{proof}
    Before we proceed, let us comment that the proof will be based on estimates in Proposition~\ref{prop:DL.rescaled} together with properties of the frames in Lemma~\ref{lem:basic.double.null} and Lemma~\ref{lem:commutators}. When using Proposition~\ref{prop:DL.rescaled}, we will in particular use the observation in Remark~\ref{rmk:norm.equivalence} to freely change the norms on the spheres and as well as to change between the background and the dynamical derivatives.

    \pfstep{Step~1: Proof of \eqref{eq:RC.double.null.main.1}} We show that \eqref{eq:RC.double.null.main.1} vanishes because each term vanishes; we only consider the terms in the dynamical spacetime, the background terms can be treated in the same manner. We first consider the case where $\nu = \lambda$ so that
    \begin{equation}\label{eq:Ricci.total.derivative}
        g(\nabla_{e'^{(\cdot)}_\mu} e'^{(\cdot)}_\nu, e'^{(\cdot)}_\nu ) = \f 12 e'^{(\cdot)}_\mu \Big(g( e'^{(\cdot)}_\nu, e'^{(\cdot)}_\nu )\Big) = 0,
    \end{equation}
    since $g( e'^{(\cdot)}_\nu, e'^{(\cdot)}_\nu )$ is a constant (either $0$, $1$ or $-2$).
    
    For the case where $(\mu,\nu,\lambda)$ contains two $3$. If $\mu = \nu = 3$, then since $e'_3$ is chosen to be geodesic, i.e., $\nabla_{e'_3} e'_3 = 0$, the term $=0$. If $\mu = \lambda = 3$, then 
    $$g(\nabla_{e_3'^{(\cdot)}} e'^{(\cdot)}_\nu, e_3'^{(\cdot)} )= -g( e'^{(\cdot)}_\nu, \nabla_{e_3'^{(\cdot)}} e_3'^{(\cdot)} )=0$$
    for the same reason as before. Finally, if $\nu = \lambda = 3$, this vanishes due to \eqref{eq:Ricci.total.derivative}.

    It remains to consider the case where $(\mu,\nu,\lambda)$ is a permutation of $(4,4,A)$. If $\nu = \lambda =4$, then the term $=0$ by \eqref{eq:Ricci.total.derivative}. If $(\mu,\nu,\lambda) = (4,4,A)$ we observe that $\nabla_{e_4'^{(\cdot)}}e_4'^{(\cdot)} = -2\om e_4'^{(\cdot)}$ (by \eqref{eq:Ricci.def}) and so $g(\nabla_{e'^{(\cdot)}_\mu} e'^{(\cdot)}_\nu, e'^{(\cdot)}_\lambda ) = -2\om g(e_4'^{(\cdot)}, e'^{(\cdot)}_A )=0$. Finally, if $(\mu,\nu,\lambda) = (4,A,4)$, we note that
    $$g(\nabla_{e'^{(\cdot)}_4} e'^{(\cdot)}_A, e'^{(\cdot)}_4 )= -g( e'^{(\cdot)}_A, \nabla_{e'^{(\cdot)}_4} e'^{(\cdot)}_4 )$$
    so that it is also vanishing.
    

    \pfstep{Step~2: Proof of \eqref{eq:RC.double.null.main.2}} For each of the cases, we write out the corresponding terms. Some terms directly correspond to connection coefficients, for which bound by comparing with Proposition~\ref{prop:DL.rescaled}. There are other terms that involve the quantities in Lemma~\ref{lem:computation.nablaslashed} so that we will estimate using Lemma~\ref{lem:commutators}. 
    
    \pfstep{Step~2(a): The case $(\mu,\nu,\lambda) = (A,B,C)$} When $(\mu,\nu,\lambda) = (1,1,1), (1,2,2), (2,1,1), (2,2,2)$, the terms vanish identically (since $\nu = \lambda$). The cases $(\mu,\nu,\lambda) = (1,2,1), (1,1,2), (2,1,2), (2,2,1)$ are similar to each other; we only give the details in the specific case of $(1,2,1)$. Using \eqref{eq:Ricci.total.derivative} and that the connections are torsion free, we can write
    \begin{equation}\label{eq:Ricci.nab.slashed.terms}
        \begin{split}
            &\: \Big|g(\nabla_{e_1'^{(\cdot)}} e_2'^{(\cdot)}, e_1'^{(\cdot)} ) - \pmb{g}(\pmb{\nabla}_{\pmb{e}'^{(\cdot)}_1} \pmb{e}'^{(\cdot)}_2, \pmb{e}'^{(\cdot)}_1 ) \Big| \\
            = &\: \Big|g(e_1'^{(\cdot)},[e_1'^{(\cdot)}, e_2'^{(\cdot)}] ) - \pmb{g}(\pmb{e}'^{(\cdot)}_1,[\pmb{e}_1'^{(\cdot)}, \pmb{e}_2'^{(\cdot)}] ) \Big| \\
            \leq &\: \Big| g(e_1'^{(\cdot)} - \pmb{e}'^{(\cdot)}_1,[e_1'^{(\cdot)}, e_2'^{(\cdot)}] )\Big| + \Big| (g - \pmb{g})(\pmb{e}'^{(\cdot)}_1,[e_1'^{(\cdot)}, e_2'^{(\cdot)}] ) \Big|\\
            &\: + \Big|\pmb{g}(\pmb{e}'^{(\cdot)}_1,[e_1'^{(\cdot)}, e_2'^{(\cdot)}] - [\pmb{e}_1'^{(\cdot)}, \pmb{e}_2'^{(\cdot)}] )\Big| =:I+II +III.
        \end{split}
    \end{equation}
    We can now bound each of the terms. We can express the dynamical frames and their commutators in terms of the background double null frame $\pmb{e}_\mu'^{(\cdot)}$ using Lemma~\ref{lem:basic.double.null} and Lemma~\ref{lem:commutators}. Using the bounds in these two lemmas for the differences $e_1'^{(\cdot)} - \pmb{e}'^{(\cdot)}_1$ and $[e_1'^{(\cdot)}, e_2'^{(\cdot)}] - [\pmb{e}_1'^{(\cdot)}, \pmb{e}_2'^{(\cdot)}]$, we thus obtain the desired bounds for $I$ and $III$. The $(g-\pmb{g})$ factor in $II$ can be replaced by $\gamma - \pmb{\gamma}$ since it is evaluated on $S_{u,\ub}$-tangent vector fields. Thus this term can be controlled using Proposition~\ref{prop:DL.rescaled}.

    \pfstep{Step~2(b): The case $(\mu,\nu,\lambda)$ is a permutation of $(3,4,A)$} If $(\mu,\nu,\lambda) = (3,4,A)$, this corresponds to $\eta$ (see \eqref{eq:Ricci.def}). We give this as an example for controlling terms which correspond to connection coefficients:
    \begin{equation}\label{eq:Ricci.eta.diff.term}
        g(\nabla_{e_3'^{(\cdot)}} e_4'^{(\cdot)}, e_A'^{(\cdot)} ) - \pmb{g}(\pmb{\nabla}_{\pmb{e}'^{(\cdot)}_3} \pmb{e}'^{(\cdot)}_4, \pmb{e}'^{(\cdot)}_A ) = 2 (\eta - \pmb{\eta})(e_A'^{(\cdot)}) + 2 \pmb{\eta}(e_A^{(\cdot)} -\pmb{e}_A'^{(\cdot)}),
    \end{equation}
    which can then be controlled in the $\opnorm{\cdot}$ norm (without degeneration in $|\Delta|$) using Proposition~\ref{prop:DL.rescaled} and Lemma~\ref{lem:basic.double.null}.
    
    For $(\mu,\nu,\lambda) = (3,A,4)$, we use
    \begin{equation}\label{eq:stupid}
    g(\nabla_{e'^{(\cdot)}_\mu} e'^{(\cdot)}_\nu, e'^{(\cdot)}_\lambda) = - g( e'^{(\cdot)}_\nu, \nabla_{e'^{(\cdot)}_\mu} e'^{(\cdot)}_\lambda),
    \end{equation}
    to see that it also corresponds to $\eta$ (see \eqref{eq:Ricci.def}). Similarly, $(\mu,\nu,\lambda) = (4,A,3), (4,3,A)$ both correspond to $\etab$, while $(\mu,\nu,\lambda) = (A,4,3),(A,3,4)$ both correspond to $\zeta$ (see \eqref{eq:Ricci.def}). All these terms can be bounded using Proposition~\ref{prop:DL.rescaled} and Lemma~\ref{lem:basic.double.null}.

    \pfstep{Step~2(c): The case $(\mu,\nu,\lambda)$ is a permutation of $(4,A,B)$} The cases $(A,4,B)$, $(B,4,A)$ correspond directly to $\chi$, and $(A,B,4)$, $(B,A,4)$ also correspond to $\chi$ after a manipulation similar to \eqref{eq:stupid}. Both can be dealt with as \eqref{eq:Ricci.eta.diff.term} using Proposition~\ref{prop:DL.rescaled} and Lemma~\ref{lem:basic.double.null}. For the remaining cases $(4,A,B)$ and $(4,B,A)$, we use \eqref{eq:nablaslashed.chi} in Lemma~\ref{lem:computation.nablaslashed} to write
    \begin{equation}\label{eq:chi.diff.expanded}
        \begin{split}
            &\:  g(\nabla_{e_4'^{(\cdot)}} e_A'^{(\cdot)}, e_B'^{(\cdot)} ) - \pmb{g}(\pmb{\nabla}_{\pmb{e}'^{(\cdot)}_4} \pmb{e}'^{(\cdot)}_A, \pmb{e}'^{(\cdot)}_B ) \\
            = &\: g([e_4'^{(\cdot)}, e_A'^{(\cdot)}], e_B'^{(\cdot)}) + \chi(e_A'^{(\cdot)}, e_B'^{(\cdot)}) - \pmb{g}([\pmb{e}_4'^{(\cdot)}, \pmb{e}_A'^{(\cdot)}], \pmb{e}_B'^{(\cdot)}) - \pmb{\chi}(\pmb{e}_A'^{(\cdot)}, \pmb{e}_B'^{(\cdot)}).        
        \end{split}
    \end{equation}
    We have thus reduced to two types of terms that we have dealt with before. The term $g([e_4'^{(\cdot)}, e_A'^{(\cdot)}], e_B'^{(\cdot)})- \pmb{g}([\pmb{e}_4'^{(\cdot)}, \pmb{e}_A'^{(\cdot)}], \pmb{e}_B'^{(\cdot)})$ can be handled in the same manner as
    \eqref{eq:Ricci.nab.slashed.terms}, using Lemma~\ref{lem:basic.double.null}, Lemma~\ref{lem:commutators}, and Proposition~\ref{prop:DL.rescaled}. The term $\chi(e_A'^{(\cdot)}, e_B'^{(\cdot)}) - \pmb{\chi}(\pmb{e}_A'^{(\cdot)}, \pmb{e}_B'^{(\cdot)})$ can be controlled as in \eqref{eq:Ricci.eta.diff.term}.
    
    \pfstep{Step~2(d): The case $(\mu,\nu,\lambda)$ is a permutation of $(4,4,3)$} The cases $(\mu,\nu,\lambda) = (4,3,4)$ or $(4,4,3)$ correspond to $\om$, which can be treated as in \eqref{eq:Ricci.eta.diff.term} using Proposition~\ref{prop:DL.rescaled} and Lemma~\ref{lem:basic.double.null}. In the remaining case $(3,4,4)$, the quantity vanishes (by \eqref{eq:Ricci.total.derivative}).

    \pfstep{Step~3: Proof of \eqref{eq:RC.double.null.main.3}} The cases $(A,3,B)$ and $(B,3,A)$ correspond to $\chib$. Arguing as in \eqref{eq:stupid}, the cases $(A,B,3)$ and $(B,A,3)$ also can be rewritten as $\chib$. Now we argue in a similar manner as \eqref{eq:Ricci.eta.diff.term}, except for noticing that the estimate for $\chib$ from Proposition~\ref{prop:DL.rescaled} has a $|\Delta|$ degeneration. This therefore results in a $|\Delta|$ degeneration in \eqref{eq:RC.double.null.main.3}.
    
    Finally, the case $(3,A,B)$ is similar to \eqref{eq:chi.diff.expanded}: After using \eqref{eq:nablaslashed.chib} in Lemma~\ref{lem:computation.nablaslashed}, we obtain
    \begin{equation}\label{eq:chib.diff.expanded}
        \begin{split}
            &\:  g(\nabla_{e_3'^{(\cdot)}} e_A'^{(\cdot)}, e_B'^{(\cdot)} ) - \pmb{g}(\pmb{\nabla}_{\pmb{e}'^{(\cdot)}_3} \pmb{e}'^{(\cdot)}_A, \pmb{e}'^{(\cdot)}_B ) \\
            = &\: g([e_3'^{(\cdot)}, e_A'^{(\cdot)}], e_B'^{(\cdot)}) + \chib(e_A'^{(\cdot)}, e_B'^{(\cdot)}) - \pmb{g}([\pmb{e}_4'^{(\cdot)}, \pmb{e}_A'^{(\cdot)}], \pmb{e}_B'^{(\cdot)}) - \pmb{\chib}(\pmb{e}_A'^{(\cdot)}, \pmb{e}_B'^{(\cdot)}).        
        \end{split}
    \end{equation}
    We can then conclude as in Step~2(c), except noticing that both $\widetilde{\chib}$ in Proposition~\ref{prop:DL.rescaled} and $(\mathfrak{c}_{A3}^C, \overline{\mathfrak{c}}_{A3}^C)$ in Lemma~\ref{lem:commutators} have degenerations in $|\Delta|$, resulting in the $|\Delta|$ degeneration in \eqref{eq:RC.double.null.main.3}.

    \pfstep{Step~4: Higher derivatives} We now show that for all the quantities considered above, the derivatives with respect to $\pmb{e}_1'^{(\cdot)}$, $\pmb{e}_2'^{(\cdot)}$, $|\Delta| \pmb{e}_3'$, or $\pmb{e}_4'$ obey the same bounds. There are two types of terms that need to consider: (1) terms coming from the connection coefficients (i.e.,~terms such as \eqref{eq:Ricci.eta.diff.term}) and (2) terms coming from the commutators in Lemma~\ref{lem:commutators} (i.e.,~terms such as \eqref{eq:Ricci.nab.slashed.terms}).

    \pfstep{Step~4(a): Contributions from the connection coefficients} For the connection coefficients terms, notice that we already have control of the $\Omg^2\nab_3$, $\nab_4$ and the $\nab$ derivatives. This, together with the bounds that we have already obtained in Steps~1-3, give the desired estimate. We take again \eqref{eq:Ricci.eta.diff.term} as an example and consider its derivatives:
    \begin{equation}
    \begin{split}
        &\: e'_4 [\eta(e_A'^{(\cdot)}) - \pmb{\eta}(\pmb{e}_A'^{(\cdot)})] \\
        = &\: \underbrace{(\nab_{e_4'} (\eta - \pmb{\eta}))(e_A'^{(\cdot)})}_{=:I} + \underbrace{(\nab_{e_4'} \pmb{\eta})( e_A'^{(\cdot)} - \pmb{e}_A'^{(\cdot)})}_{=:II} + \underbrace{g(\nab_{e_4'}e_A'^{(\cdot)}, e_B'^{(\cdot)}) (\eta - \pmb{\eta})(e_B'^{(\cdot)})}_{=:III} \\
        &\: + \underbrace{g(\nab_{e_4'}(e_A'^{(\cdot)} - \pmb{e}_A'^{(\cdot)}), e_B'^{(\cdot)}) \pmb{\eta}(e_B'^{(\cdot)})}_{=:IV}.
    \end{split}
\end{equation}
    For term $I$, we use the bound for $\nab_{e_4'} (\eta - \pmb{\eta})$ in Proposition~\ref{prop:DL.rescaled}. For the term $II$, we first control $\opnorm{e_A'^{(\cdot)} - \pmb{e}_A'^{(\cdot)}}$ by \eqref{eq:frame.diff.1} and then notice that $\nab_{e_4'} \pmb{\eta}$ can be controlled by $\pmb{\nab}_{\pmb{e}_4'} \pmb{\eta}$ (see Remark~\ref{rmk:norm.equivalence}), which is a bounded background quantity. For the term $III$, $\eta - \pmb{\eta}$ can be controlled using Proposition~\ref{prop:DL.rescaled}, while $g(\nab_{e_4'}e_A'^{(\cdot)}, e_B'^{(\cdot)}) = g([e_4'^{(\cdot)}, e_A'^{(\cdot)}], e_B'^{(\cdot)}) + \chi(e_A'^{(\cdot)}, e_B'^{(\cdot)})$ can be controlled using Lemma~\ref{lem:basic.double.null}, Lemma~\ref{lem:commutators}, Proposition~\ref{prop:DL.rescaled} (and the estimate for the background $\pmb{\chi}$). Finally, for term $IV$, we use \eqref{eq:double.null.transform} to write 
    \begin{equation}
        \begin{split}
            g(\nab_{e_4'}(e_A'^{(\cdot)} - \pmb{e}_A'^{(\cdot)}), e_B'^{(\cdot)}) = (\calA^{(\cdot)})_4^\mu (\calA^{(\cdot)})_B^\lambda g(\nab_{\pmb{e}_\mu'}(((\calA^{(\cdot)})_A^\nu - \de_A^\nu) \pmb{e}_\nu'^{(\cdot)}), \pmb{e}_\lambda'^{(\cdot)}).
        \end{split}
    \end{equation}
    By Lemma~\ref{lem:basic.double.null}, none of $\mu$, $\nu$, $\lambda$ can $=3$. Thus the needed estimates follows from Lemma~\ref{lem:basic.double.null}.

\pfstep{Step~4(b): Contributions from the commutators} We take as an example the term in \eqref{eq:Ricci.nab.slashed.terms}. As in Step~3, we split $g(\nabla_{e_1'^{(\cdot)}} e_2'^{(\cdot)}, e_1'^{(\cdot)} ) - \pmb{g}(\pmb{\nabla}_{\pmb{e}'^{(\cdot)}_1} \pmb{e}'^{(\cdot)}_2, \pmb{e}'^{(\cdot)}_1 )$ into three terms. We now observe that the application of Lemma~\ref{lem:basic.double.null} and Lemma~\ref{lem:commutators} also allows us to bound the higher derivatives.
\end{proof}

We now turn to the estimates for the curvature. Notice that the cases we consider in the proposition below exhaust all possibilities due to the symmetries of the Riemann curvature tensor. (In particular, when the same vector appears three times, the term vanishes.)
\begin{proposition}\label{prop:curvature.double.null.main}
    Denote by $R$ and $\pmb{R}$ the Riemann curvature tensors of $g$ and $\pmb{g}$, respectively, on $\os{\calM}{2}\setminus (\calU\cap \{s \leq \f{s_f}2\}) s$. The following estimates hold in $\os{\calM}{2}\setminus (\calU\cap \{s \leq \f{s_f}2\})$: 
    \begin{enumerate}
        \item If $(\mu_1,\mu_2,\mu_3,\mu_4)$ is a permutation of $(4,4,A,B)$, $(4,4,3,A)$, or $(4,A,B,C)$, then
        \begin{equation}\label{eq:main.curvature.DN.1}
            \begin{split}
                &\: \sum_{i_1+i_2+i_3+i_4\leq I} \opnorm[\Big]{(\pmb{e}_1'^{(\cdot)})^{i_1} (\pmb{e}_2'^{(\cdot)})^{i_2} (|\Delta| \pmb{e}_3')^{i_3} (\pmb{e}_4')^{i_4}\Big(R(e'_{\mu_1}, e'_{\mu_2}, e'_{\mu_3}, e'_{\mu_4}) - \pmb{R}(\pmb{e}'_{\mu_1}, \pmb{e}'_{\mu_2}, \pmb{e}'_{\mu_3}, \pmb{e}'_{\mu_4})\Big)}_* \ls \ep. 
            \end{split}            
        \end{equation}
        \item If $(\mu_1,\mu_2,\mu_3,\mu_4)$ is a permutation of $(4,4,3,3)$, $(4,3,A,B)$, $(A,B,C,D)$, $(4,3,3,A)$, or $(3,A,B,C)$, then
        \begin{equation}\label{eq:main.curvature.DN.2}
            \begin{split}
                &\: \sum_{i_1+i_2+i_3+i_4\leq I} \opnorm[\Big]{|\Delta| (\pmb{e}_1'^{(\cdot)})^{i_1} (\pmb{e}_2'^{(\cdot)})^{i_2} (|\Delta| \pmb{e}_3')^{i_3} (\pmb{e}_4')^{i_4}\Big(R(e'_{\mu_1}, e'_{\mu_2}, e'_{\mu_3}, e'_{\mu_4}) - \pmb{R}(\pmb{e}'_{\mu_1}, \pmb{e}'_{\mu_2}, \pmb{e}'_{\mu_3}, \pmb{e}'_{\mu_4})\Big)}_* \ls \ep .
            \end{split}            
        \end{equation}
        \item If $(\mu_1,\mu_2,\mu_3,\mu_4)$ is a permutation of $(3,3,A,B)$, then
        \begin{equation}\label{eq:main.curvature.DN.3}
            \begin{split}
                &\: \sum_{i_1+i_2+i_3+i_4\leq I} \opnorm[\Big]{|\Delta|^2 (\pmb{e}_1'^{(\cdot)})^{i_1} (\pmb{e}_2'^{(\cdot)})^{i_2} (|\Delta| \pmb{e}_3')^{i_3} (\pmb{e}_4')^{i_4}\Big(R(e'_{\mu_1}, e'_{\mu_2}, e'_{\mu_3}, e'_{\mu_4}) - \pmb{R}(\pmb{e}'_{\mu_1}, \pmb{e}'_{\mu_2}, \pmb{e}'_{\mu_3}, \pmb{e}'_{\mu_4})\Big)}_* \ls \ep .
            \end{split}            
        \end{equation}
    \end{enumerate}

\end{proposition}
\begin{proof}    
    We first consider the bounds without any derivatives. Given $(\mu_1,\mu_2,\mu_3,\mu_4)$, we express the differences in terms of the differences of the null-decomposed curvature $S$-tensors and the differences of the frames. This is possible since both $(e'^{(\cdot)}_1,e'^{(\cdot)}_2)$ and $(\pmb{e}'^{(\cdot)}_1,\pmb{e}'^{(\cdot)}_1)$ are $S$-tangent.

    As an example, suppose $(\mu_1,\mu_2,\mu_3,\mu_4)$ is a permutation of $(4,4,A,B)$. Without loss of generality using the symmetry of the curvature tensor, we can assume $(\mu_1,\mu_2,\mu_3,\mu_4)= (4,A,4,B)$. Then
    \begin{equation}\label{eq:R.diff.alp}
        \begin{split}
            &\: R(e'^{(\cdot)}_{\mu_1}, e'^{(\cdot)}_{\mu_2}, e'^{(\cdot)}_{\mu_3}, e'^{(\cdot)}_{\mu_4} - \pmb{R}(\pmb{e}'^{(\cdot)}_{\mu_1}, \pmb{e}'^{(\cdot)}_{\mu_2}, \pmb{e}'^{(\cdot)}_{\mu_3}, \pmb{e}'^{(\cdot)}_{\mu_4}) \\
            = &\: \alp(e'^{(\cdot)}_A,e'^{(\cdot)}_B) - \pmb{\alp}(\pmb{e}'^{(\cdot)}_A,\pmb{e}'^{(\cdot)}_B) \\
            = &\: (\alp - \pmb{\alp})(e'^{(\cdot)}_A,e'^{(\cdot)}_B) + \pmb{\alp}(e'^{(\cdot)}_A- \pmb{e}'^{(\cdot)}_A,e'^{(\cdot)}_B) + \pmb{\alp}(\pmb{e}'^{(\cdot)}_A,e'^{(\cdot)}_B- \pmb{e}'^{(\cdot)}_B).
        \end{split}
    \end{equation}
    Therefore, controlling $\alp - \pmb{\alp}$ by Proposition~\ref{prop:DL.rescaled}, and the differences of the frames by Lemma~\ref{lem:basic.double.null}, we obtain the desired bound. 

    The same estimate applies to the other curvature components. The components in \eqref{eq:main.curvature.DN.2} correspond to $\rho$, $\sigma$, and $\betab$, which therefore grows as $|\Delta|^{-1}$ by Proposition~\ref{prop:DL.rescaled} (note that $\pmb{\Omg}^2 \sim |\Delta|$), while the components in \eqref{eq:main.curvature.DN.3} correspond to $\alphab$ and need an additional $|\Delta|$ weight.

    We now control the derivatives. Notice that Proposition~\ref{prop:DL.rescaled} controls the derivatives of the null decomposed components of the curvature, except that the derivatives are taken with respect to $\pmb{\nab}_{\pmb{e}_\mu^{(\cdot)}}$. Using product rule, we can rewrite a $\pmb{e}_\mu$ as a $\pmb{\nab}_{\pmb{e}_\mu^{(\cdot)}}$ derivative, up to terms that we have already controlled. Again take \eqref{eq:R.diff.alp} as an example.
    \begin{equation}\label{eq:R.diff.alp.2}
        \begin{split}
            &\: \pmb{e}_\mu'^{(\cdot)} \Big(R(e'^{(\cdot)}_{\mu_1}, e'^{(\cdot)}_{\mu_2}, e'^{(\cdot)}_{\mu_3}, e'^{(\cdot)}_{\mu_4} - \pmb{R}(\pmb{e}'^{(\cdot)}_{\mu_1}, \pmb{e}'^{(\cdot)}_{\mu_2}, \pmb{e}'^{(\cdot)}_{\mu_3}, \pmb{e}'^{(\cdot)}_{\mu_4}) \Big)\\
            = &\: \pmb{e}_\mu'^{(\cdot)} \Big((\alp - \pmb{\alp})(e'^{(\cdot)}_A,e'^{(\cdot)}_B) + \pmb{\alp}(e'^{(\cdot)}_A- \pmb{e}'^{(\cdot)}_A,e'^{(\cdot)}_B) + \pmb{\alp}(\pmb{e}'^{(\cdot)}_A,e'^{(\cdot)}_B- \pmb{e}'^{(\cdot)}_B)\Big).
        \end{split}
    \end{equation}
    We can distribute the derivatives to control each term: For instance, the term $\pmb{e}_\mu^{(\cdot)} \Big((\alp - \pmb{\alp})(e'^{(\cdot)}_A,e'^{(\cdot)}_B)\Big)$ can be estimated by
    \begin{equation}
        \begin{split}
            &\: \pmb{e}_\mu'^{(\cdot)} \Big((\alp - \pmb{\alp})(e'^{(\cdot)}_A,e'^{(\cdot)}_B)\Big) \\
            = &\: \pmb{\nab}_{\pmb{e}_\mu'^{(\cdot)}} (\alp - \pmb{\alp})(e'^{(\cdot)}_A,e'^{(\cdot)}_B) + (\alp - \pmb{\alp})(\pmb{\nab}_{\pmb{e}_\mu'{(\cdot)}} e'^{(\cdot)}_A,e'^{(\cdot)}_B)+ (\alp - \pmb{\alp})(e'^{(\cdot)}_A,\pmb{\nab}_{\pmb{e}_\mu'^{(\cdot)}} e'^{(\cdot)}_B),
        \end{split}
    \end{equation}
    which obeys the desired bound by Proposition~\ref{prop:DL.rescaled} and Lemma~\ref{lem:basic.double.null}. Notice that in the process, we need to bound the vector fields $\pmb{\nab}_{\pmb{e}_\mu'{(\cdot)}} e'^{(\cdot)}_A$ (and $\pmb{\nab}_{\pmb{e}_\mu'^{(\cdot)}} e'^{(\cdot)}_B$), saying that when expressing $\pmb{\nab}_{\pmb{e}_\mu'{(\cdot)}} e'^{(\cdot)}_A$ in the $(\pmb{e}_1'^{(\cdot)}, \pmb{e}_2'^{(\cdot)})$ basis, the coefficients are bounded if $\mu \neq 3$ and are bounded above pointwise by $|\Delta|^{-1}$. This in turn follows from Lemma~\ref{lem:basic.double.null} after writing $e_A'^{(\cdot)} = (\calA^{(\cdot)})_A^{\nu} \pmb{e}_\nu'^{(\cdot)}$ ((see \eqref{eq:double.null.transform}) and using the boundedness of the background quantities. The other terms in \eqref{eq:R.diff.alp.2} can be treated similarly. \qedhere
\end{proof}

\subsubsection{Estimates in the blue-shift region in terms of the principal null frame}


\begin{proposition}\label{prop:Ricci.principal}
In $\os{\calM}{2}\setminus\{\calU \cap \{s \leq \f{s_f}{2}\})$, the following holds in the principal null frame of the dynamical background: 
    \begin{enumerate}
        \item If $\nu = \lambda$, then 
        \begin{equation}\label{eq:RC.principal.main.1}
        \begin{split}
            g(\nabla_{e_\mu^{(\cdot)}} e_\nu^{(\cdot)}, e_\lambda^{(\cdot)} ) - \pmb{g}(\pmb{\nabla}_{\pmb{e}_\mu^{(\cdot)}} \pmb{e}_\nu^{(\cdot)}, \pmb{e}_\lambda^{(\cdot)} )  =0.
        \end{split}
    \end{equation}
    \item If $(\mu,\nu,\lambda)$ is a permutation of $(A,B,C)$, $(3,4,A)$, $(A,B,4)$, $(4,4,3)$, or\footnote{With more work it seems that in the case where $(\mu,\nu,\lambda)$ is a permutation of $(4,4,A)$, one can obtain a further improvement. However, this will not be needed later.} $(4,4,A)$, then
        \begin{equation}\label{eq:RC.principal.main.2}
            \begin{split}
                \sum_{i_1+i_2+i_3+i_4\leq I} \opnorm[\Big]{ (\pmb{e}_1^{(\cdot)})^{i_1} (\pmb{e}_2^{(\cdot)})^{i_2} (|\Delta| \pmb{e}_3)^{i_3} (\pmb{e}_4)^{i_4} \Big( g(\nabla_{e_\mu^{(\cdot)}} e_\nu^{(\cdot)}, e_\lambda^{(\cdot)} ) - \pmb{g}(\pmb{\nabla}_{\pmb{e}_\mu^{(\cdot)}} \pmb{e}_\nu^{(\cdot)}, \pmb{e}_\lambda^{(\cdot)} ) \Big) } \ls \ep.
            \end{split}
        \end{equation}
    \item If $(\mu,\nu,\lambda)$ is a permutation of $(3,A,B)$, $(3,3,A)$, or $(3,3,4)$, then
        \begin{equation}\label{eq:RC.principal.main.3}
            \begin{split}
                \sum_{i_1+i_2+i_3+i_4\leq I} \opnorm[\Big]{|\Delta| (\pmb{e}_1^{(\cdot)})^{i_1} (\pmb{e}_2^{(\cdot)})^{i_2} (|\Delta| \pmb{e}_3)^{i_3} (\pmb{e}_4)^{i_4}\Big( g(\nabla_{e_\mu^{(\cdot)}} e_\nu^{(\cdot)}, e_\lambda^{(\cdot)} ) - \pmb{g}(\pmb{\nabla}_{\pmb{e}_\mu^{(\cdot)}} \pmb{e}_\nu^{(\cdot)}, \pmb{e}_\lambda^{(\cdot)} ) \Big)  } \ls \ep.
            \end{split}
        \end{equation}
    \end{enumerate}
\end{proposition}
\begin{proof}
The equation \eqref{eq:RC.principal.main.1} is clear as each of the two terms vanishes by a similar computation as \eqref{eq:Ricci.total.derivative}.

For the other terms, we compute the transformation:
    \begin{equation}\label{eq:RC.main.transformation}
        \begin{split}
            &\: g(\nabla_{e_\mu^{(\cdot)}} e_\nu^{(\cdot)}, e_\lambda^{(\cdot)} ) - \pmb{g}(\pmb{\nabla}_{\pmb{e}_\mu^{(\cdot)}} \pmb{e}_\nu^{(\cdot)}, \pmb{e}_\lambda^{(\cdot)} ) \\
            = &\: \underbrace{(\pmb{\calB}^{(\cdot)})_\mu^{\mu'} (\pmb{\calB}^{(\cdot)})_\lambda^{\lambda'}  (\pmb{\calB}^{(\cdot)})_\nu^{\nu'} \Big( g(\nabla_{e_{\mu'}'^{(\cdot)}} e_{\nu'}'^{(\cdot)}, e_{\lambda'}'^{(\cdot)} ) - \pmb{g}(\pmb{\nabla}_{\pmb{e}_{\mu'}'^{(\cdot)}} \pmb{e}_{\nu'}'^{(\cdot)}, \pmb{e}_{\lambda'}'^{(\cdot)} )\Big)}_{=:I} \\ 
            &\: + \underbrace{(\pmb{\calB}^{(\cdot)})_\mu^{\mu'} (\pmb{\calB}^{(\cdot)})_\lambda^{\lambda'}\Big( (e_{\mu'}'^{(\cdot)} - \pmb{e}_{\mu'}'^{(\cdot)}) (\pmb{\calB}^{(\cdot)})_\nu^{\nu'} \Big) g(e_{\nu'}'^{(\cdot)}, e_{\lambda'}'^{(\cdot)})}_{=:II} ,
        \end{split}
    \end{equation}
    where we used $g(e_{\nu'}^{(\cdot)}, e_{\lambda'}^{(\cdot)}) = \pmb{g}(\pmb{e}_{\nu'}^{(\cdot)}, \pmb{e}_{\lambda'}^{(\cdot)})$. 

    In the proposition, we need to bound \eqref{eq:RC.main.transformation} and its derivatives. Nonetheless, in view of the fact that all the estimates we will use (Proposition~\ref{prop:Ricci.double.null}, Lemma~\ref{lem:basic.double.null}, Proposition~\ref{prop:B}) allow for derivatives, once we bound the term \eqref{eq:RC.main.transformation}, its derivatives can be bounded similarly. Moreover, notice that while we took $(\pmb{e}_1'^{(\cdot)})^{i_1} (\pmb{e}_2'^{(\cdot)})^{i_2} (|\Delta| \pmb{e}_3'^{(\cdot)})^{i_3} (\pmb{e}_4'^{(\cdot)})^{i_4}$ derivatives in Proposition~\ref{prop:Ricci.double.null}, Lemma~\ref{lem:basic.double.null}, Proposition~\ref{prop:B}, we can equivalently take derivatives in $(\pmb{e}_1^{(\cdot)})^{i_1} (\pmb{e}_2^{(\cdot)})^{i_2} (|\Delta| \pmb{e}_3)^{i_3} (\pmb{e}_4)^{i_4}$ after using Proposition~\ref{prop:B}.

    In order to prove the desired estimate, we need the powers of $|\Delta|$ in $(\pmb{\calB}^{(\cdot)})_{\mu}^\nu$ and its derivatives. In particular, we use the estimates in Proposition~\ref{prop:B}. Moreover, using also Lemma~\ref{lem:basic.double.null}, we obtain
        \begin{equation}\label{eq:diff.frame.on.B}
            \opnorm{(e_{\mu'}'^{(\cdot)} - \pmb{e}_{\mu'}'^{(\cdot)}) (\pmb{\calB}^{(\cdot)})_\nu^{\nu'}} \ls \ep. 
        \end{equation}

    \pfstep{Step~1: Bounds for term $II$ in \eqref{eq:RC.main.transformation}} We first control term $II$ corresponding to the cases in \eqref{eq:RC.principal.main.2}--\eqref{eq:RC.principal.main.3}. By Proposition~\ref{prop:B}, \eqref{eq:diff.frame.on.B}, and Proposition~\ref{prop:Ricci.double.null}, $\opnorm{II} \ls \ep$. In particular, this satisfies the bound needed for \eqref{eq:RC.principal.main.2}--\eqref{eq:RC.principal.main.3}. 
    

    \pfstep{Step~2: Bounds for term $I$ in \eqref{eq:RC.main.transformation}} We then turn to the term $I$ in \eqref{eq:RC.main.transformation}.

    \pfstep{Step~2(a): Components in \eqref{eq:RC.principal.main.3}} Since all components of $\pmb{\calB}^{(\cdot)}$ are bounded, it follows from \eqref{eq:RC.double.null.main.2}--\eqref{eq:RC.double.null.main.3} that no terms can be worse than \eqref{eq:RC.principal.main.3}. As a result, when $(\mu,\nu,\lambda)$ is a permutation of $(3,A,B)$, $(3,3,A)$, or $(3,3,4)$, term $I$ in \eqref{eq:RC.main.transformation} satisfies the desired estimate in \eqref{eq:RC.principal.main.3}.

    \pfstep{Step~2(b): Components in \eqref{eq:RC.principal.main.2}} Next, we consider the estimate \eqref{eq:RC.principal.main.2} for term $I$ in \eqref{eq:RC.main.transformation} when $(\mu,\nu,\lambda)$ is a permutation of $(A,B,C)$, $(3,4,A)$, $(A,B,4)$, $(4,4,3)$ or $(4,4,A)$. Suppose $(\mu',\nu',\lambda')$ is a permutation of $(3,D,E)$ (which by Proposition~\ref{prop:Ricci.double.null} is the only way that one has a term of size $O(|\Delta|^{-1})$). Then we must have either $(\pmb{\calB}^{(\cdot)})_A^3$, $(\pmb{\calB}^{(\cdot)})_4^A$, or $(\pmb{\calB}^{(\cdot)})_4^3$. Any of these would give a factor of $|\Delta|$ (by Proposition~\ref{prop:B}) and hence the term is acceptable. \qedhere
    
\end{proof}

The following proposition gives the main estimates on the curvature components and their derivatives in the dynamical principal null frame. Observe that we do not explicitly write down estimates for all curvature components. This can be carried out but is not needed for the proof of the main theorem. Observe also that (as in Proposition~\ref{prop:curvature.double.null.main}) \eqref{eq:main.curvature.1} and \eqref{eq:main.curvature.2} involve the (stronger) $\opnorm{\cdot}_*$ norm.
\begin{proposition}\label{prop:curvature.principal}
    In $\os{\calM}{2}\setminus\{\calU \cap \{s \leq \f{s_f}{2}\})$, the following estimates hold in the dynamical principal null frame:
    \begin{enumerate}
        \item If $(\mu_1,\mu_2,\mu_3,\mu_4)$ is a permutation of $(4,4,A,B)$ or $(4,4,3,A)$, then
        \begin{equation}\label{eq:main.curvature.1}
            \begin{split}
                &\: \sum_{i_1+i_2+i_3+i_4\leq I} \opnorm[\Big]{(\pmb{e}_1^{(\cdot)})^{i_1} (\pmb{e}_2^{(\cdot)})^{i_2} (|\Delta| \pmb{e}_3)^{i_3} (\pmb{e}_4)^{i_4} \Big( R(e_{\mu_1}^{(\cdot)}, e_{\mu_2}^{(\cdot)}, e_{\mu_3}^{(\cdot)}, e_{\mu_4}^{(\cdot)}) - \pmb{R}(\pmb{e}_{\mu_1}^{(\cdot)}, \pmb{e}_{\mu_2}^{(\cdot)}, \pmb{e}_{\mu_3}^{(\cdot)}, \pmb{e}_{\mu_4}^{(\cdot)})\Big) }_* \ls \ep. 
            \end{split}            
        \end{equation}
        \item If $(\mu_1,\mu_2,\mu_3,\mu_4)$ is a permutation of $(4,4,3,3)$ or $(4,3,A,B)$, then
        \begin{equation}\label{eq:main.curvature.2}
            \begin{split}
                &\: \sum_{i_1+i_2+i_3+i_4\leq I} \opnorm[\Big]{|\Delta| (\pmb{e}_1^{(\cdot)})^{i_1} (\pmb{e}_2^{(\cdot)})^{i_2} (|\Delta| \pmb{e}_3)^{i_3} (\pmb{e}_4)^{i_4}\Big(  R(e_{\mu_1}^{(\cdot)}, e_{\mu_2}^{(\cdot)}, e_{\mu_3}^{(\cdot)}, e_{\mu_4}^{(\cdot)}) - \pmb{R}(\pmb{e}_{\mu_1}^{(\cdot)}, \pmb{e}_{\mu_2}^{(\cdot)}, \pmb{e}_{\mu_3}^{(\cdot)}, \pmb{e}_{\mu_4}^{(\cdot)})\Big)}_* \ls \ep .
            \end{split}            
        \end{equation}
    \end{enumerate}

\end{proposition}
\begin{proof}
We write
    \begin{equation}\label{eq:curvature.diff}
        \begin{split}
            &\: R(e_{\mu_1}^{(\cdot)}, e_{\mu_2}^{(\cdot)}, e_{\mu_3}^{(\cdot)}, e_{\mu_4}^{(\cdot)}) - \pmb{R}(\pmb{e}_{\mu_1}^{(\cdot)}, \pmb{e}_{\mu_2}^{(\cdot)}, \pmb{e}_{\mu_3}^{(\cdot)}, \pmb{e}_{\mu_4}^{(\cdot)}) \\
            = &\: (\pmb{\calB}^{(\cdot)})^{\nu_1}_{\mu_1} (\pmb{\calB}^{(\cdot)})^{\nu_2}_{\mu_2} (\pmb{\calB}^{(\cdot)})^{\nu_3}_{\mu_3} (\pmb{\calB}^{(\cdot)})^{\nu_4}_{\mu_4}\Big(R(e_{\nu_1}'^{(\cdot)}, e_{\nu_2}'^{(\cdot)}, e_{\nu_3}'^{(\cdot)}, e_{\nu_4}'^{(\cdot)}) - \pmb{R}(\pmb{e}_{\nu_1}'^{(\cdot)}, \pmb{e}_{\nu_2}'^{(\cdot)}, \pmb{e}_{\nu_3}'^{(\cdot)}, \pmb{e}_{\nu_4}'^{(\cdot)})\Big).
        \end{split}
    \end{equation}
    We now use the bounds in Proposition~\ref{prop:curvature.double.null.main} to control terms on the right-hand side; the key now is to understand the degeneration in $|\Delta|$. Similarly to the proof of Proposition~\ref{prop:Ricci.principal}, we will only consider the term in \eqref{eq:curvature.diff} without derivatives; the derivative bounds follow similarly since Proposition~\ref{prop:curvature.double.null.main} and Proposition~\ref{prop:B} also provide estimates for the derivatives.
    
    \pfstep{Step~1: Proof of \eqref{eq:main.curvature.1}} In order to prove \eqref{eq:main.curvature.1}, suppose we start with $(\mu_1,\mu_2,\mu_3,\mu_4)$ being a permutation of $(4,4,A,B)$ or $(4,4,3,A)$, we need to show that whenever $(\nu_1,\nu_2,\nu_3,\nu_4)$ is not itself also a permutation of $(4,4,A,B)$, $(4,4,3,A)$, or $(4,A,B,C)$, then there must be sufficient degeneration in the prefactor $(\pmb{\calB}^{(\cdot)})^{\nu_1}_{\mu_1} (\pmb{\calB}^{(\cdot)})^{\nu_2}_{\mu_2} (\pmb{\calB}^{(\cdot)})^{\nu_3}_{\mu_3} (\pmb{\calB}^{(\cdot)})^{\nu_4}_{\mu_4}$. We consider various cases.
    \begin{itemize}
        \item $(\nu_1,\nu_2,\nu_3,\nu_4)$ is a permutation of $(3,3,C,4)$. This is the harder case as according to \eqref{eq:main.curvature.DN.3}, we need to get a factor of $|\Delta|^2$ (instead of $|\Delta|$). 
        \begin{itemize}
            \item First observe that if there is a factor of $(\pmb{\calB}^{(\cdot)})_4^3$, then because it gives a factor of $|\Delta|^2$ (by Proposition~\ref{prop:B}), the term would be acceptable. From now on we can assume that there is not a factor of $(\pmb{\calB}^{(\cdot)})_4^3$. We further divide into two subcases:
            \begin{itemize}
                \item In the case $(\mu_1,\mu_2,\mu_3,\mu_4)$ is a permutation of $(4,4,A,B)$, since there are two $3$'s in the output and no $(\pmb{\calB}^{(\cdot)})_4^3$'s, the product $(\pmb{\calB}^{(\cdot)})_A^3 (\pmb{\calB}^{(\cdot)})_B^3$ must show up, which is acceptable by Proposition~\ref{prop:B}.
                \item In the case $(\mu_1,\mu_2,\mu_3,\mu_4)$ is a permutation of $(4,4,3,A)$, the condition that there are no $(\pmb{\calB}^{(\cdot)})_4^3$'s forces us to have $(\pmb{\calB}^{(\cdot)})_A^3 (\pmb{\calB}^{(\cdot)})_4^C$, which is acceptable by Proposition~\ref{prop:B}.
            \end{itemize}
        \end{itemize}
        \item $(\nu_1,\nu_2,\nu_3,\nu_4)$ is a permutation of $(4,4,3,3)$, $(4,3,A,B)$, $(A,B,C,D)$, $(4,3,3,A)$, or $(3,A,B,C)$. In this case, by \eqref{eq:main.curvature.DN.2}, we need one factor of $|\Delta|$.
        \begin{itemize}
            \item $(\mu_1,\mu_2,\mu_3,\mu_4)$ is a permutation of $(4,4,A,B)$, and $(\nu_1,\nu_2,\nu_3,\nu_4)$ is a permutation of $(4,4,3,3)$, $(4,3,A,B)$, $(4,3,3,A)$, or $(3,A,B,C)$: Since the input has no $3$ and the output has at least one $3$, there must be at least one factor of $(\pmb{\calB}^{(\cdot)})^3_4$ or $(\pmb{\calB}^{(\cdot)})^3_D$. (We use the notation $(\pmb{\calB}^{(\cdot)})^3_D$ instead of $(\pmb{\calB}^{(\cdot)})^3_A$ to emphasize the angular component could be $A$, $B$ or $C$.) Either of these (at least) gives the necessary $|\Delta|$ factor by Proposition~\ref{prop:B}.
            \item $(\mu_1,\mu_2,\mu_3,\mu_4)$ is a permutation of $(4,4,A,B)$ or $(4,4,3,A)$, and $(\nu_1,\nu_2,\nu_3,\nu_4)$ is a permutation of $(A,B,C,D)$: Since the input has a $4$ and the output are all angular, there must be a factor of $(\pmb{\calB}^{(\cdot)})^E_4$, which gives $|\Delta|$ by Proposition~\ref{prop:B}.
            \item $(\mu_1,\mu_2,\mu_3,\mu_4)$ is a permutation of $(4,4,3,A)$, and $(\nu_1,\nu_2,\nu_3,\nu_4)$ is a permutation of $(4,3,A,B)$, $(4,3,3,A)$, or $(3,A,B,C)$: The input has two $4$'s and the output has at most one $4$. Hence, there must be a factor of $(\pmb{\calB}^{(\cdot)})^D_4$ or $(\pmb{\calB}^{(\cdot)})^3_4$, which gives $|\Delta|$ by Proposition~\ref{prop:B}.
            \item $(\mu_1,\mu_2,\mu_3,\mu_4)$ is a permutation of $(4,4,3,A)$, and $(\nu_1,\nu_2,\nu_3,\nu_4)$ is a permutation of $(4,4,3,3)$: There are more $3$'s in the output than in the input. Hence, there must be a factor of $(\pmb{\calB}^{(\cdot)})_A^3$ or $(\pmb{\calB}^{(\cdot)})_4^3$, which is acceptable by Proposition~\ref{prop:B}.
        \end{itemize}
    \end{itemize}
    \pfstep{Step~2: Proof of \eqref{eq:main.curvature.2}} This is slightly easier than Step~1 since we allow for a degeneration of $|\Delta|$ in \eqref{eq:main.curvature.2}. Take $(\mu_1,\mu_2,\mu_3,\mu_4)$ to be a permutation of $(4,4,3,3)$ or $(4,3,A,B)$, by Proposition~\ref{prop:curvature.double.null.main}, we only check the case where the output $(\nu_1,\nu_2,\nu_3,\nu_4)$ is a permutation of $(3,3,A,B)$.  In this case, since the output has at least one fewer $4$ than the input, there must be a factor of $(\pmb{\calB}^{(\cdot)})^D_4$ or $(\pmb{\calB}^{(\cdot)})^3_4$, which gives $|\Delta|$ by Proposition~\ref{prop:B}.
\end{proof}

\section{Propagation from $\mathcal{H}^+$ to $\Gamma$}\label{sec:prop.up.to.Gamma}

 In this section we propagate our main instability estimate \eqref{EqAsInst2}, combined with the faster decay of the $l>2$ mode, from the event horizon $\Hp$ to the hypersurface $\Gamma$. We do this by using the nonlinear Teukolsky equation in the NP-formalism from Section \ref{SecNonLinearTeukolsky}. To make contact with the NP-formalism we now define the complex NP frame $(l,n,m, \overline{m})$ on the dynamical spacetime $\calM$ by 
\begin{equation} \label{EqDynamicalNPFrameDef}
l:= e_4, \qquad  n := \f 12 e_3, \qquad \textnormal{ and }\quad  m:=  \frac{1}{\sqrt{2}}\cdot \frac{\sqrt{\Sigma}}{r + i a \cos \theta}(e_1 + i \cdot e_2)\;,
\end{equation}
where $e_\mu$ are as in \eqref{EqDefDynamicalPNFrame}. This defines the dynamical NP connection coefficients \eqref{EqRicciCoeff} and curvature scalars \eqref{EqCurvatureNP} on $\calM$. Furthermore, by Proposition \ref{PropDeriNonLinearTeuk} and \eqref{EqShortHandNonLinearTeuk} the nonlinear Teukolsky equation $0 = \mathfrak{T}_{[2]} \psi_0 + \mathfrak{N}$ is satisfied.

We also recall the Kerr principal null frame $\pmb{e}_\mu$ on $\calM_{\mathrm{Kerr}}$ defined in \eqref{eq:Teukolsky.in.BL}, and define 
the corresponding background NP frame by 
\begin{equation} \label{EqBackgroundNPFrameDef}
    \pmb{l}= \pmb{e}_4, \qquad  \pmb{n} = \f 12 \pmb{e}_3, \qquad \textnormal{ and } \quad  \pmb{m}:=  \frac{1}{\sqrt{2}}\cdot \frac{\sqrt{\Sigma}}{r + i a \cos \theta}(\pmb{e}_1 + i \cdot \pmb{e}_2). 
\end{equation}
This defines the background NP connection coefficients, curvature scalars, and the Teukolsky operator $\pmb{\mathfrak{T}}_{[2]} = \frac{1}{2 \Sigma} \cdot \pmb{\mathcal{T}}_{[2]}$ on $\calM_{\mathrm{Kerr}}$. In Section \ref{SecDefLinearTeukolskyField} we construct a particular reference solution $$\pmb{\mathcal{T}}_{[2]}\psil = 0$$ on $\calM_{\mathrm{Kerr}}$ of the background linear Teukolsky equation, for which we have a linear instability result. Defining $\psi_0$ on $\calM_{\mathrm{Kerr}}$ via the identification $\Phi$, we then set 
$\psi :=  \psi_0 - \psil$, which satisfies the equation $\pmb{\mathfrak{T}}_{[2]} \psi = (\pmb{\mathfrak{T}}_{[2]} -\mathfrak{T}_{[2]}) \psi_0 -   \mathfrak{N} $. Recalling $\pmb{\calT}_{[2]} = 2 \Sigma \pmb{\mathfrak{T}}_{[2]}$, we have
\begin{equation} \label{EqDiffTeuk1}
    \pmb{\mathcal{T}}_{[2]} \psi = \underbrace{2 \Sigma \cdot \big[(\pmb{\mathfrak{T}}_{[2]} -\mathfrak{T}_{[2]}) \psi_0 -   \mathfrak{N} \big]}_{=: F}\;.
\end{equation}
The right-hand side (which is independent of $\psil$) is being estimated in Section \ref{SecEstNP}. The smallness of $\psi$, i.e., the closeness of $\psi_0$ and $\psil$, is then concluded in Theorem \ref{ThmPropagationToGamma} in Section \ref{SecConcludingDifference}, using the energy estimate for the background linear Teukolsky equation from Section \ref{SecInhomLinTeuk}. In Section \ref{SecPrelFrameRot} we begin with laying out the impact of rotating the frame fields $m$ and $\pmb{m}$. We emphasize that in this section all our notation corresponds to the NP formalism, not to the CK formalism.

\subsection{Preliminary rotations of the frames} \label{SecPrelFrameRot}

We begin with the following

\begin{lemma}\label{Lem*+PhiSmooth}
    $\overline{r} + \mathfrak h (r_*, \th_*)$ is a smooth function up to the event horizon.
\end{lemma}

\begin{proof}
    We first compute
\begin{equation*}
    \frac{\rd}{\rd r}\Big|_+ \big( \overline{r} + \mathfrak h (r_*, \th_*)\big) = \f a{\Delta} + \rd_{r_*} \mathfrak{h} \f{\rd r_*}{\rd r} + \rd_{\th_*} \mathfrak h \f{\rd \th_*}{\rd r}\,,
\end{equation*}
where $\f{\rd r_*}{\rd r}$ and $\f{\rd \th_*}{\rd r}$ can be computed from taking the inverse of \eqref{eq:partials.1}--\eqref{eq:partials.4}. In particular,
$$\f{\rd r_*}{\rd r} = \f{\sqrt{(r^2+a^2)^2 - a^2 \sin^2\th_*\Delta}}{\Delta}$$
Using also $\f{\rd \mathfrak h}{\rd r_*}= -\f{2Mar}{\Sigma R^2}$ (see Definition~\ref{def:Kerr.double.null.coordinates}), we have 
\begin{equation}
    \f{\rd \mathfrak h}{\rd r_*} \f{\rd r_*}{\rd r} = -\f{2Mar}{\Sigma R^2} \f{\sqrt{(r^2+a^2)^2 - a^2 \sin^2\th_*\Delta}}{\Delta} = -\f{2Mar}{\Delta(r^2+a^2)} + O(1)
\end{equation} 
by \eqref{EqMagicFormula}, where $O(1)$ denotes a bounded smooth function. The smoothness of $\frac{1}{\sin \theta_*}\rd_{\th_*} \mathfrak h$ follows from \cite[Proposition A.12]{DafLuk17} and the smoothness of $\sin \theta_*\frac{\rd \theta_*}{\rd r}$ from the explicit expression below \cite[(A.20)]{DafLuk17}.

We now compute using \cite[(A.9), (A.20)]{DafLuk17}
\begin{equation*}
\begin{split}
    \frac{1}{\sin \theta}\frac{\rd}{\rd\theta}\Big|_+ \big( \overline{r} + \mathfrak h (r_*, \th_*)\big) &= \frac{1}{\sin \theta}\rd_{r_*} \mathfrak{h} \frac{\rd r_*}{\rd \theta} + \frac{1}{\sin \theta}\rd_{\theta_*} \mathfrak{h} \frac{\rd \theta_*}{\rd \theta} \\
    &=  - \frac{ 2M a r}{\Sigma R^2} a \frac{\sqrt{\sin^2 \theta - \sin^2 \theta_*}}{\sin \theta} - \frac{1}{\sin \theta}\rd_{\theta_*}\mathfrak{h} \frac{\sin \theta_*}{a \sin \theta_*G \sqrt{\sin^2 \theta_* - \sin^2 \theta}} \,,
\end{split}
\end{equation*}
such that the smoothness is seen by invoking \cite[(A.39), (A.40)]{DafLuk17}.
\end{proof}

\begin{corollary}
    $e^{ 2 i \varphi_+}m$ is a smooth vector field away from the south pole $\{\theta = \pi\}$ and $e^{-2i \varphi_+}m$ is smooth away from the north pole $\{\theta = 0\}$. In particular, $\psi$ and $F$ defined in \eqref{EqDiffTeuk1} are spin $2$-weighted functions in the sense of Definition \ref{DefSpinWeightedFunction} in agreement with Section \ref{SecInhomLinTeuk}.
\end{corollary}

\begin{proof}
    The first statement follows directly from  Lemmas \ref{lem:basic.double.null.RS} and \ref{lem:basic.double.null} and writing $\varphi_+ = \varphi_* + \overline{r} + \mathfrak{h}$. The second statement follows from noting that $F = 2 \Sigma \cdot \pmb{\mathfrak{T}}_{[2]} \psi_0$.
\end{proof}

The estimate on $\rd_r|_+^2 F$ in Proposition \ref{PropInhomTeukI} will be reduced to the following estimates:
\begin{lemma}\label{lem:rotate.F.with.derivatives}
The following both hold:
    \begin{align}
        | (\rd_r |_+)^2 F| \ls &\: \sum_{j\leq 2}|\pmb{e}_3^j (e^{2i\varphi_*} F)|, \label{eq:rotate.for.derivative.NP} \\
        | (\rd_r |_+)^2 F| \ls &\: \sum_{j\leq 2}|\pmb{e}_3^j (e^{-2i\varphi_*} F)|.\label{eq:rotate.for.derivative.SP}
    \end{align}
\end{lemma}
\begin{proof}
    Note that by \eqref{DefPlusCoordVectorFields} and \eqref{eq:Teukolsky.in.BL} we have $\rd_r|_+ = - \Sigma \pmb{e}_3$. Thus it suffices to show the above with $\pmb{e}_3$ replaced by $\rd_r|_+$. We then compute, using $\varphi_+ = \varphi_* + \overline{r} + \mathfrak{h}$ and Lemma \ref{Lem*+PhiSmooth}
    \begin{equation*}
        \big|\rd_r|_+^2 F\big| = \big| \rd_r|_+^2 (e^{\pm 2i \varphi_+} F)\big| = \big|\rd_r|_+^2 
        \big(e^{\pm 2i (\overline{r} + \mathfrak{h})} e^{\pm 2i \varphi_*} F\big)\big| \ls \sum_{0 \leq j \leq 2} \big| \rd_r|_+^j (e^{\pm 2i \varphi_{*}} F)\big| \;.\qedhere
    \end{equation*}
\end{proof}

In addition to the dynamical and background NP frames \eqref{EqDynamicalNPFrameDef}, \eqref{EqBackgroundNPFrameDef} we introduce dynamical and background versions labeled with an $^{(N)}$ which are smooth away from the south pole
\begin{equation*}
    l^{(N)} := l, \quad n^{(N)} := n, \quad m^{(N)}:= e^{i \varphi_*} m \qquad \textnormal{ and } \qquad \pmb{l}^{(N)} := \pmb{l}, \quad \pmb{n}^{(N)} := \pmb{n}, \quad \pmb{m}^{(N)}:= e^{i \varphi_*} \pmb{m}
\end{equation*}
and versions labeled with an $^{(S)}$ which are smooth away from the north pole
\begin{equation*}
    l^{(S)} := l, \quad n^{(S)} := n, \quad m^{(S)}:= e^{-i \varphi_*} m \qquad \textnormal{ and } \qquad \pmb{l}^{(S)} := \pmb{l}, \quad \pmb{n}^{(S)} := \pmb{n}, \quad \pmb{m}^{(S)}:= e^{-i \varphi_*} \pmb{m}
\end{equation*}
The Newman--Penrose quantities with respect to these frames will be labeled by a superscript ${}^{(N)}$ or ${}^{(S)}$, respectively. For example we have $\pmb{\rho}^{(N)} = - \pmb{g}(\pmb{\nabla}_{\overline{\pmb{m}^{(N)}}} \pmb{l}, \pmb{m}^{(N)})$ and $\psi_0^{(S)} := R({l}, {m}^{(S)}, {l}, {m}^{(S)})$. The following important transformation law holds for the right-hand side of \eqref{EqDiffTeuk1} (i.e., of $F$):

\begin{lemma}\label{LemTrafoNPFrames.2}
The following holds:
    \begin{align}
        e^{2i\varphi_*} \cdot (\pmb{\mathfrak{T}}_{[2]} -\mathfrak{T}_{[2]})\psi_0 = (\pmb{\mathfrak{T}}_{[2]}^{(N)} -\mathfrak{T}_{[2]}^{(N)}) \psi_0^{(N)},\quad e^{2i\varphi_*} \mathfrak{N} = \mathfrak{N}^{(N)}, \label{EqLemTrafoFirstLine}\\
        e^{-2i\varphi_*} \cdot (\pmb{\mathfrak{T}}_{[2]} -\mathfrak{T}_{[2]}) \psi_0 = (\pmb{\mathfrak{T}}_{[2]}^{(S)} -\mathfrak{T}_{[2]}^{(S)}) \psi_0^{(S)},\quad e^{-2i\varphi_*} \mathfrak{N} = \mathfrak{N}^{(S)},
    \end{align}
    where $\mathfrak{T}_{[2]}^{(\cdot)}$, $\mathfrak{N}^{(\cdot)}$ are defined in the obvious manner with each Newman--Penrose quantity in \eqref{eq:mathfrak.T.def} and \eqref{eq:mathfrak.N.def} replaced by its northern/southern equivalent.
\end{lemma}

\begin{proof}
Consider for example
    \begin{equation*}
    \begin{split}
        e^{2 i \varphi_*} \big(\pmb{\mathfrak{T}}_{[2]} {\psi}_0\big) =&   e^{2 i \varphi_*} \Big[\Big(\pmb{\ddel}_1 - 2 (\pmb{\beta} + \ov{\pmb{\alpha}}) + \ov{\pmb{\pi}} - 4 \pmb{\tau}\Big)\Big(\pmb{\ov{\ddel}}_2 - 2 (\pmb{\alpha} + \ov{\pmb{\beta}}) + \pmb{\pi}\Big) \\
    &\quad -\Big(\pmb{\dD}_2 -  (\pmb{\varepsilon} + \ov{\pmb{\varepsilon}}) - 4 \pmb{\rho} - \ov{\pmb{\rho}}\Big)\Big(\pmb{\dDel}_2 - 2( \pmb{\gamma} + \ov{\pmb{\gamma}}) + \pmb{\mu}\Big) 
     + 3 \pmb{\psi}_2  \Big] R({l}, {m}, {l}, {m}) \;.
     \end{split}
    \end{equation*}
    The proof follows from Proposition \ref{PropTrafoNPQuant} together with the trivial transformation properties of the remaining NP quantities, cf.\ \eqref{EqRicciCoeff}, \eqref{EqCurvatureNP}, and \eqref{EqRelRicci}. The other terms in \eqref{EqLemTrafoFirstLine} follow analogously.
\end{proof}


\subsection{Estimates in the dynamical spacetime in terms of the Newman--Penrose formalism} \label{SecEstNP}

In this section we reduce the estimates on $e^{\pm 2 i \varphi_*}F$ in the north and south NP frames to the estimates obtained in Sections \ref{SecEstRedShiftPrincipalNull}, \ref{SecEstBlueShiftPN} in the frames  $(e_1^{(\cdot)}, e_2^{(\cdot)}, e_3^{(\cdot)}, e_4^{(\cdot)})$ with $^{(\cdot)} = ^{(N)}, ^{(S)}$ using Lemma \ref{LemTrafoNPFrames.2}.

As before, the estimates proven for the $^{(\cdot)} = ^{(N)}$ variant are to hold for $\th_* \in [0,\f{3\pi}4]$, and those for the $^{(\cdot)} = ^{(S)}$ variant are to hold for $\th_* \in [\f{\pi}4,\pi]$. It will be convenient to introduce for the remainder of the section the notation that
$$\os{\calM}{1}{}^{(N)} = \os{\calM}{1}\cap \{\th_* \in [0, \f{3\pi}{4}]\},\quad \os{\calM}{1}{}^{(S)} = \os{\calM}{1}\cap \{\th_* \in [\f{\pi}4, \pi]\},$$
and similarly for $\os{\calM}{2}{}^{(N)}$, $\os{\calM}{2}{}^{(S)}$, $\calM^{(N)}$ and $\calM^{(S)}$.


\subsubsection{Estimates in the red-shift region}

We first write the estimates we have already obtained with respect to the dynamical principal null frame in the Newman--Penrose formalism.
\begin{proposition}\label{prop:RS.NP.final}
    The following estimates hold in $\os{\calM}{1} \cap \{s \leq \f{3s_f}{4}\}$ (with conventions about $^{(\cdot)}$ defined in the beginning of the subsection). In what follows, denote $\pmb{\calD} \in \{ \pmb{\dD}, \pmb{\dDel}, \pmb{\ddel}^{(\cdot)}, \ov{\pmb{\ddel}}^{(\cdot)} \}$, and take $i$ to be a multi-index. Then
    \begin{equation}
        \sum_{|i|\leq I} |\pmb{\calD}^i \widetilde{\phi}| \ls \ep \mathfrak w(\ub),
    \end{equation}
    where 
    \begin{equation}\label{eq:terms.in.RS.NP.final}
        \begin{split}
            \widetilde{\phi} \in \{ &\:  \rho^{(\cdot)} - \pmb{\rho}^{(\cdot)}, \mu^{(\cdot)} - \pmb{\mu}^{(\cdot)}, \sigma^{(\cdot)}, \kappa^{(\cdot)}, \tau^{(\cdot)} - \pmb{\tau}^{(\cdot)}, \pi^{(\cdot)} - \pmb{\pi}^{(\cdot)},\\
            &\: \varepsilon^{(\cdot)} - \pmb{\varepsilon}^{(\cdot)},  \alp^{(\cdot)} - \pmb{\alp}^{(\cdot)}, \beta^{(\cdot)} - \pmb{\beta}^{(\cdot)}, \lambda^{(\cdot)}, \nu^{(\cdot)}, \gamma^{(\cdot)} \ - \pmb{\gamma} ^{(\cdot)}\}.
        \end{split}
    \end{equation}
    Moreover, the following estimates for the curvature components hold:
    \begin{equation}
        \sum_{|i|\leq I} |\pmb{\calD}^i (\psi_0^{(\cdot)}, \psi_1^{(\cdot)}, \psi_2^{(\cdot)} - \pmb{\psi}_2^{(\cdot)}) |\ls \ep \mathfrak w(\ub)\;.
    \end{equation}  
\end{proposition}
\begin{proof}
    The estimates for the Ricci coefficients represented by $\widetilde{\phi}$ is an immediate consequence of Proposition~\ref{prop:RS.principal.null}.
In the process, we noted that the background values $\pmb{\kappa} = \pmb{\sigma} = \pmb{\lambda} = \pmb{\nu} = 0$; see \eqref{eq:vanishing.background.NP.quantities}. 
    As a consequence, bounding $\kappa^{(\cdot)}$ is the same as bounding $\kappa^{(\cdot)} - \pmb{\kappa}^{(\cdot)}$. Similarly for $\sigma$, $\lambda$ and $\nu$. Hence, for those terms we do not need to write the difference.
    Similarly, for the curvature estimates, we used Proposition~\ref{prop:RS.curvature.principal.null} together with the fact that $\pmb{\psi}_0$, $\pmb{\psi}_1$ vanish (see \eqref{eq:background.NP.curvature}). \qedhere
\end{proof}

We now look at the precise terms in the Teukolsky equation.

\begin{proposition}\label{prop:Teukolsy.error.RS}
    In what follows, denote $\pmb{\calD} \in \{ \pmb{\dD}, \pmb{\dDel}, \pmb{\ddel}^{(\cdot)}, \ov{\pmb{\ddel}}^{(\cdot)} \}$, and take $i$ to be a multi-index. Then the following holds:
    \begin{equation}\label{eq:Teukolsky.error.RS}
        \sum_{|i|\leq I} \int_{\os{\calM}{1}{}^{(\cdot)}\cap \{s \leq \f{3s_f}{4}\}} \ub^q \Big( |\pmb{\calD}^i ((\pmb{\mathfrak{T}}_{[2]}^{(\cdot)} -\mathfrak{T}_{[2]}^{(\cdot)}) \psi_0^{(\cdot)})|^2 + |\pmb{\calD}^i \mathfrak{N}^{(\cdot)}|^2 \Big) \, \vol_{\gamma} \ud s \ud \ub \ls \ep^4
    \end{equation}
\end{proposition}
\begin{proof}
    We consider separately the term $(\pmb{\mathfrak{T}}_{[2]}^{(\cdot)} -\mathfrak{T}_{[2]}^{(\cdot)}) \psi_0^{(\cdot)}$ and $\mathfrak{N}^{(\cdot)}$. 
    
    We begin with $\mathfrak{N}^{(\cdot)}$. Each term in $\mathfrak{N}^{(\cdot)}$ is at least quadratic in the quantities in Proposition~\ref{prop:RS.NP.final}. As a result, we have
    \begin{equation}\label{EqUsedForDefLinear.1}
        \sum_{|i|\leq I} |\pmb{\calD}^i \mathfrak{N}^{(\cdot)}| \ls \ep^2 \mathfrak w^2(\ub).
    \end{equation}
    We will show that $(\pmb{\mathfrak{T}}_{[2]}^{(\cdot)} -\mathfrak{T}_{[2]}^{(\cdot)}) \psi_0^{(\cdot)}$ satisfies a similar bound. After writing out the derivatives
    \begin{align*}
		\ddel_1^{(\cdot)} = \ddel^{(\cdot)} - (\beta^{(\cdot)} - \ov{\alpha}^{(\cdot)}), \enspace \ov{\ddel}_2^{(\cdot)} = \ov{\ddel}^{(\cdot)} - 2(\alpha^{(\cdot)} - \ov{\beta}^{(\cdot)}),\enspace \dD_2^{(\cdot)} = \dD^{(\cdot)} - 2(\varepsilon^{(\cdot)} - \ov{\varepsilon}^{(\cdot)}),\enspace \dDel_2^{(\cdot)} = \dDel^{(\cdot)} - 2(\gamma^{(\cdot)} - \ov{\gamma}^{(\cdot)}),
	\end{align*}
    we have four types of terms:
    \begin{align}
		\hbox{difference of the second derivatives of $\psi_0^{(\cdot)}$}, \label{eq:diff.T.1}\\
		\hbox{difference of terms of the form } (\hbox{connection coefficient}) \times (\hbox{derivative of $\psi_0^{(\cdot)}$}), \label{eq:diff.T.2}\\
		\hbox{difference of terms of the form } (\hbox{first derivative of connection coefficient}) \times \psi_0^{(\cdot)},  \label{eq:diff.T.3}\\
		\hbox{difference of terms of the form } (\hbox{connection coefficient}) \times (\hbox{connection coefficient}) \times \psi_0^{(\cdot)}. \label{eq:diff.T.4}
	\end{align}
    Thus, obtaining the desired estimate reduces to bounding one of the following two types of terms:
    \begin{itemize}
        \item Terms where (the derivatives of) $\psi_0^{(\cdot)}$ is multiplied by (the derivatives of) the difference of the connection coefficients (i.e., one of the $\widetilde{\phi}$ terms in \eqref{eq:terms.in.RS.NP.final} in Proposition~\ref{prop:RS.NP.final}). These terms can be bounded using Proposition~\ref{prop:RS.NP.final} similarly to the $\mathfrak{N}^{(\cdot)}$
        \item Terms where (the derivatives of) $\psi_0^{(\cdot)}$ is multiplied by (the derivatives of) the difference of the frame fields. This can be estimated using Proposition~\ref{prop:RS.NP.final} (for $\psi_0^{(\cdot)}$) and \eqref{EqTrafoFrame1}--\eqref{EqTraforFrame2} (for the difference of the frame fields). 
    \end{itemize} 
    Altogether we thus have
    \begin{align} \label{EqUsedForDefLinear.2}
        \sum_{|i|\leq I} |\pmb{\calD}^i ((\pmb{\mathfrak{T}}_{[2]}^{(\cdot)} -\mathfrak{T}_{[2]}^{(\cdot)}) \psi_0^{(\cdot)})|\ls \ep^2 \mathfrak w^2(\ub).
    \end{align}
    
    Using \eqref{EqUsedForDefLinear.1} and \eqref{EqUsedForDefLinear.2}, we now integrate and recall the definition of the weight function in Definition~\ref{def:RS.weight}. Notice that the $s$-integration is on a finite interval $[0,s_f)$, and that the spheres have finite volume. Hence, the term on the left-hand side of \eqref{eq:Teukolsky.error.RS} is bounded by
    $$\ep^4 \int_{1}^{\infty} \ub^q \mathfrak w^4(\ub) \ud \ub \ls \ep^4 \Big( \sup_{\ub \in [1,\infty)} \ub^{q- 2(q_--2)} \Big) \| \ub^{\f{q_--2}2} \mathfrak w(\ub)\|_{L^\i_{\ub}}^2 \int_{1}^{\infty}  \ub^{q_--2} \mathfrak w^2(\ub) \ud \ub \ls \ep^4 \Big( \sup_{\ub \in [1,\infty)} \ub^{q- 2(q_--2)} \Big).$$
    Since $q\geq 7$, the final term is bounded by $\ep^4$ as long as $q-q_- \leq \f 32$. \qedhere
\end{proof}

\subsubsection{Estimates in the blue-shift region}

\begin{proposition}\label{prop:NP}
    The following holds in $\os{\calM}{2} \setminus (\calU \cap \{s \leq \f{s_f}{2}\})$ (with conventions about $^{(\cdot)}$ defined in the beginning of the subsection). In what follows, denote $\pmb{\calD} \in \{ \pmb{\dD}, |\Delta|\pmb{\dDel}, \pmb{\ddel}^{(\cdot)}, \ov{\pmb{\ddel}}^{(\cdot)} \}$, and take $i$ to be a multi-index. Then
    \begin{align}
        \sum_{|i|\leq 3}\opnorm{\pmb{\calD}^i (\rho^{(\cdot)} - \pmb{\rho}^{(\cdot)},  \sigma^{(\cdot)}, \kappa^{(\cdot)}, \tau^{(\cdot)} - \pmb{\tau}^{(\cdot)}, \pi^{(\cdot)} - \pmb{\pi}^{(\cdot)}, \varepsilon^{(\cdot)} - \pmb{\varepsilon}^{(\cdot)},  \alp^{(\cdot)} - \pmb{\alp}^{(\cdot)}, \beta^{(\cdot)} - \pmb{\beta}^{(\cdot)}) } \ls &\: \ep, \label{eq:Ricci.NP.1}\\
        \sum_{|i|\leq 3}\opnorm{|\Delta| \pmb{\calD}^i (\lambda^{(\cdot)}, \nu^{(\cdot)}, \mu^{(\cdot)} - \pmb{\mu}^{(\cdot)}, \gamma^{(\cdot)} \ - \pmb{\gamma} ^{(\cdot)})} \ls &\: \ep. \label{eq:Ricci.NP.2}
    \end{align}
    Moreover, the following estimates for the curvature components hold:
    \begin{align}
    	\sum_{|i|\leq 4} \opnorm{\pmb{\calD}^i (\psi_0^{(\cdot)}, \psi_1^{(\cdot)}) }_* \ls &\: \ep, \label{eq:curvature.NP.1}  \\
	\sum_{|i|\leq 4} \opnorm{|\Delta| \pmb{\calD}^i (\psi_2^{(\cdot)} - \pmb{\psi}_2^{(\cdot)}) }_* \ls &\: \ep. \label{eq:curvature.NP.2}
	\end{align}
	
\end{proposition}
\begin{proof}

	The estimates \eqref{eq:Ricci.NP.1}--\eqref{eq:Ricci.NP.2} amounts to rewriting the estimates in Proposition~\ref{prop:Ricci.principal} using the definition of \eqref{EqRicciCoeff}, $\rho$, $\mu$, etc.~in the Newman--Penrose formalism. In the process, we noted that the background values $\pmb{\kappa} = \pmb{\sigma} = \pmb{\lambda} = \pmb{\nu} = 0$; see \eqref{eq:vanishing.background.NP.quantities}. 
    As a consequence, bounding $\kappa^{(\cdot)}$ is the same as bounding $\kappa^{(\cdot)} - \pmb{\kappa}^{(\cdot)}$. Similarly for $\sigma$, $\lambda$ and $\nu$. Hence, for those terms we do not need to write the difference.
	
	Similarly, the estimates \eqref{eq:curvature.NP.1}--\eqref{eq:curvature.NP.2} follow from Proposition~\ref{prop:curvature.principal}. \qedhere

\end{proof}

We now look at the precise terms that arise in the Teukolsky equation. We notice that the main difficulty in the blue-shift region, which is not present in the red-shift region, is that some of the norms have $|\Delta|$-weight degeneration towards the Cauchy horizon. Nonetheless, if we restrict to the subregion where $u+\ub$ is bounded above, then the term can be treated only with slight modification of the proof of Proposition~\ref{prop:Teukolsy.error.RS}:

\begin{proposition}\label{prop:Teukolsy.error.BS.easy}
    For any fixed $C_{\mathrm{red}} \in \bbR$, the following holds (with an implicit constant depending on $C_{\mathrm{red}}$): 
    \begin{align}
        \sum_{j\leq 2} \int_{\os{\calM}{2}{}^{(\cdot)}\cap \{u+\ub\leq C_{\mathrm{red}}\}} \ub^q \Big( |\pmb{\dDel}^j ((\pmb{\mathfrak{T}}_{[2]}^{(\cdot)} -\mathfrak{T}_{[2]}^{(\cdot)}) \psi_0^{(\cdot)})|^2 + |\pmb{\dDel}^j \mathfrak{N}^{(\cdot)}|^2 \Big) \, |\Delta| \vol_{\gamma} \ud u \ud \ub\ls \ep^4.
    \end{align}
\end{proposition}
\begin{proof}
    Since we are restricted to $u+\ub\leq C_{\mathrm{red}}$, we can ignore the degeneration of weights towards the Cauchy horizon in Proposition~\ref{prop:NP} (see \eqref{EqAsympDelta}, Definition~\ref{def:Kerr.double.null.coordinates}, Lemma~\ref{LemRelationStarAndNullCoord}). Recall now that the norm $\opnorm{\cdot}$ (see Definition~\ref{def:BS.general.norm}) controls both the spacetime $L^2$ and the pointwise $L^\i$ norm, and that every term is at least quadratic in a similar manner as Proposition~\ref{prop:Teukolsy.error.RS}. (Note that the quadratic terms could be quadratic terms in quantities in Proposition~\ref{prop:NP}, or quadratic terms in $\psi_0$ and the frame differences. All these terms can be controlled using Proposition~\ref{prop:NP} and Lemma~\ref{lem:basic.double.null}.) Schematically denote the quadratic terms by $\phi \cdot \phi$, we thus have
    \begin{equation}
        \begin{split}
            &\: \int_{\os{\calM}{2}\cap \{u+\ub\leq C_{\mathrm{red}}\}} \ub^q |\phi \cdot \phi|^2 \vol_{\gamma} \ud u\ud \ub \\
            \ls &\: \Big(\sup_{\ub \geq 1} \ub^{q-(q_--3)-(q_--2)} \Big)\Big( \sup_{\os{\calM}{2}\cap \{u+\ub\leq C_{\mathrm{red}}\}} \ub^{\f{q_--3}2} |\phi|\Big)^2 \int_{\os{\calM}{2}\cap \{u+\ub\leq C_{\mathrm{red}}\}} \ub^{q_--2} |\phi|^2 \vol_{\gamma} \ud u\ud \ub \\
            \ls &\: \ep^4 \Big(\sup_{\ub \geq 1} \ub^{q-(q_--3)-(q_--2)}\Big).
        \end{split}
    \end{equation}
    Since $q \geq 7$, we obtain the conclusion as long as $q-q_- \leq 1$. \qedhere
\end{proof}

We now turn to the estimates in $\os{\calM}{2}$ without the restriction of $u+\ub\leq C_{\mathrm{red}}$, but instead all the way up to the hypersurface $\Gamma$. We make three observations related to the fact that the integral is restricted to the past of $\Gamma$:
\begin{enumerate}
    \item To the past of $\Gamma$, $|u|$ and $\ub$ are comparable (see \eqref{EqUVFutureGamma} and Lemma~\ref{LemRelationStarAndNullCoord}). As a result, all the $|u|$ weights in the norms in Definition~\ref{def:BS.general.norm} can be replaced by $\ub$ weights.
    \item We can use the $L^\i$ estimate in Definition~\ref{def:BS.general.norm} that is restricted to $\{f_\Gamma \leq 0\}$, which means that in the $L^\i$ norm we can have $|u|^{\f{q_{--}-2}{2}}$ weights (or $\ub^{\f{q_{--}-2}{2}}$ weights) instead of $|u|^{\f{q_{-}-3}{2}}$ weights (or $\ub^{\f{q_{-}-3}{2}}$ weights).
    \item For the curvature components, we have the improved bound where we can use $\opnorm{\cdot}_*$ instead of $\opnorm{\cdot}$. Moreover, to the past of $\Gamma$, the degenerate weight $\brk{u+\ub}^{-\f 34}$ satisfies  $\brk{u+\ub} \ls \log \ub$ by \eqref{eq:Gamma.in.u.ub}. Hence, after absorbing the power log-loss in the $\ub$-power, we have the inequality
    \begin{equation}\label{eq:good.norm.for.curvature!}
        \| \ub^{\f{q_{--}-2}2}  (\phi - \pmb{\phi}) \|_{L^2(\os{\calM}{2}\cap \{f_\Gamma \leq 0\}, \vol_{\gamma}\ud u\ud \ub)} \ls \opnorm{\phi - \pmb{\phi}}_*.
    \end{equation}
\end{enumerate}
The upshot now is that since in every quadratic error term, at least one of the factors is a curvature term, we can alway put the curvature term in $L^2$. As a result, we do not need an additional $|\Delta|$ weight in the spacetime norms. 

We now proceed to the proof in earnest, starting with the $\mathfrak N^{(\cdot)}$ term.

\begin{proposition}\label{prop:nonlinear.Teukolsky.error.1}
	The following holds:
		\begin{equation}
			\int_{\os{\calM}{2}{}^{(\cdot)}\cap \{f_\Gamma \leq 0\}} \ub^q |\mathfrak N^{(\cdot)}|^2 |\Delta| \vol_{\gamma} \ud u \ud \ub \ls \ep^4. 
		\end{equation}
\end{proposition}
\begin{proof}
	We recall the terms in $\mathfrak{N}$ from \eqref{eq:mathfrak.N.def}:
	\begin{equation*}
    \mathfrak{N} := 4  \ov{\ddel}_3(\sigma \psi_1)   - 4 \dDel_2 ( \kappa \psi_1) + \sigma \psi_1 \big[ -8( \alpha + \ov{\beta}) - 4 \ov{\tau} \big]  + \kappa \psi_1 \big[ 12(\gamma + \ov{\gamma}) - 4 \ov{\mu} \big]   + 3\psi_0 (\nu \kappa - \lambda \sigma)  - 10 \psi_1^2.
\end{equation*}
	Now we estimate $\mathfrak{N}^{(\cdot)}$. 
	We separate out the terms which have vanishing background values:
	\begin{equation}\label{eq:Teu.N.van}
			\psi_0^{(\cdot)}, \psi_1^{(\cdot)}, \sigma^{(\cdot)}, \kappa^{(\cdot)}, \lambda^{(\cdot)}, \nu^{(\cdot)}, 
	\end{equation}
	and the terms which have non-vanishing, but bounded, background values:
	\begin{equation}\label{eq:Teu.N.nonvan}
			\psi_2^{(\cdot)}, \rho^{(\cdot)},\mu^{(\cdot)}, \tau^{(\cdot)}, \pi^{(\cdot)}, \varepsilon^{(\cdot)}, \alp^{(\cdot)}, \bt^{(\cdot)}, \gamma^{(\cdot)}.
	\end{equation}
	There are two important observations:
	\begin{enumerate}
		\item Every term is either quadratic or cubic, as must be at least quadratic in the quantities \eqref{eq:Teu.N.van}.
		\item In every term, there is at most one factor that has to be controlled by \eqref{eq:Ricci.NP.2} or \eqref{eq:curvature.NP.2}. 
	\end{enumerate}
	
	All cubic terms will be treated as if they were quadratic after suitably putting one factor in $L^\i$. As all the terms can be treated similarly after making these observations, we will only consider one example, namely $\kappa^{(\cdot)} \psi_1^{(\cdot)} \gamma^{(\cdot)}$. All other terms are either similar or simpler. 
	
	For the term $\kappa^{(\cdot)} \psi_1^{(\cdot)} \gamma^{(\cdot)}$, observe that $\gamma^{(\cdot)}$ (is the only factor that) belongs to the group \eqref{eq:Teu.N.nonvan}, which has a non-vanishing, but bounded, background value. Hence, we write $\gamma^{(\cdot)} = \pmb{\gamma}^{(\cdot)} + (\gamma^{(\cdot)} - \pmb{\gamma}^{(\cdot)})$. For the contribution from $\pmb{\gamma}^{(\cdot)}$, we simply use that it is bounded and so we need to estimate $\psi_1^{(\cdot)}\kappa^{(\cdot)}$, which can be controlled as follows:
	\begin{equation}\label{eq:Teus.N.1}
		\begin{split}
			&\: \int_{\os{\calM}{2}{}^{(\cdot)}\cap \{f_\Gamma \leq 0\}} \ub^q |\psi_1^{(\cdot)} \kappa^{(\cdot)}|^2 |\Delta| \vol_{\gamma}\ud u\ud \ub \\
			\ls &\: \| \ub^{\f{q_{--}-2}{2}}  \psi_1^{(\cdot)}  \|_{L^2(\os{\calM}{2}{}^{(\cdot)}\cap \{f_\Gamma \leq 0\}, \vol_{\gamma}\ud u\ud \ub)}^2 \| \ub^{\f{q_{--}-2}{2}} \kappa^{(\cdot)} \|_{L^\i(\os{\calM}{2}{}^{(\cdot)}\cap \{f_\Gamma \leq 0\}, \vol_{\gamma}\ud u\ud \ub)}^2 \Big( \sup_{\ub \geq 1} \ub^{q-2q_{--}+4} \Big),
		\end{split}
	\end{equation}
	where we used $|\Delta| \ls 1$. (As mentioned before the proof, we always put the curvature term in $L^2$ so that we can use \eqref{eq:good.norm.for.curvature!}.) Now using the bounds for $\psi_1^{(\cdot)}$ and $\kappa^{(\cdot)}$ in \eqref{eq:curvature.NP.1} and \eqref{eq:Ricci.NP.1}, respectively, the term \eqref{eq:Teus.N.1} is therefore bounded above by $\ep^4$ after noting $\ub^{q-2q_{--}+4} \ls 1$ for $q \geq 7$ and $q_{--}$ sufficiently close to $q$.
	
	The contribution from $\gamma^{(\cdot)} - \pmb{\gamma}^{(\cdot)}$ is harder because of the degeneration in $|\Delta|$-weight and so we will need to be precise with the $\ub$-weights. Now since $\gamma^{(\cdot)} - \pmb{\gamma}^{(\cdot)}$ only obeys the degenerate estimate \eqref{eq:Ricci.NP.2}, we will instead put $\kappa^{(\cdot)}$ in $L^\i$ using \eqref{eq:Ricci.NP.1}. We then need to estimate $\psi_1^{(\cdot)} (\gamma^{(\cdot)} - \pmb{\gamma}^{(\cdot)})$, which is bounded as follows:
	\begin{equation}\label{eq:Teus.N.2}
		\begin{split}
			&\: \int_{\os{\calM}{2}{}^{(\cdot)}\cap \{f_\Gamma \leq 0\}} \ub^q |\psi_1^{(\cdot)} (\gamma^{(\cdot)} - \pmb{\gamma}^{(\cdot)})|^2 |\Delta| \vol_{\gamma} \ud u \ud \ub \\
			\ls &\: \| \ub^{\f{q_{--}-2}{2}}  \psi_1^{(\cdot)}  \|_{L^2(\os{\calM}{2}{}^{(\cdot)}\cap \{f_\Gamma \leq 0\}, \vol_{\gamma}\ud u\ud \ub)}^2 \| \ub^{\f{q_{--}-2}{2}} |\Delta| (\gamma^{(\cdot)} - \pmb{\gamma}^{(\cdot)}) \|_{L^{\i}(\os{\calM}{2}{}^{(\cdot)}\cap \{f_\Gamma \leq 0\}, \vol_{\gamma}\ud u\ud \ub)}^2 \\
            &\: \qquad \times \Big( \sup_{\os{\calM}{2}{}^{(\cdot)}\cap \{f_\Gamma \leq 0\}} \ub^{q-2q_{--}+4} |\Delta|^{-1} \Big).
		\end{split}
	\end{equation}
	As in \eqref{eq:Teus.N.1}, $\| \ub^{\f{q_{--}-2}{2}}  \psi_1^{(\cdot)}  \|_{L^2(\os{\calM}{2}{}^{(\cdot)}\cap \{f_\Gamma \leq 0\}, \vol_{\gamma}\ud u\ud \ub)}$ and  $\| \ub^{\f{q_{--}-2}{2}} |\Delta| (\gamma^{(\cdot)} - \pmb{\gamma}^{(\cdot)}) \|_{L^{\i}(\os{\calM}{2}{}^{(\cdot)}\cap \{f_\Gamma \leq 0\}, \vol_{\gamma}\ud u\ud \ub)}$ are each bounded by $\ep$ using Proposition~\ref{prop:NP}. Since we are to the past of $\Gamma$, by \eqref{EqDeltaPastGamma}, we have
	\begin{equation}\label{eq:polynomial.check}
		\sup_{\os{\calM}{2}{}^{(\cdot)}\cap \{f_\Gamma \leq 0\}} \ub^{q-2q_{--}+ 4} |\Delta|^{-1} \ls \ub^{q-2q_{--}+ 4} \ub^{\sigma_q} = \ub^{q-2q_{--}+ 4} \ub^{\f 14 (q+3)}.
	\end{equation}
	Since $q \geq 7$, we have $q-4 \geq \f{q+3}{4} + \f 12$. Thus, as long as $q_{--}$ is sufficiently close to $q$ (precisely, with $q - q_{--} \leq \f 14$), we have $\hbox{\eqref{eq:polynomial.check}}\ls 1$, which then gives the desired bound for this term. 
	
	Let us also remark that there are terms involving derivatives $\ov{\ddel}_3$ or $\dDel_2$, which can be treated similarly using the bounds for the derivatives in Proposition~\ref{prop:NP}. Consider, say $\dDel_2^{(\cdot)} ( \kappa^{(\cdot)} \psi_1^{(\cdot)}) = \dDel^{(\cdot)} ( \kappa^{(\cdot)} \psi_1^{(\cdot)}) -2 (\gamma^{(\cdot)} - \ov{\gamma}^{(\cdot)})\kappa^{(\cdot)} \psi_1^{(\cdot)} $ (see Definition~\ref{def:Dr}). The terms $\gamma^{(\cdot)} \kappa^{(\cdot)} \psi_1^{(\cdot)}$ and $\ov{\gamma}^{(\cdot)} \kappa^{(\cdot)} \psi_1^{(\cdot)}$ can be treated in the same manner as above. For the derivative term,
    $$ \dDel^{(\cdot)} ( \kappa^{(\cdot)} \psi_1^{(\cdot)})= \kappa^{(\cdot)} \dDel^{(\cdot)}  \psi_1^{(\cdot)} + \psi_1^{(\cdot)} \dDel^{(\cdot)}  \kappa^{(\cdot)}. $$
	We write $\dDel^{(\cdot)} = (\dDel^{(\cdot)} - \pmb{\dDel}^{(\cdot)}) + \pmb{\dDel}^{(\cdot)}$. For the difference of frames, we use \eqref{eq:double.null.transform}, \eqref{EqDefBackgroundPNFrame} and \eqref{eq:B().def} to obtain 
    \begin{equation}\label{eq:diff.of.frame.in.Teukolsky}
        e_\mu^{(\cdot)} - \pmb{e}_\mu^{(\cdot)} = (\pmb{\calB}^{(\cdot)})_\mu^\nu ((\calA^{(\cdot)})_{\nu}^\sigma \pmb{e}'^{(\cdot)}_\sigma - \pmb{e}'^{(\cdot)}_\nu).
    \end{equation}
    Thus the difference $\dDel^{(\cdot)} - \pmb{\dDel}^{(\cdot)}$ can be controlled using Proposition~\ref{prop:B} and Lemma~\ref{lem:basic.double.null}, and as a result, we use Proposition~\ref{prop:NP} to deduce that
    $$\| \ub^{\f{q_{--}-2}{2}} |\Delta| \dDel^{(\cdot)}  \psi_1^{(\cdot)} \|_{L^2(\os{\calM}{2}{}^{(\cdot)}\cap \{f_\Gamma \leq 0\}, \vol_{\gamma}\ud u\ud \ub)} + \| \ub^{\f{q_{--}-2}{2}} |\Delta| \dDel^{(\cdot)}  \kappa^{(\cdot)}\|_{L^\i(\os{\calM}{2}{}^{(\cdot)}\cap \{f_\Gamma \leq 0\}, \vol_{\gamma}\ud u\ud \ub)} \ls \ep.$$
	Thus, we have reduced to a quadratic expression where exactly one of the two terms has a $|\Delta|$-degeneration in the estimate, which can then be controlled as above using \eqref{eq:Teus.N.2} and \eqref{eq:polynomial.check}. \qedhere

\end{proof}

\begin{proposition}\label{prop:nonlinear.Teukolsky.error.2}
	The following holds: 
		\begin{equation}
			\int_{\os{\calM}{2}{}^{(\cdot)}\cap \{f_\Gamma \leq 0\}} \ub^q |(\pmb{\mathfrak{T}}_{[2]}^{(\cdot)} -\mathfrak{T}_{[2]}^{(\cdot)}) \psi_0^{(\cdot)}|^2 |\Delta| \vol_{\gamma} \ud u \ud \ub \ls \ep^4. 
		\end{equation}
\end{proposition}
\begin{proof}
	Recall from \eqref{eq:mathfrak.N.def} that
	\begin{equation*}
    \mathfrak{T}_{[2]} := \Big(\ddel_1 - 2 (\beta + \ov{\alpha}) + \ov{\pi} - 4 \tau\Big)\Big(\ov{\ddel}_2 - 2 (\alpha + \ov{\beta}) + \pi\Big)  -\Big(\dD_2 -  (\varepsilon + \ov{\varepsilon}) - 4 \rho - \ov{\rho}\Big)\Big(\dDel_2 - 2( \gamma + \ov{\gamma}) + \mu\Big)   + 3  \psi_2.
\end{equation*}
	
	We write this in the $^{(N)}$ and $^{(S)}$ versions, and expand using Definition~\ref{def:Dr},
	\begin{align*}
		\ddel_1^{(\cdot)} = \ddel^{(\cdot)} - (\beta^{(\cdot)} - \ov{\alpha}^{(\cdot)}), \enspace \ov{\ddel}_2^{(\cdot)} = \ov{\ddel}^{(\cdot)} - 2(\alpha^{(\cdot)} - \ov{\beta}^{(\cdot)}),\enspace \dD_2^{(\cdot)} = \dD^{(\cdot)} - 2(\varepsilon^{(\cdot)} - \ov{\varepsilon}^{(\cdot)}),\enspace \dDel_2^{(\cdot)} = \dDel^{(\cdot)} - 2(\gamma^{(\cdot)} - \ov{\gamma}^{(\cdot)}).
	\end{align*}
	As in the proof of Proposition~\ref{prop:Teukolsy.error.RS}, we observe that every term can be written as one of \eqref{eq:diff.T.1}--\eqref{eq:diff.T.4}.
	We consider each of these types of terms in the four steps below, highlighting the structure of the terms.
    
    \pfstep{Step~1: Terms \eqref{eq:diff.T.1}} These are terms 
    \begin{equation}\label{eq:Teus.diff.second.order}
        \Big(\ddel^{(\cdot)} \ov{\ddel}^{(\cdot)} - \pmb{\ddel}^{(\cdot)}\ov{\pmb{\ddel}}^{(\cdot)} - \dD^{(\cdot)}\dDel^{(\cdot)} - \pmb{\dD}^{(\cdot)}\pmb{\dDel}^{(\cdot)} \Big) \psi_0.    
    \end{equation}
    For these we need to compute the difference of the dynamical principal null frame and the background principal null frame, given by \eqref{eq:diff.of.frame.in.Teukolsky}. 
    Now the key is to notice that since there is only one factor of $\dDel^{(\cdot)}$ or $\pmb{\dDel}^{(\cdot)}$, after expanding \eqref{eq:Teus.diff.second.order} by \eqref{eq:diff.of.frame.in.Teukolsky}, there can be at most one copy of $\pmb{e}_3$ that is not multiplied by a factor of $|\Delta|$. In other words, using Proposition~\ref{prop:B} and Lemma~\ref{lem:basic.double.null}, we have a schematic expansion of the following form:
    \begin{equation}\label{eq:Teus.diff.second.order.sche}
        \begin{split}
            \hbox{\eqref{eq:Teus.diff.second.order}} = &\: \sum_{\mu,\nu\neq 3} O(\ep\ub^{-\f{q_{--}-2}{2}}) \pmb{e}^{(\cdot)}_\mu\pmb{e}^{(\cdot)}_\nu \psi_0 + \sum_{\mu \neq 3} O(\ep\ub^{-\f{q_{--}-2}{2}}) \pmb{e}^{(\cdot)}_\mu\pmb{e}_3 \psi_0 + \sum_{\mu \neq 3} O(\ep\ub^{-\f{q_{--}-2}{2}}) \pmb{e}_3\pmb{e}^{(\cdot)}_\mu \psi_0 \\
        &\: + O(\ep|\Delta|\ub^{-\f{q_{--}-2}{2}}) \pmb{e}_3\pmb{e}_3 \psi_0 + \sum_{\mu} O(\ep\ub^{-\f{q_{--}-2}{2}}) \pmb{e}^{(\cdot)}_\mu \psi_0.
        \end{split}
    \end{equation}
    Thus,
    \begin{equation}\label{eq:Teus.diff.second.order.est}
        \begin{split}
            &\: \int_{\os{\calM}{2}{}^{(\cdot)}\cap \{f_\Gamma \leq 0\}} \ub^q |\hbox{\eqref{eq:Teus.diff.second.order}}|^2 |\Delta| \vol_{\gamma} \ud u \ud \ub \\
            \ls &\: \ep^2 \int_{\os{\calM}{2}{}^{(\cdot)}\cap \{f_\Gamma \leq 0\}} \ub^{q-q_{--}+2} \Big(\sum_{\mu,\nu\neq 3} |\pmb{e}^{(\cdot)}_\mu\pmb{e}^{(\cdot)}_\nu \psi_0|^2 + \sum_{\mu \neq 3}(|\pmb{e}^{(\cdot)}_\mu\pmb{e}_3 \psi_0|^2+ |\pmb{e}_3\pmb{e}^{(\cdot)}_\mu \psi_0|^2) \Big) |\Delta| \vol_{\gamma} \ud u \ud \ub\\
            &\: + \ep^2 \int_{\os{\calM}{2}{}^{(\cdot)}\cap \{f_\Gamma \leq 0\}} \ub^{q-q_{--}+2} \Big(|\Delta|^2 |\pmb{e}_3\pmb{e}_3 \psi_0|^2 + \sum_\mu |\pmb{e}^{(\cdot)}_\mu \psi_0|^2 \Big) |\Delta| \vol_{\gamma} \ud u \ud \ub.
        \end{split}
		\end{equation}
    We now bound \eqref{eq:Teus.diff.second.order.est} using \eqref{eq:curvature.NP.1}, noting that we can use \eqref{eq:good.norm.for.curvature!} to the past of $\Gamma$. When no $\pmb{e}_3$ appears, the required estimate is immediate from \eqref{eq:curvature.NP.1}. When there is an $\pmb{e}_3$, the application of \eqref{eq:curvature.NP.1} requires putting in $|\Delta|$ factors, and thus we have
    \begin{equation}\label{eq:Teus.diff.second.order.est.2}
        \begin{split}
            &\: \ep^2 \int_{\os{\calM}{2}{}^{(\cdot)}\cap \{f_\Gamma \leq 0\}} \ub^{q-q_{--}+2} \Big(\sum_{\mu \neq 3}(|\pmb{e}^{(\cdot)}_\mu\pmb{e}_3 \psi_0|^2+ |\pmb{e}_3\pmb{e}^{(\cdot)}_\mu \psi_0|^2) +|\Delta|^2 |\pmb{e}_3\pmb{e}_3 \psi_0|^2 + \sum_\mu |\pmb{e}^{(\cdot)}_\mu \psi_0|^2 \Big) |\Delta| \vol_{\gamma} \ud u \ud \ub \\
            \ls &\: \ep^4 \Big( \sup_{\os{\calM}{2}{}^{(\cdot)}\cap \{f_\Gamma \leq 0\}} \ub^{q-2q_{--}+ 4} |\Delta|^{-1} \Big).
        \end{split}
		\end{equation}
    (For the $\pmb{e}^{(\cdot)}_\mu\pmb{e}_3 \psi_0$ term, we can write $\pmb{e}^{(\cdot)}_\mu\pmb{e}_3 \psi_0 = |\Delta|^{-1} \pmb{e}^{(\cdot)}_\mu(|\Delta|\pmb{e}_3) \psi_0 - (\pmb{e}^{(\cdot)}_\mu \log |\Delta|)\pmb{e}_3 \psi_0$. Noting that $(\pmb{e}^{(\cdot)}_\mu \log |\Delta|)$ is bounded, we can then apply the bound in \eqref{eq:curvature.NP.1}.) Using \eqref{eq:polynomial.check}, we thus conclude that this term satisfies the desired bound for $q_{--}$ sufficiently close to $q$. 

    \pfstep{Step~2: Terms \eqref{eq:diff.T.2}} For these terms, the structure we need is that when the derivative is given by $\dDel$ or $\pmb{\dDel}$, then it is necessarily multiplied by connection coefficients that are not $\lambda^{(\cdot)}$, $\nu^{(\cdot)}$, $\mu^{(\cdot)}$ or $\gamma^{(\cdot)}$ (so that we do not need to use the degenerate estimate \eqref{eq:Ricci.NP.2}). For these terms, we have
    \begin{equation}\label{eq:Teus.diff.0.1}
        \begin{split}
            \Big((\varepsilon^{(\cdot)} + \bar{\varepsilon}^{(\cdot)}+4\rho^{(\cdot)} + \bar{\rho}^{(\cdot)})\dDel^{(\cdot)} - (\pmb{\varepsilon}^{(\cdot)} + \bar{\pmb{\varepsilon}}^{(\cdot)}+4\pmb{\rho}^{(\cdot)} + \bar{\pmb{\rho}}^{(\cdot)})\pmb{\dDel}^{(\cdot)}\Big) \psi_0 = \sum_{\mu} O(\ep\ub^{-\f{q_{--}-2}{2}}) \pmb{e}_\mu^{(\cdot)} \psi_0,
        \end{split}
    \end{equation}
    where we used Proposition~\ref{prop:NP} to control the differences of the connection coefficients and bounded the differences of the frames as in Step~1. 
    
    Now the right-hand side of \eqref{eq:Teus.diff.0.1} also appeared as one of the terms in \eqref{eq:Teus.diff.second.order.sche} and can therefore be estimated in the same manner.

    The remaining terms in \eqref{eq:diff.T.2} may have the connection coefficients $\lambda^{(\cdot)}$, $\nu^{(\cdot)}$, $\mu^{(\cdot)}$ or $\gamma^{(\cdot)}$, but they must be multiplied by the $\dD$, $\ddel^{(\cdot)}$, or $\ov{\ddel}^{(\cdot)}$ (or $\pmb{\dD}$, $\pmb{\ddel}^{(\cdot)}$, or $\ov{\pmb{\ddel}}^{(\cdot)}$) derivatives of $\psi_0$. As an example, we have
    \begin{equation}\label{eq:Teus.diff.0.1.2}
        \begin{split}
            &\: (\gamma^{(\cdot)} \dD^{(\cdot)} - \pmb{\gamma}^{(\cdot)} \pmb{\dD}^{(\cdot)})\psi_0 = \sum_{\mu \neq 3} O(\ep|\Delta|^{-1} \ub^{-\f{q_{--}-2}{2}}) \pmb{e}_\mu^{(\cdot)} \psi_0 + O(\ep\ub^{-\f{q_{--}-2}{2}}) \pmb{e}_3 \psi_0.
        \end{split}
    \end{equation}
    The $O(\ep\ub^{-\f{q_{--}-2}{2}}) \pmb{e}_3 \psi_0$ term in \eqref{eq:Teus.diff.0.1.2} has already been treated previously. For the other term, we use that \eqref{eq:curvature.NP.1} in Proposition~\ref{prop:NP} gives better bounds for the $\pmb{e}_1^{(\cdot)}$, $\pmb{e}_2^{(\cdot)}$, or $\pmb{e}_4$ derivatives of $\psi_0$ (without the need to $|\Delta|^{-1}$ degeneration, i.e., we have
    \begin{equation}
        \begin{split}
            &\: \ep^2\sum_{\mu \neq 3} \int_{\os{\calM}{2}{}^{(\cdot)}\cap \{f_\Gamma \leq 0\}} \ub^q ||\Delta|^{-1} \ub^{-\f{q_{--}-2}{2}}\pmb{e}_\mu \psi_0|^2 |\Delta| \vol_{\gamma} \ud u \ud \ub 
            \ls \ep^4 \Big( \sup_{\os{\calM}{2}{}^{(\cdot)}\cap \{f_\Gamma \leq 0\}} \ub^{q-2q_{--}+4} |\Delta|^{-1} \Big).
        \end{split}
		\end{equation}
    We then conclude with \eqref{eq:polynomial.check} as before.

    \pfstep{Step~3: Terms \eqref{eq:diff.T.3}} For these terms, notice that for the first derivatives of the connection coefficients, the derivatives must be $\dD$, $\ddel^{(\cdot)}$, or $\ov{\ddel}^{(\cdot)}$ (or $\pmb{\dD}$, $\pmb{\ddel}^{(\cdot)}$, or $\ov{\pmb{\ddel}}^{(\cdot)}$), which does not give an extra $|\Delta|^{-1}$. By Proposition~\ref{prop:NP}, the worst contribution from the derivative of the connection coefficient is thus given by the difference of the derivative of $\gamma$. Hence, by Proposition~\ref{prop:NP}, \eqref{eq:diff.of.frame.in.Teukolsky}, Proposition~\ref{prop:B} and Lemma~\ref{lem:basic.double.null}, the term can be reduced to 
    $$O(\ep|\Delta|^{-1} \ub^{-\f{q_{--}-2}{2}}) \psi_0.$$
    Using \eqref{eq:curvature.NP.1} in Proposition~\ref{prop:NP}, and \eqref{eq:good.norm.for.curvature!}, we thus bound
    \begin{equation}
        \begin{split}
            &\: \ep^2 \int_{\os{\calM}{2}{}^{(\cdot)}\cap \{f_\Gamma \leq 0\}} \ub^q ||\Delta|^{-1} \ub^{-\f{q_{--}-2}{2}} \psi_0|^2 |\Delta| \vol_{\gamma} \ud u \ud \ub 
            \ls \ep^4 \Big( \sup_{\os{\calM}{2}{}^{(\cdot)}\cap \{f_\Gamma \leq 0\}} \ub^{q-2q_{--}+4} |\Delta|^{-1} \Big).
        \end{split}
		\end{equation}
    and conclude as before.
    
    \pfstep{Step~4: Terms \eqref{eq:diff.T.4}} This is similar to Step~3. The key is to notice that in the cubic terms, at most one of the connection coefficients belongs to the set $\{\lambda^{(\cdot)}, \nu^{(\cdot)}, \mu^{(\cdot)}, \gamma^{(\cdot)}\}$. Therefore, all the relevant terms take the form
    $$O(\ep|\Delta|^{-1} \ub^{-\f{q_{--}-2}{2}}) \psi_0,$$
    and can thus be treated as in Step~3. \qedhere
\end{proof}

\subsection{Definition of the reference linear Teukolsky field on the background} \label{SecDefLinearTeukolskyField}

We begin with two elementary lemmas.
\begin{lemma} \label{LemBlueDecayPoly}
    Let $c>0$, $p \geq 0$, and $\phi, \mathfrak{f}$  spin $2$-weighted functions on $[1, \infty) \times \Sp^2$ related by 
    $$\phi(v, \theta) = \int_{v}^\infty e^{-c(v' - v)} \mathfrak{f}(v', \vartheta) \, \ud v' \;.$$ Then $\int_1^\infty \int_{\Sp^2} v^p |\mathfrak{f}(v, \vartheta)|^2 \vols 
    \ud v < \infty$ implies $\int_1^\infty \int_{\Sp^2} v^p |\phi(v, \vartheta)|^2 \vols 
    \ud v < \infty$.
\end{lemma}
\begin{proof}
    We first assume that $\mathfrak{f}$ is compactly supported. Then 
    \begin{equation*}
        \begin{split}
            \int_1^\infty \int_{\Sp^2}v^p& \Big| \int_v^\infty e^{-c(v'- v)} \mathfrak{f}(v', \vartheta) \, \ud v' \Big|^2 \vols \ud v \leq 
            \int_1^\infty \int_{\Sp^2}v^p \Big( \int_v^\infty e^{-c(v'- v)} |\mathfrak{f}(v', \vartheta)| \, \ud v' \Big)^2 \vols \ud v \\
            &=\frac{1}{2c}\int_1^\infty \int_{\Sp^2}v^p (\rd_ve^{2cv})\Big( \underbrace{\int_v^\infty e^{-cv'} |\mathfrak{f}(v', \vartheta)| \, \ud v'}_{\textnormal{compactly supported}} \Big)^2 \vols \ud v \\
            &\leq \frac{1}{2c}2  \int_1^\infty \int_{\Sp^2} v^p e^{2cv} e^{-cv} |\mathfrak{f}(v,\vartheta)| \Big(\int_v^\infty e^{-cv'} |\mathfrak{f}(v', \vartheta)| \, \ud v'\Big)\vols \ud v \\
            & \leq \frac{1}{c}  \Big( \int_1^\infty \int_{\Sp^2} v^p |\mathfrak{f}(v,\vartheta)|^2 \vols \ud v \Big)^{\frac{1}{2}} \Big( \int_1^\infty \int_{\Sp^2}v^p \Big( \int_v^\infty e^{-c(v'- v)} |\mathfrak{f}(v', \vartheta)| \, \ud v' \Big)^2 \vols \ud v\Big)^{\frac{1}{2}}\;,
        \end{split}
    \end{equation*}
    where in the second inequality we have dropped negative boundary and bulk terms from the integration by parts and in the second inequality just Cauchy--Schwarz. Dividing by the second factor proves the claim. If $\mathfrak{f}$ is not compactly supported, the claim follows by approximation with compactly supported functions.
\end{proof}

\begin{lemma}\label{LemNoShiftDecay}
    Let $p \geq 2$ and  $\phi, \mathfrak{f}$  spin $2$-weighted functions on $[1, \infty) \times \Sp^2$ related by 
    $$\phi(v, \theta) =  \int_{v}^\infty \mathfrak{f}(v', \vartheta) \ud v' \;.$$ Then $\int_1^\infty \int_{\Sp^2} v^p |\mathfrak{f}(v, \vartheta)|^2 \vols \ud v < \infty$ implies $\int_1^\infty \int_{\Sp^2} v^{p-2} |\phi(v, \vartheta)|^2 \vols 
    \ud v < \infty$.
\end{lemma}
\begin{proof}
    Assuming again in addition that $\mathfrak{f}$ is compactly supported, we compute in a similar fashion as before
    \begin{equation*}
        \begin{split}
            \int_1^\infty \int_{\Sp^2} v^{p-2} & \Big| \int_{v}^\infty \mathfrak{f}(v', \vartheta) \ud v' \Big|^2 \vols \ud v \leq 
            \int_1^\infty \int_{\Sp^2} v^{p-2}  \Big( \int_{v}^\infty |\mathfrak{f}(v', \vartheta)| \ud v' \Big)^2 \vols \ud v \\
            &=\int_1^\infty \int_{\Sp^2}\frac{1}{p-1} ( \rd_v v^{p-1}) \Big( \int_{v}^\infty |\mathfrak{f}(v', \vartheta)| \ud v' \Big)^2 \vols \ud v \\
            &\leq \frac{2}{p-1} \int_1^\infty \int_{\Sp^2} v^{p-1} |\mathfrak{f}(v, \vartheta)| \Big( \int_{v}^\infty |\mathfrak{f}(v', \vartheta)| \ud v' \Big) \vols \ud v \\
            &\leq \frac{2}{p-1} \Big(\int_1^\infty \int_{\Sp^2} v^p |\mathfrak{f}(v, \vartheta)|^2 \vols \ud v\Big)^{\f{1}{2}} \Big( \int_1^\infty \int_{\Sp^2} v^{p-2}  \Big( \int_{v}^\infty |\mathfrak{f}(v', \vartheta)| \ud v' \Big)^2 \vols \ud v \Big)^{\f{1}{2}} \;.
        \end{split}
    \end{equation*}
    The proof concludes as before.
\end{proof}

We define the following two smooth spin $2$-weighted functions on $\Hp$ which impact the dynamical behavior of  derivatives of $\psi_0 - \psil$ transversal to $\Hp$:
\begin{align*}
    & \mathfrak{f}_0:= \frac{\Sigma}{r_+^2 + a^2} \big[(\pmb{\mathfrak{T}}_{[2]} -\mathfrak{T}_{[2]}) \psi_0 -   \mathfrak{N} \big]\Big|_{\Hp} \\
    & \mathfrak{f}_1 := \frac{1}{r_+^2 +a^2} \rd_r \Big( \Sigma \cdot \big[(\pmb{\mathfrak{T}}_{[2]} -\mathfrak{T}_{[2]}) \psi_0 -   \mathfrak{N} \big] \Big)\Big|_{\Hp} \;.
\end{align*}
Here, $\rd_r$ is with respect to the $(v_+, r, \theta, \varphi_+)$ coordinates.

\begin{lemma} \label{LemDecayBoundsf0}
    We have $\sum\limits_{0 \leq i_0 + i_1 + i_2 + i_3 \leq I}\int_1^\infty \int_{\Sp^2} v_+^{q+2} |\rd_{v_+}^{i_0} \widetilde{Z}_{1,+}^{i_1}\widetilde{Z}_{2,+}^{i_2}\widetilde{Z}_{3,+}^{i_3} \mathfrak{f}_a|^2 \, \vols \ud v_+ \ls 1$ for $a = 0,1$.
\end{lemma}
\begin{proof}
   We show
   \begin{equation} \label{Eqggh}
       \sum\limits_{0 \leq i_0 + i_1 + i_2 + i_3 + i_4 \leq I} \int_1^\infty \int_{\Sp^2_{\pm}} v_+^{q+2} |\rd_{v_+}^{i_0} \rd_r^{i_4}{Z}_{1,+}^{i_1}{Z}_{2,+}^{i_2}{Z}_{3,+}^{i_3} \big(e^{\pm2i\varphi_+}\mathfrak{f}_0\big)|^2 \, \vols \ud  v_+ \ls 1\;,
   \end{equation}
where $\Sp_{+}$ denotes the northern hemisphere $\{0 \leq \theta \leq \frac{\pi}{2}\} \subseteq \Sp^2$ and $\Sp_-$ the southern hemisphere $\{\frac{\pi}{2} \leq \theta \leq \pi\} \subseteq \Sp^2$. Recalling \eqref{EqCommuteExpZ} this then establishes the claim.

   We further note that $e^{\pm 2 i \varphi_+} \mathfrak{f}_0 = e^{\pm 2i(\overline{r} + \mathfrak{h})} e^{\pm 2i \varphi_*}\mathfrak{f}_0$. Since $\overline{r} + \mathfrak{h}$ is a smooth function up to the event horizon by Lemma \ref{Lem*+PhiSmooth}, it suffices to show \eqref{Eqggh} with $e^{\pm2i\varphi_+}\mathfrak{f}_0$ replaced by $e^{\pm2i\varphi_*}\mathfrak{f}_0$. We now recall \eqref{EqUsedForDefLinear.1}, \eqref{EqUsedForDefLinear.2} from the proof of Proposition \ref{prop:Teukolsy.error.RS}. Noting the equivalence between the sets of derivatives $\{\rd_{v_+}, \rd_r, Z_{i,+}\}$ and $\{\pmb{\dDel}, \pmb{\dD}, \pmb{\ddel}^{(\cdot)}, \overline{\pmb{\ddel}}^{(\cdot)}\}$ this gives 
   \begin{equation*} \begin{split}
        \sum\limits_{0 \leq i_0 + i_1 + i_2 + i_3 + i_4 \leq I} &\int_1^\infty \int_{\Sp^2_{\pm}} v_+^{q+2} |\rd_{v_+}^{i_0} \rd_r^{i_4}{Z}_{1,+}^{i_1}{Z}_{2,+}^{i_2}{Z}_{3,+}^{i_3} \big(e^{\pm2i\varphi_*}\mathfrak{f}_0\big)|^2 \, \vols \ud  v_+  \\
        &\lesssim \int_1^\infty v_+^{q+2} \mathfrak{w}^4(v_+) \, \ud v_+ \ls ||v_+^{\frac{q_- - 2}{2}} \mathfrak{w}(v_+)||^2_{L^\infty(v_+)} \int_1^\infty v_+^{q+2 - q_- + 2} \mathfrak{w}^2(v_+) \, \ud v_+ \ls 1\;,
        \end{split}
   \end{equation*}
   where we have used $q - q_- + 4 \leq q_- -2$, since $q \geq 7$. This concludes the proof.
\end{proof}

We want to define $\psil$ by prescribing initial data $\psil|_{\Hp} := \psi_0|_{\Hp}$ on $\Hp$ as well as on $\{f_+ = v_1\}$, where $v_1 \geq 1$ is large. The initial data prescribed on $\{f_+ = v_1\}$ determines the asymptotic behavior of $\rd_r \psil$ and $\rd_r^2 \psil$ along the event horizon $\Hp$. Since the linear Teukolsky equation exhibits a blue shift effect along $\Hp$ (which turns into a no-shift after one commutation with $\rd_r$), there exists exactly one prescription of $\rd_r \psil|_{\Hp \cap \{v_+ = v_1\}}$ and $\rd_r^2 \psil|_{\Hp \cap \{v_+ = v_1\}}$ such that transversal derivatives to $\Hp$ of $\psil$ decay towards the future. The angular regularity of this prescription depends on the number of derivatives of the dynamical geometry for which decay bounds along $\Hp$ are assumed. Since we only assume decay for a finite number of derivatives along $\Hp$ in \eqref{EqAsDecRest}, the  regularity of the prescription $\rd_r \psil|_{\Hp \cap \{v_+ = v_1\}}$ and $\rd_r^2 \psil|_{\Hp \cap \{v_+ = v_1\}}$ will also only be finite. This then translates into finite regularity of the linear Teukolsky field $\psil$. We will now give the precise construction. Partial derivatives will be with respect to the $(v_+, r, \theta, \varphi_+)$-coordinate system in this section unless explicitly stated otherwise. 

\textbf{Assumption 1:} assume that $\psil$ is a $C^I$-regular spin $2$-weighted solution of $\pmb{\mathfrak{T}}_{[2]}\psil = 0$ in $(\MK \cup \Hp) \cap \{f_+ \geq v_1\}$ satisfying $\psil|_{\Hp} = \psi_0|_{\Hp}$. We set $$\psi := \psi_0 - \psil \;.$$ 
Recall $ 2 \Sigma \cdot \pmb{\mathfrak{T}}_{[2]} = \pmb{\mathcal{T}}_{[2]}$ and thus we  rewrite \eqref{EqDiffTeuk1} as
\begin{equation}\label{EqTeukForTransversalDerivative}
    \pmb{\mathcal{T}}_{[2]} \psi = 2 \Sigma \cdot \big[ (\pmb{\mathfrak{T}}_{[2]} -\mathfrak{T}_{[2]}) \psi_0 -   \mathfrak{N} \big] \;.
\end{equation}
Since we have $\psi|_{\Hp} = 0$, it follows from the expression \eqref{EqTeukolskyStarCoordinates} that $$\pmb{\mathcal{T}}_{[2]} \psi|_{\Hp} = 2 [ (r_+^2 + a^2) \rd_{v_+} + a \rd_{\varphi_+}] \rd_r \psi|_{\Hp} - 2(r_+ - M) \rd_r \psi|_{\Hp} \;.$$ 
We introduce $\hat{\varphi}_+ := \varphi_+ + \frac{a}{r_+^2 + a^2} v_+$. 
We now introduce the vector field $X:=\rd_{v_+} + \frac{a}{r_+^2 + a^2} \rd_{\varphi_+} = \f{\rd}{\rd v_+}\Big|_{(v_+, \theta, \hat{\varphi}_+)}$ tangent to $\Hp$ and also note that $\kappa_+ = \frac{r_+ - M}{r_+^2 + a^2}$.
Thus, restricting \eqref{EqTeukForTransversalDerivative} to $\Hp$ gives the ODE
\begin{equation}\label{EqXEqFirst}
    X \rd_r \psi|_{\Hp} - \kappa_+ \rd_r \psi|_{\Hp} = \mathfrak{f}_0 \;.
\end{equation}

The solution of the ODE \eqref{EqXEqFirst} is given, in $(v_+,\theta, \hat{\varphi}_+)$ coordinates\footnote{But partial derivatives are still with respect to the $(v_+,r, \theta, \varphi_+)$ coordinate system by our convention in this section!}, by
\begin{equation} \label{EqSolFirstDerPsi}
        \rd_r \psi (v_+, \theta, \hat{\varphi}_+) ) = e^{\kappa_+ (v_+ - v_1)} \Big[  \rd_r \psi (v_1, \theta, \hat{\varphi}_+ ) + \int_{v_1}^{v_+} e^{- \kappa_+ (v' - v_1)} \mathfrak{f}_0( v', \theta, \hat{\varphi}_+) \, \ud v'\Big]\;. 
\end{equation}
Note that in general the solution will blow up.

\textbf{Assumption 2:} assume that the initial value of $\rd_r \psil$ is given by 
\begin{equation*}
    \rd_r \psil(v_1,\theta, \hat{\varphi}_+) = \rd_r \psi_0 (v_1, \theta, \hat{\varphi}_+) + \int_{v_1}^\infty e^{-\kappa_+ (v' - v_1)} \mathfrak{f}_0(v', \theta, \hat{\varphi}_+) \ud v'\;.
\end{equation*}
Then the solution \eqref{EqSolFirstDerPsi} becomes
\begin{equation} \label{EqSolFirstDerPsiWithIV}
     \rd_r \psi \big(v_+, \theta, \hat{\varphi}_+ \big) = - \int_{v_+}^\infty e^{-\kappa_+ (v'-v_+)} \mathfrak{f}_0(v', \theta, \hat{\varphi}_+)  \, \ud v' \;.
\end{equation}
\begin{lemma} \label{LemDecayRdRPsi}
    Let $\rd_r \psi$ be defined on $\Hp \cap \{v_+ \geq v_1\}$ by \eqref{EqSolFirstDerPsiWithIV}. We then have $\rd_r \psi \in \mathscr{I}^I_{[2]}(\Hp \cap \{v_+ \geq v_1\})$ and
    $$\sum\limits_{0 \leq i_0 + i_1 + i_2 + i_3 \leq I} \int_{v_1}^\infty \int_{\Sp^2} v_+^{q+2}|\rd_{v_+}^{i_0} \widetilde{Z}_{1_+}^{i_1}\widetilde{Z}_{2_+}^{i_2}\widetilde{Z}_{3_+}^{i_3} (\rd_r \psi) |^2 \vols \ud v_+ \lesssim 1 \;.$$
\end{lemma}

\begin{proof}
This follows easily from Lemmas \ref{LemBlueDecayPoly} and \ref{LemDecayBoundsf0} by replacing the set of derivatives $\rd_{v_+}$, $\wtZ_{i,+}$ by $\frac{\rd}{\rd v_+}\Big|_{(v_+, \theta, \hat{\varphi}_+)}$ and $\wtZ_{i,\hat{+}}\Big|_{(v_+, \theta, \hat{\varphi}_+)}$ and noting that these sets of derivatives are comparable. For the first claim one also uses a standard Sobolev embedding. \qedhere

\end{proof}

To derive the propagation equation for $\rd_r^2 \psi$ along $\Hp$ we differentiate \eqref{EqTeukolskyStarCoordinates} with respect to $\rd_r$ to obtain
\begin{equation*}
\begin{split}
\partial_r \pmb{\mathcal{T}}_{[s]} \psi &= a^2 \sin^2 \theta \rd_{v_+}^2 \rd_r \psi + 2a \rd_{v_+} \rd_{\varphi_+} \rd_r \psi + 2(r^2 + a^2) \rd_{v_+} \rd_r^2 \psi + 2a\rd_{\varphi_+} \rd_r^2 \psi + \Delta \rd_r^3 \psi\\
&\quad  + \swl \rd_r \psi +2\big(r(3-2s) - isa \cos \theta\big) \rd_{v_+} \rd_r \psi + 2(r-M)(2-s) \rd_r^2 \psi \\
&\quad +2(1-2s) \rd_r \psi +2(1-2s)\rd_{v_+} \psi \;.
\end{split}
\end{equation*}
Restricting this expression to $\Hp$ we get
\begin{equation*}
        \frac{1}{2(r_+^2 + a^2)} \partial_r \pmb{\mathcal{T}}_{[2]} \psi = X \rd_r^2 \psi|_{\Hp} + \mathfrak{f}_2 \;,
\end{equation*}
where
\begin{equation}\label{EqDefMathfrakF2}
    \mathfrak{f}_2 := \frac{1}{2(r_+^2 + a^2)} \Big( a^2 \sin^2 \theta \rd_{v_+}^2 \rd_r \psi|_{\Hp} + 2a \rd_{v_+} \rd_{\varphi_+} \rd_r \psi|_{\Hp} + \swlt \rd_r \psi|_{\Hp} -2(r_+ + 2ia \cos \theta) \rd_{v_+} \rd_r \psi|_{\Hp} - 6 \rd_r \psi|_{\Hp} \Big) \;.
\end{equation}
Hence, differentiating \eqref{EqTeukForTransversalDerivative} in $\rd_r$ and restricting it to $\Hp$ gives the ODE
\begin{equation*} 
    X \rd_r^2 \psi|_{\Hp} + \mathfrak{f}_2 = \mathfrak{f}_1 \;.
\end{equation*}
The solution, in $(v_+, \theta, \hat{\varphi}_+)$ coordinates is given by
\begin{equation} \label{EqSolXSecDer}
    \rd_r^2\psi|_{\Hp} (v_+, \theta, \hat{\varphi}_+) = \rd_r^2\psi|_{\Hp} (v_1, \theta, \hat{\varphi}_+) + \int_{v_1}^{v_+} (\mathfrak{f}_1 - \mathfrak{f}_2) (v', \theta, \hat{\varphi}_+) \, \ud v' \;.
\end{equation}
\textbf{Assumption 3:} assume that the initial value for $\rd_r^2 \psil$ is given by
$$\rd_r^2 \psil (v_1,  \theta, \hat{\varphi}_+) = \rd_r^2 \psi_0 (v_1, \theta, \hat{\varphi}_+) + \int_{v_1}^{\infty} (\mathfrak{f}_1 - \mathfrak{f}_2) (v', \theta ,\hat{\varphi}_+) \, \ud v' \;.$$
Then the solution \eqref{EqSolXSecDer} becomes
\begin{equation} \label{EqSecondSolWithIV}
     \rd_r^2\psi|_{\Hp} \big(v_+, \theta, \hat{\varphi}_+\big) =-\int_{v_+}^{\infty} (\mathfrak{f}_1 - \mathfrak{f}_2)  (v', \theta, \hat{\varphi}_+) \, \ud v' \;.
\end{equation}

\begin{lemma} \label{LemDecaySecondRdRPsi}
    Let $\rd_r \psi$ be defined on $\Hp \cap \{v_+ \geq v_1\}$ by \eqref{EqSolFirstDerPsiWithIV} and $\mathfrak{f}_2$ be defined in terms of this $\rd_r 
\psi$ according to \eqref{EqDefMathfrakF2}. Moreover, let $\rd_r^2 \psi$ be defined on $\Hp \cap \{v_+ \geq v_1\}$ by \eqref{EqSecondSolWithIV} with this choice of $\mathfrak{f}_2$. We then have $\rd_r^2 \psi \in \mathscr{I}^{I}_{[2]}(\Hp \cap \{v_+ \geq v_1\})$ and
$$\sum\limits_{0 \leq i_0 + i_1 + i_2 + i_3 \leq I} \int_{v_1}^\infty \int_{\Sp^2} v_+^q|\rd_{v_+}^{i_0} \widetilde{Z}_{1_+}^{i_1}\widetilde{Z}_{2_+}^{i_2}\widetilde{Z}_{3_+}^{i_3}  (\rd_r^2 \psi)|^2 \vols \ud v_+ \lesssim 1 \;.$$
\end{lemma}

\begin{proof}
 This follows similarly to the proof of Lemma \ref{LemDecayRdRPsi}, this time from Lemmas \ref{LemNoShiftDecay}, \ref{LemDecayBoundsf0}, and Lemma \ref{LemDecayRdRPsi} combined with \eqref{EqDefMathfrakF2}.
\end{proof}

Finally, we need the following
\begin{lemma} \label{LemGoOverToDynamicalFrame}
    The bounds \eqref{EqAsDecAlpha}, \eqref{EqAsInst2}, \eqref{EqAsInst3}, \eqref{EqAsInst4}  hold with $R(\pmb{e}_4, \pmb{m}, \pmb{e}_4, \pmb{m})$ replaced by $\psi_0$ and $I_0$ replaced by some $10^{10} \leq I \leq I_0$.
\end{lemma}

\begin{proof}
The statement reduces in a similar manner as in the proof of Lemma \ref{LemDecayBoundsf0} to the statement 
\begin{equation} \label{EqForTransitionIDPsi0}
    \sum_{|j| \leq I}  \int\limits_1^\infty \int\limits_{\Sp^2_\pm} \ub^{q} |\pmb{\calD}^j\big[e^{\pm 2i\varphi_*}\big(R(\pmb{e}_4, \pmb{m}, \pmb{e}_4, \pmb{m}) - R(e_4,m,e_4,m)\big)\big]|^2 \volg \ud \ub \ls 1 \;,
\end{equation}
where $\Sp^2_{\pm}$ again denote the two hemispheres and $\pmb{\calD} \in \{ \pmb{\dD}, \pmb{\dDel}, \pmb{\ddel}^{(\cdot)}, \ov{\pmb{\ddel}}^{(\cdot)} \}$. 
We now note that
\begin{equation*}
\begin{split}
    \big|\pmb{\calD}^j\big( R(e_4, & m^{(\cdot )}, e_4, m^{(\cdot)}) - R(\pmb{e}_4, \pmb{m}^{(\cdot)}, \pmb{e}_4, \pmb{m}^{(\cdot)})\big) \big| \\
    &\lesssim \big|\pmb{\calD}^j R(e_4 - \pmb{e}_4,  m^{(\cdot )}, e_4, m^{(\cdot)})| + \big|\pmb{\calD}^jR(e_4, m^{(\cdot)}
     - \pmb{m}^{(\cdot)}, e_4, m^{(\cdot)})\big| + \mathfrak{w}^2(\ub) \\
     &\ls \big| \pmb{\calD}^j \big( (\calE^{(\cdot)})_4^\beta R(e_\beta^{(\cdot)}, m^{(\cdot )}, e_4, m^{(\cdot)}) \big) \big| + \big|\pmb{\calD}^j\big(\frac{\mathfrak{c}}{\sqrt{2}} [ (\calE^{(\cdot)})_1^\beta + i ((\calE^{(\cdot)})_2^\beta ] R(e_4, e_\beta^{(\cdot)}, e_4, m^{(\cdot)})\big)\big| + \mathfrak{w}^2(\ub)\;,
     \end{split}
\end{equation*}
where we have used $e_\alpha^{(\cdot)} - \pmb{e}_\alpha^{(\cdot)} = \big[ \delta_\alpha^\beta - (\pmb{\calE}^{(\cdot)})_\alpha^\beta \big]  e_\beta^{(\cdot)}$ with $|\delta_\alpha^\beta - (\pmb{\calE}^{(\cdot)})_\alpha^\beta| \ls \mathfrak{w}(\ub)$ by Proposition \ref{prop:RS.principal.null}. Furthermore, we note that the corresponding background quantities $\pmb{R}(\pmb{e}_\beta^{(\cdot)}, \pmb{m}^{(\cdot)}, \pmb{e}_4, \pmb{m}^{(\cdot)})$ and  $\pmb{R}(\pmb{e}_4, \pmb{e}_\beta^{(\cdot)}, \pmb{e}_4, \pmb{m}^{(\cdot)})$ vanish so that also the first two summands above are $\ls \mathfrak{w}^2(\ub)$ by Proposition \ref{prop:RS.curvature.principal.null}. The claim \eqref{EqForTransitionIDPsi0} now follows as in the last line of the proof of Lemma \ref{LemDecayBoundsf0}.
\end{proof}

We are now ready to formally define the linear Teukolsky field $\psil$ used in this paper:

\begin{definition} \label{DefLinearTeukField}
    We define the linear Teukolsky field $\psil$ in $(\MK \cup \Hp) \cap \{f_+ \geq v_1\}$ by solving the mixed characteristic initial value problem for the linear Teukolsky equation $\pmb{\mathfrak{T}}_{[2]}\psil = 0$ (see \cite[Appendix A]{Sbie23}) with initial data given on $\Hp \cap \{f_+ \geq v_1\}$ by $\psil|_{\Hp} := \psi_0|_{\Hp}$ and an arbitrary choice of $C^{I}$-regular spin $2$-weighted initial data on $\{f_+ = v_1\}$ such that $\rd_r \psil|_{\Hp \cap \{\{f_+ = v_1\}}$ and $\rd_r^2 \psil|_{\Hp \cap \{\{f_+ = v_1\}}$ are given by
\begin{align}
    \rd_r \psil(v_1,r_+,\theta, \hat{\varphi}_+) &= \rd_r \psi_0 (v_1, r_+, \theta, \hat{\varphi}_+) +  \int_{v_1}^\infty e^{-\kappa_+ (v'- v_1)} \mathfrak{f}_0(v', \theta, \hat{\varphi}_+) \, \ud v'  \label{EqDefFirstTransversalDerivativePsi}\\
    \rd^2_r \psil(v_1,r_+,\theta, \hat{\varphi}_+) &=   \rd_r^2 \psi_0 (v_1, r_+, \theta, \hat{\varphi}_+) + \int_{v_1}^{\infty} (\mathfrak{f}_1 - \mathfrak{f}_2) (v', \theta, \hat{\varphi}_+) \, \ud v' \;, \label{EqDefSecondTransversalDerivativePsi}
\end{align}
where in \eqref{EqDefSecondTransversalDerivativePsi} the function $\mathfrak{f}_2$ is given by \eqref{EqDefMathfrakF2} with $\rd_r\psi|_{\Hp}$ in this expression defined by \eqref{EqSolFirstDerPsiWithIV}. Note that in this way the right-hand side of \eqref{EqDefSecondTransversalDerivativePsi} is defined just in terms of $\mathfrak{f}_0$ and $\mathfrak{f}_1$.
\end{definition}
\begin{remark}
    Note that in general one cannot set $\psil|_{\{f_+ = v_1\}} := \psi_0|_{\{f_+ = v_1\}}$. This would in general lead to a solution to the linear Teukolsky equation whose derivatives transversal to $\Hp$ blow up in time.
\end{remark}
In order to see that Definition~\ref{DefLinearTeukField} is  well-defined, we  check that the right-hand sides of \eqref{EqDefFirstTransversalDerivativePsi} and \eqref{EqDefSecondTransversalDerivativePsi} are $C^{I}$-regular spin $2$-weighted functions on $\Hp \cap \{f_+ = v_1\}$ such that the construction of an arbitrary $C^{I}$-regular spin $2$-weighted function of $\{f_+ = v_1\}$ can be carried out by an elementary finite power series construction. For \eqref{EqDefFirstTransversalDerivativePsi} this follows from  Lemma \ref{LemDecayBoundsf0} combined with Sobolev embedding. For the integral in \eqref{EqDefSecondTransversalDerivativePsi} this follows from  Lemma \ref{LemDecayBoundsf0} combined with Lemma \ref{LemDecayRdRPsi}.

Having now defined $\psil$, we take stock of the behavior of $\psil$ and $\psi$ (and their transversal derivatives) on $\Hp$ in the next proposition.

\begin{proposition} \label{PropPropLinTeukField}
   The solution $\psil$ of $\pmb{\mathfrak{T}}_{[2]}\psil = 0$ defined in Definition \ref{DefLinearTeukField} is a $C^I$-regular spin $2$-weighted function in $(\MK \cup \Hp) \cap \{f_+ \geq v_1\}$ which satisfies  
\begin{align}
    &\int\limits_{\Hp \cap \{f_+ \geq v_1\}} v_+^q |(\psil)_{l=2}|^2 \vols \ud v_+ = \infty  \label{EqPropLinearTeukHorizon1}\\
   \sum_{\substack{0 \leq i_0 + i_1 + i_2 + i_3 \leq 5 \\ 0 \leq k \leq 2}} \;&\int\limits_{\Hp \cap \{f_+ \geq v_1\}} v_+^{q_-} |\rd_{v_+}^{i_0} \widetilde{Z}^{i_1}_{1,+} \widetilde{Z}^{i_2}_{2,+} \widetilde{Z}^{i_3}_{3,+} \rd_r^k \psil|^2 \vols \ud v_+ \leq C \label{EqPropLinearTeukHorizon2} \\
   \sum_{\substack{0 \leq i_0 + i_1 + i_2 + i_3 \leq 5 \\ 0 \leq k \leq 2}} \;&\int\limits_{\Hp \cap \{f_+ \geq v_1\}} v_+^{q} |\rd_{v_+}^{i_0} \widetilde{Z}^{i_1}_{1,+} \widetilde{Z}^{i_2}_{2,+} \widetilde{Z}^{i_3}_{3,+} \rd_r^k (\rd_{v_+}\psil)|^2 \vols \ud v_+ \leq C \label{EqPropLinearTeukHorizon3}\\
   \sum_{\substack{0 \leq i_0 + i_1 + i_2 + i_3 \leq 5 \\ 0 \leq k \leq 2}} \;&\int\limits_{\Hp \cap \{f_+ \geq v_1\}} v_+^{q} |\rd_{v_+}^{i_0} \widetilde{Z}^{i_1}_{1,+} \widetilde{Z}^{i_2}_{2,+} \widetilde{Z}^{i_3}_{3,+} \rd_r^k (\psil)_{l>2}|^2 \vols \ud v_+ \leq C \;. \label{EqPropLinearTeukHorizon4}
\end{align}
   Moreover, for the difference $\psi = \psi_0 - \psil$ the following bound holds
   \begin{equation} \label{EqPropPsiHorizon} 
   \sum_{k=0}^2 \int\limits_{\mathcal{H}^+ \cap \{f_+ \geq v_1\}}  v_+^{q} d_\Delta(\rd_r^k \psi)\, \vols \ud v_+ \leq C.
   \end{equation}
\end{proposition}
Here, we have introduced the shorthand $(\psil)_{l=2}$ for $\Sp_{(2)} \psil$ and similarly for $(\psil)_{l > 2}$.

\begin{proof} We begin by recalling that by Lemma \ref{LemGoOverToDynamicalFrame} the assumptions \eqref{EqAsDecAlpha} - \eqref{EqAsInst4} hold with $\psi_0$ instead of $R(\pmb{e}_4, \pmb{m}, \pmb{e}_4, \pmb{m})$ (and with $I_0$ replaced by $I$).
   Since we have $\psil|_{\Hp} = \psi_0|_{\Hp}$, \eqref{EqPropLinearTeukHorizon1} follows directly from assumption \eqref{EqAsInst2}. 
    The other statements follow directly once 
   \begin{equation}\label{EqClaimDefRefField}
   \sum\limits_{\substack{0 \leq i_0 + i_1 + i_2 + i_3 \leq I \\ 0 \leq k \leq 2}} \int_{v_1}^\infty \int_{\Sp^2} v_+^q|\rd_{v_+}^{i_0} \widetilde{Z}_{1_+}^{i_1}\widetilde{Z}_{2_+}^{i_2}\widetilde{Z}_{3_+}^{i_3}  (\rd_r^k \psi)|^2 \vols \ud v_+ \lesssim 1 
   \end{equation}
   is established: \eqref{EqPropPsiHorizon} follows directly and \eqref{EqPropLinearTeukHorizon2}, \eqref{EqPropLinearTeukHorizon3}, \eqref{EqPropLinearTeukHorizon4} follow since their equivalent statements with $\psil$ replaced by $\psi_0$ are satisfied by Lemma \ref{LemGoOverToDynamicalFrame}.\footnote{For \eqref{EqPropLinearTeukHorizon4} we  use $$\widetilde{Z}_{1,+}^{i_1} \widetilde{Z}_{2,+}^{i_2} \widetilde{Z}_{3,+}^{i_3} (\rd_{v_+}^{i_0} \rd_r^k \psi)_{l>2} = \widetilde{Z}_{1,+}^{i_1} \widetilde{Z}_{2,+}^{i_2} \widetilde{Z}_{3,+}^{i_3} (\rd_{v_+}^{i_0} \rd_r^k \psi) - \widetilde{Z}_{1,+}^{i_1} \widetilde{Z}_{2,+}^{i_2} \widetilde{Z}_{3,+}^{i_3} (\rd_{v_+}^{i_0} \rd_r^k \psi)_{l=2}$$
    and $|| \widetilde{Z}_{1,+}^{i_1} \widetilde{Z}_{2,+}^{i_2} \widetilde{Z}_{3,+}^{i_3} (\rd_{v_+}^{i_0} \rd_r^k \psi)_{l=2}||_{L^2(\Sp^2)} \lesssim ||\rd_{v_+}^{i_0} \rd_r^k \psi||_{L^2(\Sp^2)}$, where we have used the smoothness of the spin $2$-weighted spherical harmonics.}

   To show \eqref{EqClaimDefRefField}, we first note that it trivially holds for $k=0$, since $\psi|_{\Hp} = 0$. For $k=1,2$ we revisit the derivation of the bounds for $\rd_r^k \psi$ along $\Hp$ from earlier in this section. By Definition \ref{DefLinearTeukField} Assumption 1 is satisfied and Assumption 2 is satisfied by \eqref{EqDefFirstTransversalDerivativePsi}. Thus Lemma \ref{LemDecayRdRPsi} applies which shows \eqref{EqClaimDefRefField} for $k=1$. Furthermore, Assumption 3 is met by \eqref{EqDefSecondTransversalDerivativePsi} so that Lemma \ref{LemDecaySecondRdRPsi} covers the case $k=2$. 
\end{proof}

\subsubsection{Blow-up result for the linear Teukolsky field from \cite{Sbie23}, \cite{Sbie26}}

\begin{theorem}
    For $\psil$ defined in Definition \ref{DefLinearTeukField} and $\Gamma$ defined in Section \ref{sec:Dafermos} the following bounds hold:
    \begin{align}
        &\int\limits_{\Gamma \cap \{f_+ \geq v_1\}} v_+^q |(\psil)_{l = 2}|^2 \vols \ud v_+ = \infty  \label{EqPropEstLinearL2}\\
        &\int\limits_{\Gamma \cap \{f_+ \geq v_1\}} v_+^q |(\psil)_{l > 2}|^2 \vols \ud v_+ < \infty \label{EqPropEstLinearL>2} \;.
    \end{align}
\end{theorem}

\begin{proof}
    Given the bounds on $\Hp$ stated in Proposition \ref{PropPropLinTeukField}, this follows directly from Theorem 2.2 in \cite{Sbie26}.
\end{proof}

\subsection{Putting everything together} \label{SecConcludingDifference}

\begin{lemma}\label{lem:choice.of.v_1}
    There exists $v_1 \geq 1$ sufficient large such that
    \begin{equation}\label{eq:set.to.show.v_1.large}
        \{f_+ \geq v_1 \}\cap \{f_\Gamma \leq 0\} \subset \Phi(\calM).
    \end{equation}
    In fact, we have the stronger statement that
    \begin{equation}\label{eq:set.to.show.v_1.large.2}
        \{f_+ \geq v_1 \}\cap \{f_\Gamma \leq 0\} \subset \Phi(\os{\calM}{1} \cap \{s \leq \f{3 s_f}{4}\}) \cup \Phi(\os{\calM}{2} \setminus\{U \cap \{s \leq \f{s_f}{2}\}).
    \end{equation}
\end{lemma}
\begin{proof}
    Denote the set on the left-hand side of \eqref{eq:set.to.show.v_1.large} by $D$. For either claim, it suffices to show that for $v_1$ sufficiently large, the following two statements hold:
    \begin{itemize}
        \item $\ub > 1$ on $D$: Recall from the beginning of Section~\ref{SecInhomLinTeuk} that $f_+ := v_+ -r + r_+$. Since $r$ is  bounded, by Lemma~\ref{LemRelationStarAndNullCoord}, we can choose $v_1$ large so that $\ub \geq 1$ on $D$.
        \item $u < u_f$  on $D$: Recall $f_\Gamma(v_+, v_-) := v_+ + v_- - \frac{\sigma_q}{\kappa_-} \log (v_+)$ from the beginning of Section~\ref{sec:Dafermos}. Using the definitions of $f_+$ and $f_\Gamma$, we have that $f_\Gamma \leq 0$ implies
        $$v_- \leq -f_++r-r_+ - \f{\sigma_q}{\kappa_-} \log (f_++r-r_+),$$
        with the right-hand side $\to -\infty$ as $f_+\to \infty$. Thus, choosing $v_1$ sufficiently large, we have $v_-$ arbitrarily negative and we can conclude with Lemma~\ref{LemRelationStarAndNullCoord}. \qedhere
    \end{itemize} 
\end{proof}

\begin{theorem} \label{ThmPropagationToGamma}
    Let $\psi_0$ be the pull-back of the corresponding dynamical Teukolsky field on $\calM$ to $\calM_{\mathrm{Kerr}}$ with the map $\Phi$, and $\psil$ be as in Definition~\ref{DefLinearTeukField}. Then, the following estimates hold for $v_1$ chosen as in Lemma~\ref{lem:choice.of.v_1}:
    \begin{equation} \label{EqThmPropToGamma}
        \int\limits_{\Gamma \cap \{f_+ \geq v_1\}} v_+^q |\psi_0 - \psil|^2 \vols \ud v_+ \leq C \;.
    \end{equation}
\end{theorem}

\begin{proof}
    We apply Proposition~\ref{PropInhomTeukI} and Proposition~\ref{PropInhomTeukII} to the equation $\pmb{\mathcal{T}}_{[2]}\psi =:  F$ for $\psi = \psi_0 - \psil$, where $F$ is given by \eqref{EqDiffTeuk1}. When applying Proposition~\ref{PropInhomTeukI}, the first term
    $$\sum_{k=0}^2 \int\limits_{\Hp \cap \{f_+ \geq v_1\}}  v_+^{q} d_\Delta(\rd_r^k \psi)\, \vols \ud v_+$$
    on the right-hand side of \eqref{EqPropRightH} is bounded in \eqref{EqPropPsiHorizon}. When applying Proposition~\ref{PropInhomTeukII}, the first term 
    $$\int\limits_{\{r= \rred\} \cap \{f_+ \geq v_1\}} v_+^q d(\psi) \, \vols \ud v_+$$
    on the right-hand side of \eqref{EqPropGamma} is controlled by the conclusion of Proposition~\ref{PropInhomTeukI}. Thus, in order to obtain the bound \eqref{EqThmPropToGamma}, it suffices to estimate the term
    \begin{equation}\label{eq:final.main.term.1}
        \int\limits_{\{\rred \leq r \leq r_+\} \cap \{f_+ \geq v_1\}} v_+^{q} |\rd_r^2F|^2 \,\vols\, \ud v_+ \ud r
    \end{equation}
    from \eqref{EqPropRightH} and the term
    \begin{equation}\label{eq:final.main.term.2}
        \int\limits_{\substack{\{f_\Gamma \leq 0\} \cap \{r \leq \rred\} \\ \cap \{f_+ \geq v_1\}} }v_+^q|F|^2 \, \vols \ud v_+ \ud r 
    \end{equation}
    from \eqref{EqPropGamma}.

    We have essentially already bounded these terms. We now make three observations that allow us to use the previously established estimates to bound the terms \eqref{eq:final.main.term.1}--\eqref{eq:final.main.term.2}.
    \begin{enumerate}
        \item Regions of integration: 
        Define the integration regions by $\os{\calR}{1} := \{\rred \leq r \leq r_+\} \cap \{f_+ \geq v_1\}$ and $\os{\calR}{2} := \{f_\Gamma \leq 0\} \cap \{r \leq \rred\} \cap \{f_+ \geq v_1\} $.
        By Lemma~\ref{lem:choice.of.v_1}, we know that $\os{\calR}{1}\cup \os{\calR}{2}\subset \Phi(\os{\calM}{1}\cap \{s \leq \f{3 s_f}{4}\})\cup \Phi(\os{\calM}{2} \setminus(\calU \cap \{ s \leq \f{s_f}{2}\}))$. We thus split the integration region in \eqref{eq:final.main.term.1} into (the overlapping regions) $\Phi(\os{\calM}{1}\cap \{s \leq \f{3 s_f}{4}\}) \cap \os{\calR}{1}$ and $\Phi(\os{\calM}{2} \setminus(\calU \cap \{ s \leq \f{s_f}{2}\}))\cap \os{\calR}{1}$, and the integration region in \eqref{eq:final.main.term.2} into $\Phi(\os{\calM}{1}\cap \{s \leq \f{3 s_f}{4}\})\cap \os{\calR}{2}$ and $\Phi(\os{\calM}{2} \setminus(\calU \cap \{ s \leq \f{s_f}{2}\})) \cap \os{\calR}{2}$. An important observation is the following:
        \begin{equation}\label{eq:red.blue.intersection}
            \hbox{In the intersection $\Phi(\os{\calM}{2})\cap \os{\calR}{1}$, we must have $u+\ub \leq C_{\mathrm{red}}$ for some $C_{\mathrm{red}} \in \mathbb R$.}
        \end{equation}
        This is true because of translation invariance: Notice that the sets $\{r \leq \hbox{const}\}$, $\{r \geq \hbox{const}\}$, $\{u+\ub \leq \hbox{const}\}$, $\{u+\ub \geq \hbox{const}\}$ are all invariant sets for the vector field $\f{\rd}{\rd u}\Big|_{DN} + \f{\rd}{\rd \ub}\Big|_{DN}$. The desired conclusion thus follows a compactness argument as in Lemma~\ref{lem:can.initialize}.

        \item Volume forms: For the integration in $\Phi(\os{\calM}{1}\cap \{s \leq \f{3 s_f}{4}\})$, we use that
        \begin{equation}\label{eq:vol.form.comp.1}
            \,\vols\, \ud v_+ \ud r \ls \,\vol_{\pmb{\gamma}}\, \ud \ub \ud s \ls \,\vol_{\gamma}\, \ud \ub \ud s.    
        \end{equation}

        For the integration in $\Phi(\os{\calM}{2}\setminus(\calU \cap \{ s \leq \f{s_f}{2}\}))$, we use that
        \begin{equation}\label{eq:vol.form.comp.2}
            \,\vols\, \ud v_+ \ud r \ls |\Delta|\,\vol_{\pmb{\gamma}}\, \ud u\ud \ub \ls |\Delta|\,\vol_{\gamma}\, \ud u\ud \ub.    
        \end{equation}

        The first inequalities in \eqref{eq:vol.form.comp.1} and \eqref{eq:vol.form.comp.2} are explicit Kerr computations that can be carried out using the definition of the coordinate functions. The second inequalities follow from the estimates of $\gamma - \pmb{\gamma}$ in Theorem~\ref{thm:DL.2} and Proposition~\ref{prop:DL.rescaled}.

        \item Weights: In view of Lemma~\ref{LemRelationStarAndNullCoord}, the $v_+^q$ weights in the integrals \eqref{eq:final.main.term.1} and \eqref{eq:final.main.term.2} can be replaced by $\ub^q$ weights.
        
        \item Rotation of frames: We have
        \begin{align}
            |F|^2 = |e^{2i\varphi_*} F|^2 = 4\Sigma^2 | (\pmb{\mathfrak{T}}_{[2]}^{(N)} -\mathfrak{T}_{[2]}^{(N)}) \psi_0^{(N)} + \mathfrak{N}^{(N)}|^2\ls |\pmb{\mathfrak{T}}_{[2]}^{(N)} -\mathfrak{T}_{[2]}^{(N)}) \psi_0^{(N)}|^2 + | \mathfrak{N}^{(N)}|^2,\label{eq:final.Teukolsky.F.rotate.1} \\
            |(\rd_r |_+)^2 F|^2 \ls \sum_{j\leq 2} \Big( \Big| \pmb{\dDel}^j \Big((\pmb{\mathfrak{T}}_{[2]}^{(N)} -\mathfrak{T}_{[2]}^{(N)}) \psi_0^{(N)}\Big) \Big|^2 + \Big| \pmb{\dDel}^j \mathfrak{N}^{(N)}\Big|^2 \Big), \label{eq:final.Teukolsky.F.rotate.2}
        \end{align}
        where in \eqref{eq:final.Teukolsky.F.rotate.1}, we used the equation \eqref{EqDiffTeuk1} and Lemma~\ref{LemTrafoNPFrames.2}, and in \eqref{eq:final.Teukolsky.F.rotate.2}, we additionally used Lemma~\ref{lem:rotate.F.with.derivatives}.
        

        We also obtain a similar bound where $^{(N)}$ on the right-hand sides are replaced by $^{(S)}$.
        \item Covering by northern and southern portions: By definition (see beginning of Section~\ref{SecEstNP}), $\os{\calM}{1}{}^{(N)}\cup \os{\calM}{1}{}^{(S)} = \os{\calM}{1}$ and $\os{\calM}{2}{}^{(N)}\cup \os{\calM}{2}{}^{(S)} = \os{\calM}{2}$. We can thus further split the integrals into the northern and southern portions and use \eqref{eq:final.Teukolsky.F.rotate.1}--\eqref{eq:final.Teukolsky.F.rotate.2}.
        \end{enumerate}

        We now put everything together. With the above considerations, in order to bound \eqref{eq:final.main.term.1} and \eqref{eq:final.main.term.2}, it suffices to estimate
        \begin{equation*}
            \begin{split}
                I:=&\: \int_{\Phi(\os{\calM}{1}{}^{(\cdot)} \cap \{s \leq \f{3 s_f}{4}\})\cap \os{\calR}{1}} \ub^q |(\rd_r|_+)^2 F|^2 \,\vol_{\gamma}\, \ud \ub \ud s, \\
                II:=&\: \int_{\Phi(\os{\calM}{1}{}^{(\cdot)}\cap \{s \leq \f{3 s_f}{4}\}) \cap \os{\calR}{2}} \ub^q|F|^2 \,\vol_{\gamma}\, \ud \ub \ud s,\\
                III :=&\: \int_{\Phi(\os{\calM}{2}{}^{(\cdot)} \cap (\calU \setminus\{s\leq \f{s_f}{2} \}))\cap \os{\calR}{1}} \ub^q|(\rd_r|_+)^2 F|^2 |\Delta|\,\vol_{\gamma}\, \ud u\ud \ub,\\ IV := &\:\int_{\Phi(\os{\calM}{2}{}^{(\cdot)} \cap (\calU \setminus\{s\leq \f{s_f}{2} \}))\cap \os{\calR}{2}} \ub^q |F|^2|\Delta|\,\vol_{\gamma}\, \ud u\ud \ub,
            \end{split}
        \end{equation*}
        for $^{(\cdot)} = ^{(N)}, ^{(S)}$ separately.

        We bound first $F$ and $\rd_r|_+ F$ by \eqref{eq:final.Teukolsky.F.rotate.1}--\eqref{eq:final.Teukolsky.F.rotate.2} or its southern equivalent depending on whether we are integrating on $\Phi(\calM^{(N)})$ or $\Phi(\calM^{(S)})$. Then after pulling back to $\calM$, the terms $I$ and $II$ are bounded using Proposition~\ref{prop:Teukolsy.error.RS}, the term $III$ is bounded by Proposition~\ref{prop:Teukolsy.error.BS.easy} after using \eqref{eq:red.blue.intersection}, and finally the term $IV$ is bounded using Proposition~\ref{prop:nonlinear.Teukolsky.error.1} and Proposition~\ref{prop:nonlinear.Teukolsky.error.2}.  \qedhere
\end{proof}

\section{Propagation from $\Gamma$ to the Cauchy horizon}\label{sec:final}

In this final section we propagate the  blow-up bound from Theorem \ref{ThmPropagationToGamma} on the hypersurface $\Gamma$  to the Cauchy horizon $\CH^+$ to prove Theorem \ref{ThmInextConditionVerified}. The propagation is carried out along integral curves of $e_3'$.  Theorem \ref{ThmInextConditionVerified} is phrased in terms of the $(u, \ub_{\CH^+}, \vartheta_{\CH^+})$ coordinates. Propagating a neighborhood in this coordinate system back to $\Gamma$ along $e_3'$ leads to a winding behavior in $\varphi_+$ on $\Gamma$ in terms of the $(v_+, r, \theta, \varphi_+)$ coordinates. This behavior is described in Section \ref{SecWinding} in terms of a new coordinate $\varphi_=$. The actual propagation of the blow-up bound is then carried out in Section \ref{SecPropagation} and the proof of Theorem \ref{ThmInextConditionVerified} in Section \ref{SecProofMainThm}.


\subsection{Projecting a subset of $\{u = \mathrm{const}\}$ in $\{\ub, \vartheta_{\CH^+}\}$ coordinates along the integral curves of $e_3'$ onto $\Gamma$ in $(v_+, \vartheta_=)$ coordinates} \label{SecWinding}

Recall that $\Gamma = \{v_+ + v_- - \frac{\sigma_q}{\kappa_-} \log v_+ = 0\}$ with $\sigma_q = \frac{1}{4}(q+3)$. Since $v_+ + v_- = 2r^*$, the intersection of $\Gamma$ with a level set of $v_+$ has constant value of $r$. It thus follows that this intersection agrees with a Boyer--Lindquist sphere. It can be parameterized by the $\vartheta_+ := (\theta, \varphi_+)$-coordinates but also by the $\vartheta_= :=(\theta, \varphi_=)$-coordinates, where 
\begin{equation} \label{EqDefTildePhiMinus}
\varphi_= := \varphi_+ - \frac{a}{r_-^2 + a^2} v_+ = \varphi + \overline{r} - \frac{a}{r_-^2 + a^2} (t+ r^*) \quad \textrm{mod}\;  2 \pi\,.
\end{equation}
We can thus parameterize $\Gamma \cap \{f_+ \geq v_1\}$ by $(v_+, \vartheta_+)$ coordinates such that $\Gamma \cap \{f_+ \geq v_1\} \simeq [v_{1+}, \infty) \times \Sp^2$ or by $(v_+, \vartheta_=)$ coordinates such that $\Gamma \cap \{f_+ \geq v_1\} \simeq [v_{1+}, \infty) \times \Sp^2$. 
The main result of this subsection is Proposition \ref{PropProjectionUToGamma}, which is needed for the proof of Theorem \ref{ThmBlowUpCH}. In the following we introduce various maps which, when composed, give the projection map from $\{u = \mathrm{const}\}$ in $\{\ub, \vartheta_{\CH^+}\}$ coordinates along the integral curves of $e_3'$ onto $\Gamma$ in $(v_+, \vartheta_=)$ coordinates.

\subsubsection{Relation between Pretorius-Israel coordinates $(u, \ub, \theta_*, \varphi_*)$ and $(v_-, v_+,  \theta, \varphi_=)$ coordinates: the map $\mathscr{T}$}

We denote partial derivatives with respect to the $(u, \ub, \theta_*, \varphi_*)$ coordinate system by $|_{DN}$, partial derivatives with respect to the $(t, r_*, \theta_*, \varphi_*)$ coordinate system by $|_*$, partial derivatives with respect to the Boyer--Lindquist coordinate system by $|_{BL}$, and partial derivatives with respect to the $( v_-, v_+, \theta, \varphi_=)$ coordinate system by $|_\pm$.

Since we have $\ub = \frac{1}{2}(r_* +t)$ and $u = \frac{1}{2}(r_* -t)$, it follows that
\begin{equation*}
    \frac{\rd}{\rd \ub}\Big|_{DN} = \frac{\rd}{\rd r_*}\Big|_* + \frac{\rd}{\rd t}\Big|_*\,, \qquad \frac{\rd}{\rd u}\Big|_{DN} = \frac{\rd}{\rd r_*}\Big|_* - \frac{\rd}{\rd t}\Big|_*\,, \qquad \frac{\rd}{\rd \theta_*}\Big|_{DN} = \frac{\rd}{\rd \theta_*}\Big|_*\,, \qquad \frac{\rd}{\rd \varphi_*}\Big|_{DN} = \frac{\rd}{\rd \varphi_*}\Big|_* \;.
\end{equation*}
Furthermore, using $r(r_*, \theta_*)$, $\theta(r_*, \theta_*)$, and $\varphi = \varphi_* + h(r_*, \theta_*)$, we obtain
\begin{equation*}
\begin{aligned}
    &\frac{\rd}{\rd t}\Big|_* = \frac{\rd}{\rd t}\Big|_{BL}\,, \qquad && \frac{\rd}{\rd r_*}\Big|_* = \frac{\rd r}{\rd r_*}\Big|_* \frac{\rd}{\rd r}\Big|_{BL} + \frac{\rd \theta}{\rd r_*}\Big|_* \frac{\rd}{\rd \theta}\Big|_{BL} + \frac{\rd \mathfrak h}{\rd r_*}\Big|_* \frac{\rd}{\rd \varphi}\Big|_{BL} \,, \\
    &\frac{\rd}{\rd \varphi_*}\Big|_* = \frac{\rd}{\rd \varphi}\Big|_{BL}\,, && \frac{\rd}{\rd \theta_*}\Big|_* = \frac{\rd r}{\rd \theta_*}\Big|_* \frac{\rd}{\rd r}\Big|_{BL} + \frac{\rd \theta}{\rd \theta_*}\Big|_* \frac{\rd}{\rd \theta}\Big|_{BL} + \frac{\rd \mathfrak h}{\rd \theta_*}\Big|_* \frac{\rd}{\rd \varphi}\Big|_{BL} \;.
    \end{aligned}
\end{equation*}
And recalling \eqref{EqDefKerrStarCoord} and \eqref{EqDefTildePhiMinus} we obtain
\begin{equation*}
    \begin{aligned}
        &\frac{\rd}{\rd t}\Big|_{BL} = \frac{\rd}{\rd v_+} \Big|_\pm - \frac{\rd}{\rd v_-} \Big|_\pm  - \frac{a}{r_-^2 + a^2} \frac{\rd}{\rd \varphi_=}\Big|_\pm\,, \qquad && \frac{\rd}{\rd \theta}\Big|_{BL} = \frac{\rd}{\rd \theta}\Big|_\pm\,,\\
        &\frac{\rd}{\rd r}\Big|_{BL} = \frac{r^2 + a^2}{\Delta} \frac{\rd}{\rd v_+}\Big|_\pm + \frac{r^2 + a^2}{\Delta} \frac{\rd}{\rd v_-} \Big|_\pm + \frac{a}{\Delta}\Big(1 - \frac{r^2 + a^2}{r_-^2 + a^2}\Big) \frac{\rd}{\rd \varphi_=} \Big|_\pm \,, \qquad  && \frac{\rd}{\rd \varphi}\Big|_{BL} = \frac{\rd}{\rd \varphi_=}\Big|_\pm \;.
    \end{aligned}
\end{equation*}
Combining those we obtain
\begin{equation} \label{EqDNPM}
    \begin{aligned}
        \frac{\rd}{\rd \ub}\Big|_{DN} &= \Big( \frac{\rd r}{\rd r_*}\Big|_* \frac{r^2 + a^2}{\Delta} +1 \Big) \frac{\rd}{\rd v_+} \Big|_\pm + \Big( \frac{\rd r}{\rd r_*}\Big|_* \frac{r^2 + a^2}{\Delta} -1 \Big) \frac{\rd}{\rd v_-} \Big|_\pm  \\
        &\qquad + \Big(  \frac{\rd r}{\rd r_*} \Big|_* \frac{a}{\Delta}\big( 1 - \frac{r^2 + a^2}{r_-^2 + a^2} \big)  + \frac{\rd \mathfrak h}{\rd r_*} \Big|_* - \frac{a}{r_-^2 + a^2} \Big) \frac{\rd}{\rd \varphi_=}\Big|_\pm + \frac{\rd \theta}{\rd r_*}\Big|_* \frac{\rd}{\rd \theta}\Big|_\pm \;, \\
        \frac{\rd}{\rd u}\Big|_{DN} &= \Big( \frac{\rd r}{\rd r_*}\Big|_* \frac{r^2 + a^2}{\Delta} -1 \Big) \frac{\rd}{\rd v_+} \Big|_\pm + \Big( \frac{\rd r}{\rd r_*}\Big|_* \frac{r^2 + a^2}{\Delta} +1 \Big) \frac{\rd}{\rd v_-} \Big|_\pm  \\
        &\qquad + \Big(  \frac{\rd r}{\rd r_*} \Big|_* \frac{a}{\Delta} \big(1 - \frac{r^2 + a^2}{r_-^2 + a^2}\big) + \frac{\rd \mathfrak h}{\rd r_*} \Big|_* + \frac{a}{r_-^2 + a^2}\Big) \frac{\rd}{\rd \varphi_=}\Big|_\pm + \frac{\rd \theta}{\rd r_*}\Big|_* \frac{\rd}{\rd \theta}\Big|_\pm \;, \\
        \frac{\rd}{\rd \theta_*}\Big|_{DN} &= \frac{\rd r}{\rd \theta_*}\Big|_* \frac{r^2 +a^2}{\Delta} \frac{\rd}{\rd v_+}\Big|_\pm + \frac{\rd r}{\rd \theta_*} \Big|_* \frac{r^2 + a^2}{\Delta} \frac{\rd}{\rd v_-}\Big|_\pm + \frac{\rd \theta}{\rd \theta_*} \Big|_* \frac{\rd}{\rd \theta}\Big|_\pm \\
        &\qquad + \Big(  \frac{\rd r}{\rd \theta_*}\Big|_* \frac{a}{\Delta}\big(1 - \frac{r^2 + a^2}{r_-^2 + a^2} \big) + \frac{\rd \mathfrak h}{\rd \theta_*} \Big|_* \Big) \frac{\rd}{\rd \varphi_=}\Big|_\pm\,, \\
        \frac{\rd}{\rd \varphi_*}\Big|_{DN} &= \frac{\rd}{\rd \varphi_=}\Big|_\pm \,.
    \end{aligned}
\end{equation}

Defining now the coordinate transformation $\mathscr{T}(u, \ub, \theta_*, \varphi_*) = ( v_-, v_+, \theta, \varphi_=)$, we can read off its derivative from \eqref{EqDNPM}:
\begin{equation} 
 \label{EqDerivativeT}
 \begin{split}
 D\mathscr{T} &= \begin{pmatrix}
    \frac{\rd v_-}{\rd u} & \frac{\rd v_-}{\rd \ub} & \frac{\rd v_-}{\rd \theta_*} & \frac{\rd v_-}{\rd \varphi_*} \\
     \frac{\rd v_+}{\rd u} & \frac{\rd v_+}{\rd \ub} & \frac{\rd v_+}{\rd \theta_*} & \frac{\rd v_+}{\rd \varphi_*} \\  \frac{\rd \theta}{\rd u}   &\frac{\rd \theta}{\rd \ub}   &\frac{\rd \theta}{\rd \theta_*} &\frac{\rd \theta}{\rd \varphi_*} \\
     \frac{\rd \varphi_=}{\rd u}  &\frac{\rd \varphi_=}{\rd \ub}    &\frac{\rd \varphi_=}{\rd \theta_*} & \frac{\rd \varphi_=}{\rd \varphi_*} 
 \end{pmatrix} \\
 &\text{\tiny{$= \begin{pmatrix}  \frac{\rd r}{\rd r_*}\Big|_* \frac{r^2 + a^2}{\Delta} +1  &\frac{\rd r}{\rd r_*}\Big|_* \frac{r^2 + a^2}{\Delta} -1  &\frac{\rd r}{\rd \theta_*}\Big|_* \frac{r^2 +a^2}{\Delta} &0 \\
\frac{\rd r}{\rd r_*}\Big|_* \frac{r^2 + a^2}{\Delta} -1 &\frac{\rd r}{\rd r_*}\Big|_* \frac{r^2 + a^2}{\Delta} +1 &\frac{\rd r}{\rd \theta_*}\Big|_* \frac{r^2 +a^2}{\Delta} &0 \\
\frac{\rd \theta}{\rd r_*}\Big|_* &\frac{\rd \theta}{\rd r_*}\Big|_* &\frac{\rd \theta}{\rd \theta_*} \Big|_* &0 \\
  \frac{\rd r}{\rd r_*} \Big|_* \frac{a}{\Delta} \big(1 - \frac{r^2 + a^2}{r_-^2 + a^2}\big) + \frac{\rd \mathfrak h}{\rd r_*} \Big|_* + \frac{a}{r_-^2 + a^2} &\frac{\rd r}{\rd r_*} \Big|_* \frac{a}{\Delta}\big( 1 - \frac{r^2 + a^2}{r_-^2 + a^2} \big)  + \frac{\rd \mathfrak h}{\rd r_*} \Big|_* - \frac{a}{r_-^2 + a^2}  &\frac{\rd r}{\rd \theta_*}\Big|_* \frac{a}{\Delta}\big(1 - \frac{r^2 + a^2}{r_-^2 + a^2} \big) + \frac{\rd \mathfrak h}{\rd \theta_*} \Big|_* &1 \\
 \end{pmatrix}$}}
 \end{split}
\end{equation}

\subsubsection{Projection onto $\Gamma$ along integral curves of $e_3'$: the map $\mathscr{G}$}

The following lemma shows that the hypersurface $\Gamma$ can be written as a graph over the $(\ub, \theta_*, \varphi_*)$ coordinates.
\begin{lemma} \label{LemGammaGraph}
    There exists a $\ub_0 \geq 1$ and a smooth function $u_\Gamma : [\ub_0, \infty) \times \Sp^2 \to \R$ such that
    \begin{equation} \label{EqLemmaGammaGraph}
    \Gamma \cap \{\ub \geq \ub_0\} = \Big{\{} \big( u_\Gamma(\ub, \vartheta_*), \ub, \vartheta_*\big) \; \Big| \; (\ub, \vartheta_*) \in [\ub_0, \infty) \times \Sp^2\Big{\}} \;.
    \end{equation}
\end{lemma}

\begin{proof}
It follows from \eqref{EqDerivativeT} that $$\frac{\rd v_+}{\rd u}\Big|_{DN} = \frac{\rd r}{\rd r_*} \Big|_* \frac{r^2 + a^2}{\Delta} -1 \,, \qquad \frac{\rd v_-}{\rd u}\Big|_{DN} =  \frac{\rd r}{\rd r_*}\Big|_* \frac{r^2 + a^2}{\Delta} + 1\;.$$
Hence, Lemma \ref{LemPartialsLimit} gives $\frac{\rd v_+}{\rd u}\Big|_{DN} \to 0$ and $\frac{\rd v_-}{\rd u}\Big|_{DN} \to 2$ for $r \to r_-$. Recalling the definition $f_\Gamma(v_+, v_-) = v_+ + v_- - \frac{\sigma_q}{\kappa_-} \log v_+$ we get 
\begin{equation*}
\frac{\rd}{\rd u}\Big|_{DN} f_\Gamma = \frac{\rd v_+}{\rd u}\Big|_{DN} +   \frac{\rd v_-}{\rd u}\Big|_{DN} - \frac{\sigma_q}{\kappa_-} \cdot \frac{1}{v_+} \frac{\rd v_+}{\rd u}\Big|_{DN} \;.
\end{equation*}
By Lemma \ref{LemRelationStarAndNullCoord} we can choose $\ub_0 \geq 1$ so large such that in $\{\ub \geq \ub_0\}$ we have $v_+ \geq 1$. Then choose $r_0 > r_-$ sufficiently close to $r_-$ such that for $r_- < r < r_0$ and $\ub \geq \ub_0$ we have 
\begin{equation}\label{EqDerFGammaPos}
    \frac{\rd}{\rd u}\Big|_{DN} f_\Gamma \geq 1 \;.
\end{equation}

Furthermore, since $v_+ + v_- = 2r^*$ (and using again Lemma \ref{LemRelationStarAndNullCoord}), it is clear that we can choose $\ub_0$ even bigger such that $f_\Gamma$ is negative on $\{r = r_0\} \cap \{\ub \geq \ub_0\}$. Again, by Lemma \ref{LemRelationStarAndNullCoord}, the function $f_\Gamma$ is positive on $\{u = u_f\}$ for $\ub_0$ large enough. Hence, for each $(\ub, \vartheta_*) \in [\ub_0, \infty) \times \Sp^2$ there exists a $u_\Gamma(\ub, \vartheta_*) \in \R$ with $\big( u_\Gamma(\ub, \vartheta_*), \ub, \vartheta_*\big) \in \Gamma$. The smoothness of $u_\Gamma$ follows from the smoothness of $f_\Gamma$ by the implicit function theorem. 

Finally, $f_\Gamma$ being negative on $\{r = r_0\} \cap \{\ub \geq \ub_0\}$ implies that $r$ is smaller than $r_0$ on $\Gamma \cap \{\ub \geq \ub_0\}$. Since $-r$ is a time function and $\frac{\rd}{\rd u}\Big|_{DN}$ is null, it follows that the past directed integral curve of $\frac{\rd}{\rd u}\Big|_{DN}$, once it has reached $\{r=r_0\}$, can never reach smaller $r$-values  again and thus cannot intersect $\Gamma \cap \{\ub \geq \ub_0\}$ again. This shows the equality in \eqref{EqLemmaGammaGraph}.
\end{proof}

We define the function $\mathscr{G} : [ \ub_0, \infty) \times \Sp^2 \to \Gamma$  by $\mathscr{G}(\ub, \vartheta_*) := \big(u_\Gamma(\ub, \vartheta_*), \ub, \vartheta_*\big)$, which, by Lemma \ref{LemGammaGraph}, is clearly a diffeomorphism onto its image. By the implicit function theorem, its derivative is given by
\begin{equation}
    \label{EqDerivativeG}
    \begin{split}
    D \mathscr{G} &= \begin{pmatrix}
        \frac{\rd u_\Gamma}{\rd \ub} & \frac{\rd u_\Gamma}{\rd \theta_*} & \frac{\rd u_\Gamma}{\rd \varphi_*} \\
        \frac{\rd \ub}{\rd \ub} & \frac{\rd \ub}{\rd \theta_*} & \frac{\rd \ub}{\rd \varphi_*} \\
        \frac{\rd \theta_*}{\rd \ub} & \frac{\rd \theta_*}{\rd \theta_*} & \frac{\rd \theta_*}{\rd \varphi_*} \\
        \frac{\rd \varphi_*}{\rd \ub} & \frac{\rd \varphi_*}{\rd \theta_*} & \frac{\rd \varphi_*}{\rd \varphi_*}
    \end{pmatrix} = \begin{pmatrix}
        - \frac{\frac{\rd f_\Gamma}{\rd \ub}}{\frac{\rd f_\Gamma}{\rd u}} & - \frac{\frac{\rd f_\Gamma}{\rd \theta_*}}{\frac{\rd f_\Gamma}{\rd u}} & - \frac{\frac{\rd f_\Gamma}{\rd \varphi_*}}{\frac{\rd f_\Gamma}{\rd u}}  \\
        1 & 0 & 0\\
        0 & 1 & 0 \\
        0 & 0 & 1
    \end{pmatrix} \\
    &= \begin{pmatrix}
        - \frac{\frac{\rd r}{\rd r_*}\Big|_* \frac{r^2+a^2}{\Delta}(2 - \frac{\sigma_q}{\kappa_- v_+}) - \frac{\sigma_q}{\kappa_- v_+}}{\frac{\rd r}{\rd r_*}\Big|_* \frac{r^2+a^2}{\Delta}(2 - \frac{\sigma_q}{\kappa_- v_+}) + \frac{\sigma_q}{\kappa_- v_+}} & - \frac{\frac{\rd r}{\rd \theta_*}\Big|_* \frac{r^2 + a^2}{\Delta}(2 - \frac{\sigma_q}{\kappa_- v_+})}{\frac{\rd r}{\rd r_*}\Big|_* \frac{r^2+a^2}{\Delta}(2 - \frac{\sigma_q}{\kappa_- v_+}) + \frac{\sigma_q}{\kappa_- v_+}} & 0 \\
        1 & 0 & 0\\
        0 & 1 & 0 \\
        0 & 0 & 1
    \end{pmatrix} = \begin{pmatrix}
        \calO(1) & \calO(\sin \theta_*) & 0 \\
        1 & 0 & 0\\
        0 & 1 & 0 \\
        0 & 0 & 1
    \end{pmatrix}\;,
    \end{split}
\end{equation}
where we have also used \eqref{EqDerivativeT} in the third equality and Lemma \ref{LemPartialsLimit} in the last equality.

\subsubsection{$\Gamma$ as a graph over $(v_+, \theta, \varphi_=)$: the maps $\mathscr{H}$ and $\mathscr{P}$}

In $(v_-, v_+, \theta, \varphi_=)$ coordinates the hypersurface $\Gamma$ can trivially be written as a graph $$\mathscr{H}(v_+, \theta, \varphi_=) = (v_{-, \Gamma}(v_+), v_+, \theta, \varphi_=) = (\frac{\sigma_q}{\kappa_-} \log v_+ - v_+, v_+, \theta, \varphi_=)\;.$$
If, again in $(v_-, v_+, \theta, \varphi_=)$ coordinates, $\mathscr{P}$ denotes the projection map $\mathscr{P}(v_-,v_+, \theta, \varphi_=) = (v_+, \theta, \varphi_=)$, then it is easy to see that we have
\begin{equation}
\label{EqHGP}
    \mathscr{H}^{-1} \circ \mathscr{G} = \mathscr{P} \circ \mathscr{T} \circ \mathscr{G}\;.
\end{equation}

\subsubsection{The concatenation $\mathscr{P} \circ \mathscr{T} \circ \mathscr{G}$}

Using $$D\mathscr{P} = \begin{pmatrix}
    0 & 1 & 0 & 0 \\ 0 & 0 & 1 & 0 \\ 0 & 0 & 0 & 1
\end{pmatrix}$$
we compute by matrix multiplication 
\begin{equation}
\label{EqDerivativePTG}
\begin{split}
    D(\mathscr{P} \circ \mathscr{T} \circ \mathscr{G}) &= \begin{pmatrix}
        - \frac{\frac{\rd f_\Gamma}{\rd \ub}}{\frac{\rd f_\Gamma}{\rd u}} \cdot \frac{\rd v_+}{\rd u} + \frac{\rd v_+}{\rd \ub} &     - \frac{\frac{\rd f_\Gamma}{\rd \theta_*}}{\frac{\rd f_\Gamma}{\rd u}} \cdot \frac{\rd v_+}{\rd u} + \frac{\rd v_+}{\rd \theta_*} & 0       \\
        - \frac{\frac{\rd f_\Gamma}{\rd \ub}}{\frac{\rd f_\Gamma}{\rd u}} \cdot \frac{\rd \theta}{\rd u} + \frac{\rd \theta}{\rd \ub} &    - \frac{\frac{\rd f_\Gamma}{\rd \theta_*}}{\frac{\rd f_\Gamma}{\rd u}} \cdot \frac{\rd \theta}{\rd u} + \frac{\rd \theta}{\rd \theta_*} & 0      \\
        - \frac{\frac{\rd f_\Gamma}{\rd \ub}}{\frac{\rd f_\Gamma}{\rd u}} \cdot \frac{\rd \varphi_=}{\rd u} + \frac{\rd \varphi_=}{\rd \ub} &   - \frac{\frac{\rd f_\Gamma}{\rd \theta_*}}{\frac{\rd f_\Gamma}{\rd u}} \cdot \frac{\rd \varphi_=}{\rd u} + \frac{\rd \varphi_=}{\rd \theta_*} & 1
    \end{pmatrix} = \begin{pmatrix}
        \calO(|\Delta|)  + \frac{\rd v_+}{\rd \ub} & \calO(|\Delta|) + \frac{\rd v_+}{\rd \theta_*} & 0 \\
        \calO(|\Delta|) + \frac{\rd \theta}{\rd \ub} & \calO(|\Delta|) + \frac{\rd \theta}{\rd \theta_*} & 0 \\
        \calO(|\Delta|) + \frac{\rd \varphi_=}{\rd \ub} & \calO(|\Delta|) + \frac{\rd \varphi_=}{\rd \theta_*} & 1
    \end{pmatrix} \\
    &= \begin{pmatrix}
        2+\calO(|\Delta|)  & \calO(1) & 0 \\
        \calO(|\Delta|) &  \underbrace{- a \sqrt{\sin^2 \theta_* - \sin^2 \theta} G + \calO(|\Delta|)}_{= \calO(1)} & 0 \\
        -\frac{2a}{r_-^2 + a^2} + \calO(|\Delta|) & \calO(1) & 1
    \end{pmatrix}\;,
    \end{split}
\end{equation}
where all partial derivatives are with respect to $(u, \ub, \theta_*, \varphi_*)$ (i.e., $\Big|_{DN}$) and the evaluation of the matrix entries is at $\mathscr{G}(\ub, \theta_*, \varphi_*)$. For the first equality  we have used the structure of $D \mathscr{T}$ and $D \mathscr{G}$ given by their vanishing components and the components that are equal to $1$ as given \eqref{EqDerivativeT} and \eqref{EqDerivativeG}, but have otherwise kept the shorthand for their entries to keep the notation compact. For the second equality we have used \eqref{EqDerivativeG}, \eqref{EqDerivativeT}, and Lemma \ref{LemPartialsLimit}. For the third equality we used again \eqref{EqDerivativeT} and Lemma \ref{LemPartialsLimit}. It follows from \cite[Propositions A.1 and A.3]{DafLuk17} that the middle entry is bounded away from $0$ and $\infty$ which directly gives
\begin{equation}
    \label{EqDetPTG}
    \det D(\mathscr{P} \circ \mathscr{T} \circ \mathscr{G}) \sim 1 \;.
\end{equation}

\subsubsection{Stereographic projection: the maps $\mathscr{S}_{(i)}$}

Consider the stereographic projections maps $\mathscr{S}_{(2)}$ and $\mathscr{S}_{(1)}$ defined in \eqref{eq:def.scrS.2}--\eqref{eq:def.scrS.1}. 
The derivative of $\mathscr{S}_{(2)}$ is computed to be
\begin{equation}
    \label{EqDerivativeS}
    D \mathscr{S}_{(2)} = \begin{pmatrix}
        \frac{\rd \theta^1_{(2)}}{\rd \theta} & \frac{\rd \theta^1_{(2)}}{\rd \varphi} \\ \frac{\rd \theta^2_{(2)}}{\rd \theta} & \frac{\rd \theta^2_{(2)}}{\rd \varphi}
    \end{pmatrix} = \begin{pmatrix}
        \frac{\cos \varphi}{2 \cos^2 \frac{\theta}{2}} & - \tan \frac{\theta}{2} \sin \varphi \\ \frac{\sin \varphi}{2 \cos^2 \frac{\theta}{2}} & \tan \frac{\theta}{2} \cos \varphi 
    \end{pmatrix} \overset{\calV_2''}{=} \begin{pmatrix}
        \mathcal{O}(1) & \mathcal{O}( \sin \theta) \\ 
        \mathcal{O}(1) & \mathcal{O}( \sin \theta) \\ 
    \end{pmatrix}  \;,
\end{equation}
where the last equality holds uniformly in $\calV_2''$ (for $\calV_2''$ as in Section~\ref{sec:CH.coord}). The determinant is computed to be
\begin{equation}
    \label{EqDeterminantDS}
    \det D \mathscr{S}_{(2)} = \frac{\tan \frac{\theta}{2}}{2 \cos^2 \frac{\theta}{2}} \overset{\calV_2''}{\sim} \sin \theta \;,
\end{equation}
where again the overset notation denotes the domain on which the uniform equivalence holds.
Inverting the matrix gives
\begin{equation}
    \label{EqDerivativeSInverse}
    D \mathscr{S}^{-1}_{(2)} = \begin{pmatrix}
        \frac{\rd \theta}{\rd \theta^1_{(2)}} & \frac{\rd \theta}{\rd \theta^2_{(2)}} \\ \frac{\rd \varphi}{\rd \theta^1_{(2)}} & \frac{\rd \varphi}{\rd \theta^2_{(2)}}
    \end{pmatrix} = \begin{pmatrix}
        2 \cos^2 \frac{\theta}{2} \cos \varphi & 2 \cos^2 \frac{\theta}{2} \sin \varphi \\ - \frac{\sin \varphi}{\tan \frac{\theta}{2}} & \frac{\cos \varphi}{\tan \frac{\theta}{2}}
    \end{pmatrix} \overset{\calV_2''}{=} \begin{pmatrix}
        \mathcal{O}(1) & \calO(1) \\
        \calO(\frac{1}{\sin \theta}) & \calO(\frac{1}{\sin \theta})
    \end{pmatrix}\;.
\end{equation}
Similarly for the stereographic projection $\mathscr{S}_{(1)}$, the derivative is computed analogously.

In the following the stereographic projection will be used twice: once for $(\theta, \varphi_=)$ coordinates and once for $(\theta_*, \varphi_*)$ coordinates.

\subsubsection{Mapping to angular coordinates $\vartheta_{\CH^+}$ which are regular at the Cauchy horizon: the maps $\mathscr{C}_{(i)}$}

We first define the background maps (see \eqref{eq:Kerr.CH.coord}, \eqref{eq:Kerr.CH.stereo}) $$\pmb{\mathscr{C}}_{(i)}(\ub, \theta^1_{(i)}, \theta^2_{(i)}) = \big(\ub, \pmb{\theta}^1_{(i), \CH^+} = \theta^1_{(i)} \cos(\omega_- \ub) + \theta^2_{(i)} \sin ( \omega_- \ub), \pmb{\theta}^2_{(i), \CH^+} = \theta^2_{(i)} \cos( \omega_- \ub) - \theta^1_{(i)} \sin (\omega_- \ub) \big)\;,$$ where $\omega_- := \frac{4Ma r_-}{(r_-^2 + a^2)^2} = \frac{2a}{r_-^2 + a^2}$.
The derivative is
\begin{equation}
    \label{EqDerivativeCK}
    \begin{split}
    D \pmb{\mathscr{C}}_{(i)} &= \begin{pmatrix}
        \frac{\rd \ub}{\rd \ub} & \frac{\rd \ub}{\rd \theta^1_{(i)}} & \frac{\rd \ub}{\rd \theta^2_{(i)}} \\
        \frac{\rd \pmb{\theta}^1_{(i), \CH^+}}{\rd \ub} & \frac{\rd \pmb{\theta}^1_{(i), \CH^+}}{\rd\theta^1_{(i)}} & \frac{\rd \pmb{\theta}^1_{(i),\CH^+}}{\rd \theta^2_{(i)}} \\
        \frac{\rd \pmb{\theta}^2_{(i), \CH^+}}{\rd \ub} & \frac{\rd \pmb{\theta}^2_{(i),\CH^+}}{\rd \theta^1_{(i)}} & \frac{\rd \pmb{\theta}^2_{(i), \CH^+}}{\rd \theta^2_{(i)}}
    \end{pmatrix} \\
    &= \begin{pmatrix}
        1 & 0 & 0 \\
         \omega_-\big( - \theta^1_{(i)} \sin(\omega_- \ub) + \theta^2_{(i)} \cos( \omega_- \ub) \big)& \cos( \omega_- \ub) & \sin(\omega_- \ub) \\
        - \omega_-\big( \theta^2_{(i)} \sin ( \omega_- \ub) + \theta^1_{(i)} \cos ( \omega_- \ub) \big) & - \sin(\omega_- \ub) & \cos( \omega_- \ub)
    \end{pmatrix}\;.
    \end{split}
\end{equation}

Recall that the maps $\mathscr{C}_{(i)}(\ub, \theta^1_{(i)}, \theta^2_{(i)}) = (\ub, \theta^1_{(i), \CH^+}, \theta^2_{(i), \CH^+})$ are defined in \eqref{eq:reg.coord.def.1} and \eqref{eq:reg.coord.def.2}. The derivative is of the form
\begin{equation}
    \label{EqDerivativeC}
    D \mathscr{C}_{(i)} = \begin{pmatrix}
        \frac{\rd \ub}{\rd \ub} & \frac{\rd \ub}{\rd \theta^1_{(i)}} & \frac{\rd \ub}{\rd \theta^2_{(i)}} \\
        \frac{\rd \theta^1_{(i), \CH^+}}{\rd \ub} & \frac{\rd \theta^1_{(i), \CH^+}}{\rd\theta^1_{(i)}} & \frac{\rd \theta^1_{(i),\CH^+}}{\rd \theta^2_{(i)}} \\
        \frac{\rd \theta^2_{(i), \CH^+}}{\rd \ub} & \frac{\rd \theta^2_{(i),\CH^+}}{\rd \theta^1_{(i)}} & \frac{\rd \theta^2_{(i), \CH^+}}{\rd \theta^2_{(i)}}
    \end{pmatrix} = \begin{pmatrix}
        1 & 0 & 0 \\
        -b^B_{(i)}|_{u = u_f} \frac{\rd \theta^1_{(i), \CH^+}}{\rd \theta^B_{(i)}} + \underline{\pmb{b}}^1_{(i)}|_{u = u_f}& \frac{\rd \theta^1_{(i), \CH^+}}{\rd\theta^1_{(i)}} & \frac{\rd \theta^1_{(i),\CH^+}}{\rd \theta^2_{(i)}} \\
        -b^B_{(i)}|_{u = u_f} \frac{\rd \theta^2_{(i), \CH^+}}{\rd \theta^B_{(i)}} + \underline{\pmb{b}}^2_{(i)}|_{u = u_f} & \frac{\rd \theta^2_{(i),\CH^+}}{\rd \theta^1_{(i)}} & \frac{\rd \theta^2_{(i), \CH^+}}{\rd \theta^2_{(i)}}
    \end{pmatrix}\;.
\end{equation}
The bottom right $2 \times 2$ submatrix is $\epsilon$-close to the bottom right $2 \times 2$ submatrix of \eqref{EqDerivativeCK} by \cite[Proposition 16.11]{DafLuk17} and thus, for $\epsilon>0$ sufficiently small, is uniformly bounded and invertible with uniformly bounded inverse.
In particular this gives
\begin{equation}
    \label{EqDeterminantDC}
    \det D \mathscr{C}_{(i)} \sim 1 \;.
\end{equation}
Using the block form of \eqref{EqDerivativeC} we can easily invert $D \mathscr{C}_{(i)}$ to obtain
\begin{equation}
    \label{EqDerivativeCInverse}
    \begin{split}
    D \mathscr{C}^{-1}_{(i)} &= \begin{pmatrix}
        \frac{\rd \ub}{\rd \ub} & \frac{\rd \ub}{\rd \theta^1_{(i), \CH^+}} & \frac{\rd \ub}{\rd \theta^2_{(i), \CH^+}} \\
        \frac{\rd \theta^1_{(i)}}{\rd \ub} & \frac{\rd \theta^1_{(i)}}{\rd\theta^1_{(i), \CH^+}} & \frac{\rd \theta^1_{(i)}}{\rd \theta^2_{(i), \CH^+}} \\
        \frac{\rd \theta^2_{(i)}}{\rd \ub} & \frac{\rd \theta^2_{(i)}}{\rd \theta^1_{(i), \CH^+}} & \frac{\rd \theta^2_{(i)}}{\rd \theta^2_{(i), \CH^+}}
    \end{pmatrix} = \begin{pmatrix}
        1 & 0 & 0 \\
        b^1_{(i)}|_{u = u_f} - \frac{\rd \theta^1_{(i)}}{\rd \theta^B_{(i), \CH^+}}  \underline{\pmb{b}}^B_{(i)}|_{u = u_f}& \frac{\rd \theta^1_{(i)}}{\rd\theta^1_{(i), \CH^+}} & \frac{\rd \theta^1_{(i)}}{\rd \theta^2_{(i), \CH^+}} \\
        b^2_{(i)}|_{u = u_f} - \frac{\rd \theta^2_{(i)}}{\rd \theta^B_{(i), \CH^+}}  \underline{\pmb{b}}^B_{(i)}|_{u = u_f} & \frac{\rd \theta^2_{(i)}}{\rd \theta^1_{(i), \CH^+}} & \frac{\rd \theta^2_{(i)}}{\rd \theta^2_{(i), \CH^+}}
    \end{pmatrix} \\
    &\overset{\calV_2'}{=}\begin{pmatrix}
        1 & 0 & 0 \\
        b^1_{(i)}|_{u = u_f} + \calO(e^{-2 \kappa_- \ub}) & \calO(1) & \calO(1) \\
        b^2_{(i)}|_{u = u_f} + \calO(e^{-2 \kappa_- \ub}) & \calO(1) & \calO(1) 
    \end{pmatrix}\;,
    \end{split}
\end{equation}
we have used \cite[Proposition A.15]{DafLuk17} to estimate the $\underline{\pmb{b}}$ terms.

\subsubsection{The concatenation $( \id_1 \times \mathscr{S}^{-1}_{(2)}) \circ \mathscr{C}_{(2)}^{-1}$}

We begin by computing $D \big( ( \id_1 \times \mathscr{S}^{-1}_{(2)}) \circ \mathscr{C}_{(2)}^{-1}\big)$ and we note that in \eqref{EqDerivativeSInverse} we need to replace $\theta$ by $\theta_*$ and $\varphi$ by $\varphi_*$:
\begin{equation}
    \label{EqDerivativeSCInv}
    \begin{split}
   D \big( ( \id_1 \times \mathscr{S}^{-1}_{(2)}) \circ \mathscr{C}_{(2)}^{-1}\big) &= \begin{pmatrix}
       1 & 0 & 0 \\
       b^{\theta_*}|_{u = u_f} - \frac{\rd \theta_*}{\rd \theta^A_{(2)}} \frac{\rd \theta^A_{(2)}}{\rd \theta^B_{(2),\CH^+}} ( \underline{\pmb{b}})^B_{(2)}|_{u = u_f} & \frac{\rd \theta_*}{\rd \theta^A_{(2)}} \frac{\rd \theta^A_{(2)}}{\rd \theta^1_{(2), \CH^+}} & \frac{\rd \theta_*}{\rd \theta^A_{(2)}} \frac{\rd \theta^A_{(2)}}{\rd \theta^2_{(2), \CH^+}} \\
       b^{\varphi_*}|_{u = u_f} - \frac{\rd \varphi_*}{\rd \theta^A_{(2)}} \frac{\rd \theta^A_{(2)}}{\rd \theta^B_{(2),\CH^+}} ( \underline{\pmb{b}})^B_{(2)}|_{u = u_f} & \frac{\rd \varphi_*}{\rd \theta^A_{(2)}} \frac{\rd \theta^A_{(2)}}{\rd \theta^1_{(2), \CH^+}}  & \frac{\rd \varphi_*}{\rd \theta^A_{(2)}} \frac{\rd \theta^A_{(2)}}{\rd \theta^2_{(2), \CH^+}} 
   \end{pmatrix} \\
   &\overset{\calV_2'}{=}\begin{pmatrix}
       1 & 0 & 0 \\
      b^{\theta_*}|_{u = u_f} + \calO(e^{-2 \kappa_- \ub}) &\mathcal{O}(1) & \calO(1) \\
      b^{\varphi_*}|_{u = u_f} + \calO(\frac{e^{-2 \kappa_- \ub}}{\sin \theta_*}) & \calO(\frac{1}{\sin \theta_*}) & \calO(\frac{1}{\sin \theta_*})
   \end{pmatrix}
   \end{split}
\end{equation}
where   we have used that $\theta_{(2), \CH^+} \in \calV_2'$ implies $\theta_{(2)} \in \calV_2''$ (see Section~\ref{sec:CH.coord}). 
Moreover, from \eqref{EqDeterminantDS} and \eqref{EqDeterminantDC} it follows that 
\begin{equation}
    \label{EqDeterminantDCSInv}
    \det D \big( ( \id_1 \times \mathscr{S}^{-1}_{(2)}) \circ \mathscr{C}_{(2)}^{-1}\big) \overset{[\ub_0, \infty) \times \calV_2''}{\sim} \frac{1}{\sin \theta_*} \;.
\end{equation}

\subsubsection{Concatenating the individual maps}

For $i=1,2$ we define the maps
\begin{equation*}
    \Pi_{(i)}:=(\id_1 \times \mathscr{S}_{(i)}) \circ \mathscr{H}^{-1} \circ \mathscr{G} \circ (\id_1 \times \mathscr{S}){(i)}^{-1}) \circ \mathscr{C}_{(i)}^{-1} = (\id_1 \times \mathscr{S}_{(i)}) \circ \mathscr{P} \circ \mathscr{T} \circ \mathscr{G} \circ (\id_1 \times \mathscr{S}_{(i)}^{-1}) \circ \mathscr{C}_{(i)}^{-1} \;.
\end{equation*}
Here, we have used \eqref{EqHGP} in the last equality.
It follows from \cite[Lemma 16.8]{DafLuk17} that the coordinate maps $\Pi_{(1)}$ and $\Pi_{(2)}$ patch together to yield a well-defined map $\Pi : [\ub_0, \infty) \times \Sp^2 \to \R \times \Sp^2$, where the differential structure on the domain is given by the $(\ub, \vartheta_{\CH^+})$ coordinates and on the target space by $(v_+, \vartheta_=)$ coordinates. Note that since $\mathscr{G}$ and $\mathscr{H}$ are diffeomorphisms (onto their images), the map $\Pi$ is a diffeomorphism onto its image.

We now compute $D\Pi_{(2)}$ (the computation of $D\Pi_{(1)}$ is analogous) \underline{in $[\ub_0, \infty) \times \calV_2'$} by matrix multiplication, using \eqref{EqDerivativeS}, \eqref{EqDerivativePTG}, \eqref{EqDerivativeSCInv} and also $b^{\theta_*}|_{u = u_f} = \tilde{b}^{\theta_*}|_{u = u_f}$ and $b^{\varphi_*}|_{u = u_f} = \frac{2a}{r_-^2 + a^2} + \calO(e^{-2 \kappa_- \ub}) + \tilde{b}^{\varphi_*}|_{u = u_f}$:
\begin{equation}
    \label{EqDerP}
    \begin{split}
        &D\Pi_{(2)} = D\big((\id_1 \times \mathscr{S}_{(i)})\big) D (\mathscr{P} \circ \mathscr{T} \circ \mathscr{G} ) D \big((\id_1 \times \mathscr{S}_{(i)}^{-1}) \circ \mathscr{C}_{(i)}^{-1} \big) \\
        &\;= \text{\footnotesize{$\begin{pmatrix}
            1 & 0 & 0 \\
            0 & \calO(1) & \calO(\sin \theta) \\
            0 & \calO(1) & \calO(\sin \theta)
        \end{pmatrix}
        \begin{pmatrix}
            2+\calO(|\Delta|) & \calO(1) & 0 \\
            \calO(|\Delta|) & \calO(1) & 0 \\
            -\frac{2a}{r_-^2 + a^2} + \calO(|\Delta|) & \calO(1) & 1 
        \end{pmatrix}
        \begin{pmatrix}
            1 & 0 & 0 \\
            \tilde{b}^{\theta_*}|_{u = u_f} + \calO(e^{-2 \kappa_- \ub}) & \calO(1) & \calO(1) \\
            \frac{2a}{r_-^2 + a^2} + \calO(\frac{e^{- 2\kappa_-\ub}}{\sin \theta_*}) + \tilde{b}^{\varphi_*}|_{u = u_f} & \calO(\frac{1}{\sin \theta_*}) & \calO( \frac{1}{\sin \theta_*})
        \end{pmatrix}$} }\\
        &\;=\begin{pmatrix}
            1 & 0 & 0 \\
            0 & \calO(1) & \calO(\sin \theta) \\
            0 & \calO(1) & \calO(\sin \theta)
        \end{pmatrix} \begin{pmatrix}
            2 + \calO(|\Delta|) + \calO(1)\big( \tilde{b}^{\theta_*}|_{u = u_f} + \calO(e^{-2 \kappa_- \ub}) \big) & \calO(1) & \calO(1) \\
             \calO(|\Delta|) + \calO(1) \tilde{b}^{\theta_*}|_{u = u_f} + \calO(e^{-2 \kappa_- \ub})    & \calO(1) & \calO(1) \\
             \calO(\frac{|\Delta| + e^{-2\kappa_- \ub}}{\sin \theta_*}) + \tilde{b}^{\varphi_*}|_{u = u_f} + \calO(1)\tilde{b}^{\theta_*}|_{u = u_f}       & \calO(\frac{1}{\sin \theta_*}) & \calO(\frac{1}{\sin \theta_*})
        \end{pmatrix} \\
        &\;= \begin{pmatrix}
           2 + \calO(|\Delta|) + \calO(1)\big( \tilde{b}^{\theta_*}|_{u = u_f} + \calO(e^{-2 \kappa_- \ub}) \big) & \calO(1) &  \calO(1) \\
             \calO(|\Delta|) + \calO(1) \tilde{b}^{\theta_*}|_{u = u_f} + \calO(e^{-2 \kappa_- \ub}) + \calO(\sin \theta) \tilde{b}^{\varphi_*}|_{u = u_f}  &  \calO(1) &  \calO(1)\\
             \calO(|\Delta|) + \calO(1) \tilde{b}^{\theta_*}|_{u = u_f} + \calO(e^{-2 \kappa_- \ub}) + \calO(\sin \theta) \tilde{b}^{\varphi_*}|_{u = u_f}  &  \calO(1) &  \calO(1)
        \end{pmatrix}\;,
    \end{split}
\end{equation}
where in the last equality we have used again \cite[Proposition A.1]{DafLuk17}. 

Recall $\Pi_{(2)}(\ub, \theta_{(2), \CH^+}^1, \theta_{(2), \CH^+}^2) = (v_+, \theta_{(2), =}^1, \theta_{(2), =}^2)$. It then follows from \eqref{EqDerP} in combination with \eqref{EqDeltaFutureGamma}, Lemma \ref{LemRelationStarAndNullCoord}, and Lemma~\ref{lem:b.improve} that
\begin{equation}
    \label{EqIntegratedBoundP}
    \max_{A=1,2}\int_{\ub_0}^\infty \sup_{\theta_{(2), \CH^+} \in \calV_2'} \Big|\frac{\rd \theta_{(2), =}^A}{\rd \ub} \Big| \, \ud\ub  < \infty \;.
\end{equation}
Furthermore note that it follows directly from \eqref{EqDeterminantDS}, \eqref{EqDetPTG}, \eqref{EqDeterminantDCSInv} together with \cite[Proposition A.1]{DafLuk17} that
\begin{equation*}
    \det \Pi_{(2)} \overset{[\ub_0, \infty) \times \calV_2'}{\sim} 1 \;.
\end{equation*}
Since by \cite[(4.24)]{DafLuk17} all entries in \eqref{EqDerP} are bounded, it thus follows that all entries in the inverse matrix are bounded as well. Hence we conclude for the inverse map $\Pi_{(2)}^{-1}(v_+, \theta_{(2), =}^1, \theta_{(2), =}^2) = (\ub, \theta_{(2), \CH^+}^1, \theta_{(2), \CH^+}^2)$ in particular the following bounds on the partial derivatives:
\begin{equation}
    \label{EqPointwiseBoundsPInverse}
    \max_{A,B \in \{1,2\}}\sup_{(v_+, \theta_{(2), =}) \in \Pi_{(2)}\big([\ub_0, \infty) \times \calV_2'\big)} \Big| \frac{\rd \theta^A_{(2), \CH^+}}{\rd \theta^B_{(2), =}} \Big| \lesssim 1 \;.
\end{equation}
Furthermore, by choosing $\ub_0$ larger and $\epsilon>0$ smaller if needed it also follows from \eqref{EqDerP} that
\begin{equation}
    \label{EqMonP}
    \frac{\rd v_+}{\rd \ub} >0 \;.
\end{equation}
The bounds corresponding to \eqref{EqIntegratedBoundP}, \eqref{EqPointwiseBoundsPInverse}, and \eqref{EqMonP} for $\Pi_{(1)}$ are being proved analogously.

We can now prove the main result of this subsection:

\begin{proposition} \label{PropProjectionUToGamma}
    Let $W_{\CH^+} \subseteq \Sp^2$ be an open subset with respect to the $(\ub, \vartheta_{\CH^+})$ coordinates on $ \R \times \Sp^2$ and let $\ub_1 \geq \ub_0$ with $\ub_0$ as in Lemma \ref{LemGammaGraph}. Then there exists an open subset $W_\Gamma \subseteq \Sp^2$ with respect to $(v_+, \vartheta_=)$ coordinates on $\R \times \Sp^2$ and $v_{2+} >0$ such that 
    \begin{equation}
        \label{EqPropProjGamma}
        \Pi \big([\ub_1, \infty) \times W_{\CH^+} \big) \supseteq [v_{2+}, \infty) \times W_\Gamma\;.
    \end{equation}
\end{proposition}

\begin{proof}
Consider a point $\mathring{\vartheta}_{\CH^+} \in W_{\CH^+} \subseteq \Sp^2$ and then the curve 
\begin{equation}\label{EqPfRingCH}
  \ub \mapsto \Pi(\ub, \mathring{\vartheta}_{\CH^+}) = \big(v_+(\ub), \vartheta_{=}(\ub)\big)  \;.
\end{equation}
It follows from \eqref{EqIntegratedBoundP} (and its analogue for $\Pi_{(1)})$ that the limit
\begin{equation}\label{EqPfRingConvergence}
    \lim_{\ub \to \infty} \mathbb{P}_{\Sp^2} \circ \Pi (\ub, \mathring{\vartheta}_{\CH^+}) = \lim_{\ub \to \infty}\vartheta_{=}(\ub) =: \mathring{\vartheta}_{=} \in \Sp^2
\end{equation}
exists, where $\mathbb{P}_{\Sp^2} : \R \times \Sp^2 \to \Sp^2$ denotes the projection onto the sphere. 
We now claim that
\begin{equation}
    \label{EqClaimProjectionGamma}
    \lim_{v_+ \to \infty} \mathbb{P}_{\Sp^2} \circ \Pi^{-1}(v_+, \mathring{\vartheta}_{=}) = \mathring{\vartheta}_{\CH^+} \;.
\end{equation}
To see this, we first note that in light of \eqref{EqMonP} the curve \eqref{EqPfRingCH} can be reparameterized by $v_+$, i.e., $v_+ \mapsto \big(v_+, \vartheta_{=}(v_+) \big)$, and by Lemma \ref{LemRelationStarAndNullCoord}  we have $v_+(\ub, \mathring{\vartheta}_{\CH^+}) \to \infty$ for $\ub \to \infty$.   
But then, using the uniform angular Lipschitz property \eqref{EqPointwiseBoundsPInverse} of $\Pi^{-1}$, we conclude
\begin{equation*}
    d_{\Sp^2} \big( \mathbb{P}_{\Sp^2} \circ \Pi^{-1} (v_+, \mathring{\vartheta}_{=}), \mathring{\vartheta}_{\CH^+}\big) = d_{\Sp^2} \Big( \mathbb{P}_{\Sp^2} \circ \Pi^{-1} (v_+, \mathring{\vartheta}_{=}), \mathbb{P}_{\Sp^2} \circ \Pi^{-1} \big(v_+, \vartheta_{=}(v_+)\big)\Big) \lesssim d_{\Sp^2}(\mathring{\vartheta}_{=}, \vartheta_{=}(v_+) \to 0
\end{equation*}
 for $v_+ \to \infty$ by \eqref{EqPfRingConvergence}. Here, $d_{\Sp^2}$ denotes the canonical angular distance function on $\Sp^2$.

Having established \eqref{EqClaimProjectionGamma}, we can now choose $v_{2+}$ large enough and $W_{\Gamma}$ to be a small enough neighborhood of $\mathring{\vartheta}_{=}$ such that \eqref{EqPropProjGamma} holds, where one uses again \eqref{EqPointwiseBoundsPInverse} and \eqref{EqClaimProjectionGamma}.
\end{proof}

\subsection{The propagation} \label{SecPropagation}

\begin{proposition} \label{PropWindingCurveGammaL2}
    Consider the parameterisation of $\Gamma \cap \{f_+ \geq v_1\} \simeq [v_{1+}, \infty) \times \Sp^2$ in terms of $(v_+, \theta, \varphi_=)$-coordinates. For every open set $W_{\Gamma} \subseteq \Sp^2$ in these coordinates there exists an open subset $V_{\Gamma} \subseteq W_{\Gamma}$ such that
    \begin{equation} \label{EqPropOnGammaForPsi}
        \int_{v_{1+}}^\infty v_+^q \Big| \iint_{V_\Gamma} \psi_0 \, \sin \theta \ud  \theta \ud  \varphi_= \Big|^2 \, \ud v_+ = \infty \;.
    \end{equation}
\end{proposition}

For the proof we will need the following two elementary lemmas.

\begin{lemma} \label{LemNormedSpace}
    Let $V$ denote a normed vector space based on functions $f : [1, \infty) \to \C$.\footnote{For example below we will consider $V$ to be a weighted $L^2$-space.} Let $N \in \N$ and for $i \in \{1, \ldots, N\}$ let  $f_i : [1, \infty) \to \C$ be functions which do not belong to $V$. Then there exists an $\vec{n} \in \C^N$ such that if for $\vec{a} \in \C^N$ we have $\vec{a} \cdot \vec{f} := \sum_{i=1}^N a_i f_i \in V$, then $\vec{n} \cdot \vec{a} =0$.
\end{lemma}
In other words, the set of linear combinations of $f_i$ which lie in $V$ form at most an $N-1$-dimensional hyperplane in $\C^N$.
\begin{proof}
    There exist at most $N-1$ linearly independent $\vec{a}_k \in \C^N$ such that $\vec{a}_k \cdot \vec{f} \in V$. If there existed $N$ linearly independent $\vec{a}_k \in C^N$ with $\vec{a}_k \cdot \vec{f} \in V$, then there exists $c_k \in \C$, $k = 1, \ldots, N$ with $f_1 = (1, 0, \ldots, 0) \cdot \vec{f} = (c_1 \vec{a}_1 + \ldots + c_N \vec{a}_N) \cdot \vec{f} \in V$, which is a contradiction. But this implies the existence of an $\vec{n} \in \C^N$ as in the statement of the lemma.
\end{proof}

\begin{lemma} \label{LemLinIndSpherHarm}
    For $N \in \N$ and $i \in \{1, \ldots, N\}$ let $Y_i := Y^{[2]}_{m(i)l(i)}$ denote a finite number of different spin $2$-weighted spherical harmonics and let $W \subseteq \Sp^2 \setminus\{\theta = 0, \pi\}$. Then there does not exist an $ 0 \neq \vec{n} \in \C^N $ such that $\vec{n} \cdot \vec{Y} := \sum_{i=1}^N n_i Y_i $ vanishes identically on $W$.
\end{lemma}

\begin{proof}
    Assuming the existence of such an $\vec{n}$, since the spin $2$-weighted spherical harmonics are analytic on $\Sp^2 \setminus\{\theta = 0, \pi\}$, it would follow that $\vec{n} \cdot \vec{Y} =0$ on all of $\Sp^2 \setminus\{\theta = 0, \pi\}$. But this contradicts the fact that different spin $2$-weighted spherical harmonics are $L^2$-orthogonal to each other, in particular linearly independent. 
\end{proof}

\begin{proof}[Proof of Proposition \ref{PropWindingCurveGammaL2}]
    Since the spin $2$-weighted spherical harmonics $Y^{[2]}_{ml}$ form an orthonormal basis of $L^2(\Sp^2)$, and since the Boyer--Lindquist spheres are contained in $\Gamma$, we directly obtain from \eqref{EqThmPropToGamma}
    \begin{align*}
        &\int\limits_{\Gamma \cap \{f_+ \geq v_1\}} v_+^q |(\psi_0)_{l=2} - (\psil)_{l=2}|^2 \vols \ud v_+ \leq C  \\
        &\int\limits_{\Gamma \cap \{f_+ \geq v_1\}} v_+^q |(\psi_0)_{l>2} - (\psil)_{l>2}|^2 \vols \ud v_+ \leq C \;.
    \end{align*}
    Together with \eqref{EqPropEstLinearL2}, \eqref{EqPropEstLinearL>2} this gives
    \begin{align}
        &\infty = \int\limits_{\Gamma \cap \{f_+ \geq v_1\}} v_+^q |(\psi_0)_{l=2}|^2 \vols \ud v_+ = \int_{v_{1+}}^\infty v_+^q \sum_{|m| \leq 2}\big|(\psi_0)_{m2}|_{\Gamma}\big|^2 \, \ud v_+ \label{EqPfL=2} \\
        &\infty >  \int\limits_{\Gamma \cap \{f_+ \geq v_1\}} v_+^q |(\psi_0)_{l>2}|^2 \vols \ud v_+ \label{EqPfL>2} \;.
    \end{align}
We now consider the $(v_+, \theta, \varphi_=)$-coordinates on $\Gamma$ and claim that for every open set $W_{\Gamma} \subseteq \Sp^2$ in these coordinates there exists an open subset $V_{\Gamma} \subseteq W_{\Gamma}$ such that
    \begin{equation} \label{EqClaimL=2}
        \int_{v_{1+}}^\infty v_+^q \Big| \iint_{V_\Gamma} (\psi_0)_{l=2} \, \sin \theta \ud  \theta \ud  \varphi_= \Big|^2 \, \ud v_+ = \infty \;.
    \end{equation}
Once \eqref{EqClaimL=2} is established, \eqref{EqPropOnGammaForPsi} follows by contradiction from first using $(\psi_0)_{l=2} = \psi_0 - (\psi_0)_{l>2}$ and the Minkowski inequality which gives
\begin{equation*}
    \Big(\int_{v_{1+}}^\infty v_+^q \Big| \iint_{V_\Gamma} (\psi_0)_{l=2}) \vols \Big|^2 \ud v_+\Big)^{\frac{1}{2}} \leq \Big(\int_{v_{1+}}^\infty v_+^q \Big| \iint_{V_\Gamma} \psi_0 \vols \Big|^2 \ud v_+\Big)^{\frac{1}{2}} + \Big(\int_{v_{1+}}^\infty v_+^q \Big| \iint_{V_\Gamma} (\psi_0)_{l>2}) \vols \Big|^2 \ud v_+\Big)^{\frac{1}{2}}
\end{equation*}
    and then concluding with H\"older on $V_\Gamma \subseteq \Sp^2$ and \eqref{EqPfL>2} for the last term. Thus, it remains to prove \eqref{EqClaimL=2}.

Let now $W_{\Gamma} \subseteq \Sp^2$ be given. Recall that we have $\varphi_+ = \varphi_= + \frac{a}{r_-^2 + a^2} v_+$. Thus, for $V_{\Gamma} \subseteq W_\Gamma$ to be determined momentarily, we have
\begin{equation} \label{Eqfm}
    \int\limits_{v_{1+}}^\infty v_+^q \Big| \iint\limits_{V_\Gamma} (\psi_0)_{l=2} \, \sin \theta \ud  \theta \ud  \varphi_= \Big|^2 \, \ud v_+ = \int\limits_{v_{1+}}^\infty v_+^q \Big| \sum_{|m| \leq 2} \iint\limits_{V_\Gamma} Y^{[2]}_{m2}(\theta, \varphi_=) \, \sin \theta \ud \theta \ud  \varphi_= \cdot \underbrace{e^{im\frac{a}{r_-^2 + a^2} v_+} (\psi_0)_{m2}|_\Gamma (v_+)}_{=: f_m} \Big|^2 \, \ud v_+
\end{equation}
Let $V$ denote the space of measurable functions $f : [v_{1+}, \infty) \to \C$ such that $\int_{v_{1+}}^\infty v_+^q |f|^2 \, dv_+ < \infty$. It then follows from \eqref{EqPfL=2} that there is at least one $f_m$ which is not contained in $V$ and at most $5$. Let $1 \leq N \leq 5 $ denote the number of such $f_m$ and relabel those which are not contained in $V$ by $f_{m(i)}$ with $i \in \{1, \ldots, N\}$. Note that those remaining $f_m$ which are contained in $V$ do not impact on whether the sum inside the absolute value on the right-hand side of \eqref{Eqfm} is contained in $V$ or not. Let $\vec{f}$ denote $(f_{m(1)}, \ldots, f_{m(N)})$.

By Lemma \ref{LemNormedSpace} there exists an $\vec{n} \in \C^N$ with $\vec{a} \cdot  \vec{f} \in V$ implying $\vec{n} \cdot \vec{a} =0$.  By Lemma \ref{LemLinIndSpherHarm} there exists a point $\mathring{\vartheta}_= \in W_\Gamma \subseteq \Sp^2$ with $\vec{n} \cdot \vec{Y}(\mathring{\vartheta}_=) := \sum_{i = 1} n_i Y^{[2]}_{m(i)2}(\mathring{\vartheta}_=) \neq 0$. This, however, implies the existence of a small neighborhood $V_{\Gamma} \subseteq W_\Gamma$ of $\mathring{\vartheta}_=$ such that $\vec{n} \cdot \iint_{V_\Gamma} \vec{Y} (\theta, \varphi_=) \, \sin \theta \ud  \theta \ud  \varphi_= \neq 0$, which concludes the proof of the claim \eqref{EqClaimL=2}.
\end{proof}

\begin{corollary}\label{CorLocalGamma}
     Consider the parameterisation of $\Gamma \cap \{f_+ \geq v_1\} \simeq [v_{1+}, \infty) \times \Sp^2$ in terms of $(v_+, \theta, \varphi_=)$-coordinates. For every open set $W_{\Gamma} \subseteq \Sp^2$ in these coordinates we have
    \begin{equation} \label{EqCorLocalGamma}
        \int_{v_{1+}}^\infty  \iint_{W_\Gamma} v_+^q  |\psi_0|^2\, \vols \ud v_+ = \infty \;.
    \end{equation}
\end{corollary}

\begin{proof}
    This follows trivially by contradiction: if \eqref{EqCorLocalGamma} were finite for some $W_{\Gamma} \subseteq \Sp^2$, then by Proposition \ref{PropWindingCurveGammaL2} there exists $V_{\Gamma} \subseteq W_{\Gamma}$ such that
    $$\infty = \int_{v_{1+}}^\infty v_+^q \Big| \iint_{V_\Gamma} \psi_0 \, \sin \theta \ud  \theta \ud  \varphi_= \Big|^2 \, \ud v_+ \leq \vols(V_{\Gamma})  \cdot \int_{v_{1+}}^\infty v_+^q \iint_{V_\Gamma} |\psi_0|^2 \, \vols \ud v_+ <\infty \;,$$
    where we have used H\"older's inequality.
\end{proof}

We now translate the blow-up \eqref{EqCorLocalGamma} of the curvature component $\psi_0$ on $\Gamma$, which is with respect to the dynamical principal null frame, to blow-up of curvature with respect to the dynamical double null frame on $\Gamma$. The reason for this is that it is slightly more convenient to propagate the blow-up of curvature from $\Gamma$ to $\CH^+$ in the dynamical double null gauge. However, note that when translating \eqref{EqCorLocalGamma} into $\alpha$, background curvature terms enter; it is here that the $q$-dependent choice of $\Gamma$ becomes important in generating enough decay from the degenerate $e_4$ vector field to beat the $v_+^q$-weight in \eqref{EqCorLocalGamma}. In particular, from now on we stop considering difference quantities.

\begin{proposition} \label{PropTransitionToAlpha}
    Consider the parameterisation of $\Gamma \cap \{f_+ \geq v_1\} \simeq [v_{1+}, \infty) \times \Sp^2$ in terms of $(v_+, \theta, \varphi_=)$-coordinates. For every open set $W_{\Gamma} \subseteq \Sp^2$ in these coordinates we have
    \begin{equation} \label{EqPropLocalGammaAlpha}
        \int_{v_{1+}}^\infty  \iint_{W_\Gamma} v_+^q   |\Omega^4\alpha|_\gamma^2\, \vols \ud v_+ = \infty \;.
    \end{equation}
\end{proposition}
Recall that $\alpha$, as in Section \ref{sec:DL.refined}, is with respect to $\os{\ee}{2}_4$ which is non-degenerate at $\CH^+$.

\begin{proof}
   
Recall that $m' = \frac{1}{\sqrt{2}} \frac{ \sqrt{\Sigma}}{r + ia \cos \theta}(e_1' + ie_2') =: \frac{1}{\sqrt{2}}\mathfrak{c} (e_1' + i e_2')$, where we have defined the complex function $\mathfrak{c}$ of norm one. Also recall that $m = \frac{1}{\sqrt{2}}\mathfrak{c}(e_1 + i e_2)$. Furthermore, the double null vector fields $\os{\ee}{2}_3$, $\os{\ee}{2}_4$ which are regular at $\CH^+$, are related to $e_3'$ and $e_4'$ by $e_3' = \frac{1}{\Omega^2} \ee_3'$ and $e_4' = \Omega^2 \ee_4'$. We obtain from \eqref{EqDefDynamicalPNFrame} together with \eqref{eq:principal.null.in.terms.of.double.null.1}, \eqref{eq:principal.null.in.terms.of.double.null.2}
\begin{equation} \label{EqAsympm}
    \begin{split}
        m &= \underbrace{\frac{1}{2} \Big( \frac{R}{\ell} \frac{\rd \theta}{\rd \theta_*} \sqrt{\Sigma} + \frac{r^2 + a^2}{R \sqrt{\Sigma}}}_{\to 1} \Big) m' + \underbrace{ \frac{1}{2} \frac{\mathfrak{c}}{\overline{\mathfrak{c}}} \Big(\frac{R}{\ell} \frac{\rd \theta}{\rd \theta_*} \sqrt{\Sigma} - \frac{r^2 + a^2}{R \sqrt{\Sigma}} \Big)}_{\sim |\Delta|} \overline{m}' \\
        & \quad + \underbrace{ \frac{\mathfrak{c}}{2 \sqrt{2} \Omega^2} \Big( \frac{i a \Si \Delta}{\sqrt{\Sigma} R^2} - \sqrt{\Sigma} \frac{\rd \theta}{\rd r_*} \Big)}_{\sim 1} \ee_3' + \underbrace{\frac{\mathfrak{c} \Omega^2}{2 \sqrt{2}} \Big( \frac{ ia \Si}{\sqrt{\Sigma}} + \frac{R^2 \sqrt{\Sigma}}{\Delta} \frac{\rd \theta}{\rd r_*} \Big)}_{\sim |\Delta|} \ee_4' \;,
    \end{split}
\end{equation}
where we have also given the asymptotics of the coefficients for $r \to r_-$. For the asymptotics, we have used Lemma \ref{LemPartialsLimit}, \eqref{EqMagicFormula}, $\frac{\Omega^2}{{\boldsymbol \Omega}^2} \sim 1$ (from \eqref{NS.def}) and \eqref{Kerr.metric.comp}. We emphasize here that $\Omega^2$ is the component of the \emph{dynamical} metric, while all other quantities in the underbraced coefficients are background quantities (which, however, do not have a dynamical analogue and for this reason are not in bold).  Similarly we obtain from \eqref{eq:principal.null.in.terms.of.double.null.4}
\begin{equation} \label{EqAsympe4}
    \begin{split}
        e_4 &= \underbrace{ \frac{\sqrt{2}}{2 \mathfrak{c}} \Big( \frac{\Sigma R}{\ell} \frac{\rd r}{\rd \theta_*} + \frac{a \Delta \Si}{i R}\Big)}_{\sim |\Delta|} m' + \underbrace{\frac{\sqrt{2}}{2 \overline{\mathfrak{c}}} \Big( \frac{\Sigma R}{\ell} \frac{\rd r}{\rd \theta_*} - \frac{a \Delta \Si}{R i} \Big)}_{\sim |\Delta|} \overline{m}' \\
        &\quad + \underbrace{\Big( \frac{\Delta (r^2 + a^2)}{2 R^2} - \frac{\Sigma}{2} \frac{\rd r}{\rd r_*} \Big) \frac{1}{\Omega^2}}_{\sim |\Delta|} \ee_3' + \underbrace{\underbrace{\Big( \frac{r^2 + a^2}{2} + \frac{\Sigma R^2}{2 \Delta} \frac{\rd r}{\rd r_*} \Big)}_{\to r_-^2 + a^2} \Omega^2}_{ \sim |\Delta|} \ee_4' \;.
    \end{split}
\end{equation}
We now use \eqref{EqAsympm} and \eqref{EqAsympe4} to write $\psi_0 = R(e_4,m,e_4,m)$ in terms of the double null curvature components $\alpha, \beta, \rho, \sigma, \betab, \alphab$ (all with respect to $\os{\ee}{2}_3$, $\os{\ee}{2}_4$) to obtain
\begin{equation} \label{EqPsi0ToAlpha}
\begin{split}
    |\psi_0| &= |R(e_4,m,e_4,m)| \\
    &\leq \big[ (r_-^2 + a^2)^2 + \mathcal{O}(|\Delta|) \big] \cdot \Omega^4  \frac{1}{\sqrt{2}}|\alpha|_\gamma + \underbrace{\mathcal{O}(|\Delta|^2) \big( |\beta|_\gamma + |\rho| + |\sigma| + |\betab|_\gamma + |\alphab|_\gamma \big)}\;.
    \end{split}
\end{equation}
This is seen as follows: first we recall that by virtue of the vacuum Einstein equations any curvature component with respect to the frame $(m', \overline{m}', \os{\ee}{2}_3, \os{\ee}{2}_4)$ can be expressed in terms of $\alpha, \beta, \rho, \sigma, \betab, \alphab$ and $\gamma, \epsilon$ evaluated on $m'$ and/or $\overline{m}'$.  Hence, $R(e_4,m, e_4,m)$ can be written as a linear combination  of those null curvature components. Since each coefficient in \eqref{EqAsympe4} is of order $\mathcal{O}(|\Delta|)$ -- and the coefficients in \eqref{EqAsympm} are regular -- each coefficient in the linear combination of those null curvature components is at least $\mathcal{O}(|\Delta|^2)$. This in particular yields the underbraced terms in \eqref{EqPsi0ToAlpha}. The leading order coefficient of $\alpha(m', m')$ in the linear combination comes from the $m'$ terms in \eqref{EqAsympm} and the $\os{\ee}{2}_4$ terms in \eqref{EqAsympe4} -- all other terms are subleading. We also use $|\alpha(m',m')| = \frac{1}{\sqrt{2}}|\alpha|_\gamma = |\alpha(\mb',\mb')|$ and $\alpha(m',\mb') = 0$. This shows \eqref{EqPsi0ToAlpha}.  

Furthermore we have
\begin{equation*}
\begin{split}
    \int_{v_{1+}}^\infty  \iint_{W_\Gamma} v_+^q \Delta^4 \big( |\beta|^2_\gamma + |\rho|^2 + |\sigma|^2 + |\betab|^2_\gamma + |\alphab|^2_\gamma \big) \vols \ud v_+    \leq  \int_{v_{1+}}^\infty v_+^q v_+^{-2 \sigma_q} v_+^{-2 \min\{\frac{q_- - 3}{2}, \sigma_q\}} \ud v_+ < \infty\;,
    \end{split}
\end{equation*}
where we have used \eqref{EqDeltaFutureGamma} together with the first two points of Proposition \ref{PropEstimatesNonDifference} below and $q - 2 \sigma_q - 2 \frac{q_- - 3}{2} = q - q_- + \frac{3}{2} - \frac{q}{2} <-1$ for $q \geq 7$ and $q - 4 \sigma_q = -3 <-1$.
Hence, Proposition \ref{PropTransitionToAlpha} now follows in conjunction with \eqref{EqPsi0ToAlpha} and Corollary \ref{CorLocalGamma}. \qedhere


\end{proof}

We have the following bounds on the dynamical (non-difference) quantities in particular on and to the future of $\Gamma$. We use here a schematic notation as in \cite{DafLuk17}, similar to \eqref{S.def}, except for non-difference quantities. In particular, we write
\begin{equation}
\mathcal S_{\psi} := \{\eta,\,\etab,\, \zeta\},\quad\mathcal S_{\psi_{\Hb}} := \{\slashed{tr}\chib,\,\chibh\}, \quad \mathcal S_{\psi_H}:= \{\slashed{tr}\chi,\,\chih\}
\end{equation}
so that $\psi \in \mathcal S_{\psi}$, $\psi_{\Hb} \in \mathcal S_{\psi_{\Hb}}$ and $\psi_H \in \mathcal S_{\psi_H}$.
\begin{proposition} \label{PropEstimatesNonDifference} We have  on and to the future of $\Gamma$
\begin{multicols}{2}
   \begin{enumerate}[leftmargin=*]
    \item $||(\ub^{\f{q_--3}{2}}\wedge \ub^{\sigma_q})(\pmb{\Omega}^2 \beta, \pmb{\Omega}^2 \rho, \pmb{\Omega}^2 \sigma)||_{L^\infty_uL^\infty_{\ub}L^\infty(S)} \lesssim 1 $
    \item $||(|u|^{\f{q_--3}{2}}\wedge \ub^{\sigma_q}) (\betab, \alphab)||_{L^\infty_uL^\infty_{\ub}L^\infty(S)} \lesssim 1$ 
   \item $||(|u|^{\f{q_--3}{2}}\wedge |u|^{\sigma_q}) \pHb ||_{L^\infty_uL^\infty_{\ub}L^\infty(S)} \lesssim 1 $  
   \item $|| (\ub^{\f{q_{--}-2}2} \wedge \ub^{\sigma_{q_-}+\f 12}) \pmb{\Omega}^2 \pH ||_{L^\infty_u L^2_{\ub} L^2(S)}  \lesssim 1$ 
   \item  $||(\ub^{\f{q_--3}{2}}\wedge \ub^{\sigma_q}) \pmb{\Omega}^2 \pH||_{L^\infty_u L^\infty_{\ub} L^\infty(S)} \lesssim 1$. 
\item $\sum_{i=0}^1 || (\ub^{\f{q_{--}-2}2}\wedge \ub^{\sigma_{q_-}+\f 12}) \pmb{\Omega}^2 \nab^i \beta||_{L^\infty_u L^2_{\ub}L^2(S)} \lesssim 1$
\item $\sum_{i=0}^1 ||\nab^i \psi||_{L^\infty_u L^\infty_{\ub} L^\infty(S)} \lesssim 1$
\item $||K||_{L^\infty_uL^\infty_{\ub}L^\infty(S)} \lesssim 1$,
  \end{enumerate}
  \end{multicols}
  where $\sigma_q= \f 14(q+3)$ as before, $\sigma_{q_-} = \f 14(q_-+3)$, $q_{--} < q_-$ but arbitrarily close to $q_-$, and we used $\wedge$ to denote minimums.
\end{proposition}

\begin{proof}
    For each of the geometric quantities in question, we split into the background part and the difference, i.e., we write $\bt = \pmb{\bt} + \widetilde{\bt}$, etc., and we will prove that all the stated bounds hold separately for the background and the difference (which explains taking the minimums of two weights). It is straightforward to check that for the differences, the estimates follow from Proposition~\ref{prop:derivatives.all.directions} and Proposition~\ref{prop:curvature.original} (where the weights correspond to the first in the minimums and we integrate in $\ub$ for points 4 and 6), using also Sobolev embedding \eqref{eq:Sobolev.BS} on the spheres.

    We now turn to the background quantities. For this, we will freely use that to the future of $\Gamma$, 
\begin{equation}\label{eq:Delta.after.Gamma.again}
\Omg^2 \sim \pmb{\Omg}^2 \sim |\Delta|\ls \ub^{-\sigma_q}\quad \hbox{with $\sigma_q= \f 14(q+3)$}
\end{equation}
which follows from by combining \eqref{EqDeltaFutureGamma} and Lemma~\ref{LemRelationStarAndNullCoord}.
    \begin{enumerate}
        \item We first use that all the written background quantities (without $\pmb{\Omg}^2$ weights), i.e., $\pmb{\bt}$, $\pmb{\rho}$, $\pmb{\sigma}$, $\pmb{\betab}$, $\pmb{\alphab}$, $\pmb{\psi_{\Hb}}$, $\pmb{\psi_H}$, $\pmb{\psi}$, $\pmb{K}$, are bounded. We claim that this gives the desired bounds for the background contributions in points 1, 4, 5, 6, 7, 8. Indeed, this is obvious for 7, 8, which only requires an $L^\i_uL^\i_{\ub}L^\i(S)$ bound. For points 1 and 5, we use additionally \eqref{eq:Delta.after.Gamma.again} so that we can put in $\ub^{\f{q+3}{4}}$ weights. For points 4, 6, we again use \eqref{eq:Delta.after.Gamma.again} and note that $\| \ub^{-\f 12+\f 14(q-q_-)} \|_{L^2_{\ub}}$ is finite. 
        \item We now turn to the background contributions for points 2 and 3. For this we need that these components are better, and that $|\pmb{\alphab}|_\gamma, \, |\pmb{\betab}|_\gamma,\,|\pmb{\nabla} \pmb{\betab}|, |\pmb{\psi_{\Hb}}| \ls |\Delta|$. The bounds for $\pmb{\betab}$, $\pmb{\nabla} \pmb{\betab}$, and  $\pmb{\psi_{\Hb}}$ were proven in \cite[Propositions~A.19, A.23]{DafLuk17}. That for $\pmb{\alphab}$ is not contained in \cite{DafLuk17}, but can be proven similarly by taking another derivative of $\pmb{\chib}$. With the extra power of $|\Delta|$, the estimates in points 2 and 3 are then a consequence of \eqref{eq:Delta.after.Gamma.again} and \eqref{EqUVFutureGamma}. \qedhere
    \end{enumerate}

\end{proof}

We now consider the coordinate system $(u, \ub_{\CH^+}, \theta^1_{(i), \CH^+}, \theta^2_{(i), \CH^+})$ from Section~\ref{sec:CH.coord}, $i=1,2$, where $(\theta^1_{(i), \CH^+}, \theta^2_{(i), \CH^+})$ are coordinates on $\calV_i' \subseteq \Sp^2$, with $\calV_1' \cup \calV_2' = \Sp^2$. Recall that the metric extends continuously to the Cauchy horizon $\{\ub_{\CH^+} = 0\}$ in these coordinates.

\begin{theorem} \label{ThmBlowUpCH}
    Let $\overline{p} =(u_0, 0, \mathring{\vartheta}_{\CH^+}) \in \CH^+ $ for some $-\infty < u_0 < u_f$ and $\mathring{\vartheta}_{\CH^+} \in \Sp^2$. For any $\lambda_0>0$ so small that $u_0 + \lambda_0 <u_f$ and any open neighborhood $W_{\CH^+} \subseteq \Sp^2$ of $\mathring{\vartheta}_{\CH^+}$, we have
    \begin{equation} \label{EqThmIntegratedBlowUpWeightedAlpha}
        \int\limits_{\ub_{0,\CH^+}}^0  \int\limits_{u_0 - \lambda_0}^{u_0 + \lambda_0} \;\iint\limits_{W_{\CH^+}} \ub^q  \Omega^{-2}|\Omega^4\alpha|_{\gamma}^2  \, \volg \,\ud u \ud \ub_{\CH^+} = \infty \;,
    \end{equation}
    where $\ub_{0,\CH^+} <0$ and the integration is with respect to $(u, \ub_{\CH^+}, \theta_{\CH^+})$ coordinates.
\end{theorem}

\begin{remark} \label{RemThmBlowUpCH}
    By $\Omega \sim \boldsymbol{\Omega}$, the finite range of the $u$-coordinate in the integration, and the definition of $\ub_{\CH^+}$, it follows that \eqref{EqThmIntegratedBlowUpWeightedAlpha} is equivalent to
    \begin{equation} \label{EqThmIntegratedBlowUpWeightedAlpha2}
        \int\limits_{\ub_0}^\infty\int\limits_{u_0 - \lambda_0}^{u_0 + \lambda_0} \;\iint\limits_{W_{\CH^+}} \ub^q |\Omega^4\alpha|_{\gamma}^2  \, \volg \,\ud u \ud \ub = \infty \;.
    \end{equation}
\end{remark}

\begin{proof}
Indeed, we will prove Theorem \ref{ThmBlowUpCH} by proving \eqref{EqThmIntegratedBlowUpWeightedAlpha2}.
Recall the Bianchi equation (see \cite[(3.6)]{DafLuk17})
\begin{equation} \label{EqBianchiAlpha}
    \nab_3 \alpha = - \frac{1}{2} \trchb \alpha + \nab \hot \beta + 4 \omegab \alpha - 3(\chih \rho + {}^*\chih \sigma) + (\zeta + 4 \eta) \hot \beta \;.
\end{equation}
All quantities here are with respect to the regular double null frame $\os{\ee}{2}_3$, $\os{\ee}{2}_4$. Also recall that $\os{\ee}{2}_3 = \frac{\rd}{\rd u}$ in the $(u, \ub_{\CH^+}, \theta^1_{(i), \CH^+}, \theta^2_{(i), \CH^+})$ coordinates. We now use $\omegab = - \nab_3(\log \Omega) = - \frac{\rd_u \Omega}{\Omega}$ (see \cite[Proposition 2.3]{DafLuk17}) and  $\rd_u (\det \gamma_{\CH^+})^{\frac{1}{4}} = \frac{1}{2}(\det \gamma_{\CH^+})^{\frac{1}{4}} \cdot \trchb$ (see \cite[Proposition 2.4]{DafLuk17}), where in the latter $\gamma_{\CH^+}$ stands for the components of $\gamma$ with respect to the $\theta_{\CH^+}$ coordinates. This gives
$$\nab_3\big( (\det \gamma_{\CH^+})^{\frac{1}{4}} \cdot \Omega^4 \alpha  \big) = - (\det \gamma_{\CH^+})^{\frac{1}{4}} 4 \omegab \Omega^4 \alpha  + \frac{1}{2} (\det \gamma_{\CH^+})^{\frac{1}{4}} \trchb \Omega^4 \alpha + (\det \gamma_{\CH^+})^{\frac{1}{4}} \Omega^4 \nab_3 \alpha \;.$$ Together with \eqref{EqBianchiAlpha}, \eqref{eq:Bianchi.for.curvature.3}, \eqref{eq:Bianchi.for.curvature.4}, and $\zeta = \eta$ in the given gauge -- see \cite[Proposition 2.3]{DafLuk17} -- we obtain
\begin{equation*}
    \begin{split}
        \nab_3 ( (\det \gamma_{\CH^+})^{\frac{1}{4}} \cdot \Omega^4 \alpha) = (\det \gamma_{\CH^+})^{\frac{1}{4}} \Omega^4 \Big[ \nab \hot \beta + 3 \chih K - &\frac{3}{2} \chih (\chih \cdot \chibh) + \frac{3}{4} \chih \trch \trchb \\
        &- 3 {}^*\chih \curl \eta  + \frac{3}{2} {}^* \chih (\chibh \wedge \chih) +5 \eta \hot \beta \Big] \,.
    \end{split}
\end{equation*}
Using the schematic notation of \cite[Section 3.2]{DafLuk17} we get
\begin{equation} \label{EqBianchiAlphaSchematic}
    \nab_3 ((\det \gamma_{\CH^+})^{\frac{1}{4}} \Omega^4 \alpha) \eqs   (\det \gamma_{\CH^+})^{\frac{1}{4}} \Omega^4 \Big[ (\nab, \psi) \beta + \pH K + \pH \nab \psi + \pH^2 \pHb \Big]
\end{equation}
We now recall from Lemma \ref{LemGammaGraph} that for given $(\ub,\vartheta_*) \in [\ub_0, \infty) \times \Sp^2$ the value $u_\Gamma(\ub,\vartheta_*)$ is such that $(u_\Gamma(\ub,\theta_*), \ub, \vartheta_*) \in \Gamma$. Recall also that the map $(\ub, \vartheta_{\CH^+}) \overset{\calC}{\mapsto} (\ub, \vartheta_{*})$ is a diffeomorphism.
We then have
\begin{equation*}
    \begin{split}
        |\Omega^4 \alpha|^2_\gamma &(\det \gamma_{\CH^+})^{\frac{1}{2}}\big(u_\Gamma \circ \calC (\ub, \vartheta_{\CH^+}), \ub, \vartheta_{\CH^+}\big) - |\Omega^4 \alpha|_\gamma^2 (\det \gamma_{\CH^+})^{\frac{1}{2}}(u, \ub, \vartheta_{\CH^+}) \\
        &\leq \int_{u_\Gamma}^u \Big| 2 \langle \Omega^4 \alpha (\det \gamma_{\CH^+})^{\frac{1}{4}}, \nab_3\big( (\det \gamma_{\CH^+})^{\frac{1}{4}} \Omega^4 \alpha\big)\rangle_\gamma \Big| (u', \ub, \vartheta_{\CH^+}) \ud u' \\
        &\lesssim \int_{u_\Gamma}^u \Big| \langle \Omega^4 \alpha,  \Omega^4 \pH^2 \pHb + u^{-\f 58} u^{\f 58} \Omg^4 \big[(\nab, \psi) \beta + \pH K + \pH \nab \psi \big] \rangle_\gamma \Big|  (\det \gamma_{\CH^+})^{\frac{1}{2}}     (u', \ub, \vartheta_{\CH^+}) \ud u' \\
        &\lesssim \int_{u_\Gamma}^u \Big( |\pHb|_\gamma |\Omega^4 \alpha|_\gamma^2 + |\pHb|_\gamma |\Omega^4 \pH^2|_\gamma^2 + u^{-\f 54}  |\Omega^4 \alpha|_\gamma^2  \\
        &\hspace*{4cm} + u^{\f 54}\Omg^4 |\Omega^2 \big[(\nab, \psi) \beta + \pH K + \pH \nab\psi\big]|_\gamma^2 \Big) (\det \gamma_{\CH^+})^{\frac{1}{2}}(u', \ub, \vartheta_{\CH^+}) \ud u'\;,
    \end{split}
\end{equation*}
where we have used Cauchy-Schwarz in the last inequality and where the implicit constants only depend on the numerical constants dropped in the schematic notation \eqref{EqBianchiAlphaSchematic}. 
Observe that the weight $|u|^{-\f 54}$ is integrable in $u$ (since $u_f \leq -1$). Furthermore by Proposition \ref{PropEstimatesNonDifference} $|\pHb|_\gamma  $ is uniformly (in $\ub$ and $\theta_{\CH^+}$) integrable in $u$ so that Gr\"onwall's inequality gives
\begin{equation*}
\begin{split}
    |\Omega^4 \alpha|^2_\gamma &(\det \gamma_{\CH^+})^{\frac{1}{2}}\big(u_\Gamma\circ \calC (\ub, \vartheta_{\CH^+}), \ub, \vartheta_{\CH^+}\big) \lesssim |\Omega^4 \alpha|_\gamma^2 (\det \gamma_{\CH^+})^{\frac{1}{2}}(u, \ub, \vartheta_{\CH^+}) \\ 
    &+ \int_{u_\Gamma}^{u_f} \Big(|\pHb|_\gamma |\Omega^4 \pH^2|_\gamma^2 + u^{\f 54} \Omg^4 |\Omega^2 \big[(\nab, \psi) \beta + \pH K + \pH \nab\psi\big]|_\gamma^2 \Big)(\det \gamma_{\CH^+})^{\frac{1}{2}} (u', \ub, \vartheta_{\CH^+}) \ud u'
    \end{split}
\end{equation*}
with the implicit constant being independent of $\ub \geq \ub_0$, $\vartheta_{\CH^+} \in \Sp^2$, and $ u_{\Gamma}<u< u_f$.

We now multiply both sides by $\ub^{q}$ and integrate in $\iint\limits_{W_{\CH^+}} \ud  \theta_{\CH^+}^1 \ud \theta_{\CH^+}^2$, $\int\limits_{u_0 - \lambda_0}^{u_0 + \lambda_0} \ud u$, and $\int\limits_{\ub_0}^\infty \ud  \ub$ to obtain
\begin{equation}\label{EqPfThmPropCauchy1}
\begin{split}
   \int\limits_{\ub_0}^\infty& \,\iint\limits_{W_{\CH^+}}  \ub^{q}|\Omega^4 \alpha|^2_\gamma(u_\Gamma\circ \calC(\ub, \vartheta_{\CH^+}), \ub, \vartheta_{\CH^+}) \volg \ud  \ub \\
   &\lesssim \int\limits_{\ub_0}^\infty \int\limits_{u_0 - \lambda_0}^{u_0 + \lambda_0} \,\iint\limits_{W_{\CH^+}} \ub^{q} |\Omega^4 \alpha|_\gamma^2 (u, \ub, \vartheta_{\CH^+}) \volg \ud u \ud  \ub\\ 
    &\; \;+ \int\limits_{\ub_0}^\infty \int\limits_{u_\Gamma}^{u_f} \,\iint\limits_{W_{\CH^+}} \ub^{q} \Big(\underbrace{|\pHb|_\gamma |\Omega^4 \pH^2|_\gamma^2 }_{=:I}+ \underbrace{u^{\f 54}\Omg^4 |\Omega^2 \big[(\nab, \psi) \beta + \pH K + \pH \nab\psi\big]|_\gamma^2 }_{=:II}\Big) (u', \ub, \vartheta_{\CH^+}) \volg \ud u'  \ud  \ub
    \end{split}
\end{equation}
We now show that the second integral on the right-hand side is finite.  The term $I$ is estimated by 
$$||\pHb||_{L^1_u L^\infty_{\ub} L^\infty(S)} \cdot || \ub^{\min\{q_{--} -2, 2 \sigma_{q_-} +1\}} \Omega^4 \pH^2 ||_{L^\infty_u L^1_{\ub} L^1(S)} \cdot || \ub^{q -\min\{q_{--} -2, 2 \sigma_{q_-} +1\}} \Omega^4 \pH^2||_{L^\infty_u L^\infty_{\ub} L^\infty(S)}\;,$$
where we use $q - q_{--} +2 <q_- -3$ and  $q - 2 \sigma_{q_-} -1 < 2 \sigma_q $ and Proposition \ref{PropEstimatesNonDifference}.
For the second term we take another $|u|^{-\f 54}$ for the $L^1_u$ integral and estimate $\Omg^4 = (\Omg^2)^2 \ls \ub^{-2\sigma_q}$. The term $II$ is then estimated by
\begin{equation*}
    \begin{split}
 |||u|^{- \f 54}||_{L^1_uL^\infty_{\ub}L^\infty(S)} \cdot \Big( &|| \ub^{q +\f 52- 2\sigma_q} \Omega^4 ( \nab \beta)^2 ||_{L^\infty_u L^1_{\ub} L^1(S)} + ||\ub^{q +\f 52- 2\sigma_q} \Omega^4 \beta^2 ||_{L^\infty_u L^1_{\ub} L^1(S)} \cdot ||\psi||^2_{L^\infty_u L^\infty_{\ub}L^\infty(S)}      \\
 &+ ||\ub^{q +\f 52- 2\sigma_q} \Omega^4 \pH^2 ||_{L^\infty_u L^1_{\ub} L^1(S)} \cdot \big[ ||K||^2_{L^\infty_uL^\infty_{\ub}L^\infty(S)} + ||\nab \psi||^2_{L^\infty_u L^\infty_{\ub} L^\infty(S)}\big]\Big)
    \end{split}
\end{equation*}
which we use in conjunction with $q + \f 52- 2\sigma_q <\min\{q_{--} -2, 2\sigma_{q_-} +1\}$ (which holds since $q\geq 7$) and Proposition \ref{PropEstimatesNonDifference}.
Hence, \eqref{EqPfThmPropCauchy1} gives 
\begin{equation*}
    \underbrace{\int\limits_{\ub_0}^\infty \,\iint\limits_{W_{\CH^+}}  \ub^{q}|\Omega^4 \alpha|^2_\gamma\big(u_\Gamma \circ \calC(\ub, \vartheta_{\CH^+}), \ub, \vartheta_{\CH^+}\big) \volg \ud  \ub }_{=:III}
   \lesssim 1+ \int\limits_{\ub_0}^\infty \int\limits_{u_0 - \lambda_0}^{u_0 + \lambda_0} \,\iint\limits_{W_{\CH^+}} \ub^{q} |\Omega^4 \alpha|_\gamma^2 (u, \ub, \vartheta_{\CH^+}) \volg \ud u \ud  \ub \;.
\end{equation*}
We now lower-bound $III$. By Lemma \ref{LemRelationStarAndNullCoord} we have $\ub \sim v_+$. For the domain of integration we note that by Proposition \ref{PropProjectionUToGamma} there exist $v_{2+} \geq v_{1+}$ and an open subset $W_{\Gamma} \subseteq \Sp^2$ such that
$\Pi \big([\ub_1, \infty) \times W_{\CH^+} \big) \supseteq [v_{2+}, \infty) \times W_\Gamma$.  And for the measure we first use that $\volg = \sqrt{\det \gamma} \,\ud \theta_*\ud  \varphi_* \sim \sqrt{\det \boldsymbol{\gamma} }\, \ud  \theta_* \ud  \varphi_*$ by \eqref{NS.def} for $\epsilon >0$ small enough. Furthermore, from \cite[A.43]{DafLuk17} one directly computes $\det \boldsymbol{\gamma} = \ell^2 \sin^2 \theta$ and \cite[Proposition A.3 and A.1]{DafLuk17} give $\ell \sim 1$. Hence, we have $\det \gamma \sim \sin^2 \theta$.
Together with \eqref{EqDetPTG} this gives
\begin{equation*}
    \int\limits_{v_{2+}}^\infty \, \iint\limits_{W_\Gamma} v_+^q |\Omega^4 \alpha|^2_\gamma\big(v_{-, \Gamma}(v_+),v_+, \vartheta_=\big) \, \vols \ud v_+ \lesssim III\;.
\end{equation*}
But now \eqref{EqThmIntegratedBlowUpWeightedAlpha2} follows from Proposition \ref{PropTransitionToAlpha}, which concludes the proof.
\end{proof}

\subsection{Proof of Theorem \ref{ThmInextConditionVerified}} \label{SecProofMainThm}

We collate a few more bounds on curvature and connection coefficients which hold to the future of $\Gamma$ and are needed in the following:

\begin{proposition} \label{PropFinalBounds}
The following bounds hold to the future of $\Gamma$, where all curvature components are with respect to the dynamical double null frame $\os{\ee}{2}_3 = \Omg^2 e_3'$, $\os{\ee}{2}_4 = \frac{1}{\Omega^2} e_4'$:
    \begin{enumerate}
        \item $||\nab^k\nab_{e_4'}^j(\Omega^4\alpha)||_{L^2_{\ub} L^\infty_{u}L^\infty(S)} \lesssim 1$  for $0 \leq j+k \leq 1$
        \item $||\pmb{\Omega}^2(\beta, \rho, \sigma, \betab, \alphab)||_{L^1_{\ub}L^\infty_uL^\infty(S)} \lesssim 1$
        \item $|| e^{-2 \kappa_- \ub} b_{\CH^+}||_{L^\infty_{\ub} L^\infty_u L^\infty(S)} \lesssim 1$
    \end{enumerate}
\end{proposition}

\begin{proof}
We will freely use that \eqref{eq:Delta.after.Gamma.again} holds to the future of $\Gamma$ (see the proof of Proposition~\ref{PropEstimatesNonDifference}).

For all of the curvature components, we need to control the component without taking differences with the background. We will write $\alp = (\alp - \pmb{\alp}) + \pmb{\alp}$, etc. For the background quantities $\pmb{\alp},\pmb{\beta},\pmb{\rho}, \pmb{\sigma}, \pmb{\betab}, \pmb{\alphab}$, we use that they are bounded (without $\pmb{\Omg}^2$ weights) in $L^\i_{\ub} L^\i_u L^\i(S)$ (since they are background quantities) so that the corresponding bounds in points 1 and 2 follow from \eqref{eq:Delta.after.Gamma.again} and that $\ub^{-\sigma_q}$ is $L^1_{\ub}$ (and hence also $L^2_{\ub}$) bounded (since $\sigma_q \geq \f {10}4 >1$).

For the differences, the estimates needed for $\alp$, $\bt$, $\rho$ and $\sigma$ in the first two points are immediate from \eqref{eq:curvature.double.null.final.pointwise}, after using Sobolev embedding on the spheres \eqref{eq:Sobolev.BS} and that $\ub^{-\f{q_--3}2}$ is $L^1_{\ub}$ (and hence also $L^2_{\ub}$) integrable.

For the difference estimates for $\betab$ and $\alphab$, however, \eqref{eq:curvature.double.null.final.pointwise} only gives $|u|$-decay. Instead we use that \eqref{eq:curvature.double.null.final.pointwise} gives $L^\i_{\ub}L^\i_u L^\i(S)$ boundedness and then obtain the $L^1_{\ub}$ integrability in $\ub$ by using the $\pmb{\Omg}^2$ weight and \eqref{eq:Delta.after.Gamma.again}.

The third point  follows directly from \cite[(16.60) and Lemma 16.12]{DafLuk17}. \qedhere 


\end{proof}

The final ingredient in the proof of our main theorem  is the following

\begin{lemma} \label{LemFinal}
    Let $K \subset \mathbb R^n$ be a compact set and $\kappa>0$ be a constant. Let $f : [1, \infty)_{\ub} \times K_y \to \C$ with $f, \f{\rd f}{\rd {\ub}} \in L^2_{\ub} L^\i_y$ and let $\mathfrak{a}, \mathfrak{b} \in C^0 \big( [1, \infty) \times K ; \C\big)$ with $\Rea (\mathfrak{a}) -|\mathfrak{b}| \gtrsim 1$.

    If $$ \int_1^\infty \Big| \int_{K} \int_1^{{\ub}} e^{\kappa{\ub}'} \big[\mathfrak{a}({\ub}',y) f({\ub}',y) +\mathfrak{b}({\ub}',y) \overline{f}({\ub}',y)\big] \, \ud {\ub}' \ud y \Big|^2 e^{-\kappa{\ub}} \, \ud {\ub} < \infty$$ then also $\int_K \int_1^\infty {\ub}^p |f|^2({\ub},y) \, \ud {\ub} \ud y < \infty$ for all $p \geq 0$.
\end{lemma}

\begin{proof}
    By the assumption on $\mathfrak{a}$ and $\mathfrak{b}$ we have $|f|^2 \leq c' \cdot \Rea \big( \overline{f}(\mathfrak{a}f + \mathfrak{b} \overline{f} )\big)$ for some $c'>0$. Thus, we estimate
    \begin{equation*}
        \begin{split}
            &\: \int_K \int_1^\infty {\ub}^p |f|^2({\ub},y) \, \ud {\ub} \ud y \\
            \lesssim &\:\Big| \int_K \int_1^\infty  {\ub}^p \overline{f}({\ub},y) \big[\mathfrak{a}({\ub},y) f({\ub},y) + \mathfrak{b}({\ub},y) \overline{f}({\ub},y) \big]\, \ud {\ub} \ud y\Big| \\
            = &\: \Big| \int_K \int_1^\infty {\ub}^p e^{-\kappa{\ub}} \overline{f({\ub},y)}  \Big( \f{\rd}{\rd {\ub}} \int_1^{\ub} e^{\kappa{\ub}'} \big[ \mathfrak{a}({\ub}',y)  f({\ub}',y) + \mathfrak{b}({\ub}',y) \overline{f}({\ub}',y)\big] \, \ud {\ub}' \Big) \, \ud {\ub} \ud y \Big| \\
            = &\:\Big| \int_K \int_1^\infty \f{\rd}{\rd {\ub}}\big[ {\ub}^p e^{-\kappa{\ub}} \overline{f({\ub},y)} \big]  \Big(  \int_1^{\ub} e^{\kappa{\ub}'}  \big[ \mathfrak{a}({\ub}',y)  f({\ub}',y) + \mathfrak{b}({\ub}',y) \overline{f}({\ub}',y)\big] \, \ud {\ub}' \Big) \, \ud {\ub} \ud y \Big| \\
            \leq &\: \Big( \int_1^\infty \Big| \int_K \int_1^{\ub} e^{\kappa{\ub}'}  \big[ \mathfrak{a}({\ub}',y)  f({\ub}',y) + \mathfrak{b}({\ub}',y) \overline{f}({\ub}',y)\big] \, \ud {\ub}' \ud y \Big|^2 e^{-\kappa{\ub}} \ud {\ub}\Big)^{\f 12} \\
            &\qquad \qquad \cdot \Big( ||f||_{L^2_{\ub}L^\i_y } ||({\ub}^p e^{-\kappa{\ub}})' e^{\f{\kappa{\ub}}2}||_{L^\i_{\ub}} + ||\f{\rd f}{\rd {\ub}}||_{L^2_{\ub}L^\i_y} ||{\ub}^p e^{-\f{\kappa{\ub}}2}||_{L^\i_{\ub}}  + ||{\ub}^p e^{-\f{\kappa{\ub}}2}||_{L^\i_{\ub}} ||f||_{L^2_{\ub} L^\i_y}  \Big),
        \end{split}
    \end{equation*}
    where in the last line we used the Cauchy--Schwarz inequality. 
\end{proof}

\begin{remark}
    The lemma still holds with $f, \f{\rd f}{\rd \ub} \in L^\i_{\ub,y}$.
\end{remark}

\begin{proof}[Proof of Theorem \ref{ThmInextConditionVerified}.] 
    Let $K= K_u \times K_{\Sp^2} \subseteq (-\infty, u_f) \times \Sp^2 \simeq \{\ub_{\CH^+}=0\}$ be given and assume $K_{\Sp^2} \subseteq \calV_2'$; the other case follows analogously. 
    
    Let $\hat{\varepsilon}>0$; the exact value will be fixed later on in the proof. For the convenience of indexing, we now set here in the proof
        $E_4 := \os{\ee}{2}_4$ and   $E_3 := \os{\ee}{2}_3$ and furthermore define via Gram--Schmidt in the $(u, \ub_{\CH^+}, \vartheta_{(2),\CH^+})$ coordinates
        \begin{equation*}
         E_1 := \frac{1}{\sqrt{(\gamma_{(2),\CH^+})_{11}}} \f{\rd}{\rd \theta^1_{(2), \CH^+}}  , \qquad  E_2 :=  \f{1}{\sqrt{(\gamma_{(2),\CH^+})_{22} - \f{(\gamma_{(2),\CH^+})_{21}^2}{(\gamma_{(2),\CH^+})_{11}}}} \Big(\f{\rd}{\rd \theta^2_{(2), \CH^+}} - \f{(\gamma_{(2),\CH^+})_{21}}{(\gamma_{(2),\CH^+})_{11}} \f{\rd}{\rd \theta^1_{(2), \CH^+}} \Big)\;.
    \end{equation*}
    Note that $E_j$ form a normalized and smooth double null frame in $(-\delta_0,0) \times K $ which extends continuously to the Cauchy horizon by \eqref{Eqe4ExtCont} and Theorem \ref{thm:C0.extendibility}. Similarly, the $\hat{\varepsilon}$-perturbations of those vector fields from the statement of Theorem \ref{ThmInextConditionVerified} will be denoted with a hat. 
It follows from  $||\hat{E}_\mu - E_\mu||_h < \hat{\varepsilon}$ for $\mu \in \{1,2,4\}$
that we can write
\begin{align*}
\hat{E}_4 &= (1 + \calO(\hat{\varepsilon})) E_4 + \calO(\hat{\varepsilon}) E_3 + \calO(\hat{\varepsilon}) E_2 + \calO(\hat{\varepsilon}) E_1 \\
\hat{E}_2 &= \calO(\hat{\varepsilon})) E_4 + \calO(\hat{\varepsilon}) E_3 + (1+\calO(\hat{\varepsilon})) E_2 + \calO(\hat{\varepsilon}) E_1 \\
\hat{E}_1 &= \calO(\hat{\varepsilon}) E_4 + \calO(\hat{\varepsilon}) E_3 + \calO(\hat{\varepsilon}) E_2 + (1+\calO(\hat{\varepsilon})) E_1 \;,
\end{align*}
where $\calO(\hat{\varepsilon})$ stands for functions which are, in absolute value, uniformly bounded by $C \cdot \hat{\varepsilon}$ on $(\ub_0, \infty) \times K$, where the constant $C>0$ can be explicitly determined. Then using the symmetry and trace-freeness of the Weyl curvature tensor we compute
\begin{equation} \label{EqCurvatureHat}
    \begin{split}
R(\hat{E}_4, &\hat{E}_1 , \hat{E}_4, \hat{E}_1 ) + i R(\hat{E}_4, \hat{E}_1, \hat{E}_4, \hat{E}_2) = R(E_4,E_1,E_4,E_1) + iR(E_4,E_1,E_4,E_2)  \\
&\qquad +  \sum_{j_1, j_2, j_3, j_4=1}^4\mathcal{O}(\hat{\varepsilon}) R(E_{j_1}, E_{j_2}, E_{j_3}, E_{j_4}) + i  \sum_{j_1, j_2, j_3, j_4=1}^4\mathcal{O}(\hat{\varepsilon}) R(E_{j_1}, E_{j_2}, E_{j_3}, E_{j_4}) \\
&=(1+ \mathcal{O}(\hat{\varepsilon})) R(E_4,E_1,E_4,E_1) + \mathcal{O}(\hat{\varepsilon}) R(E_4,E_1,E_4,E_2) \\
&\qquad +i \Big[\mathcal{O}(\hat{\varepsilon})R(E_4,E_1,E_4,E_1) + (1+ \mathcal{O}(\hat{\varepsilon}))R(E_4,E_1,E_4,E_2)\Big]  \\
&\qquad + \textnormal{integrable terms }
\end{split}
\end{equation}
Here, `integrable terms' stands for a complex linear combination of $$\alphab(E_1, E_1), \quad \alphab(E_1,  E_2), \quad \betab(E_1), \quad \betab(E_2), \quad \rho,\quad  \sigma, \quad \beta(E_1),\quad  \beta(E_2)$$
with uniformly bounded coefficients.
Furthermore, we set 
$$F:= \alpha\big(\frac{1}{\sqrt{2}}(E_1 + iE_2),\frac{1}{\sqrt{2}}(E_1 + iE_2)\big) = \alpha(E_1,E_1) + i \alpha (E_1,E_2)\;,$$
where we have used the trace-freeness and symmetry of $\alpha$. Then \eqref{EqCurvatureHat} reads
\begin{equation*}
    R(\hat{E}_4, \hat{E}_1 , \hat{E}_4, \hat{E}_1 ) + i R(\hat{E}_4, \hat{E}_1, \hat{E}_4, \hat{E}_2) = \mathring{\mathfrak{a}} F + \mathring{\mathfrak{b}} \overline{F} + \textnormal{integrable terms }\;,
\end{equation*}
for some $\mathring{\mathfrak{a}}$ and $\mathring{\mathfrak{b}}$ complex-valued functions with $|\mathring{\mathfrak{a}} -1| = \mathcal{O}(\hat{\varepsilon})$ and $|\mathring{\mathfrak{b}}| = \mathcal{O}(\hat{\varepsilon})$. In particular we now fix $\hat{\varepsilon}>0$ small enough such that $\Rea(\mathring{\mathfrak{a}}) -|\mathring{\mathfrak{b}}| \geq \frac{1}{2}$.

It follows from \eqref{EqDetG} that $\vol_g = 2 \Omega^2_{\CH^+} \sqrt{\det \gamma_{\CH^+}} \ud u \wedge \ud \ub_{\CH^+} \wedge \ud  \theta^1_{(2), \CH^+} \wedge \ud  \theta^2_{(2), \CH^+}$. By continuity we have $\Omega^2_{\CH^+} \sim 1$ on $(-\delta_0,0] \times K$  so that it follows from the second point of Proposition \ref{PropFinalBounds}  that
\begin{align}
    \int\limits_{-\delta_0}^0 &\Big| \int\limits_{-\delta_0}^{\ub_{\CH^+}'}\int_{K} \big( R(\hat{E}_4, \hat{E}_1 , \hat{E}_4, \hat{E}_1 ) + i R(\hat{E}_4, \hat{E}_1, \hat{E}_4, \hat{E}_2) \big) \vol_g  \Big|^2 \ud \ub_{\CH^+}' \nonumber \\
    &= \int\limits_{-\delta_0}^0 \Big| \int\limits_{-\delta_0}^{\ub_{\CH^+}'}\int_{K} \big(\mathring{\mathfrak{a}} F + \mathring{\mathfrak{b}} \overline{F} \big) 2 \Omega^2_{\CH^+}\volg \ud u \ud \ub_{\CH^+}  \Big|^2 \ud \ub_{\CH^+}' + \calO(1) \nonumber\\
    &=\int\limits_{\ub_{\delta_0}}^\infty\Big| \int\limits_{\ub_{\delta_0}}^{\ub'}\int_{K} e^{4 \kappa_-(u + \ub)}\frac{e^{-4 \kappa_-(u + \ub)}}{\pmb{\Omega}^4}\frac{\pmb{\Omega}^4}{\Omega^4}\big(\mathring{\mathfrak{a}} \Omega^4F + \mathring{\mathfrak{b}} \overline{ \Omega^4 F} \big) \nonumber \\
    &\qquad \qquad \qquad \qquad \cdot 2 \Omega^2_{\CH^+}\det\gamma_{(2),\CH^+} \ud \theta^1_{(2),\CH^+} \ud \theta^2_{(2),\CH^+} \ud u\, e^{-2 \kappa_- \ub}\,\ud \ub  \Big|^2 e^{-2 \kappa_- \ub'} \,\ud \ub'  + \calO(1)  \nonumber\\
    &= \int\limits_{\ub_{\delta_0}}^\infty\Big| \int\limits_{\ub_{\delta_0}}^{\ub'}\int_{K} e^{2 \kappa_- \ub}\big(\mathfrak{a} f + \mathfrak{b} \overline{ f} \big)  \ud \theta^1_{(2),\CH^+} \ud \theta^2_{(2),\CH^+} \ud u \ud \ub  \Big|^2 e^{-2 \kappa_- \ub'} \ud \ub'  + \calO(1) \;,\label{EqToShowIntInf}
\end{align}
where in the second equality we have just changed the integration from $\ub$ to $\ub_{\CH^+}$ coordinates and added factors whose product  equals $1$, and in the third equality we have defined $$f:= \Omega^4 F, \quad \mathfrak{a} := e^{4 \kappa_- u}\frac{e^{-4 \kappa_-(u + \ub)}}{\pmb{\Omega}^4}\frac{\pmb{\Omega}^4}{\Omega^4}  \cdot 2 \Omega^2_{\CH^+}\det\gamma_{(2),\CH^+} \mathring{\mathfrak{a}}, \quad \mathfrak{b} := e^{4 \kappa_- u}\frac{e^{-4 \kappa_-(u + \ub)}}{\pmb{\Omega}^4}\frac{\pmb{\Omega}^4}{\Omega^4} \cdot 2 \Omega^2_{\CH^+}\det\gamma_{(2),\CH^+} \mathring{\mathfrak{b}}\;.$$
Note that since $K_u$ is compact and $K_{\Sp^2}$ is compactly contained in $\calV_2'$, each of the factors multiplying $\mathring{\mathfrak{a}}$ and $\mathring{\mathfrak{b}}$ is $\sim 1$, so that we obtain $\Rea(\mathfrak{a}) - |\mathfrak{b}| \geq c' >0$ in $(\ub_1, \infty) \times K_u \times K_{\Sp^2}$. We will now apply Lemma \ref{LemFinal} to show the infinitude of the integral in \eqref{EqToShowIntInf}.

    In the setting of Theorem \ref{ThmBlowUpCH} and Remark \ref{RemThmBlowUpCH}  let $\overline{p} = (u_0, 0, \mathring{\vartheta}_{\CH^+}) \in K$ and choose $\lambda_0 >0$ so small that $(u_0 - \lambda_0, u_0 + \lambda_0) \subseteq K_u$ and $W_{\CH^+} \subseteq K_{\Sp^2}$. Then by \eqref{EqThmIntegratedBlowUpWeightedAlpha2} we have 
    \begin{equation*} 
        \int_{\ub_{\delta_0}}^\infty \iiint_K \ub^q |f|^2 \, d\theta^1_{(2),\CH^+} d\theta^2_{(2),\CH^+}\ud u \ud\ub = \infty\;,
    \end{equation*}
    where we have again used $\det \gamma_{(2), \CH^+} \sim 1$ in $K_u \times (\ub_0, \infty) \times K_{\Sp^2}$.
    Note that it remains to show
    \begin{equation} \label{EqLastThingTS}
        \int_{\ub_{\delta_0}}^\infty \Big(\sup\limits_{(u, \theta_{\CH^+}) \in K_u \times K_{\Sp^2}} |\frac{\rd}{\rd \ub}\Big|_{(u, \ub, \theta_{(2),\CH^+})}f(u,\ub, \theta_{\CH^+})| \Big)^2 \, \ud\ub < \infty \;.
    \end{equation}
     Once this is established, we can apply Lemma \ref{LemFinal} with $\kappa = 2 \kappa_-$, since  $f \in L^2_{\ub} L^\infty_K$ by the first point of Proposition \ref{PropFinalBounds}.

     We claim that
     \begin{equation} \label{EqLastClaim12}
         |g(\nab_{E_A} E_B, E_C)| + |g(\nab_{e_4'}E_A,E_B)| \ls 1 \textnormal{ in } (-\delta_0,0) \times K \;.
     \end{equation}
      For the first summand we note that (cf.\ \cite[(16.59)]{DafLuk17})
      $$(\gamma_{(2),\CH^+})_{AB} = \gamma_{A'B'} \Big( \f{\rd \theta_{(2),\CH^+}}{\rd \theta_{(2)}}\Big)^{-1}{\strut}_A^{\;\,A'}\Big( \f{\rd \theta_{(2),\CH^+}}{\rd \theta_{(2)}}\Big)^{-1}{\strut}_B^{\; \,B'} $$
so that the claim for the first summand follows from \eqref{eq:NS.higher} together with \cite[(16.45)]{DafLuk17} and \eqref{eq:angular.second.derivative.bound}. For the second summand we first recall
 \begin{equation}\label{EqExpre4p}
     \begin{split}
         e_4' &= \Omega^2 \os{\ee}{2}_4 = \frac{\rd}{\rd \ub}\Big|_{DN} + b^A \frac{\rd}{\rd \vartheta_*^A}\Big|_{DN} \\ 
         &= \frac{\rd}{\rd \ub}\Big|_{(u, \ub, \theta_{(2), \CH^+})} + \underbrace{\Big( \frac{\rd \theta^A_{(2), \CH^+}}{\rd \ub} \Big|_{DN} + b^B \frac{\rd \theta_{(2), \CH^+}^A}{\rd \vartheta_*^B} \Big|_{DN} \Big) }_{=e^{- 2 \kappa_- \ub}b_{\CH^+}^A \textnormal{ by \cite[(16.60)]{DafLuk17}}} \frac{\rd}{\rd \theta_{(2), \CH^+}^A}\Big|_{(u, \ub, \theta_{(2), \CH^+})}
     \end{split}
 \end{equation}
 so that  the claim for the second summand follows from \eqref{eq:nablaslashed.chi} together with \eqref{eq:sbi.cond.goal} and \eqref{eq:NS.higher}. Having established \eqref{EqLastClaim12} we now use \eqref{EqExpre4p} to compute
  $$2 \frac{\rd}{\rd \ub}\Big|_{(u, \ub, \theta_{(2),\CH^+})} f = \big[\nab_{e_4'} - e^{- 2 \kappa_- \ub}b^A_{\CH^+}\nab_A\big]\big( \Omega^4 \alpha (E_1 + i E_2, E_1 + i E_2) \big) $$
  so that \eqref{EqLastThingTS} follows from Proposition \ref{PropFinalBounds} together with \eqref{EqLastClaim12}. This proves Theorem \ref{ThmInextConditionVerified}.
\end{proof}

\appendix

\section{Commutator computations for the linear Teukolsky equation} 

The second order terms of $ \pmb{\mathcal{T}}_{[s]}$  in $\{v_+, r, \theta, \varphi_+\}$ coordinates are 
\begin{equation}\label{EqSecOrderTermsLinTeuk}
II := a^2 \sin^2 \theta \,\partial_{v_+}^2\psi + 2a \,\partial_{v_+}\partial_{\varphi_+} \psi + 2(r^2 + a^2)\, \partial_{v_+}\partial_r \psi +2 a\, \partial_{\varphi_+}\partial_r \psi 
 + \Delta \,\partial_r^2 \psi + \mathring{\slashed{\Delta}}_{[s]} \psi 
\end{equation}
We use $\chi(v_+) e^{\lambda r} (\partial_{v_+} - \frac{M}{r} \rd_r)\overline{\psi}$ as a multiplier with $\lambda >0$ and compute the commutator expressions in the following individually for each term in \eqref{EqSecOrderTermsLinTeuk}.
Again we use the notation $\underset{a.i.}{=}$ to denote equality after integration over the spheres with respect to $\vols$. We also use Proposition 2.26 and equation (2.32) from \cite{Sbie23}.

\subsection{Computing $II \cdot \chi(v_+) e^{\lambda r} (\partial_{v_+} - \frac{M}{r} \rd_r)\overline{\psi}$} \label{AppendixCommutator}

\begin{flalign*}
    &\chi(v_+)  e^{\lambda r} a^2 \sin^2 \theta \Rea \big( \rd_{v^+}^2 \psi [ \overline{\rd_{v_+}\psi} - \frac{M}{r} \overline{\rd_r \psi}] \big) & \\
    &= \rd_{v_+} \big[ \frac{1}{2} \chi(v_+) e^{\lambda r} a^2 \sin^2 \theta |\rd_{v^+} \psi|^2 \big] - \frac{1}{2} \chi'(v_+) e^{\lambda r} a^2 \sin^2 \theta |\rd_{v_+} \psi|^2 - \rd_{v_+} \big[ \chi(v_+) e^{\lambda r} a^2 \sin^2 \theta \frac{M}{r} \Rea(\rd_{v_+} \psi \overline{\rd_r \psi})\big] & \\
    & \quad + \chi'(v_+) e^{\lambda r} a^2 \sin^2 \theta \frac{M}{r} \Rea(\rd_{v_+} \psi \overline{\rd_r \psi}) + \rd_r \big[ \frac{1}{2} \chi(v_+) e^{\lambda r} a^2 \sin^2 \theta \frac{M}{r} | \rd_{v_+} \psi|^2 \big] - \uwave{\frac{1}{2} \chi(v_+) e^{\lambda r} a^2 \sin^2 \theta \frac{M \lambda}{r} |\rd_{v_+} \psi|^2 } \\
    &\quad  + \frac{1}{2} \chi(v_+) e^{\lambda r} a^2 \sin^2 \theta \frac{M}{r^2} |\rd_{v_+} \psi|^2 &
    \end{flalign*}
    \begin{flalign*}
    &\chi(v_+)  e^{\lambda r} 2a \Rea \big(\rd_{v_+} \rd_{\varphi_+} \psi [ \overline{\rd_{v_+} \psi} - \frac{M}{r} \overline{\rd_r \psi}] \big) &\\
    &\eai - \rd_{v_+} \big[ \chi(v_+)  e^{\lambda r} \frac{aM}{r} \Rea(\rd_{\varphi_+} \psi \overline{\rd_r \psi}) \big] + \chi'(v_+) e^{\lambda r} \frac{aM}{r} \Rea(\rd_{\varphi_+} \psi \overline{\rd_r \psi}) + \rd_r \big[ \chi(v_+)  e^{\lambda r} \frac{aM}{r} \Rea(\rd_{\varphi_+} \psi \overline{\rd_{v_+} \psi}) \big] & \\
    &\quad - \uwave{\chi(v_+)  e^{\lambda r} \lambda \frac{aM}{r} \Rea(\rd_{\varphi_+} \psi \overline{\rd_{v_+} \psi})} + \chi(v_+)  e^{\lambda r} \frac{aM}{r^2} \Rea(\rd_{\varphi_+} \psi \overline{\rd_{v_+} \psi}) &
    \end{flalign*}
    \begin{flalign*}
    &\chi(v_+)  e^{\lambda r} 2 (r^2 + a^2) \Rea (\rd_{v_+} \rd_r \psi [ \overline{\rd_{v_+} \psi} - \frac{M}{r} \overline{\rd_r \psi} ]) &\\
    &=\rd_r \big[\chi(v_+)  e^{\lambda r}  (r^2 + a^2) |\rd_{v_+} \psi|^2 \big] - \uwave{\chi(v_+)  e^{\lambda r} \lambda (r^2 + a^2) |\rd_{v_+} \psi|^2} - \chi(v_+)  e^{\lambda r}  2r |\rd_{v_+} \psi|^2 &\\
    &\quad - \rd_{v_+} \big[ \chi(v_+) e^{\lambda r} \frac{M (r^2 + a^2)}{r}|\rd_r \psi|^2 \big] +  \chi'(v_+) e^{\lambda r} \frac{M (r^2 + a^2)}{r}|\rd_r \psi|^2 &
    \end{flalign*}
    \begin{flalign*}
    &\chi(v_+)  e^{\lambda r} 2a \Rea(\rd_{\varphi_+} \rd_r \psi [ \overline{\rd_{v_+} \psi} - \frac{M}{r} \overline{ \rd_r \psi}]) &\\
    &\eai \rd_r \big[ \chi(v_+)  e^{\lambda r} a \Rea(\rd_{\varphi_+} \psi \overline{\rd_{v_+} \psi}) \big] - \uwave{ \chi(v_+)  e^{\lambda r} \lambda a \Rea( \rd_{\varphi_+} \psi \overline{\rd_{v_+} \psi})} - \rd_{v_+} \big[\chi(v_+)  e^{\lambda r} a \Rea(\rd_{\varphi_+} \psi \overline{\rd_{r} \psi}) \big] &\\
    &\quad + \chi'(v_+)  e^{\lambda r} a \Rea(\rd_{\varphi_+} \psi \overline{\rd_{r} \psi}) &
    \end{flalign*}
    \begin{flalign*}
    &\chi(v_+)  e^{\lambda r} \Delta \Rea(\rd_{r}^2 \psi [ \overline{\rd_{v_+} \psi} - \frac{M}{r} \overline{ \rd_r \psi}]) &\\
    &= \rd_r \Big[\chi(v_+)  e^{\lambda r}  \Delta \big( \Rea( \rd_r \psi \overline{\rd_{v_+} \psi}) - \frac{M}{2r}|\rd_r \psi|^2 \big) \Big] - \uwave{ \chi(v_+)  e^{\lambda r}  \lambda \Delta \Rea( \rd_r \psi \overline{\rd_{v_+} \psi}) } - \chi(v_+)  e^{\lambda r}  2(r-M) \Rea( \rd_r \psi \overline{\rd_{v_+} \psi}) & \\
    &\quad + \uwave{\chi(v_+) e^{\lambda r} \lambda \frac{\Delta M}{2r} | \rd_r \psi|^2} + \dashuline{\chi(v_+) e^{\lambda r}  \frac{(r-M) M}{r} | \rd_r \psi|^2 } - \chi(v_+) e^{\lambda r} \frac{\Delta M}{2r^2} | \rd_r \psi|^2 & \\
    &\quad - \rd_{v_+} \big[ \frac{1}{2} \chi(v_+) e^{\lambda r}\Delta | \rd_r \psi|^2 \big] +  \frac{1}{2} \chi'(v_+) e^{\lambda r}\Delta | \rd_r \psi|^2  &
    \end{flalign*}
    \begin{flalign*}
    & \chi(v_+) e^{\lambda r} \Rea(\mathring{\slashed{\Delta}}_{[s]} \psi [ \overline{\rd_{v_+} \psi} - \frac{M}{r} \overline{ \rd_r \psi}]) & \\
    &\eai \rd_{v_+} \Big[ \frac{1}{2} \chi(v_+) e^{\lambda r} \big((s+s^2)|\psi|^2 - \sum_{i=1}^3 |\wtZ_{i,+} \psi|^2 \big) \Big] - \frac{1}{2} \chi'(v_+) e^{\lambda r} \big((s+s^2)|\psi|^2 - \sum_{i=1}^3 |\wtZ_{i,+} \psi|^2 \big) & \\
    &\quad + \rd_r \Big[ \frac{1}{2} \chi(v_+) e^{\lambda r}\frac{M}{r} \big(-(s+s^2)|\psi|^2 + \sum_{i=1}^3 |\wtZ_{i,+} \psi|^2 \big)\Big] - \uwave{ \frac{1}{2} \chi(v_+) e^{\lambda r} \lambda \frac{M}{r} \big(-(s+s^2)|\psi|^2 + \sum_{i=1}^3 |\wtZ_{i,+} \psi|^2 \big)} & \\
    &\quad 
    + \frac{1}{2} \chi(v_+) e^{\lambda r}  \frac{M}{r^2} \big(-(s+s^2)|\psi|^2 + \sum_{i=1}^3 |\wtZ_{i,+} \psi|^2 \big) &
\end{flalign*}

\subsection{Estimates on the main bulk term and the boundary terms} \label{AppendixEstimates}

After integration over the spheres, $II \cdot \chi(v_+) e^{\lambda r} (\partial_{v_+} - \frac{M}{r} \rd_r)\overline{\psi}$ is then of the form
\begin{equation*}
    II \cdot \chi(v_+) e^{\lambda r} (\partial_{v_+} - \frac{M}{r} \rd_r)\overline{\psi} \eai \rd_{v_+} [A(\psi)] + \rd_r [B(\psi)] + \uwave{D(
    \psi)} + D_{\mathrm{rem}}(\psi) \;,
\end{equation*}
where $A(\psi)$ comprises all the terms above with a $\rd_{v_+}$ in front, $B(\psi)$ all those terms with a $\rd_r$ in front, $\uwave{D(\psi)}$ all the terms that are wavily underlined (these are all the terms that contain a factor of $\lambda$) , and $D_{\mathrm{rem}}(\psi)$ the remaining terms. In the following we will usually just write $A$ for $A(\psi)$, etc. We also introduce the notation $\Phi:= \frac{1}{\sin \theta} (is \cos \theta + \rd_{\varphi_+})$.

\begin{lemma}\label{LemmaAppendixDUWave}
    There exist constants $0< c, C$  such that $$\uwave{D} \leq -c \lambda \chi(v_+) e^{\lambda r} \Big( - \Delta |\rd_r \psi|^2 + |\rd_{v_+} \psi|^2 + |\rd_\theta \psi|^2 + |\Phi \psi|^2 \Big) + C \lambda \chi(v_+) e^{\lambda r} |\psi|^2$$ for $r \in [r_-, r_+]$.
\end{lemma}

\begin{proof}
  We have
\begin{align}
   \uwave{D} &=\lambda \chi(v_+) e^{ \lambda r} \Big[ - \big( \frac{1}{2} a^2 \sin^2 \theta \frac{M}{r} + r^2 + a^2\big) |\rd_{v_+} \psi|^2 - a(1 + \frac{M}{r}) \Rea ( \rd_{\varphi_{+}} \psi \overline{\rd_{v_+} \psi} ) - \Delta \Rea(\rd_r \psi \overline{\rd_{v_+} \psi}) \notag \\
    &\qquad \qquad \qquad + \frac{\Delta M}{2r} |\rd_r \psi|^2 - \frac{1}{2} \frac{M}{r} \big(\sum_{i = 1}^3 | \wtZ_{i,+} \psi|^2 -(s+s^2)|\psi|^2 \big) \Big]  \label{EqUpBoundB}\\
    &= \lambda \chi(v_+) e^{ \lambda r} \Big[ - \big( \frac{1}{2} a^2 \sin^2 \theta \frac{M}{r} + r^2 + a^2\big) |\rd_{v_+} \psi|^2 - a(1 + \frac{M}{r}) \sin \theta \Rea \big(  \Phi \psi \overline{\rd_{v_+} \psi} \big)  \notag \\
    &\qquad \qquad \qquad - \Delta \Rea(\rd_r \psi \overline{\rd_{v_+} \psi}) + \frac{\Delta M}{2r} |\rd_r \psi|^2 - \frac{1}{2} \frac{M}{r} \big( |\rd_\theta \psi|^2 + |\Phi \psi|^2 - \underbrace{s |\psi|^2}\big) \Big] \notag \\
    &\qquad \qquad \qquad + \underbrace{\lambda \chi_n(v_+) e^{ \lambda r} a(1 + \frac{M}{r}) \Rea( is \cos \theta \cdot \psi \overline{\rd_{v_+} \psi})}\;, \notag
\end{align}
where we have used Lemma 2.33 from \cite{Sbie23} in the second equality. The underbraced terms are considered as error terms. With the exception of the $|\rd_\theta \phi|^2$ term, which is already manifestly negative definite, the non-underbraced terms can be written as a quadratic form in $(\sqrt{- \Delta} \rd_r \psi, \Phi \psi , \rd_{v_+} \psi)$ with matrix $\lambda \chi(v_+) e^{ \lambda r}  \mathbb{Q}_1 $, where
\begin{equation*}
    \mathbb{Q}_1 = \begin{pmatrix}
        - \frac{M}{2r} & 0 & \frac{1}{2} \sqrt{- \Delta} \\
        0 & - \frac{1}{2} \frac{M}{r} & - \frac{a}{2} (1 + \frac{M}{r}) \sin \theta \\
        \frac{1}{2} \sqrt{- \Delta} & - \frac{a}{2} ( 1+ \frac{M}{r}) \sin \theta & - ( \frac{1}{2} a^2 \sin^2 \theta \frac{M}{r} + r^2 + a^2)
    \end{pmatrix}
\end{equation*}
The first two principal sub-matrices obviously have negative determinant and a computation gives $$\det \mathbb{Q}_1 = - \frac{M\big[ r^3 +(2M+r) a^2 \cos^2 \theta\big]}{8r^2} \;,$$
so that Sylvester's criterion gives that $\mathbb{Q}_1$ is uniformly negative definite for all $r \in [r_-, r_+]$. Finally, applying a weighted Cauchy--Schwarz inequality to the second underbraced term yields the result.
\end{proof}

\begin{lemma} \label{LemAppendixB}
    There exist constants $0<c,C$ such that
    $$B \geq c \chi(v_+) e^{\lambda r} \Big( - \Delta |\rd_r \psi|^2 + |\rd_{v_+} \psi|^2 + |\rd_\theta \psi|^2 + |\Phi \psi|^2 \Big) - C \chi(v_+) e^{\lambda r} |\psi|^2  $$ for $r \in [r_-, r_+]$.
\end{lemma}

\begin{proof}
  This follows directly from Lemma \ref{LemmaAppendixDUWave} by observing that we have $B = - \frac{1}{\lambda} \uwave{D}$.
\end{proof}

For the proof of the next two lemmas we note that we have 
\begin{equation*}
    \begin{split}
        A&= \chi(v_+) e^{\lambda r} \Big[ \frac{1}{2} a^2 \sin^2 \theta |\rd_{v_+} \psi|^2 - a^2 \sin^2 \theta \frac{M}{r} \Rea(\rd_{v_+} \psi \overline{\rd_{r}\psi}) -a(1+ \frac{M}{r}) \Rea(\rd_{\varphi_+} \psi \ov{\rd_r \psi}) \\
        &\qquad - \big(\frac{1}{2} \Delta + \frac{M(r^2 + a^2)}{r} \big) |\rd_r \psi|^2 + \frac{1}{2} \big( (s+s^2) |\psi|^2 - \sum_{i = 1}^3 |\wtZ_{i,+} \psi|^2 \big)\Big] \\
        &= \chi(v_+) e^{\lambda r} \Big[ \frac{1}{2} a^2 \sin^2 \theta |\rd_{v_+} \psi|^2 - a^2 \sin^2 \theta \frac{M}{r} \Rea(\rd_{v_+} \psi \overline{\rd_{r}\psi}) -a(1+ \frac{M}{r}) \sin \theta \Rea \big( \Phi \psi \ov{\rd_r \psi}\big) \\
        &\qquad - \big(\frac{1}{2} \Delta + \frac{M(r^2 + a^2)}{r} \big) |\rd_r \psi|^2  - \frac{1}{2} |\rd_{\theta} \psi|^2 - \frac{1}{2 } |\Phi \psi|^2 \\
        &\qquad + \underbrace{a(1 + \frac{M}{r}) \Rea(is \cos \theta \cdot \psi \ov{\rd_r \psi}) + \frac{1}{2}s |\psi|^2} \;,
    \end{split}
\end{equation*}
where we have used again Lemma 2.33 from \cite{Sbie23} in the second equality. The underbraced terms will be considered again as error terms.

\begin{lemma} \label{LemAppendixBA}
    There exist constants $0<c,C$ such that
    $$B - A \geq c \chi(v_+) e^{\lambda r} \Big( |\rd_r \psi|^2 + |\rd_{v_+} \psi|^2 + |\rd_\theta \psi|^2 + |\Phi \psi|^2 \Big) - C \chi(v_+) e^{\lambda r} |\psi|^2  $$ for $r \in [r_-, r_+]$.
\end{lemma}

\begin{proof}
    Recalling the expression of $B = - \frac{1}{\lambda}\uwave{D}$ from the proof of Lemma \ref{LemmaAppendixDUWave} and the form of $A$ from above we get
    \begin{equation*}
        \begin{split}
           B-A &= \chi(v_+) e^{\lambda r} \Big[\big(\frac{1}{2} a^2 \sin^2 \theta ( \frac{M}{r} - 1) + r^2 + a^2\big) | \rd_{v_+} \psi|^2 + a(1 + \frac{M}{r}) \sin \theta \Rea \big( \Phi \psi \ov{\rd_{v_+} \psi} \big) \\
           &\qquad + ( \Delta + a^2 \sin^2 \theta \frac{M}{r}) \Rea(\rd_{v_+} \ov{\rd_r \psi}) + a(1 + \frac{M}{r}) \sin \theta \Rea \big( \Phi \psi \ov{\rd_r \psi} \big) \\
           &\qquad + \big( \frac{\Delta}{2}(1 - \frac{M}{r}) + \frac{M(r^2 + a^2)}{r} \big) |\rd_r \psi|^2  + \frac{1}{2} (1 + \frac{M}{r}) |\Phi \psi|^2 + \frac{1}{2}(1 + \frac{M}{r})|\rd_\theta \psi|^2 \\
           &\qquad \underbrace{- \frac{1}{2} s(1+\frac{M}{r})|\psi|^2 -a(1 + \frac{M}{r}) \Rea\big( is \cos \theta \cdot \psi [\ov{\rd_{v_+} \psi + \rd_r \psi}]}\big)\Big]\;.
        \end{split}
    \end{equation*}
    Disregarding the $|\rd_\theta \psi|^2$ term, which is manifestly positive definite, and the underbraced error terms, the remaining terms can be written as a quadratic form in $(\rd_r \psi, \Phi \psi, \rd_{v_+} \psi)$ with matrix $\chi(v_+) e^{\lambda r} \mathbb{Q}_2$, where
    \begin{equation*}
        \mathbb{Q}_2 = \begin{pmatrix}
            \frac{1}{2r} \big[r(r^2 -rM + 2M^2) + a^2(M+r)\big] & \frac{1}{2} a \sin \theta (1 + \frac{M}{r}) & \frac{1}{2} ( \Delta + a^2 \sin^2 \theta \frac{M}{r}) \\
           \frac{1}{2} a \sin \theta (1 + \frac{M}{r})  & \frac{1}{2}(1 + \frac{M}{r}) & \frac{1}{2} a \sin \theta (1 + \frac{M}{r}) \\
           \frac{1}{2} ( \Delta + a^2 \sin^2 \theta \frac{M}{r}) &  \frac{1}{2} a \sin \theta (1 + \frac{M}{r}) & \frac{1}{2} a^2 \sin^2 \theta (\frac{M}{r} -1) + r^2 + a^2
        \end{pmatrix} \;.
    \end{equation*}
For the first principal submatrix we observe that $r^2 - rM + 2M^2$ takes on its minimum at $r=\frac{M}{2}$, for which the expression evaluates to $\frac{7}{4}M >0$. This shows that the first principal submatrix is strictly positive.  The determinant of the second principal submatrix is computed to be
$$\frac{(M+r)\big[a^2(M+r) \cos^2 \theta + r( r^2- rM + 2M^2  )\big]}{4r^2}$$
which is seen to be uniformly positive in the same way. Finally we compute
$$ \det \mathbb{Q}_2 = \frac{M+r}{8r^2}\big(a^2\cos^2 \theta + r(2M + r)\big) \big( r^3 + a^2(2M + r) \cos^2 \theta\big)\;,$$
    which is again uniformly positive. A weighted Cauchy--Schwarz inequality on the last error term concludes the proof.
\end{proof}

\begin{lemma} \label{LemApendixGamma}
There exists $r_G \in (r_-, r_+)$ and constants $0 < c,C$ such that
\begin{equation*}
    B + \frac{\sigma_q}{\kappa_- v_+}\cdot \frac{|\Delta|}{2(r^2 + a^2)} A \geq  c \chi(v_+) e^{\lambda r} \Big( - \Delta |\rd_r \psi|^2 + |\rd_{v_+} \psi|^2 + |\rd_\theta \psi|^2 + |\Phi \psi|^2 \Big) - C \chi(v_+) e^{\lambda r} |\psi|^2 
\end{equation*}
on $\Gamma \cap \{r_- \leq r \leq r_G\}$.
\end{lemma}

\begin{proof}
    The expressions for $B$ and $A$ have been computed above. Modulo the underbraced error terms and the $|\partial_\theta \psi|^2$ terms, we again consider $B + \frac{\sigma_q}{\kappa_- v_+}\cdot \frac{|\Delta|}{2(r^2 + a^2)} A$ as a quadratic form in $(\sqrt{-\Delta} \rd_r \psi, \Phi \psi, \rd_{v_+} \psi)$ with matrix $\chi(v_+) e^{\lambda r} \mathbb{Q}_3$. Recall that on $\Gamma$ we have $2 r^* = \frac{\sigma_q}{\kappa_-} \log (v_+)$ so that, on $\Gamma$, $v_+(r) \to \infty$ for $r \to r_-$. Hence, it suffices to show that the matrix $\mathbb{Q}_3$ is positive definite at $r=r_-$ (uniformly in $\theta \in [0,\pi]$) since then, by continuous dependence, there will be $r_G > r_-$ close enough to $r_-$ such that $\mathbb{Q}_3$ is uniformly positive definite on $\Gamma \cap \{r_- \leq r \leq r_G\}$.
    Since $\Delta(r_-) = 0$ and $\frac{1}{v_+(r)} \to 0$ for $r \to r_-$, we observe that all of the terms in $\mathbb{Q}_3|_{r = r_-}$ coming from $\frac{\sigma_q}{\kappa_- v_+}\cdot \frac{|\Delta|}{2(r^2 + a^2)} A$ vanish, i.e., $\mathbb{Q}_3 (r=r_-) = -\mathbb{Q}_1|_{r = r_-}$. Hence, the claim follows from the proof of Lemma \ref{LemmaAppendixDUWave}.
\end{proof}

\section{Glossary}

\subsection{Frame fields used in the paper}\label{sec:glossary.frame}

For the convenience of the reader we provide a list of the different frame fields used in the paper and where they are defined.

\begin{enumerate}
    \item In the \textbf{red-shift region} $\os{\calM}{1}$ using the $(s, \ub, \vartheta_*)$ coordinates we have the background frame field (see \eqref{eq:background.RS.ee.3.4.def}, \eqref{eq:fat.ee.1}, \eqref{eq:fat.ee.2})
\begin{equation*}
        \os{\pmb{\ee}}{1}_3 = \f{\rd}{\rd s}\Big|_s,\quad \os{\pmb{\ee}}{1}_4 = \f{\rd}{\rd \ub}\Big|_s + \pmb{f} \f{\rd}{\rd s}\Big|_s + \pmb{h}^A \f{\rd}{\rd\vartheta_*^A}\Big|_s, \quad  \os{\pmb{\ee}}{1}_1 = \f{R}{\ell} \Big( \f{\rd}{\rd{\th_*}}\Big|_s - (\f{\rd \mathfrak h}{\rd\th_*}) \f{\rd}{\rd {\varphi_*}}\Big|_s  \Big), \quad   \os{\pmb{\ee}}{1}_2=  \f{1}{R\Si} \f{\rd}{\rd {\varphi_*}}\Big|_s.
\end{equation*}
    and the dynamical frame field (see \eqref{eq:dynamical.ee.3.4.def}, \eqref{eq:fat.ee.1}, \eqref{eq:fat.ee.2})
    \begin{equation*}
         \overset{\scriptscriptstyle{\text{[1]}}}{\ee}_3 = \f{\rd}{\rd s}\Big|_s, \quad \overset{\scriptscriptstyle{\text{[1]}}}{\ee}_4 = \f{\rd}{\rd \ub}\Big|_s + f\f{\rd}{\rd s}\Big|_s + h^A\f{\rd}{\rd\vartheta_*^A}\Big|_s, \quad  \os{\ee}{1}_1= \os{\pmb{\ee}}{1}_1, \quad 
            \os{\ee}{1}_2 =   \os{\pmb{\ee}}{1}_2.
    \end{equation*}
The frame fields are regular at the event horizon $\Hp$. The background frame field is a properly normalized null frame with respect to the background metric. For the dynamical null frame we note that while $\os{\ee}{1}_A$ are orthogonal to $\os{\ee}{1}_3$ and $\os{\ee}{1}_4$, the $\os{\ee}{1}_A$ only form an approximate orthonormal basis with respect to the dynamical metric, since they are defined with respect to the background coordinate quantities.

    \item In the \textbf{blue-shift region} $\os{\calM}{2}$ using the $(u, \ub, \vartheta_*)$ coordinates we have the double null background frame fields (see \eqref{eq:background.ee.BS}, \eqref{def:Kerr.double.null})
\begin{equation*}
    \os{\pmb{\ee}}{2}_3  =  \frac{\rd}{\rd u}\Big|_{DN}, \quad \os{\pmb{\ee}}{2}_4 = \pmb{\Omg}^{-2}\Big(\frac{\rd}{\rd {\ub}}\Big|_{DN} + \f{4Mar}{\Sigma R^2} \frac{\rd}{\rd {\varphi_*}}\Big|_{DN}\Big), \quad 
\end{equation*}
    and the dynamical double null frame fields (see \eqref{eq:dynamical.ee.2.3.4.def})
    \begin{equation*}
        \overset{\scriptscriptstyle{\text{[2]}}}{\ee}_3 = \f{\rd}{\rd u}\Big|_{DN}, \quad \overset{\scriptscriptstyle{\text{[2]}}}{\ee}_4 = \Omg^{-2} \Big(\f{\rd}{\rd \ub}\Big|_{DN} + b^A\f{\rd}{\rd\vartheta_*^A}\Big|_{DN}\Big)\;.
    \end{equation*}
These frame fields are regular at the Cauchy horizon $\CH^+$.

\item In the \textbf{smaller blue-shift region} $\os{\calM}{2}\setminus (\calU \cap \{s \leq \f{s_f}{2}\})$ we have the rescaled, non-regular  background frame fields 
(see \eqref{def:Kerr.double.null})
\begin{equation*}
    \pmb{e}'_3 = \frac{1}{\pmb{\Omega}^2} \os{\pmb{\ee}}{2}_3, \quad \pmb{e}_4' = \pmb{\Omega}^2 \os{\pmb{\ee}}{2}_4\;.
\end{equation*}
and the rescaled, non-regular dynamical frame fields
(see \eqref{eq:e3'.e4'.def.restricted})
\begin{equation*}
    e_3' = \frac{1}{\Omg^{2}} \os{\ee}{2}_3,\quad e_4' = \Omg^2 \os{\ee}{2}_4\;.
\end{equation*}

\item On the \textbf{global spacetime} $\calM$ using the $(u', \ub, \vartheta_*)$ coordinates of Definition \ref{eq:u'.def} we have the background double null frame field (see \eqref{eq:pmbe3e4.def}, \eqref{EqDefBolde_A})
\begin{equation*}
     \pmb{e}_3' = \pmb{\Omg}^{-2} \f{\rd}{\rd u'}\Big|_{DN'}, \quad 
        \pmb{e}_4' = \f{\rd}{\rd \ub}\Big|_{DN'} + \pmb{b}^A \f{\rd}{\rd \vartheta_*^A}\Big|_{DN'}, \quad 
        \pmb{e}_1' = \f{R }{\ell} \Big( \rd_{\th_*}\Big|_{DN'} - (\f{\rd \mathfrak h}{\rd\th_*}) \rd_{\varphi_*} \Big|_{DN'} \Big),\quad \pmb{e}_2' = \f{1}{R\Si} \rd_{\varphi_*} \Big|_{DN'}, 
\end{equation*}
and the dynamical frame field (see Definition \ref{def:e3'.e4'} and \ref{DefDefe_A'}, in particular \eqref{eq:e'})
\begin{equation*}
    e_3' = -2(g^{-1})^{\alp\bt} \rd_\alp \ub \rd_\bt, \quad e_4', \quad e_1', \quad e_2' \;.
\end{equation*}
In the smaller blue-shift region we in particular have   $ e_3' = \Omg^{-2} \os{\ee}{2}_3$ and  $e_4' = \Omg^2 \os{\ee}{2}_4$. These frame fields are regular at $\Hp$ but non-regular at $\CH^+$.

We also have the background principal null frame (see \eqref{eq:Teukolsky.in.BL}, \eqref{EqDefBackgroundPNFrame})
\begin{equation*}
\begin{aligned}
    &\pmb{e}_4 =\Delta \frac{\rd}{\rd r}\Big|_{BL} + (r^2 + a^2)\frac{\rd}{\rd t}\Big|_{BL} + a \frac{\rd}{\rd {\varphi}}\Big|_{BL}, \quad  &&\pmb{e}_3 =- \frac{1}{ \Sigma} \frac{\rd}{\rd r}\Big|_{BL} + \frac{r^2 + a^2}{ \Delta \Sigma} \frac{\rd}{\rd t}\Big|_{BL} + \frac{a}{ \Delta \Sigma} \frac{\rd}{\rd {\varphi}}\Big|_{BL},\\
    &\pmb{e}_1 = \frac{1}{\sqrt{\Sigma}} \frac{\partial}{\rd \theta}\Big|_{BL}, \quad &&\pmb{e}_2 = \frac{1}{\sqrt{\Sigma} \Si}(\frac{\partial}{\rd \varphi}\Big|_{BL} + a \Si^2 \frac{\partial}{\rd t}\Big|_{BL})\;,
    \end{aligned}
\end{equation*}
which is related to the background double null frame by $\pmb{e}_\mu = \pmb{\calB}_\mu^\nu \pmb{e}'_\nu$. The dynamical principal null frame (see \eqref{EqDefDynamicalPNFrame})
\begin{equation*}
    e_4, \quad e_3, \quad e_1, \quad e_2
\end{equation*}
is defined via $e_\mu = \pmb{\calB}_\mu^\nu e'_\nu$. Again, those frame fields are regular at $\Hp$ and non-regular at $\CH^+$.

\end{enumerate}

\bibliographystyle{acm}
\bibliography{Bibly}

@article{LukSbi15,
	author = {J. Luk and J. Sbierski},
	journal = {J. Funct. Anal.},
	number = {7},
	pages = {1948-1995},
	title = {{Instability results for the wave equation in the interior of Kerr black holes}},
	volume = {271},
	year = {2016}}

@article{NP62,
  title={An approach to gravitational radiation by a method of spin coefficients},
  author={Newman, Ezra and Penrose, Roger},
  journal={Journal of Mathematical Physics},
  volume={3},
  number={3},
  pages={566--578},
  year={1962},
  publisher={American Institute of Physics}
}

@article{Teu73,
  title={Perturbations of a rotating black hole. I. Fundamental equations for gravitational, electromagnetic, and neutrino-field perturbations},
  author={Teukolsky, Saul A},
  journal={Astrophysical Journal, Vol. 185, pp. 635-648 (1973)},
  volume={185},
  pages={635--648},
  year={1973}
}

@article{GHP73,
  title={A space-time calculus based on pairs of null directions},
  author={Geroch, Robert and Held, Alan and Penrose, Roger},
  journal={Journal of Mathematical Physics},
  volume={14},
  number={7},
  pages={874--881},
  year={1973},
  publisher={American Institute of Physics}
}

@article{IoKlai09,
  title={On the uniqueness of smooth, stationary black holes in vacuum},
  author={Ionescu, Alexandru D and Klainerman, Sergiu},
  journal={Inventiones mathematicae},
  volume={175},
  number={1},
  pages={35--102},
  year={2009},
  publisher={Springer}
}

@article{Sbie23,
  title={Instability of the {K}err {C}auchy horizon under linearised gravitational perturbations},
  author={Sbierski, Jan},
  journal={Annals of PDE},
  volume={9},
  number={1},
  pages={7},
  year={2023},
  publisher={Springer}
}

@article{Sbie26,
  title={{Supplement to: Instability of the Kerr Cauchy horizon under linearised gravitational perturbations}},
  author={Sbierski, Jan},
  year={preprint},
}

@article{GalLinSbi17,
	author = {G. Galloway and E. Ling and J. Sbierski},
	journal = {Comm. Math. Phys.},
	number = {3},
	pages = {937-949},
	title = {{Timelike completeness as an obstruction to $C^0$-extensions}},
	volume = {359},
	year = {2018}}

@article{Le25a,
	author = {Pengyu Le},
	journal = {arXiv:2508.03263v2},
	title = {{Volume-distance-ratio asymptote and spacetime inextendibility for spatially flat and hyperbolic FLRW spacetimes}},
	year = {2025}}

@article{Le25,
	author = {Pengyu Le},
	journal = {arXiv:2507.23097v2},
	title = {{Volume-Distance-Ratio Asymptote and Spacetime Inextendibility}},
	year = {2025}}

@article{GraKuSa19,
	author = {J. Grant and M. Kunzinger and C. S\"amann},
	journal = {Ann. Glob. Anal. Geom.},
	pages = {133-147},
	title = {{Inextendibility of spacetimes and Lorentzian length spaces}},
	volume = {55},
	year = {2019}}

@article{Racz10,
	author = {R{\'a}cz, Istv{\'a}n},
	date = {2010/06/17},
	doi = {10.1088/0264-9381/27/15/155007},
	isbn = {0264-9381},
	journal = {Classical and Quantum Gravity},
	number = {15},
	pages = {155007},
	title = {Spacetime extensions {II}},
	url = {https://dx.doi.org/10.1088/0264-9381/27/15/155007},
	volume = {27},
	year = {2010}}

@article{CaSbi25,
	author = {Peter Cameron and Jan Sbierski},
	journal = {arXiv:2511.13422},
	title = {{On the uniqueness of continuous spacetime extensions in $1+1$ dimensions with applications to weak null singularities}},
	year = {2025}}

@article{Sbie24a,
	author = {Sbierski, Jan},
	doi = {10.1093/imrn/rnae194},
	eprint = {https://academic.oup.com/imrn/article-pdf/2024/20/13221/59813913/rnae194.pdf},
	issn = {1073-7928},
	journal = {International Mathematics Research Notices},
	month = {09},
	number = {20},
	pages = {13221-13254},
	title = {{Uniqueness and Non-Uniqueness Results for Spacetime Extensions}},
	url = {https://doi.org/10.1093/imrn/rnae194},
	volume = {2024},
	year = {2024}}

@article{MinSuhr19,
	author = {E. Minguzzi and S. Suhr},
	journal = {Ann. Glob. Anal. Geom.},
	pages = {597-611},
	title = {{Some regularity results for Lorentz-Finsler spaces}},
	volume = {56},
	year = {2019}}

@article{Sbie24,
  AUTHOR = {Sbierski, Jan},
     TITLE = {Lipschitz inextendibility of weak null singularities from
              curvature blow-up},
   JOURNAL = {Invent. Math.},
  FJOURNAL = {Inventiones Mathematicae},
    VOLUME = {243},
      YEAR = {2026},
    NUMBER = {3},
     PAGES = {961--991},
      ISSN = {0020-9910,1432-1297},
   MRCLASS = {99-06},
  MRNUMBER = {5008158},
       DOI = {10.1007/s00222-025-01387-0},
       URL = {https://doi-org.stanford.idm.oclc.org/10.1007/s00222-025-01387-0},
}

@article{DafLuk17,
    AUTHOR = {Dafermos, Mihalis and Luk, Jonathan},
     TITLE = {The interior of dynamical vacuum black holes {I}: {T}he
              {$C^0$}-stability of the {K}err {C}auchy horizon},
   JOURNAL = {Ann. of Math. (2)},
  FJOURNAL = {Annals of Mathematics. Second Series},
    VOLUME = {202},
      YEAR = {2025},
    NUMBER = {2},
     PAGES = {309--630},
      ISSN = {0003-486X,1939-8980},
   MRCLASS = {83C57 (35Q76 83C05 83C75)},
  MRNUMBER = {4964220},
       DOI = {10.4007/annals.2025.202.2.1},
       URL = {https://doi.org/10.4007/annals.2025.202.2.1},
}

@article{DafLuk26,
	author = {M. Dafermos and J. Luk},
	journal = {in preparation},
	title = {{The interior of dynamical vacuum black holes II: Event horizon data and the stability of the red-shift region}},
	year = {2026}}

@article{McN,
     author    = "J. McNamara",
     title     = "{Instability of black hole inner horizons}",
     year      = "1978",
     journal   = "Proc. Roy. Soc. Lon. A",
		 volume		 = "358",
		 pages		 = "499-517",		 
}

@article {McN.2,
    AUTHOR = {McNamara, J. M.},
     TITLE = {Behavior of scalar perturbations of a {R}eissner-{N}ordstr\"om
              black hole inside the event horizon},
   JOURNAL = {Proc. Roy. Soc. London Ser. A},
  FJOURNAL = {Proceedings of the Royal Society. London. Series A.
              Mathematical, Physical and Engineering Sciences},
    VOLUME = {364},
      YEAR = {1978},
    NUMBER = {1716},
     PAGES = {121--134},
      ISSN = {0962-8444,2053-9169},
   MRCLASS = {83D05},
  MRNUMBER = {521510},
MRREVIEWER = {Alan\ Barnes},
       DOI = {10.1098/rspa.1978.0191},
       URL = {https://doi-org.stanford.idm.oclc.org/10.1098/rspa.1978.0191},
}

@book{ChandBlackHole,
	author = {S. Chandrasekhar},
	publisher = {Clarendon Press},
	title = {{The Mathematical Theory of Black Holes}},
	year = {1998}}

@article{CH,
  title = {On Crossing the {C}auchy Horizon of a {R}eissner-{N}\"ordstrom Black-Hole},
  author = {Chandrasekhar, S. and Hartle, J. B.},
  journal = {Proc. Royal Society of London A},
  volume = {384},
  number = {1787},
  pages = {301--315},
  year = {1982},
  }

@article{SP,
	author = {Simpson, M. and Penrose, R.},
	journal = {Internat. J. Theoret. Phys.},
	pages = {183--197},
	title = {Internal instability in a {R}eissner-{N}ordstr\"om black hole},
	volume = {7},
	year = {1973}}

@article{GSNS,
	author = {Gursel, Y. and Sandberg, V. and Novikov, I. and Starobinsky, A.},
	journal = {Phys. Rev. D},
	pages = {413--420},
	title = {Evolution of scalar perturbations near the {C}auchy horizon of a charged black hole},
	volume = {19},
	year = {1979}}

@article{PI2,
  adsurl   = {http://adsabs.harvard.edu/abs/1990PhRvD..41.1796P},
  author   = {Poisson, E. and Israel, W.},
  doi      = {10.1103/PhysRevD.41.1796},
  journal  = {Phys. Rev. D},
  pages    = {1796--1809},
  refdb_id = {http://relativity.livingreviews.org/refdb/record/1797},
  title    = {Internal structure of black holes},
  volume   = {41},
  year     = {1990}
}

@article{PI1,
  author   = {Poisson, E. and Israel, W.},
  journal  = {Phys. Rev. Lett.},
  pages    = {1663--1666},
  title    = {Inner-horizon instability and mass inflation in black holes},
  volume   = {63},
  year     = {1989}
}

@article{Hiscock,
  author   = {Hiscock, W. A.},
  journal  = {Phys. Rev. Lett.},
  pages    = {110--112},
  title    = {Evolution of the interior of a charged black hole},
  volume   = {83A},
  year     = {1981}
}

@article {LOSR,
    AUTHOR = {Luk, Jonathan and Oh, Sung-Jin and Shlapentokh-Rothman, Yakov},
     TITLE = {A scattering theory approach to {C}auchy horizon instability
              and applications to mass inflation},
   JOURNAL = {Ann. Henri Poincar\'e},
  FJOURNAL = {Annales Henri Poincar\'e. A Journal of Theoretical and
              Mathematical Physics},
    VOLUME = {24},
      YEAR = {2023},
    NUMBER = {2},
     PAGES = {363--411},
      ISSN = {1424-0637,1424-0661},
   MRCLASS = {35L05},
  MRNUMBER = {4548525},
       DOI = {10.1007/s00023-022-01216-7},
       URL = {https://doi-org.stanford.idm.oclc.org/10.1007/s00023-022-01216-7},
}

@article {Gurriaran.linear,
    AUTHOR = {Gurriaran, Sebastian},
     TITLE = {Precise asymptotics of the spin +2 {T}eukolsky field in the
              {K}err black hole interior},
   JOURNAL = {Comm. Math. Phys.},
  FJOURNAL = {Communications in Mathematical Physics},
    VOLUME = {406},
      YEAR = {2025},
    NUMBER = {7},
     PAGES = {Paper No. 152, 71},
      ISSN = {0010-3616,1432-0916},
   MRCLASS = {83C57 (35B44 58J90 83C75)},
  MRNUMBER = {4915730},
MRREVIEWER = {Roberto\ Giamb\`o},
       DOI = {10.1007/s00220-025-05332-3},
       URL = {https://doi-org.stanford.idm.oclc.org/10.1007/s00220-025-05332-3},
}

@article {cK2020,
    AUTHOR = {Kehle, Christoph},
     TITLE = {Uniform boundedness and continuity at the {C}auchy horizon for
              linear waves on {R}eissner-{N}ordstr\"om-{A}d{S} black holes},
   JOURNAL = {Comm. Math. Phys.},
  FJOURNAL = {Communications in Mathematical Physics},
    VOLUME = {376},
      YEAR = {2020},
    NUMBER = {1},
     PAGES = {145--200},
      ISSN = {0010-3616,1432-0916},
   MRCLASS = {81T20 (35Q75 35R01 83C57 83C75)},
  MRNUMBER = {4093859},
MRREVIEWER = {Mustafa\ Salti},
       DOI = {10.1007/s00220-019-03529-x},
       URL = {https://doi-org.stanford.idm.oclc.org/10.1007/s00220-019-03529-x},
}

@article {cK2021,
    AUTHOR = {Kehle, Christoph},
     TITLE = {Blowup of the local energy of linear waves at the
              {R}eissner-{N}ordstr\"om-{A}d{S} {C}auchy horizon},
   JOURNAL = {Classical Quantum Gravity},
  FJOURNAL = {Classical and Quantum Gravity},
    VOLUME = {38},
      YEAR = {2021},
    NUMBER = {21},
     PAGES = {Paper No. 214001, 22},
      ISSN = {0264-9381,1361-6382},
   MRCLASS = {83C75 (81T20 83C57)},
  MRNUMBER = {4327717},
MRREVIEWER = {Farhang\ Loran},
       DOI = {10.1088/1361-6382/ac28e3},
       URL = {https://doi-org.stanford.idm.oclc.org/10.1088/1361-6382/ac28e3},
}

@article{cK2022,
    AUTHOR = {Kehle, Christoph},
     TITLE = {Diophantine approximation as cosmic censor for {K}err-{A}d{S}
              black holes},
   JOURNAL = {Invent. Math.},
  FJOURNAL = {Inventiones Mathematicae},
    VOLUME = {227},
      YEAR = {2022},
    NUMBER = {3},
     PAGES = {1169--1321},
      ISSN = {0020-9910,1432-1297},
   MRCLASS = {83C75 (11J04 81T20 83C57)},
  MRNUMBER = {4384195},
MRREVIEWER = {Moritz\ Reintjes},
       DOI = {10.1007/s00222-021-01078-6},
       URL = {https://doi-org.stanford.idm.oclc.org/10.1007/s00222-021-01078-6},
}

@article {pHaV2017,
    AUTHOR = {Hintz, Peter and Vasy, Andr\'as},
     TITLE = {Analysis of linear waves near the {C}auchy horizon of
              cosmological black holes},
   JOURNAL = {J. Math. Phys.},
  FJOURNAL = {Journal of Mathematical Physics},
    VOLUME = {58},
      YEAR = {2017},
    NUMBER = {8},
     PAGES = {081509, 45},
      ISSN = {0022-2488,1089-7658},
   MRCLASS = {83C57 (81T20)},
  MRNUMBER = {3687649},
MRREVIEWER = {Vladimir\ Dzhunushaliev},
       DOI = {10.1063/1.4996575},
       URL = {https://doi-org.stanford.idm.oclc.org/10.1063/1.4996575},
}

@article{Song,
      author         = "Song, Yuefeng",
      title          = {Weak null singularity for the {E}instein--{E}uler system},
      year           = "2025",
	journal        = "arXiv:2506.16635, preprint",
      eprint         = "2506.16635",
      archivePrefix  = "arXiv",

}

@article{Mancheva,
      author         = "Mancheva, R.~V.",
      title          = {Contrasting behaviour of two spherically symmetric perfect fluids near a weak null singularity in a spherically symmetric black hole},
      year           = "2026",
	journal        = "arXiv:2603.18340, preprint",
      eprint         = "2603.18340",
      archivePrefix  = "arXiv",

}

@Article{LukWeakNull,
  AUTHOR = {Luk, Jonathan},
     TITLE = {Weak null singularities in general relativity},
   JOURNAL = {J. Amer. Math. Soc.},
  FJOURNAL = {Journal of the American Mathematical Society},
    VOLUME = {31},
      YEAR = {2018},
    NUMBER = {1},
     PAGES = {1--63},
      ISSN = {0894-0347,1088-6834},
   MRCLASS = {83C75 (35L65 35L67 35Q75 83C57)},
  MRNUMBER = {3718450},
MRREVIEWER = {Jos\'e\ Nat\'ario},
       DOI = {10.1090/jams/888},
       URL = {https://doi-org.stanford.idm.oclc.org/10.1090/jams/888},
}

@article {syMlZ2023,
    AUTHOR = {Ma, Siyuan and Zhang, Lin},
     TITLE = {Precise late-time asymptotics of scalar field in the interior
              of a subextreme {K}err black hole and its application in
              strong cosmic censorship conjecture},
   JOURNAL = {Trans. Amer. Math. Soc.},
  FJOURNAL = {Transactions of the American Mathematical Society},
    VOLUME = {376},
      YEAR = {2023},
    NUMBER = {11},
     PAGES = {7815--7856},
      ISSN = {0002-9947,1088-6850},
   MRCLASS = {83C75 (35Q75 58J45 83C57)},
  MRNUMBER = {4657222},
MRREVIEWER = {Wolfgang\ Hasse},
       DOI = {10.1090/tran/8957},
       URL = {https://doi-org.stanford.idm.oclc.org/10.1090/tran/8957},
}

@article {wLmVdM2025,
    AUTHOR = {Li, Warren and Van de Moortel, Maxime},
     TITLE = {Kasner bounces and fluctuating collapse inside hairy black
              holes with charged matter},
   JOURNAL = {Ann. PDE},
  FJOURNAL = {Annals of PDE. Journal Dedicated to the Analysis of Problems
              from Physical Sciences},
    VOLUME = {11},
      YEAR = {2025},
    NUMBER = {1},
     PAGES = {Paper No. 3, 122},
      ISSN = {2524-5317,2199-2576},
   MRCLASS = {83C75 (83C20 83C25 83C57)},
  MRNUMBER = {4846256},
MRREVIEWER = {Roberto\ Giamb\`o},
       DOI = {10.1007/s40818-024-00192-x},
       URL = {https://doi-org.stanford.idm.oclc.org/10.1007/s40818-024-00192-x},
}

@article {mVdM2024,
    AUTHOR = {Van de Moortel, Maxime},
     TITLE = {Violent nonlinear collapse in the interior of charged hairy
              black holes},
   JOURNAL = {Arch. Ration. Mech. Anal.},
  FJOURNAL = {Archive for Rational Mechanics and Analysis},
    VOLUME = {248},
      YEAR = {2024},
    NUMBER = {5},
     PAGES = {Paper No. 89, 62},
      ISSN = {0003-9527,1432-0673},
   MRCLASS = {83C20 (35Q75 35Q76 83C57 83D05)},
  MRNUMBER = {4798963},
MRREVIEWER = {Theophanes\ Grammenos},
       DOI = {10.1007/s00205-024-02038-z},
       URL = {https://doi-org.stanford.idm.oclc.org/10.1007/s00205-024-02038-z},
}

@article {cKmVdM2024,
    AUTHOR = {Kehle, Christoph and Van de Moortel, Maxime},
     TITLE = {Strong cosmic censorship in the presence of matter: the
              decisive effect of horizon oscillations on the black hole
              interior geometry},
   JOURNAL = {Anal. PDE},
  FJOURNAL = {Analysis \& PDE},
    VOLUME = {17},
      YEAR = {2024},
    NUMBER = {5},
     PAGES = {1501--1592},
      ISSN = {2157-5045,1948-206X},
   MRCLASS = {35Q75 (35Q76 83C05 83C57 83C75)},
  MRNUMBER = {4761250},
MRREVIEWER = {Sari\ Ghanem},
       DOI = {10.2140/apde.2024.17.1501},
       URL = {https://doi-org.stanford.idm.oclc.org/10.2140/apde.2024.17.1501},
}

@article{Maxime.survey,
	author = {Van de Moortel, Maxime},
	journal = {Comptes Rendus. Mécanique},
	pages = {415-454},
	title = {{The Strong Cosmic Censorship conjecture}},
	volume = {353},
	year = {2025}}

@article{D1,
     author    = "M. Dafermos",
     title     = "{Stability and instability of the {C}auchy horizon for the spherically symmetric {E}instein-{M}axwell-scalar field equations}",
     year      = "2003",
     journal   = "Annals of Math.",
		 volume	 	 = "158",
		 number    = "3",
		 pages		 = "875-928",

}

@article{D2,
     author    = "M. Dafermos",
     title     = "{The interior of charged black holes and the problem of uniqueness in general relativity}",
     year      = "2005",
     journal   = "Comm. Pure Appl. Math.",
		 volume	 	 = "58",
		 number    = "4",
		 pages		 = "445-504",

}

@article{D3,
     author    = "M. Dafermos",
     title     = "{Black holes without spacelike singularities}",
     year      = "2014",
     journal   = "Comm. Math. Phys.",
		 volume	 	 = "332",
		 pages		 = "729-757",

}

@mastersthesis{Gleeson,
      author         = "Gleeson, Eavan",
      title          = "{Linear Instability of the Reissner-Nordstr\"om Cauchy
                        Horizon}",
      url            = "https://inspirehep.net/record/1510420/files/arXiv:1701.06668.pdf",
      year           = "2017",
        school       = {University of Cambridge, arXiv:1701.06668},
      eprint         = "1701.06668",
      archivePrefix  = "arXiv",
      primaryClass   = "gr-qc",
      SLACcitation   = "%%CITATION = ARXIV:1701.06668;%%"
}

@article {gFjS2020,
    AUTHOR = {Fournodavlos, Grigorios and Sbierski, Jan},
     TITLE = {Generic blow-up results for the wave equation in the interior
              of a {S}chwarzschild black hole},
   JOURNAL = {Arch. Ration. Mech. Anal.},
  FJOURNAL = {Archive for Rational Mechanics and Analysis},
    VOLUME = {235},
      YEAR = {2020},
    NUMBER = {2},
     PAGES = {927--971},
      ISSN = {0003-9527,1432-0673},
   MRCLASS = {81T20 (35B44 35Q75 83C57)},
  MRNUMBER = {4064191},
MRREVIEWER = {Farhang\ Loran},
       DOI = {10.1007/s00205-019-01434-0},
       URL = {https://doi-org.stanford.idm.oclc.org/10.1007/s00205-019-01434-0},
}

@article {Gajic:2015csa,
    AUTHOR = {Gajic, Dejan},
     TITLE = {Linear Waves in the Interior of Extremal Black Holes
              {I}},
   JOURNAL = {Commun. Math. Phys.},
  FJOURNAL = {Communications in Mathematical Physics},
    VOLUME = {353},
      YEAR = {2017},
    NUMBER = {2},
     PAGES = {717--770},
      ISSN = {0010-3616},
   MRCLASS = {83C57 (83C05)},
  MRNUMBER = {3649484},
       DOI = {10.1007/s00220-016-2800-y},
       URL = {http://dx.doi.org/10.1007/s00220-016-2800-y},
}

@article{Gajic:2015hyu,
    AUTHOR = {Gajic, Dejan},
     TITLE = {Linear waves in the interior of extremal black holes {II}},
   JOURNAL = {Ann. Henri Poincar\'e},
  FJOURNAL = {Annales Henri Poincar\'e. A Journal of Theoretical and
              Mathematical Physics},
    VOLUME = {18},
      YEAR = {2017},
    NUMBER = {12},
     PAGES = {4005--4081},
      ISSN = {1424-0637,1424-0661},
   MRCLASS = {83C57 (35L10 35Q75 35R01 83C05)},
  MRNUMBER = {3723347},
MRREVIEWER = {Stefanos\ Aretakis},
       DOI = {10.1007/s00023-017-0614-x},
       URL = {https://doi-org.stanford.idm.oclc.org/10.1007/s00023-017-0614-x},
}

@article{LukOh2017one,
      author = {Luk, Jonathan and Oh, Sung-Jin},
	doi = {10.4007/annals.2019.190.1.1},
	fjournal = {Annals of Mathematics. Second Series},
	issn = {0003-486X},
	journal = {Ann. of Math. (2)},
	mrclass = {83C75 (35Q75)},
	mrnumber = {3990601},
	number = {1},
	pages = {1--111},
	title = {Strong cosmic censorship in spherical symmetry for two-ended asymptotically flat initial data {I}. {T}he interior of the black hole region},
	url = {https://doi-org.stanford.idm.oclc.org/10.4007/annals.2019.190.1.1},
	volume = {190},
	year = {2019}}

@article{LukOh2017two,
      author = {Luk, Jonathan and Oh, Sung-Jin},
	doi = {10.1007/s40818-019-0062-7},
	fjournal = {Annals of PDE. Journal Dedicated to the Analysis of Problems from Physical Sciences},
	issn = {2524-5317},
	journal = {Ann. PDE},
	mrclass = {83C75 (35Q75 35R01 83C57)},
	mrnumber = {3969149},
	mrreviewer = {Theophanes Grammenos},
	number = {1},
	pages = {Paper No. 6, 194},
	title = {Strong cosmic censorship in spherical symmetry for two-ended asymptotically flat initial data {II}: the exterior of the black hole region},
	url = {https://doi-org.stanford.idm.oclc.org/10.1007/s40818-019-0062-7},
	volume = {5},
	year = {2019}}

@article {LukOhpub,
    AUTHOR = {Luk, Jonathan and Oh, Sung-Jin},
     TITLE = {Proof of linear instability of the {R}eissner--{N}ordstr\"om
              {C}auchy horizon under scalar perturbations},
   JOURNAL = {Duke Math. J.},
  FJOURNAL = {Duke Mathematical Journal},
    VOLUME = {166},
      YEAR = {2017},
    NUMBER = {3},
     PAGES = {437--493},
      ISSN = {0012-7094},
   MRCLASS = {35Q75 (83C57)},
  MRNUMBER = {3606723},
       DOI = {10.1215/00127094-3715189},
       URL = {http://dx.doi.org/10.1215/00127094-3715189},
}

@article{LO,
	author = {Luk, Jonathan and Oh, Sung-Jin},
	journal = {arXiv:2404.02220, preprint},
	title = {Late time tail of waves on dynamic asymptotically flat spacetimes of odd space dimensions},
	year = {2024}}

@article{Gurriaran.nonlinear,
	author = {Gurriaran, S.},
	journal = {arXiv:2603.17911, preprint},
	title = {Non-linear instability of the {K}err {C}auchy horizon near $i_+$},
	year = {2026}}

@article{Gurriaran.mass,
	author = {Gurriaran, S.},
	journal = {arXiv:2503.24114, preprint},
	title = {Generic linearized curvature singularity at the perturbed {K}err {C}auchy horizon},
	year = {2025}}

@article{Gau,
	author = {Onyx Gautam},
    journal = {arXiv:2412.17927, preprint},
	eprint = {2412.17927},
	month = {12},
	title = {Late-time tails and mass inflation for the spherically symmetric {E}instein-{M}axwell-scalar field system},
	url = {https://arxiv.org/pdf/2412.17927.pdf},
	year = {2024}}

@article{FO,
  title = {How generic are null spacetime singularities?},
  author = {Ori, Amos and Flanagan, \'Eanna \'E.},
  journal = {Phys. Rev. D},
  volume = {53},
  pages = {1754--1758},
  year = {1996},
  }

@incollection {Penroseunsolv,
    AUTHOR = {Penrose, R.},
     TITLE = {Some unsolved problems in classical general relativity},
 BOOKTITLE = {Seminar on {D}ifferential {G}eometry},
    SERIES = {Ann. of Math. Stud.},
    editor= {S.-T. Yau},
    VOLUME = {102},
     PAGES = {631--668},
 PUBLISHER = {Princeton Univ. Press, Princeton, N.J.},
      YEAR = {1982},
   MRCLASS = {83-02 (83C30)},
  MRNUMBER = {645761 (83c:83001)},
MRREVIEWER = {Richard Hansen},
}

@article{DRPL,
     author    = "M. Dafermos and I. Rodnianski",
     title     = "{A proof of Price's law for the collapse of a self-gravitating scalar field}",
     year      = "2005",
     journal   = "Invent. Math.",
		 volume	 	 = "162",
		 number    = "2",
		 pages		 = "381-457",
}

@Article{Franzen1,
    AUTHOR = {Franzen, Anne T.},
     TITLE = {Boundedness of massless scalar waves on {R}eissner-{N}ordstr\"om
              interior backgrounds},
   JOURNAL = {Commun. Math. Phys.},
  FJOURNAL = {Communications in Mathematical Physics},
    VOLUME = {343},
      YEAR = {2016},
    NUMBER = {2},
     PAGES = {601--650},
      ISSN = {0010-3616},
   MRCLASS = {83C57 (35Q75 35R01 83C75)},
  MRNUMBER = {3477348},
       DOI = {10.1007/s00220-015-2440-7},
       URL = {http://dx.doi.org/10.1007/s00220-015-2440-7},
}

@article{Franzen2,
    AUTHOR = {Franzen, Anne T.},
     TITLE = {Boundedness of massless scalar waves on {K}err interior
              backgrounds},
   JOURNAL = {Ann. Henri Poincar\'{e}},
  FJOURNAL = {Annales Henri Poincar\'{e}. A Journal of Theoretical and
              Mathematical Physics},
    VOLUME = {21},
      YEAR = {2020},
    NUMBER = {4},
     PAGES = {1045--1111},
      ISSN = {1424-0637},
   MRCLASS = {81T20 (35Q75 35R01)},
  MRNUMBER = {4078277},
       DOI = {10.1007/s00023-020-00900-w},
       URL = {https://doi-org.stanford.idm.oclc.org/10.1007/s00023-020-00900-w},
}

@article {pH2017,
    AUTHOR = {Hintz, Peter},
     TITLE = {Boundedness and decay of scalar waves at the {C}auchy horizon
              of the {K}err spacetime},
   JOURNAL = {Comment. Math. Helv.},
  FJOURNAL = {Commentarii Mathematici Helvetici. A Journal of the Swiss
              Mathematical Society},
    VOLUME = {92},
      YEAR = {2017},
    NUMBER = {4},
     PAGES = {801--837},
      ISSN = {0010-2571,1420-8946},
   MRCLASS = {58J47 (35B40 35L10 35R01 83C57)},
  MRNUMBER = {3718488},
MRREVIEWER = {Davide\ Batic},
       DOI = {10.4171/CMH/425},
       URL = {https://doi-org.stanford.idm.oclc.org/10.4171/CMH/425},
}

@article {dGlmaK2022,
    AUTHOR = {Gajic, Dejan and Kehrberger, Leonhard M. A.},
     TITLE = {On the relation between asymptotic charges, the failure of
              peeling and late-time tails},
   JOURNAL = {Classical Quantum Gravity},
  FJOURNAL = {Classical and Quantum Gravity},
    VOLUME = {39},
      YEAR = {2022},
    NUMBER = {19},
     PAGES = {Paper No. 195006, 26},
      ISSN = {0264-9381,1361-6382},
   MRCLASS = {83C30 (83C35 83C40)},
  MRNUMBER = {4476718},
MRREVIEWER = {Wolfgang\ Hasse},
}

@article{DHRT,
	archiveprefix = {arXiv},
	author = {Dafermos, Mihalis and Holzegel, Gustav and Rodnianski, Igor and Taylor, Martin},
	eprint = {2104.08222},
	journal = {arXiv:2104.08222, preprint},
	title = {The non-linear stability of the {S}chwarzschild family of black holes},
	year = {2021}}

@article{GKS,
   AUTHOR = {Giorgi, Elena and Klainerman, Sergiu and Szeftel, J\'er\'emie},
     TITLE = {Wave equations estimates and the nonlinear stability of slowly
              rotating {K}err black holes},
   JOURNAL = {Pure Appl. Math. Q.},
  FJOURNAL = {Pure and Applied Mathematics Quarterly},
    VOLUME = {20},
      YEAR = {2024},
    NUMBER = {7},
     PAGES = {2865--3849},
      ISSN = {1558-8599,1558-8602},
   MRCLASS = {58J45 (83C05 83C10 83C57)},
  MRNUMBER = {4835104},
MRREVIEWER = {Pierre\ Noundjeu},
       DOI = {10.4310/pamq.241128023033},
       URL = {https://doi-org.stanford.idm.oclc.org/10.4310/pamq.241128023033},
}

@article{KS.Kerr,
    AUTHOR = {Klainerman, Sergiu and Szeftel, J\'er\'emie},
     TITLE = {Kerr stability for small angular momentum},
   JOURNAL = {Pure Appl. Math. Q.},
  FJOURNAL = {Pure and Applied Mathematics Quarterly},
    VOLUME = {19},
      YEAR = {2023},
    NUMBER = {3},
     PAGES = {791--1678},
      ISSN = {1558-8599,1558-8602},
   MRCLASS = {58J90 (58J45 83C05 83C10)},
  MRNUMBER = {4621379},
MRREVIEWER = {Daniele\ Gregoris},
       DOI = {10.4310/pamq.2023.v19.n3.a1},
       URL = {https://doi-org.stanford.idm.oclc.org/10.4310/pamq.2023.v19.n3.a1},
}

@inproceedings {Dafermos.ICM,
    AUTHOR = {Dafermos, Mihalis},
     TITLE = {The mathematical analysis of black holes in general
              relativity},
 BOOKTITLE = {Proceedings of the {I}nternational {C}ongress of
              {M}athematicians---{S}eoul 2014. {V}ol. {III}},
     PAGES = {747--772},
 PUBLISHER = {Kyung Moon Sa, Seoul},
      YEAR = {2014},
      ISBN = {978-89-6105-806-3; 978-89-6105-803-2},
   MRCLASS = {83C57 (83C75)},
  MRNUMBER = {3729050},
}

@article {Dafermos.extremal,
    AUTHOR = {Dafermos, Mihalis},
     TITLE = {The stability problem for extremal black holes},
   JOURNAL = {Gen. Relativity Gravitation},
  FJOURNAL = {General Relativity and Gravitation},
    VOLUME = {57},
      YEAR = {2025},
    NUMBER = {3},
     PAGES = {Paper No. 60, 20},
      ISSN = {0001-7701,1572-9532},
   MRCLASS = {83C57 (83C05)},
  MRNUMBER = {4880678},
       DOI = {10.1007/s10714-025-03394-1},
       URL = {https://doi-org.stanford.idm.oclc.org/10.1007/s10714-025-03394-1},
}

@article {gjGeL2017,
    AUTHOR = {Galloway, Gregory J. and Ling, Eric},
     TITLE = {Some remarks on the {$C^0$}-(in)extendibility of spacetimes},
   JOURNAL = {Ann. Henri Poincar\'e},
  FJOURNAL = {Annales Henri Poincar\'e. A Journal of Theoretical and
              Mathematical Physics},
    VOLUME = {18},
      YEAR = {2017},
    NUMBER = {10},
     PAGES = {3427--3447},
      ISSN = {1424-0637,1424-0661},
   MRCLASS = {83C75 (53C50 81T20)},
  MRNUMBER = {3697199},
MRREVIEWER = {Clemens\ Saemann},
       DOI = {10.1007/s00023-017-0602-1},
       URL = {https://doi-org.stanford.idm.oclc.org/10.1007/s00023-017-0602-1},
}

@article {pCpK2018,
    AUTHOR = {Chru\'sciel, Piotr T. and Klinger, Paul},
     TITLE = {The annoying null boundaries},
   JOURNAL = {J. Phys. Conf. Ser.},
  FJOURNAL = {Journal of Physics. Conference Series},
    VOLUME = {968},
      YEAR = {2018},
     PAGES = {012003, 15},
      ISSN = {1742-6588,1742-6596},
   MRCLASS = {83C75 (53C20 53C50 83C05 83C20)},
  MRNUMBER = {3919946},
MRREVIEWER = {S.\ Timothy\ Swift},
       DOI = {10.1088/1742-6596/968/1/012003},
       URL = {https://doi-org.stanford.idm.oclc.org/10.1088/1742-6596/968/1/012003},
}

@incollection {mGmvdBS2025,
    AUTHOR = {Graf, Melanie and van den Beld-Serrano, Marco},
     TITLE = {{$C^0$}-inextendibility of {FLRW} spacetimes within a subclass
              of axisymmetric spacetimes},
 BOOKTITLE = {Progress in {L}orentzian geometry},
    SERIES = {Springer Proc. Math. Stat.},
    VOLUME = {512},
     PAGES = {163--187},
 PUBLISHER = {Springer, Cham},
      YEAR = {[2025] \copyright 2025},
      ISBN = {978-3-031-99211-7; 978-3-031-99212-4},
   MRCLASS = {53C50 (83C05)},
  MRNUMBER = {4995415},
       DOI = {10.1007/978-3-031-99212-4\_9},
       URL = {https://doi-org.stanford.idm.oclc.org/10.1007/978-3-031-99212-4_9},
}

@article {Sbi.Schwarzschild,
    AUTHOR = {Sbierski, Jan},
     TITLE = {The {$C^0$}-inextendibility of the {S}chwarzschild spacetime
              and the spacelike diameter in {L}orentzian geometry},
   JOURNAL = {J. Differential Geom.},
  FJOURNAL = {Journal of Differential Geometry},
    VOLUME = {108},
      YEAR = {2018},
    NUMBER = {2},
     PAGES = {319--378},
      ISSN = {0022-040X,1945-743X},
   MRCLASS = {53C50 (83C75)},
  MRNUMBER = {3763070},
MRREVIEWER = {Clemens\ Saemann},
       DOI = {10.4310/jdg/1518490820},
       URL = {https://doi-org.stanford.idm.oclc.org/10.4310/jdg/1518490820},
}

@article{bM2024,
	archiveprefix = {arXiv},
	author = {Miethke, B.},
	eprint = {2408.05257},
	journal = {arXiv:2408.05257, preprint},
	title = {{$C^0$}-inextendibility of the {K}asner spacetime},
	year = {2024}}

@article{eL2024,
	archiveprefix = {arXiv},
	author = {Ling, E.},
	eprint = {2404.08257},
	journal = {arXiv:2404.08257, preprint},
	title = {The {$C^0$}-inextendibility of some spatially flat {FLRW} spacetimes},
	year = {2024}}

@article{jS2023,
	archiveprefix = {arXiv},
	author = {Sbierski, J.},
	eprint = {2312.07443},
	journal = {arXiv:2312.07443, preprint},
	title = {The {$C^0$}-inextendibility of a class of {FLRW} spacetimes.},
	year = {2023}}

@article {fPwI1998,
    AUTHOR = {Pretorius, Frans and Israel, Werner},
     TITLE = {Quasi-spherical light cones of the {K}err geometry},
   JOURNAL = {Classical Quantum Gravity},
  FJOURNAL = {Classical and Quantum Gravity},
    VOLUME = {15},
      YEAR = {1998},
    NUMBER = {8},
     PAGES = {2289--2301},
      ISSN = {0264-9381,1361-6382},
   MRCLASS = {83C15},
  MRNUMBER = {1645565},
MRREVIEWER = {Juan\ Antonio\ S\'aez},
       DOI = {10.1088/0264-9381/15/8/012},
       URL = {https://doi-org.stanford.idm.oclc.org/10.1088/0264-9381/15/8/012},
}

@article {dwS2023,
    AUTHOR = {Shen, Dawei},
     TITLE = {Construction of {GCM} hypersurfaces in perturbations of
              {K}err},
   JOURNAL = {Ann. PDE},
  FJOURNAL = {Annals of PDE. Journal Dedicated to the Analysis of Problems
              from Physical Sciences},
    VOLUME = {9},
      YEAR = {2023},
    NUMBER = {1},
     PAGES = {Paper No. 11, 112},
      ISSN = {2524-5317,2199-2576},
   MRCLASS = {53C50 (53C40 53C80 83C57)},
  MRNUMBER = {4598042},
MRREVIEWER = {Rossella\ Bartolo},
       DOI = {10.1007/s40818-023-00152-x},
       URL = {https://doi-org.stanford.idm.oclc.org/10.1007/s40818-023-00152-x},
}

@article{VDM,
	author = {Van de Moortel, Maxime},
	doi = {10.1007/s00220-017-3079-3},
	fjournal = {Communications in Mathematical Physics},
	issn = {0010-3616},
	journal = {Comm. Math. Phys.},
	mrclass = {83C22 (83C57 83C75)},
	mrnumber = {3795189},
	number = {1},
	pages = {103--168},
	title = {Stability and instability of the sub-extremal {R}eissner-{N}ordstr\"{o}m black hole interior for the {E}instein-{M}axwell-{K}lein-{G}ordon equations in spherical symmetry},
	url = {https://doi-org.stanford.idm.oclc.org/10.1007/s00220-017-3079-3},
	volume = {360},
	year = {2018}}

@article{VDM3,
    AUTHOR = {Van de Moortel, Maxime},
     TITLE = {The breakdown of weak null singularities inside black holes},
   JOURNAL = {Duke Math. J.},
  FJOURNAL = {Duke Mathematical Journal},
    VOLUME = {172},
      YEAR = {2023},
    NUMBER = {15},
     PAGES = {2957--3012},
      ISSN = {0012-7094,1547-7398},
   MRCLASS = {35Q75 (35Q76 83C57 83C75)},
  MRNUMBER = {4675045},
MRREVIEWER = {Theophanes\ Grammenos},
       DOI = {10.1215/00127094-2022-0096},
       URL = {https://doi-org.stanford.idm.oclc.org/10.1215/00127094-2022-0096},
}

@article{VDM4,
    AUTHOR = {Van de Moortel, Maxime},
     TITLE = {Mass inflation and the {$C^2$}-inextendibility of spherically
              symmetric charged scalar field dynamical black holes},
   JOURNAL = {Comm. Math. Phys.},
  FJOURNAL = {Communications in Mathematical Physics},
    VOLUME = {382},
      YEAR = {2021},
    NUMBER = {2},
     PAGES = {1263--1341},
      ISSN = {0010-3616,1432-0916},
   MRCLASS = {83C05 (83C57 83C75)},
  MRNUMBER = {4227173},
MRREVIEWER = {Mihai\ Tohaneanu},
       DOI = {10.1007/s00220-020-03923-w},
       URL = {https://doi-org.stanford.idm.oclc.org/10.1007/s00220-020-03923-w},
}

@article{VDM.coexistence,
	archiveprefix = {arXiv},
	author = {Van de Moortel, Maxime},
	eprint = {2504.12370},
	journal = {arXiv:2504.12370, preprint},
	title = {The coexistence of null and spacelike singularities inside spherically symmetric black holes},
	year = {2025}}

@article{VDM.coexistence.global,
	archiveprefix = {arXiv},
	author = {Van de Moortel, Maxime},
	eprint = {2510.07431},
	journal = {arXiv:2510.07431, preprint},
	title = {Asymptotically flat black holes with a singular {C}auchy horizon and a spacelike singularity},
	year = {2025}}

@article{Sbierski.C1,
	author = {Sbierski, Jan},
	doi = {10.1215/00127094-2022-0040},
	fjournal = {Duke Mathematical Journal},
	issn = {0012-7094},
	journal = {Duke Math. J.},
	mrclass = {83C75 (83C22)},
	mrnumber = {4491709},
	number = {14},
	pages = {2881--2942},
	title = {On holonomy singularities in general relativity and the {$C_{\rm loc}^{0,1}$}-inextendibility of space-times},
	url = {https://doi-org.stanford.idm.oclc.org/10.1215/00127094-2022-0040},
	volume = {171},
	year = {2022}}

@article{HintzPriceLaw,
	author = {Hintz, Peter},
	doi = {10.1007/s00220-021-04276-8},
	fjournal = {Communications in Mathematical Physics},
	issn = {0010-3616,1432-0916},
	journal = {Comm. Math. Phys.},
	mrclass = {58J37 (35Q83 81T20 83C57)},
	mrnumber = {4365146},
	mrreviewer = {Siyuan\ Ma},
	number = {1},
	pages = {491--542},
	title = {A sharp version of {P}rice's law for wave decay on asymptotically flat spacetimes},
	url = {https://doi.org/10.1007/s00220-021-04276-8},
	volume = {389},
	year = {2022}}

@article {mDySR2018,
    AUTHOR = {Dafermos, Mihalis and Shlapentokh-Rothman, Yakov},
     TITLE = {Rough initial data and the strength of the blue-shift
              instability on cosmological black holes with {$\Lambda>0$}},
   JOURNAL = {Classical Quantum Gravity},
  FJOURNAL = {Classical and Quantum Gravity},
    VOLUME = {35},
      YEAR = {2018},
    NUMBER = {19},
     PAGES = {195010, 28},
      ISSN = {0264-9381,1361-6382},
   MRCLASS = {83C20},
  MRNUMBER = {3876012},
MRREVIEWER = {H\'ector\ H.\ Hern\'andez},
       DOI = {10.1088/1361-6382/aadbcf},
       URL = {https://doi-org.stanford.idm.oclc.org/10.1088/1361-6382/aadbcf},
}

@article{DSS1,
	author = {Donninger, Roland and Schlag, Wilhelm and Soffer, Avy},
	doi = {10.1016/j.aim.2010.06.026},
	fjournal = {Advances in Mathematics},
	issn = {0001-8708},
	journal = {Adv. Math.},
	mrclass = {58J45 (35P25 35Q75 58J90 83C57)},
	mrnumber = {2735767},
	mrreviewer = {Alan D. Rendall},
	number = {1},
	pages = {484--540},
	title = {A proof of {P}rice's law on {S}chwarzschild black hole manifolds for all angular momenta},
	url = {https://doi-org.stanford.idm.oclc.org/10.1016/j.aim.2010.06.026},
	volume = {226},
	year = {2011}}

@article{dGlK2025,
	archiveprefix = {arXiv},
	author = {Gajic, Dejan and Kehrberger, Lionor},
	eprint = {2511.23242},
	journal = {arXiv:2511.23242, preprint},
	title = {Linear and nonlinear late-time tails on dynamical black hole spacetimes via time integrals},
	year = {2025}}

@article{Ta,
	author = {D. Tataru},
	eprint = {0910.5290},
	journal = {Amer. J. Math.},
	number = {2},
	pages = {361-401},
	title = {Local decay of waves on asymptotically flat stationary space-times},
	url = {http://www.citebase.org/abstract?id=oai:arXiv.org:0910.5290},
	volume = {135},
	year = {2013}}

@article{AAGPrice,
	archiveprefix = {arXiv},
	author = {Angelopoulos, Yannis and Aretakis, Stefanos and Gajic, Dejan},
	eprint = {2102.11888},
	journal = {arXiv:2102.11888, preprint},
	title = {Price's law and precise late-time asymptotics for subextremal {R}eissner-{N}ordstr{\"o}m black holes},
	year = {2021}}

@article{syMlZ2022.2,
	author = {Ma, Siyuan and Zhang, Lin},
	doi = {10.1007/s40818-022-00139-0},
	fjournal = {Annals of PDE. Journal Dedicated to the Analysis of Problems from Physical Sciences},
	issn = {2524-5317},
	journal = {Ann. PDE},
	mrclass = {58K55 (35Q83 53C27 83C50 83C57 83C60)},
	mrnumber = {4510626},
	number = {2},
	pages = {Paper No. 25, 100},
	title = {Price's law for spin fields on a {S}chwarzschild background},
	url = {https://doi-org.stanford.idm.oclc.org/10.1007/s40818-022-00139-0},
	volume = {8},
	year = {2022}}

@article{Price,
	author = {Price, Richard H.},
	doi = {10.1103/PhysRevD.5.2419},
	fjournal = {Physical Review. D. Particles and Fields. Third Series},
	issn = {0556-2821},
	journal = {Phys. Rev. D (3)},
	mrclass = {85.35},
	mrnumber = {376103},
	pages = {2419--2438},
	title = {Nonspherical perturbations of relativistic gravitational collapse. {I}. {S}calar and gravitational perturbations},
	url = {https://doi-org.stanford.idm.oclc.org/10.1103/PhysRevD.5.2419},
	volume = {5},
	year = {1972}}

@article{AAGKerr,
	archiveprefix = {arXiv},
	author = {Angelopoulos, Yannis and Aretakis, Stefanos and Gajic, Dejan},
	eprint = {2102.11884},
	journal = {arXiv:2102.11884, preprint},
	title = {Late-time tails and mode coupling of linear waves on {K}err spacetimes},
	year = {2021}}

@article{MTT,
	archiveprefix = {arXiv},
	author = {Metcalfe, J. and Tataru, D. and Tohaneanu, M.},
	eprint = {arXiv:1104.5437},
	journal = {Adv. Math.},
	number = {3},
	pages = {995-1028},
	title = {{Price's law on nonstationary spacetimes}},
	volume = {230},
	year = {2012}}

@article{DSS2,
	author = {Donninger, Roland and Schlag, Wilhelm and Soffer, Avy},
	doi = {10.1007/s00220-011-1393-8},
	fjournal = {Communications in Mathematical Physics},
	issn = {0010-3616},
	journal = {Comm. Math. Phys.},
	mrclass = {58J45 (58J90 83C57)},
	mrnumber = {2864787},
	mrreviewer = {Atanas G. Stefanov},
	number = {1},
	pages = {51--86},
	title = {On pointwise decay of linear waves on a {S}chwarzschild black hole background},
	url = {https://doi-org.stanford.idm.oclc.org/10.1007/s00220-011-1393-8},
	volume = {309},
	year = {2012}}

@article{AAG2020,
	author = {Angelopoulos, Y. and Aretakis, S. and Gajic, D.},
	doi = {10.1016/j.aim.2020.107363},
	fjournal = {Advances in Mathematics},
	issn = {0001-8708},
	journal = {Adv. Math.},
	mrclass = {81T20 (35L05 35Q75 58J45 83C30 83C57)},
	mrnumber = {4135420},
	mrreviewer = {Sergey I. Tertychniy},
	pages = {107363, 139},
	title = {Late-time asymptotics for the wave equation on extremal {R}eissner-{N}ordstr\"{o}m backgrounds},
	url = {https://doi-org.stanford.idm.oclc.org/10.1016/j.aim.2020.107363},
	volume = {375},
	year = {2020}}

@article {lmaK2022,
    AUTHOR = {Kehrberger, Leonhard M. A.},
     TITLE = {The case against smooth null infinity {II}: a logarithmically
              modified {P}rice's law},
   JOURNAL = {Adv. Theor. Math. Phys.},
  FJOURNAL = {Advances in Theoretical and Mathematical Physics},
    VOLUME = {26},
      YEAR = {2022},
    NUMBER = {10},
     PAGES = {3633--3676},
      ISSN = {1095-0761,1095-0753},
   MRCLASS = {83C30 (81T20 83C20 83C35)},
  MRNUMBER = {4753927},
       DOI = {10.4310/atmp.2022.v26.n10.a6},
       URL = {https://doi-org.stanford.idm.oclc.org/10.4310/atmp.2022.v26.n10.a6},
}

@article {mDySR2017,
    AUTHOR = {Dafermos, Mihalis and Shlapentokh-Rothman, Yakov},
     TITLE = {Time-translation invariance of scattering maps and blue-shift
              instabilities on {K}err black hole spacetimes},
   JOURNAL = {Comm. Math. Phys.},
  FJOURNAL = {Communications in Mathematical Physics},
    VOLUME = {350},
      YEAR = {2017},
    NUMBER = {3},
     PAGES = {985--1016},
      ISSN = {0010-3616,1432-0916},
   MRCLASS = {83C57},
  MRNUMBER = {3607468},
MRREVIEWER = {Vladimir\ Dzhunushaliev},
       DOI = {10.1007/s00220-016-2771-z},
}

@article {jS2015,
    AUTHOR = {Sbierski, Jan},
     TITLE = {Characterisation of the energy of {G}aussian beams on
              {L}orentzian manifolds: with applications to black hole
              spacetimes},
   JOURNAL = {Anal. PDE},
  FJOURNAL = {Analysis \& PDE},
    VOLUME = {8},
      YEAR = {2015},
    NUMBER = {6},
     PAGES = {1379--1420},
      ISSN = {2157-5045,1948-206X},
   MRCLASS = {58J45 (35R01 58Z05 83C57)},
  MRNUMBER = {3397001},
MRREVIEWER = {Stefanos\ Aretakis},
       DOI = {10.2140/apde.2015.8.1379},
       URL = {https://doi-org.stanford.idm.oclc.org/10.2140/apde.2015.8.1379},
}

@article{pM2023,
	archiveprefix = {arXiv},
	author = {Millet, Pascal},
	eprint = {2302.06946},
	journal = {arXiv:2302.06946, preprint},
	title = {Optimal decay for solutions of the {T}eukolsky equation on the {K}err metric for the full subextremal range $|a| < M$},
	year = {2023}}

@book {CK,
    AUTHOR = {Christodoulou, Demetrios and Klainerman, Sergiu},
     TITLE = {The global nonlinear stability of the {M}inkowski space},
    SERIES = {Princeton Mathematical Series},
    VOLUME = {41},
 PUBLISHER = {Princeton University Press, Princeton, NJ},
      YEAR = {1993},
     PAGES = {x+514},
      ISBN = {0-691-08777-6},
   MRCLASS = {83C05 (35Q75 58G16 83C35)},
  MRNUMBER = {1316662},
MRREVIEWER = {Alan\ D.\ Rendall},
}

@article {MR4776522,
    AUTHOR = {Davey, Alex and Dias, \'Oscar J. C. and Sola Gil, David},
     TITLE = {Strong cosmic censorship in {K}err--{N}ewman--de {S}itter},
   JOURNAL = {J. High Energy Phys.},
  FJOURNAL = {Journal of High Energy Physics},
      YEAR = {2024},
    NUMBER = {7},
     PAGES = {Paper No. 113, 53},
      ISSN = {1126-6708,1029-8479},
   MRCLASS = {83C75 (83C57)},
  MRNUMBER = {4776522},
MRREVIEWER = {Stefano\ Baiguera},
       DOI = {10.1007/jhep07(2024)113},
       URL = {https://doi-org.stanford.idm.oclc.org/10.1007/jhep07(2024)113},
}

@article {MR4487911,
    AUTHOR = {Casals, Marc and Marinho, C\'assio I. S.},
     TITLE = {Glimpses of violation of strong cosmic censorship in rotating
              black holes},
   JOURNAL = {Phys. Rev. D},
  FJOURNAL = {Physical Review D},
    VOLUME = {106},
      YEAR = {2022},
    NUMBER = {4},
     PAGES = {Paper No. 044060, 17},
      ISSN = {2470-0010,2470-0029},
   MRCLASS = {83C75 (83C20 83C25 83C57)},
  MRNUMBER = {4487911},
       DOI = {10.1103/physrevd.106.044060},
       URL = {https://doi-org.stanford.idm.oclc.org/10.1103/physrevd.106.044060},
}

@article {MR3882684,
    AUTHOR = {Dias, Oscar J. C. and Eperon, Felicity C. and Reall, Harvey S.
              and Santos, Jorge E.},
     TITLE = {Strong cosmic censorship in de {S}itter space},
   JOURNAL = {Phys. Rev. D},
  FJOURNAL = {Physical Review D},
    VOLUME = {97},
      YEAR = {2018},
    NUMBER = {10},
     PAGES = {104060, 13},
      ISSN = {2470-0010,2470-0029},
   MRCLASS = {83C75},
  MRNUMBER = {3882684},
       DOI = {10.1103/physrevd.97.104060},
       URL = {https://doi-org.stanford.idm.oclc.org/10.1103/physrevd.97.104060},
}

@article {MR3952830,
    AUTHOR = {Cardoso, Vitor and Costa, Jo\~ao L. and Destounis, Kyriakos
              and Hintz, Peter and Jansen, Aron},
     TITLE = {Strong cosmic censorship in charged black-hole spacetimes:
              still subtle},
   JOURNAL = {Phys. Rev. D},
  FJOURNAL = {Physical Review D},
    VOLUME = {98},
      YEAR = {2018},
    NUMBER = {10},
     PAGES = {104007, 7},
      ISSN = {2470-0010,2470-0029},
   MRCLASS = {83C75},
  MRNUMBER = {3952830},
       DOI = {10.1103/physrevd.98.104007},
       URL = {https://doi-org.stanford.idm.oclc.org/10.1103/physrevd.98.104007},
}

@article{PhysRevLett.120.031103,
  title = {Quasinormal Modes and Strong Cosmic Censorship},
  author = {Cardoso, Vitor and Costa, Jo\~ao L. and Destounis, Kyriakos and Hintz, Peter and Jansen, Aron},
  journal = {Phys. Rev. Lett.},
  volume = {120},
  issue = {3},
  pages = {031103},
  numpages = {6},
  year = {2018},
  month = {Jan},
  publisher = {American Physical Society},
  doi = {10.1103/PhysRevLett.120.031103},
  url = {https://link.aps.org/doi/10.1103/PhysRevLett.120.031103}
}

@article {MR3892262,
    AUTHOR = {Dias, Oscar J. C. and Reall, Harvey S. and Santos, Jorge E.},
     TITLE = {Strong cosmic censorship: taking the rough with the smooth},
   JOURNAL = {J. High Energy Phys.},
  FJOURNAL = {Journal of High Energy Physics},
      YEAR = {2018},
    NUMBER = {10},
     PAGES = {001, front matter+53},
      ISSN = {1126-6708,1029-8479},
   MRCLASS = {83C75},
  MRNUMBER = {3892262},
       DOI = {10.1007/jhep10(2018)001},
       URL = {https://doi-org.stanford.idm.oclc.org/10.1007/jhep10(2018)001},
}

@article {aO1992,
    AUTHOR = {Ori, Amos},
     TITLE = {Structure of the singularity inside a realistic rotating black
              hole},
   JOURNAL = {Phys. Rev. Lett.},
  FJOURNAL = {Physical Review Letters},
    VOLUME = {68},
      YEAR = {1992},
    NUMBER = {14},
     PAGES = {2117--2120},
      ISSN = {0031-9007,1079-7114},
   MRCLASS = {83C75 (83C57)},
  MRNUMBER = {1158251},
MRREVIEWER = {Mikhail\ Katanaev},
       DOI = {10.1103/PhysRevLett.68.2117},
       URL = {https://doi-org.stanford.idm.oclc.org/10.1103/PhysRevLett.68.2117},
}

@article {prBcmC1995,
    AUTHOR = {Brady, Patrick R. and Chambers, Chris M.},
     TITLE = {Nonlinear instability of {K}err-type {C}auchy horizons},
   JOURNAL = {Phys. Rev. D (3)},
  FJOURNAL = {Physical Review. D. Third Series},
    VOLUME = {51},
      YEAR = {1995},
    NUMBER = {8},
     PAGES = {4177--4186},
      ISSN = {0556-2821},
   MRCLASS = {83C57 (83C20)},
  MRNUMBER = {1326575},
MRREVIEWER = {Jorge\ A.\ Pullin},
       DOI = {10.1103/PhysRevD.51.4177},
       URL = {https://doi-org.stanford.idm.oclc.org/10.1103/PhysRevD.51.4177},
}

@article {aO1997,
    AUTHOR = {Ori, Amos},
     TITLE = {Perturbative approach to the inner structure of a rotating
              black hole},
   JOURNAL = {Gen. Relativity Gravitation},
  FJOURNAL = {General Relativity and Gravitation},
    VOLUME = {29},
      YEAR = {1997},
    NUMBER = {7},
     PAGES = {881--929},
      ISSN = {0001-7701,1572-9532},
   MRCLASS = {83C57},
  MRNUMBER = {1459327},
       DOI = {10.1023/A:1018887317656},
       URL = {https://doi-org.stanford.idm.oclc.org/10.1023/A:1018887317656},
}

@book {Chr,
    AUTHOR = {Christodoulou, Demetrios},
     TITLE = {The formation of black holes in general relativity},
    SERIES = {EMS Monographs in Mathematics},
 PUBLISHER = {European Mathematical Society (EMS), Z\"urich},
      YEAR = {2009},
     PAGES = {x+589},
      ISBN = {978-3-03719-068-5},
   MRCLASS = {83C05 (35L70 35Q75 83C75)},
  MRNUMBER = {2488976},
MRREVIEWER = {Piotr\ T.\ Chru\'sciel},
       DOI = {10.4171/068},
       URL = {https://doi-org.stanford.idm.oclc.org/10.4171/068},
}

@article {pHaV2018,
    AUTHOR = {Hintz, Peter and Vasy, Andr\'as},
     TITLE = {The global non-linear stability of the {K}err--de {S}itter
              family of black holes},
   JOURNAL = {Acta Math.},
  FJOURNAL = {Acta Mathematica},
    VOLUME = {220},
      YEAR = {2018},
    NUMBER = {1},
     PAGES = {1--206},
      ISSN = {0001-5962,1871-2509},
   MRCLASS = {83C57 (35B35 35Q75 35R01 58Z05)},
  MRNUMBER = {3816427},
MRREVIEWER = {Mihai\ Tohaneanu},
       DOI = {10.4310/ACTA.2018.v220.n1.a1},
       URL = {https://doi-org.stanford.idm.oclc.org/10.4310/ACTA.2018.v220.n1.a1},
}

@article {mDiR2009,
    AUTHOR = {Dafermos, Mihalis and Rodnianski, Igor},
     TITLE = {The red-shift effect and radiation decay on black hole
              spacetimes},
   JOURNAL = {Comm. Pure Appl. Math.},
  FJOURNAL = {Communications on Pure and Applied Mathematics},
    VOLUME = {62},
      YEAR = {2009},
    NUMBER = {7},
     PAGES = {859--919},
      ISSN = {0010-3640,1097-0312},
   MRCLASS = {83C57 (35B40 35L05)},
  MRNUMBER = {2527808},
MRREVIEWER = {Guilherme\ De Berredo-Peixoto},
       DOI = {10.1002/cpa.20281},
       URL = {https://doi.org/10.1002/cpa.20281},
}

@article{xtCsK2024,
	archiveprefix = {arXiv},
	author = {Chen, Xuantao and Klainerman, Sergiu},
	eprint = {2409.05700},
	journal = {arXiv:2409.05700, to appear in {D}uke {M}ath {J}.},
	title = {Regularity of the Future Event Horizon in Perturbations of {K}err},
	year = {2024}}

@article{pH2024,
	archiveprefix = {arXiv},
	author = {Hintz, Peter},
	eprint = {2411.12568},
	journal = {arXiv:2411.12568, to appear in {D}uke {M}ath {J}.},
	title = {Horizons of some asymptotically stationary spacetimes},
	year = {2024}}

@article {mDiRySR2018,
    AUTHOR = {Dafermos, Mihalis and Rodnianski, Igor and
              Shlapentokh-Rothman, Yakov},
     TITLE = {A scattering theory for the wave equation on {K}err black hole
              exteriors},
   JOURNAL = {Ann. Sci. \'Ec. Norm. Sup\'er. (4)},
  FJOURNAL = {Annales Scientifiques de l'\'Ecole Normale Sup\'erieure.
              Quatri\`eme S\'erie},
    VOLUME = {51},
      YEAR = {2018},
    NUMBER = {2},
     PAGES = {371--486},
      ISSN = {0012-9593,1873-2151},
   MRCLASS = {58J45 (35L10 35P25 35R01 83C30 83C57)},
  MRNUMBER = {3798305},
MRREVIEWER = {Jiaju\ Zhang},
       DOI = {10.24033/asens.2358},
       URL = {https://doi-org.stanford.idm.oclc.org/10.24033/asens.2358},
}

\end{document}